\documentclass[12pt,a4paper,twoside,openright]{book}
\usepackage[nottoc,numbib]{tocbibind}
\usepackage{amsmath}
\usepackage{amssymb}
\usepackage[italian,english]{babel}
\usepackage[utf8]{inputenc}
\usepackage[font=small, format=hang, labelfont={bf}]{caption}
\usepackage{graphicx}
\usepackage{siunitx}	
\usepackage{tikz}	
\usepackage{pgfplots}
\usetikzlibrary{decorations}
\usepackage{pdfpages}		
\usepackage{float}
\usepackage{cite}
\usepackage{fancyhdr} 
\graphicspath{{Immagini/}}
\usepackage{longtable}
\DeclareGraphicsExtensions{.eps}
\usepackage{psfrag}
\usepackage{appendix}
\usepackage{subfig}
\usepackage[T1]{fontenc}
\usepackage{xcolor}
\usepackage{listings}	
\usepackage[absolute,overlay]{textpos}	 
\usepackage[a4paper,top=3.3cm,bottom=3.5cm,left=3.3cm,right=3.3cm]{geometry}
\usepackage[
citecolor=black,%
filecolor=black,%
linkcolor=black,%
urlcolor=black,%
colorlinks=true]{hyperref}
\usepackage{xspace}
\usepackage{tablefootnote}
\usepackage{multirow}
\usepackage{adjustbox}
\usepackage{bigints}
\usepackage{marginnote}

\newcommand{\centchapter}[1]{\chapter*{\centering{#1}}}
\newcommand{\Likeli}{\mathcal{L}}
\newcommand{\W}{WIMP\xspace}
\newcommand{\Ws}{WIMPs\xspace}
\newcommand{\SN}{SNDM\xspace}
\newcommand{\Gevc}{GeV/c$^2$\xspace}
\newcommand{\Mevc}{MeV/c$^2$\xspace}
\newcommand{\SF}{SF$_6$\xspace}
\newcommand{\fe}{$^{55}$Fe\xspace}
\newcommand{\kap}{Kapton\texttrademark\xspace}
\newcommand{\myl}{Mylar\textregistered\xspace}

\begin{document}

\begin{center}

	\begin{Large}\textsc{Gran Sasso Science Institute}\end{Large}\vspace{0.5cm}\\
	\begin{large}\textsc{\vspace{0.5cm}\\Astroparticle Physics}
	\end{large}
	
	\medskip\medskip\medskip\medskip\medskip\medskip
	
	\begin{large}
		\begin{center}
			{PhD in Astroparticle Physics \\PhD Cycle XXXIV 2018-2022}
		\end{center}
	\end{large}
	\medskip\medskip
	\begin{Large}
		\begin{center}
			\begin{figure}[!hbtp]
				\centering
				\includegraphics[height=2.5 cm]{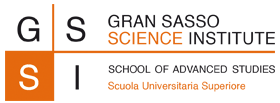}
			\end{figure}
		\end{center}
		11$^{th}$ February 2023
	\end{Large}
		\begin{center}
	\begin{Huge}
		\textbf{Optimisation of amplification and gas mixture for directional Dark Matter searches with the CYGNO/INITIUM project}
		
	\end{Huge}
		\end{center}

	\begin{huge}\textbf{}\end{huge}
	\medskip\medskip
	
	\begin{flushleft}
		\textit{Supervisor:
			Prof.ssa Elisabetta Baracchini\vspace{0.5cm}	}
	\end{flushleft}
	
	\begin{flushright}
		\textit{Candidate:
			Giorgio Dho
		}
	\end{flushright}

\end{center}

\thispagestyle{empty} 
\newpage
\thispagestyle{empty}
\pagenumbering{roman}
\setcounter{page}{0}
\centchapter{Abstract}
\addcontentsline{toc}{chapter}{Abstract}
Astrophysical and cosmological observations performed within the last century provide extensive evidences that the comprehension of the Universe is not complete and that an ingredient known as dark matter (DM) is required to interpret the measurements.
One of the most supported class of theories suggests that dark matter is composed of weakly interactive massive particles (\Ws). As the Milky Way is expected to reside in a halo of these particles, a possible experimental methodology to observe them is to exploit the potential weak interaction between \Ws and standard matter, which results in the recoil of the latter. Due to the strength of this interaction, direct detection experiments for \Ws are searching for ultra rare events, by exposing large volume of instrumented mass to detect recoils of regular matter.
The motion of the Sun and Earth with respect to the Galactic Rest Frame and the foreseen non relativistic nature of the \Ws, induces an apparent wind of DM particles coming from the Cygnus constellation, which imprints a strong directional dependence in the recoil spectrum.
Direct detection experiments capable of measuring the angular features of the recoils gain access to a wide range of advantages. These include the possibility to positively claim a discovery of DM, along with the ability to constrain DM properties.
The CYGNO project sets into this context, with the aim to deploy a large directional detector for rare event searches as DM. It comprises a gaseous time projection chamber filled with a He:CF$_4$ gas mixture, an amplification stage based on a triple stack of gas electron multiplier (GEM) optically readout employing scientific CMOS cameras and photomultiplier tubes (PMTs). This setup allows to perform the imaging of the recoil tracks, measuring on an event by event basis the energy, the direction and the topology of the tracks. 

The advantage of the optical readout through sCMOS cameras is that it allows to image large areas with an high granularity but a reduced amount of sensors. The major drawback is that the resulting small solid angle coverage of the optical setup ends up limiting the number of photons that reach the light detectors, effectively affecting the energy threshold. 
In this thesis, it is presented the work carried out with small CYGNO prototypes in order to maximise the light yield without degrading spatial and energy resolution. Different number of GEMs of various thicknesses are tested together with the addition of an extra electrode to produce strong electric fields below the last GEM. The latter addition proved to be an innovative way to enhance the light production below the GEM holes for the He:CF$_4$ gas mixture, which allowed to improve the light yield by a factor close to 2 and reducing the intrinsic diffusion of the amplification structure by tens of micrometers.\\
Minimising the electron cloud diffusion while drifting towards the amplification stage is very relevant to precisely measure the topological information of the recoil tracks.
The addition of highly electronegative gases induces the electrons freed by the passage of an ionising particle to be captured by them in few micrometres. The negative ions produced are drifted in place of the electrons and their larger mass allows them to remain thermal with the gas, strongly reducing the diffusion. This technique is usually referred to as Negative Ion Drift  (NID) operation. Using a CYGNO prototype, the addition of 1.6\% of SF$_6$ to the He:CF$_4$ gas mixture is demonstrated to produce negative ions drift operation reaching gas gains of the order of 10$^4$. A strong reduction in diffusion is measured, with diffusion coefficient as low as 45 $\mu$m/$\sqrt{\text{cm}}$, among the smallest ever measured in a gas detector. 

The potential performances of directional detectors in the context of a direct DM search are analysed with the use of rigorous statistical tools. The improvement in setting limits in the \W to nucleon and DM mass parameter space with a directional detector are evaluated employing the future expected performances of the CYGNO experiment. In addition, the improvement in the capability of discerning two different DM models exploiting a directional detector in comparison to a non directional one is analysed. It is found that, taking advantage of the angular feature, the two models considered can be distinguished with orders of magnitude less events than conventional non directional detectors.
\chapter*{Acknowledgements}
\addcontentsline{toc}{chapter}{Acknowledgements}
Part of this project is funded by the European Union's Horizon 2020 research and innovation programme under the ERC Consolidator Grant Agreement No 818744.\\
In questo percorso di quattro anni abbondanti, tantissime persone mi sono state accanto in vari momenti e mi hanno aiutato ancora a crescere e maturare.
Innanzi tutto ringrazio la Prof.ssa Baracchini, mia advisor. Mi ha insegnato tanto, sia dentro sia fuori dal laboratorio, dalla prima settimana dentro il \textit{bunker} a Frascati fino all'ultima settimana nell'ormai collaudato e attrezzato laboratorio al Gran Sasso. Inoltre sono grato anche per la sua pazienza e aiuto nella stesura della tesi!\\
Una persona molto importante da ringraziare per questi anni è sicuramente Cecilia che, nonostante la lontananza, mi ha aiutato e insieme abbiamo condiviso gioie, feste, viaggi, sogni, insomma tutto. Il suo senso pratico mi ha spesso direzionato saldamente verso le decisioni prese e mi ha su/opportato durante la scrittura della tesi.\\
Ringrazio anche tanto la mia famiglia, mamma, papà e mio fratello, per tutte le quasi regolari chiamate skype domenicali e per farmi sentire sempre a casa. Il loro supporto si è fatto sentire da sempre, quando mi hanno accompagnato a L'Aquila e in questi anni ogni volta ne avessi bisogno.\\
Ringrazio il GSSI che permette a noi studenti di fruire di un percorso di dottorato di alto livello fornendo un ambiente altamente stimolante e pieno di persone che diventano amici e che alla fine si trasformano in una nuova famiglia da cui imparare tradizioni, conoscere luoghi e soprattutto mangiare cibi da tutto il mondo.\\
In modo particolare una menzione importante va alla mia \textit{classe} del 34$^{\circ}$ Asish, $\Delta\eta\mu\eta\tau\rho\eta\sigma$ (I know the last letter is wrong!), Teena, Francesca, Tomislav, Benedikt, Vittoria, Priyanka, Samuele, Mattia (sempre adottato dai fisici!) e Sig. Pilo per le mille cene e avventure di questi anni. In particolare ringrazio Dimitris per la compagnia nel basket in tutte le salse e Francesca e Vittoria, tra le altre cose, per avermi involontariamente spinto a suonare uno strumento.\\
Un ringraziamento particolare va a Benedikt in quanto mio primo storico coinquilino. Insieme abbiamo passato la quarantena, il trasloco, gli insetti e un sacco di altre avventure. Inoltre gli sono grato per avermi riavvicinato a Age of Empires!\\
Un altro ringraziamento speciale va a Mattia Seagull. Dopo esserci conosciuti a Torino e ritrovati a L'Aquila abbiamo trascorso 4 anni di follie e anche coinquilinaggio pazzesco, dalle feste, alla bici con una mano rotta, alle sciate, al basket e anche alle briciole sul tavolo.\\
Un'altra persona importante è Giulia con cui mi sono divertito tantissimo e ho condiviso moltissimo specie nell'ultimo anno, dallo sport alla vita di tutti i giorni.\\
Ringrazio anche Carlo con cui le lunghe chiacchierate e riflessioni hanno aggiunto tantissime sfaccettature alla mia esperienza a L'Aquila oltre a fornirmi, come Giulia, uno sguardo in un campo di ricerca completamente lontano dal mio. E non può mancare anche Cecilia con tutte le serate e gli aperitivi matti al bar del corso.\\
Vorrei sottolineare il grande ruolo dei \textit{cygni} del GSSI, Samuele, Flaminia, Atul, David e André con cui abbiamo condiviso molto apertamente il lavoro, la frustrazione (siamo sperimentali) e il divertimento (siamo sperimentali) in questi anni. In particolare, Samuele e David, mastro e compagno di laboratorio fin dal principio, per le feste, compagnia e supporto continuo.\\
Una menzione particolare va a tutti i Cygners con cui ho potuto in qualche modo lavorare soprattutto Davide, Giovanni e Stefano. Inoltre sono grato a William per il lavoro svolto insieme, prima collaborazione personale intercontinentale del mio dottorato.\\
Per la facilità, immediatezza e naturalezza con cui sono stato accolto e incluso voglio ringraziare i due gruppi di basket e in particolare Fabio e Moreno grazie a cui non mi sono spiaggiato e ho finalmente riportato la mia prima frattura.\\
Inoltre ringrazio i miei amici a Torino che riesco a vedere ogni volta che torno su e con cui mi confronto sempre, Giorgio, Bob, Luke e Cocco.\\
Tutte queste persone insieme alla moltitudine di cui non ho modo di scrivere, come i nuovi amici del \textit{Del Nuovo}, sono state ingredienti per questa tempesta perfetta che è stato il mio dottorato.

\newpage
\setcounter{page}{3}
\tableofcontents

\mainmatter
\pagenumbering{arabic}
\chapter*{Introduction}
\addcontentsline{toc}{chapter}{Introduction}
\label{intro}
The last sixty years saw the flourishing of fundamental particle physics, with the development and experimental verification of the Standard Model (SM), capped with the discovery of the Higgs Boson in 2012. Recent years also witnessed the first detection of gravitational waves, phenomena predicted by the fundamental theory of gravity, the General Relativity. This detection, followed by many others, marked an extreme agreement of experimental data with the theoretical predictions. Yet, the SM is known to have various limitations and issues, such as the complete lack of the description of the gravitational force. Moreover, at the same time, numerous astrophysical and cosmological observations suggest the knowledge of the Universe is still incomplete, with the leading cosmological model, the $\Lambda$CDM, advocating for the existence of new constituents of the Universe, the dark matter (DM) and the dark energy. In particular, DM is observed to behave gravitationally as a mass, challenging the current knowledge of either gravity or particle physics. One of the most supported hypotheses speculates that DM is composed of one or more types of particles different from the ones described in the Standard Model. Indeed, theories such as the supersymmetric model or the QCD axions were developed in order to tackle some issues of the Standard Model and ended up predicting the existence of a particle which turned out to be a perfect candidate for DM. Among these, the Weakly Interactive Massive Particles (\Ws) are a class of DM matter candidates in which DM consists in a non relativistic, with \Gevc to TeV/c$^2$ masses, that weakly interacts with Standard Model particles. Indeed, a stable, weakly-interacting particle in thermal equilibrium with the early Universe would be able to reproduce the observed relic DM density.\\
The measurements of the rotation curve of the Milky Way suggest the presence of high concentrations of DM at the Galactic radius of the Sun, with an halo embedding the entire Galaxy.
Assuming DM can interact with SM matter other than gravitationally, in order to detect it, it is possible to exploit the weak interaction with regular matter and measure the recoils induced by a DM interaction. This is known as direct detection and the fundamental strategy is to expose large amount of instrumented mass and wait for DM to produce recoils in it. Direct detection experiments look for nuclear recoils of low energy, 1-100 keV, with an expected rate below 1 event per kg year. Such very low expected rates imply detectors with extremely challenging requirements on the allowed experimental backgrounds. Classical background minimisation techniques are deep underground operation (to suppress cosmic rays), use of radio-pure components (to avoid natural radioactivity) and active or passive shielding of the detector.\\
The motion of the Sun and Earth around the centre of the Galaxy induces an apparent wind of \Ws coming from a fixed direction in the frame of reference in which the Galaxy is at rest. This motion imprints a strong directional feature in the distribution of the recoils, which cannot be mimicked by any form of background, also thanks to the rotation of the Earth around its own axis. Thus, differently from the energy distribution of the recoils, which is expected to possess a generic falling exponential shape, the access to the angular distribution opens the possibility for a positive claim for a DM discovery and even constrain some characteristics of the DM models.\\
The CYGNO project intends to construct a large, $\mathcal{O}$(30) m$^3$, directional detector for rare event searches, such as DM. The detector is a gaseous time projection chamber, a detector intrinsically sensitive to the topology of the recoil tracks which allows to measure their direction. A helium and fluorine rich gas mixture is employed, in order to be sensitive to both spin-depedent and spin-independent interactions. CYGNO will be equipped with an amplification stage composed of three gas electron multipliers (GEMs) and will be optically readout with photomultiplier tubes and scientific CMOS cameras. The latter allows to image large readout areas with high sensitivity while maintaining extremely good granularity. This, in turn, grants the capability to measure the direction of the recoils. Various CYGNO prototypes have been built and tested both overground and underground, to determine their performances and their viability in a realistic environment for the DM search.\\
This work has been carried out in the context of the CYGNO experiment with the goal of optimising the light yield and diffusion properties of the detector. These are extremely relevant parameters as they directly characterise the quality direction and energy can be measured with. In addition, an assessment of the potential of such directional detectors in the directional search is dealt with in this work. The structure of the thesis is outlined in the following.

In Chapter \ref{chap1}, a brief overview of the theoretical arguments and experimental observations which lead to consider the existence of DM a paradigm of modern physics is presented. The Chapter also addresses the most relevant theoretical models which are proposed to explain its nature.

In Chapter \ref{chap2}, the phenomenology of the \W scattering with nuclei is given in the context of the direct detection of DM. The basic principles, the observables and the challenges of the direct detection technique are also discussed and the current status of the direct detection is presented. In addition, large focus is dedicated to the directional search, presenting its numerous advantages and portraying the state of the art of directional detectors for DM searches.

In Chapter \ref{chap3}, an extensive description of the CYGNO experiment is provided. The concept of the detector is thoroughly described with particular attention to the main feature of the CYGNO project, the optical readout. The most important results obtained with several prototypes are also depicted and a glimpse to the future prospects is furnished.

In Chapter \ref{chap4}, the issue of the analysis of scientific CMOS camera images is described, detailing the algorithm developed and its optimisation. Moreover, it is detailed the analysis strategy to assess the diffusion characteristics of the CYGNO detectors employed in the studies presented in the subsequent Chapters.

In Chapter \ref{chap5}, the study of the optimisation of the amplification stage is detailed. Different types and number of GEMs are tested together with a varying percentage of helium in the gas mixture. Moreover, the addition of an extra electrode below the last GEM to induce strong electric fields is examined, also with the help of electric field simulations. The performances of the different amplification structures are analysed and compared in terms of light yield, energy resolution and spatial diffusion.

In Chapter \ref{chap6}, the study of the reduction of the diffusion in the gas by means of the addition of a high electronegative gas is described. The principle of the negative ion drift operation for the SF$_6$ gas is given and the experimental results in the context of the CYGNO optical readout are presented.

In Chapter \ref{chap7}, a detailed study of the directional performances of a CYGNO-like detector is given. The advantages obtained by using a directional detector in the determination of the upper bound limits for a DM search with a CYGNO future detector is analysed in a rigorous statistical framework. In addition, the capabilities of a directional detector in discerning different DM hypotheses are worked out.

Finally, the Conclusions report the final comments and a summary of the results obtained in this work.
\chapter{Dark matter}
\label{chap1}
In the last decades, the fact that a broad number of astrophysical and cosmological observations is suggesting that something is lacking in the description of the Universe has become a well established paradigm of modern physics. The measurements indicate that an ingredient different from the known forces and particles is required to account for the experimental findings. As it behaves like a mass, it is also known as dark matter (DM). Even though experimental efforts in the years unveiled details of its phenomenology, its nature is still elusive and unknown. One of the most credited possible explanations assumes that this is indeed a form of matter which makes up the 84\% of total \cite{bib:Plank_2020} and plays a fundamental role in the formation of the organised structures of the Universe \cite{Frenk_2012}. A plethora of theoretical models have being trying to correctly describe this elusive phenomena, while experimental efforts have being relentlessly aimed at detecting clear signatures of DM, pushing the current research activities to become very active and relevant in the context of present and future research. In the following, Section \ref{sec:evidenceDM} is dedicated to a short overview on the astrophysical and cosmological measurements which lead to DM hypothesis, whilst Section \ref{sec:candidateDM} focuses on the description of the most supported theoretical explanation of this phenomenon.
\section{Evidence for dark matter}
\label{sec:evidenceDM}
Evidences for the presence of an ulterior ingredient to the Universe, commonly referred to as dark matter (DM), are present is many astronomical observations spanning all the scales of the Universe, from single galaxies to the largest structures. In the following Sections, the most critical and decisive will be briefly illustrated, starting from the galactic scale up to the cosmological one. For a more comprehensive review, see the reference \cite{bib:Bertone:2004pz}.
\subsection{Galactic rotation curves}
\label{subsec:galaxyrot}
One of the most compelling and established evidences of DM derives from the observation of the rotation curves of spiral galaxies, including the Milky Way\cite{Strigari_2013,Begeman_1991}. In these galaxies, most of the stars are located in the central part, called bulge, around which largely flat spiral arms composed of stars and gas clouds are moving in approximately circular orbits. The rotation curve is the profile of the circular velocities of the stars and gas clouds of the galaxy along the radial distance from the centre. When considering a rotating galaxy as a closed system, an object of mass $m$ at a radius $R$ from the  centre is subject to a centripetal force equal to the gravitational force of the overall mass present within the same radius $M(R)$:
\begin{equation}
\label{eq:centripetal}
m\frac{v^2}{R}= \frac{GM(R)m}{R^2},
\end{equation}
with $G$ the gravitational constant and $v$ the circular velocity of the object with mass $m$.
Hence, the rotational velocity of the constituents of a galaxy can be expressed as a function of $R$ and of the total mass $M$ as:
\begin{equation}
\label{eq:rotcurve}
v= \sqrt{\frac{GM(R)}{R}}.
\end{equation}
It is reasonable to assume that the large majority of the mass of a galaxy is located within a characteristic radius $R_c$, estimated by measurements to fall between 5 and 10 kpc \cite{Strigari_2013}. So, for $R<R_c$ the mass density can be adopted as uniform, therefore  $M(R)$ grows linearly with the volume and the velocity rises linearly with $R$. Instead, for $R>R_c$ the the mass M is constant and the velocity decreases as $\sqrt{R}$.
\begin{figure*}[t]
	\centering
	\includegraphics[width=0.75\textwidth]{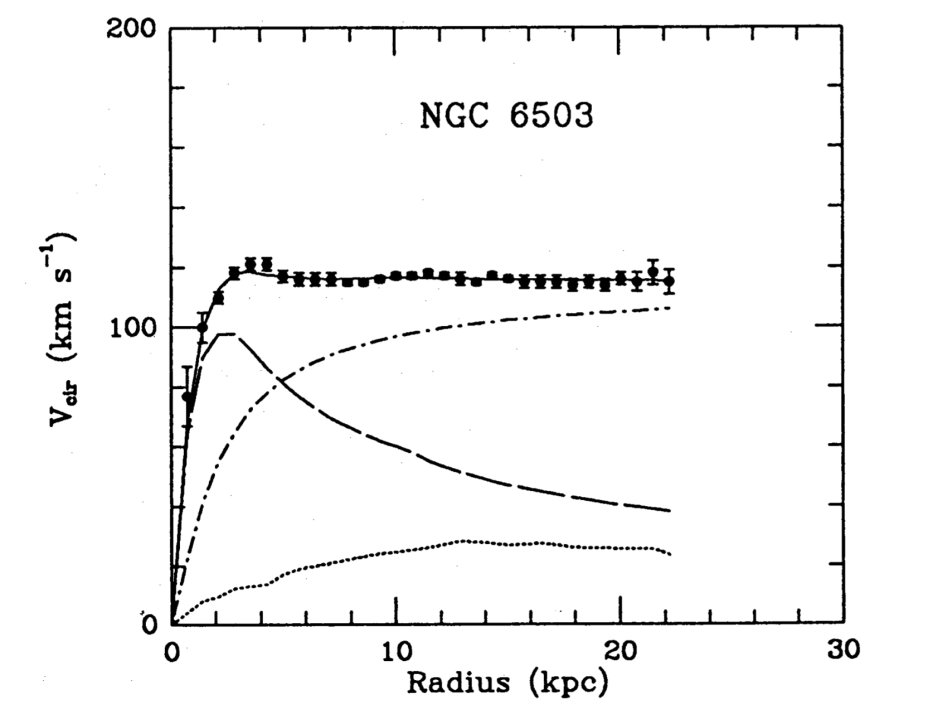}
	\caption{Example of the rotation curve measurement of the galaxy NGC6503, taken from \cite{Begeman_1991}. The solid line represents the combined fit of the data, while the dashed refers to the gas component, the dotted to the visible one, and the dash-dotted to the DM.}
	\label{fig:rotcurves}
\end{figure*}
The rotation curve can be measured via the Doppler shift of the 21 cm emission line of the of the HI. A large set of rotation curves, extending well beyond the optical limit and down 200 kpc, has been measured, exhibiting similar results. Figure \ref{fig:rotcurves} shows, as an example, the experimental findings on the rotation curve of the galaxy NGC6503 \cite{Begeman_1991}. The measured orbital velocity increases linearly as expected at small radii but stays constant at large radii, contrary to expectation. This flatness suggested the existence of an additional and invisible mass in the form of a dark galactic halo around any spiral galaxy, which does not interact electromagnetically. This halo of matter appears to be consistent at first order to a spherical distribution with a density proportional to $R^{-2}$.
\subsection{Galaxy clusters velocity dispersion}
\label{subsec:galaxyclusvir}
Galaxy clusters are systems composed of thousands of galaxies and gas clouds gravitationally bound to each other. By applying the Virial theorem, it is possible to relate their time average kinetic energy to the total gravitational potential energy of the system \cite{sanders_2010}:
\begin{equation}
\label{eq:virial0}
\sum_{i}m_iv_i^2= \sum_{i}\sum_{j\neq i} \frac{Gm_im_j}{r_{ij}} 
\end{equation}
with  $i$ and $j$ representing the index on the galaxies of the cluster, $m$ the mass of a galaxy, $v$ the time average velocity of the galaxy, $G$ the gravitational constant and $r_{ij}$ the distance between the galaxies $i$ and $j$.
By performing an average on the galaxies, a relation can be obtained that correlates the total mass of the cluster to the dispersion of the velocities of its galaxies:
\begin{equation}
\label{eq:virial}
M \simeq \frac{5}{3}R_{G}\frac{\left\langle v^2\right\rangle}{G},
\end{equation}
where it is assumed that the galaxies are uniformly distributed in a sphere of radius $R_G$.\\
The application of the Virial theorem to the Coma Cluster by Zwicky in 1933 provided the first experimental observation of a missing mass component in the Universe \cite{sanders_2010,Zwicky_1937}. By measuring the line of sight velocities of the galaxies of this globular cluster via redshift and assuming a spherical distribution, Zwicky obtained the velocity dispersion of the Coma cluster and its gravitational mass through the Virial theorem. In parallel, Zwicky estimated the cluster mass from its luminous component assuming an average luminosity to mass ratio of galaxies. The mass estimated from the gravitational component resulted $\mathcal{O}$(10$^2$) larger than measured luminous mass, advocating for the presence of matter in the galaxy cluster structure that is not interacting electromagnetically. Many other measurements on different galaxy clusters soon followed with consistent results \cite{Smith_1936}.  Basically, the velocities of the galaxies appear too large to be explained by the gravitational potential caused by the luminous matter of the cluster.
\subsection{X-ray emission from galaxy clusters}
\label{subsec:galaxyclusxray}
The large amount of gas present in between galaxies can be used to assess the galaxy cluster mass by studying their X-rays emission, whose intensity is strongly related to the gravitational potential experienced due to the surrounding mass. Assuming the gas plasma to be in hydrostatic equilibrium, it is possible to relate gas density $\rho$ and pressure $P$ at a given distance $r$ from the centre of the system to the mass $M$ of the cluster as \cite{Bertone_2018}:
\begin{equation}
\label{eq:hydroeq}
\frac{1}{\rho(r)}\frac{dP(r)}{dr}=-\frac{GM(r)}{r^2},
\end{equation}
Inserting the perfect gas law to make the dependence on the temperature of the gas explicit, the relation can be transformed in:
\begin{equation}
\label{eq:gasT0}
\frac{d \log\rho}{d \log r} + \frac{d \log\ T}{d \log r} = -\frac{r}{T}\left(\frac{\mu m_p}{k_B}\right)\frac{GM(r)}{r^2},
\end{equation}
where $\mu$ is the average molecular weight of the primordial gas of about 0.6, and $m_p$ is the mass of the proton, $T$ is the temperature of the gas and $k_B$ the Boltzmann constant. Direct measurements suggest that $\rho$ is a power law in $r$ with a spectral index between -1.5 and -2 \cite{Bertone_2018}. By adding this density model, the temperature of the gas can be parametrised as follows:
\begin{equation}
\label{eq:gasT}
k_BT \sim (1.3 \text{--}1.8)\text{keV}\left(\frac{M_r}{10^{14}M_{\odot}}\right)\left(\frac{1 Mpc}{r}\right),
\end{equation}
where a typical mass of a cluster of 10$^{14}$ solar masses and a radius of 1 Mpc were assumed in the calculation. Typical measured temperatures of galaxy clusters are of the order of 10 keV, pointing at the existence of non-luminous matter which adds gravitational potential to the visible one which is required in order to reproduce the observations under these assumptions.
\subsection{Gravitational lensing}
\label{subsec:gravlensing}
\begin{figure*}[t]
	\centering
	\includegraphics[width=0.55\textwidth]{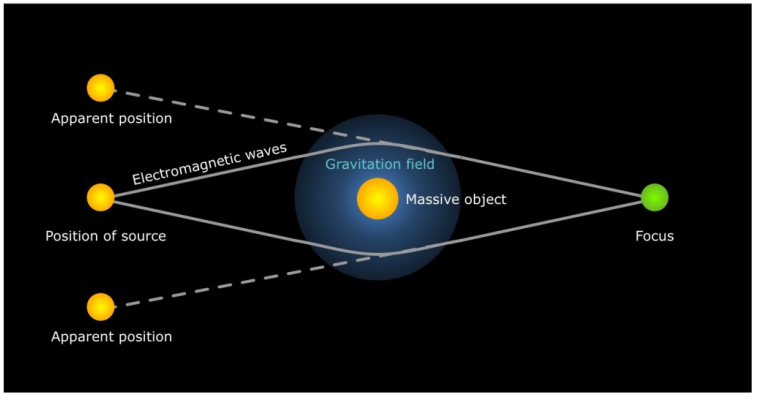}
	\includegraphics[width=0.40\textwidth]{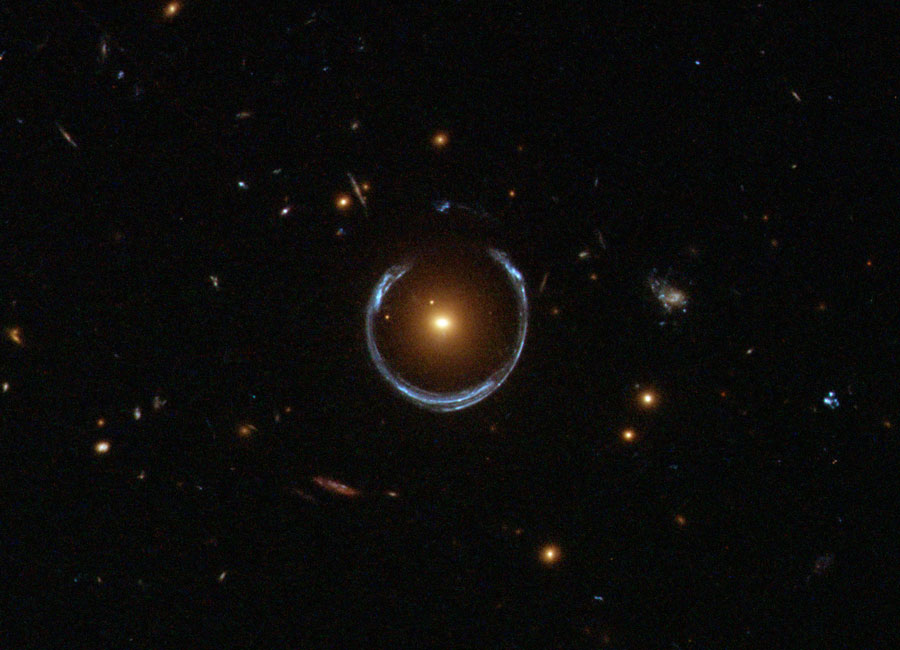}
	\caption{On the left, a scheme of the working principle of the gravitational lensing (© Horst Frank/Designergold). On the right, an example of a horseshoe Einstein ring due to gravitational lensing taken from a Hubble image \cite{Bandiera:2003qgt}.}
	\label{fig:lensing}
\end{figure*}
The gravitational lensing  provides additional evidences of the presence of DM. This phenomenon is predicted by General Relativity and is caused by the space-time distortions induced by the energy and mass distributed within it. In a flat region of space, particles and light travel on a straight path in the 3D space. Yet, in presence of a large mass concentrated in a small region, the space-time around it is distorted so that the straight paths are deflected in the direction of the mass. A schematic illustration of the gravitational lensing principle is shown in left panel of Figure \ref{fig:lensing}. When a massive object is present between an observer and a source of photons, its gravitational potential modifies the space-time, effectively deflecting the light coming from the source and generating multiple images of the original object at the source, as shown in right panel of  Figure \ref{fig:lensing}. In this context, the mass distribution between the observer and the photon source is called \emph{lens}.
The angle $\theta$ of deflection from the straight path of a light ray passing at a distance $r$ from a lens of mass $M$ is expressed in the simple Schawrzschild approximation by \cite{GRWald}:
\begin{equation}
\label{eq:lensing}
\theta=\sqrt{\frac{4GM}{rc^2}},
\end{equation}
\begin{figure*}[t]
	\centering
	\includegraphics[width=0.65\textwidth]{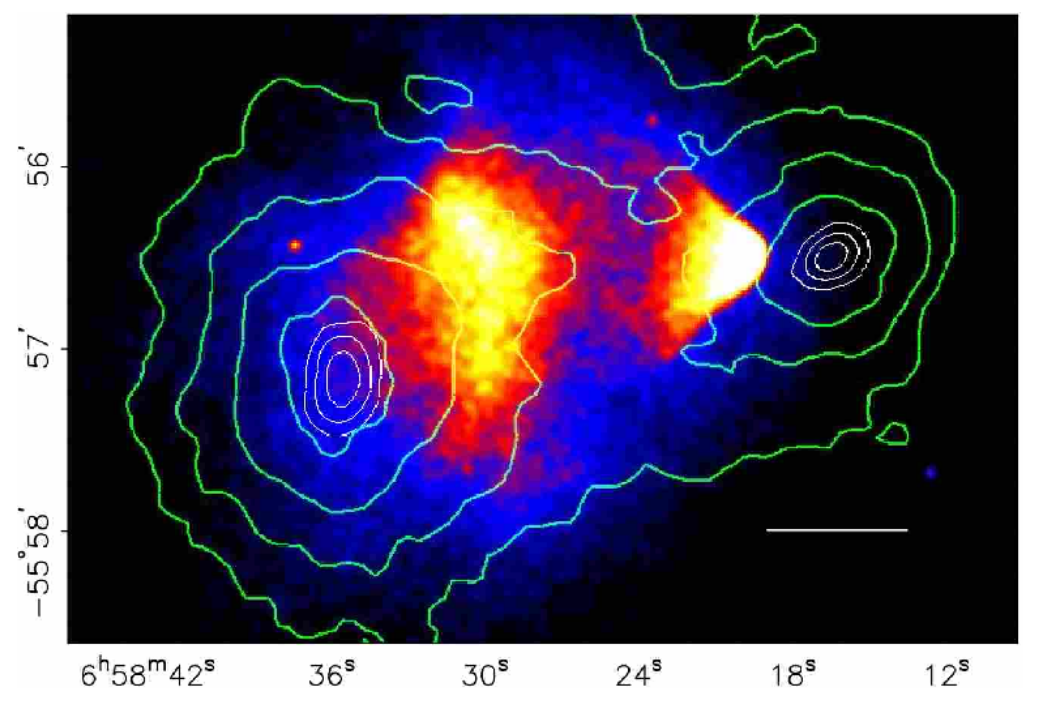}
	\caption{Observation of the bullet cluster performed with the Chandra X-ray telescope \cite{Clowe_2006}. Superimposed in green are the contour lines representing the gravitational potential measured with gravitational lensing.}
	\label{fig:galclust}
\end{figure*}
with $c$ the speed of light in vacuum. As a result, measuring the deflection allows to infer the mass of the object causing it.\\
Depending on the mass and the mass distribution of a lens, the effects of the gravitational lensing are different. Strong lensing happens when mass of the lens is large and its position change in the sky is small compared to Earth's velocity. The effects are resolvable with telescopes in single images\cite{Clowe_2006}. Weak lensing is produced by smaller masses and results in a distortion of galaxy images due to the bending of light \cite{Huterer_2010}. Finally, when the mass of a lens is so small that it does not produce visible deformation, but only affects the light intensity of the luminous sources, it is called microlensing \cite{Wegg_2016}. Different experimental and analysis techniques allow to retrieve evaluation of the masses distributions in between the observer and the luminous sources.\\
An example of a  strong proof of the presence of missing mass in galaxy clusters can be obtained by studying the ones that are merging one into the other, comparing the X-rays emission of the hot gases to the mass distribution inferred from gravitational lensing. In this context, while stars are not expected to collide due to their small size with respect to the dimension of the cluster, the intergalactic hot gas interacts mainly electromagnetically. As a consequence, the gas slows down in the direction of the motion of the galaxies, while its temperature rises and the X-rays emission becomes more intense. In Figure \ref{fig:galclust}, the example of the Bullet cluster is presented. The red halo highlights where the X-ray emission is the strongest and the green contour lines show the gravitational potential estimated with lensing. The observed displacement between these two measurement strongly support the hypothesis that more mass than the one visible has to be present, along with the hint that DM is not collisional, being unscathed by the encounter of the two clusters.
\subsection{Cosmological scale}
\label{subsec:cosmoscale}
One of the most compelling indications of DM existence is inferred by measurements of cosmological-related quantities. The currently most supported theory of cosmology is the Big Bang model, which states that the Universe was once very small, dense and hot, and later expanded to reach the shape it has today\cite{Kolb:1990vq}. 
The Friedmann equations describe the expansion of space-time in a homogeneous and isotropic Universe within the context of General Relativity. These assumptions originate from the cosmological principle, or else the notion that the Universe is homogeneous and isotropic when viewed on a large enough scale. The cosmological principle implies a Friedmann–Lema\^{i}tre–Robertson–Walker metric, that, when applied to General Relativity, allows to express the space-time evolution and its geometry as a function of the energy density of some constituents fluid. The velocity of the expansion of the Universe can be then expressed:
\begin{equation}
\label{eq:Friedmann}
H^2:=\left(\frac{\dot{a}}{a}\right)^2 = \frac{8\pi G}{3}\rho(a) -\frac{k}{a^2},
\end{equation}
with $H$ the Hubble parameter, $a$ the scale factor of the Universe, $\rho$ the energy density of the fluid, $k$ the curvature of the 4-dimensional space-time. The scale factor of the Universe is a normalisation term which represents the dimension of the Universe. In general, its value at the time $t=0$ is taken as zero and subsequently grows with time due to the Universe adiabatic expansion. The fluid which constitutes the Universe can be described as a composed by different components. The most supported theoretical framework which depicts the composition and evolution of the Universe is the $\Lambda$CDM model\cite{OZER1987776}. The $\Lambda$CDM assumes that the energy density $\rho$ is, in fact, the sum of the energy densities of fluids with different properties:
\begin{itemize}
	\item \textbf{Radiation} $\rho_r$ which comprises ultra-relativistic particles such as photons and neutrinos for which  $\rho_r\propto a^{-4}$ is valid;
	\item \textbf{Matter} $\rho_m$ which includes the non-relativistic matter among which the baryonic matter and also the non-baryonic matter like the DM. In this case $\rho_m\propto a^{-3}$;
	\item \textbf{Cosmological constant} $\rho_{\Lambda}$ that accounts for the dark energy (DE), an unknown and mysterious form of energy which is responsible for the acceleration of the Universe's expansion. For the DE, $\rho_{\Lambda}$ is independent of $a$.
\end{itemize}
All of these constituents of the Universe combined with Equation \ref{eq:Friedmann}, describe the evolution of the Universe.\\
A common notation used to express all the Universe constituents into the Friedmann equation is to relate each one's energy density with the critical density $\rho_c$, the density required for the Universe to be perfectly flat:
\begin{equation}
\label{eq:omega}
\Omega_{i}=\frac{\rho_i}{\rho_c},
\end{equation}
Friedmann equation can be rewritten as:
\begin{equation}
\label{eq:omegafried}
H^2 = H_0^2\left(\Omega_{r,0}(1+z)^4 + \Omega_{m,0}(1+z)^3 + \Omega_{\Lambda,0} + \Omega_{k,0}(1+z)^2\right),
\end{equation}
with $z$ the redshift which replaces $a$ following $z=\frac{a_0}{a}-1$, the subscript $0$ denotes that the variable is evaluated now, and $\Omega_{k,0}$ does not represent a constituent of the Universe, but is the curvature term of Equation \ref{eq:Friedmann}. If the sum of all the $\Omega$ components is 1, then the total energy density of the Universe is equal to the critical one and the Universe is flat. The term $\Omega_m$ can be further split into the baryonic part and the DM one. Any measurement able to constrain $\Omega$ parameters provides information on the structure of the Universe and help quantify the amount of DM.\\

\subsubsection{Cosmic microwave background (CMB)} 
\label{subsubsec:CMB}
\begin{figure*}[t]
	\centering
	\includegraphics[width=0.8\textwidth]{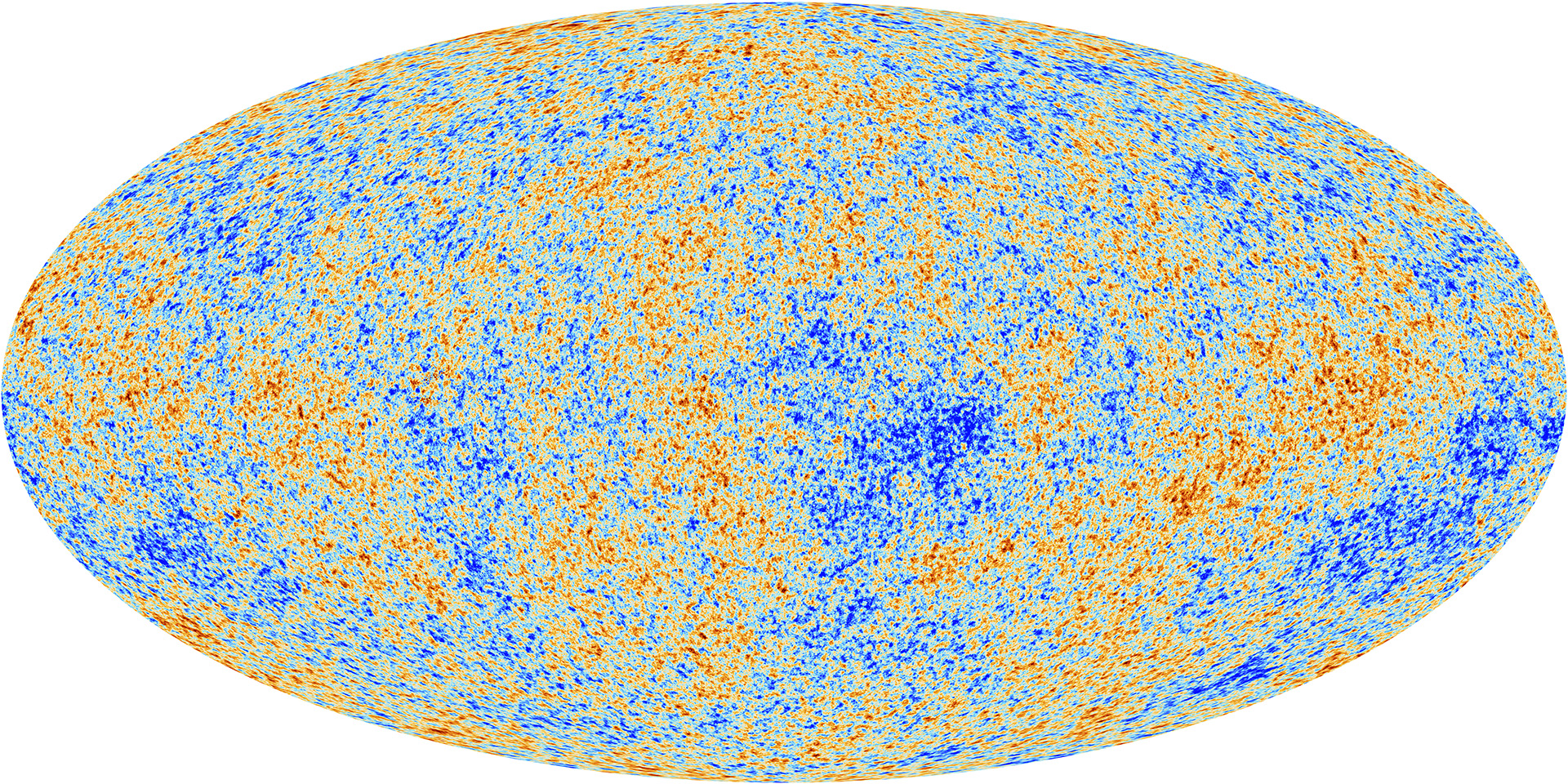}
	\caption{CMB anisotropies measured by the PLANCK satellite. Figure by ESA \url{https://www.esa.int/ESA_Multimedia/Images/2013/03/Planck_CMB}.}
	\label{fig:CMB}
\end{figure*}
\begin{figure*}[!t]
	\centering
	\includegraphics[width=0.8\textwidth]{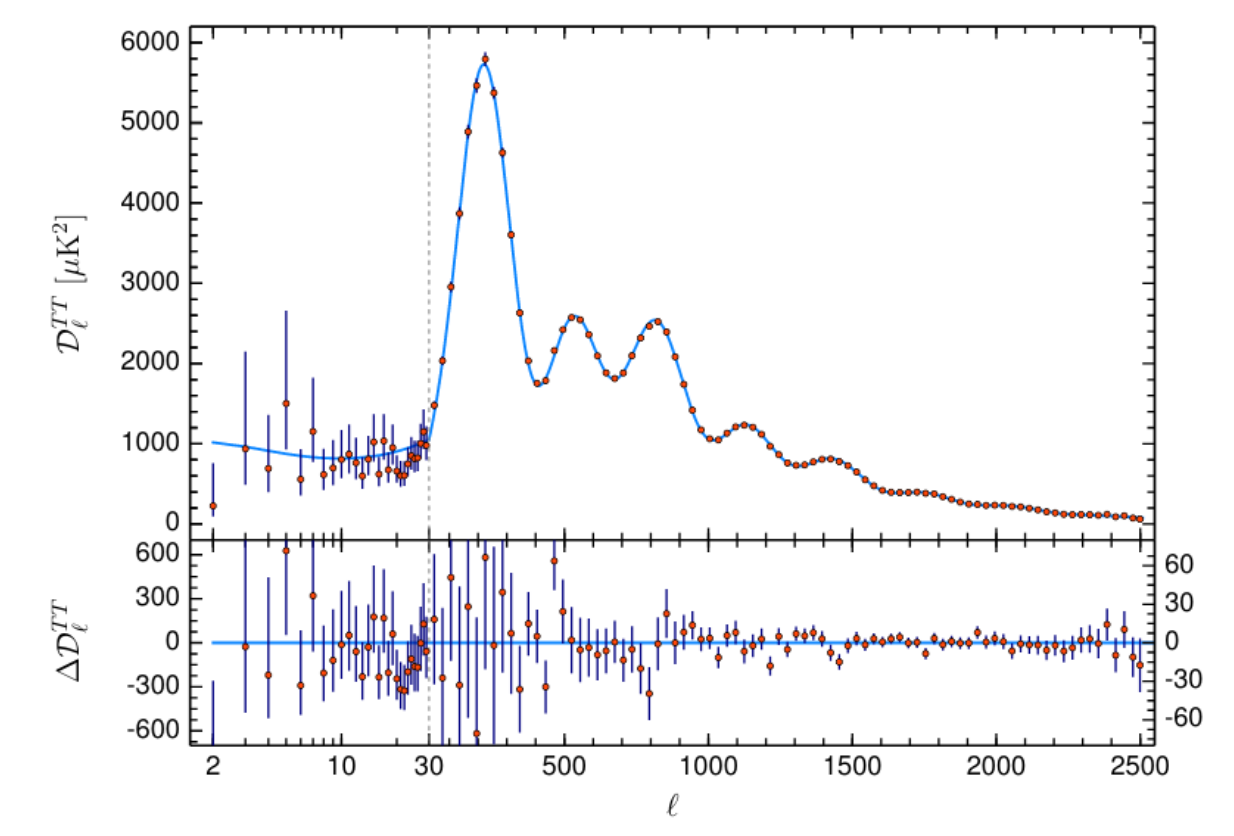}
	\caption{Temperature anisotropies spectrum in spherical harmonics presented by PLANCK\cite{bib:Plank_2020}. The blue line is the $\Lambda$CDM model fit of the experimental data.}
	\label{fig:Plank_spect}
\end{figure*}

A possible method to constrain the $\Omega$ parameters is to exploit the measurement of the cosmic microwave background (CMB), a relic radiation originated during a period called \emph{recombination}. At the very early stages of the Universe, few seconds after the Big Bang, the temperature, inversely proportional to the adiabatic expansion  of the Universe\cite{Arbey_2021}, was of the order of 100 MeV. The cosmic plasma comprised photons, neutrinos, electrons, positrons and baryons in thermal equilibrium. Every photon emitted at those times is lost as they interacted soon after with other particles, being in thermal equilibrium. With the expansion of the Universe, the temperature dropped resulting in reduction of the intensity of the interactions. When the temperature reached values around 1 eV, the energy carried by photons started to be not enough to prevent protons and electrons from combining and forming the first atoms. This is known as the \emph{recombination} period. As a consequence, the photons emitted in that era were not absorbed and can be still detected today as the CMB. This relic radiation possesses a spectrum which follows the black body radiation spectrum, and its isotropy reaches levels of $\frac{\Delta T}{T}\sim 10^{-5}$ \cite{bib:Plank_2020}. Nevertheless, small anisotropies exist and are shown in Figure \ref{fig:CMB}, from the measurement of the PLANCK collaboration. These anisotropies are related to fluctuations in the local gravitational potential and any kind of phenomenon which mirrors its behaviour on the energy density \cite{ELWright}. Examining them carefully can provide constraints on various scenarios hypothesised for the early Universe. The temperature anisotropies are usually decomposed in the spherical harmonic functions
\begin{equation}
\label{eq:sferarm}
\frac{\Delta T}{T}=\sum_{l}\sum_{m=-l}^{m=l}a_{lm}Y_l^m(\theta, \phi),
\end{equation}
and the power spectrum is computed from:
\begin{equation}
\label{eq:sferarmpowerspect}
C_l=\left\langle \vert a_{lm}\vert^2\right\rangle_m
\end{equation}
Figure \ref{fig:Plank_spect} shows the temperature anisotropies as presented by PLANCK collaboration as a function of the spherical harmonics. The blue line represents the fit of the $\Lambda$CDM model, the simplest parametric model of the Big Bang which includes DM and DE. The temperature fluctuations follow the oscillation of gravitational wells, matter, DM and photon plasma. The first intense peak is related to the base frequency of the energy and matter density oscillation. Its position is strongly related to the value of total curvature of the Universe at the recombination time. The other peaks are overtones of the fundamental frequency. The difference in intensity is related to the amount of baryonic matter (for the second peak) and non-baryonic one (for the subsequent peaks) as plasma oscillation can interact destructively with the fluctuation of the gravitational potential wells. The model fits very well the data and allows to measure the $\Omega$ components from the relative intensity of the peaks and their position, even though more complete and precise analyses which include other measurable quantities is described in \cite{bib:Plank_2020}.
The CMB anisotropy description is one of the possible cosmological measurements which can provide information on the $\Omega$ components of the $\Lambda$CDM model. The art on the $\Omega$ relative densities which are summarised in Table \ref{tab:planck}, taken from \cite{bib:Plank_2020}:
\begin{table}[!t]
	\centering
	\begin{tabular}{|c|c|c|}
		\hline
		Component & Density parameter & Planck results with 68\% CL    \\ \hline
		Radiation & $\Omega_r$ & $\sim 9 \times 10^{-5}$ \\ \hline
		Baryonic matter & $\Omega_b$ & $0.0489 \pm 0.0003$ \\ \hline
		Non-baryonic matter & $\Omega_{nb}$ & $0.2607 \pm 0.0019$ \\ \hline
		Dark energy & $\Omega_{\Lambda}$ & $0.6889 \pm 0.0056$ \\ \hline
		Total & $\Omega_{tot}$ & $0.9993 \pm 0.0019$ \\ \hline
	\end{tabular}
	\caption{Table summarising the Planck combined results on the $\Omega$ parameters of the $\Lambda$CDM model, taken from \cite{bib:Plank_2020}.}
\label{tab:planck}
\end{table}
The $\Omega_{tot}$ is perfectly consistent with 1, suggesting that the Universe is flat. The non-baryonic matter which is considered to mostly be DM is a required constituent of the Universe. On top of this, its abundance is about 26\% of the total energy density, comprising 84\% of total matter.
\subsubsection{Structure formation} 
\label{subsubsec:strutform}
\begin{figure*}[t]
	\centering
	\includegraphics[width=0.35\textwidth]{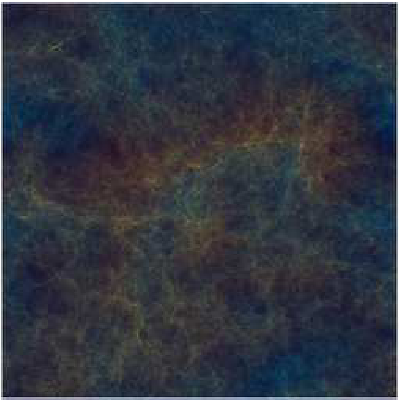}
	\includegraphics[width=0.35\textwidth]{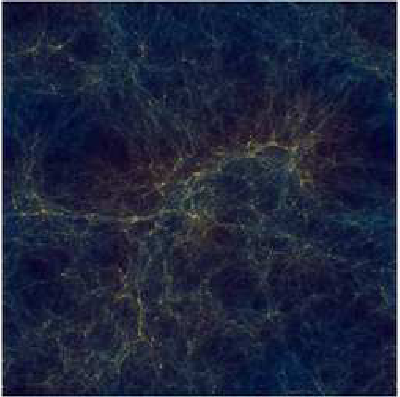}
	\includegraphics[width=0.35\textwidth]{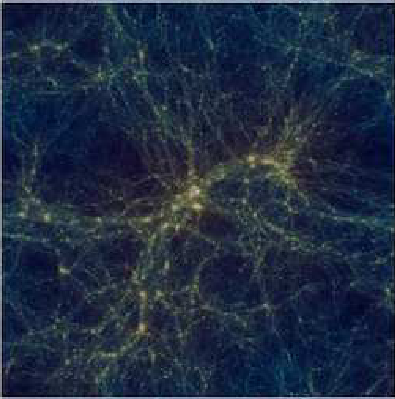}
	\includegraphics[width=0.35\textwidth]{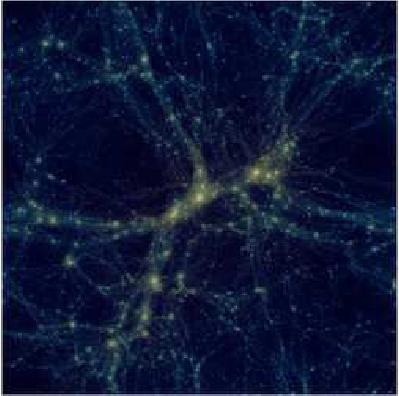}
	\caption{Example of the N-body simulation result for the Illustris \cite{Sparre_2015}. Different redshift are displayed: top left z=7, top right z=3, bottom left z= 1, bottom right z=0.}
	\label{fig:largescalesim}
\end{figure*}
The formation of large scale structures, such as galaxies and clusters of galaxies, is a very relevant test bench for the DM characteristics and evidence of its presence.  Since today's highly inhomogeneous Universe evolved from a very uniform and smooth state at the time of recombination (when density fluctuations were of the order 10$^{-5}$), the analysis of the large structures can provide constraint on what influenced such evolution. The homogeneity of the Universe can be expressed as the fluctuation intensity of the fluid energy density of the Universe $\rho$ defined in the previous Section. As a simple approximation, until the Universe is radiation dominated ($\rho_r>>\rho_m,\rho_{\Lambda})$, $\frac{\delta \rho}{\rho}$ can grow as the logarithm of the scale factor $a$. Only in a matter dominated epoch, the perturbations can inflate linearly. However, if only baryonic matter existed, the linear growth of the perturbations would start only after recombination. Scaling them to the present day, $\frac{\delta\rho}{\rho}$ would still be of the order of 10$^{-2}$, a quite uniform Universe. A non linear growth of the perturbations can be induced if a different matter species that decouples from photon before recombination is introduced in this picture.  This would generate potential wells where baryonic matter falls into immediately after the recombination, provoking non linear enhancement of the perturbations. This different, non baryonic form of matter would perfectly fit as DM.\\
In order to better detail the peculiarities of this DM assumption in the evolution of these large structures, complex N-body simulations are used. A set quantity of particles belonging to the primordial fluids (radiation, DM and baryonic matter) is prepared in the abundance present at the very early stages of the Universe, inferred by the PLANCK results. The system is left to evolve following the gravitational attraction of the constituents and the interactions between the fluids as derived from particle physics. The evolution is continued until the present day and the large structure formed and formation processes are checked with the astronomical observations. Various N-body simulations like Millennium\cite{Springel_2005}, $\nu^2$GC \cite{nu2GC} and Illustris\cite{Sparre_2015} are employed and typically utilise billions of particles in order to generate a modern universe few hundreds of light-years wide. Figure \ref{fig:largescalesim} exemplifies the outcome of some of these simulations.
For example, in the hypothesis that DM is made of particles (see next Section), some physical characteristics can modify the result of the simulation. If DM were made of relativistic particles at the moment of decoupling from baryonic matter, the pressure they induce would be enough to eliminate the small perturbations and large scale gas-structure would form first. The galaxies would be generated only afterwards from the collapse of these gaseous systems. Recent observation rule out the hypothesis of hot DM as they are too different from the simulation outputs \cite{White}. As a consequence, the most supported model endorses a cold or warm DM hypothesis, with non-relativistic or semi-relativistic particles at the decoupling time, as simulations results succeed in reproducing the current measurements.
\section{Candidates for dark matter}
\label{sec:candidateDM}
Astronomical and cosmological observations based on gravitational probes are not consistent with the current picture of Universe particle and forces content. Thus, an ulterior ingredient is believed to be necessary to explain the measurements performed. Nevertheless, all of these hints derive from gravitational probes of observables. Various theories to explain these anomalies exist, and in the following, the most established will be reviewed.
\subsection{Modified Newtonian Dynamics}
\label{subsec:MOND}
All the evidences of the lack of complete understanding of the Universe derived from the astrophysical and cosmological measurements are gravitational probes which assume the application of the Newtonian dynamics and General Relativity. Alternative models try to explain these observed discrepancies by modifying the underlying hypotheses on the laws of gravitation themself. These theories go under the name of Modified Newtonian Dynamics (MOND) and attempt to explain the observations by modifying Newtonian gravitational force to correctly predict the anomalies measured \cite{Milgrom_2015,Famaey_2012}. The core idea is that it exists a new fundamental acceleration constant $a_0$ that marks the transition between Newton and deep-MOND regime. When the acceleration of the system $a$ is smaller than $a_0$, the gravitational force scales differently than standard Newtonian gravity describes. On the contrary, at higher accelerations than $a_0$, the standard dynamics is recovered. With $a_0$ estimated from the measurements of the galactic rotation curves as $10^{-10}$ cms$^{-2}$, the dynamics of galaxies and cluster of galaxies is affected, leaving untouched the Solar system scale. The theories are accurate and predictive at a galactic scale. In particular, the foreseen dependence of the rotation curve on the baryonic mass content of the galaxy is matching the expectation \cite{Mcgaugh_1,Milgrom_2020}. However, the difficulties in the MOND hypothesis resides in the capability to explain the discrepancies along the varied and different scales of the Universe. Large structure behaviours present in galaxy clusters and the CMB inhomogeneities require advanced extensions of the MOND in a relativistic form \cite{Bertone_2018}. These theories become less predictive and more functions and parameters have to added to correctly describe the measurements \cite{Famaey_2012}. The MOND theories largely succeeded in describing the galactic and galaxy cluster scale\cite{Famaey_2012}, but fail with more complicated structures and phenomena, such as the Bullet cluster and the CMB \cite{Skordis_2021,Angus_1}. While MOND theories propose an interesting view to solve the observation discrepancies, a complete theory is still missing\cite{McGaugh_1999}.
\subsection{Primordial black holes}
\label{subsec:PBH}
Gravitational observation of a missing component in the Universe can be explained without the need to invoke a modification of the known laws of gravity, but rather  assuming the existence of astrophysical objects that have yet to be resolved and identified. The possibility that black holes could be formed by the collapse of matter in the early Universe was proposed by Zeldovich, Novikov and Hawking in the late sixties \cite{Zeldovich,Hawking}. These primordial black holes (PBHs) are expected to be produced before the period of the Big bang nucleosynthesis (BBN).
The BBN is a period in which the temperature and density of the Universe are allowing the nuclear fusion of light elements, with the products of such reaction not being destroyed by the energetic photons thermally coupled to matter. This epoch permitted the generation of nuclei heavier than hydrogen and it is used to constrain the baryonic matter content of the Universe\cite{Coc_2017,Zyla:2020zbs}. Being these massive objects produced before the BBN, they are decoupled from the rest of the baryonic matter and are not accounted for in the evaluation of its abundance\cite{Carr_2020}. For these reasons, since these massive objects do not enter in the CMB predictions, they could constitute the origin of the missing mass of the Universe and still be consistent with the evidences presented in this Chapter\cite{Chapline}. Indeed, for PBH masses greater than 5 $\times$ 10$^{14}$ g, their lifetime is longer than the age of the Universe \cite{Billard_2022}. In fact, since PBH do not originate from collapse of massive star, their mass is unconstrained and could span many orders of magnitude. No positive discovery of a PBH has ever been claimed, but measurements of microlensing, gravitational waves, and CMB are putting stringent limits on the available parameter space of fraction of DM as PBH versus PBH mass\cite{Green_2021,PrimordialBH}, leaving wide open only a region of masses between 10$^{17}$ and 10$^{21}$ g. 
\subsection{Dark matter as a particle}
\label{subsec:DMpart}
The last decades have seen the completion of the discovery of the particles predicted by the most complete and known model of particle physics, the Standard Model (SM), with the observation of the Higgs boson at LHC in 2012. Nonetheless, the SM  is known to possess several limitations, from the Higgs mass correction to the neutrino masses, to the strong CP violation, to the matter to antimatter asymmetry and the lack of the inclusion of gravity\cite{limitsSM,Lee_2021}. Extensions of the SM to tackle these problems predict the existence of new forces or particles, some of which have the characteristics to be a DM candidate. From the observation discussed in Section \ref{sec:evidenceDM}, a series of characteristics that any model of DM as a particle has to respect can be obtained:
\begin{itemize}
	\item \textbf{Non baryonic}. Planck measurements of the CMB imply a non baryonic nature of these particles;
	\item \textbf{Abundant}: Planck measurements found a $\Omega_{nb}\sim$0.26, a much larger density than baryonic matter. The combination of number density of DM particles and their mass depends on the model in order to correctly match this abundance and the gravitational effects observed;
	\item \textbf{Neutral and colour-free}. Electromagnetic interactions are excluded by the lack of interaction with photons which would have been otherwise detected. Strong interaction is also excluded since DM would lose energy and concentrate more in the galactic centres than what is observed \cite{Fan_2013}. 
	\item \textbf{Weakly interacting}. While at the moment experimental observations supporting the hypothesis that DM interacts other than gravitationally do not exist\cite{Billard_2022}, a form of weak interaction between DM and SM particles is hoped for in order to obtain a positive detection. As this non baryonic matter needs to be decoupled from the radiation-baryonic matter fluid already at the epoch of BBN, the intensity of interaction must be weak. This weak interaction does not refer to the SM electro-weak interaction, but can be any form of sub-weak strength type of interaction \cite{Billard_2022}.
	\item \textbf{Stable} or extremely long lived. Due to the weak interaction, its presence in the early Universe and the measurements of its gravitational influence today, the DM particles are required to be preserved in large amount.
	\item \textbf{Warm or cold}. DM relics can be classified according to their energy at the cosmological moment of decoupling from the baryonic matter fluid. Hot DM models require relativistic particles at the decoupling and typically masses are of the order $\mathcal{O}$(10) eV to match the abundance measured. Warm and cold DM models have need larger masses and smaller kinetic energies, with cold DM being completely non-relativistic. Large scale structure simulations are consistent with the existence of a hot population of DM, but only if its abundance is small, as it is shown to be incompatible with the formation of the large scale structures. Warm and cold DM allow the formation of the galaxies as observed, but the two modify the galactic structure. The information available on the galactic satellites are not conclusive enough to determine the type of DM population\cite{Lovell_2012}.
\end{itemize}
As a consequence of these characteristics, the modelling of a DM particle necessarily requires to extend the SM, because no SM particle satisfies the requirements listed above, not even neutrinos. Indeed, neutrinos are non baryonic neutral only weakly interactive particles, but the evaluation of the expected relic density does not match the observations. Moreover, being a warm-hot candidate the formation of small structures like dwarf galaxies is strongly hindered if neutrinos accounted for 100\% of DM \cite{1992ApJ}.
\subsubsection{WIMP}
\label{subsubsec:WIMP}
\begin{figure*}[t]
	\centering
	\includegraphics[width=0.8\textwidth]{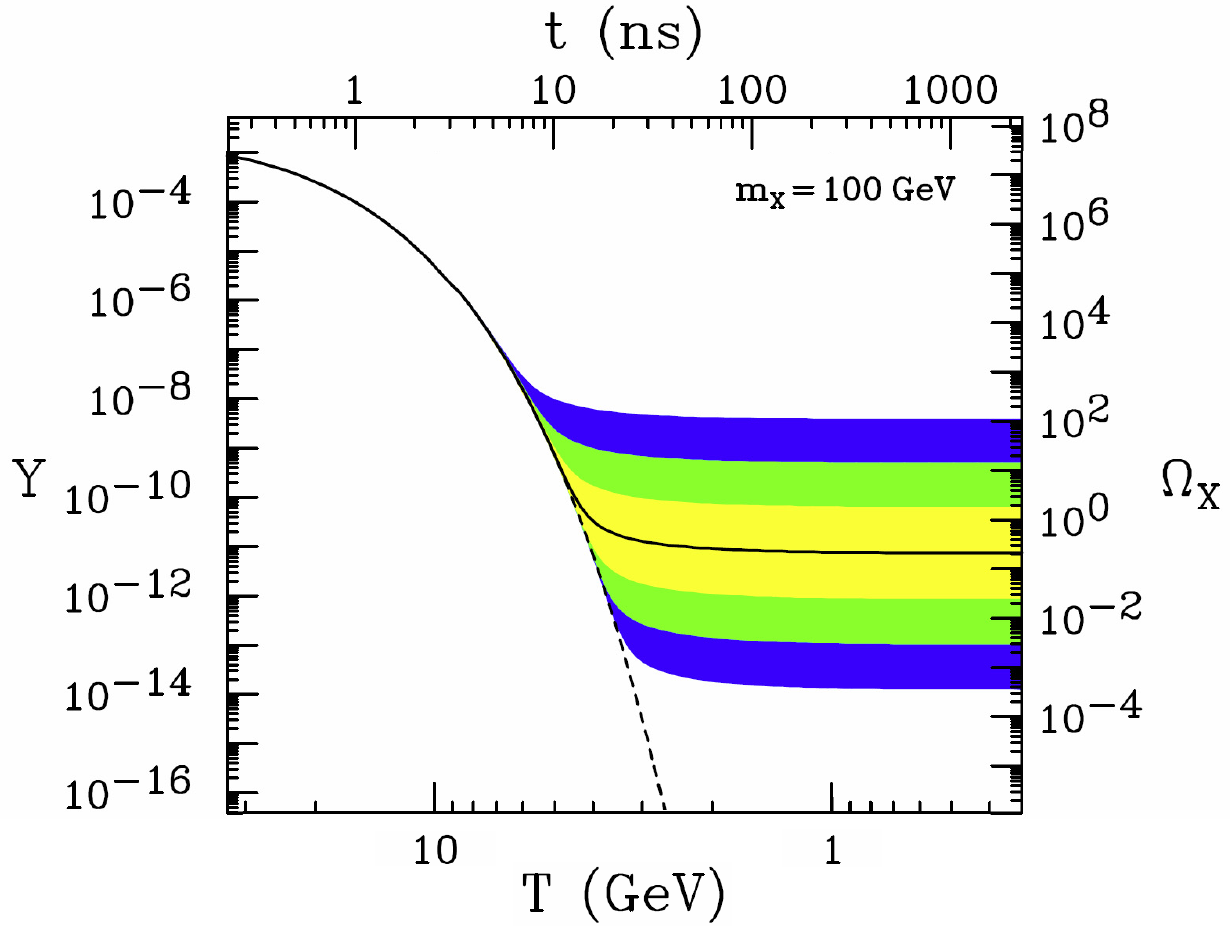}
	\caption{Comoving number density $Y$ as a function of the temperature along the evolution of the Universe for a WIMP mass of 100 GeV/c$^2$. The solid black line is the solution to the Boltzmann equation that result in the measured DM density parameter $\Omega_{nb}$. The coloured bands show the diverse evolution for cross sections that differ of a factor 10, 100 and 1000 \cite{Feng_2010}.}
	\label{fig:freezeout}
\end{figure*}

One of the most studied and supported candidates are the Weakly Interactive Massive Particles (WIMPs), because these SM extensions predict type of particles which can naturally be produced in the correct abundance as thermal relics. The basic assumption behind WIMPs is that they are neutral massive particles with interaction with regular matter at the order of the weak scale or below. Starting from this assumptions, it is possible to relate the WIMP mass and cross section to today's relic density\cite{PhysRevD.33.1585,Bernal_2017}.  Naming $\chi$ a WIMP and $\phi$ a SM one, and assuming that \Ws can annihilate into SM ones, different epochs can be defined in the history of the Universe. In the hot early Universe, WIMPs were in thermal and dynamic equilibrium with SM, where the following expressions were valid:
\begin{equation}
\label{eq:chi-phi_termal}
\phi+\chi \leftrightarrow  \phi + \chi
\end{equation}
\begin{equation}
\label{eq:chi-phi_coupling}
\chi+\bar{\chi} \leftrightarrow \phi + \phi*.
\end{equation}
With the expansion of the Universe, the density and the temperature drop so that the DM decouples from the SM. The number of DM particles is therefore fixed and this thermal relic abundance survives until now. This mechanism is called freeze-out.\\
The number density of WIMPs $n_{\chi}$ can be described by the Boltzmann equation
\begin{equation}
\label{eq:boltzmanWIMP}
\frac{dn_{\chi}}{dt}=-Hn_{\chi}-\left\langle\sigma_Av\right\rangle(n_{\chi}^2-n_{\chi,eq}^2),
\end{equation}
with $H$ the Hubble parameter, $n_{\chi,eq}$ the number density in equilibrium and $\left\langle\sigma_Av\right\rangle$ the thermal average of the annihilation cross section flux. The equation shows how the number density diminishes in time  due to the spatial expansion of the Universe and is modified by the annihilation and generation of DM particles. In cosmology, the total entropy of the Universe is conserved during thermal equilibrium and its density $s$ is often used as a variable to characterise the comoving volume of the Universe. As a consequence, the comoving number density $Y_{\chi}$ of DM can be rewritten as:
\begin{equation}
\label{eq:comoving}
Y_{\chi}:=\frac{n_{\chi}}{s}
\end{equation}
The moment of freeze-out is very relevant to determine the abundance of WIMPs now. It is defined as when
\begin{equation}
\label{eq:freezeout_moment}
n_{\chi,FO}\left\langle\sigma_Av\right\rangle \sim H
\end{equation}
Moreover, the $\Omega_{nb}$ parameter can be formulated as function of comoving quantities. Then, assuming that the comoving density now is the same that was present at the freeze out, it possible to derive: 
\begin{equation}
\label{eq:omegafreezeout}
\Omega_{nb}=\frac{\rho_{\chi}^0}{\rho_c}= \left(\frac{s^0}{\rho_c}\right)m_{\chi}Y_{\chi}^0=\left(\frac{s^0}{\rho_c}\right)m_{\chi}Y_{\chi}^{FO}=\left(\frac{s^0}{\rho_c}\right)m_{\chi}\frac{H^{FO}}{s^{FO}\left\langle\sigma_Av\right\rangle}\propto\frac{m_{\chi}}{\left\langle\sigma_Av\right\rangle},
\end{equation}
with the superscripts $0$ and $FO$ referring to the time in the history of the Universe when the quantities are evaluated, respectively now and at the time of freeze out, $m_{\chi}$ the \W mass, $\rho_c$ the critical density, and  $\Omega_{nb}$ the density parameter of non baryonic matter (see Section \ref{subsec:cosmoscale}).
Since the PLANCK measurements show that $\Omega_{nb}$ is $\mathcal{O}$(0.1), Equation \ref{eq:omegafreezeout} implies that the ratio of the interaction strength and the \W mass are of the same order. As the interaction is expected to be weak, the \W mass is also found at the weak scale. The fact that theories conceived to tackle SM issues naturally predicted DM candidate at the weak scale was considered extremely fascinating and takes the name of "WIMP miracle". Figure \ref{fig:freezeout} shows the $Y$ comoving number density of DM in the freeze out mechanism as a function of the temperature of the evolving Universe.
It is possible to formulate hypotheses in which the constraints on the DM mass and cross section interaction is loosen. The coloured bands represent the effects on the thermal relic density of the cross section intensity variation of various order of magnitude.\\

Particles with the characteristics of the WIMPs discussed above can emerge in many scenarios beyond the SM. Here, the two most popular classes of theories will be discussed, that are supersymmetry (SUSY) and extra dimensions (EDi).

The SUSY \cite{Fox:20196D,Roszkowski_2018} is an extension of the SM which tries to address the gauge coupling unification and the Higgs mass corrections\cite{Roszkowski_2018} by introducing a copy of each of the SM particle with opposite parity with respect to a new gauge transformation. The calculation of the Higgs boson mass includes corrections which are proportional to the mass of all the SM particles. In the SM, the series of corrections does not converge and the Higgs mass diverges. The new symmetry added by SUSY links bosons and fermions so that each boson of the SM has a so-called superpartner which is fermionic, and for each SM fermion there is a boson superpartner. These superpartners possess a different parity to this new gauge transformation which causes their correction to the Higgs mass to have opposite sign. Therefore, only if the superpartner masses are close enough the the SM ones, the overall contributions to the Higgs mass calculation are strongly reduced so that the series of correction converges. 
In a subclass of SUSY, the Minimal Supersymmetric Standard Model (MSSM), another ingredient is added, the R-parity conservation. R is a quantum number, combination of the baryonic number, the leptonic number and the spin. SM particles have R=1, while SUSY ones have R=$-$1. The R-parity conservation implies that the decay of a SM particle can happen only with an odd number of SM particles. This was introduced to suppress the decay of the proton in supersymmetric particles. Akin, also supersymmetric particles can only decay in an odd number of themselves.\\
In the SM, the gauge bosons of the electroweak symmetry (B, W$^{1}$, W$^{2}$, W$^{3}$) combine with Higgs fields in the electroweak symmetry breaking to generate the mediator bosons of (A, W$^{+}$, W$^{-}$, Z). Similarly, in SUSY, their superpartners ($\tilde{B}$, $\tilde{W}^{1}$, $\tilde{W}^{2}$, $\tilde{W}^{3}$) combine and produce neutral, colourless, weakly interacting particles, the neutralinos.  Thanks to the R-parity, the lightest of the neutralinos is a stable particle which can disappear only by means of annihilation, a perfect WIMP candidate.

It is possible to accommodate the WIMP hypothesis also in the EDi framework \cite{Bertone_2018}.  The basic idea behind EDi is to try to unify electromagnetism with gravity by introducing additional dimensions to reconcile the weakness of the gravity interaction with respect to the other forces. In this class of models, the strong and electroweak forces are confined in the 3+1 dimensions of the Universe, called \emph{brane}. Instead, the gravitational force is described in a 3+1+$n$ dimension space named \emph{bulk}, which embeds the brane, and thus appears feeble in the 3+1 brane. All of these extra dimensions are \emph{compactified} on circles of size $R$, because if their dimension were big, they would have been detected \cite{Andriot_2017,PDG}. The compactification forces all of the fields propagating in the bulk have their momentum quantised. The result is that for each bulk field, a set of Fourier expanded modes, called Kaluza–Klein (KK) states, appears. From the point of view of the four dimensional world, these KK states manifest as a series, \emph{tower}, of states with masses $m_n = n/R$, where $n$ labels the mode number. When the extra dimensions are small, it is possible for these fields to propagate throughout the bulk and not being confined only to the single brane. This is known as the Universal extra dimensions \cite{Appelquist_2001}. In these models, new particles coming from these KK states can be produced and the lightest ones have masses of the order of hundreds of \Gevc \cite{Servant_2003}. Again, a theoretical model whose the main goal is solving known issues of the SM returns a perfect WIMP candidate.
\subsubsection{Axions and ALPs}
\label{subsubsec:ALPS}
The axions are particles predicted by an extension of the SM formulated by Peccei and Quinn \cite{Peccei:1977hh} to tackle the strong CP problem. The Lagrangian of the Quantum ChromoDynamics (QCD) contains a term which causes the violation of the charge parity symmetry (CP). The parameter responsible for the intensity of the symmetry violation, $\theta$, is not constrained by any theoretical reason. Such CP violating term can give an arbitrarily large electric dipole moment to the neutron. Since no electric dipole moment for the neutron has ever been observed \cite{Abel:2020pzs}, $\theta$ has to be less than 10$^{-11}$ \cite{Pendlebury:2015lrz}. Nonetheless, the consequences of a CP violating term are not measured, leading to the necessity of a new symmetry which prevents the CP violation. The Peccei-Quinn solution is to introduce a $U$(1) chiral symmetry to prevent $\theta$ from being $\mathcal{O}$(1). Similar to the Higgs potential, this symmetry is spontaneously broken at a scale $f_a$, leading to the generation of Goldstone bosons that carry the degrees of freedom. Specifically, the boson generated is a pseudo-Nambu-Goldstone boson which is called axion, $a$ \cite{PhysRevLett.40.279}. When the Lagrangian is rewritten in terms of this newly generated boson, the expectation value of the axion term is capable of compensating for the CP violating component of the QCD, solving the fine tuning issue. This also implies that the axion can couple to gluons and photons. In particular, the former allows to estimate the expected mass of the axion \cite{di_Cortona_2016} as:
\begin{equation}
\label{eq:axionmass}
m_a= 5.70(6) \mu\text{eV}\left(\frac{10^{12}\text{GeV}}{f_a}\right)
\end{equation}
connecting the mass of the particle to mixing strength with gluons. As a matter of fact, with a large enough $f_a$, the axion is stable and weakly interacting with matter, enough to be considered as a genuine DM candidate \cite{axions_astro}. The axions are light, but being bosons, a group of them acting as a fluid can be demonstrated to recover all the relevant properties to explain the cold DM hypothesis\cite{Sikivie_2008}. The coupling with photons allows to thermally obtain the correct relic density, converting photons into axions in the early Universe \cite{Mass__2002}, or with an oscillation mechanism around the minimum of the potential \cite{Garcia_Irastorza_2022}. Moreover, the coupling with photons permits different phenomena which can result in the destruction of an axion in favour of a SM particle, such as the axion decay in a couple of photons. This provides channels through which axion particles can be detected. Current constraints on the $f_a$ parameter due to cosmological measurements are \cite{Raffelt_2007}:
\begin{equation}
\label{eq:axionlimitfa}
10^9\text{GeV} \lesssim f_a \lesssim 10^{12}\text{GeV}
\end{equation}

Theoretical frameworks which predict the existence of particles similar to axions have been elaborated. These models keep the coupling of these axion-like particles (ALPs) with photons, but relax the relation which binds $f_a$ to its mass. Though these theories do not resolve the SM issues, these particles still have the characteristics to be a DM candidate.

\subsubsection{Sterile neutrinos}
\label{subsubsec:sterile}
The SM predicts the existence of 3 types of neutrino with left handed chirality and zero mass. However, the observation of the oscillation of neutrino flavours implies that the neutrinos possess a mass\cite{nuoscillation}. A possible solution to the SM neutrino mass problem is found by introducing a new particle, the sterile neutrino. This is a singlet in the standard gauge group and thus it is interacting very weakly with SM matter only through a mixing mass matrix. Exploiting the seesaw mechanism, the mixing matrix between the sterile neutrino and the SM ones can be expressed as a symmetric matrix with one term much larger than the others. This causes the masses of the mixed neutrinos to be correlated so that when one is very large the other is very small. The tiny SM neutrino mass is explained by the high value of the mass of the sterile neutrino. If its mass is larger than $\sim$ 1 keV/c$^2$ and the mixing angles of just mentioned mass matrix are small, it can be shown that it becomes non-relativistic early enough to produce a thermal relic density consistent with a cold DM hypothesis \cite{Dodelson_1994}. The coupling through the mixing mass matrix allows a tiny interaction between SM particles and the sterile neutrino.
The extremely weak strength of mixing with the SM guarantees a stability which matches the requirements to be an example of a class of theories known as Feebly Interacting Massive Particles (FIMPs). In order to obtain the correct relic density with such small interaction the freeze in mechanism is preferred to the freeze out  \cite{Hall_2010}. In this case the cross section of the interaction of DM with SM is so low that even at the early stages of the Universe DM is decoupled from the SM. The amount of DM is very small, but slowly increases by means of the annihilation of SM couples. The thermal relic density is reached when the production of DM is suppressed by the expansion of the Universe.\\ The signature sought to look for these particles is the decay of the sterile neutrino in a monochromatic photon, on which research is still ongoing \cite{SHROCK1982359}.

\subsubsection{Dark sector}
\label{subsubsec:Dph}
In the context of DM searches, numerous models in the direction of a \textit{dark sector} have become popular. In these frameworks, the SM is just one sector of the  particles and forces present in the Universe, next to which a dark sector composed by a collection of particles that are not charged directly under the SM strong, weak, or electromagnetic forces exists. The dark sector can contain from few particles up to entire new interaction forces and structure similar to the SM one. The dark sector scenario is well motivated by the ease with which SM limitations can be overcome \cite{darksector,Billard_2022}. Particles belonging to this sector are assumed to interact gravitationally, but may also possess a messenger sector, called \textit{portal}, which contains one or more states that mediate SM–dark sector interactions. The gravitational interaction along with the suppression of strong or electromagnetic interaction make the dark sector particles optimal DM candidates\cite{Fabbrichesi_2021}, whose relic density can be thermally obtained both via freeze out and freeze in scenarios \cite{darksector}. \\
One of the most renowned class of dark sector models is the dark photon portal in which a light thermal WIMP, in the \Mevc mass range, interacts with the SM sector via a dark photon (a dark sector gauge boson) which mixes with the SM photon. The dark photon can be detected because of its kinetic mixing with the ordinary, visible photon. This kinetic mixing is always possible because the field strengths of two Abelian gauge fields can be multiplied together to give a dimension four operator. The existence of such an operator means that the two gauge bosons can go into each other as they propagate. This kinetic mixing provides the portal linking the dark and visible sectors\cite{Fabbrichesi_2021}.\\
The generic Lagrangian of these models includes a kinetic term of the dark photon:
\begin{equation}
\label{eq:lagrdph}
\mathcal{L}_0= -\frac{1}{4}F_{a\mu\nu}F_{a}^{\mu\nu}-\frac{1}{4}F_{b\mu\nu}F_{b}^{\mu\nu}-\frac{\epsilon}{2}F_{a\mu\nu}F_{b}^{\mu\nu}
\end{equation}
where $a$ and $b$ denote respectively the dark and the SM photon components, while the last block is the kinetic mixing of the two gauge bosons. In some cases the dark photon can be a massless gauge boson and the DM is identified with \Gevc or TeV/c$^2$ scale dark fermions\cite{PhysRevD.70.083501}, in some other the dark photon is massive and light (\Mevc scale) and it is the DM candidate itself \cite{Boehm_2004}. \\
The portal a dark sector exploits to interact with the SM can also be modelled employing dark fermions in place of dark photons\cite{DeRocco_2019}. A generic dark fermion term can be added to a Lagrangian as follows:
\begin{equation}
\label{eq:lagrdfer}
\mathcal{L}=eJ_{\mu}A_b^{\mu} + e'J_{\mu}^{'}A_b^{'\mu}
\end{equation}
with the $'$ indicating the dark sector terms.\\
Possible constraints can be found in the analyses of the BBN, in the structure formation theory, but are also sought at accelerators experiments \cite{Fabbrichesi_2021}.
\chapter{Direct detection of dark matter}
\label{chap2}
As illustrated in Chapter \ref{chap1}, astronomical and cosmological measurements indicate that the description of the fundamental particles and forces of the Universe is still incomplete. All the unexplained observed effects are compatible with a missing mass problem.\\
In the assumption that dark matter (DM) is composed by particles which can couple to Standard Model (SM) ones not only gravitationally (see Section \ref{subsec:DMpart}), this interaction can be exploited in different ways to search for DM signatures. DM can be indirectly probed by studying potential unexplained excesses of SM particles produced by DM annihilation in space. Unpredicted fluxes of SM particles from space, like positrons, antiprotons or $\gamma$s, are looked for by dedicated detectors \cite{Conrad2014}. Alternatively, the ability to create new particles, typical of experiments at high energy colliders, can be employed to search for DM signature either in direct production \cite{Buchmueller_2017} or via beam dump approaches \cite{Battaglieri_2022}. \\
The direct DM detection techniques discussed in this thesis aim at observing the elastic scatters of Galactic DM particles with the detector material. The measurements of the rotation curve of the Milky Way imply, in fact, that the Galaxy is embedded in a DM halo, like all spiral galaxies (see Subsection \ref{subsec:galaxyrot}). The peculiar motion of the Earth and Sun with respect to the centre of the Galaxy imprints a preferred direction of arrival of these DM particles with respect to the laboratory rest frame, which can be exploited to measure and characterise DM interactions. Unfortunately, astrophysical constraints imply that the coupling between DM and SM particles is very weak, making the expected signal rate very low. Given the large number of phenomena able to produce interactions that can mimic a DM signal, direct DM searches are experimentally extremely challenging. \\
This thesis, from now on, will address specifically the direct detection of \Ws through nuclear recoil signatures, and refers the reader to  \cite{Conrad2014,Buchmueller_2017,snowall} for a recent comprehensive review of all other approaches. In Section \ref{sec:WIMPparadigm}, the phenomenology of the WIMP interaction for the direct detection is described. In Section \ref{sec:direct_detectors}, a summary of the working principles of the direct detectors for DM is presented. Finally, Section \ref{sec:directional} discusses the features and advantages of the directional detection.
\section{WIMPs-nuclei scattering}
\label{sec:WIMPparadigm}
In the \W assumption, the Milky Way resides in a halo composed of these \emph{cold}, non relativistic, DM particles. The peculiar motion of the Sun around the centre of the Galaxy, along with the rotation of the Earth around the Sun, produces an apparent wind of DM particles coming from a preferential direction. This phenomenon boosts the energy of these DM particles from Earth's frame of reference and fosters the possibility of detectable interaction between \Ws and SM particles, in the assumption of the existence of an interaction between the two. The core idea of a direct detection experiment is to position a volume of material able to measure recoils of regular matter after a WIMP interaction occurs. The most common SM constituents available in a material are electrons and nuclei. DM can potentially interact with both of them, but large focus is on the nuclear scatter. Indeed, as \Ws are expected to possess a mass of $\mathcal{O}$(10) \Gevc, a better kinematic coupling is foreseen for nuclei than electrons. It should also be noted that electron recoils can be more easily generated by other interactions, such as $\gamma$ rays with electrons, than nuclear recoils.
\subsection{Expected event rate}
\label{subsec:WIMP_rate}
Following \cite{Gondolo_2002,LEWIN199687,SCHNEE_2011}, the expected rate per unit mass for an interaction between Galactic \Ws and atomic nuclei can be defined as:
\begin{equation}
\label{eq:rate1}
dR = \frac{N_0}{A_{mol}}\sigma v \frac{\rho_0}{m_{\chi}}f(\vec{v})d^3v,
\end{equation} 
with $N_0$ the Avogadro number, $\rho_0$ the DM density at Earth, $A_{mol}$ the molar mass of the atom, $\sigma$ the cross section between the \W and the nucleus, $m_{\chi}$ the mass of the \W, and $f(\vec{v})$ the distribution of the velocities of the \Ws in the Galactic Rest Frame, with the Galactic Rest Frame being the reference frame in which the Galactic centre is at rest. The differential rate in the recoiling nucleus momentum $q$ and solid angle $\Omega$ can then be obtained by integrating on all the possible velocities which can induce detectable signatures on the detector as: 
\begin{equation}
\label{eq:rate2}
\frac{dR}{dq^2d\Omega} = \frac{N_0}{A_{mol}}\frac{\rho_0}{m_{\chi}}\int_{v_{min}(q)}\frac{d\sigma}{dq^2d\Omega} vf(\vec{v})d^3v
\end{equation}
where the integral is performed on the domain of the velocity distribution of the DM particles above the minimum velocity, $v_{min}$, which is required to induce a recoil of momentum $q$. 
Equation \ref{eq:rate2} explicitly highlights how predicted rates assume a certain mass and scattering cross section, as well as a set of astrophysical parameters, material target and experimental performances. Each of these will be discussed in details in the following.
\subsection{DM halo}
\label{subsec:Astro}
The DM halo of the Milky Way is usually described by the Standard Halo Model (SHM) \cite{Standard_halo}. The SHM assumes an isotropic, isothermal sphere of DM particles with a density profile of $\rho(r)\propto r^{-2}$. Under the assumption of collisionless particles and isothermal system, the velocity density distribution can be expressed as a Maxwell-Boltzmann distribution:
\begin{equation}
\label{eq:fv}
f(\vec{v}) =
\begin{cases}
\alpha e^{-\frac{v^2}{v_p^2}} & \text{if  } |\vec{v}|<v_{esc}\\
0 & \text{if  }|\vec{v}|>v_{esc}\\
\end{cases}
\end{equation}
with $\alpha$ the normalisation factor and $v_p$ the most probable value of the velocity distribution close to the Earth and also represents the velocity dispersion of the \Ws. When assuming a perfectly flat rotation curve for the Galaxy, $v_p$ can be measured at various distances and is found to be about 230 km/s at the Earth radius \cite{bib:Baxter_2021,Reid_2019,Eilers_2019}. Yet, more precise calculations of the rotation curve predict a non perfectly flat Milky Way and, depending on its modelling, systematic uncertainties on the $v_p$ arise leading to a value span from 200 up to 279 km/s \cite{bib:Mayet_2016zxu}. While the Boltzmann distribution of the velocity extends mathematically up to infinity, extremely fast particles are bound to leave the Galaxy and escape, effectively posing an upper  limit on the available velocities for \Ws detection. In the SHM, this is implemented by manually truncating the velocity distribution at the local escape speed $v_{esc}$. This varies according to the distance $R$ from the centre of the Galaxy, following the amount of matter inside $R$. At the Solar position, 8.2 kpc from the centre, and assuming a smooth halo in equilibrium, the escape velocity $v_{esc}$ is measured by the Gaia telescope and RAVE survey \cite{Piffl_2014,Gaia_escape} and it is taken as 544 km/s.\\
Another interesting astrophysical parameter is the DM density at the Solar distance from the centre of the Galaxy, $\rho_0$. Despite some systematic uncertainties, the most commonly used value is 0.3 GeV cm$^{-3}$ c$^{-2}$ \cite{bib:Baxter_2021}. Recent measurements which followed the hypotheses of the Frenk Navarro White density profiles \cite{Navarro_1996} set constraints on the $\rho_0$ at slightly higher values, about 0.4 GeV cm$^{-3}$ c$^{-2}$ . Other local estimations which do not require mass-modelling of the Milky Way \cite{refId0}, are in agreement with 0.3 GeV cm$^{-3}$ c$^{-2}$, but with large uncertainties.\\
In the \W theoretical framework, the DM is cold, non relativistic. As a consequence, the speed of the DM particles is comparable to the peculiar velocity of the Sun and the Earth. For this to be taken into account when calculating the expected rate of interaction of \Ws with SM particles on Earth, a coordinate transformation is needed. A Galilean transformation is sufficient because the velocities are well below the relativistic limit, and the gravitational potential of the Sun and the Earth are weak enough to affect only \Ws with less than 40 km/s \cite{PhysRevD.37.2703}, which represent a very small fraction. In this transformation of coordinates, the laboratory velocity $v_{lab}$ is the sum of the Sun's velocity with respect to the Galactic Rest Frame and Earth's motion around the Sun. In some cases, the direct detector is approximated as positioned in the centre of the Sun, neglecting Earth's motion. While this is limiting for phenomena in the context of annual (see Section \ref{subsubsec:timesignature}) and daily modulation searches, it does not affect the recoil distribution or the rate above few percents \cite{Froborg_2020,bib:Mayet_2016zxu}. When only the peculiar velocity of the Sun is taken into account when evaluating the  laboratory velocity $v_{lab}$, it is valid that $v_{lab}= 242$ km/s \cite{bib:Baxter_2021}.\\
The SHM may not be the best approximation of the DM halo of the Milky Way as the isothermal and isotropic hypotheses are widely considered too simplistic. In cold DM cosmologies,
structures form hierarchically, leading to DM halos which are triaxial, anisotropic and contain phase-space substructure \cite{triaxialhalo}. The correct estimation of the DM halo is still debated with measurements and various simulation pointing at different structures and velocity distributions \cite{Gaia,halosim1,Pillepich_2014,Purcell_2009}. While the SHM might not be the best modelling of the real DM distribution of \Ws, it is generally used as benchmark for the computation of the \Ws rate of interaction with SM particles in the direct detection context. The guidelines for the astrophysical quantities regarding the halo model can be found in \cite{bib:Baxter_2021}.
\subsection{WIMP-nucleon elastic cross section}
\label{subsec:xsection}
The type and model of interaction between the \Ws and SM matter is unknown, hence an operative approximation can be performed by employing the Fermi's Golden rule. Thus, the factor $\frac{d\sigma}{dq^2}$ can be approximated splitting the dependence on the energy of the differential cross section in an independent term $\sigma_{WA}$ and another one, $S(q)$, which includes the entire dependence on the transfer momentum $q$, the form factor (which will be discussed later).
\begin{equation}
\label{eq:Fermi_rule}
\frac{d\sigma}{dq^2} = \frac{1}{\pi v^2}\vert\mathcal{M}\vert^2 = \frac{\sigma_{WA}S(q)}{4\mu_A^2 v^2}
\end{equation}
where $\mu_A$ is the reduced mass of the nucleus $A$ and the \W $\chi$, defined as:
\begin{equation}
\label{eq:red_mass}
\mu_A =\left(\frac{m_{\chi}m_A}{m_{\chi}+m_A}\right)
\end{equation}
The nuclei present in common materials only posses thermal energy and are non relativistic. At the same time, the SHM and the cold DM hypothesis need the \Ws to be non-relativistic particles. With the peak of the local \W velocity being $\sim$ 230 km/s, the De Broglie wavelength associated to the DM particle of roughly 10 \Gevc mass is:
\begin{equation}
\label{eq:debroglie}
\lambda_{DM}= \frac{h}{p}\sim  160 fm
\end{equation}
On the other hand, the typical nuclear dimension is of the order of few fm, at least one order of magnitude less than the $\lambda_{DM}$. As a consequence, it is possible to assume that the WIMP perceives the nucleus as a whole and interacts with it coherently. Moreover, due to the small expected energy transfers the scattering is considered elastic. 
While the actual nature of DM coupling is not known, it is possible to make a simplified standard assumption that it can possess spin-independent (SI) and spin-dependent (SD) interactions, which couple to the charge, the mass and spin of the nucleus, respectively. From this and \cite{Kurylov_2004}, it is possible to write a generic expression of the cross section as:
\begin{equation}
\label{eq:sigmawa}
\sigma_{WA}= \sigma_{WA,SI}+  \sigma_{WA,SD}
\end{equation}

The SI term can be parameterised as a function of the target nature as:
\begin{equation}
\label{eq:sigmawa_si}
\sigma_{WA}= \frac{4\mu_A^2}{\pi}\left[Zf_p+(A-Z)f_n\right]^2 
\end{equation}
Here $f_p$ and $f_n$ are the effective spin independent couplings to the proton ($p$) and neutron ($n$), while $A$ and $Z$ are the atomic mass and number of the nucleus. A reasonable assumption, supported by various models, is that WIMPs have the same coupling to proton and neutrons, resulting in $f_p$ = $f_n$. Thus:
\begin{equation}
\label{eq:sigmasi}
\sigma_{WA,SI} \simeq \frac{4\mu_A^2}{\pi}A^2f_n^2 = \sigma_{n,SI}\frac{\mu_A^2}{\mu_n^2}A^2,
\end{equation}
with $\mu_n$ the reduced mass of $m_{\chi}$ with the mass of a nucleon. This way, the cross section with a generic nucleus can be expressed as a function of the cross section with a nucleon $\sigma_{n,SI}$. The Equation also shows that the SI cross section is enhanced by a factor $A^2$, favouring heavy nuclei. This is consistent with the expectation of a coherent scattering where the contribution from all the nucleons are summed up together.

The SD cross section can be expressed as:
\begin{equation}
\label{eq:sigmawa_sd}
\sigma_{WA,SD}=  \frac{32G_F^2\mu_A^2}{\pi}\frac{J+1}{J}\left(a_p\left\langle S_p\right\rangle + a_n\left\langle S_n\right\rangle\right)
\end{equation}
with $a_p$ and $a_n$ the effective spin dependent couplings to the proton ($p$) and neutron ($n$), $J$ the nuclear spin, while $\left\langle S_p\right\rangle$ (and $\left\langle S_n\right\rangle$) is the expectation value of proton (neutron) spin inside the nucleus. The cross section strongly depends on the spin state of the nucleus. An asymmetry in the number of protons or neutrons is required to have a large expectation value of spin and therefore  sensitivity to SD interaction, as the contributions of the different nucleons generally cancel out\cite{Tovey_2000}. Table \ref{tab:spindep} summarises some of the elements with largest expected spin value, extracted from \cite{Tovey_2000}. Each of these nuclei is sensitive mainly to proton or neutron interaction depending on the which nucleon is odd in the nuclear structure.
\begin{table}[!t]
	\centering
	\begin{tabular}{|c|c|c|c|c|c|c|c|}
		\hline
		Nucleus & $Z$ & Odd Nuc. & $J$ & $\left\langle S_p\right\rangle^2$ & $\left\langle S_n\right\rangle$ & $\frac{4\left\langle S_p\right\rangle(J+1)}{3J}$ & $\frac{4\left\langle S_n\right\rangle(J+1)}{3J}$ \\ \hline
		$^{1}$H &1 & p & 1/2 & 0.500 & 0.0 & 1.0 & 0  \\
		$^{19}$F &9 & p & 1/2 & 0.447 & -0.004 & 9.1 $\times 10^{-1}$ & 6.4 $\times 10^{-5}$ \\
		$^{73}$Ge &32 & n & 9/2 & 0.030 & 0.378 & 1.5 $\times 10^{-3}$ & 2.3 $\times 10^{-1}$ \\
		$^{129}$Xe &54 & n & 1/2 & 0.028 & 0.359 & 3.1 $\times 10^{-3}$ & 5.2 $\times 10^{-1}$ \\ \hline
	\end{tabular}
	\caption{Table taken from \cite{Tovey_2000} summarising the spin nuclear characteristic of few of the most relevant nuclei in regards to the sensitivity to the SD interaction.}
	\label{tab:spindep}
\end{table}
As reported in \cite{SCHNEE_2011}, the SD cross section to proton and to neutron can be defined as:
\begin{equation}
\label{eq:sigmasd}
\sigma_{p,SD} = \frac{24G_F^2\mu_p^2a_p^2}{\pi} 
\end{equation}
$$\sigma_{n,SD} = \frac{24G_F^2\mu_n^2a_n^2}{\pi}.$$
Now, combining Equations \ref{eq:sigmawa}, \ref{eq:sigmasi}, \ref{eq:sigmawa_sd} and \ref{eq:sigmasd}, the total cross section for a nucleus containing an odd number of protons can be written as:
\begin{equation}
\label{eq:sigmawa2}
\sigma_{WA}= \sigma_{n,SI}\frac{\mu_A^2}{\mu_p^2}A^2 + \sigma_{p,SD}\frac{\mu_A^2}{\mu_p^2}\frac{4\left\langle S_p\right\rangle^2(J+1)}{3J}
\end{equation}
With respect to Equation \ref{eq:sigmawa2}, it is possible to notice that the SD term $\frac{4\left\langle S_p\right\rangle(J+1)}{3J}$, generally less than 1, is much smaller than the SI term. Thus, in case the $\sigma_{n,SI}$ is of the same order as $\sigma_{p,SD}$, for the majority of the targets the SI interaction dominates, and the SD component becomes negligible. For this reason, the experiments which set limits on the \W to nucleon cross section separate the SI scenario from the SD one. As a convention, the DM limit plots are put on the parameter space generated by $\sigma_{n,SI}$ and $m_{\chi}$ for the SI, whilst on $\sigma_{p,SD}$ ($\sigma_{n,SD}$) and $m_{\chi}$ for the SD coupling to the proton (neutron).
\subsubsection{Form factor}
\label{subsubsec:formfactor}
$S(q)$ is the nuclear form factor which carries the dependence of the differential cross section on the momentum transfer of the collision, and it depends on the structural form of the nucleus treated in the Fermi's approximation. This describes the nucleus in a spherical symmetry with a radius $R_n$ and a skin thickness \cite{SCHNEE_2011}. The form factor for the SI takes into account the fact that with larger momentum transfers the nucleus appears less and less homogeneous and the interaction loses coherence. It can be evaluated as the Fourier transformation of the nucleus radius function found in \cite{Gondolo_2002} and results in:
\begin{equation}
\label{eq:SIformfact}
S(q)= \vert F(q)\vert^2 = \biggr\rvert \frac{9[\sin(qR_n)+qR_n\cos(qR_n)]^2}{(qR_n)^6}\biggr\rvert^2
\end{equation}
with the radius of the nucleus approximated as:
\begin{equation}
\label{eq:nuclearradius}
R_n \simeq\left[0.91A^{1/3}+0.3\right] fm.
\end{equation}
The form factor for SD interactions embeds corrections due to the spin structure function, and thus its calculation results much more complex\cite{SCHNEE_2011}. Nonetheless, at first order it can be expressed as:
\begin{equation}
\label{eq:SDformfact}
S(q)= \vert F(q)\vert^2 = \biggr\rvert \frac{\sin(qR_n)}{qR_n}\biggr\rvert^2.
\end{equation}
\subsection{Kinematic}
\label{subsec:kinematics}
\begin{figure*}[t]
	\centering
	\includegraphics[width=0.8\textwidth]{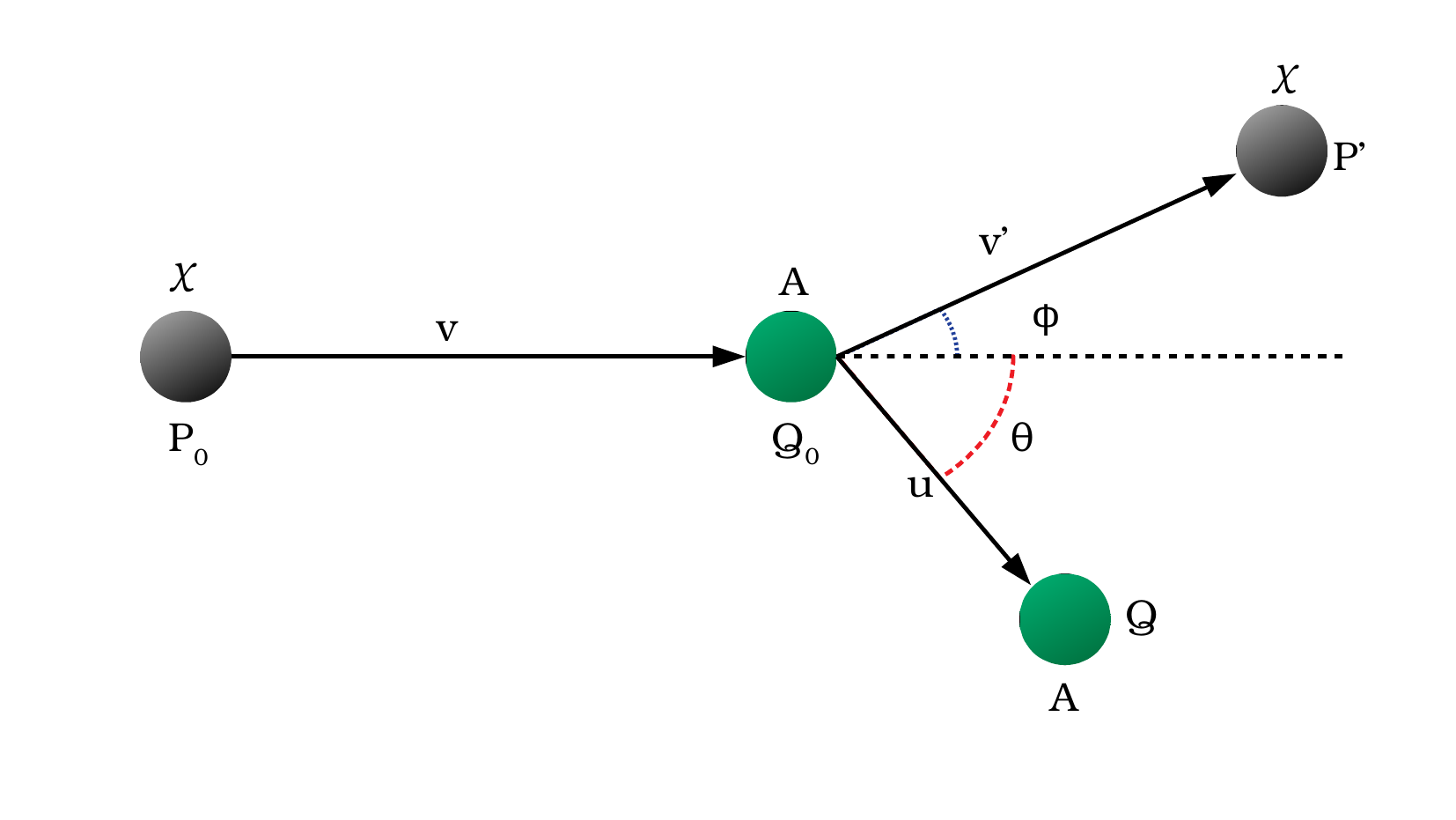}
	\caption{Schematics of the scatter between a DM particle $\chi$ with momentum $p_0$ and a nucleus $A$ at rest. The nucleus scatters with momentum $q$ at an angle $\theta$.}
	\label{fig:kinematics_scheme}
\end{figure*}
The kinematics of the elastic scattering between two particles can be evaluated in general terms as a relativistic collision. Under the SHM assumption, \Ws are expected to have an average velocity of the order of 10$^{-3}$c, whilst the thermal energy of a nucleus of a $\mathcal{O}$(10) GeV/c$^{2}$ at 300 K is about 10$^{-5}$c. Therefore, as displayed in Figure \ref{fig:kinematics_scheme}, it is reasonable to consider a DM particle $\chi$ with a momentum $p_0$ and velocity $v$ which collides with a nucleus $A$ at rest. The latter recoils with momentum $q$ at an angle $\theta$ as a result of the elastic collision. The laws of conservation of energy and momentum require that:
\begin{equation}
\label{eq:conservation}
P_0^{\mu}+Q_0^{\mu} = P'^{\mu} + Q^{\mu}
\end{equation}
The 3-dimensional momentum $q$ of the recoiling nucleus can be evaluated in natural units as:
\begin{equation}
\label{eq:relativ_kinematics}
q = \frac{2 m_A (\sqrt{p_0^2 + m_{\chi}^2} + m_A) p_0 \cos\theta}{(\sqrt{p_0^2 + m_{\chi}^2} + m_A)^2 - p_0^2 \cos^2 \theta}.
\end{equation}
Given the non-relativistic nature of the particles involved in the \Ws to nuclei scattering, the following approximation is valid $m_{\chi}, m_A >> p_0$. Thus, Equation \ref{eq:relativ_kinematics} can be reduced to:
\begin{equation}
\label{eq:nonrelativ_kinematics}
q = 2v\mu_A\cos\theta,
\end{equation}

Since the expected \W mass spans over several orders of magnitude, the efficiency in transfer momentum to the nucleus depends on the masses of the two particles. The maximum energy the recoil can attain is:
\begin{equation}
\label{eq:max_energy}
E_{max}=\frac{q_{max}^2}{2m_A}=\frac{2\mu_A^2v^2}{m_A}\equiv r\frac{1}{2}m_{\chi}v^2,
\end{equation}
where $r$ is the adimensional efficiency in the momentum transfer defined as 
\begin{equation}
\label{eq:r}
r =4\frac{m_{\chi}m_A}{(m_{\chi}+m_A)^2}.
\end{equation}
When $m_A=m_{\chi}$, the efficiency is maximal, $r=1$. This shows how the maximal sensitivity of a given target is for WIMPs of the same mass of the atoms composing the target. For instance, to optimise the sensitivity to \W masses close to 1 \Gevc, light targets as helium or hydrogen represent the best choices.\\
Another relevant relation, is the minimum velocity a \W needs to have in order to cause a recoil of energy $E$.  Indeed, relative to the angle $\theta$ of scattering, the same $E$ can be induced by different $v$ of a DM particle. The minimum velocity required to impinge a recoil energy $E$ is:
\begin{equation}
\label{eq:minvel}
v_{min}=\frac{\sqrt{2Em_A}}{2\mu_A}.
\end{equation}

\subsection{Experimental signatures}
\label{subsec:ratecalculation}
Combining the expression derived in the Subsection \ref{subsec:kinematics}, which binds the angle of the recoil to its energy, with Equation \ref{eq:rate2}, one obtains:
\begin{equation}
\label{eq:rate2.2}
\frac{dR}{dq^2d\Omega}  = \frac{N_0}{A_{mol}}\frac{\rho_0}{m_{\chi}} \frac{d\sigma}{dq^2}\frac{1}{2\pi}\int \delta\left(\cos\theta-\frac{q}{2\mu_A v}\right) v f(\vec{v})d^3v,
\end{equation}
This way, the double differential cross section is expressed in terms of the differential cross section on momentum, while the dependence on the angular part is made explicit in the Dirac's delta.\\
Introducing the expression for the SI cross section discussed in Subsection \ref{subsec:xsection}, the expected rate for SI interactions per unit of mass for a mono-atomic molecule becomes:
\begin{equation}
\label{eq:rate3}
\frac{dR}{dEd\Omega} = \frac{2N_0}{A_{mol}}\frac{\rho_0\sigma_{n,SI}S(E)}{m_{\chi}^2r}\frac{\mu_A^2}{\mu_n^2}A^2   \frac{1}{2\pi}\int \delta\left(\cos\theta-\frac{q}{2\mu_A v}\right) \frac{f(\vec{v})}{v}d^3v .
\end{equation}
It is now useful to define the integral on the velocity distribution, which is independent on the type of interaction between \W and nucleus, as:
\begin{equation}
\label{eq:rate_I}
I= \frac{1}{2\pi}\int \delta\left(\cos\theta-\frac{q}{2\mu_A v}\right) \frac{f(\vec{v})}{v}d^3v .
\end{equation}

A possible method to solve the integral of Equation \ref{eq:rate_I} is to rely on the Radon transform \cite{Gondolo_2002, Radon_dean,Radon_radon}, that takes into account the proper coordinate transformation to express the Equation \ref{eq:rate3} in the laboratory rest frame.
The velocity employed for the change of  reference frame is $\vec{v}_{lab}=-v_{lab}\hat{u_z}$, with $\hat{u_z}$ defined in the opposite direction of the motion of the laboratory in the Galactic rest frame. Following the calculation in Appendix \ref{app:calcrateWIMP}, the double differential cross section for a \W scattering as a function of the recoil energy $E$ and the angle $\gamma$ between the recoil direction and $\vec{v}_{lab}$ can be expressed as:
\begin{eqnarray}
\label{eq:rateWIMP}
\frac{dR(t)}{dEd\cos\gamma} &= \frac{ N_0}{A_{mol}}\frac{2\rho_0\sigma_{n,SI}S(E)}{m_{\chi}^2r}\frac{\mu_A^2}{\mu_n^2}A^2 \pi \frac{v_p^3}{v_{lab}(t)}\alpha'\times&    \\
&\times \left(e^{-\frac{\left(\frac{\sqrt{2m_AE}}{2\mu_A}-v_{lab}(t)\cos\gamma\right)^2}{v_p^2}} - e^{-\frac{v_{esc}^2}{v_p^2}}\right)&\Theta\left(\cos\gamma -\frac{\frac{\sqrt{2m_AE}}{2\mu_A}-v_{esc}}{v_{lab}(t)}\right) \nonumber,
\end{eqnarray}
with $\alpha'$ the normalisation factor of the velocity distribution, and where a dependence on the time $t$ has been added to the $v_{lab}$ as it includes the component of the Earth's velocity. The role of $v_{esc}$ is critical as it basically limits the integral for large velocities, indirectly restricting the energy a recoil can possess. Indeed, for each DM mass, there is a maximum energy $E_{max}$ which can be induced to the nucleus $A$ when a \W is at the maximum velocity allowed:
\begin{equation}
\label{eq:maxenergyrec}
E_{max}= \frac{1}{2}m_{\chi} r (v_{lab}(t)\cos\gamma +v_{esc})^2 
\end{equation}
This Equation shows explicitly how the energy threshold of an experiment effectively limits the mass parameter space that can be studied depending on the mass of the target, as a $m_{\chi}$ will be undetectable if $E_{max}$ is below the threshold. Overall, the differential rate is a function of three variables which can be experimentally measured: recoiling energy $E$, time $t$, and angle of scatter $\gamma$.
\subsubsection{Recoil rate energy dependence}
\label{subsubsec:energysignature}
\begin{figure*}[!t]
	\centering
	\includegraphics[width=0.5\textwidth]{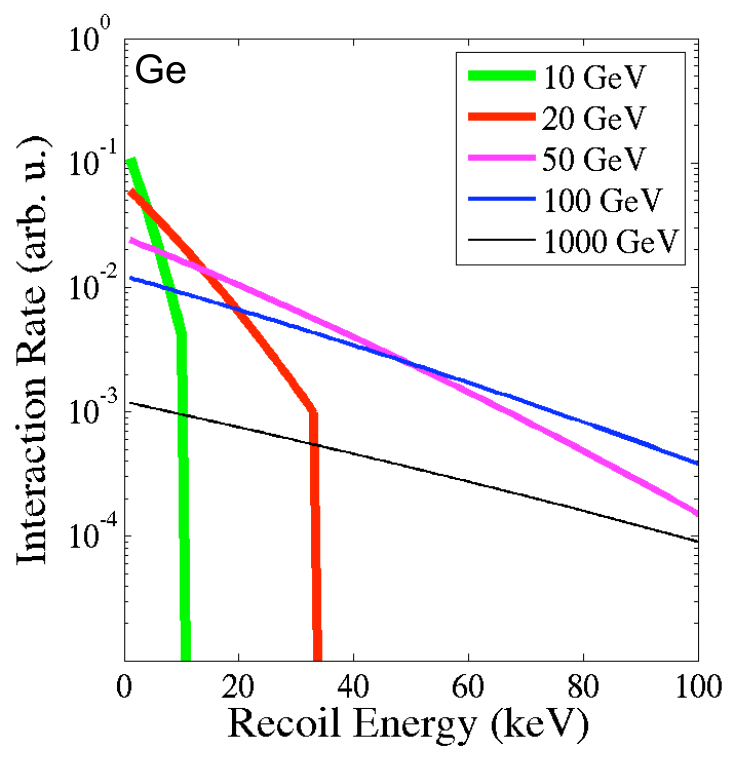}
	\caption{The energetic recoil spectra of a Ge target for different \W masses, taken from \cite{SCHNEE_2011}. With lower \W mass the spectrum is sharply cut at high energies due to the escape velocity effect on the recoil maximum energy.}
	\label{fig:Gespectr}
\end{figure*}
The expected energy spectrum of WIMP induced nuclear recoil, as a function of the WIMP mass, is obtained from Equation \ref{eq:rateWIMP} by averaging on time the $v_{lab}$ and integrating over the $\cos\gamma$ domain from -1 to 1. The spectrum is limited at high energies by the $E_{max}$ and in intensity by the dumping effect of the form factor, especially for large $A$ targets. The energy distribution is constrained at low values by the energy threshold $E_{thr}$ which depends on the characteristics of the specific detector employed for the direct search.
At first order of approximation, the energy spectra have a falling exponential behaviour without any particular feature nor peak. Figure \ref{fig:Gespectr} from \cite{SCHNEE_2011} shows examples of the energetic recoil spectra of a Ge target for different \W masses. The effect of the escape velocity on the spectrum is highly visible as with lower \W mass it results sharply cut at high energies.  
\subsubsection{Recoil rate temporal dependence}
\label{subsubsec:timesignature}
The motion of the Earth around the Sun modifies the relative velocity of \Ws with respect to the target nuclei during the year. In December, in the northern hemisphere, Earth's orbital velocity has a component anti-parallel to the motion of the Sun around the Galactic centre, and parallel during June.  This effectively induces a temporal dependence in the expected interaction rate of a few percent over one year, since the orbital velocity about 30 km/s. Thus, a diverse strategy to seek for \W signature is to exploit this annual modulation, which can be expressed as \cite{RevModPhys.85.1561}:
\begin{equation}
\label{eq:annualmod}
\frac{dR}{dE} \simeq R_0 + R_m\cos(\omega(t-t_0))
\end{equation}
where the differential rate depends on the time averaged signal $R_0$ and the modulated one $R_m$ whose dependence on time is expressed by a sinusoidal function of frequency $\omega= 2\pi/$year and by the phase of the modulation $t_0$. The observation of an annual modulation of the nuclear recoil rate in the detector could therefore be used to test the DM hypothesis. In order be sensitive to this signature, an experiment needs a large exposure with a highly stability in the response over an $\mathcal{O}$(1) year of time.
\subsubsection{Recoil rate angular dependence}
\label{subsubsec:anglesignature}
Earth's co-motion with the Sun introduces a directional dependence of the rate, that peaks in the mean radial direction of the Sun (or else roughly towards the Cygnus constellation), with a diurnal temporal modulation due Earth's rotation around its own axis. The angular spectrum of the recoils can be obtained by averaging on time the $\vec{v}_{lab}$ and integrating over all the range of energies $\left[E_{thr},E_{max}\right]$. From Equation \ref{eq:rateWIMP}, the most relevant term for the angular dependence can be extracted:
\begin{equation}
\label{eq:angsimpl}
\frac{dR}{d\cos\gamma} \propto \int_{E_{thr}}^{E_{max}}e^{-\frac{\left(v_{lab}\cos\gamma-v_{min}\right)^2}{v_p^2}},
\end{equation}
\begin{figure*}[!t]
	\centering
	\includegraphics[width=0.4\textwidth]{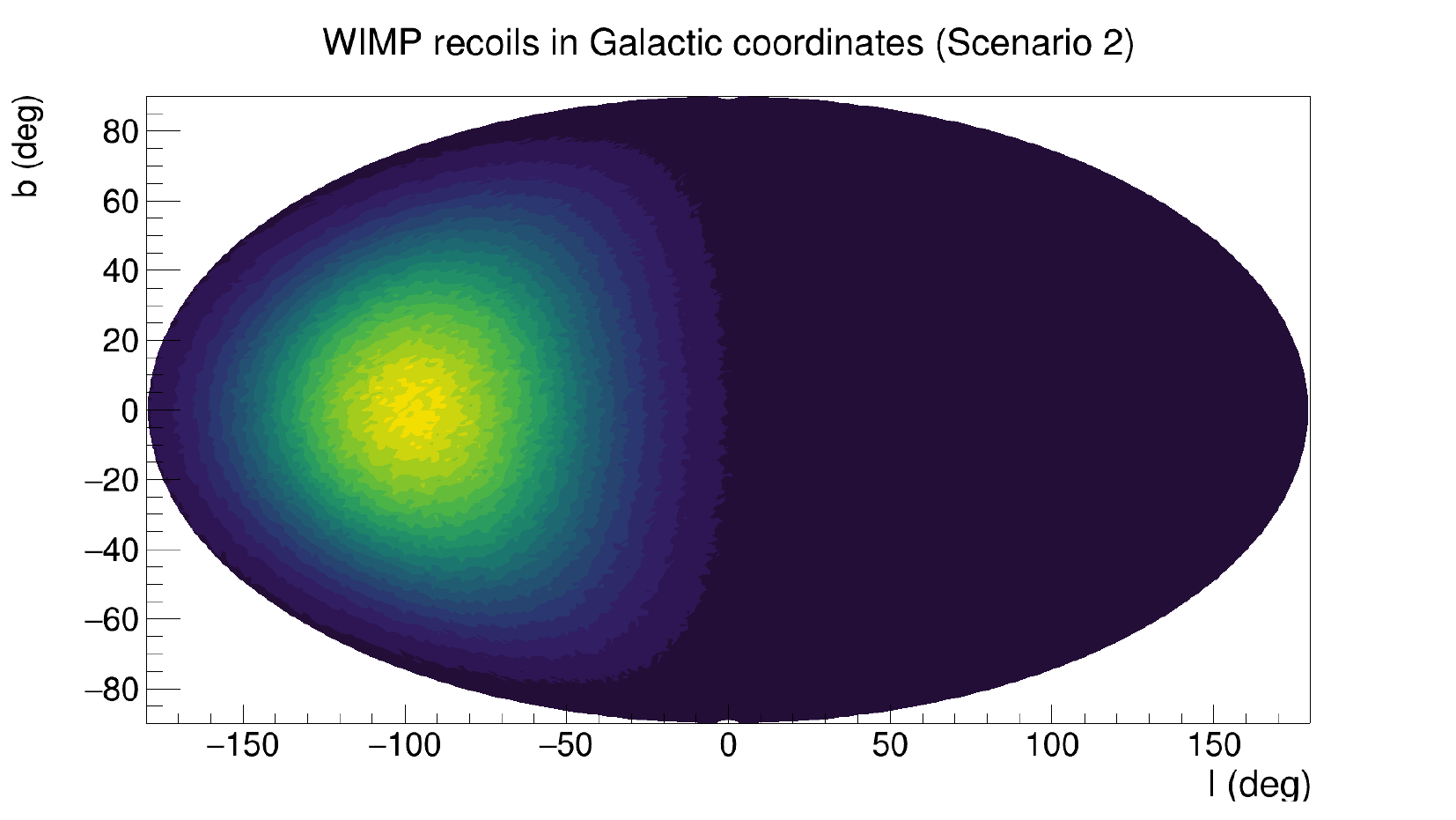}
	\includegraphics[width=0.4\textwidth]{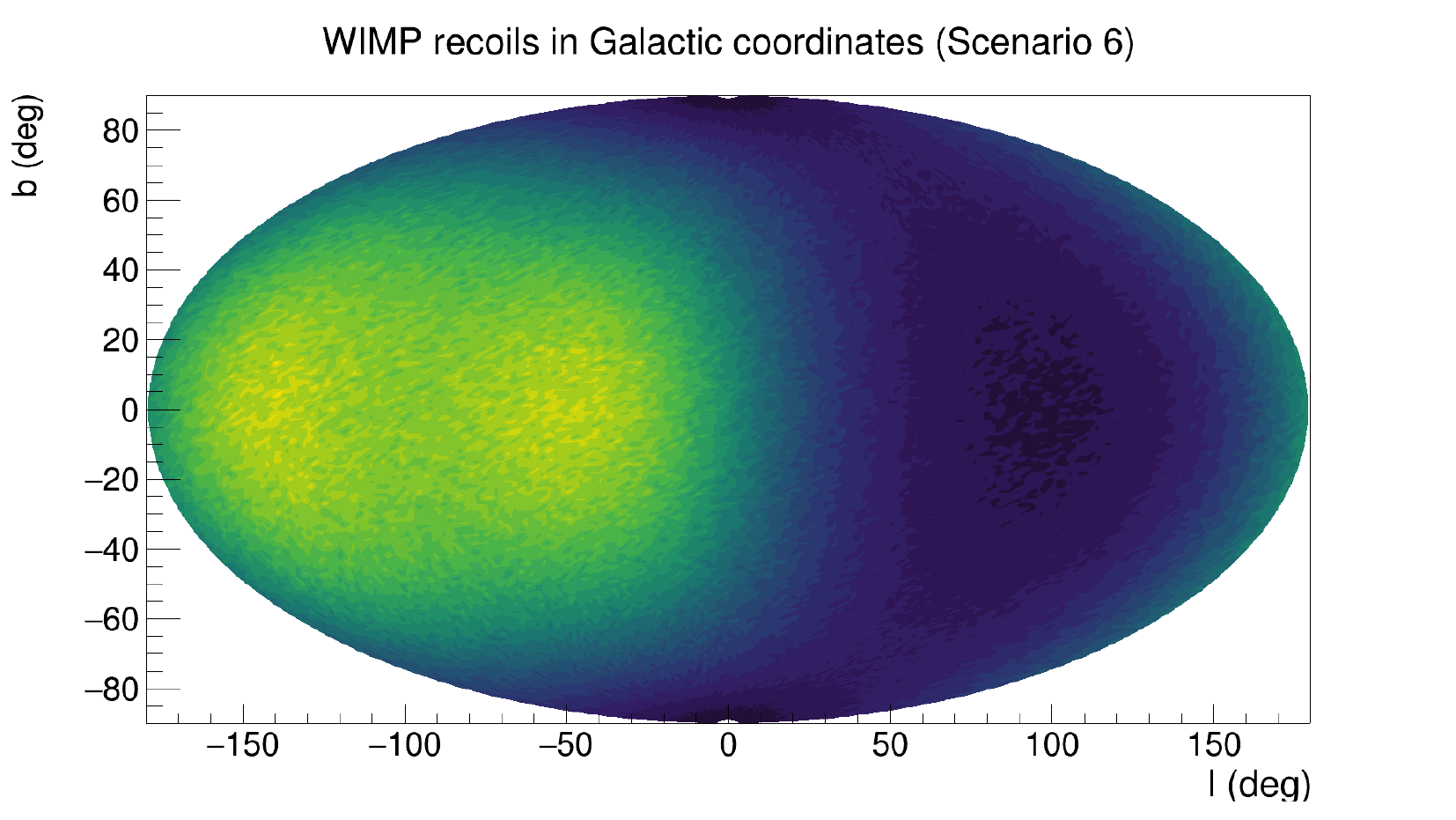}
	\caption{Examples of angular distribution of the recoils induced by \W interaction calculated following the SHM assumptions (for more details see Section \ref{sec:discr_2models}). On the left, the distribution is evaluated using a \W mass of 10 \Gevc on fluorine target which exhibits the typical dipole structure. On the right, the \W mass used is 100 \Gevc with xenon target. In this case the recoil follow the ring-like shape due to the masses combination.}
	\label{fig:example_angul} 
\end{figure*}
where the dependence of the exponential on the $\cos\gamma$ remarks the anisotropic nature of the angular distribution of the recoils, which is clear once displayed in Galactic coordinates. The Galactic coordinate system is a right-handed celestial coordinate system in spherical coordinates centred on the Sun. The x-axis points at the centre of the Milky Way, while the y-axis approximately is parallel to the peculiar motion of the Sun with respect to the Galactic Rest Frame. This way, the Galactic plane corresponds to the equatorial plane of the coordinate system. The coordinates employed are the Galactic longitude $l$ and the Galactic latitude $b$, whose behaviour mirrors the one of terrestrial longitude and latitude. In this coordinate system, the constellation of Cygnus is located at $(l,b) \simeq (81, 0.5)$. Examples of the angular distribution of recoils in Galactic coordinates are shown in Figure \ref{fig:example_angul} (for more details on the plot assumptions, see Section \ref{sec:discr_2models}). These clearly display the anisotropic nature of the angular distribution of WIMP-induced recoils, with an excess of events at negative longitude, opposite to the motion of the Sun. This signature present a directional correlation with an astrophysical source that no background whatsoever can mimic (also thanks to the rotation of Earth's rotation around its own axis) and can hence offer the possibility for a positive claim of a DM signal.
A more detailed description of the features of the angular distribution of the recoils, along with the advantages of the directional search will be presented in Section \ref{sec:directional}.
\section{Experimental challenges and detection techniques}
\label{sec:direct_detectors}
The direct detection experiments search for signatures which are induced by a DM interaction. A comprehensive report of the direct detection approach and status can be found in \cite{Billard_2022}. Taking into account the kinematic at play and the constraints from Equation \ref{eq:maxenergyrec}, the expected recoil energies for \W interactions are of the order of few keV, while the rate, estimated from Equation \ref{eq:rateWIMP} and current limits, is foreseen to be below the $\mathcal{O}$(1) event kg$^{-1}$ y$^{-1}$, making the direct detection a search for very rare events. In addition, many sources of backgrounds can mimic DM induced signal, further complicating its quest. Cosmic rays and natural environmental and material radioactivity produces electromagnetic radiation with rates typically 10$^6$-10$^8$  or more times larger than an expected WIMP signals, and need therefore to be properly passively shielded or actively suppressed in the analysis.  Neutrons and neutrinos from ambient and space, moreover, induce a detector response nearly identical to WIMPs, unless some additional topology or directional handle is employed for further discrimination. Direct DM searches experimental techniques are based on the possibility of measuring the energy deposited by a particle interacting with the instrumented target material in the form of ionisation, scintillation or lattice vibrations (i.e. heat). Most of the experiments measure more than one of these channels, a feature that allows to distinguish an electron recoil (ER) from a nuclear recoil (NR), due to the different energy release mechanism of such particles. In large detectors with heavy dense targets, fast neutrons with $\mathcal{O}$(1) cm mean free path can be suppressed by looking for multiple scatterings or defining an internal fiducial volume, at the price of reducing the active material sometimes even by 50\%. Neutrinos, conversely, can not be effectively shielded and can induce NRs by means of coherent scattering with nuclei (CE$\nu$NS)\cite{Monroe_2007} with an energy spectrum very similar to NRs expected from WIMP interaction.\\
As a consequence, the knowledge of the backgrounds and its minimisation are of the utmost importance for the success of a direct DM experiment. Ideally, an experiment aims at the complete suppression of any form of interaction that could resemble a DM signal, by either minimising the possible sources or by actively suppressing this contribution in the experiment setup or analysis. Therefore, their goal is to meticulously detail the typical features of background events in order to be able to properly account for their contribution at the analysis stage.
\subsection{Background sources}
\label{subsec:background}
The rare nature of \W induced recoils requires the detector to be in a controlled environment where these key signatures of DM interaction can be efficiently detected. The flux of secondary cosmic rays coming from the interaction of high energetic nuclei from outer space in the atmosphere can be as high as $\sim$ 100 particles per m$^2$ per s. This stream is for the majority composed of muons with energies ranging from MeV up to hundreds of GeV, which represent a quite penetrating radiation. Immense shielding facilities would be required to suppress this form of background. As a consequence, all the direct detection experiments are located in underground laboratories, below at least hundreds of metres of rock to strongly suppress the cosmic flux and reduce this background. 
Typically the underground facilities are tunnels dug in mountains or mines. Some examples are the Laboratori Nazionali del Gran Sasso (LNGS), and the Boulby mine among others.

\begin{figure*}[t]
	\centering
	\includegraphics[width=1\textwidth]{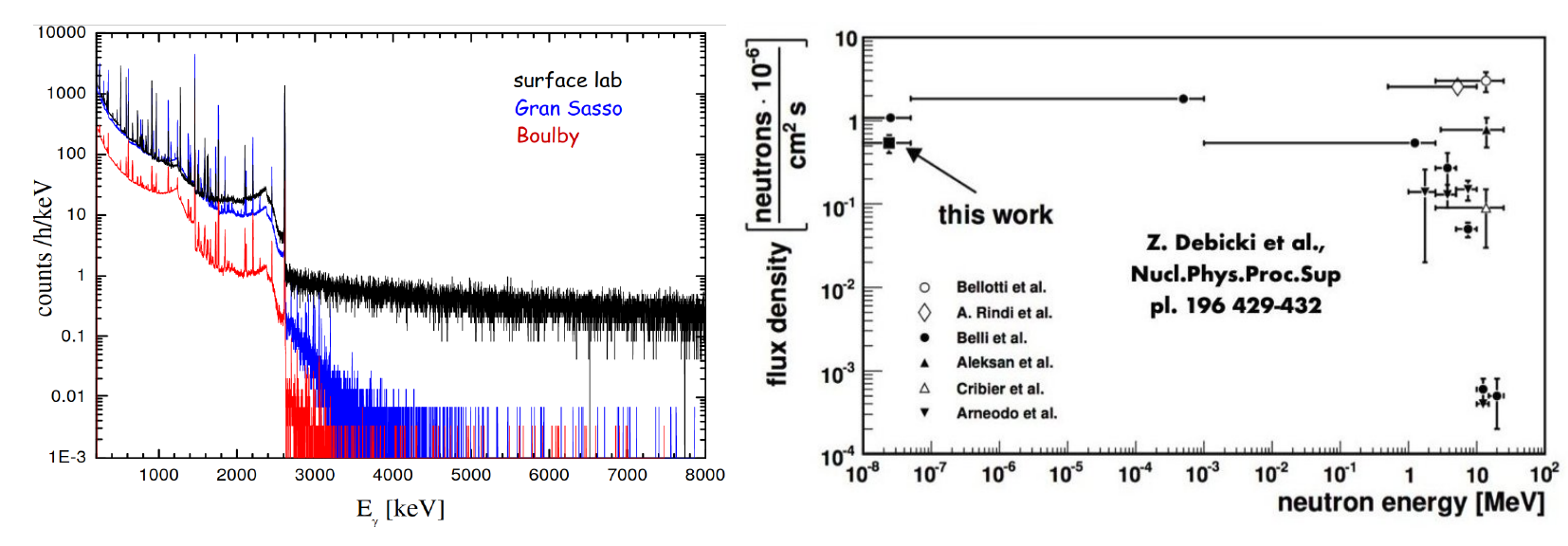}
	\caption{On the left, the $\gamma$ spectrum above ground and in two underground sites as a function of energy. On the right, the LNGS underground environmental neutron flux \cite{DEBICKI2009429}.}
	\label{fig:muons}
\end{figure*}

Underground facilities, while protected from cosmic rays, present nonetheless a natural environmental radioactivity produced by the surrounding rocks. The radioactive chains of thorium and uranium are present in almost every rock including the natural bed of rocks of the underground sites. These chains include a wide variety of elements and due to the secular equilibrium, even the isotopes with very short half-life are important contributors. Among them, the largest contributions come from $^{238}$U, $^{234}$Th, $^{232}$Th, $^{226}$Rn, $^{214}$Pb, $^{214}$Bi, $^{227}$Ac, $^{220}$Rn, $^{212}$Pb, and $^{212}$Bi. In addition, also the $^{40}$K is a radioactive element often present, especially when human activity is involved. These isotopes can decay in diverse ways with the emission of alpha particles, electrons, positrons or $\gamma$s. Alphas, positrons and electrons can be more easily stopped, as their energy loss is continuous, and generally do not exit from the rock itself. However, $\gamma$s, mostly coming from the de-excitation of a nucleus after a radioactive decay, can cross the rock unaffected and reach the active volume of an experiments, where they induce ERs. Figure \ref{fig:muons} on the left panel shows a measurements of the gamma induced background above ground and in two underground sites. A difference in the environmental gamma spectrum is visible and is due to the different rock composition of the two laboratories, Dolomite calcareous rock for the LNGS and salt and potash for the Boulby mine.
A non negligible amount of neutrons can also be emitted by natural rocks which can induce NRs. The origin of the emission of these neutrons can be separated in two categories: radiogenic, when caused by radioactivity, and cosmogenic, when induced by muons. In the former case, they can be emitted following a natural fission for very high $A$ elements, or by means of an ($\alpha$,n) reaction, when an alpha particle interacts with a high $Z$ material with the resulting emission of a neutron \cite{osti_15215}. The cosmogenic neutrons instead are produced by spallation processes subsequent to a muon collision\cite{Kudryavtsev2008}.\\
To suppress the amount of radiation that reaches the sensitive part of the detectors, underground experiments are also equipped with shielding material. The shields can be passive, when it is used merely for its stopping power properties, or active, when it is instrumented and signals can be obtained from the energy deposited in the shield material to characterise the external background and actively suppress it through vetos. Copper and lead are often used as passive shielding with high $Z$ to stop $\gamma$s and $\beta$s, whilst materials rich in hydrogen, such as water or polyethylene, are employed to slow down and absorb neutrons thanks to the large kinematic match with H. In addition, scintillators instrumented with photon detectors and water Cherenkov detectors are possible solutions for  active shielding configurations \cite{Westerdale_2016}.

Electromagnetic and neutral backgrounds can also come from the material the shielding and the detector themselves are manufactured with, and for this reason, they are typically called \emph{internal} backgrounds. As this is much harder to block due to the vicinity to the sensitive volume, the best approach is to suppress it as much as possible enhancing the radiopurity of the materials employed \cite{Schumann_2019}. 
Along with the same $^{40}$K and the chains of $^{238}$U, $^{235}$U and $^{232}$Th (which are always present as contaminants), other typical contaminating isotopes are $^{60}$Co, $^{57}$Co, $^{58}$Co and $^{54}$Mn from the activation of the copper and iron utilised as shielding and structural materials due to muon-induced neutrons interactions \cite{osti_1056763}.\\
Background can also come from the progeny of radon deposited on the internal structure of the detector. $^{222}$Rn is a radionuclide which belongs to the $^{238}$U chain and is an inert gas. As a consequence, radon can escape the rocks and be present in the air of the underground facility, from which it can easily penetrate inside the experiment structure, even in the internal layers close to the sensitive volume. When radon decays into $^{218}$Po, the unstable positively charged ion $^{218}$Po$^+$ (80\% of times) is sensitive to electric fields and can therefore deposit on detector electrodes and decay with the emission of alphas. The geometry of this process implies a high probability for the alpha to be completely embedded in the detector material, leaving only the $^{218}$Po recoiling in the fiducial volume and mimicking a WIMP interaction. In order to reject these events, the full 3D position of the events needs to be reconstructed (a technique usually referred to as “fiducialisation”), something not necessarily trivial for experiments in the DM search field, since the absolute time of interaction is not always known.
\begin{figure*}[t]
	\centering
	\includegraphics[width=0.6\textwidth]{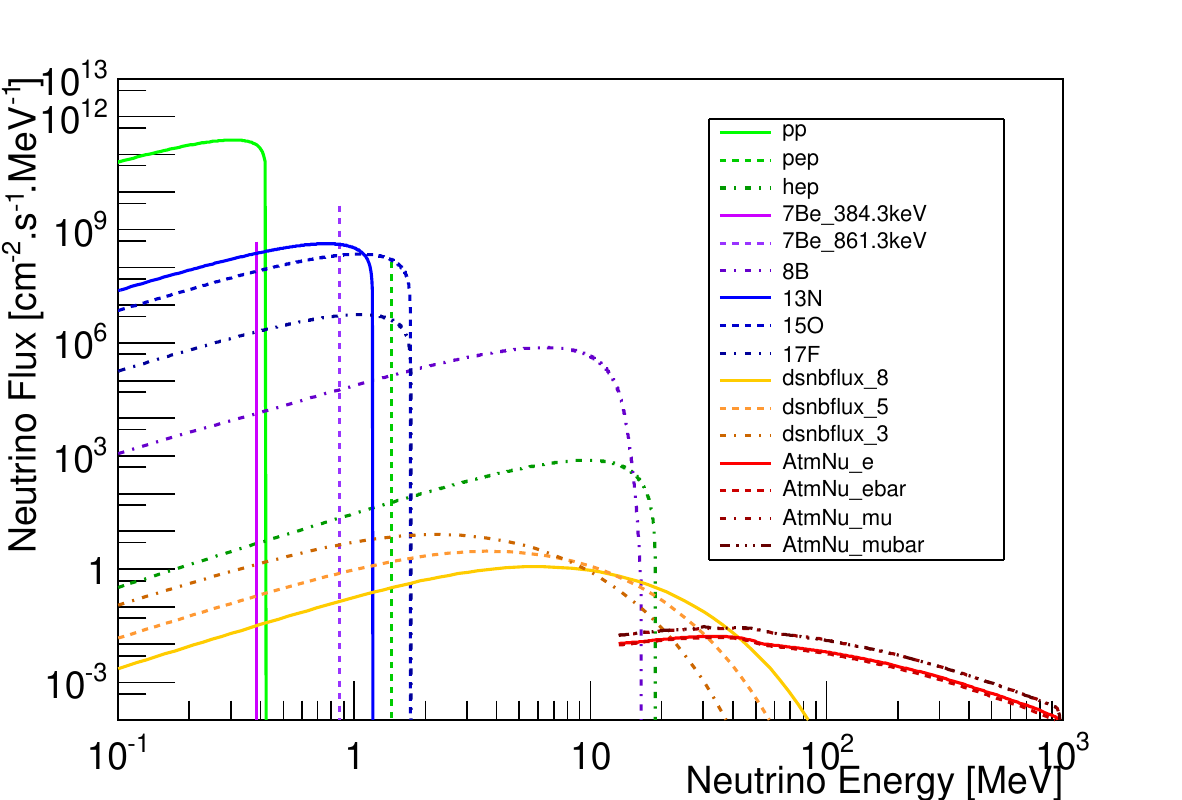}
	\caption{Neutrino energy spectrum at Earth, integrated over directions and summed over flavours for energies above 100 keV. Figure taken from \cite{Billard_2014}.}
	\label{fig:spectrneu}
\end{figure*}

Neutrinos emitted by the Sun, produced in the atmosphere by cosmic rays and coming from the diffuse flux of the Supernovae are a critical sources of background \cite{Billard_2014}.   Figure \ref{fig:spectrneu} shows the energy spectrum of the neutrinos from these different astrophysical sources. The energies and fluxes span over several orders of magnitude. The most intense are the ones coming from the Sun, especially from the \emph{pp} cycle, which mostly induce ERs of the order of few keV, and $^8$B chains, which generate NRs of the same energies, impossible to be singularly discriminated from \W induced ones. Relevant contributions in form of NRs also come from the atmospheric flux. Current Xe-based experiments (and several next generation detectors) are starting to be sensitive to the $^8$B solar component \cite{Xenon_future}.\\
Experiments at the $\mathcal{O}$(1) tonne scale are reaching a sensitivity at which Solar neutrinos are bound to affect the improvements in the cross section-\W mass limit phase space. This makes the realisation and construction of larger detectors with the goal of improving the sensitivity for DM  highly problematic and not convenient. In the past years, this has been considered a nearly hard limit (hence denominated \emph{Neutrino Floor}) where DM signals become hidden underneath a remarkably similar-looking NR background, unless a very large statistic (possibly with multiple targets) \cite{Billard_2014} or directional information \cite{bib:Mayet_2016zxu, Vahsen:2020pzb} are employed. A more recent approach \cite{O_Hare_2021}, already well accepted by the international community \cite{bib:Snowmass2022}, has refined this definition from a \emph{floor} to a \emph{fog}, that is interpreted as:
\begin{center}
	\textit{"A region of the parameter space where a clear distinction between signal and background is challenging, but not impossible \cite{O_Hare_2021}."}    
\end{center}
\begin{figure*}[!t]
	\centering
	\includegraphics[width=0.9\textwidth]{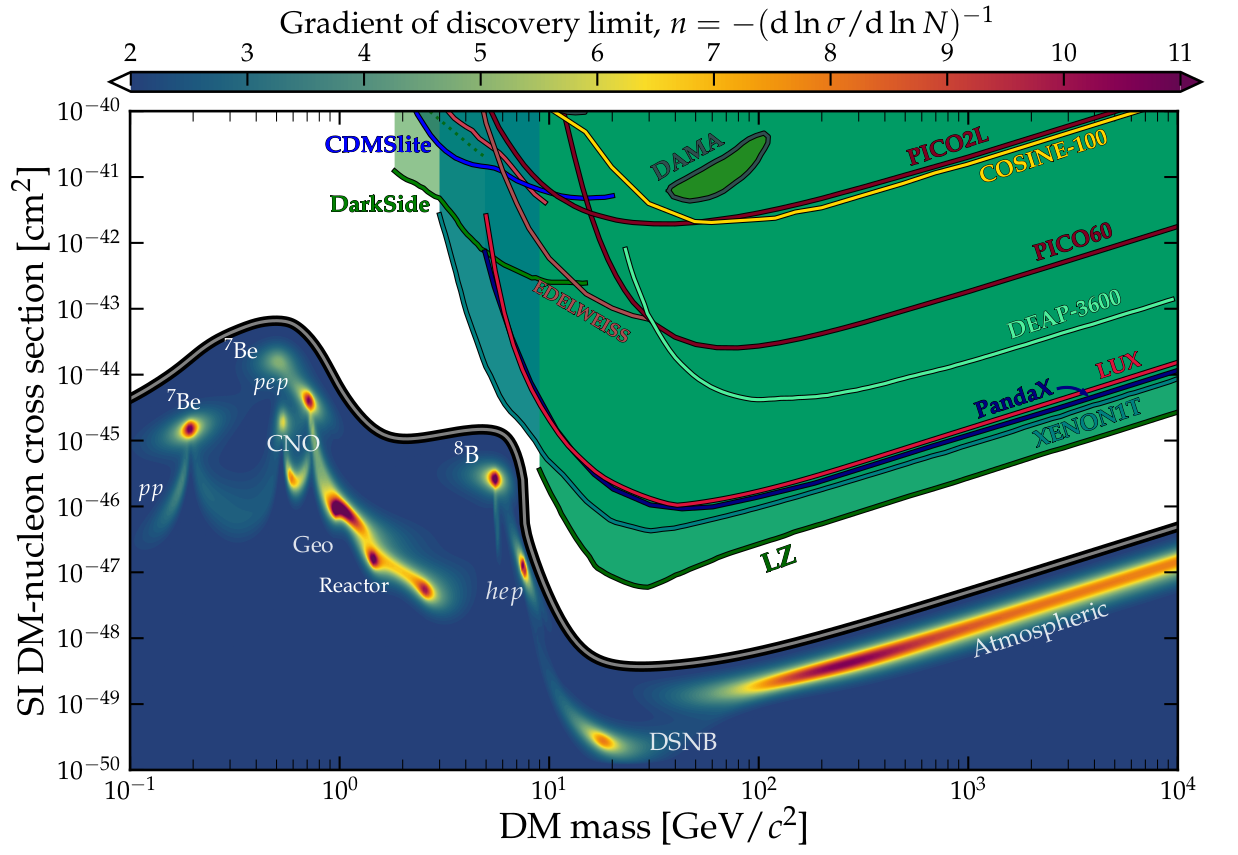}
	\caption{The neutrino fog of xenon shown along with some experimental limits on the SI \W to nucleon cross section. The colour intensity represents the $n$ index of logarithmic proportionality between the limits on the cross section and the number of CE$\nu$NS events detected. Figure extracted from \cite{O_Hare_2021}.}
	\label{fig:nufog} 
\end{figure*}
The new definition is based on the derivative of the discovery limit $\sigma$. This technique makes the \emph{Neutrino Fog} definition independent from the experimental exposure and energy threshold and highlights how the uncertainties of the various neutrino fluxes of Figure \ref{fig:spectrneu} and the shape of the recoiling spectra influence its \emph{opacity} in different \W masses and cross sections regions. In general, the discovery limit of an experiment improves when the exposure or the number of observed background neutrino events $N$ increases. Until the experiment stays background free, the limit on the cross section $\sigma$ evolves as $\sigma\propto N^{-1}$. With increasing $N$, the limit progresses into Poissonian background subtraction with sensitivity improving as $\sigma\propto 1/\sqrt{N}$. At some point, depending on the uncertainties in the background fluxes $\delta \Phi$, the limit gets stalled and becomes purely dominated by systematics as $\sigma\propto 1/\sqrt{(1 + N \delta \Phi^2 )/N}$.
More formally, let $n$ be the improvement in the cross section limit for SI with the increase of $N$, defined as \cite{bib:Snowmass2022}:
\begin{equation}
\label{eq:n_vsneutrino}
n = -\left(\frac{d\ln\sigma_{n,SI}}{d\ln N}\right)^{-1}.
\end{equation}
From what discussed above $n=1$ when an experiment has no background, whilst $n=2$ when the background can be Poissonianly subtracted. The recoils induced by neutrinos do not have exactly the same energy distribution as the ones cause by \W interaction and in large numbers can be statistically separated. Therefore, eventually, the limit can still emerge from saturation and return to a $\sigma\propto N^{-1}$ scaling, but only at an extremely large exposure not worth the cost-to-benefit investment. In this context, the \emph{Neutrino Floor} for each experiment can be consistently defined as the largest cross section for which $n$ crosses the value of 2.
Figure \ref{fig:nufog} shows the value of $n$, defined above as the index with which a discovery limit scales with the number of background events, for xenon targets, along with some experimental limits on the SI \W to nucleon cross section that will be discussed in Section \ref{subsec:status}. The colour intensity on $n$ stresses the regions of the parameter space where the neutrino events hinder more the DM search.\\
In spite of the fact that the discrimination between neutrinos and \Ws induced events is possible, a $n>2$ stresses that, employing the same detection technique, the improvement in the discovery limit does not justify the immense investments required to increase the exposure accordingly. Among the different approaches that can be used to minimise the limitations imposed by neutrinos, directionality appears to be the most effective \cite{O_Hare_2021,Ohare2,O_Hare_2015} and will be discussed in Section \ref{sec:directional}.
\subsection{Direct dark matter searches experimental techniques}
\label{subsec:status}
The experimental approaches to detect \W induced NRs are diverse and employ a wide range of technologies. These are based on the possibility of measuring the energy deposited in matter by a particle interaction through either ionisation, scintillation or heat. Since the energy partition into this three channels depends on the particle nature and energy (and target), the simultaneous measurement of more than one component provides discrimination between NRs and ERs. This is, in fact, related to the specific energy release mechanism of each particle type, which basically reflects the energy loss profile.\\

\begin{figure*}[t]
	\centering
	\includegraphics[width=0.9\textwidth]{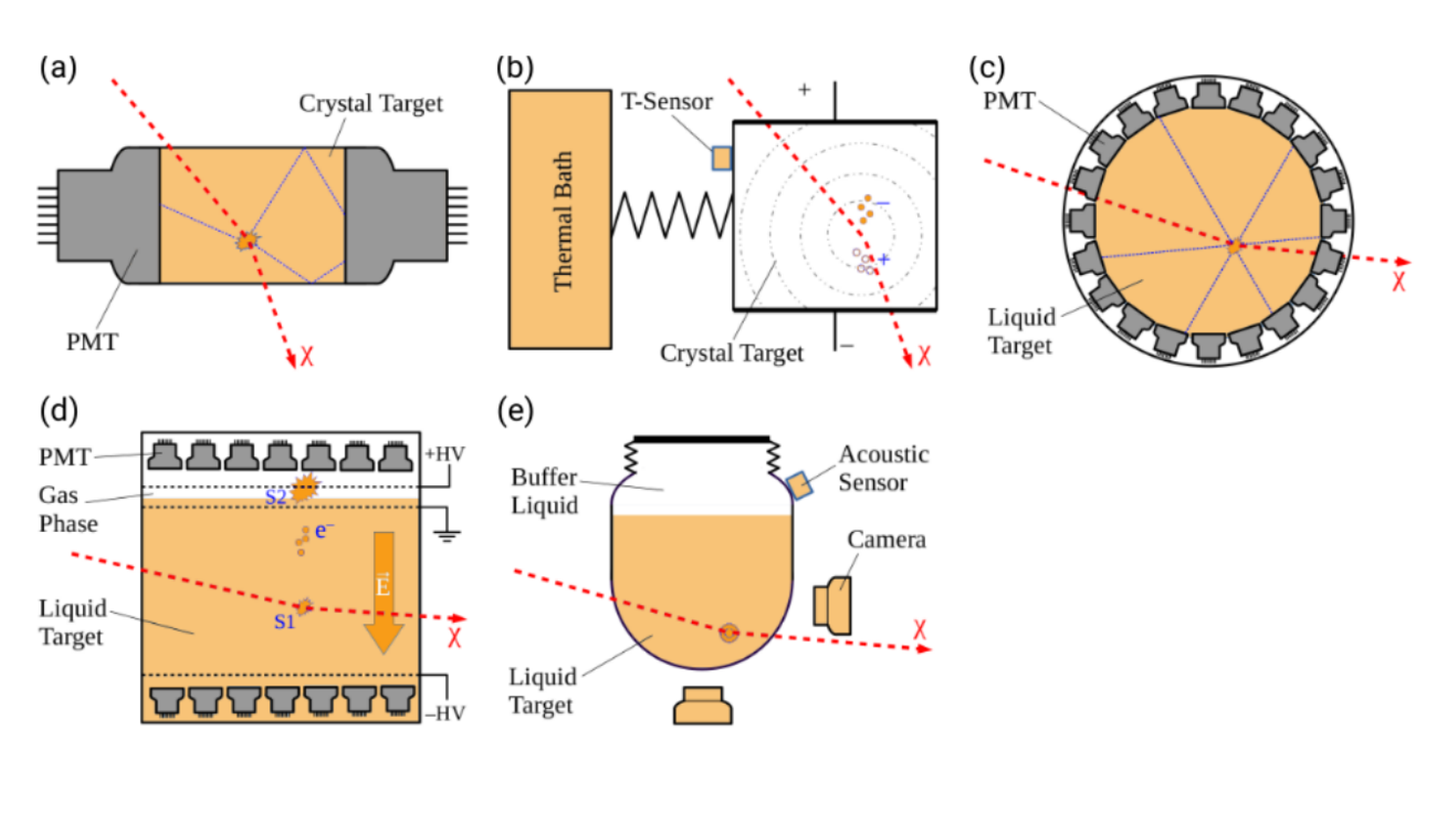}
	\caption{Schematics of the working principles of the most common detectors for direct searches: (a) scintillating crystal; (b) bolometer; (c) single-phase and (d) dual-phase liquid noble gas detectors; (e) bubble chamber. Figure taken from \cite{Billard_2022}.}
	\label{fig:detectors}
\end{figure*}

In some other cases, the response of the medium changes according to the recoiling particle. For example, the noble liquid scintillating properties allow a pulse shape discrimination. Indeed, NRs and ERs excite the atoms of the liquid differently, generating distinct populations of triplet and singlet states, whose average lifetime before decaying with the emission of photons strongly differs. Solid scintillators exploit the doping of additional elements which causes the generation of extra energy levels in the lattice, from whose decay light is emitted. Akin to noble liquids, electrons in these energy states have various lifetimes of decay and ERs and NRs populate these energy states with different lifetimes in distinctive ways. For imaging detectors, the local energy deposition can be taken advantage of, with NRs having larger and strongly different distributed $dE/dx$ along the track with respect to ERs. Finally, experiments can be built in such a way that the energy deposits of ERs do not induce a detectable response due to the low dE/dx.
Following \cite{Billard_2022}, a brief summary of some of these techniques is presented supported by Figure \ref{fig:detectors}. The experimental approaches most relevant for directional DM searches, namely nuclear emulsions and gaseous Time Projection Chambers (TPCs), will be illustrated in detail in the dedicated directional Section in \ref{sec:directional}.
\paragraph{Scintillating crystal}
Inorganic crystals like NaI(Tl) or CsI(Tl) convert energy deposited in the material into UV or optical photons, which can be readout by low background photomultiplier tubes (PMTs). They are known to have large light yield and their experimental implementation is demonstrated to be simple and stable for long periods of time. The presence of intermediate/high Z makes them sensitive to high \W masses in the SI search. Difficulties in the growth of these crystals with high radiopurity induces large background contribution, exception made for the DAMA experiment. ANAIS-112 \cite{Coarasa_2019}, DAMA/LIBRA \cite{Bernabei:2022xgg} are examples of scintillating crystal-based experiments (Figure \ref{fig:detectors}(a)).
\paragraph{Bolometers}
Bolometers are cryogenic detectors operated at extremely low temperature, usually below 50 mK, whose main detection channel is based on the exploitation of their large heat capacitance to measure the increase in temperature caused by the phonons produced by a recoil inside their volume. The excellent energy resolution, the energy threshold potentially at the eV scale, and the small impact of the quenching factor on the phonon production are the highlight parts of the bolometers. In order to reject the ER background component, bolometers are typically sensitive to another energy deposition channel, namely ionisation or scintillation. Experiments based on semiconductor crystals take advantage of the collection of the charge, thanks to the very small energy gap required for ionisation, whilst scintillating crystals profit from the light emission. Since the growth of these scintillating crystals with extremely low radioactivity is problematic, the exposure attained are still limited. Examples of experiments implementing bolometers are EDELWEISS-II \cite{G_Gerbier_2008}, CRESST-III\cite{Willers:2017vae}, COSINUS \cite{Angloher_2016}, and Super-CDMS \cite{SuperCDMS:2016wui} (Figure \ref{fig:detectors}(b)).
\paragraph{Noble liquid detectors}
Noble liquids like argon and xenon can be easily ionised and possess excellent scintillation properties. The excitation of atoms by the passage of an ionising particle causes the generation of a population of excited states in triplet and singlet states whose relaxation provokes the emission of UV photons with different decay times. The type of ionising particle affects the population of triplet and singlet states. Thus, experiments can measure the primary scintillation light by instrumenting a large volume of noble elements with light detectors. Good particle identification is obtained by analysing the shape of the scintillation signal. In liquid form at low temperatures these experiments can reach large masses of the order of tonnes. Examples of these single phase noble liquid detectors are DEAP-3600\cite{KUZNIAK2016340} or XMASS \cite{ABE201378} (Figure \ref{fig:detectors}(c)).\\
Noble elements can be used in dual phase TPCs in order to measure both the scintillation light and the ionisation charge. The primary light is collected by photo-sensors, S1 signal, whilst the charge is drifted by means of an electric field towards a grid that separates the liquid part from the gaseous one. Here, the primary ionisation electrons are strongly accelerated and thanks to electro-luminescence phenomenon their signal is amplified into photons and measured by optical sensors, S2 signal. From the combination of  S1 and S2 the energy deposited, particle identification and absolute position of the interaction can be determined. XenonnT \cite{Aprile_2022}, DarkSide \cite{Aalseth:2017fik} are examples of experiments based on this technology (Figure \ref{fig:detectors}(d)).

\paragraph{Bubble chamber}
Superheated liquids can be operated in metastable thermodynamic conditions wherein they are virtually insensitive to gammas or electrons. An energy deposition in this metastable state will result in nucleating a bubble that can grow to macroscopic scales and be optically detected. The higher the degree of superheat, the lower energy threshold for bubble nucleation. In addition, superheated fluids are uniquely sensitive to the local energy deposition, or stopping power dE/dx. Since ERs have a much smaller dE/dx with respect to NRs, they result extremely inefficient at nucleating bubbles, effectively becoming incapable to induce a signal in these kind of experimental approach.
This feature effectively implies a threshold both in energy and in energy density of the deposit, which allows to be completely insensitive to low energy ERs. Experiments based on this approach can not therefore measure the energy of an interaction and rely only on counting statistics for the \W search, being unable to characterise the DM mass in case of discovery. PICO \cite{bib:Amole_2019} is the most successful experiment of this category (Figure \ref{fig:detectors}(e)).
\subsubsection{Current limits}
\label{subsubsec:limi}
\begin{figure*}[!t]
	\centering
	\includegraphics[width=0.9\textwidth]{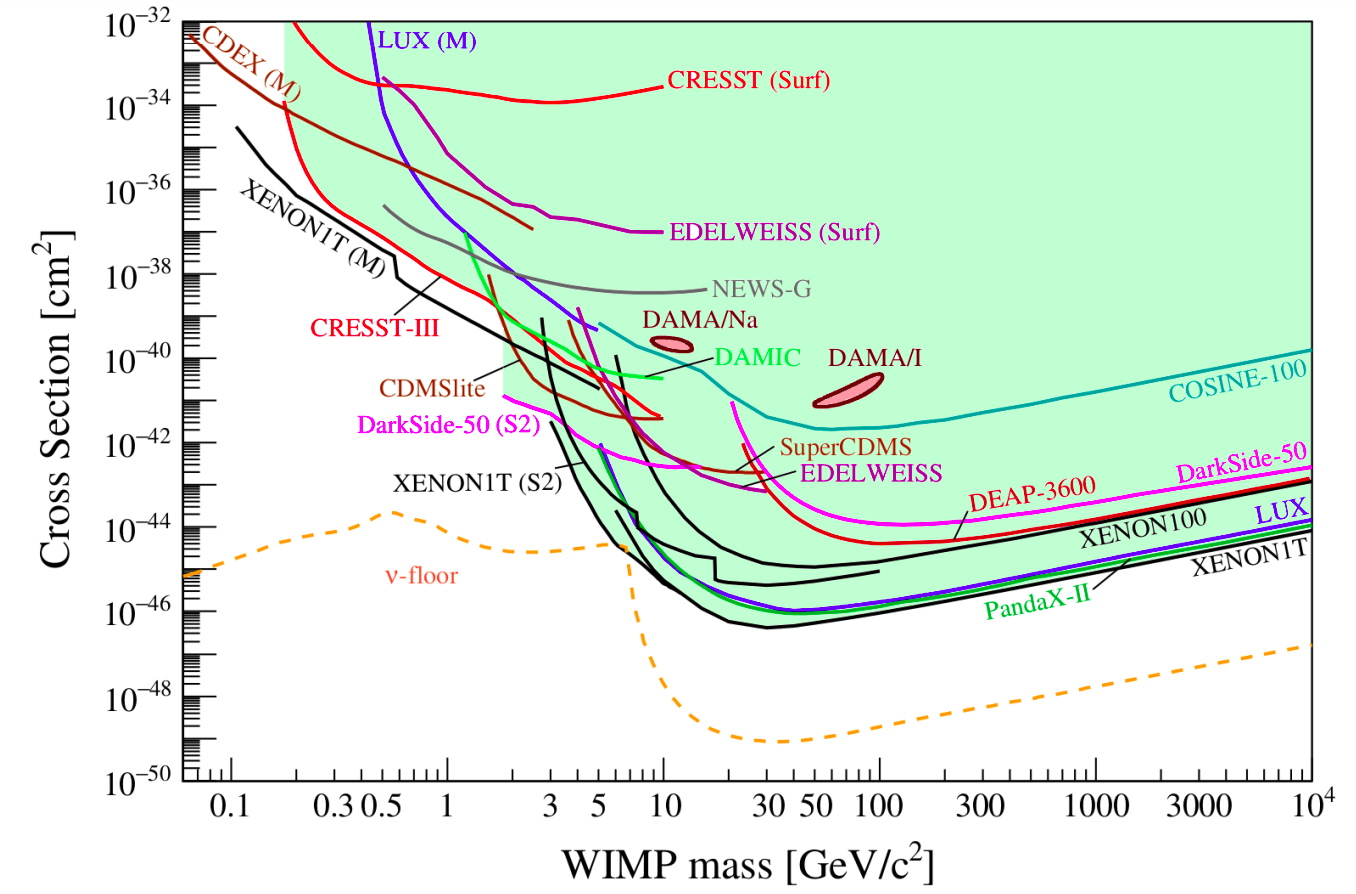}
	\caption{Current status of the SI limits on the cross section \W mass parameter space based on SHM assumption and nuclear recoils searches. Figure taken from \cite{Billard_2022}.}
	\label{fig:SI_status}
\end{figure*}

As the unknown variables of the \W model determining the nuclear recoil characteristics are two, the total \W to nucleus cross section and the \W mass, a positive detection of DM induced events will allow to define a contour in the parameter space of these two variables.
If instead no event is detected, only an upper bound limit contour can be determined.\\
Figure \ref{fig:SI_status} shows the current status of the SI limits on the cross section  versus \W mass parameter space based on nuclear recoils searches. The DAMA/LIBRA experiment, based on NaI scintillating crystals, has observed an annual modulation of NRs in the energy range between 2-6 keV with a 2.86 tonne y exposure and 13.7 $\sigma$ significance, which is consistent with \W hypothesis \cite{Bernabei:2022xgg}. Translating the annual modulation observed by DAMA into constraints in the free parameter space allows to evaluate two possible signal regions with 90\% confidence.
Despite the extremely large significance, more sensitive experiments have not found any hint of DM interaction even when probing smaller cross section than the constraints obtained from the DAMA data suggest. Therefore, and because of the small amplitude of the modulation, DAMA results could be affected by uncertainties which may mimic this modulation, from the possibility that the background itself could be modulated in time to systematics arising from the choice of the analysis technique itself \cite{Buttazzo_2020,Davis_2014,Messina_2020,Damaanalysis}. In order to solve this tension, experiments like COSINE-100 \cite{Adhikari_2021}, ANAIS \cite{Coarasa_2019} and  COSINUS \cite{Angloher_2016} have being designed or are already taking data with the same target material as DAMA/LIBRA, though they have not yet matched the crystal radiopurity achieved by DAMA collaboration. Despite the exposure needed to test the annual modulation hypothesis has not been reached yet, the 90\% upper bound limits obtained analysing the energy spectrum are already excluding the DAMA regions\cite{Adhikari_2021,Coarasa_2019}.\\
The majority of the other experiments measure only the energy spectrum of the recoils and none ever claimed any discovery. As a consequence, the 90\% confidence level exclusion limits are set by the experiments based on the assumption described in Section \ref{sec:WIMPparadigm}, in order to compare the different setups and techniques. The typical shape of a limit put by a single element target is characterised by a minimum, in the \W mass region close to the atomic mass of the target for kinematic reasons (see Section \ref{subsec:kinematics}). For larger \W masses,  the event rate is overall suppressed by 1/m$_{\chi}$. Given that the local DM density is a constant, the heavier the individual particles, the less particles are available for scattering. Conversely, at lower \W masses than the minimum, the energy threshold more steeply raises the limit value. Limits at large masses, above 3 \Gevc, are dominated by large $Z$ noble liquid TPCs as they reached monolithic tonne scale masses with relative ease thanks to the excellent scintillating properties and high density of these media. 
\begin{figure*}[!t]
	\centering
	\includegraphics[width=0.48\textwidth]{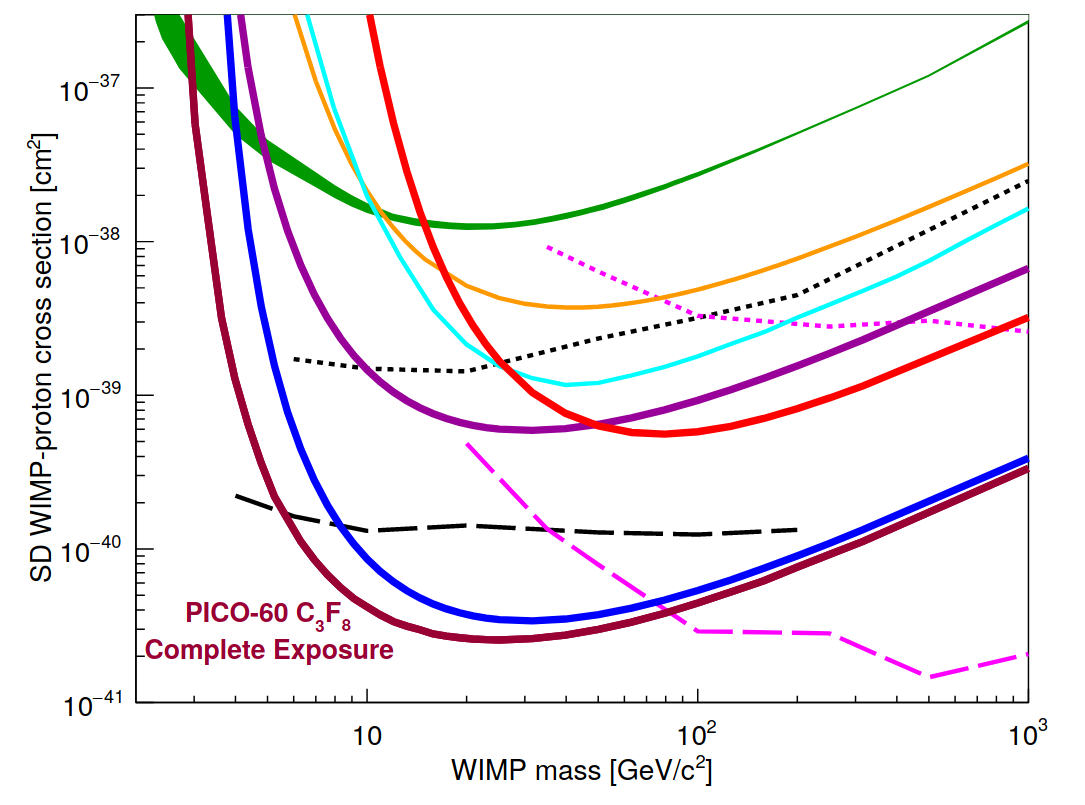}
	\includegraphics[width=0.46\textwidth]{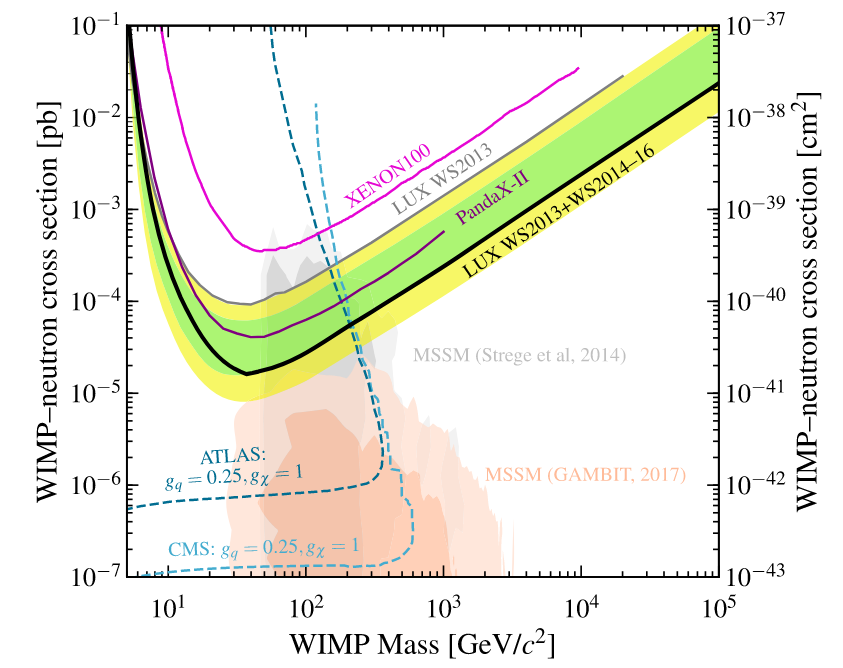}	
	\caption{On the left the current status of the SD limits on the proton cross section \W mass parameter space based on nuclear recoils searches. Figure taken from \cite{bib:Amole_2019}. The most relevant limits are from PICO-60 (blue), PICO-2L (purple), PICASSO (green), SIMPLE (orange), Panda-X (cyan). On the right, the current status of the SD limits on the neutron cross section \W mass parameter space based on nuclear recoils searches. Figure taken from \cite{PhysRevLett.118.251302}.}
	\label{fig:SD_status} 
\end{figure*}
Figure \ref{fig:SI_status} shows Xenon1T with the best limits even though recently published results by LUX-Zeplin experiment further improved the state of the art with 3$\cdot$ 10$^{-47}$ cm$^2$ at 100 \Gevc \cite{LZ}. On the other hand, improvements in low \W mass can be achieved by employing light targets and small energy thresholds. In fact, the region between 1 and 10 \Gevc WIMP mass still remains theoretically well motivated and largely unexplored to these days. Bolometer detectors with $\mathcal{O}$(10) eV energy threshold such as CREST-III and SuperCDMS published the best limits in this region, though a wide region of the parameter space is still left uncharted.\\

The sensitivities reached in the SD coupling (Figure \ref{fig:SD_status}) are between 10$^{5}$-10$^{7}$ times weaker than for the SI coupling. This is mainly due to the restricted availability of spin-odd target nuclei sensitive to axial vector currents and due to the lack of the coherent enhancement term $A^2$ in the interaction. Figure \ref{fig:SD_status} on the left shows the 90\% confidence level on the \W to proton SD interaction and on the right on the \W to neutron SD interaction. PICO is the leading experiment in the SD to proton interaction thanks to the large exposure achievable in the dense overheated liquid used in the bubble chamber and to the high fluorine content, which grants high SD coupling \cite{bib:Amole_2019}. Instead, the natural $^{129}$Xe isotope provides quite large SD coupling with the neutron and the high density of xenon allows large exposures to be attainable\cite{PhysRevLett.118.251302}. In this regard, LUX possess the best limit in the SD to neutron parameter space.\\ 

In this panorama, several experiments reported a significant number of observed excesses of unexpected low-energy events in the latest years. Deviations from expectation were detected in a variety of distinct experimental techniques and in both NR and ER signatures, such as Xenon1T \cite{Takahashi_2020}, DAMIC, SuperCDMS, CRESST-III, and EDELWEISS among the many\cite{Adari_2022}. These anomalies are being actively tackled down by the various collaborations, in some cases by demonstrating the disappearance of the excess in larger detector with reduced backgrounds\cite{Aprile_2022}, and in others by joining forces and sharing the knowledge about the individual observations \cite{Adari_2022}.
Thus, direct DM search has been (and is still)  hampered by many false promises, since all the experimental approaches discussed so far lack the capability of a positive identification of a DM signature. In addition to this, current Xe-based and several next generation experiments will be sensitive to the Solar, atmospheric and diffuse supernovae neutrinos. All these considerations consequently advocate the development of experimental approaches able to provide a positive and unambiguous identification of a DM signal, even in presence of (unknown) backgrounds. This is the goal of directional DM searches, whose methods and reasons will be discussed in the following Section.
\section{Directional dark matter searches}
\label{sec:directional}
The measurement of the direction of the recoils induced by \Ws can improve substantially the search for DM, from  an easier discrimination of the signal from the background to a potential mean of positive identification of \W. Equation \ref{eq:rateWIMP} shows the double differential rate in solid angle and energy for a Earth-based detector as a function of time, while the most relevant contribution to the directional dependence is expressed in Equation \ref{eq:angsimplbis}.
\begin{equation}
\label{eq:angsimplbis}
\frac{dR}{d\cos\gamma} \propto \int_{E_{thr}}^{E_{max}}e^{-\frac{\left(v_{lab}\cos\gamma-v_{min}\right)^2}{v_p^2}},
\end{equation}
The term $v_{lab}\cos\gamma$ is fundamental as, on one side, predicts that \Ws coming from a certain direction will have larger velocities and thus be more easily detected, and, at the same time, it induces a strong directional anisotropy, clear when the recoil spectrum is expressed in Galactic coordinates. In these coordinate system any non-\W induced recoil distribution is expected to be flat. Therefore, the anisotropic feature of the recoil spectrum cannot be mimicked by any form of background \cite{bib:Mayet_2016zxu}, which makes this feature to be considered a smoking gun of the \W DM scenario.
Figure \ref{fig:example_angul} shows two examples of angular distribution in Galactic coordinates of \W induced recoils on fluorine and xenon targets with different \W masses, obtained by integrating Equation \ref{eq:rateWIMP} on the available energy range (for operative details see Section \ref{sec:discr_2models}).\\
The anisotropic behaviour can manifest itself in various realisations. The value of $\gamma$ for the recoil depends, in fact, on the combination of the recoil energy $E$, the \W mass, the target mass and the laboratory velocity according to:
\begin{equation}
\label{eq:ringangle}
\cos\gamma = \frac{v_{min}}{v_{lab}}=\sqrt{\frac{m_AE}{2\mu_Av_{lab}^2}}.
\end{equation}
Thus, the maximum $\gamma$ can vary from 0 up to 90 degrees.\\
When $\gamma$ peaks at 0, or at very small values, the angular distribution displays a dipole in Galactic coordinates, as can be seen in Figure \ref{fig:example_angul} on the left. In this configuration, the preferred recoil direction is in the opposite direction of Earth's motion with respect to the Galactic Rest Frame, which approximately corresponds to the Cygnus constellation location. The ratio of recoil rates in the forward over the backward directions can reach differences by a factor of order 100, evaluated on the left panel of Figure \ref{fig:example_angul}. The spread of the dipolar distribution depends on the energy range of the recoils, with larger energies leading to more peaked structures. The actual distribution of the measured direction of nuclear recoils eventually depends on the experimental performances, such
as angular resolution and energy threshold.\\
Conversely, when $v_{min}<v_{lab}$, $\gamma$ peaks at values different from 0 and the maxima of the angular distribution in Galactic coordinates are distributed on a ring-like structure centred on the same direction as the dipole. An example of a ring-like shape recoil distribution is visible in Figure \ref{fig:example_angul} on the right panel. While this feature is less prominent than a pure dipole structure, of roughly a factor 2 to 6, it can still be used as a clear DM signature.\\

A directional detector is sensitive to the energy of recoils which, combined with the direction, provides an ulterior handle for the characterisation of DM. Indeed, energy and angle of recoil are bound by the kinematic of the elastic scattering and provide orthogonal information on the event nature. Measuring both enhances the power of the analysis.\\
When measuring the nuclear recoil direction, two parameters are very relevant to be addressed. One is the angular resolution which intrinsically affects the precision the recoil angle is assessed with. The angular resolution affects experimentally the spread of the anisotropy, potentially hindering the capability of discriminating it against a flat background if too large. Yet, the dipole structure is so inherently distinct from a flat distribution that a 30 $\times$ 30 degrees squared angular resolution is considered enough to efficiently detect this asymmetry\cite{bib:Mayet_2016zxu,O_Hare_2015}.\\
\begin{figure*}[!t]
	\centering
	\includegraphics[width=0.4\textwidth]{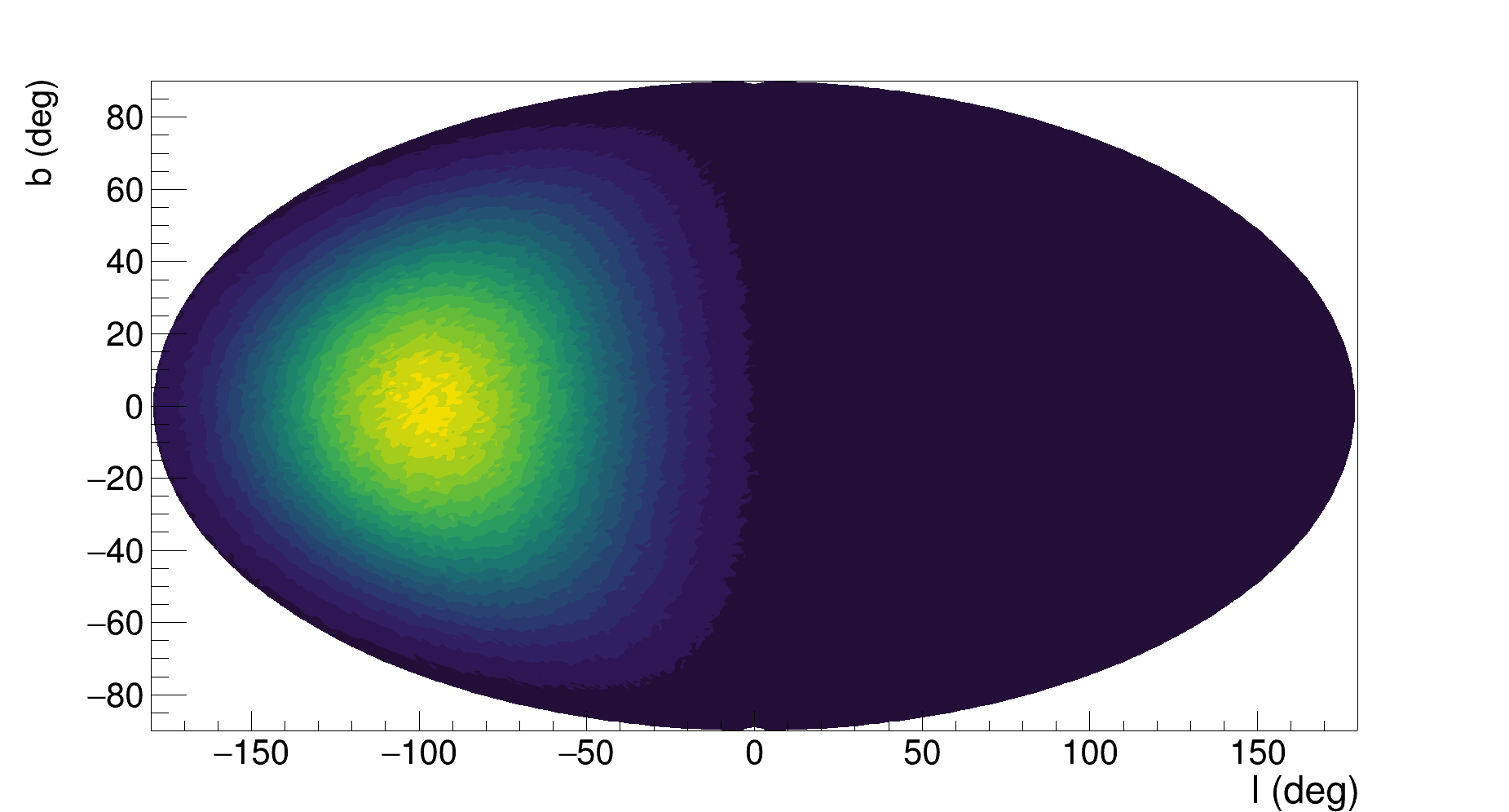}
	\includegraphics[width=0.4\textwidth]{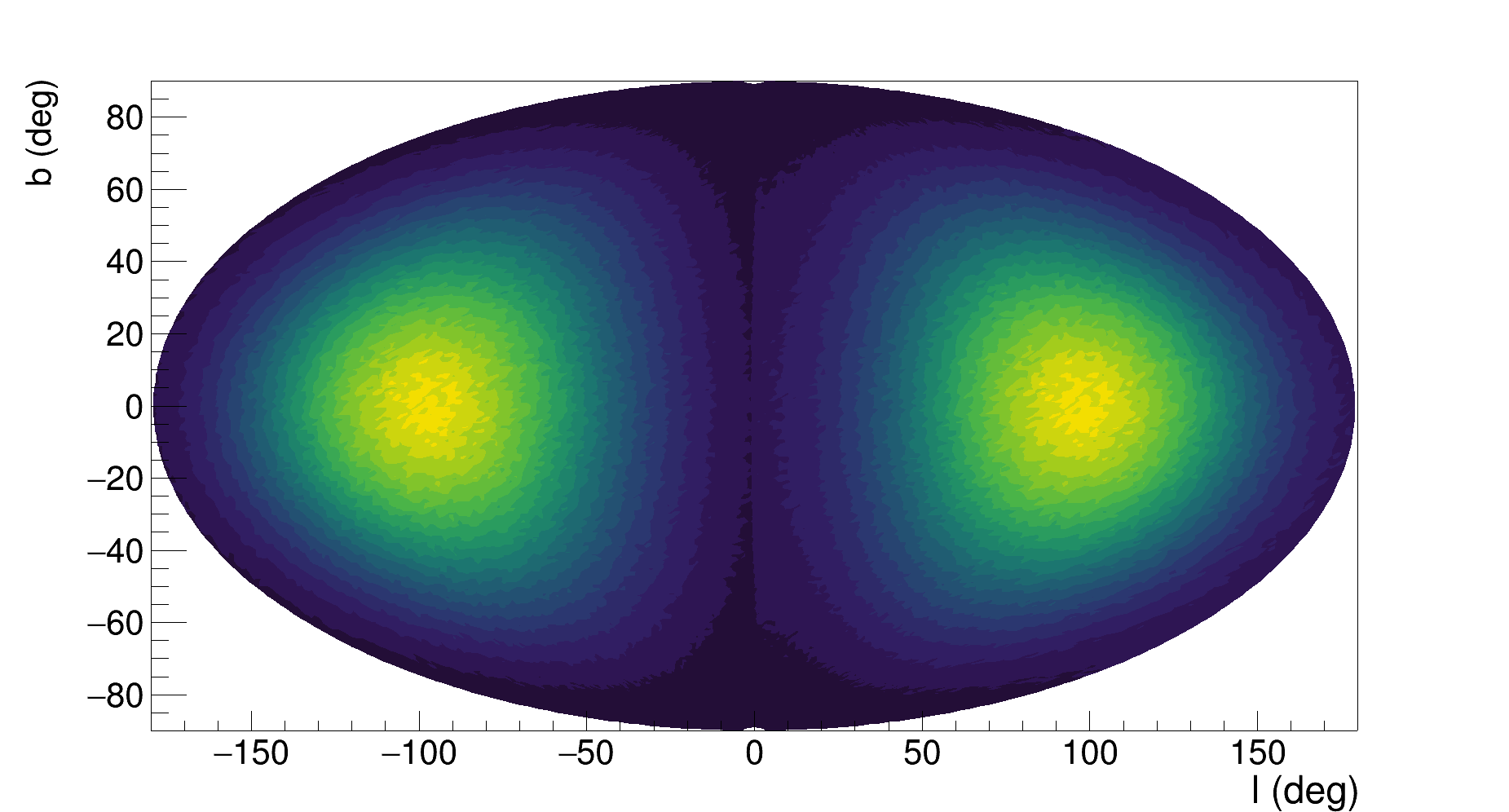}
	\caption{Examples of angular distribution of the recoils induced by \W interaction calculated following the SHM assumptions (for more details see Section \ref{sec:discr_2models}). On the left, the distribution is evaluated assuming a 100\% HT recognition, while on the right only 50\%, which corresponds to no capability in the sense reconstruction.}
	\label{fig:example_angulht} 
\end{figure*}
The other important parameter is the \emph{head-tail} recognition (HT), which represents the ability in the recognition of the sense of direction of the recoil. When the HT recognition level is at 100\%, the readout is called vectorial as it is always possible to discern the original sense of direction of the recoil. If the HT recognition drops to 50\% ,there is no sensitivity to the original orientation, as it is equivalent to randomly assess the sense of the direction, and the readout is called axial. This is an extremely relevant quantity, as the prominence of the asymmetric distribution largely depends on the recognition of the sense of the recoil. Figure \ref{fig:example_angulht} shows an example of the spectrum of the recoils in Galactic coordinates of fluorine induced by \W with 100\% (left) and 50\% (right) of HT recognition. When there is no HT discrimination capability, the dipole is mirrored in the two hemispheres, effectively strongly reducing the potential discrimination from a flat background. Many studies showed how having a vectorial readout has a larger impact than angular resolution and 3 dimensional readout requiring a factor 10 less of events to reject isotropic backgrounds\cite{Green_2007}.
\subsection{Directionality advantages}
\label{subsec:advdir}
The supplementary information provided by the measurement of the recoil direction in addition to the energy can prove to be fundamental for the search, identification and study of DM. Depending on the sensitivity of the specific detector and the experimental performances, the number of events detected can allow to obtain information on the existence and subsequently nature of DM with orders of magnitude less exposure than the one required by only energy-sensitive experiments.
The measurement of NR direction can improve the capabilities of an experiment to discriminate against background as will be discussed in Section \ref{subsubsec:dirlimits} and Section \ref{subsubsec:dirnufog}. Moreover, it can provide a correlation with an astrophysical source that represent an unique key for a positive identification of a DM signal, see Section \ref{subsubsec:dirposdisc}. Finally, it allows to infer properties of the velocity distribution and DM physical parameters as depicted in Section \ref{subsubsec:dirastronomydm}.
\subsubsection{Directional exclusion limits of a dark matter signal}
\label{subsubsec:dirlimits}
When the significance of a WIMP signal is insufficient to claim a discovery, an experiment sets an upper limit on the WIMP-nucleon interaction cross section, as discussed in Section \ref{sec:direct_detectors}. A directional detector by definition provides more information about each recoil than a non-directional detector. The additional angular information can improve the separation of signal and background. This is generally true for any background which has an angular distribution that differs from that of the WIMP recoil signal. In particular, the angular distribution of the expected internal and external backgrounds is at first order of approximation isotropic. Localised source of recoils will be washed out once the distribution is expressed in Galactic coordinates thanks to Earth's rotation around its axis. In addition, the fact that Earth's rotation axis is tilted with respect to the Galactic plane helps even more the smearing of local anisotropies in the background angular distribution so that no form of background can mimic a \W signal\cite{bib:Mayet_2016zxu}. As a result, directional detectors can set stronger exclusion limits than non-directional detectors, with the same exposure. Several works studied how the anisotropic characteristics of the \W recoil distribution can improve of $\mathcal{O}$(5) the limit setting \cite{bib:Mayet_2016zxu,Vahsen:2020pzb,COPI199943,Billard_2012}. In this work, Chapter \ref{chap7} will explicitly illustrate how the use of the directional information is practically employed to estimate the expected future sensitivity of the CYGNO experiment. 
\subsubsection{Searching for dark matter into the neutrino fog}
\label{subsubsec:dirnufog}
\begin{figure*}[!t]
	\centering
	\includegraphics[width=0.8\textwidth]{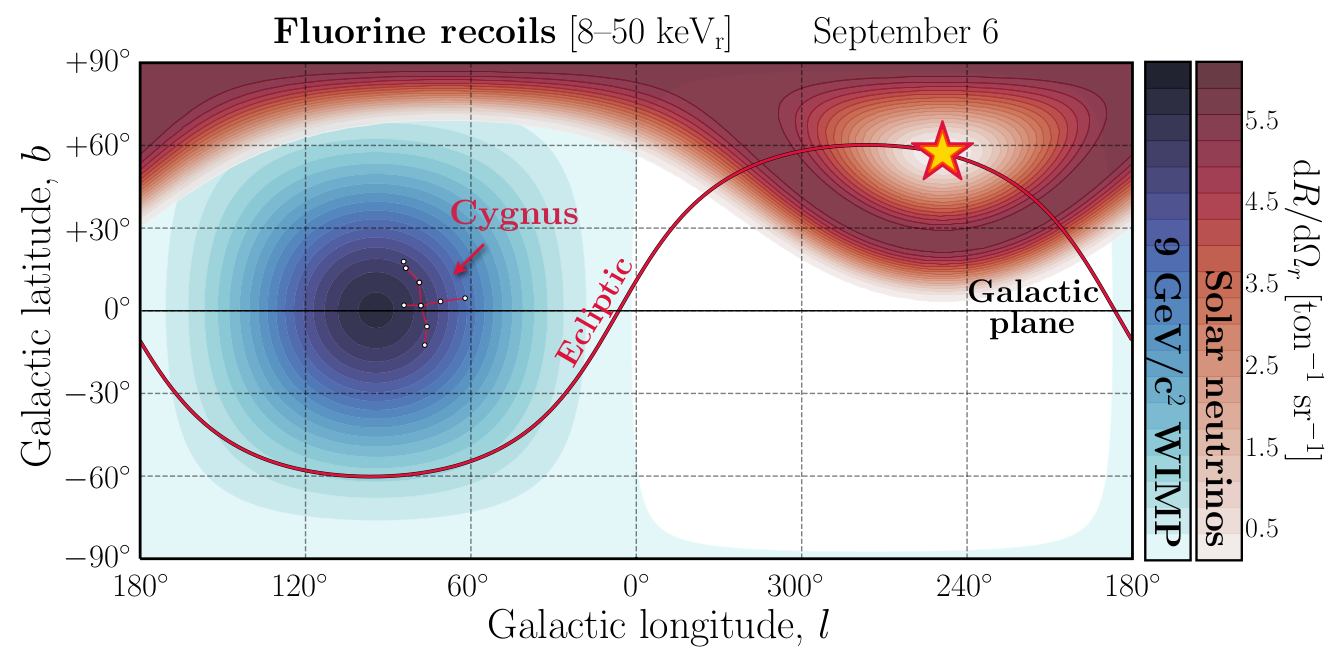}
	\caption{The angular distribution of the nuclear recoils induced by \W and Solar neutrinos in Galactic coordinates. The red line represents the ecliptic path demonstrating a very small overlapping of the two distributions in any moment of the year. Figure from \cite{Vahsen_2021}.}
	\label{fig:dmnugal} 
\end{figure*}
\begin{figure*}[!t]
	\centering
	\includegraphics[width=0.99\textwidth]{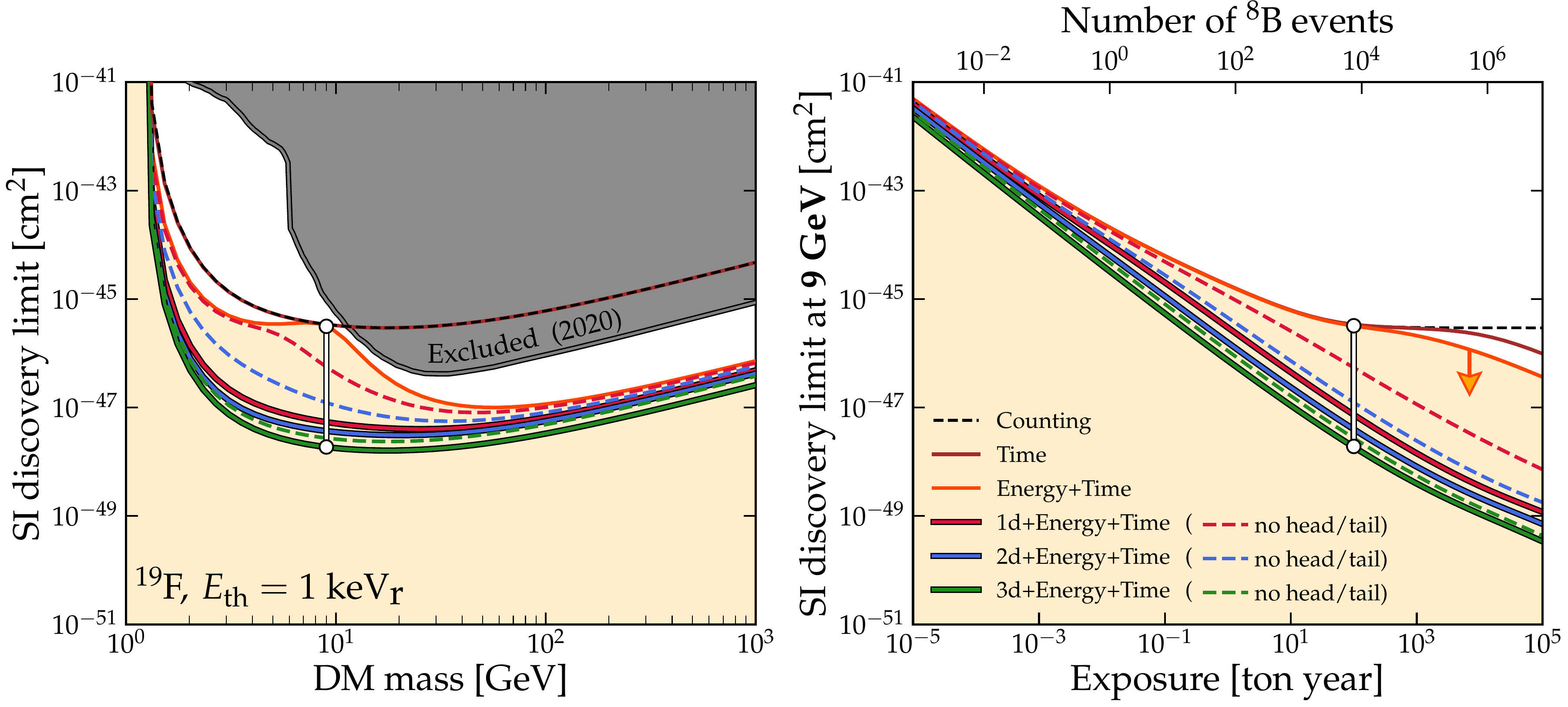}
	\caption{On the left, the discovery limits versus DM mass for a fixed detector exposure. On the right, the SI discovery limit as a function of the total detector exposure. Different curves represent, in both panels, the various characteristics these detectors can measure, as explained in the legend. With perfect head-tail recognition and angular resolution, a 9 \Gevc \W search can be performed without any worsening caused by the neutrino interaction. Figures from \cite{Vahsen_2021}}
	\label{fig:imprdir} 
\end{figure*}

Section \ref{subsec:background} discussed how neutrinos from the Sun, diffused Supernovae and atmosphere can very closely mimic a DM signal if no other handle except for the NR event energy distribution is employed to discriminate the two. Improving the knowledge on the neutrino fluxes, using multiple target elements, and detecting large amount of events allow to breach the neutrino floor into the fog as they will furnish more details on the recoil spectrum generated by the CE$\nu$NS interaction. Yet, it has been extensively demonstrated that the measurement of the direction of the recoil represents the best approach to circumvent this problem especially for Solar neutrinos, exploiting the knowledge of the arrival direction of the two different types of particles which induce recoils \cite{Vahsen:2020pzb,O_Hare_2015}. Figure \ref{fig:dmnugal} shows the angular distribution of the recoils caused by DM, peaked at the Cygnus constellation position, and the ones produced by Solar neutrinos on the 6th September (the day of maximal separation), together with the ecliptic of the Sun, assuming an energy threshold of 8 keV. With an angular resolution better than a few tens of degrees, the overlap is negligible \cite{O_Hare_2015}. Figure \ref{fig:imprdir} right, shows the discovery limit $\sigma$ from Equation \ref{eq:n_vsneutrino} as a function of the total detector exposure with different readout scenarios. The latter include readout capable of only counting events, or measure only the time, or only the energy or the energy and the direction with different performances. In the case of directional detectors, the slope of the discovery limit can be kept similar to when no background is detected even at very large exposures, effectively bypassing the neutrino induced recoil problem and showcasing the power of a directional detector. In contrast, a non-directional experiment discovery limit suffers from the neutrino background influence, with ranges of exposures where a large increase leads to no improvement. The direct effect on the discovery limit as a function of the DM mass is displayed in the left panel of Figure \ref{fig:imprdir} with a fixed exposure of 100 tonne-year. The improvement on the limits achievable with a directional detector with respect to non-directional one with the same exposure is remarkable.
As discussed in \cite{bib:Snowmass2022,Vahsen_2021,Ohare2}, the performances required on each detected event to clear the neutrino fog are:
\begin{itemize}
	\item Time resolution better than a few hours,
	\item energy resolution better than 20\%,
	\item HT above 75\%,
	\item angular resolution better than 30$^{\circ}$.
\end{itemize}
Therefore, while non-directional detectors are technically capable of overcoming the neutrino background, the requirement of exposure results not to be worth the investment for the improvement in sensitivity. Instead, the directional information provides the best solution to reduce the impact of this form of background and dive deeper into even weaker cross sections.
\subsubsection{Dark matter positive discovery}
\label{subsubsec:dirposdisc}
The strongly anisotropic structure of the angular distribution of the WIMP-induced recoils compared with the flat distribution of events originated from background sources provides a very powerful discovery potential \cite{Vahsen:2020pzb,O_Hare_2015}. The discovery of a \W signal can be reached with as few as tens of events with small background contamination and high quality directional capabilities \cite{COPI199943,Green_2007}.\\
Naturally, the experimental characteristics  such as the target material, the angular resolution, the HT recognition and the actual mass of the DM influence the number of events (hence exposure) needed for a discovery claim. The impact of different experimental characteristics on the outcome of a directional search has been evaluated to optimise the directional search in Reference \cite{Billard_2012}. The energy threshold of detection is deemed as the most important, affecting directly the amount of DM particles which can induce visible recoils in the detector. Nevertheless, this is valid for all kind of detectors, including the non-directional ones. Similarly, the background contamination, whose increment enlarges the number of events required to measure a statistically significative result, has a common effect on all kind of detectors. Instead, among the proper directional observables, the HT has the largest impact in the ability to claim for a discovery with the least possible number of events. The reduction in the amount of detected DM induced events can drop of a factor 100 when a full HT recognition is achieved with respect to no HT information, assuming an energy threshold of 20 keV, a perfect signal to background discrimination and a sulphur target \cite{Green_2007}. Finally, the the angular and energy resolutions affect only minimally the exposure needed to claim discovery.\\

Yet, the real power of directionality relies on the fact that it provides an observable for a positive identification of a DM signal and hence claim of discovery. The consistency of the expected and measured direction of the WIMP-induced recoils offers a positive mean for discovery thanks to the possibility to establish the Galactic origin of the observed events, a feature that is not achievable with the analysis of only the energy spectrum.
This highlights the fundamental difference between a measurement of the angular or energy distributions in the context of a \W search. 
\subsubsection{Dark matter identification and characterisation}
\label{subsubsec:dirastronomydm}
The ultimate goal of the research is to constrain the properties of DM, once conclusive evidence in favour of a discovery is observed by at least one experiment. A directional detector offers the chance to deeply investigate its nature by means of the employment of a single detector. The free parameters of the \W model which can be constrained by directional measurements are on one side related to its particle nature, such as the \W mass and the interaction cross section with a nucleon, while on the other hand are linked to the astrophysical structure of the DM halo. In particular, current information available on the velocity distribution of \Ws and on the structure of the DM halo, which directly determine the angular distribution of the recoils, is characterised by uncertainties and systematics in their description. In fact, as discussed in Section \ref{subsec:Astro}, the SHM is believed to be a simplistic description of the real DM halo.
Depending on the assumptions of the baryonic and non-baryonic matter content of a galaxy, simulations based on the principles discussed in \cite{halosim1,Pillepich_2014} can result in different shape of the matter profile as a function of the galactic radius, which directly affects the velocity distribution.  In addition, other phenomena like late halo merging of baryonic disks can induce the generation of co-rotating dark components, which in turn can modify the local DM density\cite{Purcell_2009}.

Directional detectors can potentially measure the DM velocity structure thanks to the information provided by the angular distribution of the recoils. Thanks to the three-dimensional nature of the latter, it is theoretically possible to access the complete structure of the DM velocity function, a feature impossible for only energy sensitive detectors whose recoil spectrum depends only on the distribution of the modulus of the velocities \cite{Lee_2012,Kavanagh_2016}. The precise measurement of the anisotropies of the velocity distribution may provide insights into the history of the Milky Way’s formation in case of halos merging, as well as the fundamental properties of DM. DM streams which generate kinematically localised structures are examples of DM halo properties particularly suited to be explored by directional detectors \cite{O_Hare_2014,Gondolo_2002}. Moreover, recent measurements from the GAIA survey observed that at a large portion of the local DM halo is likely to be part of a highly radially anisotropic feature, referred to as Gaia-Sausage \cite{gaiasausage,Kruijssen_2018,Naidu_2020}. As a result, directional detectors can prove to be much more efficient than traditional techniques in identifying tidal streams intersecting Earth position in the Galaxy\cite{Myeong_2017}.\\
Once the velocity distribution is better observed and constrained by astrophysical and DM measurements, the particle physics parameters can be also determined by directional measurements with better accuracy. The precision of the constrains on the DM mass, DM interaction cross section and the velocity dispersion of the DM halo achievable by a directional measurement is the subject of few quantitative simulations. Though these are evaluated either for specific kind of detectors \cite{bib:Mayet_2016zxu} or with perfect performances \cite{O_Hare_2015,O_Hare_2014}, they show that 10\% uncertainties on physical parameters is possible in optimistic conditions.
\subsection{Directional experimental approaches}
\label{subsec:dirdet}
The difficulty in measuring the direction of a NR induced by a WIMP scattering resides in the very low momentum exchanged in the interaction which results in recoils of less than 100 keV. While solid and liquid materials are dense and heavy, which make them excellent for exposure purposes, the length of NRs is of the order of hundreds of nm in the former and $\mu$m in the latter, requiring extremely precise detection techniques. Conversely, while a gaseous target possesses a much smaller density than a solid/liquid one, which strongly reduces the exposed mass when similar volumes of material are considered, it grants recoiling tracks of $\mathcal{O}$(1) mm whose direction can be measured with known techniques.\\
Generally speaking, two are main experimental approaches: detectors that can perform recoil imaging and detectors sensitive to a direction-dependent response of the target material to a scattering interaction. This categorisation can be further refined according to whether the detector can sense recoil directions at the event level or with only statistical distributions of events.
The different approaches and technologies developed or in R\&D to access directional information will be briefly illustrated in the following. A more comprehensive review can be found in \cite{Battat_2016}.
\subsubsection{Directional techniques in R\&D}
\label{subsubsec:dirsolid}
The following experimental efforts are worth to be mentioned due to some interesting results achieved, but have not shown full directional capabilities at the energies of interest for the DM search.
\paragraph{Anisotropic crystals} Anisotropic response of scintillating crystals has been studied in the last decade in order to obtain directional information. The two main efforts in this context are carried out by an experiment in the Kamioka mine\cite{SEKIYA_2005}, and by the ADAMO project\cite{Belli_2020}. In particular, the ZnWO$_{4}$ is a scintillating crystal with low energy threshold, multiple target and large $Z$, making it a suitable detector for DM searches. This crystal possesses one of the three axis of symmetry with a different quenching factor with respect to the others, inducing a directional dependence in the scintillation yield. This technique does not possess any HT recognition and the diversity in light yield with respect to the inclination of the crystal was measured only with high energy nuclear recoil (down to 850 keV) induced by a neutron gun.
\paragraph{Carbon nanotube} The anisotropic response of aligned single-wall Carbon NanoTubes (CNTs) has been proposed as a tool to directionally detect DM. The exploitation of a set of cylindrical CNT to build a \emph{dark PMT} is at the basis of the ANDROMEDA project \cite{carbonnano}. The basic idea is that a CNTs closed at one end and open at the other would act both as the target for a WIMP to nucleus or electron interaction and as a channel to direct the subsequent recoiling particle ejected in the process towards the free CNT end. The channeling is a phenomenon similar to total reflection in an optical fibre. By positioning many cylindrical CNTs parallel to each other and given the low energies, typical of DM induced recoils, the ions or electrons ejected from any CNT will be absorbed by the closed end or by the walls of CNTs themselves, unless they are emitted specifically in the direction which allows them to undergo channeling. The project still has to demonstrate experimentally the feasibility of such a technique.
\paragraph{Columnar recombination} Born by an idea of Nygren \cite{Nygren_2013}, the columnar recombination is a phenomenon for which the amount of primary scintillation and ionisation electrons released by a recoil in presence of an electric fields depends on the track orientation with respect to it. Indeed, after ionisation induced by a charge particle, there is a certain probability that freed electrons recombine with the positive ions. The inclination of the recoiling track with respect to the drifting electric field can affect this probability, effectively modifying the amount of charge which can reach the readout anode. Of particular interest is the possibility to employ high pressure gaseous xenon, in order to build large monolithic detectors with a high $Z$ material with directional capabilities. Measurements with xenon mixture were performed with addition of trimethylamine (TMA) to enhance the recombination probability. However, the well known absorption of TMA of the scintillation light produced by xenon, prevented the columnar recombination to be efficiently measured. Subsequent work carried out in 8 atm pure xenon searched for modifications in the shape of the scintillating signal. Indeed, the recombination in xenon gives rise to a secondary scintillation light with a larger decay time. A noticeable difference between 5.4 MeV alpha tracks orthogonal and parallel to the drift field were observed in the signal shape, though it was impossible to quantify the performances\cite{Nakamura_2018}.
\subsubsection{Nuclear Emulsions for WIMP Search with directional measurement: the NEWSdm project}
\label{subsubsec:newsdm}
NEWSdm is an experiment which aims at deploying a directional detector for DM searches with the use of nanometric nuclear emulsions (called Nano-Imaging Tracker or NIT) and a fully automated innovative scanning system, with improved optical technologies. A NIT is a super high-resolution nuclear emulsion based on a solid material composed of AgBr:I crystals dispersed in gelatin film of polyvinyl alcohol. The mean crystal size in the NIT is 40 nm (6 nm standard deviation) and can reach densities in the gelatin film of 2000 crystals/$\mu$m$^3$ \cite{NAKA2013519,Shiraishi_2023}. NEWSdm is the only experiment with solid target which demonstrated directional imaging capabilities at energies relevant for above 10 \Gevc WIMP masses DM searches.\\
When an ionising particle crosses the target, some of the crystals are modified in such a way that they are turned into grains of silver. This phenomenon is related to the band gap energy of the silver halide which is very small, 2.6 eV\cite{Ariga2020}. During the \textit{development} of the emulsion, only the silver grains are left in the gelatin and the three-dimensional trajectory of original particle can be recovered by connecting the silver grains \cite{DeLellis:20227s}. The spatial resolution and angular resolution is then set by the final silver crystal size and density after development, and by the capability of optical devices to distinguish such small images. From a SRIM\footnote{\url{http://www.srim.org/}} simulation and assuming the optical system to be 100\% efficient, the directional performance for NIT grains has about 50\% efficiency with 550 mrad angular resolution for a carbon recoil at 20 keV \cite{Newdm1}. This is the intrinsic characteristic directional performance only due to silver grain dimensions and density, that needs to be successively imaged with some optical device. Common optical microscopes have too low efficiency and too long times needed for the short tracks expected from WIMP induced nuclear recoils, and therefore an extensive dedicated R\&D was carried out in the last decade to increase the scanning speed and efficiency at large angle.\\

The scanning methodology to analyse the samples of emulsion film is based on a two-steps approach with the goal of readout as fast as possible, in order to analyse as quickly as the exposure increases, and to reach spatial resolutions of few tens of nanometres. The first step consists in a fast scanning by means of optical microscopes with the state-of-the-art nanometric resolution, with a speed of roughly hundreds of cm$^2$/h\cite{Alexandrov2019}. Due to the intrinsic resolution of these objects, the tracks appear as single ellipsoidal clusters and are recognised by means of elliptic fits. This first step already removes single isolated silver grains, whose shape more closely resembles a sphere. The second step of the analysis employs the plasmon resonance technique, originally developed to image intracellular fluorescent proteins at nanometre resolution \cite{Betzig2006-rv}, and to which was assigned the 2014 Nobel Prize in Chemistry. The plasmon resonance is a modification of the intensity and wavelength of the light reflected by metallic materials in presence of polarised light caused by the angle of polarisation and of illumination of non-spherical objects \cite{plasmonreso}. Approximating the silver grains in emulsion to ellipsoids, a larger resonant peak will be found at long wavelengths along the major axis and at short ones along the minor. Silver grains along a track do not have in general their major axes aligned, therefore, the rotation of the light polarisation produces resonant peaks at different angles for different grains. Analysing the evolution of the track while rotating the polarisation of the light allows to obtain additional information on the substructures of the tracks with a resolution beyond the optical diffraction limit \cite{DeLellis:20227s}. Spatial resolution of 80  nm have been achieved with this technique \cite{Alexandrov2020} and an extraordinary position accuracy of 10 nm has been established \cite{Di_Marco_2016,refId0}.

As any DM experiment, recoils induced by background in emulsions can be a limiting aspect for the DM search. While $\gamma$ background from environmental radioactivity can be suppressed with active or passive shielding, a significant source of background is represented by the $\beta$ decays produced by the $^{14}$C present in the light nuclei making up the gelatine. Nevertheless, such electron background is expected to be rejected by properly tuning the response of the emulsion through chemical processes, or by operating at cryogenic temperature, or by exploiting the different response of $\beta$s to polarised light \cite{DeLellis:20227s}. 
The neutron induced recoils represent the ultimate background sources, distinguishable from WIMP only for the isotropic angular distribution and for the typical track length. While appropriate shielding can reduce the external neutron flux to reasonable levels, the intrinsic emulsions radioactivity will eventually be responsible for the ultimate irreducible background through ($\alpha$,n) reactions and $^{238}$U spontaneous fission. For these reasons, each component of the emulsions (AgBr, Gelatin and PVA) have been carefully screened with both the Inductively Coupled Plasma Mass Spectrometry (ICP-MS) and high purity Ge detectors. The measured total activity of the sample is (23 $\pm$ 7) mBq/kg for the $^{238}$U chain and (5.1 $\pm$ 1.5) mBq/kg for $^{232}$Th \cite{Di_Marco_2016}. The neutron induced NRs yield determined from this activity is expected to be of the order of 0.06 recoils per year per kg\cite{DeLellis:20227s}. This finding allows the design of an emulsion detector with an exposure of about 10 kg year. A careful selection of the emulsion gelatine and production process could further increase the radio-purity, thus extending the detector mass and exposure time. 

Recently, the scanning system was improved to reach 250 g/year/machine speed in analysis of the emulsions and was employed in the environmental neutron measurements of the overground laboratories of LNGS. The large content of protons in the gelatine allows a large sensitivity to these neutral particles. The response to neutrons was calibrated with a monochromatic 880 keV source and the ambient flux was measured to be (7.6 $\pm$ 1.7) $\times$ 10$^{-3}$ cm$^{-2}$s$^{-1}$ in agreement with previous measurements. The topological analysis of the silver grains tracks of proton recoils allowed to evaluate a rejection factor for $\gamma$s of 5 $\times$ 10$^7$ \cite{Shiraishi_2023}.

Since March 2021 a NEWSdm setup for the emulsion handling and film production have been installed at underground LNGS. A facility to accommodate a 10 g mass of NITs encased in a shield of 40 cm of polyethylene and 10 cm of lead has being built close to the Borexino site in order to test the environmental background sources in an underground context\cite{DeLellis:20227s}.\\ 
The current measurements suggest the feasibility of a 10 kg year exposure experiment which is planned to explore the DAMA regions \cite{DeLellis:20227s}. Since emulsions are time insensitive, they will need to be installed on an equatorial telescope to maintain the detector with a fixed orientation towards the Cygnus constellation, achievable with an accuracy better than 1 degree.  \cite{Golovatiuk_2021}.
\subsubsection{Gaseous detectors}
\label{subsubsec:dirgas}
Although inherently challenging, gaseous TPCs potentially provide the best architecture and the best observables for a rare events search experiment, and in particular for directional DM. A TPC is  a three dimensional detector where a sensitive volume is enveloped between cathode and anode electrodes that provide a suitable electric field, maintained homogeneous by a field cage\cite{bib:tpc1,bib:tpc2,bib:tpc3}. The passage of an ionising radiation in the gas frees electrons. The presence of the electric field  prevents the recombination of ions and electrons inside the active gas volume, and drift them respectively to the cathode and the anode. When a recoil of about 10 keV releases its energy in the gas, hundreds of electrons are freed on average for a generic gas mixture. When these electrons are collected at the anode, a tiny signal of less than a fC is induced. Therefore, the electron signal needs to be amplified to be distinguished from the readout noise. When electrons are drifting inside the gas by means of an electric field, they are accelerated and exchange energy with ions by colliding with them\cite{Knoll}. The field can be strengthened enough (up to tens of kV/cm) in the so called \textit{amplification region} of the detector, so that the energy acquired by the electrons between collisions is sufficient to induce further ionisation in the gas, multiplying the charge. These secondary electrons just produced will in turn accelerate and produce more charge. The cascade of secondary electrons is called \emph{avalanche}\cite{Knoll}. With standard noble gas based mixtures avalanches can produce gains between 10$^{2}$ and 10$^{3}$ \cite{Knoll}.

The TPC is an intrinsic 3D detector which possesses the characteristics to be an excellent directional detector for DM searches. Low energy thresholds can be achieved as the typical average energy required to produce an electron-ion pair in gas is between 20 and 50 eV and modern amplification structures can reach gains of a factor 10$^6$ \cite{SAULI20162,Knoll}.
The measurement of the total ionisation indicates the energy of the recoil. Alphas and electrons can be easily identified comparing their track topology to the energy released along the path (i.e. dE/dx), providing excellent background rejection. The track itself indicates the axis of the NR and the charge (and/or dE/dx) measured along its path encodes the track orientation, providing the HT measurement.\\
In addition, the active volume is generally free of background producing materials as only the gas, which can be purified of its radon content, is present. A wide variety of gases can be utilised combining heavy and light targets with the possibility to be sensitive to both SD and SI interactions. Only one wall of the detector needs to be instrumented with an amplification and readout system leading to favourable cost to volume scaling and to reduced radioactive contamination due to the detector materials. Even though the gas density is much lower than solid and liquid targets, large TPCs up to 20000 m$^3$ of active volume have been approved for construction in the neutrino field\cite{DUNE}, showcasing that the instrumentation of a very big detector is feasible.\\

Achieving all the above mentioned 3D features at the low energy of interest for directional DM searches implies two main detector requirements. Firstly, the readout segmentation must be smaller than the recoil length of interest (0.1-1 mm), so that multiple space points along the track can be measured. While experimentally challenging, this is nowadays achievable via multiple technologies \cite{SAULI20162}. Secondly, any diffusion of the recoil trajectory must also be small compared with the recoil length, in order to not wash out its topology. This can be obtained with the use of mixtures containing cold gases (such as CO$_2$ or CF$_4$) or by exploiting Negative Ion Drift (NID) operation. Cold gases are characterised by large cross sections of the collision between electrons and the gas. This way, the energy acquired by the electrons when accelerated by electric field is more often transferred to the gas molecules, keeping the former more closely to the thermal equilibrium, effectively reducing the diffusion \cite{Blum_rolandi}.

The addition of a small amount of highly electronegative gases, like CS$_2$ and \SF, allows the primary electrons to be captured in few micrometres. This way, negative ions are drifted in place of electron, significantly reducing the diffusion they are subject to \cite{Martoff_2005}, improving the quality of the directional quantities measurable and the background rejection capabilities\cite{Vahsen_2021}. A detailed and comprehensive discussion on the features and properties of NID operation will be illustrated in the dedicated Chapter \ref{chap6}.\\

In principle, the TPC technology applied to the DM field does not allow to measure the position of an event in the drift field direction (z), as the absolute time of interaction in the sensitive volume is not known. In the case of noble liquids, the primary scintillation is so large that it can be detected by photo detectors and the fiducialisation is possible (see Section \ref{subsec:background} for the fiducialisation definition). Instead, gas-based TPCs do not possess enough scintillation yield to be able to detect it. Nevertheless, fiducialisation can be achieved by fitting the diffusion of each recoil and relate it to the distance travelled along z \cite{LEWIS201581,bib:Antochi_2021}. This is possible only when the granularity of the readout is much smaller than the length of the recoil, gaining enough details on the diffusion the track is subject to. Alternatively, in NID operation, different species of negative ions can be generated with sensibly different masses \cite{mincarry}. The unequal mass causes different times of arrival of the anions at the readout from which the original z position can be precisely determined, as will be discussed in more details in Chapter \ref{chap6}.
\begin{table}[!t]
	\centering
	\begin{adjustbox}{max width=1.01\textwidth}
		\begin{tabular}{|c|c|c|c|c|c|c|}
			\hline
			\multirow{2}{*}{Name}& \multirow{2}{*}{Readout} & \multirow{2}{*}{Directionality} & \multirow{2}{*}{Gas}  & \multirow{2}{*}{R\&D gas} & Largest  & Future  \\
			&&&&&detector&detector \\ \hline \hline
			\multirow{2}{*}{DRIFT} & \multirow{2}{*}{MWPC} & \multirow{2}{*}{1.5 D} & \textcolor{blue}{CS$_2$:CF$_4$:O$_2$} & \textcolor{blue}{\SF:(CF$_4$)} & 1 m$^3$ & 10 m$^3$  \\
			& & & \@ 0.05 bar & \@ 0.05 bar & \scriptsize{(underground}) & \scriptsize{(under study} ) \\ \hline
			\multirow{2}{*}{NEWAGE} & \multirow{2}{*}{GEM+ muPIC} & \multirow{2}{*}{3 D} & \textcolor{red}{CF$_4$} & \textcolor{blue}{\SF} & 0.04 m$^3$ & 1 m$^3$  \\
			& & & \@ 0.1 bar & \@ 0.03 bar & \scriptsize{(underground}) & \scriptsize{(vessel underground} ) \\ \hline
			\multirow{2}{*}{D$^3$} & \multirow{2}{*}{2 GEMs + pixel} & \multirow{2}{*}{3 D} & \textcolor{red}{Ar/He:CO$_2$} & \textcolor{red}{He:CF$_4$:X} & 10$^{-4}$ m$^3$ & 0.04 m$^3$  \\
			& & & \@ 1 bar & \@ 1 bar & \scriptsize{(overground}) & \scriptsize{(under construction} ) \\ \hline
			\multirow{2}{*}{CYGNO} & 3 GEMs + sCMOS & \multirow{2}{*}{3 D} & \textcolor{red}{He:CF$_4$} & \textcolor{blue}{He:CF$_4$:\SF} & 0.05 m$^3$ & 0.4 m$^3$  \\
			& + PMT & & \@ 0.8-1 bar & \@ 0.8-1 bar & \scriptsize{(underground}) & \scriptsize{(funded)}  \\ \hline
			\multirow{2}{*}{MIMAC} & \multirow{2}{*}{Micromegas + FADC} & \multirow{2}{*}{3 D} & \textcolor{red}{CF$_4$:CHF$_3$:i-C$_4$H$_ 10$} & \multirow{2}{*}{N.A.} & 0.05 m$^3$ & 1 m$^3$  \\
			& & & \@ 0.05 bar &  & \scriptsize{(underground}) & \scriptsize{(under study} ) \\ \hline
			\multirow{2}{*}{DMTPC} & Mesh chambers & \multirow{2}{*}{3 D} & \textcolor{red}{CF$_4$} & \multirow{2}{*}{N.A.} & 0.02 m$^3$ & \multirow{2}{*}{N.A.}  \\
			& + CCD & & \@ 0.04 bar & & \scriptsize{(overground}) & \\ \hline
		\end{tabular}
	\end{adjustbox}
	\caption{Table summarising the principal characteristics of the most relevant gaseous directional detectors employed in the direct search for DM. The red colour is utilised to stress that the gas is used as regular electro drift operation (ED) , while the blue underlines when the gas operates in a NID condition.}
	\label{tab:tabgasdirec}	
\end{table}
In the following a general outline of the principal directional gaseous detector, whose main features are displayed in Table \ref{tab:tabgasdirec}, is given.
\paragraph{Directional Recoil Identification From Tracks: DRIFT}
The DRIFT collaboration at Boulby has pioneered since 2001 construction and operation
of the only directional TPCs at 1 m$^3$ scale using negative ions in a gas mixture of CS$_2$:CF$_4$:O$_2$ in a proportion of 30:10:1 Torr \cite{Battat_2021}. The CS$_2$ provides the negative ion operation, CF$_4$ gives sensitivity to the SD coupling thanks to the fluorine and the O$_2$ is used to foster the generation of minority carriers, secondary negative ion species, to fiducialise along the drift direction. The DRIFT-IId detector has a total volume of 1 m$^3$, of which 0.8 are active, and is composed of two back-to-back TPCs of 50 cm drift, which share a common cathode.  Multi-Wire Proportional Counters (MWPC) are used to amplify and readout the charge reaching the grounded anode plane of 20 $\mu$m wires, positioned in-between two perpendicular grid planes of 100 $\mu$m wires. The pitch is 2 mm in both directions of the 2D readout plane \cite{DRIFT:2014bny}. The full volume fiducialisation provided by the minority carrier signal, together with the impressive gamma rejection factor of 1.98 $\cdot$ 10$^7$, granted DRIFT background-free operation \cite{DRIFT:2014bny}. The collaboration demonstrated head-tail recognition down to 38 keV$_{\rm{nr}}$ \cite{Battat_2016_HT}, on a statistical basis. The detector operated for more than a decade in the Boulby mine and has the best exclusion plots produced by a directional detector: down to 0.160 pb in the SD channel for a WIMP mass of 80 GeV and it has set constraints down to a WIMP mass of 9 \Gevc \cite{Battat_2021}.
Currently, the DRIFT collaboration is focusing on the study of innovative amplification structures and improved means of gas purification, towards the development of a larger scale detector in the context of the CYGNUS project (see below). In order to overcome the typical low gains of NID operation, they are testing an original Multi-Mesh Thick Gas Electron Multipliers device (MMTHGEM) developed at CERN, that adds additional stages of amplification to a classical thick GEM by embedding mesh layers within it, and which already demonstrated a gain larger than 6 $\times$ 10$^4$ in pure SF$_6$ at 20 Torr in an unoptimised configuration.

\paragraph{NEw generation WIMP-search with Advanced Gaseous tracking device	Experiment: NEWAGE}
NEWAGE collaboration's TPC is equipped with a gas mixture of pure CF$_4$ at 100 mbar. The chamber is 41 cm long along the drift and has an active area of about 31 $\times$ 31 cm$^2$ \cite{NEWAGE1}. Gas Electron Multipliers (GEMs) are employed to amplify the charge which is readout by a plane made of micro-pixels ($\mu$-PIC) with a pitch of 400 $\mu$m. NEWAGE measured an angular resolution of 48$^{\circ}$ for recoils in the range $\left[50, 100\right]\text{ keV}_{\text{ee}}$ with a detection threshold around 50 keV$_\text{ee}$ \cite{NEWAGE1}. After the first tests, the $\mu$-PIC was customised to suppress by a factor 100 the radioactive emission of alpha particles in the sensitive volume \cite{Hashimoto_2018}. Recently, the NEWAGE collaboration has set an exclusion limit down to 50 pb in the WIMP-proton SD channel for WIMP mass of 100 \Gevc \cite{bib:yakabe2020limits}, operating in the Laboratory B of the Kamioka Observatory. This is an extremely important limit as it is first ever by a 3D vector directional detector. The possibility of the introduction of \SF has been investigated and preliminary tests on alpha particles returned a 3D reconstruction with a spatial resolution of 130 $\mu$m with absolute fiducialisation thanks to the use of minority carriers \cite{Ikeda:2020pex}. This work is progressing in parallel to a simulation tailored to study the NID amplification processes. Meanwhile, a vessel of 1 m$^3$ has been constructed and 1/18 of it has already been instrumented.
\paragraph{Directional Dark matter Detector: D$^3$}
The D$^3$ project focused their efforts in the employment of highly segmented pixelated charge readouts to small volume TPCs. During their development different charge readout chips realised by ATLAS were utilised, from the FE-I3 up to the more modern FE-I4b \cite{LEWIS201581,VAHSEN201595}. The FE-I3 and FE-I4b chips are fabricated in 250 and 130 nm CMOS technology respectively, with pixel sizes of 50 $\times$ 400 $\mu$m$^2$ and 50 $\times$ 250 $\mu$m$^2$, and are both nominally clocked at 40 MHz. Each pixel contains an integrating amplifier, a discriminator, a shaper, and an associated digital control. Thanks to the large amplification granted by a double GEM stack (gain of $\sim$ 10$^4$ ) and the low intrinsic electronic noise of the pixels, the readout chip permits stable operation, with close to 1 single electron efficiency. With a small prototype of 4.5 cm drift filled with He:CO$_2$ (70/30) or Ar:CO$_2$ (70/30) operated at atmospheric pressure with standard 50 $\mu$m thick CERN GEM amplification, a single point resolution of $\sigma_x = (197 \pm 11)$ $\mu$m and $\sigma_y = (142 \pm 9)$ $\mu$m from cosmic ray data and an angular resolution
of $\mathcal{O}$(1) degree from alpha tracks was measured \cite{VAHSEN201595}. Utilising another prototype with 15 cm long drift distance, the absolute position along the drift direction was measured by positioning a $^{210}$Po source at various positions, reaching an accuracy of 1-2 cm by studying the diffusion profile \cite{LEWIS201581}.  A new, larger detector (CYGNUS-HD) with about 40 l active volume is under development in a back-to-back 50 cm drift configuration with shared cathode and a readout plane area of 20 $\times$ 20 cm$^2$ made of CERN Micromegas with a strip charge readout, towards the realisation of a cubic metre experiment.
\paragraph{A CYGNus module with Optical readout: CYGNO}
This thesis has been developed in the context of the CYGNO detector. The utilised gas mixture is a He:CF$_4$ (64/40) at atmospheric pressure \cite{Amaro_2022}. The amplification stage contains three 50 $\mu$m GEMs and thanks to the scintillating properties of the CF$_4$ gas, they are coupled to an optical readout comprising sCMOS cameras and PMTs to achieve a full 3D reconstruction of the tracks. The largest prototype built is a 50 l detector with 50 cm drift length which is operating in the underground facilities at LNGS. Energy threshold of 1 keV$_{\text{ee}}$ and rejection power of 10$^2$ below 10 keV has been measured. Future plans include the construction of a 0.4 m$^3$ detector in a back-to-back configuration with shared cathode, towards the realisation of a $\mathcal{O}$(30) m$^3$ one. A detailed description of the experiments is provided in Chapter \ref{chap3}.
\paragraph{MIcro-tpc MAtrix of Chambers for dark matter directional detection: MIMAC}
MIMAC utilises a gas mixture for electron drift made of CF$_4$:CHF$_3$:i-C$_4$H$_{10}$ (70:28:2) at 50 mbar. A large content of fluorine is chosen for the SD coupling and the gas mixture was specifically developed to control the gain and obtain a small velocity of the primary drifting electrons. The latter was optimised in order to cope with the sampling rate of their readout, 50 MHz, with the goal of maximising their capabilities in the reconstruction of the z coordinate. The charge amplification and readout is realised by a Micromegas of gap 512 $\mu$m whose generated avalanches are sampled at 50 MHz by a 2D pixelated anode, with a strip pitch of 424.3 $\mu$m in both directions. MIMAC achieved a ER rejection power of around 10$^5$ above 5 keV with a simulated $\sim$ 95\% nuclear recoil efficiency \cite{Riffard_2016}. The measurement was performed by comparing a data taking with only $\gamma$ interactions with another one where nuclear recoils of $\mathcal{O}$(100) keV induced by fast neutrons, generated by ($p$,$n$) resonances in $^7$Li, and ERs were present. The detection threshold is estimated to be 5 keV and angular resolution of 25$^{\circ}$ were measured on $^{19}$F ions of 6.3 keV of kinetic energy \cite{Tao:2020muh}. The MIMAC collaboration is developing a $\mathcal{O}$(1) m$^3$ detector comprising a bi-chamber modules with 25 cm drift to be constructed and installed in the future in the underground laboratory of Modane (LSM) \cite{Tao:2020muh}.
\paragraph{Dark Matter Time Projection Chamber: DMTPC}
The DMTPC experiment employs conventional electron drift with a low-pressure (30-100 Torr) CF$_4$ gas and is equipped with CCD cameras, PMTs and charge readout systems in the attempt to achieve 3D reconstruction. Large amplification and photon production are provided by a small (< 1 mm) region close to the grounded anode where high electric fields are generated. With small prototypes, with $\sim$ 10 cm drift length, a HT recognition of 75\% and angular resolution of $\sim$ 15 degrees was achieved for 100 keV recoils, with HT decreasing to 50\% (i.e. no sensitivity) at 40 keV\cite{BATTAT20146,Ahlen_2011}. The collaboration halted the operations.
\paragraph{CYGNUS}
CYGNUS is a proto-collaboration which aims at building a multi-site, multi-target Galactic Observatory at the tonne scale to probe for DM in the neutrino fog and measure Solar neutrinos \cite{bib:cygnus}. Multiple underground site installation can help reducing the dimension of the single components of the experiment and to eliminate background location systematics. The proposed target is an helium/fluorine gas mixture at atmospheric pressure, to be sensitive to light DM at the \Gevc scale and to the SD coupling. A feasibility paper of CYGNUS \cite{Vahsen:2020pzb} is public wherein the potential capabilities were simulated in the assumption of a NID operation with He:SF$_6$ gas mixture and charge readout. NID operation in this context provides a sensible reduction of the diffusion and full fiducialisation in order to obtain full 3D with HT recognition down to the keV with full background rejection. A study based on costs-benefits simulations shows that a strip charge detector with 200 $\mu$m pitch is the best suited for CYGNUS experiment. Nevertheless, combined R\&D from different groups are exploring other NID mixtures as well as ED ones coupled to both charge and optical readout. The full detector is expected to be realised through a staged approach, each one with specific reachable physics goals in the DM, neutrinos and others like Migdal effect \cite{Vahsen_2021}. With a $\mathcal{O}$(100) m$^3$ competitive search in the SD \W to proton cross section, non-\W DM searches, the study of the Solar neutrino flux would be in reach, whilst a $\mathcal{O}$(10$^5$) m$^3$ is foreseen to allow to perform DM astronomy and measure geoneutrinos from Earth.
\chapter{The CYGNO experiment}
\label{chap3}
The CYGNO experiment proposes an innovative path for the directional searches of dark matter (DM) and rare events at low energy. CYGNO employs a high resolution TPC working with a helium-fluorine based gas mixture at atmospheric temperature. The combination of Gas Electron Multipliers (GEMs) and optical sensors constitutes the amplification and readout strategy. Scientific CMOS (sCMOS) cameras  are used in order to image large areas  with high granularity and reduced number of detectors. By combining the sCMOS camera images with signals from photomultiplier tubes (PMTs), 3D tracking is achieved with directional capabilities down to $\mathcal{O}$(1) keV energy. A comprehensive overview of the CYGNO project can be found in \cite{Amaro_2022}. A handful of prototypes have been built with increasing dimensions since the project started, in order to consolidate the experimental technique and to refine its performances. The goal is to construct in underground facilities a detector of tens of cubic metres active volume to perform directional DM searches and Solar neutrinos spectroscopy. Moreover, CYGNO fits in the context of the CYGNUS project, whose goal is to create a multi-site Galactic Recoil Observatory that can probe the DM hypothesis inside the neutrino fog and measure  neutrinos from the Sun and SNe (see \ref{subsubsec:dirgas}). The CYGNO collaboration now comprise roughly 50 researchers and students from institutions from four different countries, Italy, Portugal, Great Britain and Brazil.\\
Section \ref{sec:conc_goalCYGNO} is dedicated to the description of the experiment technological approach. Section \ref{sec:timeline} describes the timeline of the CYGNO experiment. Section \ref{sec:mango}  illustrates the MANGO prototype employed to collect the data illustrated in Chapter \ref{chap5} and Chapter \ref{chap6}, while Section \ref{sec:LEMOn} contains the description and the most relevant results and measurements obtained with the LEMOn prototype. Section \ref{sec:LIME} focuses on the latest and largest prototype manufactured so far (LIME, with a 50 l active volume), its operation and performances, and Section \ref{sec:future:CYGNO} presents an outlook on the future perspectives.
\section{Detector concept}
\label{sec:conc_goalCYGNO}
The CYGNO collaboration intends to deploy a high granularity detector for direct directional DM searches. The detector is based on a gaseous TPC with an helium and fluorine rich gas mixture operated at room temperature and atmospheric pressure. An amplification stage comprising three 50 $\mu$m GEMs is readout optically by means of sCMOS cameras and PMTs. Figure \ref{fig:TPC} displays a sketch of a TPC with some of its key features. TPCs are detectors where a sensitive volume is enveloped between cathode and anode electrodes that provide a suitable electric field, maintained homogeneous by a field cage \cite{bib:tpc1,bib:tpc2,bib:tpc3}. The anode is usually instrumented, but the cathode can be as well. The passage of an ionising particle in the gas frees electron-ions pairs along its trajectory. The presence of an electric field  prevents the recombination of ions and electrons inside the active gas volume, and drifts them respectively to the cathode and the anode. Since the charge freed in the ionisation process is typically too small to be directly detected, an amplification stage is usually added at the anode to enhance the primary electron signal. The amplified signal can be readout by multiple technologies, whose characteristics, together with the gas properties, determine the overall detector performances. As discussed in Section \ref{subsubsec:dirgas}, TPCs are intrinsic 3D detectors with low energy threshold, ideal for the 3D tracking of low energy recoils. The CYGNO detector concept combines the following innovative features:
\begin{figure*}[!t]
	\centering
	\includegraphics[width=1\textwidth]{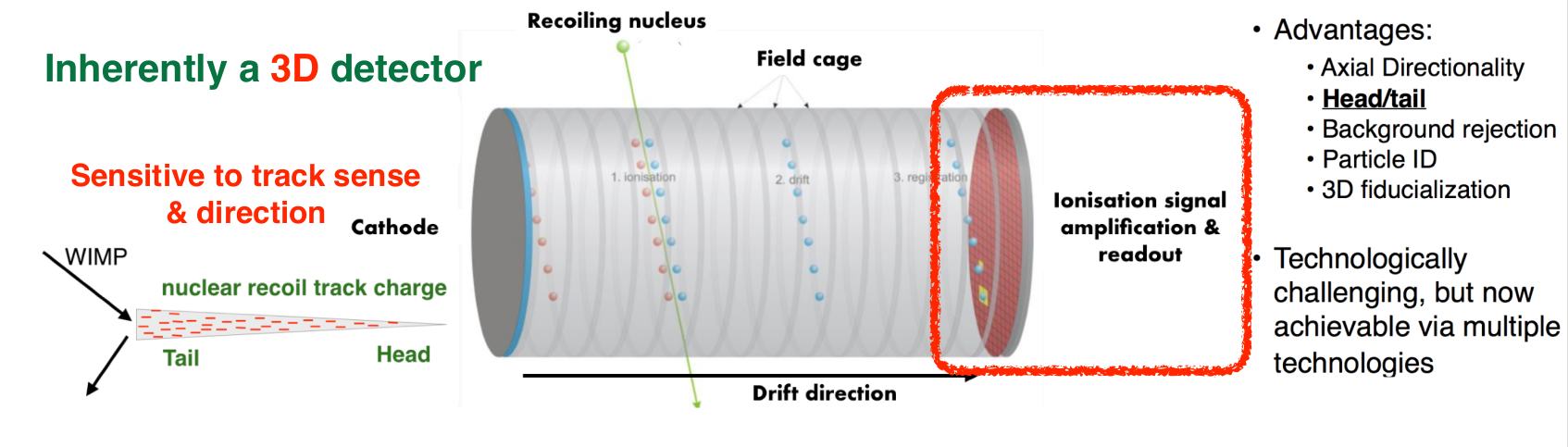}
	\caption{Sketch of a TPC with some of its feature. The intrinsic 3D nature of the detector, sensitivity to HT and particle identification through the track topology are among the key elements of a TPC based detector. Figure adapted from a drawing. Credit to Oliver Sch\"afer (DESY)}
	\label{fig:TPC} 
\end{figure*}
\begin{itemize}
	\item the gas based on a He:CF$_4$ mixture. The multi-target mixture provides simultaneous sensitivity to both SI and, thanks to fluorine, SD couplings, while the CF$_4$ scintillating properties are optimal for an optical readout. The 60/40 proportion chosen grants a small density (roughly equal to the one of air) which allows recoils to travel longer, facilitating the directional reconstruction. (See Section \ref{subsec:cyg_gas} for details.)
	\item The the amplification stage comprised by a triple 50 $\mu$m thick GEMs. GEMs are micro pattern gas detectors (MPGDs) which grants large charge amplification gain, that helps achieving low energy thresholds, and a high granularity, that nicely matches the one resulting from the coupling of the sCMOS camera sensor with the optical system implemented in the CYGNO setup\cite{SAULI20162,Knoll}. In addition, GEMs are among the few MPGDs which can be manufactured in large dimension, a necessary feature for the future stages of the experiment. (See Section \ref{subsec:cyg_GEM} for details.)
	\item The optical readout of the amplified signal by means of PMTs and sCMOS cameras. Optical sensors possess much larger signal to noise ratio than charge detectors, maximising the detection efficiency. The sCMOS cameras offer highly 2D granular readout with limited costs, large quantum efficiency in the optical band, and are able to image large areas when coupled to suitable lenses (see Section \ref{subsubsec:CMOS} for details). Due to the long acquisition time of these devices, the sCMOS camera effectively provides a 2D projection of the recoil track parallel to the plane of the GEM from which topology and directional features can be extracted. In addition, the camera furnishes a measurement of the energy deposited, proportional to the photons collected by the sensor. As a result, the combination of the sCMOS camera with PMTs allows to measure the energy of a recoil with two photo-sensors, while a full 3D reconstruction of the track can be performed by merging the 2D information granted by the sCMOS camera with the information on the orthogonal direction measured by the PMTs.  The 3D topology of the recoils is particularly interesting because it can be used as a handle to discriminate signal from background, namely NRs from ERs, and improves the directional capabilities of the detector. The solid angle coverage, which limits the amount of photons reaching the optical sensors, is balanced by the very large gain achievable with the GEM amplification structure.  (See Section \ref{subsec:cyg_readout} for details.)
	\item The possibility of operation at atmospheric pressure. This guarantees a reasonable volume-to-target-mass ratio, while, at the same time, allowing a reduction in the engineering requirements of the vessel (hence internal backgrounds). Compared with all the directional detectors discussed in Section \ref{subsubsec:dirgas}, CYGNO is the only one (except for D$^3$, whose prototypes anyway never exceed 0.3 l active volume) which operates at atmospheric pressure. This improves the exposure of the experiment, while it does not compromise the recoil track lengths thanks to the low gas density.
\end{itemize}
\subsection{The gas mixture}
\label{subsec:cyg_gas}
\begin{figure}[t]
	\centering
	\includegraphics[width=0.46\textwidth]{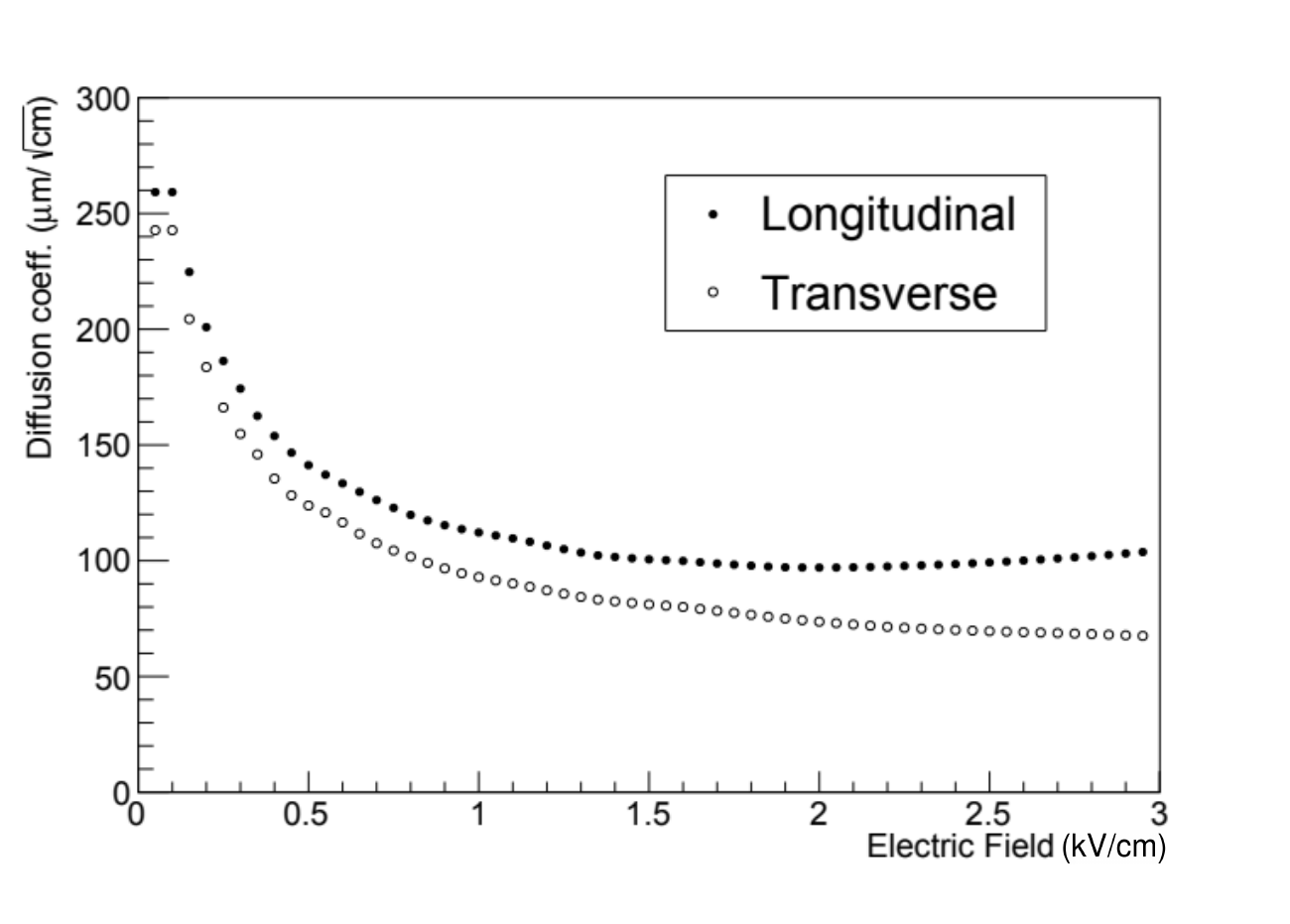}
	\includegraphics[width=0.48\textwidth]{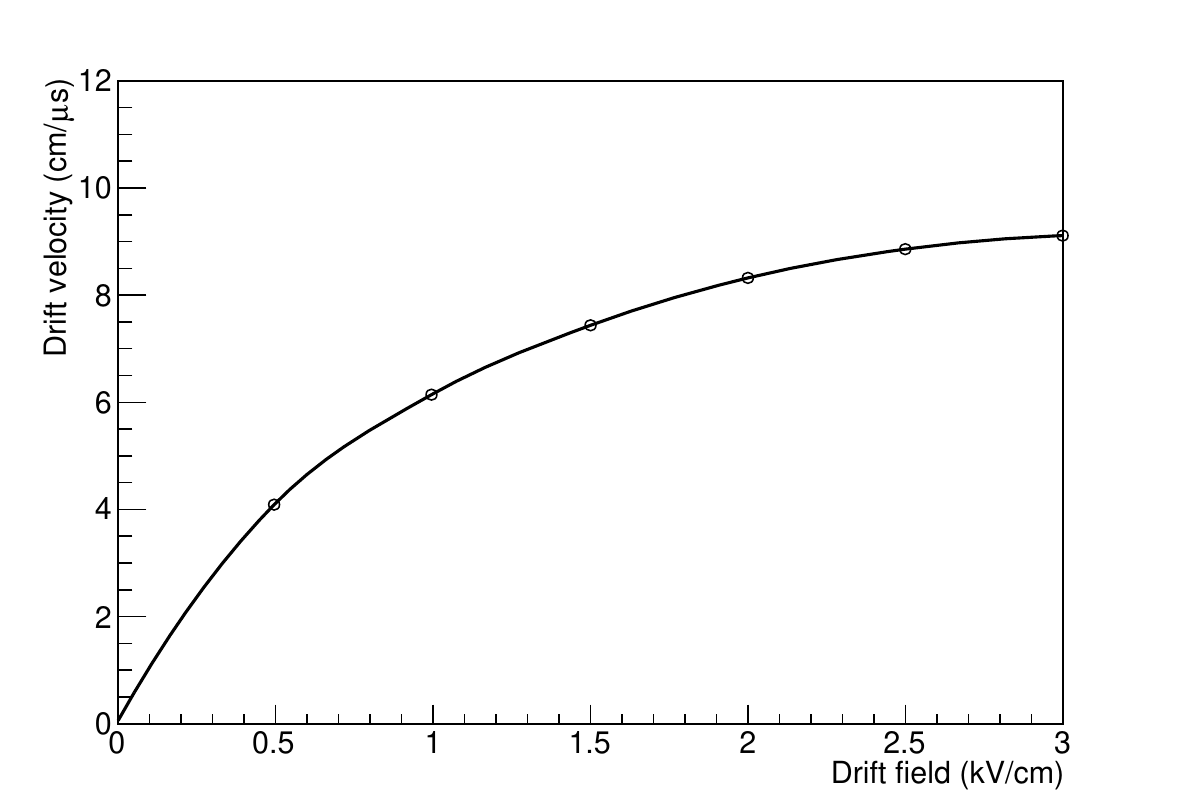}
	\caption{Transverse and longitudinal diffusion coefficients for He:CF$_{4}$ 60/40 (left) and electron drift velocity as a function of the drift field (right). Results obtained by Garfield simulations. Figure taken from \cite{bib:garfield1,bib:garfield2}.}
	\label{fig:hecf4}
\end{figure}
The choice of the gas mixture is of outstanding importance as it determines the experimental properties of the detector and defines the target for the physics searches.\\

Tetrafluoromethane, CF$_4$, is a scintillating gas that produces large light yield in the visible range, also in combination with noble elements \cite{bib:Margato_2013,bib:Fraga:2003uu,bib:Morozov_2012, bib:Kurihara}. The light emission spectrum actually comprises two continua, peaked respectively around 290 nm and 620 nm. The emission in the visible orange region results from the de-excitation from a Rydberg state of the neutral CF$_3^*$ originated from the fragmentation of CF$_4$, with an energy threshold of 12.5 eV. On the other hand, the dissociative ionisation threshold is 15.9 eV and many of the diverse ionic resulting fragmentations of the CF$_4$ end up emitting in the UV band. Gas mixtures with CF$_4$ presents also very low diffusion during drift \cite{SAULI20162}, thanks to the large characteristic scattering cross sections of electron-CF$_4$. A small diffusion coefficient improves the quality of the data as the 3D tracking in a TPC, as discussed in Section \ref{subsubsec:dirgas}.
In the CYGNO experiment, CF$_4$ is combined with He in the gas mixture. Helium is a noble gas: being chemically stable and inert, its use results practical in particle detectors, and can provide large gain especially when coupled with other molecular gases \cite{SAULI20162}. The combination of the very low helium density with CF$_4$ permits to operate CYGNO at atmospheric pressure with an overall target density only a factor 2 larger than in the NEWAGE experiment \cite{NEWAGE1}, that works with pure CF$_4$ at 100 Torr. 

As illustrated in Section \ref{subsec:kinematics}, in the context of direct DM searches, the reduced WIMP-target mass and the experimental energy threshold effectively define the minimum DM mass an experiment is sensitive to. The choice of a light target ($A$ < 10) combined with a typical $\mathcal{O}$(1) keV energy threshold, can therefore allow to extend the sensitivity to DM masses below 10 \Gevc, a parameter space still largely unconstrained as of today (see Figure \ref{fig:SI_status} and Section \ref{subsubsec:limi}). Indeed, the helium employed in the CYGNO gas mixture with $A$=4 maximises the sensitivity to \W mass of few \Gevc for SI couplings. In addition, the good kinematic match of He with low WIMP masses and its small mass produce longer nuclear recoils than heavier targets, easing the measurement of the track directional and sense. 
The large presence of fluorine atoms in the CF$_4$ provides sensitivity to SD coupling also at low WIMP masses, thanks to its relative lightness. With these choices, CYGNO results one of the few experiments in the field (see Section \ref{subsec:status}) simultaneously sensitive to both SI and SD coupling at WIMP masses below 10 \Gevc.\\
A systematic optimisation study was performed to define the He to CF$_4$ ratio that could maximise CYGNO experimental approach gain and light production, while minimising the diffusion and the detector instabilities and a 60/40 proportion was selected \cite{bib:StabilityCygno,bib:roby}.
\begin{figure}[t]
	\centering
	\includegraphics[width=0.7\textwidth]{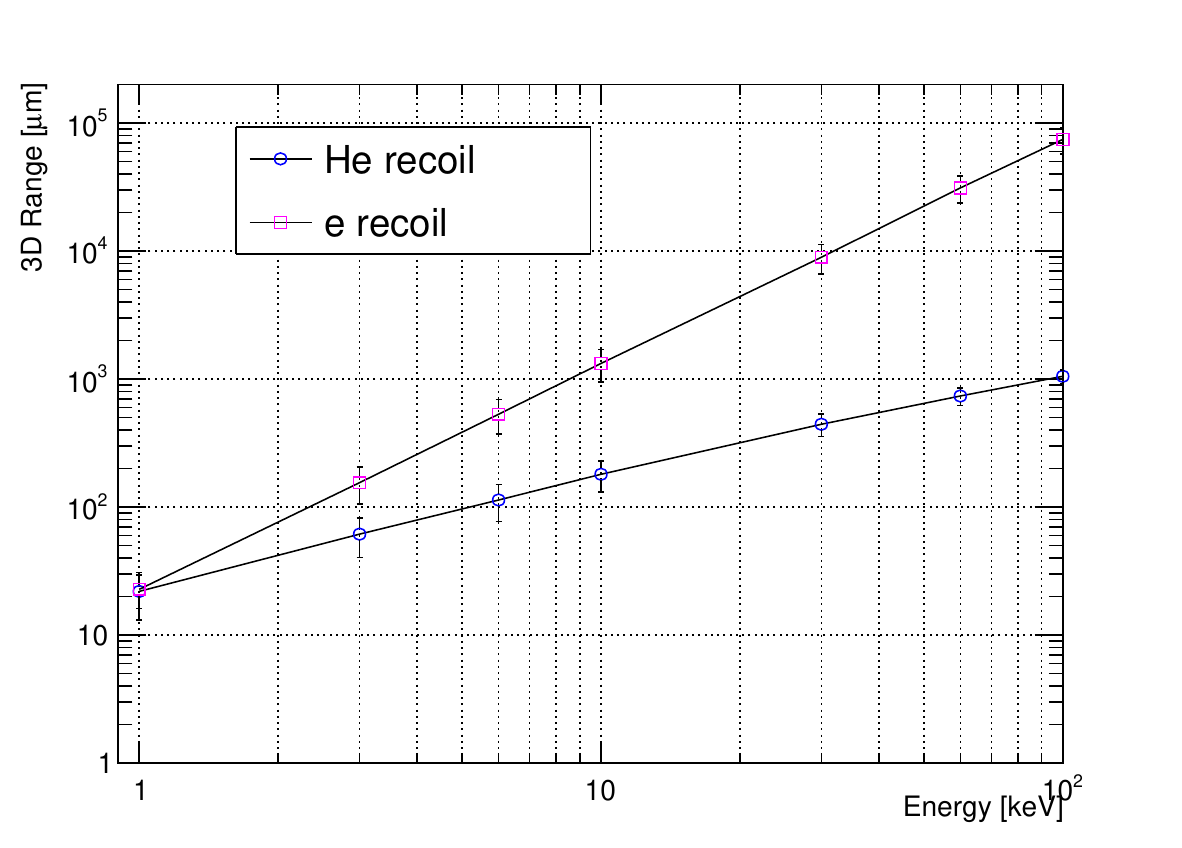}
	\caption{Average simulated 3D distance between the production and absorption point for electron and He-nucleus recoils as a function of their kinetic energy in a He:CF$_4$ (60/40) gas mixture.}
	\label{fig:sim}
\end{figure}
Figure \ref{fig:hecf4} shows the diffusion coefficient per unit of length (left) and the drift velocity (right) as a function of the applied electric field, obtained by a Garfield\footnote{\url{https://www.desy.de/~zenker/FLC/garfieldpp.html}} simulation \cite{bib:garfield1,bib:garfield2}.
The simulation confirms the expectation of the low diffusion coefficient both for transverse and longitudinal diffusion, with less than 130 $\mu m/\sqrt{cm}$ already above 0.5 kV/cm.
The drift velocity is simulated to be around 5 cm/$\mu$s between 0.5 and 1 kV/cm. The expected energy required to produce a pair of ion-electron is estimated to be $\sim$ 38 eV \cite{PDG}. For a minimum ionising particle (MIP), it translates to a loss of energy in ionisation of 250 eV per mm. In order to further minimise the diffusion, Negative Ion Drift operation within the CYGNO optical TPC approach is explored in the context of the synergic INITIUM project (see Section \ref{subsec:INITIUM}). The features of this peculiar modification of the TPC principle are illustrated in Chapter \ref{chap6}, together with the  first demonstration ever of the feasibility of this approach at atmospheric pressure in the context of the CYGNO readout strategy.
Finally, the range of the recoils of electrons and He ions were simulated respectively by means of GEANT4 \cite{bib:geant} and SRIM software. The average 3D ranges (i.e., the distance between the production and absorption point) as a function of the particle kinetic energy are shown in Figure \ref{fig:sim}. The helium recoils keep their range below the mm up to 100 keV$_{\text{nr}}$, leading to the production of dense and bright spot with most of the topology dominated by the Gaussian diffusion. On the contrary, electrons produce longer tracks which translates into less intense spots below 10 keV, whilst centimetres are foreseen at higher energies.
\subsection{Amplification stage}
\label{subsec:cyg_GEM}
The CYGNO  amplification stage is comprised of a stack of GEMs. The employment of an optical readout reduces even more the signal intensity due to the limited solid angle coverage (see Section \ref{subsubsec:CMOS}). Thus, a very large amplification is required to achieve performances suitable for a low energy deposit events search. Among the gas detectors which can furnish extremely large gains, the GEMs are excellent candidates. GEMs are one possible realisation of detectors called Micro Pattern Gas Detectors (MPGD) which were originally developed in order to improve the spatial granularity and the efficiency for high particle fluxes of wire chambers detectors \cite{Knoll,SAULI20162}.\\
\begin{figure}[t]
	\centering
	\includegraphics[width=0.5\textwidth]{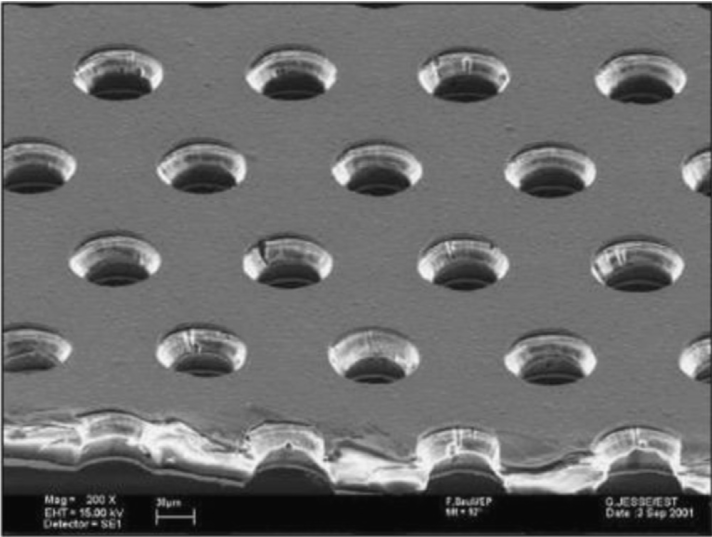}
	\caption{Electron microscope picture of a section of typical GEM electrode, 50 $\mu$m thick. The holes pitch and diameter are 140 and 70 $\mu$m, respectively. Figure taken from \cite{SAULI20162}.}
	\label{fig:GEM}
\end{figure}
The GEM was invented by Sauli in 1997 and an extensive review of their features and characteristics can be found in \cite{SAULI20162}. A GEM is constituted by a thin polymer insulating foil, coated in conducting metal on both sides and pierced with high density of holes. Once large difference of potentials is applied to the two electrodes, high fields of $\mathcal{O}$(10) kV/cm are generated in the holes, strong enough to initiate avalanches. The ions produced in the hole are mostly captured by the upper metallic coating suppressing the ion backflow in the drift region, which helps keeping stable operations even in very high rate environment \cite{Tripathy:2021puz}. The materials employed to realise the GEM can vary, but the most common are fiberglass or \kap for the insulator sheet, and copper for the metallic coating. GEMs can be realised with various thicknesses, from very thin 50 $\mu$m up to 1 mm. For the thin ones, the hole diameter is usually 70 $\mu$m wide, as smaller holes do not improve the effective gain due to a very large charge density in the hole \cite{SAULI20162}. The pitch is typically 140 $\mu$m. Such a small hole is fabricated utilising the etching technique which permits to chemically dig tiny holes in materials. The common geometry of the hole obtained via etching is double conical with the narrower diameter in its centre. With thicker GEMs, the pitch and the diameter hole are increased accordingly. For thicknesses above 300 $\mu$m the holes can be produced by mechanical drilling and then refined with wet etching, resulting in cylindrical shaped holes. The localisation of the avalanches within the holes of the GEMs allows to attain outstanding granularity, especially for the 50 $\mu$m thin ones, mostly preserving the x-y characteristics of the primary electron cloud. When applied to TPCs, a sufficiently segmented readout is able to combine the high granularity with excellent timing resolution \cite{bib:gem}.\\
The secondary electrons produced by a GEM can be collected by an anode or sent to another GEM for a second amplification stage. Indeed, multiple stacking of GEMs has been employed and studied for several applications with total gain achieved  of the order of 10$^4$/10$^6$\cite{SAULI20162,bib:lemon_btf}.
For all these reasons, CYGNO employs an amplification stage comprised of a stack of three thin 50 $\mu$m thin GEMs. R\&D work is under development to further optimise and tailor the amplification structure in the context of the CYGNO experiment, and is discussed in Chapter \ref{chap5}.
\subsection{The optical readout}
\label{subsec:cyg_readout}
Gas luminescence is a well-studied and established mechanism: charged particles travelling in the gas can ionise atoms and molecules but can also excite them. During the de-excitation processes, photons are emitted. The amount and spectrum of light produced strongly depends on the gas, on its density and on the possible presence and strength of an electric field. In most common gas mixtures, the number of emitted photons per avalanche electron can vary between 10$^{-2}$ and 10$^{-1}$ \cite{Knoll,bib:Margato_2013,bib:Fraga:2003uu,MONTEIRO201218}. The optical readout approach is based on the idea of detecting the light produced during the amplification processes rather than the charge, an idea pioneered by studies on parallel plates detector in the 90's \cite{opticalcharpak}. A charge readout is accompanied by electronic noise which can be problematic for very low energy deposits. Conversely, the optical sensors have much larger signal to noise ratios. Moreover, an optical readout allows to locate the sensors outside the sensitive volume to reduce the radioactive and gas contamination. The advantage of positioning the light sensor far from the sensitive volume carries the drawback of the reduction of the total amount of photons collected due to the solid angle coverage. Nevertheless, the very large gains attained with the triple GEM structure in the CYGNO gas mixture are able to compensate for this loss \cite{bib:roby}.
As a directional detector, CYGNO aims at measuring the direction of the recoils along with the energy. To optimally assess the direction and the HT of the recoils a full 3D analysis of the track is necessary. This is achieved by the synergic use of two optical sensors: a scientific CMOS (sCMOS) camera which measures the release of energy along the x-y projection of the tracks, and a set of PMTs which measure the integrated energy and the development of the track along the drift direction (z-axis).
\subsubsection{PMT}
\label{subsubsec:PMT}
\begin{figure}[t]
	\centering
	\includegraphics[width=0.5\textwidth]{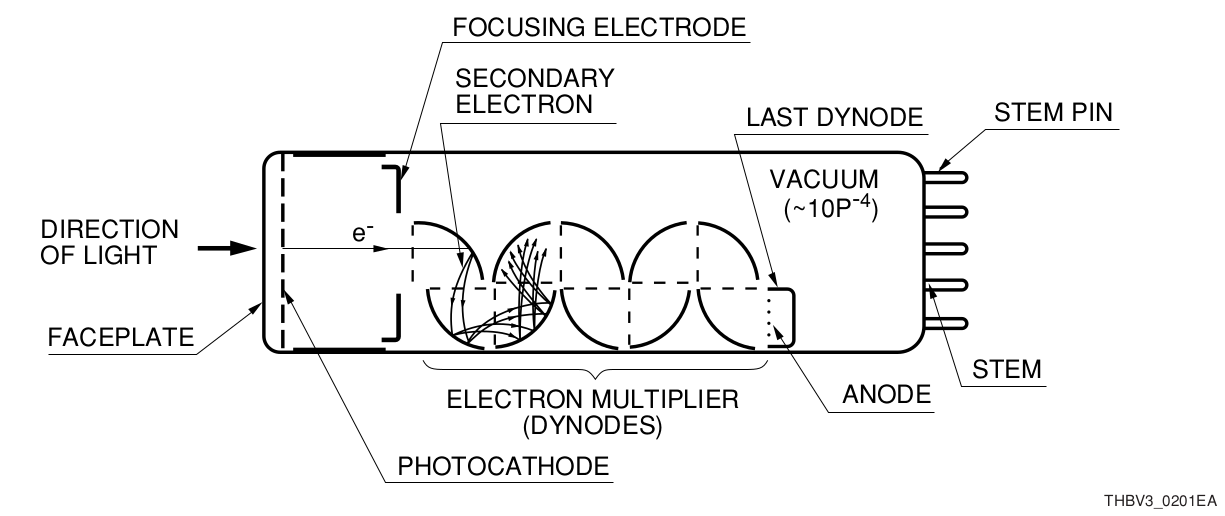}
	\caption{Sketch of a linear PMT with each component highlighted. On one side there is an input window which covers a photocathode, the sensitive material responsible to convert photons into electrons. Inside the tube there are a focusing electrodes, and a sequence of electron multipliers called dynodes. At the end of the chain an anode collects electrons and can be connected to an output cable for the transmission of the signal. Figure taken from \cite{PMT}.}
	\label{fig:PMT}
\end{figure}
The photomultiplier tube (PMT) is a photon detector, whose origin goes back to the mid 1930s \cite{PMT}, that is able to detect single photons while maintaining a large dynamic range. A sketch of a PMT is displayed in Figure \ref{fig:PMT}. A PMT is a vacuum tube usually sealed into an evacuated glass tube. On one side there is an input window which covers a photocathode, the sensitive material responsible to convert photons into electrons. Inside the tube there are a focusing electrodes, and a sequence of electron multipliers called dynodes. At the end of the chain an anode collects electrons and can be connected to an output cable for the transmission of the signal. A voltage divider partitions a high voltage so that there is a sequential voltage difference between the photocathode, each dynode and the anode.\\
Photons impinging on the photocathode can be converted into electrons by means of the photoelectric effect. The emitted electron is focused by the focusing electrode on the first dynode. These are composed by a material which can emit multiple electrons when hit by a fast electron. As a consequence, for each electron that arrived at the first dynode, three or four more are emitted by the dynode. The process is repeated for all the stages resulting in a large amplification of the signal which is finally collected by the anode circuitry. The number of dynodes, their material and the voltage divider can vary to optimise the dynamic range and the gain, which can be as large as 10$^{6}$.\\
The wavelength sensitivity and quantum efficiency (QE) of the PMT depend on the materials of the window and the photocathode. The windows are most commonly made of borosilicate glass, which permits photons above 300 nm to be transmitted and can be manufactured with relative low radioactive material. The most utilised photocathode materials are \emph{bialkali} (usually Sb-K-Cs), where the name comes from the presence of more than one alkali element. The QE of these photocathodes peaks at 400 nm, being sensitive up to 650 nm. Other materials like GaAsP(Cs) have higher QE in the visible range. Typical QE of PMTs do not exceed 25\%.\\
\begin{figure}[t]
	\centering
	\includegraphics[width=0.45\textwidth]{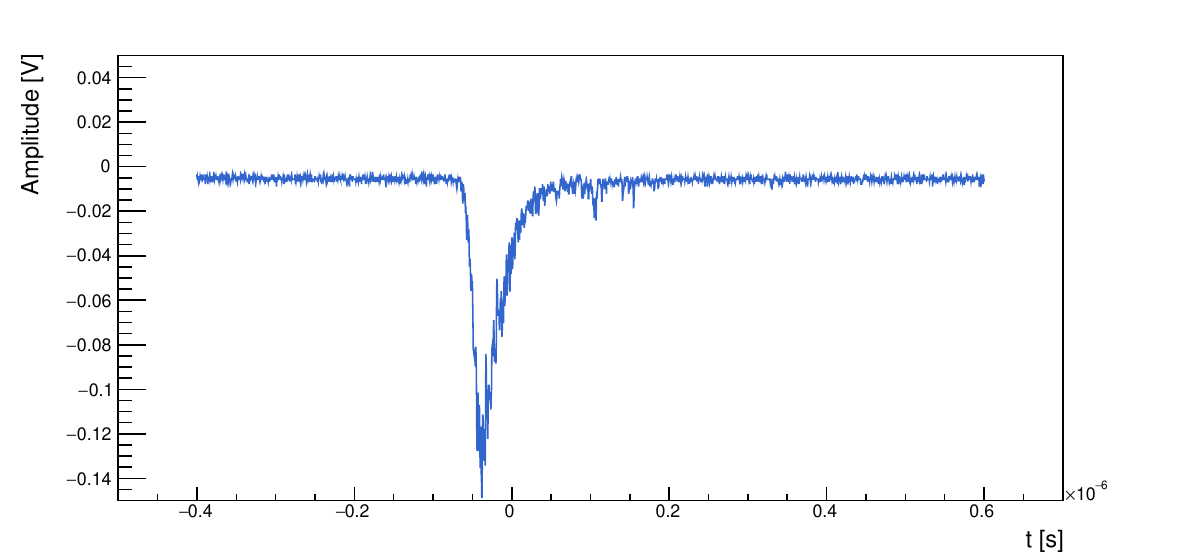}
	\includegraphics[width=0.45\textwidth]{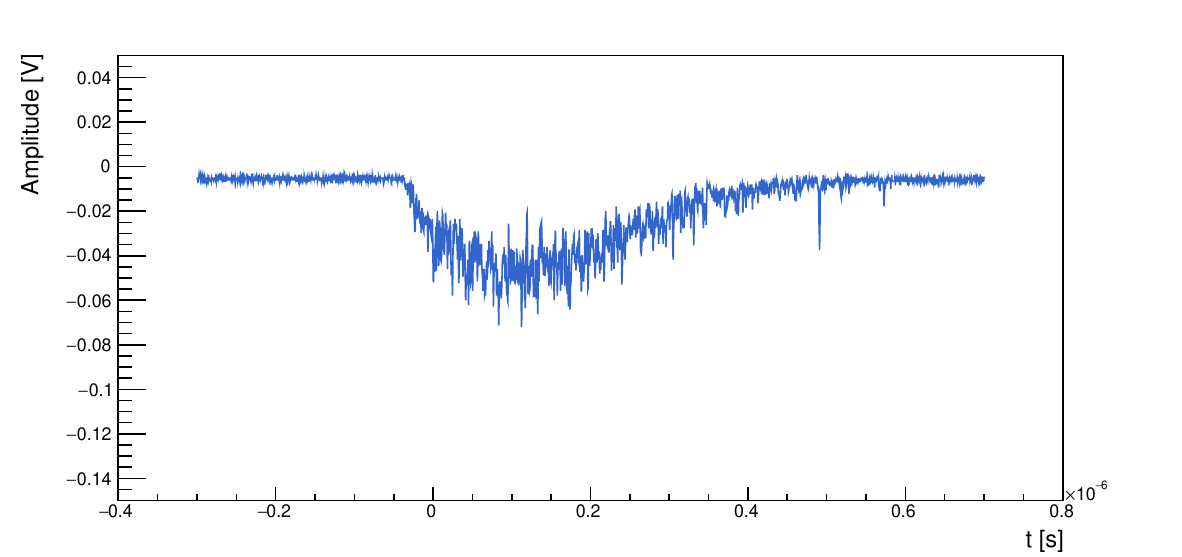}
	\caption{Examples of PMT waveforms for a track parallel to the GEM plane (left) and tilted with respect to the same plane (right). The difference in time of arrival of the primary charge in the tilted track is mirrored in the time evolution of the waveform.}
	\label{fig:PMTwave}
\end{figure}
When applied to optically readout TPCs, the measurement of the current signal collected at the anode induced by the photons detected can provide relevant pieces of information. The total charge collected by the anode is proportional to the amount of photons converted, thus it corresponds to a measurement of the energy. Depending on the QE and the dimension of the photocathode, the energy resolution can vary, but for low energy deposits, the contribution to it is typically dominated by the fluctuation of the primary electrons generated in the sensitive region of the detector. With the common rise time of a PMT of the order of ns, a sampling rate above 400 MS/s of the PMT waveforms allows to characterise the track development along the drift direction (for typical electron drift velocities). This permits not only to infer the inclination of a recoil with respect to the GEM plane, but also to attain information on the topology of the track, fundamental for direction and HT recognition. Figure \ref{fig:PMTwave} shows examples of PMT waveforms for a track parallel and tilted with respect to the GEM plane on the left and the right panel respectively. The difference in time of arrival of the primary charge in the tilted track is measured in the time evolution of the waveform. In the CYGNO gas mixture with the typical CYGNO drift field (with an electron drift velocity of 5 cm/$\mu$s), a $\mathcal{O}$(1) GS/s sampling rate of the PMT waveform grants a granularity in the z direction of roughly 50 $\mu$m. Moreover, if a set of PMTs is employed, the different intensity response of each light sensor can help to reconstruct the x-y position by utilising the barycentre of the signal intensity.
\subsubsection{sCMOS}
\label{subsubsec:CMOS}
A scientific CMOS camera (sCMOS) is a high sensitivity low noise photon detector of the family of active pixel sensors (APS) based on the complementary metal-oxide semiconductor (CMOS) pinned photodiode (PPD) technology.\\
The CMOS is the technology employed to manufacture the MOSFET transistors, the fundamental building block of modern electronics\cite{CMOStech}.\\
\begin{table}[!t]
	\centering
	\begin{adjustbox}{max width=1.01\textwidth}
		\begin{tabular}{|c|c|c|c|c|}
			\hline
			
			\large{Model} & \large{\# pixels} & \large{Pixel size ($\mu$m$^2$)}  &  \large{Noise RMS} (e$^-$)& \large{Frame rate (Hz)} \\ \hline
			ORCA Flash\tablefootnote{\url{https://www.hamamatsu.com/content/dam/hamamatsu-photonics/sites/documents/99_SALES_LIBRARY/sys/SCAS0134E_C13440-20CU_tec.pdf}} & 2048$\times$2048 & 6.5$\times$6.5  &  1.4 & 30  \\ 
			ORCA Fusion\tablefootnote{\url{ https://www.hamamatsu.com/content/dam/hamamatsu-photonics/sites/documents/99_SALES_LIBRARY/sys/SCAS0138E_C14440-20UP_tec.pdf}}& 2304$\times$2304 & 6.5$\times$6.5  &  0.7 & 5.4  \\ 
			ORCA Quest\tablefootnote{\url{ ttps://www.hamamatsu.com/content/dam/hamamatsu-photonics/sites/documents/99\_SALES\_LIBRARY/sys/SCAS0154E\_C15550-20UP\_tec.pdf}} & 4096$\times$2304 & 4.6$\times$4.6  & 0.27 & 5  \\ \hline
		\end{tabular}
	\end{adjustbox}
	\caption{Table describing the principal features of three sCMOS cameras manufactured by Hamamatsu and employed by different CYGNO prototypes. The technological advancement is clearly visible as newer cameras (lower rows) possess more pixels with reduced noise. The content of the columns \textit{Noise RMS} and \textit{Frame rate} refers to the \textit{slow scan} of the camera, the one that grants the smaller noise RMS.}
	\label{tab:hama}
\end{table}
An APS is a silicon photo-detector in which the charge collected by the conversion of the photons in the depleted region of the sensor is actively handled and readout. A PPD is an APS wherein the active handling of the charge is performed by a combination of $n$ and $p$ doped silicon architectures. PPDs can be manufactured very small, $\mathcal{O}$(10) $\mu$m per side, so that arrays and matrices of these PPDs can be arranged next to each other to form camera sensors, highly granular photo detectors. The most renowned kind of PPD based camera sensors are charge coupled devices (CCDs) and CMOS cameras\cite{PPD}.\\
In CCDs, the semiconductor structure coupled to each PPD pixel is employed to efficiently transport the charge collected by one of them to the adjacent one in a shift register technique. At every charge transfer the information carried by the pixels is shifted towards the boundary of the sensor where a circuit is located to readout the charge of a single pixel. The transport is carried out until all the pixels are readout \cite{CCDopt}.\\
CMOS cameras, instead, exploit the fact that every single PPD pixel is coupled to a dedicated CMOS circuit whose main goal is to directly readout the signal. The CMOS circuit is also responsible for the noise reduction, reset of the pixel, and in some implementations also for its exposure \cite{PPD}. The simplest architecture comprises 4 CMOS transistors,
though modern CMOS circuitry can contain more CMOS elements to provide low noise measurements\cite{PPD}.\\
CCD cameras usually possess larger light collection efficiency and higher uniformity than CMOS ones, due to the depth of the depleted region of the pixels and the different readout methodology. For long time CCDs had better performances in terms of noise level,
but recent development in the CMOS circuitries improved the CMOS features enough to surpass CCDs achievements \cite{Zhan_2017,CCDnoise,CCDcmosnoise}, especially for scientific purposes (sCMOS). Moreover, in the majority of the applications, CCD sensors require lower temperature for science operation than a CMOS, in order to limit the intrinsic noise \cite{CCDhubble,LSST}. The time required for the exposure of the pixels is usually quite high for both types of sensor, reaching levels of tens or even hundreds of ms. The CMOS cameras can be manufactured cheaper and with pixels of smaller size, granting extremely high granularity. In addition, CMOS cameras keep their power consumption lower than CCDs, thanks to the small voltages necessary to power the CMOS technology with respect to the ones required by the shift register. It is important to note that these kinds of camera have a wide variety of applications in the market, as, for example, cellphone cameras are CMOS image sensors themselves. Therefore, the technological pull driven by market industry can foster a fast improvement of these types of sensor technology. Thanks to the low noise, high granularity and large potential for technological improvement, the sCMOS camera is chosen for the CYGNO experiment.\\
\begin{figure}[t]
	\centering
	\includegraphics[width=0.4\textwidth]{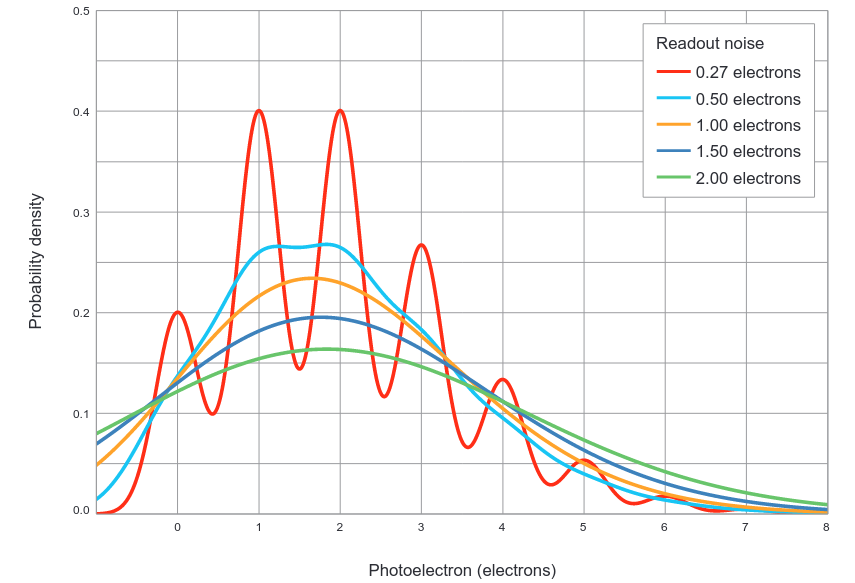}		\includegraphics[width=0.56\textwidth]{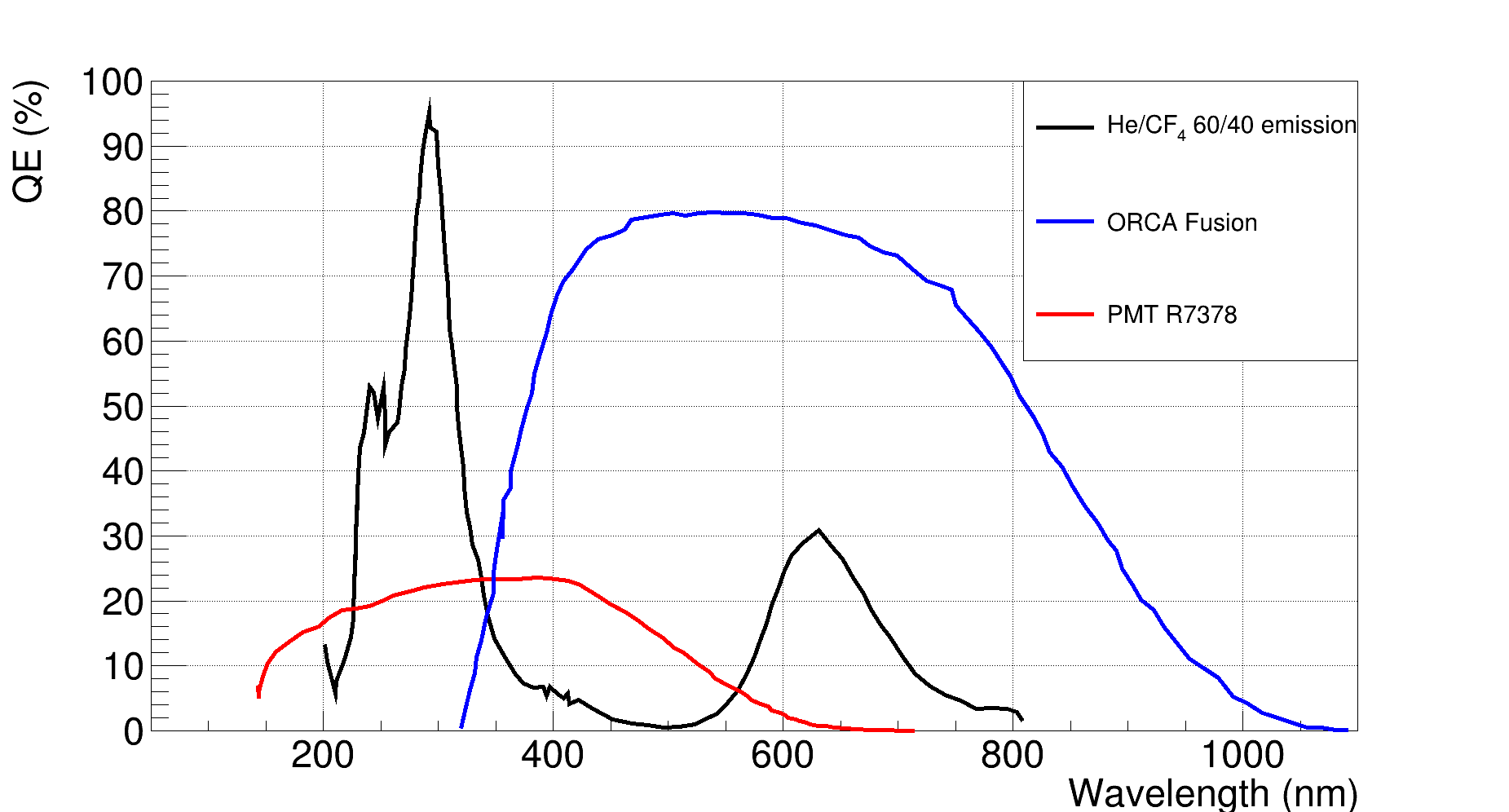}
	\caption{Some characteristics of Hamamatsu sCMOS sensors. On the left panel, the readout performances as a function of the photons converted in the sensor with different RMS noise levels in units of electrons for different camera models of different generation (taken from Hamamatsu). On the right panel, the QE of the Hamamatsu ORCA Fusion sCMOS camera and of a R7378 PMT superimposed to a renormalised emission spectrum of the CYGNO gas mixture. He:CF$_4$ spectrum taken from \cite{bib:Margato_2013}, QE of ORCA Fusion from Hamamatsu, and QE of PMT from \cite{PMT}.}
	\label{fig:CMOS1}
\end{figure}
CYGNO employs CMOS cameras manufactured by the Hamamatsu company and, in time, the collaboration tested more modern and better performing models whose main characteristics are displayed in Table \ref{tab:hama}. The technological advancement in the production of more pixels and with smaller size is clearly visible. More importantly, the noise RMS of each single pixel is reduced in the recent models, with the QUEST having improved of a factor 10 with respect to the FLASH. In particular, the QUEST model has such low noise, of 0.27 electron in RMS, that allows to count the number of photons converted in a pixel with less than 1 photon uncertainty. For this reason, they are also known as quantitative CMOS (qCMOS) cameras. This feature is displayed in Figure \ref{fig:CMOS1} on the left panel, which shows the simulation of the probability density of the photoelectron signal as a function of the number of photoelectrons generated in a pixel (thus photons converted). This probability density is the convolution of the amount of photons converted in a pixel with its intrinsic noise which results in a probability function of the expected signal. The noise level of the QUEST (red line) is the only one that allows to separate signals caused by a different number of photons, permitting direct counting.

With the bulk of the photon sensitive region made of silicon and the protective window of the sensor composed of glass, sCMOS cameras have large sensitivity in the optical range. The QE can reach peaks of 80\% at about 600 nm. This value perfectly matches the peak of emission in the visible range of the He:CF$_4$ CYGNO gas mixture. Figure \ref{fig:CMOS1} on the right shows the QE of the ORCA Fusion (second row of Table \ref{tab:hama}), along with the QE of the R7378, the bialkali borosilicate PMT \cite{PMT} employed in LIME (see Section \ref{sec:LIME}), together with the CYGNO gas mixture emission spectra, as measured in \cite{bib:Fraga:2003uu}, normalised to 1. It is important to note that the sCMOS sensor has much larger QE than common PMT and the wavelength sensitivity nicely matches the emission of the gas mixture.
\paragraph{Optics}
\begin{figure*}[t]
	\centering
	\includegraphics[height=5.8 cm]{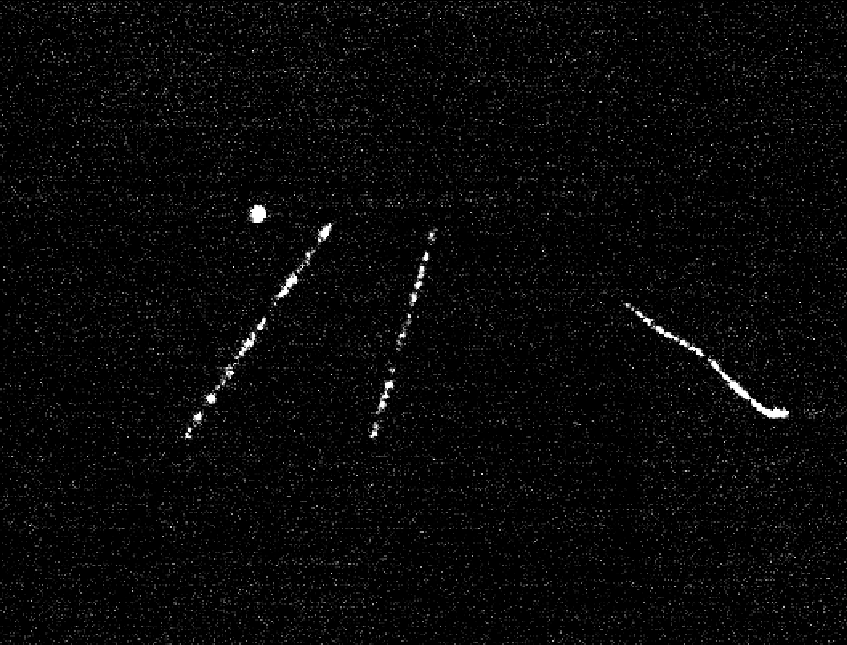}
	\includegraphics[height=5.8 cm]{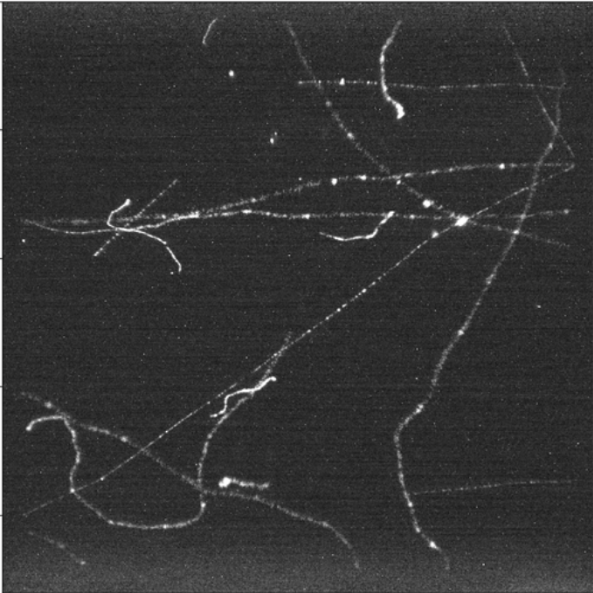}
	\caption{Two examples of sCMOS images taken with two CYGNO prototypes, MANGO (left) and LIME (right), of the natural radioactivity (details on the prototypes in Section \ref{sec:timeline}). }
	\label{fig:exCMOS}
\end{figure*}
Modern CMOS cameras with megapixel matrices allow to image very large areas if coupled to the proper optical system. Indeed, it is possible to image areas as large as $\mathcal{O}$(0.5) m$^2$ with a single camera while keeping an excellent granularity matching the segmentation of the GEM amplification stage.
The lens utilised for all the CYGNO detectors is a Schneider Xenon lens  with focal length $f=$25.6 mm, f-number or aperture ratio $N=$0.95 (see text below) and 0.85 transparency in the optical band (technical sheet in Appendix \ref{appC}). In order to image areas of 30$\times$30 cm$^2$ with this optical system, the sensor needs to be positioned $\mathcal{O}$(20) cm away from the last amplification GEM plane. Figure \ref{fig:exCMOS} shows examples of sCMOS images taken with two CYGNO prototypes, MANGO (left) and LIME (right), of the natural radioactivity (details on the prototypes in Section \ref{sec:timeline}). These images clearly display the capability of imaging the recoil tracks.
As the camera needs to be positioned far from the amplification plane, the solid angle covered is a key ingredient to evaluate the light collection properties of the sCMOS camera.\\
Figure \ref{fig:optics} shows a sketch of an optical system with a thick lens in the Gaussian approximation projected on the plane generated by the optical axis (i.e. the axis that joins the object, the sensor and crosses the lens in its centre) and an orthogonal axis (for circular lenses considered, which other direction is irrelevant)\cite{vignetting}. OP is the object plane where the source on focus is positioned, while SP is the sensor plane where the photo-sensor is located. EP and XP are the entrance and exit planes where the photons enter and exit the lens. H and H' are the hyperfocal planes, corresponding to the position of a thin lens with the same optical behaviour of a thick lens \cite{vignetting}. $D$ is the radius of opening of the lens which determines the area of it which can accept photons. Practically, it can be modified by changing the position of the stopper of a lens. The infinitesimal area emitting photons is $dA$, $u$ is the tangent of the opening angle, $s$ the distance between the object and the hyperfocal plane H, $s'$ the distance between the sensor and the hyperfocal plane H', $f_F$ is the focal distance, $F$ the focal point, and in the general case $F'=F$ and $f_R'=f_F=f$. The sketch shows the functioning of a generic thick lens with the photons in the yellow region on the left of EP that reach the lens and are focused on the sensor by the lens. 
\begin{figure*}[t]
	\centering
	\includegraphics[width=0.95\linewidth]{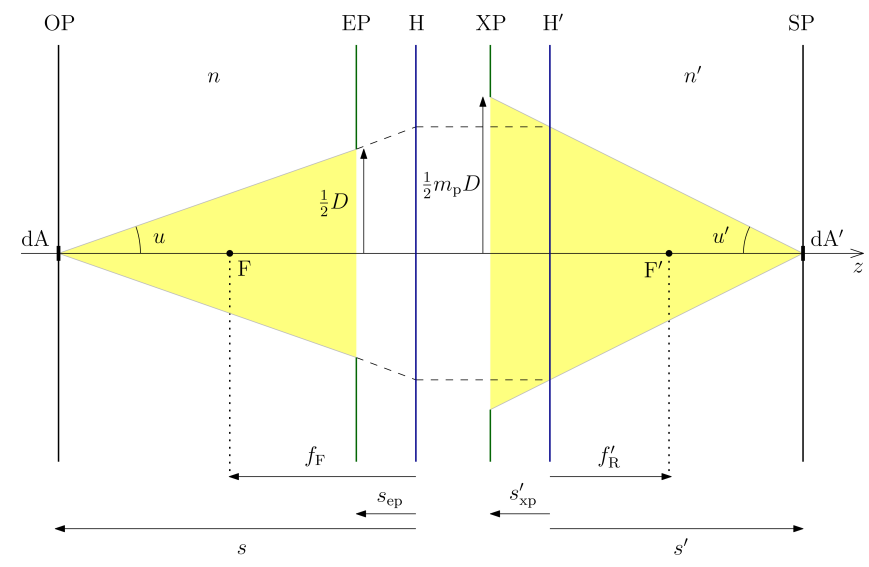}
	\caption{Schematic of a lens in the Gaussian approximation taken from \cite{vignetting}. The yellow band refers to the amount of solid angle covered that emitted from the object plane (OP) is focused on the sensor plane (SP).}
	\label{fig:optics}
\end{figure*}
When considering OP, EP, XP and SP parallel, the flux of photons $\Phi$ which reaches the lens, emitted from the area $dA$, can be evaluated as\cite{vignetting}:
\begin{equation}
\label{eq:optics1}
\Phi= \pi LdAu^2
\end{equation}
with $L$ the luminosity of the source. The fraction of solid angle covered by the lens, $\Omega_f$, can be estimated from the previous Equation and dividing by the total solid angle of the emission of the photons, $4\pi$:
\begin{equation}
\label{eq:optics2}
\Omega_f=\frac{\Phi}{LdA}\cdot\frac{1}{4\pi}=\pi u^2\frac{1}{4\pi}= \frac{\left(\frac{D}{2}\right)^2}{4 s^2}
\end{equation}
The relationship between the distances $s$ and $s'$ to the magnification $I$ can be considered identical to linear optics of thin lenses if $s>>s_{ep}$, with the latter being the distance between the hyperfocal plane and the entrance plane. In the context of the CYGNO experiment, this is true, thus, it is valid that:
\begin{equation}
\label{eq:optics4}
\frac{1}{f}=\frac{1}{s}+\frac{1}{s'}.
\end{equation}
\begin{equation}
\label{eq:optics3}
I=\frac{y'}{y}=\frac{s'}{s},
\end{equation}
with $y$ and $y'$ the object and image dimension respectively.
Combining Equation \ref{eq:optics2} with \ref{eq:optics4} and \ref{eq:optics3} it is possible to derive:
\begin{equation}
\label{eq:opticsfin}
\Omega_f=\frac{1}{\left(4N\left(\frac{1}{I}+1\right)\right)^2},
\end{equation}
with $N$ the f-number or aperture ratio, defined as $N=D/f$.\\
The typical solid angle covered with the Schneider lens and, for example, the ORCA Fusion camera (second row of Table \ref{tab:hama}) when imaging an area of 30$\times$30 cm$^2$ is of $\mathcal{O}$(10$^{-4}$), significantly reducing the number of photons that reach the active sCMOS sensor. As a result, an extremely high gain is required from the amplification stage, which leads to the employment of the stack of three GEMs (see Section \ref{subsec:cyg_GEM}).
\section{CYGNO timeline}
\label{sec:timeline}
\begin{figure*}[t]
	\centering
	\includegraphics[width=0.95\linewidth]{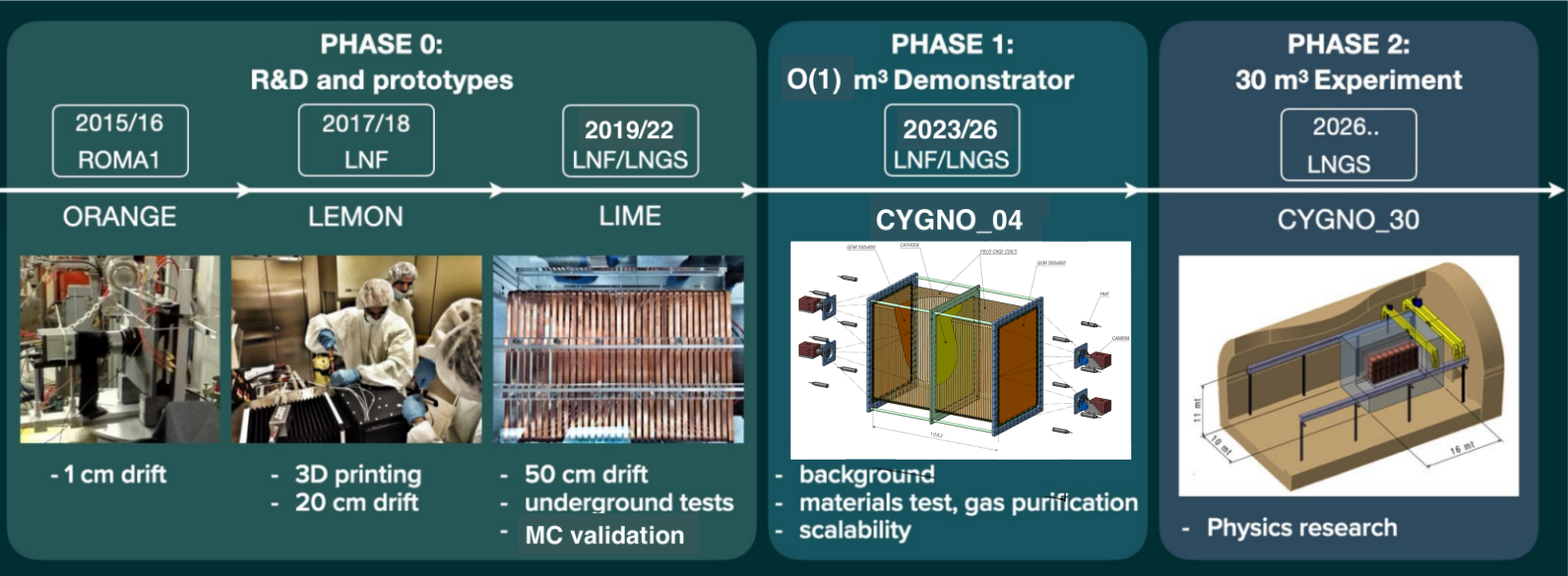}
	\caption{Schematic of the roadmap of the CYGNO experiment.}
	\label{fig:roadmap}
\end{figure*}

The CYGNO project aims to deploy a large scale detector for directional search of DM and Solar neutrino spectroscopy. In order to characterise and optimise the experimental technique, numerous prototypes were developed. Figure \ref{fig:roadmap} shows the roadmap of the experiment, while Table \ref{tab:summaryprot} shows some of the main features of the prototypes. The first prototype, ORANGE, had a 10 $\times$ 10 cm$^2$ area with a 1 cm drift gap and was used to demonstrate the possibility of detecting $\mathcal{O}$(1) keV energy deposits in the gas coupling a sCMOS camera to a GEM amplification stage, and to assess the feasibility of the 3D reconstruction with the addition of a PMT\cite{bib:roby}. ORANGE was followed by LEMOn, which possesses a 20 cm drift gap and a larger readout area (20 $\times$ 24 cm$^2$). LEMOn was utilised to deepen the knowledge on the performances of the detector as a function of a larger drift distance and area. A prototype with a 10 $\times$ 10 cm$^2$ readout area and variable drift region, MANGO (not present in Figure \ref{fig:roadmap}), was also manufactured in 2018 to perform R\&D tests on the gas properties and amplification structures. In 2019, the LIME prototype was completed, a 50 cm drift with 33 $\times$ 33 cm$^2$ readout area detector, which represents the fundamental building block of the larger scale experiment. After overground commissioning, LIME was installed in underground Laboratori Nazionali del Gran Sasso (LNGS) in June 2022, in order to characterise the detector in a realistic environment. All the mentioned prototypes constitute the CYGNO PHASE\_0, whose goal is to validate the full performances of the optical readout via APS commercial cameras and PMTs and the Monte Carlo simulation of the expected backgrounds. PHASE\_0 is to be followed by PHASE\_1, CYGNO-04, a 0.4 m$^3$ detector to be installed at the LNGS which will be realised with all the technological and material choices foreseen for the future CYGNO-30. This phase aims at demonstrating the scalability of the experimental approach and the potentialities of the large PHASE\_2 detector to reach the expected physics goals.
Finally, CYGNO-30, a $\mathcal{O}$(30) m$^3$ detector, represents the PHASE\_2 of the project. A CYGNO-30 experiment would be able to give a
significant contribution to the search and study of DM with masses below 10 \Gevc for both SI and SD coupling, with the addition of exploiting all the advantages of a directional DM search (see Section \ref{sec:directional}).
The following Sections will describe the various prototypes and the main results achieved with each of them.
\begin{table}[!t]
	\centering
	\begin{adjustbox}{max width=1.01\textwidth}
	\begin{tabular}{|c|c|c|c|c|}
		\hline
		\Large{Prototype} & \Large{Readout area (cm$^2$)} & \Large{Drift length (cm)} &\Large{Readout} & \Large{Purpose} \\ \hline
		ORANGE & 10 $\times$ 10 & 1 & 1 sCMOS + 1 PMT & Proof of technique  \\ 
		LEMOn & 20 $\times$ 24 & 20 & 1 sCMOS + 1 PMT & Stability and background studies  \\ 
		MANGO & 10 $\times$ 10 & 1-15 & 1 sCMOS + 1 PMT & R\&D for Phase\_2 \\ 
		LIME & 33 $\times$ 33 & 50 & 1 sCMOS + 4 PMT & Underground operation, MC validation  \\ 
		CYGNO-04 & 50 $\times$ 80 & 2 $\times$ 50 & 4 sCMOS + 12 PMT & Scalability test, radiopurity  \\ 
		CYGNO-30 & N.A. & N.A. & N.A. & Physics case  \\ \hline
	\end{tabular}
\end{adjustbox}
	\caption{Short summary of the main features of the various prototypes of the CYGNO project.}
	\label{tab:summaryprot}
\end{table}
\section{The MANGO detector}
\label{sec:mango}
\begin{figure}[t]
	\centering
	\includegraphics[width=0.6\textwidth]{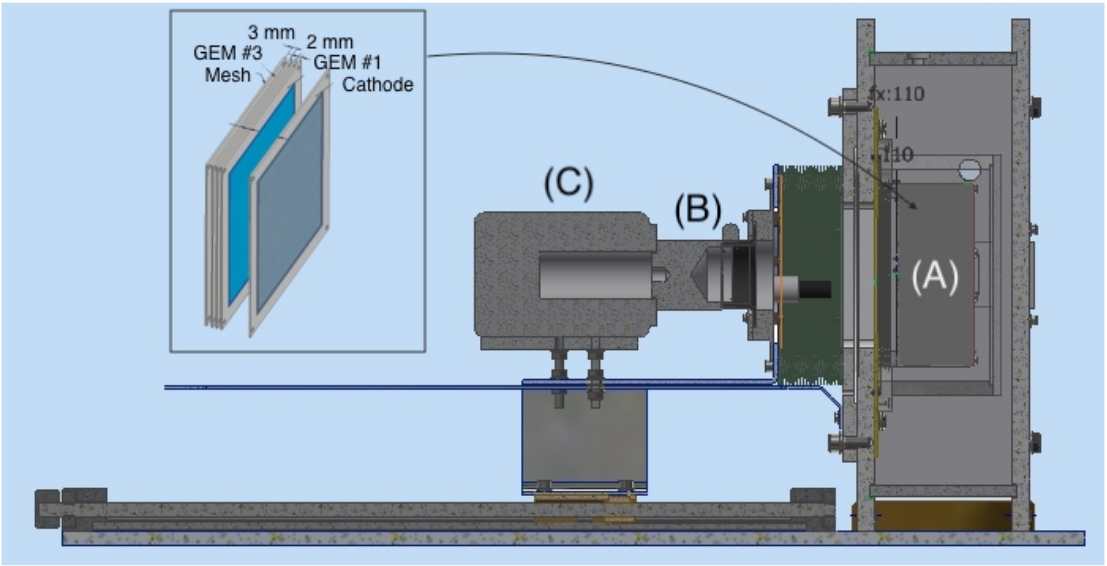}
	\caption{A simple representation of the MANGO setup with exemplified a triple GEM amplification.}\label{fig:mangosketch_0}
\end{figure}

The Multipurpose  Apparatus for Negative ions studies with GEM Optically readout (MANGO) is a small detector with a 10 $\times$ 10 cm$^2$ amplification area and a variable drift region. This detector is employed in the studies presented in Chapter \ref{chap5} and \ref{chap6}. 
A sketch of the MANGO prototype and the internal TPC structure is shown in Figure \ref{fig:mangosketch_0} and described in details \cite{bib:EL_cygno}. The amplification stage consists in a stack of multiple 10 $\times$ 10 cm$^2$ GEMs spaced 2 mm, with a transfer field of 2.5 kV/cm in between the GEMs. MANGO is typically operated with triple 50 $\mu$m thin GEM, but different thicknesses and stacking option were tested as it will be shown in Chapter \ref{chap5}.  The GEMs are numbered from 1 to 3 with GEM1 being the closest to the drift region. The GEMs are powered by the HVGEM \cite{CORRADI200796}, a dedicated high voltage supply developed by the Laboratori Nazionali di Frascati (LNF), which works as an active high voltage (HV) divider. The HVGEM is employed to power seven floating independent channels (that is the upper and bottom electrodes of the 3 GEMs plus one potential extra electrode), with each channel isolated up to 5 kV to ground and with a high sensitivity ($\sim$ 1nA) current meter.
The TPC, typically operated in continuous gas flux mode, is enclosed in a 3D printed black plastic light-tight box that contains a gas-tight acrylic internal vessel. A thin window of highly transparent (> 0.9) \myl decouples the gas detector from the optical readout, which consists in a PMT (Hamamatsu H3164-10\footnote{\url{https://www.hamamatsu.com/eu/en/product/optical-sensors/pmt/pmt-assembly/head-on-type/H3164-10.html}}) and a C14440-20UP ORCA-Fusion sCMOS camera (second row of Table \ref{tab:hama} and C in Figure \ref{fig:mangosketch_0}), placed at a distance of $(20.5 \pm 0.3)$ cm and focused on the last GEM amplification plane. The camera possesses a squared silicon sensor of 1.498 $\times$ 1.498 cm$^2$ are, segmented in 2304 $\times$ 2304 pixels. Each pixel has a QE of about 80\% at 600 nm, and a readout noise of 0.7 electrons RMS. The optical sensor is equipped with a Schneider Xenon lens (Appendix \ref{appC} and B in Figure \ref{fig:mangosketch_0} left). Within this scheme, the camera images an area of $\sim$ 11.3 $\times$ 11.3 cm$^2$, resulting in an effective pixel size of $\sim$ 49 $\times$ 49  $\mu$m$^2$. 
\section{The LEMOn detector}
\label{sec:LEMOn}
\begin{figure}[t] 
	\centering
	\includegraphics[width=0.8\linewidth]{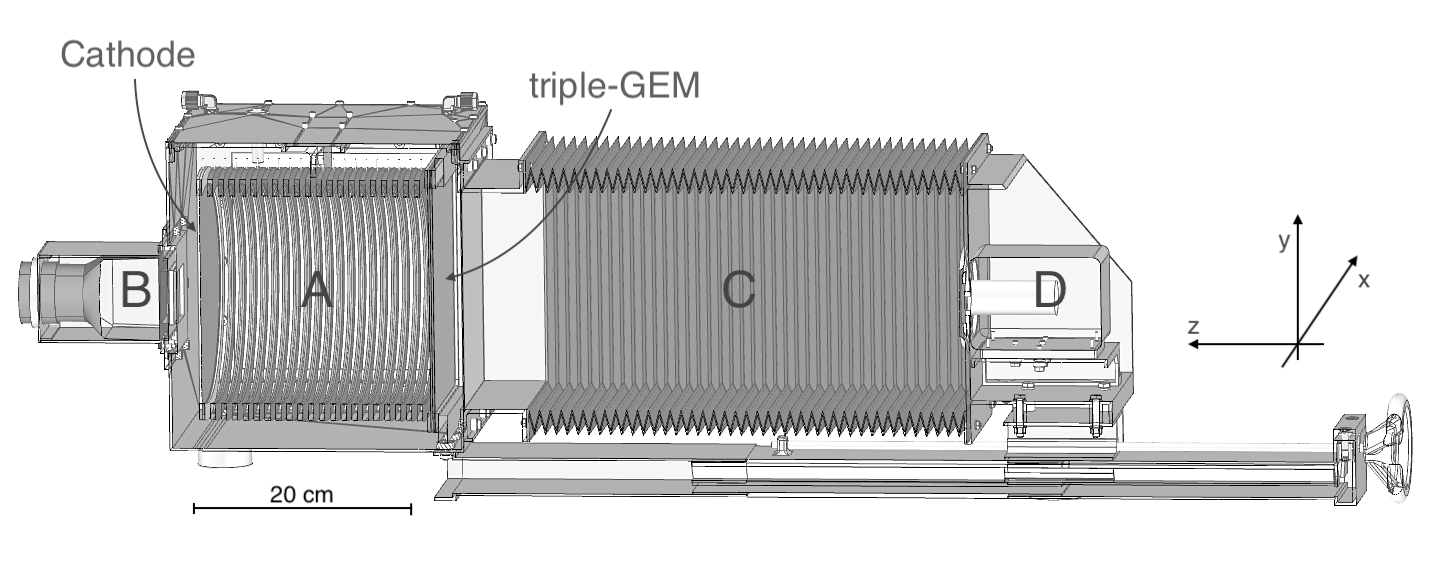}
	\caption{The LEMOn prototype \cite{bib:Antochi_2021}. The elliptical sensitive volume (\textbf{A}), the fast photomultiplier (\textbf{B}), the optical bellow (\textbf{C}) and the sCMOS-based camera (\textbf{D}) are indicated.}
	\label{fig:lemonsketch}
\end{figure}

A sketch of the Long Elliptical MOdule (LEMOn) detector is shown in Figure \ref{fig:lemonsketch}. A 7 litres active volume TPC with a 20 cm drift length and a 24 $\times$ 20 cm$^2$ readout area is enclosed in a gas-tight acrylic vessel and operated in continuous gas flux mode. An ellipsoidal field cage comprised of silver wires held by 3D printed plastic supports with 1 cm pitch guarantees drift field uniformity in the 20 cm drift gap. The cathode is made by an ATLAS MicroMegas mesh \cite{bib:Micromegas} with 30 $\mu$m diameter metallic wires with a pitch of 70 $\mu$m. Three 24 $\times$ 20 cm$^2$, 50 $\mu$m thick  GEMs spaced each other 2 mm comprise the amplification stage. The GEMs are numbered from 1 to 3 with GEM1 being the closest to the drift region. LEMOn cathode and field cage base are powered by a CAEN N1570\footnote{\url{https://www.caen.it/products/n1570/}} HV supply, while the GEMs by a CAEN A1526\footnote{\url{https://www.caen.it/products/a1526/}} power supply with 6 independent HV channels up to 15 kV and a current sensitivity of 10 nA. This powering configuration provides a very precise tool of measurement of the charge on each of the GEM electrodes.

The LEMOn detector is optically coupled to a Hamamatsu sCMOS camera ORCA FLASH 4.0 (first row of Table \ref{tab:hama}) through a TEDLAR transparent window and an adjustable plastic bellow. The camera has a calibrated response of 0.9 counts/photons \cite{bib:jinst_orange1}. The sensor is equipped with a Schneider Xenon lens (see Appendix \ref{appC}). The ORCA-Flash is positioned at $(50.6 \pm 0.1)$ cm distance from GEM3 and reads out an area of 25.6  $\times$ 25.6 cm$^2$. Therefore, each of the 2048 $\times$ 2048 pixels of the sCMOS sensor images an effective area of 125  $\times$ 125 $\mu$m$^2$. On the opposite side, behind the transparent cathode, a HZC Photonics XP3392 PMT is positioned. The drift field is set to 0.5 kV/cm and the transfer fields between GEMs at 2.5 kV/cm, while the voltage across the GEMs of 460 V was utilised in order to provide a gain of 1.5 10$^6$ \cite{bib:roby}.\\

The LEMOn prototype was utilised to validate the concept of the detector on a medium scale and to measure the energy resolution, the stability performances, to provide a preliminary estimation of the absolute position along the drift direction and the first assessment of the background rejection. With the LEMOn detector it was possible to demonstrate on a medium size TPC that the detection efficiency of a 5.9 keV electron recoil is 100\% with drift fields above 300 V/cm independently of the drift distance. A more detailed description of LEMOn and its performances can be found in \cite{bib:Antochi_2021,bib:fe55,bib:GEMoptical}. The LEMOn prototype is also utilised for the studies described in Chapter \ref{chap5}.
\subsection{Background rejection}
\label{subsec:cyg_backrej}
\begin{figure}[t]
	\centering
	\includegraphics[width=0.65\linewidth]{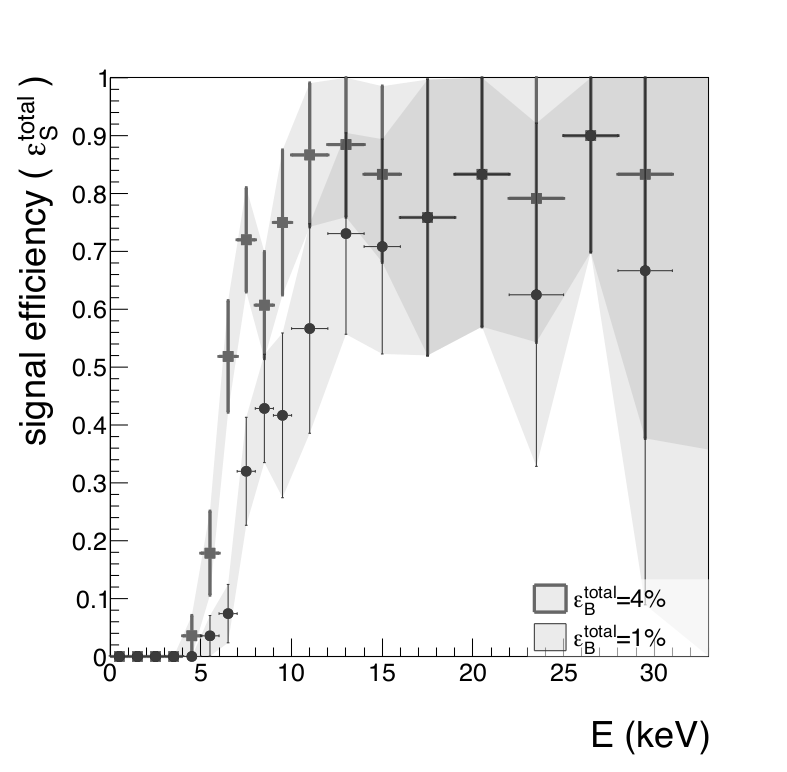}
	\caption{Detection efficiency for nuclear recoils ($\epsilon^{total}_s$) as a function of their detected energy for an efficiency on \fe electron recoils of 4\% (squares) and 1\% (circles).}
	\label{fig:blediscr}
\end{figure}
The capability to discriminate between ERs and NRs is of paramount importance for rare event searches, see Section \ref{subsec:background}. The high granularity of the optical readout allows to retrieve a large amount of information on the track shape, that can be efficiently exploited to identify its ER or NR origin.\\
In order to evaluate background rejection capability of the CYGNO experimental approach, LEMOn was exposed to a \fe source producing ERs at 5.9 keV and an AmBe source producing NRs (properly shielded with lead blocks to suppress the gamma flux emitted by the source). The gas mixture used is a He:CF$_4$ (60/40) at atmospheric pressure and room temperature with a flux of 200 cc/min. A reconstruction algorithm based on DBSCAN and discussed in detail in Chapter \ref{chap4} is employed to find and reconstruct tracks in the sCMOS images. It is based on the Density-Based Spatial Clustering  of Applications with Noise (DBSCAN) \cite{dbscan1996} which selects clusters of pixels which are believed to belong to the same group. These are furthermore grouped according to their vicinity and their derivative in the light profile, employing an algorithm based on Geodesic Active Contour (GAC \cite{gac,mgac}). More details can be found in Section \ref{sec:recocode}. The test was performed at overground facility of the Laboratori Nazionali di Frascati (LNF), implying a large contamination of cosmic rays in the data. Track shape variables were used to identify the different particles interacting in the detector. Track length and track slimness (ratio of track width over length) allow to separate the long cosmic ray tracks from the low energy ERs and NRs. Indeed, the former can cross the whole detector with reduced loss of energy per unit length producing long slim tracks, while the larger energy deposition of $\mathcal{O}$(1) keV ERs and NRs generate short thicker denser tracks. The pixel density $\delta$, defined as the ratio of the total number of photons of a track over its number of pixels, is proportional to the $dE/dx$ stopping power of an ionising particle. Thus, this quantity can be employed to discriminate NRs, characterised by continuous and dense energy depositions, from ERs, that conversely release energy in a sparse and discontinuous pattern. Figure \ref{fig:blediscr} shows the NR signal efficiency $\epsilon^{total}_S$ as a function of the measured energy for 5.9 keV ER efficiency of 4\% and 1\%. An efficiency of 18\% to detect nuclear recoils with an energy of about 6 keV is reached, while suppressing 96\% of the \fe photoelectrons, showing how this optically readout gaseous TPC is a very promising candidate for future investigations of ultra-rare events as directional direct DM searches.
Even though the selection cut is very simple and minimal, a rejection power of the order 10$^2$ is achievable below 10 keV. More sophisticated and articulated analyses which combine multiple topological features and with the use of Machine learning techniques are under study in the collaboration, and already showed the capability to improve of about one order of magnitude on these results.
\section{LIME}
\label{sec:LIME}
\begin{figure*}[!t]
	\centering
	\includegraphics[width=0.46\linewidth]{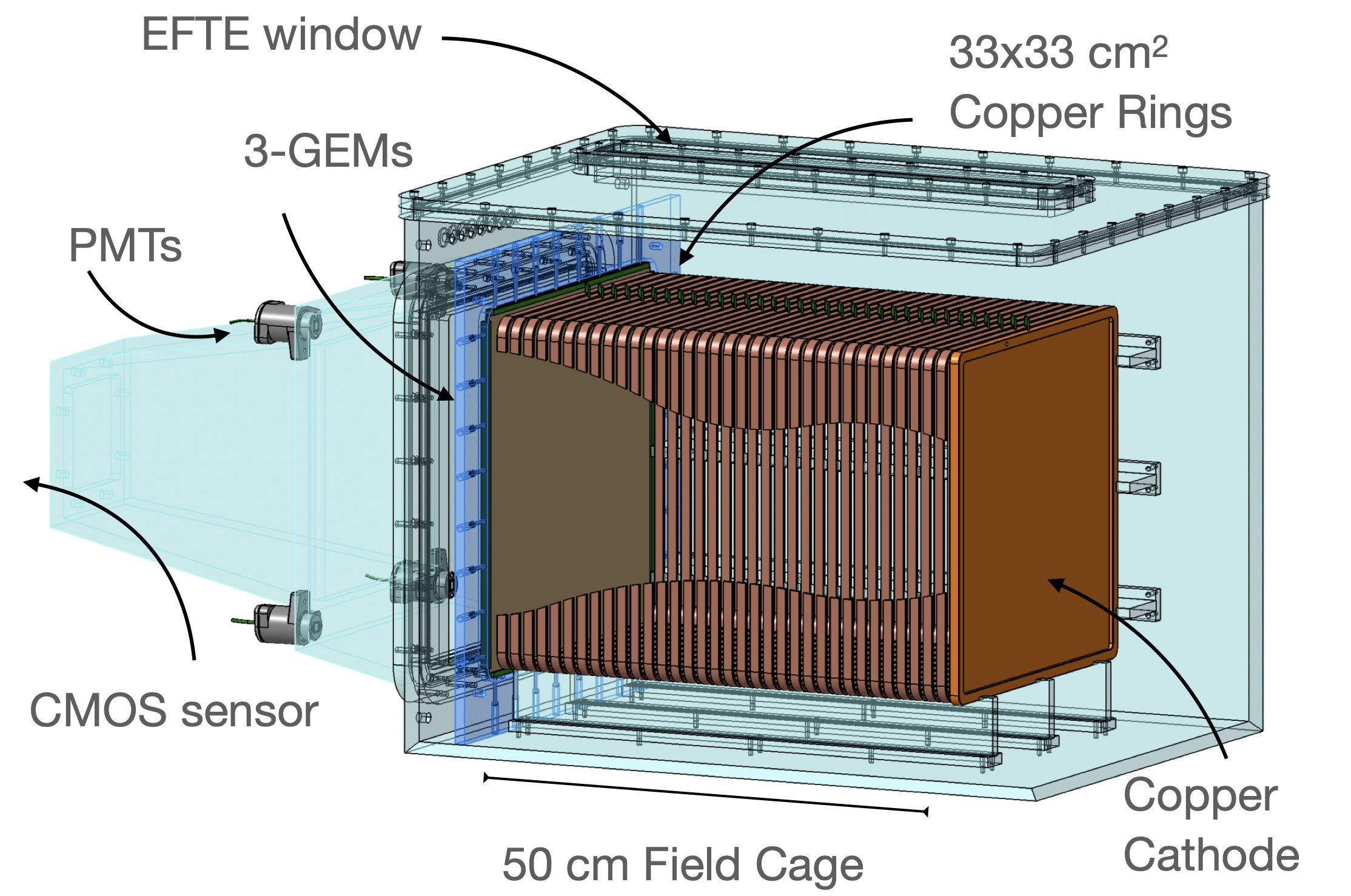}		\includegraphics[width=0.4\linewidth]{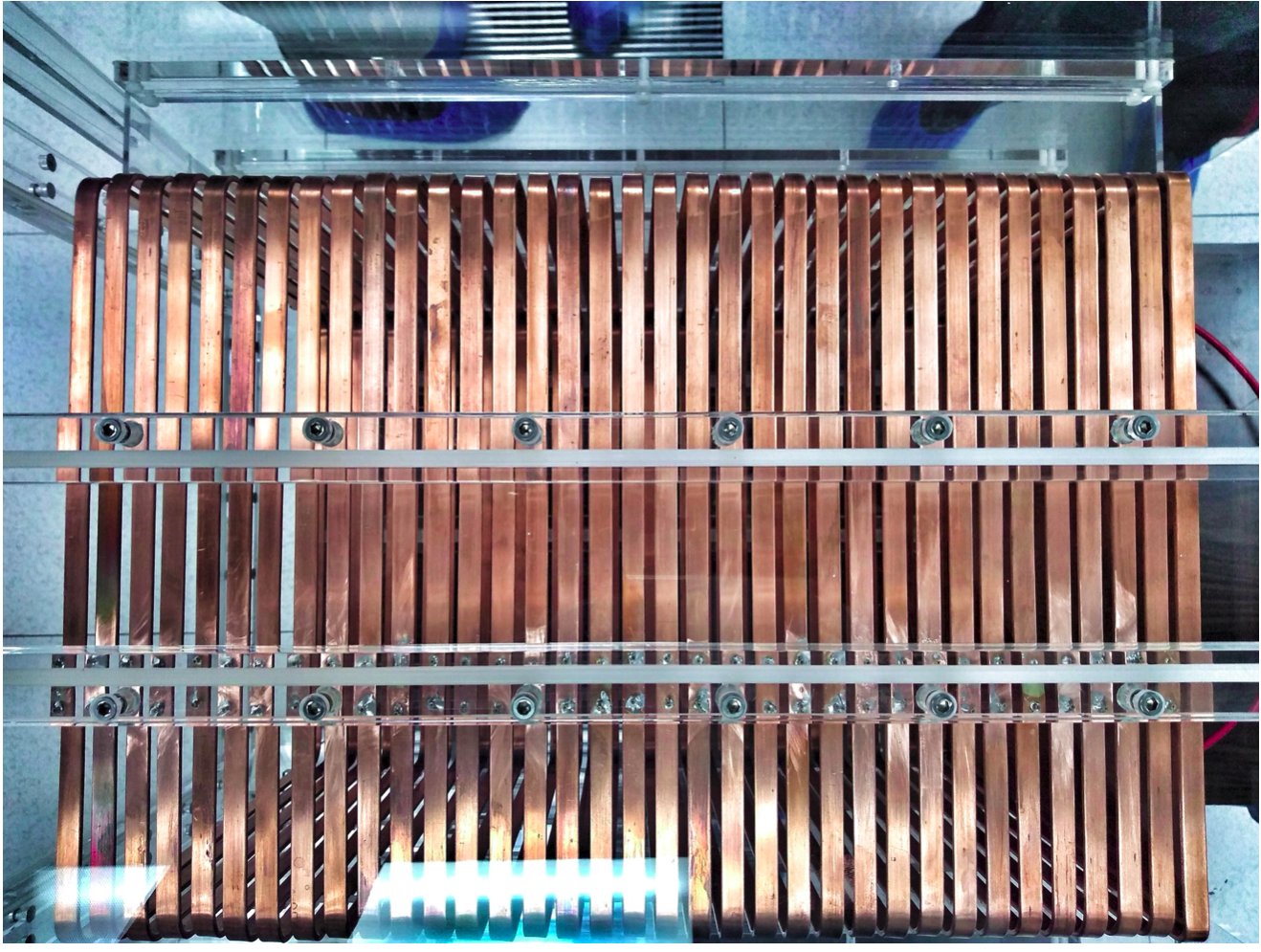}
	\caption{On the left a sketch of the LIME detector. The structure of the acrylic vessel, the copper rings and the optical readout are shown. On the right, a picture of the field cage after the installation.}
	\label{fig:limeske}
\end{figure*}

The Long Imaging ModulE (LIME) is the largest prototype built so far with the CYGNO experimental approach and it is expected to conclude the R\&D phase of the CYGNO project. A picture and a sketch of LIME are shown in Figure \ref{fig:limeske}. The TPC has a parallelepiped sensitive volume with a readout area of 33 $\times$ 33 cm$^2$ and a drift length of 50 cm, for a total of 50 l active volume. These dimensions are chosen in order to reproduce the drift length of the 0.4 m$^3$ CYGNO-04 detector, see Section \ref{sec:future:CYGNO}. The vessel is made of acrylic and an external aluminium Faraday cage shields the detector from any electromagnetic interference. The top side of the vessel has a narrow opening along the drift distance closed by a thin ETFE window in order to be able to calibrate the detector with external radioactive sources. The cathode and the squared field rings of the field cage are made of copper. The rings are positioned with a pitch of 16 mm and connected via ceramic resistor to generate the partition of the voltage which produces the drift field. The amplification stage is based on a triple stack of 50 $\mu$m thin GEMs of 33 $\times$ 33 cm$^2$ of area, each stretched on an acrylic frame rather than on conventional PCB boards to minimise the radioactivity contribution from this item. The amplification area is readout optically by a single sCMOS camera and 4 PMTs positioned towards the 4 corners in order to improve the PMT track reconstruction.
The PMTs are Hamamatsu R7378A\footnote{\url{https://www.hamamatsu.com/content/dam/hamamatsu-photonics/sites/documents/99_SALES_LIBRARY/etd/R7378A_TPMH1288E.pdf}} with 22 mm diameter bialkali photocathode sensitive from 160 up to 650 nm. The sCMOS camera is the Hamamatsu ORCA Fusion (second row of Table \ref{tab:hama}) and it is equipped with a Schneider Xenon lens (see Appendix \ref{appC}). The ORCA-Fusion is positioned at $\sim$ 62 cm distance from the farthest GEM from the sensitive volume and reads out an area of 35.7  $\times$ 35.7 cm$^2$. Therefore, each of the 2304 $\times$ 2304 pixels of the sCMOS sensor images an effective area of 155  $\times$ 155 $\mu$m$^2$. \\

After the overground commissioning, LIME was installed in the underground facility of the Laboratori Nazionali del Gran Sasso (LNGS), where it is now operating in a low background environment. The goal of LIME is to validate the full performances of the CYGNO approach in an underground environment, to verify the Monte Carlo (MC) simulation of the expected background and to test construction and operation procedures toward the realisation of PHASE\_1 (see Section \ref{subsec:cygno04}). In parallel, it will provide a precise, spectral and directional measurement of the environmental LNGS neutron flux (see Section \ref{subsubsec:cyg_measureplan}).
\subsection{Overground tests}
\label{subsec:cyg_lim}
LIME was commissioned at the overground LNF in order to characterise its performances before the underground installation at the LNGS. The obtained results on operation stability, light yield, energy response and determination of the absolute position along the drift direction are illustrated in the following.
\subsubsection{Stability}
\label{subsubsec:cyg_stability}
The long term stability of LIME was tested for a month long run, during which the detector was exposed to environmental radioactivity, cosmic rays and a \fe source. 
During the whole period, the current drawn by each of the high voltage supplier channels powering the GEMs electrodes was monitored and recorded to identify sudden and large increases that could indicate discharges or other electrostatic issues.
The main instabilities observed were:
\begin{itemize}
	\item hot-spots appearing on the GEM surface. These are likely related to localised self-sustaining micro-discharges happening in one or few GEM holes. In some cases they fade away with time, in some others they start to slowly grow up to tens of nA (on a time scale of minutes). 
	
	\item sudden discharges. High charge density due to very high ionising particles or charge accumulation on electrode imperfections can suddenly discharge across GEM holes. These events are recognisable thanks to a sudden increase in the current drawn on the supplier channels. A voltage supply can quickly restore the drop occurred during the discharge through 10 M$\Omega$ protection resistors on a few seconds time basis.  Despite these events being less frequent than hot-spots, they can be dangerous for the GEM structure and the energy released in the discharge could damage it irreversibly. 
\end{itemize}
In order to protect the detector, an automatic recovery procedure was established. Whenever the drawn currents are measured beyond a customisable threshold, this procedure decreases the voltage across all the GEMs by 100 V. Then, the voltage is restored in 5 steps of 20 V with a 30 seconds of pause between each step. This procedure lasts about 3 minutes and it demonstrated the capability of recovering both hot-spots and discharges within a couple of minutes.
An average of 16 instabilities per day was observed and the automatic recovery procedure introduced less than 4$\%$ of dead time. A detailed analysis of the time interval between two consecutive phenomena did not shown any correlation between two subsequent events, nor any increase of their rates. This study demonstrated that the detector operation is very safe and stable in satisfactory way for the purposes of the data taking.
\subsubsection{Linearity and light yield}
\label{subsubsec:cyg_lightyield}
\begin{table}[!t]
	\centering
	\begin{tabular}{|c|c|}
		\hline
		Material & Emission (keV)  \\ \hline\hline
		Calcium & 3.69  \\ 
		Titanium & 4.51  \\ 
		Iron & 5.9   \\ 
		Copper & 8.04  \\
		Rubidium & 13.37 \\
		Molybdenum & 17.44 \\
		Silver & 22.10   \\
		Barium & 32.06 \\ \hline
	\end{tabular}
	\caption{Summary of the materials and K$_{\alpha}$ emissions utilised for the X-ray calibration of LIME.}
	\label{tab:xray}
\end{table}
In order to test the response of the detector to diverse energy deposits, the emission of K$_{\alpha}$ and K$_{\beta}$ lines from different materials is exploited. A target holder was positioned on the top of the EFTE window with a 45 degrees inclination with respect to it. By irradiating different materials positioned in the holder by means of a radioactive source (\fe or $^{241}$Am), X-ray emission of K lines can be induced, with an energy dependent on the target material. The inclination of the holder is chosen in order to maximise the probability that the emitted X-rays reach LIME sensitive volume and convert in ERs into it. The \fe source was also used to directly irradiate the detector in order to provide its characteristic 5.9 keV X-ray emission. The materials utilised with their K$_{\alpha}$ line energy are reported in Table \ref{tab:xray}. The X-rays energy ranged from 3.7 keV of calcium up to 32 keV of barium for a total of 8 lines.
Images from the sCMOS camera were acquired in free running mode (i.e. without an external trigger). For each target element, additional images were acquired with the detector turned off to characterise the sensor noise that can vary along hours of data taking. The data were analysed with the reconstruction code illustrated in Section \ref{sec:recocode}. The energy spectrum of each dataset is fitted with a Gaussian, which represents the X-ray signal distribution, on top of a polynomial function modelling the background (typically composed by mis-reconstructed cosmic ray tracks and other low energy ERs).

\begin{figure*}[t]
	\centering
	\includegraphics[width=0.51\linewidth]{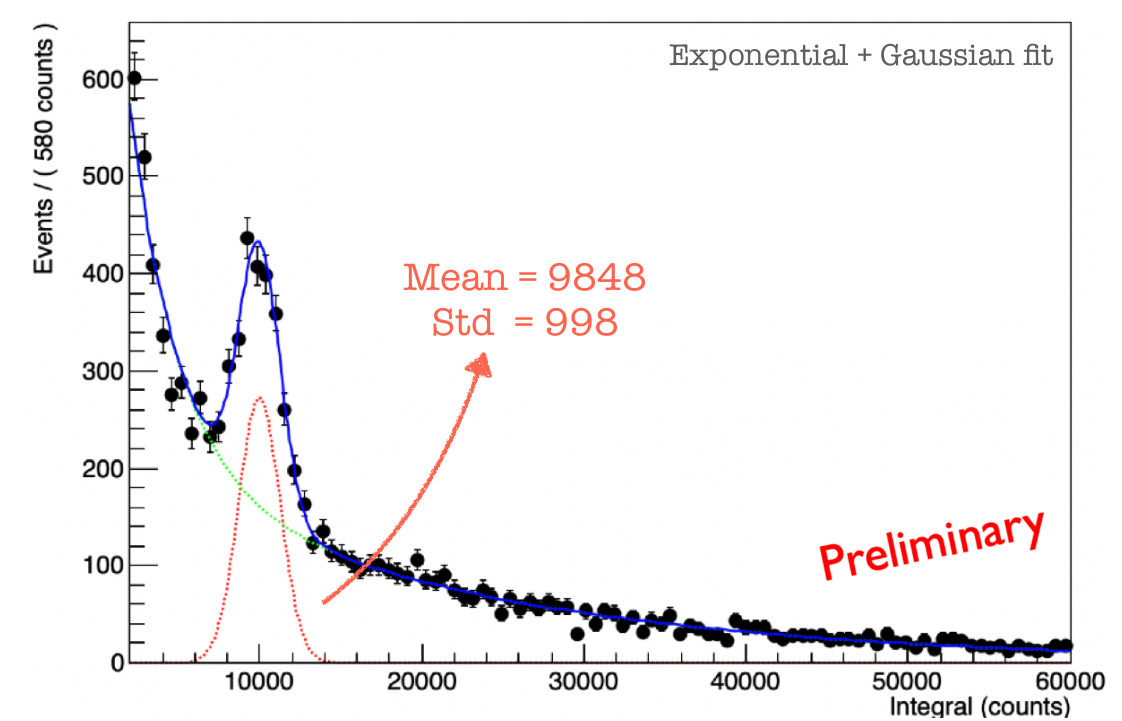}		\includegraphics[width=0.34\linewidth]{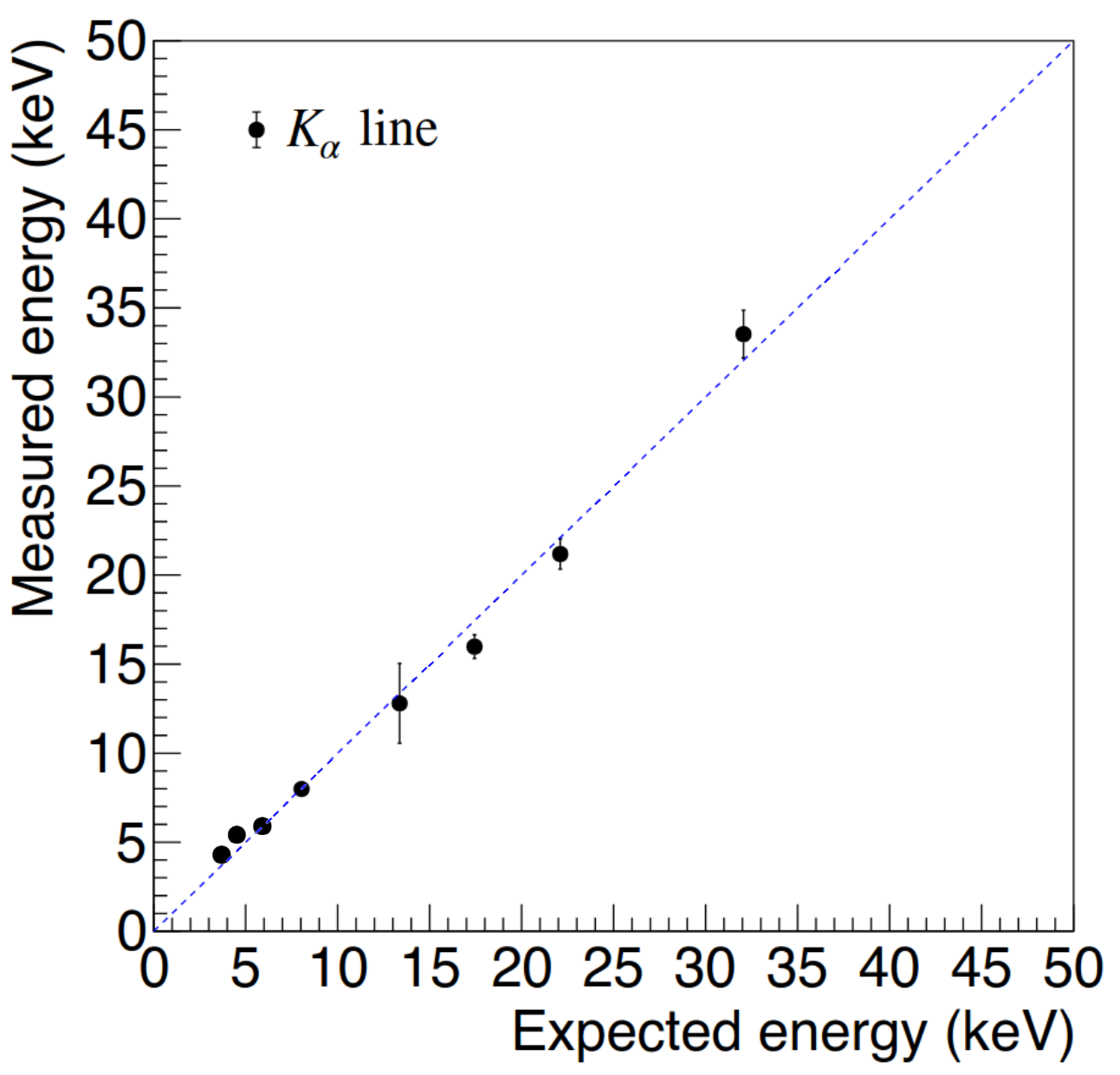}
	\caption{Preliminary results on the linearity study of the LIME detector performed at the LNF. On the left, an example of the light spectrum of the clusters found during the data taking with the 8 keV emission from a copper target material. The data are modelled as a polynomial background with a Gaussian peak on top. On the right, the measurement of the average amount of photons per X-ray emission as a function of the nominal energy. The linear fit superimposed demonstrate the linear response of the detector. Credit for this work to the fellow PhD student Samuele Torelli and Atul Prajapati.}
	\label{fig:lin}
\end{figure*}

The mean of the Gaussian fit represents the average number of counts summed on all the pixels of the track, proportional to the number of photons and hence to the track energy. The fitted sigma defines the energy resolution. 
Figure \ref{fig:lin} on the left, shows an example of an energy spectrum obtained during the data taking of the K lines of copper. Figure \ref{fig:lin} on the right shows the preliminary results on the energy measured for each X-ray emitter as a function of the nominal original energy. The superimposed linear fit shows a good linear response up to the barium emission with negligible offset. The energy resolution was found to be roughly 14\% in this range of energies, an excellent resolution for a gas detector for such low energy deposits.\\
The measurement of the average number of photons collected per track allows to estimate the light yield properties of the detector. The light yield was found to be 650 photons per keV of energy deposited in the sensitive volume, estimated from the \fe data. The knowledge of the light yield combined with the intrinsic noise of the camera leads to the determination of an effective energy threshold of 1 keV. Indeed, the sensor electronic noise represents a possible unavoidable instrumental background and it can generate \textit{ghost-clusters} in the reconstruction algorithm. The distribution of the light in each ghost-cluster found is studied in blind data sets, when a cap was placed in front of the camera to prevent light from reaching the sensor \cite{bib:fe55}. The exponential tail of this distribution is extrapolated in order to obtain the probability of having a ghost-clusters with an amount of light larger than a given threshold. It was found that the number of ghost-clusters with more than 400 photons are expected to be $\sim$ 3 10$^{-7}$ per second. With such a low rate, when the reconstruction code returns a cluster with 400 photons it is highly probable it is related to a real track. While this method does not define the energy threshold, 1 keV is believed to be a conservative estimation of it.
\subsubsection{Absolute position along the drift direction}
\label{subsubsec:cyg_fiducial}
The determination of the absolute position along the drift direction (z) is of the outermost importance because, once combined with the 2D projection provided by the sCMOS camera, it allows to localise an event in all the three dimensions. This is fundamental for the fiducialisation of the detector, as discussed in Section \ref{sec:direct_detectors}.\\
\begin{figure}[t]
	\centering
	\includegraphics[width=.4\textwidth]{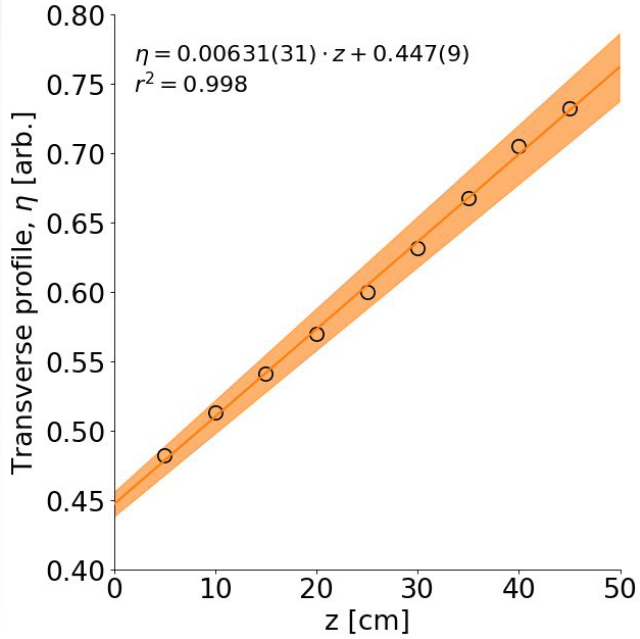}
	\caption{The dependence of the average $\eta$ as a function of the position of the \fe source away from the GEM.  Credit for this work to the fellow PhD student Rita Joana Cruz Roque.}
	\label{fig:eta}
\end{figure}
The z position determination was studied positioning the \fe source at various drift distances along the z axis of the detector. The z coordinate can be evaluated by exploiting the information on the diffusion a track is subject to \cite{bib:lemon_btf,LEWIS201581}. A 5.9 keV X-ray is absorbed by the gas producing an electron via photoelectric effect. Typical ER track length dimension at these energies is of the order of 100 $\mu$m. Given the diffusion coefficient of the CYGNO gas of about 100 $\mu$m/$\sqrt{cm}$ (see Figure \ref{fig:hecf4}), diffusion during drift will dominate over the original distribution of primary electrons. Thus, the signature of an \fe signal on a sCMOS image is a round spot. The spatial distribution of the pixels can be approximated to a Gaussian with standard deviation $\sigma$ and amplitude $A_G$. Depending on the z-coordinate of the recoil, the transverse diffusion increases enlarging $\sigma$, whilst decreasing $A_G$.
The ratio $\eta$ defined as $\sigma$/$A_G$ is expected to grow with the drift distance. Figure \ref{fig:eta} shows the average $\eta$, evaluated on all the tracks reconstructed per z position of the \fe source, as a function of the nominal depth of the track, with superimposed a linear fit. The plot demostrates that $\eta$ is sensitive to the average depth of the tracks. Yet, the $\eta$ distribution of each data set results rather large due to experimental effects, especially the not optimal collimation of the source, complicating the actual event-by-event determination of the depth in the data acquired so far. MC simulations of the non collimated geometry of the source in the LIME setup suggest a resulting spread in z of the \fe signal of $\sim$ 5.5 cm. Employing a random forest regression algorithm \cite{Randomforest} on multiple shape variables, among which $\eta$ itself, an estimation of the z coordinate is obtained with a dispersion of $\sim$ 6 cm, very close to the simulated one. Future data acquisition with a collimated \fe source will provide a better estimation of the resolution in the reconstruction of the absolute z coordinate, improving the preliminary 15\% obtained with the LEMOn detector by measuring 450 MeV electron tracks at the LNF BTF line \cite{bib:lemon_btf}.
\subsection{Underground installation}
\label{subsec:cyg_limeunder}
\begin{figure*}[t]
	\centering
	\includegraphics[width=0.35\linewidth]{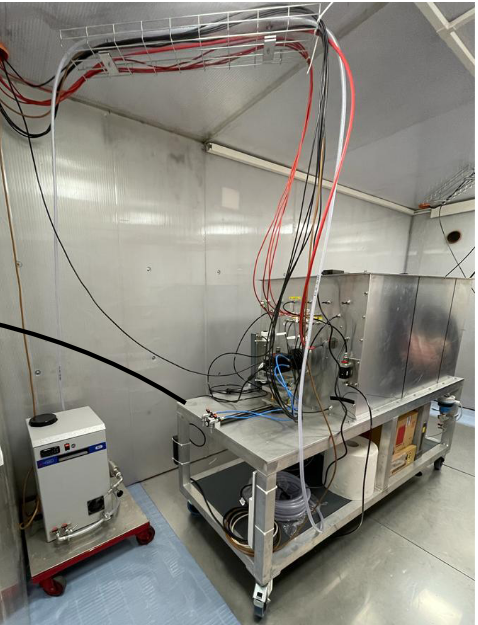} \hspace{0.3 cm}\includegraphics[width=0.36\linewidth]{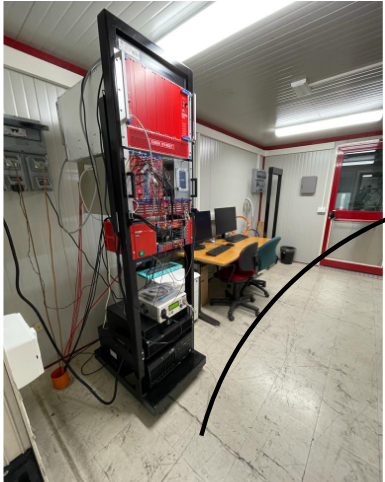}
	\caption{Pictures of the ground floor (left) and the first floor (right) with respectively LIME detector inside its Faraday cage and the control room. }
	\label{fig:limeund}
\end{figure*}

One of the main goals of the LIME prototype is to assess its performances in the low background underground environment of LNGS, where it has been operating since June 2022. The experimental area is located in a container in the corridor between Hall B and Hall A, in front of DAMA/LIBRA experiment. The container spans across two floors where the bottom one contains the detector (and the future shielding) and the top the control room, which hosts also the DAQ servers and modules, and the high voltage suppliers. Figure \ref{fig:limeund} shows two pictures of the underground LIME site. The gas system is located just outside the experimental area and is an Air Liquid GAS system which provides the flow of the mixture to the detector, purification from impurities, as well as its recirculation and its recovery for the disposal of greenhouse gases.
In the following Sections, the slow control and DAQ system LIME is equipped with will be described, as well as the background simulations and the future plan of measurement.
\subsubsection{Slow control}
\label{subsubsec:cyg_sc}
\begin{figure*}[t]
	\centering
	\includegraphics[width=0.5\linewidth]{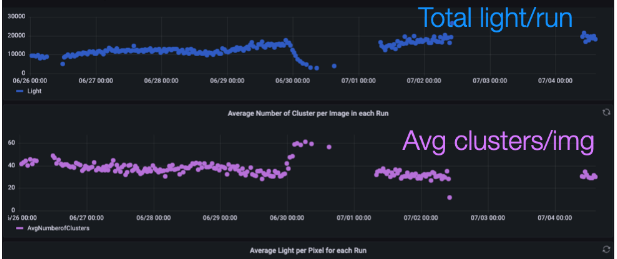}\hspace{ 0.1 cm}
	\includegraphics[width=0.27\linewidth]{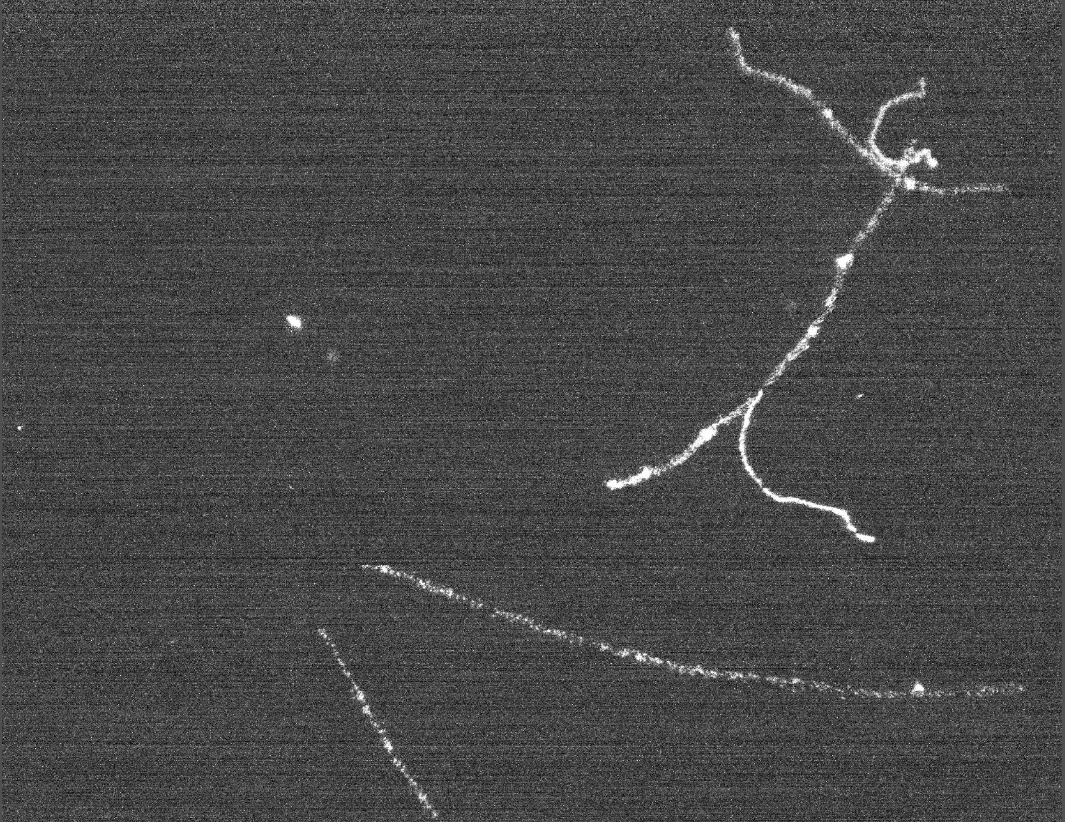}
	\caption{On the left an example of two variables of the slow control which check the quality of the data captured with the sCMOS camera. Occurrences of issues was notice by the slow control during this initial test phase. On the right, an example of an image from the sCMOS camera taken on LIME at the underground laboratories of LNGS.}
	\label{fig:sc}
\end{figure*}

During operation, LIME is handled by an open source C++ based slow control software developed within the MIDAS framework\footnote{ \url{https://daq00.triumf.ca/MidasWiki/index.php/Main_Page}}. MIDAS is a versatile software capable of interfacing with a wide range of sensors and electronic boards. The goal of the slow control is to setup the detector operating condition and periodically monitor its performances and functioning. In the CYGNO setup, this MIDAS-based software controls the high voltage suppliers of PMTs, cathode and GEMs, and records the currents drawn by the latter in order to activate recovery procedures in case of instabilities (see Section \ref{subsubsec:cyg_stability}). Furthermore, the slow control is connected to temperature and pressure sensors positioned inside the detector and in the room at the ground floor, in order to continuously monitor LIME working conditions and possible gas leakages. In parallel, it handles error messages and alarms delivered by the gas system in case of malfunctioning. Finally, it performs a first raw analysis of the PMT waveforms and of the sCMOS images to monitor the data quality during data taking. Figure \ref{fig:sc} shows on the left, an example of two variables used to check the regularity of operation of the sCMOS camera. The average number of clusters found in a set of images and total amount of photons collected are indicators of the correct gas purity and electronic stability of GEMs and camera. Figure \ref{fig:sc} exhibits, on the right, an example of a picture taken with the sCMOS camera in the underground laboratories at LNGS highlighting the reduction of the cosmic rays component in the underground environment.
\subsubsection{DAQ system}
\label{subsubsec:daq}
The Data Acquisition system (DAQ) is a set of hardware and software components responsible for the handling, acquisition and storage of the data. In the context of the CYGNO experiment, the challenging task of DAQ is to collect synchronized data from sCMOS cameras and the PMTs (i.e. detectors that have a difference in the response time of orders of magnitude), while handling the logic of the trigger. In the context of the LIME detector, the current signal waveforms induced on the bottom of GEM2 and GEM3 are also acquired by the DAQ, in order to better characterise the prototype performances and features.\\
The sCMOS camera is interfaced to the DAQ computer via USB 3.0 cable, which allows to transport the images, each of $\sim$ 5 MB, at the maximum rate allowed by the ORCA-Fusion, 5.5 Hz.\\
The typical time length of a PMT signal of short nuclear and electron recoils in a TPC filled with He:CF$_4$ depends on the inclination of the track with respect to the drift direction. On average, it ranges from few tens up to few hundreds of ns. Moreover, the usual risetime of this photosensor electrical circuit of the order of ns permits, in principle, to detail its waveform down to the ns scale. To cope with these requirements, a CAEN VME digitizer V1742\footnote{\url{https://www.caen.it/products/v1742/}} was selected to sample the PMT waveforms. It possesses 32 channels, each one with 1024 switched capacitor sampling cells at 12 bits and a sampling frequency which can vary from 750 MS/s up to 5 GS/s. In the configuration with 1 GS/s, it allows to sample every PMT signal with 1 ns step for an entire $\mu$s, enough to contain the desired waveform with a high precision both in sampling and voltage sensitivity. On the other hand, the signal induced on the GEMs is influenced by the motion of the ions produced during the amplification which causes their time extension to be much longer. As a consequence, the CAEN VME V1761\footnote{\url{https://www.caen.it/products/v1761/}} digitizer was selected for these, with two 10 bit channels equipped with a 4 GS/s Flash ADC and 7.2 MS of memory buffer. This way, a detailed sampling of the GEM waveform can be performed for a maximum time of 1.8 ms in order to include the entirety of the signal. To handle the VME digitizers and their connection to the DAQ server, a CAEN VME bridge V1718\footnote{\url{https://www.caen.it/products/v1718/}} is employed. In addition, to correctly communicate the commands received by the DAQ computer to the digitizers, these bridges also possess programmable outputs which can act as vetoes. \\

The challenge of a data acquisition with these different optical detectors lies in their imbalanced typical time of operation. Indeed, as the typical extent of a PMT waveform is of a $\mu$s, the sCMOS camera takes 80 $\mu$s just to expose one of the 2304 rows of the silicon matrix. The usual exposure time selected for the data taking is of 300 or 500 ms, 10$^5$ times longer than the PMT signals. In order to obtain a synchronised data acquisition of all the sensors, a custom front end program was developed and integrated in the MIDAS software. The acquisition is divided into runs, each composed of a variable number of events. An event is defined as a collection of a sCMOS image together with all the PMT and GEM waveforms generated during the camera acquisition window. The sCMOS is kept in continuous acquisition mode so that images are taken repeatedly without any pause. The sensor is opened to expose the pixels, kept open for the exposure time length, closed and finally the trigger signal is checked to decide whether to store the image or to discard it. The trigger signal is the logic AND of two variables: \emph{lamPMT}, coming from a logic analysis of the PMT signals, and \emph{lamCAM} coming from the camera. lamPMT is positive if two or more PMT waveforms have at least one sample below a predetermined voltage threshold (PMTs have negative polarity signals). This was fixed at $-$5 mV, in order to provide 100\% efficiency on the signal while minimising noise events from single photoelectron background. The logic signal lamCAM is positive when all the pixels of the camera are exposed together. This way, the trigger is positive when more than a PMT received a sufficient number of photons from the amplification stage while the whole sensor was exposed, in order to be certain to have the signal detected by all the sensors. A single positive trigger is enough to force the storage of the sCMOS picture. More triggers can occur during a single image, and every time the trigger signal is positive, the digitizers sample the waveforms for a maximum of 128 in total.  When the sCMOS picture is saved to disk also the buffer of the digitizers is transferred to the DAQ server and compressed along with the picture in a single event. This choice of data acquisition logic could in principle hinder the correspondence between each PMT waveform and the relative track in the CMOS image in case of high occupancy, limiting the 3D reconstruction capabilities. Nonetheless, the typical LIME rate of events in an underground environment (of few Hz in absence of shielding, see Section \ref{subsubsec:cyg_backstu}), combined with the possibility of inferring the track x-y position from the PMT waveforms light barycentre, will grant the match of the two signals. \\
During several campaigns at LNF, the DAQ system was tested and adapted to recognise different digitizers and various logic structures to maximise its flexibility. It will be thoroughly tested during the underground operation, as it represent the basic structure of the DAQ for the future CYGNO detectors.

During the last months, LIME was commissioned in the underground facility, testing the performances of DAQ, slow control and gas system utilising a \fe source. The stability of the LIME setup allowed the data taking to be undertaken from remote control.
\subsubsection{LIME background studies}
\label{subsubsec:cyg_backstu}
\begin{figure*}[t]
	\centering
	\includegraphics[width=0.9\linewidth]{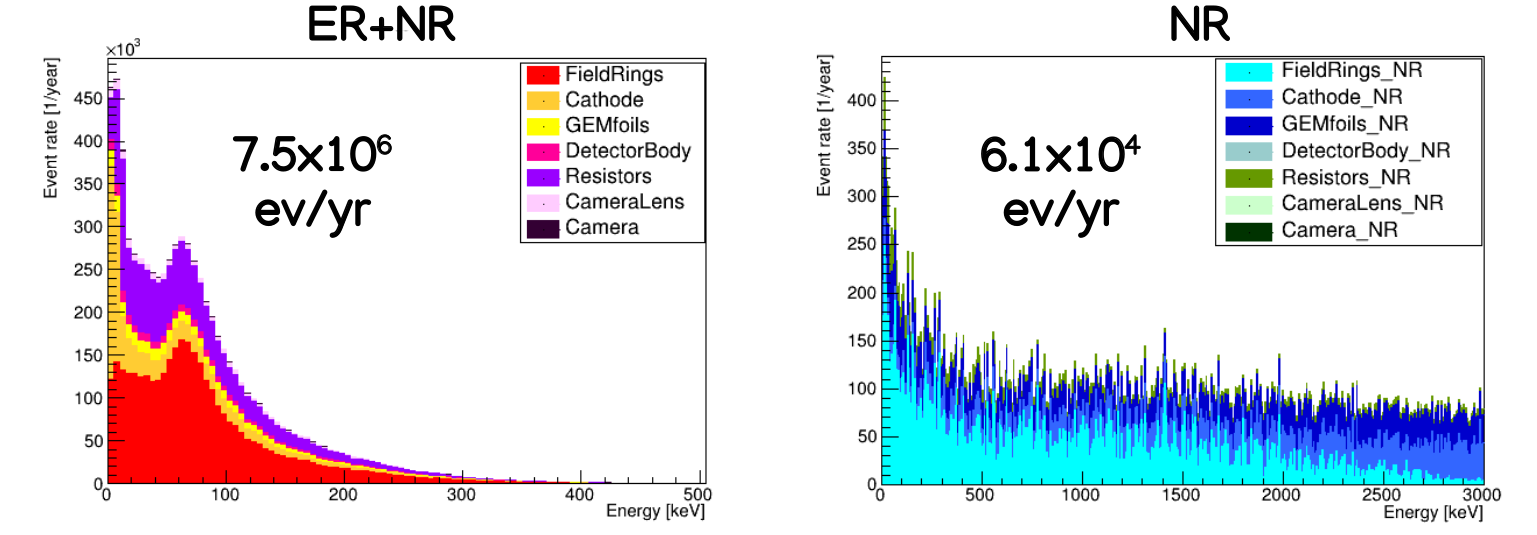}
	\caption{Result of the GEANT4 simulation of the internal radioactivity of LIME. On the left the sum of the ER and NR induced by radioactive background interactions. On the right only the NR are shown. The different detector components causing the recoils are separated by means of the colour scheme displayed in the legend. Credit for this work to the fellow PhD student Flaminia Di Giambattista.}
	\label{fig:limegeant}
\end{figure*}
The characterisation and knowledge of the radioactive background is of the utmost importance for rare events searches (see Section \ref{sec:direct_detectors}). For this reason, a thorough MC simulation of the expected background in LIME was performed with the GEANT4 package \cite{bib:geant}. The CAD engineering design of the prototype was implemented in the simulation, as well as the different shielding configurations. The radioactive background coming from outside the detector (\emph{external background}) is simulated from the spectra of neutrons and $\gamma$s fluxes measured at the LNGS by other experiments (see Figure \ref{fig:muons} and \cite{Belli1989DeepUN,DEBICKI2009429}). The intrinsic radioactivity of the main components used to manufacture LIME (cathode, copper, field rings, resistors, GEM, vessel acrylic, camera lens and body) were measured thanks to the LNGS Services with high purity Ge detectors, and used in the simulation of the  \emph{internal radioactivity}. Relevant contributions from $^{238}$U, $^{235}$U, $^{232}$Th chains and $^{40}$K were found. In particular, large 
contamination of  $^{234}$Th were measured in the resistors, cathode, field rings, camera body, lens and GEMs up to 20 Bq/kg. Resistors also contains large amount of  $^{234m}$Pa, $^{226}$Ra, as for the GEMs and cathode and field rings, and $^{210}$Pb. 
Copper activation isotopes as $^{58}$Co was found in small amount in the field rings and cathode. Relevant contribution also derives from the $^{40}$K contamination of camera sensor and lens, along with GEMs and resistors. It has to be stressed that LIME was built with standard materials, not optimised in terms of low intrinsic radioactivity.\\
In the unshielded configuration, where only the Faraday cage is present, the external background dominates inducing a total $\sim$ 10$^9$ events per year above 1 keV, a rate of $\sim$ 36 Hz. A cost-benefit study was performed in order to optimise the shielding configuration with the goal of maximising the suppression the external contribution while minimising the costs, taking into account the available space and additional radioactive contribution introduced by the shielding materials. The most cost-effective solution found was to employ a copper shield, 10 cm thick, encased in a water tank shield of 40 cm thick.
The copper component is used to stop all the $\beta$ emissions and significantly reduces the gamma $\gamma$ ones. On the other hand, the water is utilised to slow down and capture neutrons, thanks the large quantity of hydrogen present in the liquid. This shielding configuration is expected to grant that less than 2 NRs per year above 1 keV are induced by external background in the active detector volume, while the ERs are reduced to 10$^5$ per year between 1 and 20 keV. These values are deemed small enough for LIME purposes (see Section \ref{subsubsec:cyg_measureplan}), since the simulation of the internal radioactivity returns a total number of recoils induced in the sensitive volume of about $\sim$ 7 10$^6$ events per year, making it the dominant component in the full shielding configuration. Figure \ref{fig:limegeant} shows the energy spectrum of the total number of recoils induced by internal radioactivity on the left, while the right plot displays only the NRs. The various colours exhibit the contributions from each of the main detector component included in the simulation. The field rings, resistors and GEMs are among the highest contributors to the radioactivity budget, with considerable impact also from the camera lens and body.
\begin{table}[!t]
	\centering
	\begin{adjustbox}{max width=1.01\textwidth}
		\begin{tabular}{|l|c|c|c|c|c|c|}
			\hline
			\multirow{2}{*}{\Large{Shield}} & \multicolumn{2}{c|}{External} & \multicolumn{2}{c|}{Internal} &\multicolumn{2}{c|}{Total} \\
			& ER/yr & NR/yr & ER/yr & NR/yr  & ER/yr & NR/yr \\ \hline\hline
			No shield & 1.13 10$^9$ & 1450 & 7.26 10$^6$ & 6.11 10$^4$ &1.14 10$^9$ &6.25 10$^4$  \\ 
			4cm Cu & 2.64 10$^7$ & 850 & 7.26 10$^6$ & 6.11 10$^4$ &3.43 10$^7$ &6.19 10$^4$  \\ 
			10cm Cu & 1.95 10$^6$ & 915 & 7.26 10$^6$ & 6.11 10$^4$ &9.78 10$^6$ &6.20 10$^4$  \\ 
			10cm Cu + cuts & N.A. & 772 & N.A. & 16 &N.A. &788  \\ 
			40cm H$_2$O +10 Cu & 5.09 10$^5$ & 2.0 & 7.26 10$^6$ & 6.11 10$^4$ &8.34 10$^6$ &6.11 10$^4$  \\ 
			40cm H$_2$O +10 Cu + cuts & 2.0 10$^4$ & 2.0 & 2.8 10$^5$ & 17 &3.3 10$^5$ & 19  \\ \hline
		\end{tabular}
	\end{adjustbox}
	\caption{Summary the simulated background from internal and external contributions in the LIME detector with different shielding configurations. The recoils are considered above 1 keV of energy. The total column also contains the additional contribution coming from the shielding material radioactivity, which is not explicitly reported in the rest of the table.}
	\label{tab:back}
\end{table}
This is of great importance in order to tackle down the radiopurity issues for the future detectors. The MC simulation shows that it is possible to further reduce the external and internal backgrounds with fiducialisation. Applying geometrical cuts to the detector volume, namely 1 cm from the borders and GEMs (i.e. in x-y) and 4 cm from the cathode (i.e. in z), the background induced recoils can be reduced of 96\% for ERs and 99.97\% for NRs. With these preliminary cuts, below 20 keV, the total amount of recoils drops to $\sim$ 2.7 10$^5$ events per year, of which only 6 are NRs. This analysis is performed on the real events simulated without the introduction of the detector performances. The collaboration is working to include the response of the detector and to optimise the cuts and the analysis strategy.\\
Table \ref{tab:back} summarises the internal and external contributions of ERs and NRs per year in the various shielding configurations, including the fiducial cuts.
The first images taken with LIME in the underground site with only the Faraday cage suggest a rate of recoils of $\sim$ 35 Hz, in excellent agreement with the expected value from the simulation.
\subsubsection{LIME underground program }
\label{subsubsec:cyg_measureplan}
LIME underground installation and operation has multiple purposes, from the study of the background to the validation of the MC simulations, and to the measurement of the neutron flux (see beginning of Section \ref{sec:LIME}). Since each goal requires different setups and level of backgrounds, the underground LIME program will proceed through a staged approach, progressively adding shielding layers around the detector. The foreseen phases are:
\begin{itemize}
	\item \textbf{No shield} LIME was operated with only the Faraday cage in the last months, with the goal of characterising the detector response underground and study the external radioactive background, which enormously dominates over the internal one when no shield is present (see first row of Table \ref{tab:back}). After an initial commissioning dedicated to the test of the slow control and DAQ system and prototype calibration with \fe source, about one month of background data was collected (with $\sim$ 10$^8$ events per month expected) to cross check the MC simulation with real data and directly measure the backgrounds. Data analysis of the acquired events is ongoing.
	\item \textbf{4 cm Cu}. This copper shield has been recently installed in order to verify that the measured background decreases as expected from the simulations. In addition, a neutron calibration with $^{241}$AmBe is going to be performed with this configuration, to minimise the gamma background from the source. A month is anticipated to complete this step. 
	\item \textbf{10 cm Cu}. The final copper shielding configuration will be of 10 cm, a thickness able to suppress of a factor 10$^3$ the external gamma flux. As the neutron flux in the underground laboratories is a source of relevant background for all of the experiments and its measurements at LNGS are limited and with broad uncertainties \cite{Haffke2011BackgroundMI,Bruno2019FluxMO,Belli1989DeepUN}, LIME will be employed to perform a precise, spectral and directional measurement of this, which will be of paramount importance for any present or future detector at LNGS. In addition, a directional measurement of NRs in this underground environment represents an effective demonstration of the capability of the detector to search for DM via NRs. \\
	Thanks to the background studies and the fiducialisation cuts analyses, the NRs from internal background are expected to be reduced to 16 per year, a negligible contribution with respect to the 772 coming from external sources (see previous Section). As the discrimination between ERs and NRs becomes very high above 20 keV in an imaging TPC (see Section \ref{subsec:cyg_backrej} and \cite{Vahsen_2021}), it possible to obtain a highly pure sample of external neutron induced NRs. About 200 NRs above 20 keV induced by external neutrons are expected to be detected in four months of data taking.
	\item  \textbf{10 cm Cu+ 40 cm water}. To strongly suppress the external neutron interaction, a water shield of 40 cm thickness will be built around the copper layer. In this configuration, the internal radioactivity will be the dominant component of the background recoils (see last two rows of Table \ref{tab:back}). Ten months are planned to be spent in the measurement of background data, to assess in detail the internal contributions to the background and to cross-check it with the MC simulations.
\end{itemize}
\section{Future of CYGNO}
\label{sec:future:CYGNO}
The goal of the CYGNO project is to boost the advancement of gaseous TPCs for directional DM searches. Thus, the construction in the future of a large $\mathcal{O}$(30) m$^3$ detector for the DM studies is a key milestone for the directional detector community. To achieve this goal, CYGNO is proceeding through a staged approach that involves the realisation of a 0.4 m$^3$ demonstrator CYGNO-04 towards  a $\mathcal{O}$(30) m$^3$ experiment. In the following Sections, these two future CYGNO phases will be  discussed.
\subsection{Phase\_1: CYGNO-04}
\label{subsec:cygno04}
Taking advantage from the experience gained from the construction, underground installation and operation of the LIME prototype, the CYGNO collaboration is moving towards the development of a PHASE\_1 0.4 m$^3$ experiment demonstrator: CYGNO-04.
The main objectives of PHASE\_1 are to study and minimise material radioactivity on a realistic experimental layout and scale, while demonstrating the scalability and the actual potentialities of a large PHASE\_2 detector to reach the expected physics goals.
\begin{figure*}[t]
	\centering
	\includegraphics[width=0.95\linewidth]{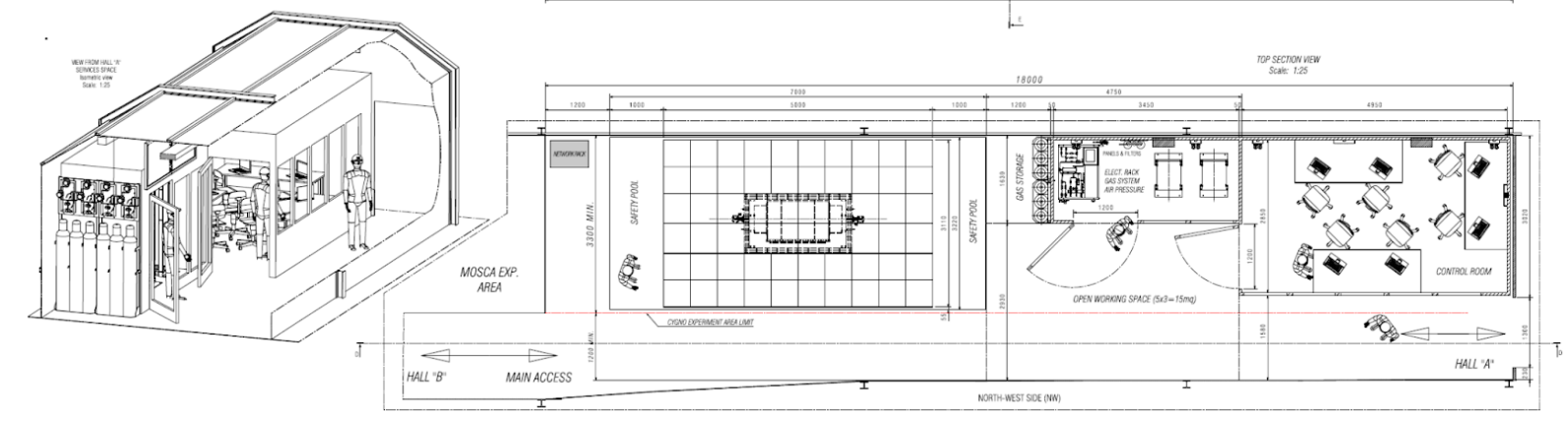}
	\caption{Technical design of CYGNO Phase\_1 adapted to the Hall F space granted to the CYGNO collaboration.}
	\label{fig:cyg04hf}
\end{figure*}
CYGNO-04 technical design report was submitted to LNGS administration and funding agencies in July 2022. It will be installed in Hall F and a technical design of the detector adapted to the experimental space is shown in Figure \ref{fig:cyg04hf}. The technical design of the internal structure of the detector is shown in Figure \ref{fig:phase1}. This will be contained in a PMMA acrylic vessel to lower the material radioactivity, to avoid the gas contamination and to guarantee the electrical insulation from the internal electrodes. As shown in Figure \ref{fig:phase1}, the gas volume is split into two chambers, each of 50 cm drift length sharing a central cathode electrode in a back-to-back configuration. Exploiting the expertise gained by the DRIFT collaboration, the cathode will be manufactured from a 0.9 $\mu$m thick aluminised \myl film \cite{bib:drift1,Battat:2015rna}. Indeed, the aluminium coating does not introduce radioactive material and the thinness of the film allows the alpha particles from radon daughter decays to enter the fiducial volume and thereby provide a means to tag and remove these events. To further reduce the internal background, the field cage will be manufactured from a \kap foil with very thin layers of copper acting as rings. This technique was employed by DRIFT \cite{bib:drift1} and removes the necessity for the highly radioactive ceramic resistors used in conventional field cages, while at the same time reducing the amount of copper. The amplification stage covers an area of 50 $\times$ 80 cm$^2$ with monolithic triple 50 $\mu$m thin GEMs. A cleaning procedure for GEMs based on DI water baths developed by T-REX collaboration will be utilised, to further suppress the radioactive contamination of this component\cite{Castel_2019}. \\
The optical readout consists in six PMTs and two sCMOS camera per side to cover all the amplification area. The model foreseen is the Hamamatsu R7378A which has a good 15\% QE at large wavelength like 550 nm \cite{Antochi_2018}. The typical radioactivity of a PMT combined with the distance these sensors will be positioned in, makes their contribution to the background budget negligible. The Hamamatsu ORCA QUEST will be used as sCMOS camera. This novel sensor displays improved noise performances with respect to the ORCA Fusion (see Table \ref{tab:hama}), about 0.27 electrons RMS. This permits each pixel to perform a single photon counting while the sensor images maintain a nice effective granularity of $\sim$ 200$\times$ 200 $\mu$m$^2$ per pixel in this setup. Given the large $^{40}$K contamination measured in the camera lens, customised ones, manufactured from ultra pure silica (SUPRASIL\textregistered), are under development for CYGNO-04 in collaboration with external companies. This will allow to reduce of a factor $\sim$10$^4$ the radioactive contribution from these objects. The collaboration is aiming at suppressing the radioactive contribution from the sCMOS camera, by customising the readout electronics which can be positioned outside the shields and by replacing the protective glass window of the sCMOS sensor with an ultra pure silica one.\\
\begin{figure*}[t]
	\centering
	\includegraphics[width=0.95\linewidth]{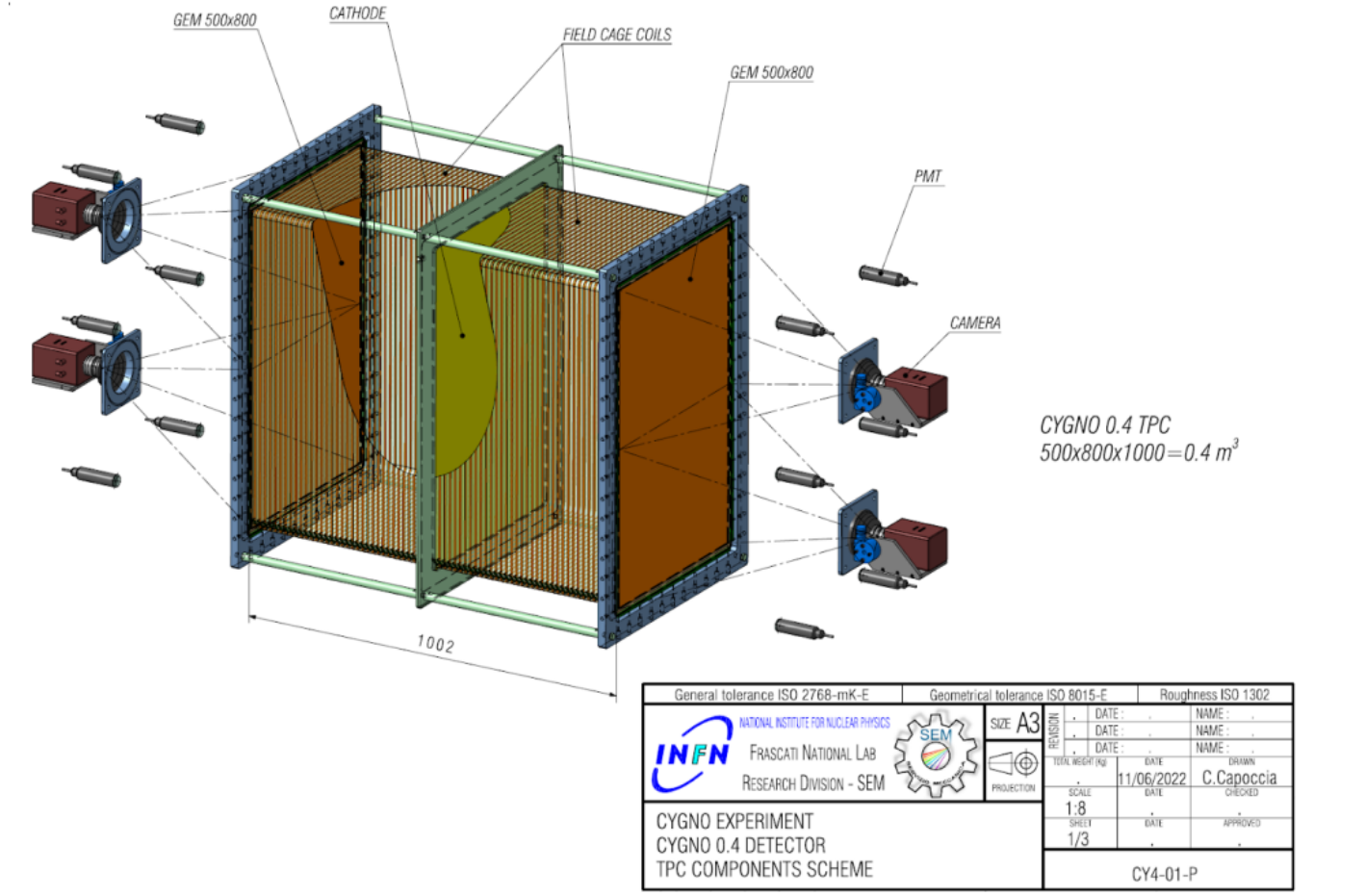}
	\caption{Technical design of CYGNO Phase\_1. The 0.4 m$^3$ volume is split in two chambers which share a common aluminised \myl cathode. Four sCMOS cameras and twelve PMTs constitute the optical detectors which image the 50 $\times$ 80 cm$^2$ readout area.}
	\label{fig:phase1}
\end{figure*}
A very preliminary GEANT4 simulation of the expected background in CYGNO-04 resulting from external environmental contribution plus internal material intrinsic radioactivity was performed in order to optimise the shielding material selection. The preliminary shield choice, which also takes into account the limited space available in Hall F, comprises 10 cm of copper surrounded by 1.1 m of water which is expected to lower the external radioactive contribution down to a factor $\sim$ 20 below the internal one. 
Given the modular design of CYGNO-04 and the results on the fiducial cuts discussed in Section \ref{subsubsec:cyg_backstu}, a similar capability of background reduction is expected to be achieved by fiducialization. To add on this, preliminary studies on gamma background rejection with Deep Neural Network algorithm based on track shape variables extracted from the sCMOS images suggests a
98.3\% rejection in the whole [1-40] keV$_{\rm{ee}}$ range with 40\% nuclear recoil efficiency. The advancement in sCMOS camera sensitivity and the addition of the PMT information can further
significantly improve these preliminary results, as well as the use of Convolutional
Neural Networks combining sCMOS and PMT information. Therefore, a level of background after all selection cuts and rejection of the order 10-100 events/year in the 1-20 keV$_{\rm{ee}}$ region for CYGNO-04 can be anticipated.\\
CYGNO-04 TDR was submitted in July 2022 and approved by LNGS management and funding agency.
\subsection{Phase\_2: CYGNO-30}
\label{subsec:cygno30}
\begin{figure*}[t]
	\centering
	\includegraphics[width=0.8\linewidth]{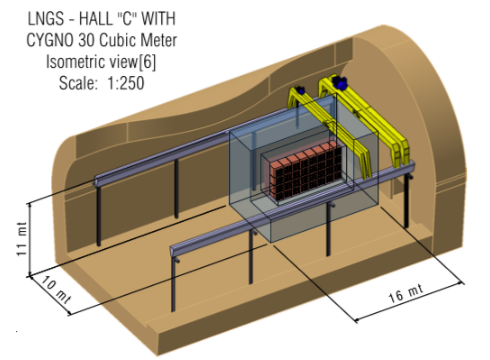}
	\caption{Preliminary technical design of a possible CYGNO Phase\_2 located in the Hall C of the underground laboratories of LNGS.}
	\label{fig:phase2}
\end{figure*}
A CYGNO-30 experiment would be able to give a significant contribution to the search and study of DM with masses below 10 GeV/c$^2$ for both SI and SD coupling. In case of a DM observation claim by other experiments, the information provided by a directional detector such as CYGNO would be fundamental to positively confirm the Galactic origin of the allegedly detected DM signal. CYGNO-30 could furthermore provide the first directional measurement via elastic scattering on electrons of solar neutrinos from the pp chain, possibly extending to lower energies the Borexino results \cite{BOREXINO:2018ohr}. A more detailed analysis of the sensitivity to WIMP DM searches of an exposure of $\mathcal{O}$(100) kg year directional detector is presented in Chapter \ref{chap7}.\\
In addition to DM, such an experiment will be relevant also for neutrino physics. The CE$\nu$NS constitute the basic interaction behind the neutrino fog, see Section \ref{sec:direct_detectors}. While the exposure of a $\mathcal{O}$(30) m$^3$ is not enough to reach the sensitivity necessary to detect Solar neutrino events through this interaction, new physics in the neutrino sector can  modify the CE$\nu$NS cross section at low energies and increase the rate of events  by orders of magnitude \cite{Boehm:2018sux,bib:bertuzzo}. Even more interesting is the scattering between neutrinos and electrons. Classical DM experiments that are not sensitive to the direction of the detected events cannot distinguish ERs induced by solar neutrino from the ones from natural radioactivity \cite{Amaro_2022}. A directional detector like CYGNO can instead exploit the directional information to identify the ERs caused by solar neutrino interaction, which display a strongly peaked asymmetry in their angular distribution, a feature not present in ERs from background sources. In fact, tonne-scale TPCs were proposed in the past with this same goal \cite{Seguinot:1992zu,Arpesella:1996uc}. Incidentally, the CF$_4$ gas present in the CYGNO mixture has a large electron density (about 10$^{21}$ cm$^{-3}$) with respect to its average low Z, which limits the multiple scattering of electron recoils in the gas. This is very important, as the multiple scattering hinders the recognition  of the original direction for low energy ERs. In the CYGNO gas mixture, 1 ER per year per m$^3$ induced by Solar neutrinos of the \emph{pp} cycle is expected above 20 keV$_{\text{ee}}$. A dedicated algorithm inspired by X-ray polarimetry studies \cite{Soffitta:2012hx} demonstrated on MC simulation that a 30$^{\circ}$ 2D angular resolution with 80\% HT recognition is achievable at this energy threshold. On top of this, the 20 keV threshold translates into 80 keV of the incoming neutrino energy. This opens a new window of opportunities on the \emph{pp} Sun processes down to low energy, unreachable to conventional neutrino detectors, such as Borexino, whose minimum neutrino energy measured is 160 keV \cite{Seguinot:1992zu}.\\
Figure \ref{fig:phase2} shows the sketch of a possible configuration of a 30 m$^3$ CYGNO Phase\_2 detector. Together with a rough estimation of 2 $\cdot$  10$^3$ m$^3$ of shields, it would fit inside the Hall C of the underground laboratories of LNGS.
\subsubsection{R\&D for the Phase\_2}
\label{subsubsec:redfuture}
A possible way to increase the sensitivity in the low \W mass region at the same energy threshold is the use of very light elements in order to enhance the momentum transfer, as discussed in Section \ref{subsec:kinematics}. Therefore, the collaboration is studying the possibility of introducing hydrogen in the CYGNO gas mixture by adding a small percentage of either iC$_4$H$_{10}$ or CH$_4$. While observed that both gases mildly reduce the photon/electron ratio production,  when methane is employed the overall amplification system results more stable and the voltages on the GEMs can be increased by $\sim$ 70 V to obtain a larger light yield without compromising the stability of the detector.\\
The optical readout needs very large gain from the amplification stage as only a small fraction of the solid angle is covered by these sensors. Maximising the light production is fundamental to lower and be sensitive to low WIMP masses, see Section \ref{sec:direct_detectors}. The possibility of introducing a strong electric field below the outermost GEM to increase the production of photons is studied. Details on the amplification stage optimisation are described in Chapter \ref{chap5}.
\subsection{INITIUM}
\label{subsec:INITIUM}
INITIUM (Innovative Negative Ion Time projection chamber for Underground dark
Matter searches) is an ERC Consolidator Grant with the goal of realising Negative Ion Drift operation within the CYGNO optical approach and results therefore highly synergic with it. The NID is a modification of the conventional TPC operation and will be extensively described in Chapter \ref{chap6}. It consists in the addition of a highly electronegative dopant which captures the primary electrons in the gas within few hundreds of $\mu$m so that anions are drifted in place of electrons, reducing the diffusion to the thermal limit. The goal of INITIUM is to develop a scintillating He:CF$_4$:\SF-based gas mixture at atmospheric pressure with a low content of \SF for NID operation in order to preserve the main features of the CYGNO experiment while minimising the diffusion and hence improve the tracking capability. If NID can be achieved within the optical approach, tracking could be even further improved by the possibility of reconstructing the track shape along the drift direction by sampling the recorded light at a kHz frame rate. At the moment, no camera exists with such a high rate which keeps high resolution and low noise. Nevertheless, given the fast development of the sCMOS technology, progress in short time is possible, which could open the door to this possibility.
\chapter{Data analysis}
\label{chap4}
The CYGNO experimental approach is based on the optical readout of the light produced by the electron avalanche process induced by the GEM amplification stage (see Chapter \ref{chap3}). This is realised by the simultaneous use of photomultiplier tubes (PMTs, sensitive to the track evolution along the drift direction $z$) and sCMOS cameras, that provide information on the track projection on the amplification plane ($x-y$). 
While a comprehensive analysis of the PMTs waveforms is still under development in the CYGNO collaboration, a dedicated preliminary algorithm has been elaborated for the analysis of the Negative Ion Drift data presented in Chapter \ref{chap6}, in order to deal with the peculiar features of the PMT signals of this original operating conditions. This will be discussed in Section \ref{subsubsec:omtanal}, to better contextualise it in the Negative Ion Drift operation context.

The sCMOS camera images represent a novel format of data in the context of Time Projection Chambers (TPCs) readout, and therefore require a tailored and original analysis strategy. A dedicated track reconstruction algorithm has been developed by the CYGNO collaboration for this task, with contributions from the work illustrated in this thesis. All the results from the analysis of sCMOS images reported in Chapters \ref{chap5} and \ref{chap6} are based on this algorithm, that represent the subject of this Chapter.
Section \ref{sec:recocode} is dedicated to the description of the methodology used to reconstruct particles tracks from the camera pictures, and to its optimisation. Section \ref{sec:difffe} describes the analysis algorithm developed for this thesis to extract the actual $x-y$ shape of 5.9 keV electron recoils from \fe X-rays absorption in the gas for the diffusion studies presented in Chapter \ref{chap5}. Section \ref{sec:diffalpha} illustrates the approach employed to measure the transversal dimension of alpha tracks for an alternative evaluation of diffusion parameters in the context of Negative Ion Drift operation (see Chapter \ref{chap6}).
\section{Reconstruction code}
\label{sec:recocode}
The ionisation patterns released by the passage of tracks inside the active gas volume of CYGNO detectors and recorded by the sCMOS camera can vary a lot in shape and photon density.
Minimum Ionising Particles (MIPs), like muons from cosmic rays, will leave long, scarcely dense trails. Electrons of tens of keV will generate curly tracks quite dense close to their end point, while nuclear recoils will produce short, straight and dense shapes due to the large energy loss per unit length. The  method to reconstruct tracks from sCMOS images needs therefore to be flexible enough to efficiently infer such diverse set of patterns. For this reason, the sCMOS images analysis procedure develops through three steps, by firstly dealing with the intrinsic sensor noise (Section \ref{subsec:noisered}), subsequently finding basic clusters from single small deposit (Section \ref{subsec:IDBSCAN}), and finally merging them into superclusters to determine the full tracks (Section \ref{subsec:supercl}).\\
The output of the reconstruction code is the collection of pixels belonging to each found track, with their original position in the CMOS image and the number of counts measured by each pixel. A quantity proportional to the energy released by the recoiling particle can be obtained by summing the content of all the pixels associated to each track. In addition, the 2D projection on the amplification $x-y$ plane of the ionisation trail topology and its $dE/dx$ along the track can be inferred from the single pixels content. This is the paramount feature of sCMOS images, which give access to the track direction and sense measurement, and to particle identification capability by the analysis of track energy loss and spatial development.
As illustrated in Section \ref{subsubsec:CMOS}, each sCMOS camera has different performances and characteristics. Since the ORCA-Fusion was employed for all of the results discussed in this thesis, it will be used as a reference for the following discussion. It is, nonetheless, important to stress that the reconstruction code is versatile with respect to this, and can be adapted to any camera model by a proper optimisation of the free parameters of the algorithm.
\subsection{Sensor noise reduction}
\label{subsec:noisered}
Every picture of the sCMOS camera is effectively a matrix of 2304 $\times$ 2304 integers  representing the raw ADC counts recorded by each pixel. The thermal energy, the manufacturing process and the electrical connections are all potential sources of noise during the readout of the charge collected in each pixel.  In order to proper deal with such issue in the images analysis, a set of pure noise images is acquired with the GEMs turned off every time physics data are collected (since the noise can change with time). As a consequence, these pictures will only contain the peculiar noise of each pixel. These are called \emph{pedestal} runs. From these noise pictures it is possible to evaluate the average noise per pixel of about 99 and the standard deviation of 3.5. The distribution of the noise counts is roughly Gaussian with a tail at large values which depends on a pixel per pixel basis, but is negligible for the vast majority of them (>98\%) and are often located at the borders of the sensor. The information of average noise ($\mu_{noise,ij}$) and standard deviation ($err_{noise,ij}$) for each pixel is stored in a matrix, called \textit{pedestal map}.\\
For each acquired image from physics run, a pixel by pixel subtraction of the pedestal map is performed. In addition, the pixels that satisfy the condition shown is Equation \ref{eq:pedsub} are removed,
\begin{equation}
\label{eq:pedsub}
C_{i,j}-\mu_{noise,i,j}< n_{\sigma n}err_{noise,ij}
\end{equation}
where $C_{ij}$ is the content of the pixel ($i,j$), and $n_{\sigma n}$ is a free parameter of the algorithm that is optimised depending on the noise condition of each data set. Sometimes, few pixels can exhibit very large amount of counts, close to the end of the dynamic range of the camera pixels. While these can be recovered by power cycling the camera, they are removed from the data analysis by sharply cutting pixels with intensities above 5000, as no physical signal can return this value. After noise subtraction and removal of outliers, the remaining pixels are rebinned by averaging the counts in 4 $\times$ 4 matrices and a median filter is applied to soften and smoothen the electronic noise fluctuations \cite{medianfilter}. 
The rebinned matrix of pixels is  further corrected to compensate for the vignetting effect. The vignetting effect is a natural geometrical reduction of the light intensity of an object imaged by a lens \cite{vignetting}. The light acceptance is, in fact, maximal when the source of light is facing the centre of the lens and decreases as 1/$R^4$, with $R$ the distance of a point on the focal plane from it. This is due to the inclination of the lens plane with respect to the emitted light cone, that effectively reduces the solid angle covered by it. In order to correct for this distortion in the light intensity, a vignetting-correction map is applied to the data. This is evaluated from images of a white paper uniformly illuminated, where the light intensity of each pixel is normalised to the value of the central one. It is important to apply this correction after having removed the noise as much as possible in order to avoid to enhance its contribution, especially on the outermost parts of the pictures.
The resulting image is analysed with the clustering algorithm illustrated in the following.
\subsection{Clustering algorithm}
\label{subsec:IDBSCAN}
Since the camera images a large area of the detector, only a fraction of the pixels of the sensor collects the photons originated from each of the recoil events occurred in a  single picture.
The goal of the clustering algorithm is to identify neighbouring pixels with counts above threshold and group them together. These are the \emph{seeds} which will be the input of the next step, the \emph{superclustering} procedure.
\begin{figure*}[t]
	\centering
	\includegraphics[width=0.45\textwidth]{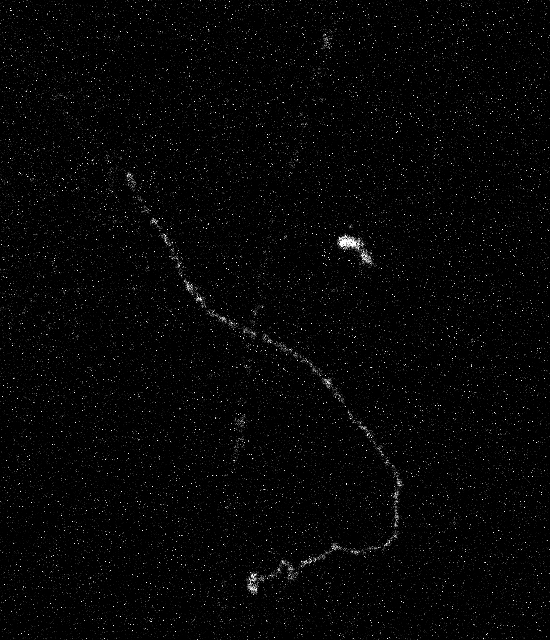}
	\includegraphics[width=0.45\textwidth]{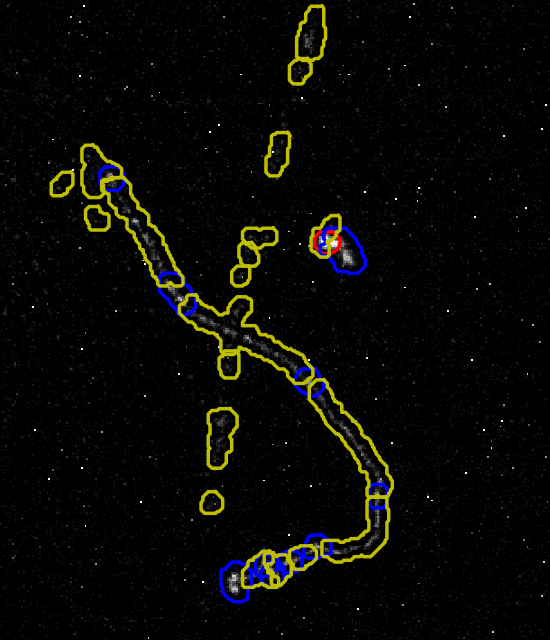}
	\caption{Example of a sCMOS picture taken with 0.5 s exposure of the natural radioactivity in MANGO. On the left, the original image with the intensity scale shown in gray scale, while on the right, the same picture with superimposed the contours found by the IDBSCAN algorithm. The first iteration (red line) recognised only the dense and intense agglomeration of pixels in the centre, the second (blue line) the brightest parts of other tracks and finally the third the remaining faded parts of the tracks (yellow line).}
	\label{fig:idbscan}
\end{figure*}
The IDBSCAN algorithm developed by the CYGNO collaboration \cite{Baracchini:2020iwg} is based on an optimised iterative application of the well-known DBSCAN method \cite{dbscan1996}. This is a non parametric, density-based clustering algorithm, which groups together pixels above a determined threshold with many neighbours in a certain phase space, within a sphere of radius $\epsilon$. This method is particularly useful for the application to sCMOS camera pictures because it is able to recognise pixels which lie isolated in low density regions and to eliminate them. Indeed, these are usually high count noise-induced pixels which survived the filters and the noise suppressions which do not represent physical energy deposits.  In the CYGNO analysis a third dimension to the phase space of the points considered is employed, adding to the pixel position (x-y coordinates) the measured number of counts in that pixel. This approach can reduce the emergence of clusters caused by combinatorial chances due to mere x-y position closeness.\\
Since the specific energy deposit (hence, the photon intensities measured by the pixels) varies with particle types and energies, and also along the path of each track, the clustering procedure is iterated three times. For the first iteration, the parameters are tuned to form clusters of spatially dense and intense pixels, characteristic of high energy deposits such as nuclear recoils, alpha particles or the Bragg peak of soft electron recoils. The density in 3D is called \emph{sparsity}. The pixels belonging to the reconstructed clusters are then removed
from the image, and the DBSCAN procedure is repeated, with a looser selection on the sparsity parameters. The second iteration is tuned to efficiently reconstruct soft very short electron tracks and slices of tracks from nuclear recoils with lower intensity. A third iteration of DBSCAN with even looser parameters is finally executed, targeting faint portions of a recoil.

Figure \ref{fig:idbscan} shows an example of a picture taken of the natural radioactivity with the MANGO prototype and processed by the IDBSCAN algorithm. The left panel displays the original picture, while on the right, the same image is shown with superimposed the contours found by the IDBSCAN algorithm. The first iteration (red line) recognised only the dense and intense agglomeration of pixels in the centre, the second (blue line) the brightest parts of other tracks and finally the third the remaining faded parts (yellow line). As it can be seen, the pixels belonging to the long and curly electron recoil are all found but segmented in various clusters also belonging to different iterations. 
Once the three iterations are terminated, the clusters found are used as input of the supercluster algorithm.
\subsection{Supercluster algorithms}
\label{subsec:supercl}
The goal of the supercluster algorithm is to identify all the pixels belonging to the each track by merging into a single cluster the seeds found by the IDBSCAN procedure. In fact, the IDBSCAN looks for neighbouring pixels with a fixed search radius $\epsilon$, a feature that is not able cope with the irregular nature of the energy deposit along a recoil path. It is not possible to substantially increase the radius search as a lot of pixels only containing noise would be added to the cluster, dramatically worsening the energy resolution and failing to separate close tracks. As a result, a more sophisticated strategy is required to follow each track along the path and group together all its pixels. Since executing any of the algorithm capable of achieving this goal on the full 2304$\times$2304 image is not manageable CPU-wise due to the huge pixel combinatorics, the seeds found by the IDBSCAN algorithm are used as the input of a subsequent method whose goal is to find objects boundaries in images.
Many algorithms exist that can perform this task, with different performances. In this thesis, the geodesic active contour method \cite{gac,mgac} is compared to the Chan-Vese algorithm in the reconstruction of \fe signals\cite{chanvese}.

In order to maximally profit from all the information available in the image, the rebinning applied to the pixels at the level of noise reduction is removed from the seeds that are employed in the above algorithms. A Gaussian filter is applied to the unbinned pixels of each seed to smoothen their response.
This is realised by dividing the full image in subsets of 16 pixels and for each of these small groups, the intensity is convolved  with a Gaussian function, whose sigma is equal to the standard deviation of the intensities of all the pixels of the subset. The pixel intensities which are outside a $n_f$ sigma window with respect to the average are replaced with the average value of the intensity of the group of pixels.
The Gaussian filter can be expressed as:
\begin{equation}
\label{eq:gausfilter}
F(N_{ph})=\alpha \left(G_{n_f\sigma}*N_{ph}\right),
\end{equation}
with $N_{ph}$ is the function representing the number of photons collected by the pixels (which is proportional to the pixel counts), $\alpha$ the relative intensity of the filter, $n_f$ the number of standard deviation $\sigma$ characteristic of the filter.
\subsubsection{Geodesic Active Contour}
\label{subsubsec:gac}
The geodesic active contour (GAC) relies on the calculation of the photon intensity gradients along the pixels to follow the track path. The borders of a cluster are found by looking for strong variations on the gradient in all directions. This algorithm is especially suited for the analysis of images where objects have sharp edges \cite{gac,mgac}.\\
The GAC algorithm applied to the Gaussian filtered images finds the contours of the tracks by minimising a function $E$ defined as:
\begin{equation}
\label{eq:gacminfunc}
E=\int_{0}^{1}g(N_{ph})C(p)\cdot\vert C_p\vert dp,
\end{equation}
where $p$ is the intrinsic coordinate along the curve $C(p)$, $\vert C_p\vert dp$ is infinitesimal arc-length along the curve, and $g(N_{ph})$ is the stopping edge function which allows to define the boundary of the cluster by weighting pixel by pixel the variation of the photon gradient. $g(N_{ph})$ is defined as:
\begin{equation}
\label{eq:gacstopping}
g(N_{ph})=\frac{1}{\sqrt{1+\left\lVert\nabla F(N_{ph})\right\rVert}}
\end{equation}
where the gradient for a generic function $f$ can be calculated pixel by pixel as:
\begin{equation}
\label{eq:gacgrad}
\left\lVert\nabla f \right\rVert= \sqrt{\left(\frac{\partial f}{\partial x}\right)^2 + \left(\frac{\partial f}{\partial y}\right)^2}
\end{equation}
In presence of a strong gradient, that is expected to mark the transition between energy deposits and low intensity sensor noise, the function $g$ is minimised, and consequently the border curve is found. A number of 300 iterations is used to evolve the supercluster contour. The GAC algorithm possesses only two free parameters that can be optimised to tune the its response to the data: $n_f$ and $\alpha$, parameters of the Gaussian filter.
This is the superclustering method employed by the CYGNO experiment to obtained the results described in Section \ref{subsec:cyg_backrej}.
\subsubsection{Chan-Vese algorithm}
\label{subsubsec:chanvese}
In this thesis, the performances of an alternative method to group clusters into a track are compared with the GAC algorithm in the reconstrution of ER produced by the absorption of X-rays produced by \fe source. This is based on an algorithm developed by Chan and Vese \cite{chanvese} to recognise objects which display blurred contours. It is a binary modification of the Mumford-Shah image approximation \cite{Mumfordshah}. If $f$ is the scalar function of domain $\Omega$ representing the gray-scale intensity of a picture, the Mumford-Shah approximation states that the best representation of $f$ is given by the function $u$ and boundary curve $\mathcal{C}$ which minimise the following expression:
\begin{equation}
\label{eq:mumfordshah}
\text{argmin}_{u,\mathcal{C}}\left[ \mu \rm{Length}(\mathcal{C}) +\lambda \int_{\Omega}(f(x)-u(x))^2dx + \int_{\Omega/\mathcal{C}} \vert\nabla u(x)\vert^2 dx\right],
\end{equation}
with $\mu$, $\lambda$ parameters. The first term penalises long boundary curves, the second assures that $u$ is as close as possible to $f$ and the third guarantees the differentiability and continuity of the function $u$ both on the domain $\Omega$ and along the curve $\mathcal{C}$. The Chan-Vese modification to the the Mumford-Shah approximation removes the differentiability requirement, imposes that $\mathcal{C}$ generates a closed set, adds a term to penalise large areas inside the curve $\mathcal{C}$, and simplifies the function $u$ to a binary value as:
\begin{eqnarray}
u(x)=
\begin{cases}
c_1  & \forall x \in inside(\mathcal{C})\\
c_2  & \forall x \in outside(\mathcal{C})\\
\end{cases}
\end{eqnarray}
Now, $c_1$,$c_2$ and $\mathcal{C}$ are the variables that better approximate the image when the following expression is minimised:
\begin{equation}
\label{eq:chanvese}
\begin{split}
\text{argmin}_{c_1,c_2,\mathcal{C}} \Bigl[ &\mu \rm{Length}(\mathcal{C}) + \nu Area(inside(\mathcal{C})) + \\
& + \lambda_1 \int_{inside(\mathcal{C})}(f(x)-c_1)^2dx 
+\lambda_2 \int_{outside(\mathcal{C})}(f(x)-c_2)^2dx \Bigr] ,
\end{split}
\end{equation}
with $\lambda_1$ and $\lambda_2$ the weight parameters of the approximation term of $f$ inside and outside the curve $\mathcal{C}$. This is particularly interesting for the application to the track finding algorithm in the sCMOS camera. Indeed, after the first clustering performed by the IDBSCAN, the interesting sets of pixels selected may contain many tracks separated each other by some low intensity noise. The Chan-Vese binary minimisation finds the borders of the tracks by separating the average value of noise ($c_2$) from the average value of pixel intensity of the tracks ($c_1$). In the Python package used for this study, the parameters $\mu$ and $\nu$ are hard coded. When applying this minimisation algorithm, the free parameters which can be tuned are $\lambda_1$, $\lambda_2$ and the $\alpha$ and $n_f$ of the Gaussian filtering.
\subsubsection{Preliminary comparison of superclustering algorithm performances}
\label{subsubsec:optimiz}
The best choice for the superclustering algorithms depends on the physics case and the analysis goal under consideration.\\
In the direct DM search context, where very low energy deposits are looked for in the images, one of the important aspects to be considered is the ability of the algorithm to correctly define the track boundaries. The goal is to maximise the inclusion in the final track of actual signal pixels, while minimising the collection of spurious noise ones. In order to investigate this aspect, images  taken with the MANGO prototype at LNGS (see Section \ref{sec:mango}) with an \fe source emitting 5.9 keV X-rays inside the gas sensitive volume of the TPC, are reconstructed with both the superclustering algorithms discussed above and the results are compared in the following. During the data taking a drift field of 1 kV/cm was utilised to minimise the charge attachment, 2.5 kV/cm transfer field between the three 50 $\mu$m thin GEMs and 420 V across each GEM were employed to guarantee high gain. More details on the operation with the \fe source are presented in Chapter \ref{chap5}. A 5.9 keV X-ray interacts by photoelectric effect inside the gas transferring all of its energy to an electron which travels for few hundreds of $\mu$m in the CYGNO gas mixture of He:CF$_4$ 60/40 before being stopped and absorbed. This recoiling electron frees about 130 pair of electrons and ions. The primary charge is drifted towards the anode where it is amplified and the light is produced. The diffusion of the primary electronic cloud during  drift and amplification induce a spatial smearing of the electron positions much larger than the original recoil track distance. As a consequence, the photons are emitted by a nearly spherical charge distribution whose shape is dominated by diffusion rather than by the initial track shape. The resulting signature of the \fe emission on the sCMOS picture is therefore an approximately round spot. Given the exposure time of the sCMOS camera of 0.5 s and the activity of the \fe source, the occupancy of these \fe spots on each image is small enough for both supercluster algorithms to distinguish and separate them.

\begin{figure*}[t]
	\centering
	\includegraphics[width=0.33\textwidth]{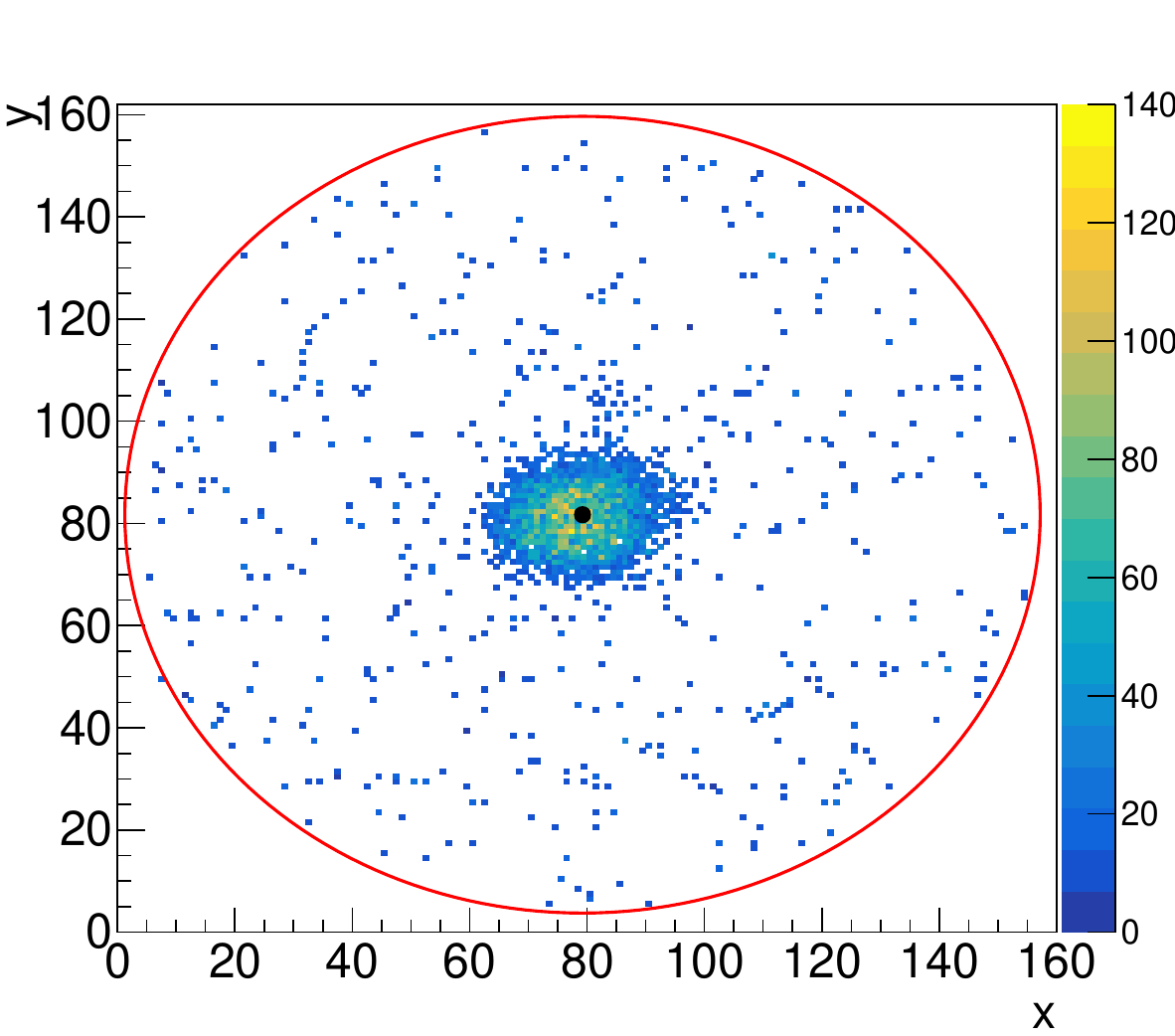}
	\includegraphics[width=0.53\textwidth]{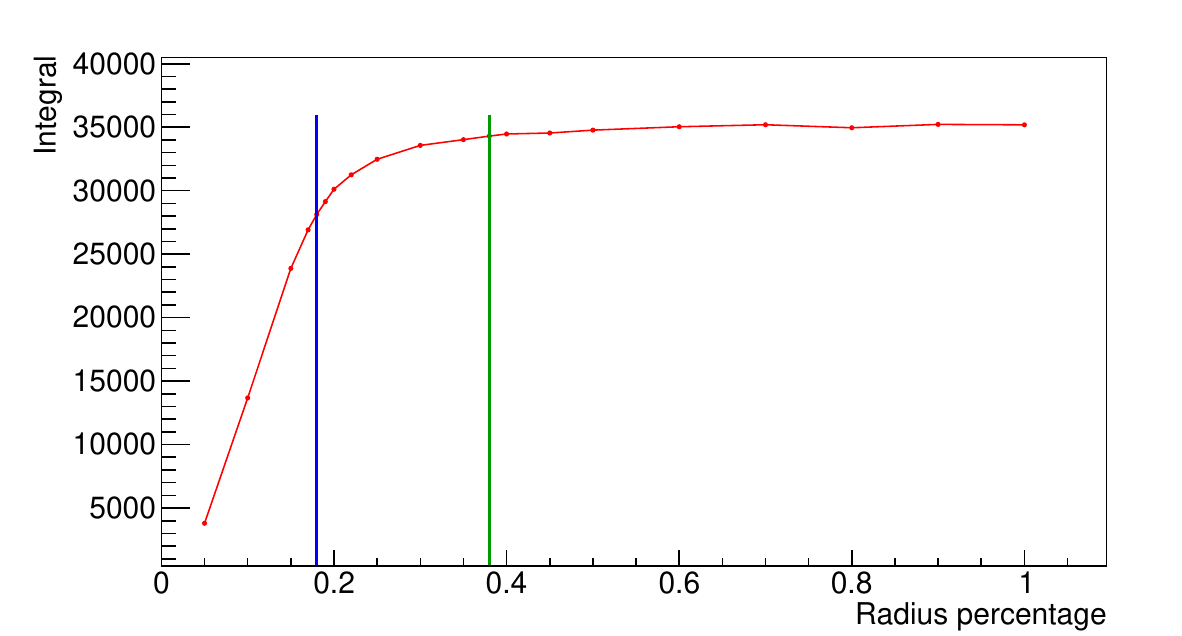}
	\caption{On the left panel an example of the pixels above the noise threshold of an \fe spot. The black dot is the evaluated barycentre and the red circle is the one with the maximum radius of $\sim$ 80 pixels in dimension. On the right panel, the average integral of the \fe tracks as a function of the percentage of the maximum radius utilised for the calculation. The blue vertical line shows the average integral of the GAC supercluster algorithm, while the green one refers to the Chan-Vese.}
	\label{fig:circleiron}
\end{figure*}
In order to evaluate the capability of each algorithm at correctly identifying the track borders, the \fe spots are selected as the IDBSCAN output with extremely loose parameters to obtain clusters much larger than the actual track. The photon barycentre is evaluated as the average x and y coordinate weighted on the pixel intensity. All the pixels enclosed by a circle with radius of $\sim$ 80 pixels and centred on the barycentre are used to calculate the \emph{integral}, proportional to the number of photons collected by the pixels of the camera and therefore to the energy released in the gas. The integral is defined as the sum of the counts of all the pixels selected, after the pedestal subtraction. This sum is evaluated repeatedly decreasing the circle dimensions and the overall integral with different radius choices is compared. Figure \ref{fig:circleiron} shows on the left an example of the pixels above the noise threshold of an \fe spot. The black dot is the light barycentre and the red circle is the maximum one with radius of $\sim$ 80 pixels. 
Figure \ref{fig:circleiron} shows on the right panel, the average integral of the \fe tracks as a function of the percentage of the maximum radius utilised in the evaluation of the integral. The \fe integral linearly depends on the radius dimension up to nearly 1/3 of the 80 pixel radius, after which it starts saturating, indicating that this represents the actual dimension of the physical \fe spot. Enlarging further the radius simply includes in the calculation more noise-containing pixels, effectively worsening the energy resolution, degrading the reconstruction of the track x-y shape, and giving a null contribution to the overall integral estimation. In order to perform the test, the parameters of the GAC and Chan-Vese algorithms are optimised to be able to reconstruct correctly a variety of tracks and then the two are applied to the same \fe data sets. The average integral obtained with GAC and Chan-Vese are superimposed on the right plot of Figure \ref{fig:circleiron}, respectively in blue and green. It can be deduced that the GAC algorithm is too strict in the determination of the borders, inducing an underestimation of \fe spot dimension and a 20\% reduction of the integral. No suitable change in the parameters of GAC is found to improve the determination of the \fe integral without causing different tracks to be wrongly merged by the algorithm.\\
\begin{figure*}[t]
	\centering
	\includegraphics[width=0.45\textwidth]{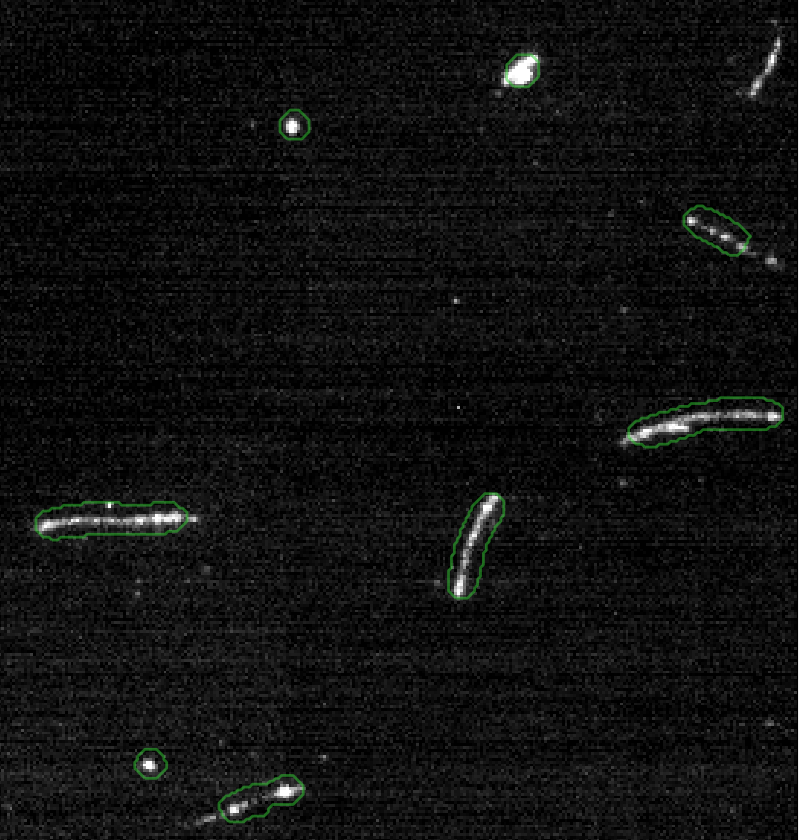}
	\includegraphics[width=0.45\textwidth]{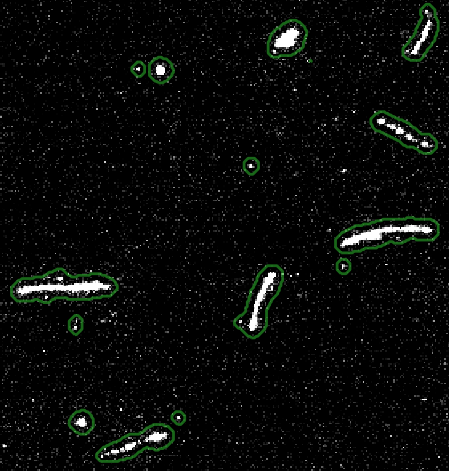}
	\caption{A sCMOS image of natural radioactivity taken with the MANGO detector. Different types of tracks are distinguishable with a wide variety of topology.  The output of the GAC and the Chan-Vese algorithms are displayed respectively on the left and the right panels.}
	\label{fig:gacvschan_clu}
\end{figure*}
Another important aspect to consider when comparing superclustering algorithms is their ability to correctly identify the totality of the track, without missing fainter portions and/or merging multiple tracks.
Moreover, when a track is fragmented into multiple clusters, not only the original information is impossible to retrieve, but it generates more background for tracks similar to the fragments generated. Figure \ref{fig:gacvschan_clu} displays an example of an image taken with the MANGO detector of the natural radioactivity. Different types of tracks are distinguishable with a wide variety of topology. The output of the GAC and the Chan-Vese algorithms are displayed respectively on the left and the right panels. It can be noted that all the tracks are well reconstructed with Chan-Vese, while the GAC algorithm sometimes fails to properly include fainter parts of a track into the superclusters. As in the analysis of \fe spot integral discussed above, the conditions on the GAC gradients result, sometimes, in a too strict requirement for the cluster to be including all the pixels of a track. Nevertheless, it also has to be noted that the Chan-Vese algorithm fails at splitting tracks which are too close to each other, as Figure \ref{fig:gacvschan_clu} on the right shows, with some round spots merged into longer straight tracks.

To determine the best superclustering algorithm for CYGNO's physics goals, several tracks reconstruction performances (such as energy resolution, directional capabilities and particle identification) need to be compared, and a detailed study on this is ongoing. For the purpose of this thesis, since the above discussion demonstrated that Chan-Vese algorithm is better suited for the reconstruction of \fe, this algorithm is employed in the analysis presented in the next Chapter. The optimised set of parameters employed henceforth in the reconstruction of the sCMOS camera pictures is the following:
\begin{equation}
\label{eq:chanveseparam}
\left[n_{\sigma n}, n_f,\alpha, \lambda_1, \lambda_2 \right] = \left[2,2,10,2,1.5 \right]
\end{equation}
\section{Analysis of the \fe spot shape}
\label{sec:difffe}
\begin{figure}[!t] 
	\centering
	\includegraphics[width=0.42\linewidth]{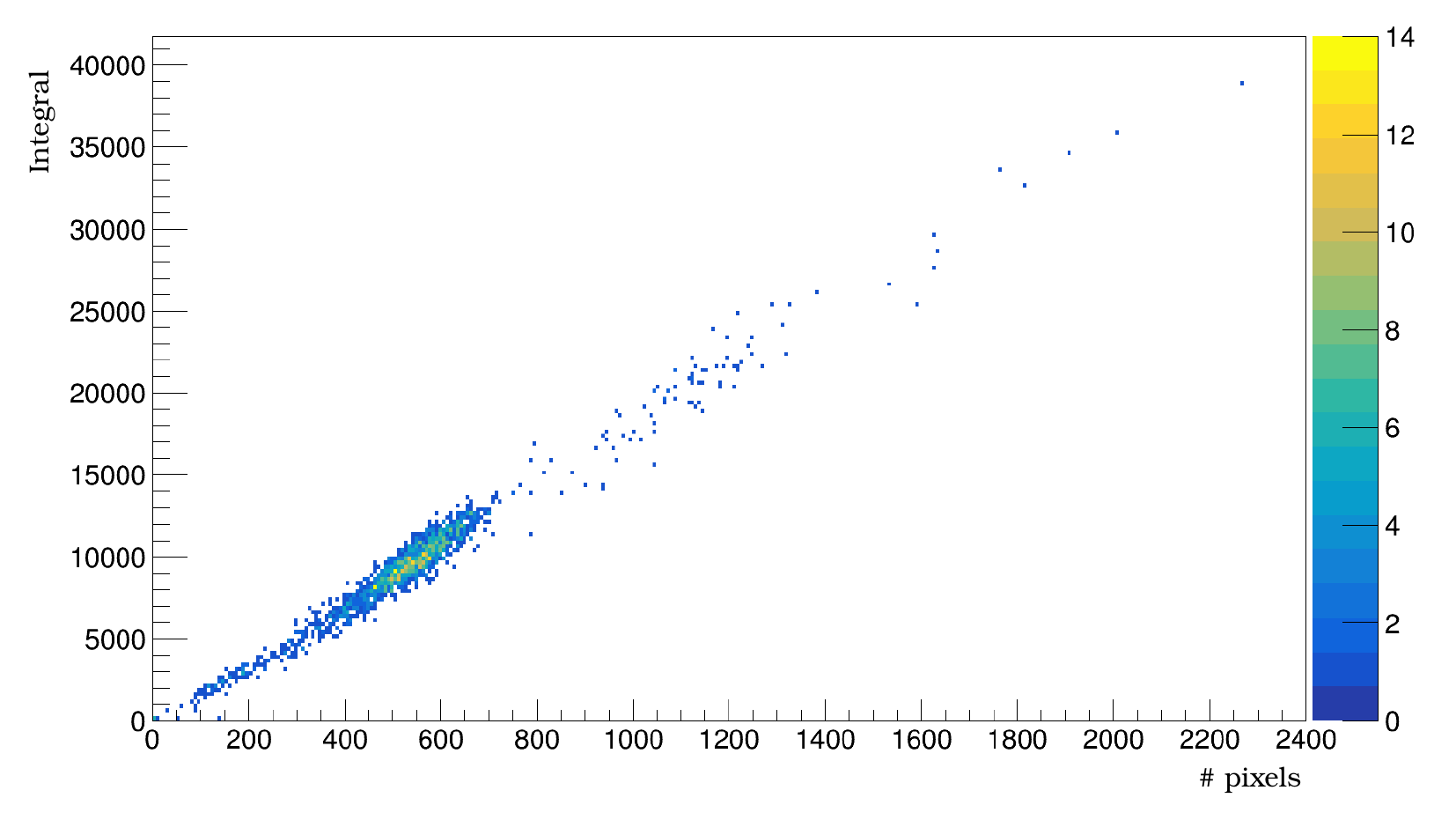}
	\includegraphics[width=0.48\linewidth]{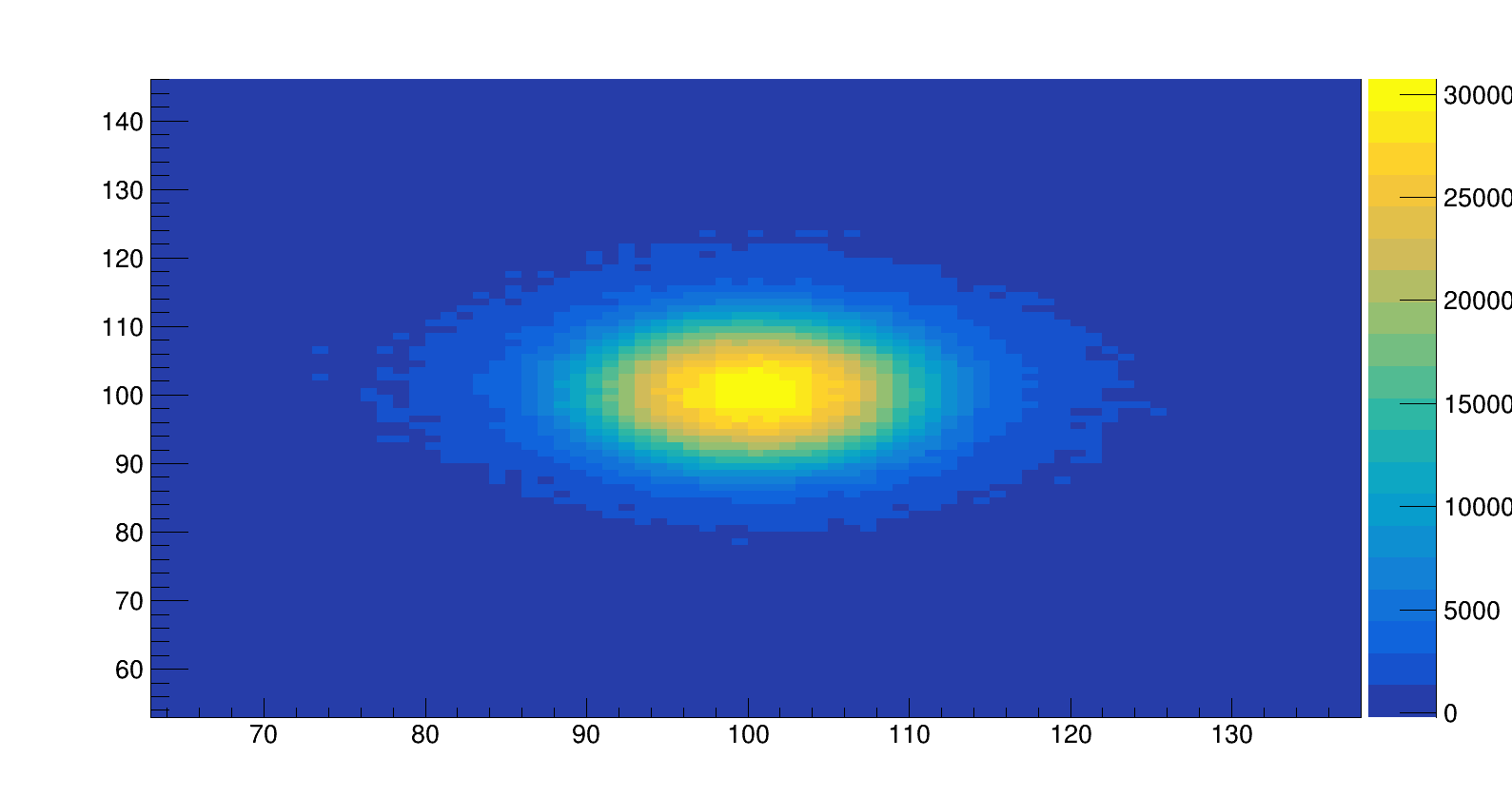}
	\caption{On the left, light integral of the selected \fe clusters versus the  number of pixels belonging to the track $n_p$. A strong linear correlation appears evident. On the right, the sum of hundreds of \fe tracks after their barycentre was aligned to the same position.}
	\label{fig:nhits0}
\end{figure}
The \fe induced recoils are an excellent tool to study the transverse diffusion of the experimental setup. The absorption of the 5.9 keV X-ray produces an electron recoil which travels about few hundreds of $\mu$m before being stopped in the CYGNO gas mixture. As shown in left panel of Figure \ref{fig:hecf4}, the diffusion of the primary charges while drifting for ten of cm in the CYGNO gas mixture is of the order of hundreds of $\mu$m in standard deviation.  The amplification stage induces an additional constant diffusion of the same order of magnitude (see Section \ref{subsec:spot}) that adds in quadrature to this term. As a consequence, the  \fe shape imaged by the sCMOS camera appears as a round spot, whose intensity roughly follows a two-dimensional Gaussian distribution when projected on the amplification plane, dominated by the diffusion the electronic cloud was subject to. Therefore, in order to infer the diffusion from the \fe data, it is necessary to precisely determine the x-y dimension of the spots. 
In this respect, it is important to notice how a simplistic definition of the spot dimension in terms of the number of pixels ($n_p$) associated to the track by the reconstruction code discussed in Section \ref{sec:recocode} results in a biased determination of spot dimension. The integral and the $n_p$ of a cluster are in fact highly correlated, as can be seen in left panel of Figure \ref{fig:nhits0}, since a larger light yield will result in a larger number of pixels with counts above the noise threshold. For this reason, a dedicated method was developed in the context of this thesis in order to define observables independent from the track energy and hence able to extract the transverse diffusion from the analysis of the \fe spot shape without introducing a bias. This is based on the analysis of the $x-y$ projection of the cluster intensity profile, since the shape is expected to be preserved even if a cluster becomes more, or less, luminous. With the goal of minimising any systematic effect and provide a robust diffusion estimation, the \fe spot cluster distributions are averaged after having aligned all their photon barycentres. An example of the 2D pixel averaged distribution is shown in right panel of Figure \ref{fig:nhits0}. 

\begin{figure}[!t] 
	\centering
	\includegraphics[width=0.45\linewidth]{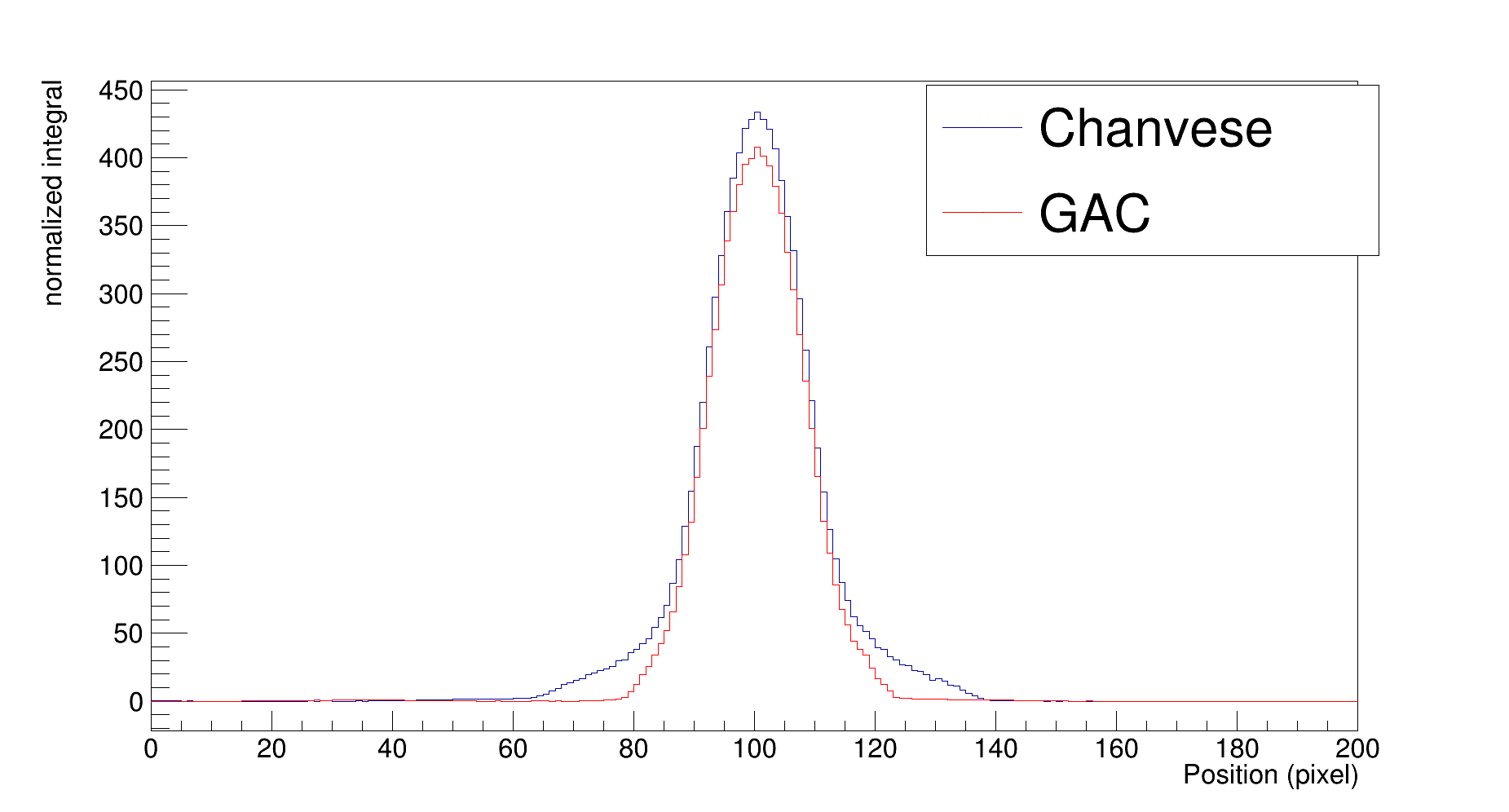}
	\includegraphics[width=0.45\linewidth]{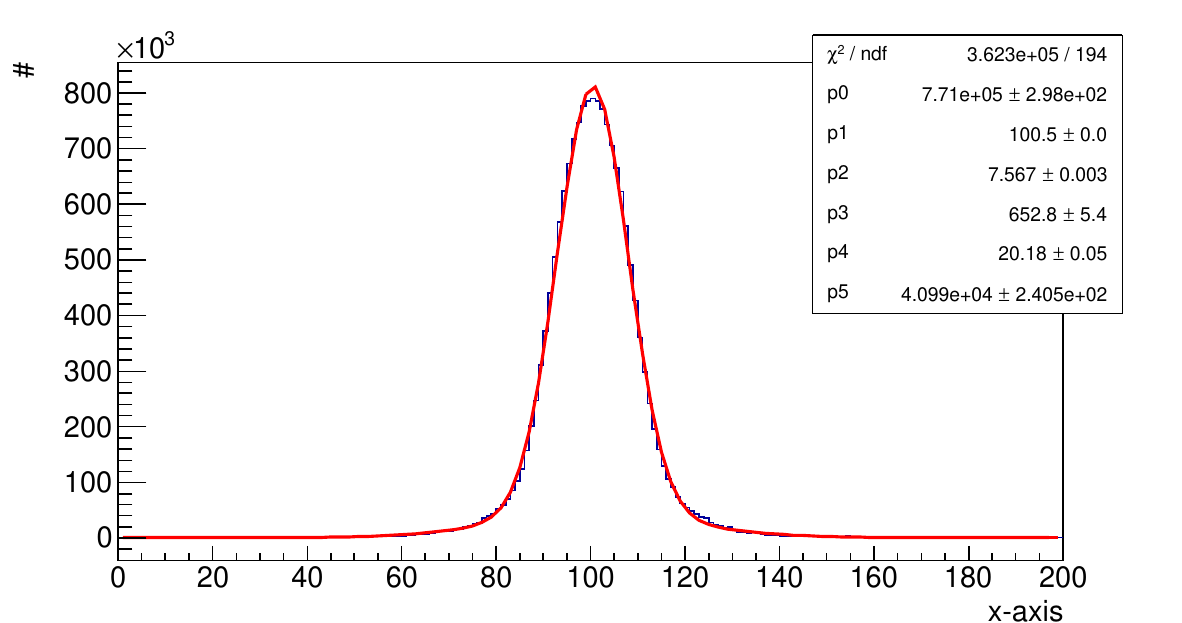}
	\caption{On the left, the light integral distribution of the x projection of the \fe centred clusters obtained with the two superclustering algorithms described in Section \ref{subsec:supercl}. On the right, the same projection distribution evaluated by the Chan-Vese algorithm with a double Gaussian fit superimposed.}
	\label{fig:fediff}
\end{figure}
The averaged distribution in x and y are fitted with a double Gaussian, in order to properly include secondary tails in the projections distributions, that anyway never account for a fraction larger than 20\%. By comparing  $^{55}$Fe spots reconstructed applying either the GAC or Chan Vese superclustering algorithms, it was verified that the observed tails in the projection distributions depend on the algorithm definition of a cluster boundary rather than by misalignment of the barycentres. This comparison is explicitly shown in Figure \ref{fig:fediff} on the left panel. The GAC algorithm has stricter requirements on the definition of the borders of a cluster that results in cutting the \fe track short of some pixels, which are instead included by the Chan-Vese algorithm. In order to not depend on the superclustering algorithm (i.e. on the definition of the cluster boundary), the diffusion is estimated by averaging only the primary sigma of the $x$ and the $y$ projection of the $^{55}$Fe spots.\\
Figure \ref{fig:fediff} shows on the right an example of an x projection of the aligned \fe tracks with a superimposed double Gaussian fit.

This method is applied to the data discussed in Chapter \ref{chap5} in the context of the optimisation of the amplification structures.
\section{Evaluation of alpha tracks transverse profile}
\label{sec:diffalpha}
Alpha particles consist in two protons and two neutrons bound together into a particle identical to a helium-4 nucleus. While also other mechanisms are possible, they are typically produced by an alpha radioactive decay. In particular, the alpha particles produced by the $^{241}$Am radioactive source employed to characterise the CYGNO experiment possess a $\sim$ 5.4 MeV energy and travel about 4 cm in the CYGNO gas mixture. Due to their large energy, alpha particles tend to travel along a mostly straight path until they have released their entire energy. As a consequence, their signature in a sCMOS picture is represented by very intense and straight tracks. Since the intrinsic transverse dimension of the primary electrons generated by the passage of an alpha in the gas is of the order of few tens of microns, and, as already discussed in Section \ref{sec:difffe}, the typical diffusion during drift and at the amplification stage is hundreds of $\mu$m, the alpha track transverse profile dimension can provide an estimation of the diffusion alternative to the analysis of the \fe spot shape.\\
\begin{figure}[!t] 
	\centering
	\includegraphics[width=1\linewidth]{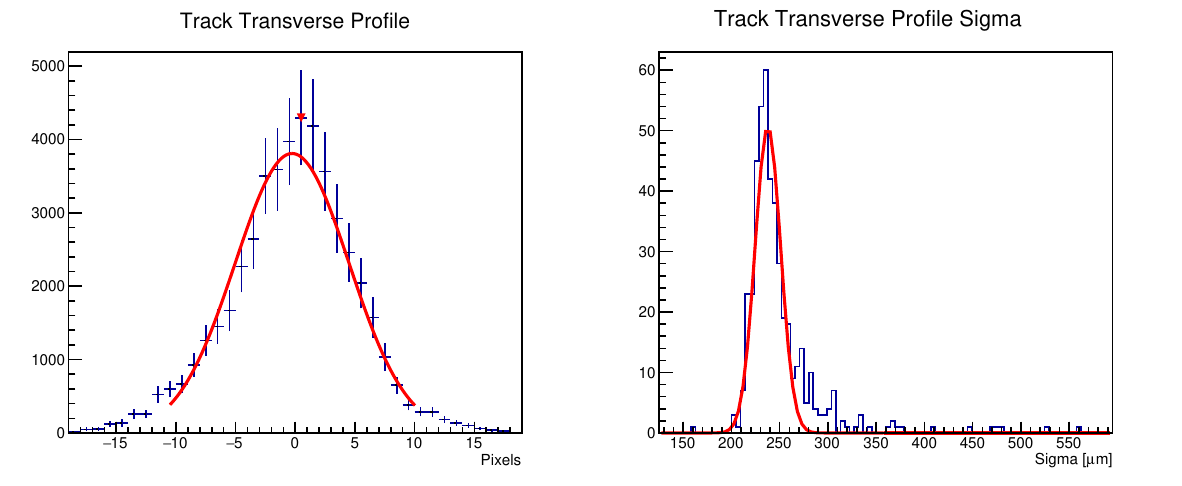}
	\caption{On the left, an example of a transverse profile of an alpha particle surviving the selection cuts described in the text with a Gaussian fit superimposed. On the right, an example of a distribution of the $\sigma$s obtained from the Gaussian fits as of the one left panel.}
	\label{fig:profalpha}
\end{figure}

The track extremes (i.e. start and end) are removed from the evaluation of the transverse profile in order to avoid including the possible multiple scattering in the gas (that can spoil the straightness of the track) and disuniformities at the boundaries of the detector. This is attained by defining a \emph{cropped} track contained inside a radius of 0.61 cm centred in the intensity barycentre as calculated from the original full track. A Principal Component Analysis (PCA) is applied to the intensity-weighted collection of pixels belonging to such cropped selection to determine the track principal axes. The cropped track pixels are then projected on the major (minor) axis, in order to produce a one-dimensional longitudinal (transverse) profile. A one-dimensional peak search function based on the TSpectrum ROOT class \cite{root} with an amplitude threshold of 10$\%$ is applied to the transverse profile of the cropped tracks and the number of found peaks is required to be equal to one in order to select single tracks, eliminating overlaps. The transverse profile of the cropped tracks satisfying this cut is fitted with a Gaussian distribution, and multiple superimposing tracks (due to pileup or mis-reconstructed images) are further rejected by requiring the $\chi^2$/nDOF of the Gaussian fit to be less than 5. Figure \ref{fig:profalpha} shows on the left panel an example of a transverse profile of an alpha particle with the Gaussian fit superimposed (details on the data taking in Chapter \ref{chap6}). The distribution of the $\sigma$s obtained from the fitting function to these selected cropped tracks is, in turn, fitted with a Gaussian whose mean is employed as an estimate of the track diffusion. An example of the distribution of $\sigma$s is displayed on the right panel of Figure \ref{fig:profalpha}.

This analysis procedure was developed to evaluate the diffusion in Negative Ion Drift operation discussed in Chapter  \ref{chap6}.
\chapter{CYGNO amplification optimisation studies}
\label{chap5}
The optical readout employed by the CYGNO experiment consists in the use of sCMOS cameras and PMTs, as discussed in Chapter \ref{chap3}. The former, coupled to suitable MPGD and optics, allows to image large areas while maintaining high granularity on the 2D projection on the amplification plane of the recoils occurred in the sensitive volume of the TPC. The drawback of this approach lies in the small solid angle covered by the sCMOS sensor, which reduces the number of collected photons per energy deposited (see Chapter \ref{chap3}).  The geometrical acceptance can be as low as 10$^{-4}$ for an imaged area of 25.6 $\times$ 25.6 cm$^2$ as in the LEMOn detector illustrated in Section \ref{sec:LEMOn}. Thus, enhancing the photon production at the amplification stage is of uttermost importance for the CYGNO experiment in particular, and for any gaseous detector exploiting optical readout in general. Increasing the voltage across the GEMs does not solve the problem as one would eventually face breakdown effects on the gas which disrupts the operation. The typical triple thin GEM amplification employed by CYGNO provides enough light yield to detect low energy recoils down to few keV. Yet, the use of multiple GEMs increases the diffusion processes that happen in the gap between the amplification foils. A large diffusion limits the topological information which can be retrieved from a recoil track, spoiling the quality of  the information relevant for directional studies (see Section \ref{sec:directional}) and particle identification and rejection (see Section \ref{subsec:cyg_backrej}). \\
Experimental R\&Ds were performed with MANGO and LEMOn to optimise the configuration of the amplification stage, in order to maximise the light gain while minimising the diffusion. The results presented in \cite{bib:EL_cygno}, showed that it is possible to enhance the light yield by introducing a strong electric field below the last GEM amplification plane. For these reasons, this possibility has been extensively explored with the CYGNO prototypes and is the subject of this Chapter. With respect to \cite{bib:EL_cygno}, the study of this phenomenon is expanded to different GEM thicknesses and stacking options also varying the He to CF$_4$ ratio.
In Section \ref{sec:lemon}, the studies of the strong electric field below the last amplification GEM performed with the LEMOn detector are presented. Section \ref{sec:mango_det} is dedicated to the description of the measurements carried out with the MANGO detector, while Section \ref{sec:mango_light} describes the results. In Section \ref{sec:maxwell}, the simulation of the electric fields close to the holes of the GEMs is presented. Finally, Section \ref{sec:disc} is dedicated to the final discussion on the obtained results.
\section{LEMOn experimental setup}
\label{sec:lemon}
\begin{figure}[t]
	\centering
	\includegraphics[width=0.6\textwidth]{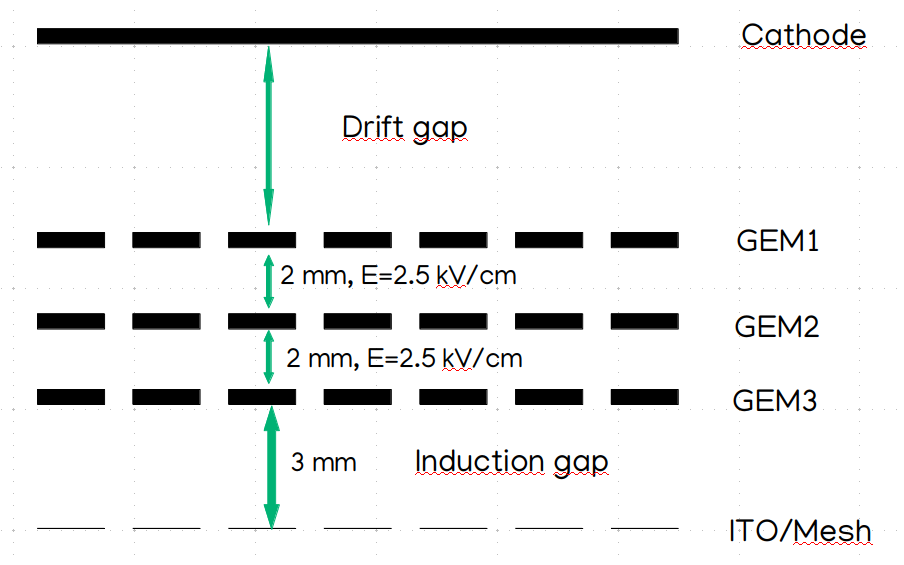}
	\caption{A sketch of the internal structure of the TPC of the CYGNO prototypes employed in this study where the addition of the ITO or mesh below the last amplification GEM plane can be appreciated.}\label{fig:tpcinternsketch}
\end{figure}
In order to further validate the findings presented in \cite{bib:EL_cygno} and to extend them to stronger applied electric fields, the larger LEMOn detector was employed.\\
A full description of the LEMOn prototype is present in Section \ref{sec:LEMOn}. With respect to that setup, for this study an ITO glass electrode with 90$\%$  transparency is placed of a distance of 3 mm below  the last GEM (GEM3), as shown in the sketch of Figure \ref{fig:tpcinternsketch}. By biasing the ITO, it is possible to produce an additional electric field in the region below GEM3 to enhance the light yield of the detector. The region between GEM3 and the ITO glass is defined in the following as the \emph{induction} region, and the field \emph{induction} field (E$_{ITO}$). In addition to the CAEN A1526 powering the LEMOn detector, a CAEN DT1470ET\footnote{\url{https://www.caen.it/products/dt1470et/}} power supply is employed for the ITO electrode as well, with a high sensitivity current-meter ($\sim$ 5 nA) able to precisely measure the continuous current signals on each channel.

For these measurements, LEMOn is optically coupled to the Hamamatsu sCMOS camera ORCA-Fusion (second row of Table \ref{tab:hama}) through a TEDLAR transparent window and an adjustable plastic bellow. The Orca Fusion is positioned at $(50.6 \pm 0.1)$ cm distance from GEM3 and reads out an area of 25.6  $\times$ 25.6 cm$^2$. Therefore, each of the 2304 $\times$ 2304 pixels of the sCMOS sensor images an effective area of 111  $\times$ 111 $\mu$m$^2$.

In this setup, LEMOn is operated with a 0.5 kV/cm drift field, a 2.5 kV/cm transfer fields between the GEMs and 400 V applied across each GEM, with a He:CF$_4$ 60/40 gas mixture at 1000 mbar, the average atmospheric pressure at Laboratori Nazionali di Frascati (LNF). A $\sim$ 115 MBq $^{55}$Fe source is placed at 5 cm distance from GEM1 in order to induce 5.9 keV energy deposits inside the detector active gas volume. The large source activity provides a detectable current signal on each of the 3 GEM electrodes and the ITO glass, also thanks to the high sensitivity of the power supplies employed. This feature, conversely, does not allow to identify each $^{55}$Fe cluster separately in the sCMOS image because of the large pileup. For this reason, and since the current on the electrodes represents an integrated information of all the $^{55}$Fe clusters produced in the gas and amplified by the GEMs, a 1 s exposure time on self-trigger pictures is used for the sCMOS camera data acquisition in LEMOn to perform a consistent light measurement. The light yield from the sCMOS camera and the charge measured on each of the seven LEMOn amplification electrodes (two for each GEM, the upper (\emph{U}) and the bottom (\emph{D}) ones) are studied by varying the induction field E$_{ITO}$ from 0 to 17 kV/cm and the results are reported in the following.	

\subsection{sCMOS images analysis}
\label{subsec:lemon_light}
\begin{figure}[t!] 
	\centering
	\includegraphics[width=0.5\linewidth]{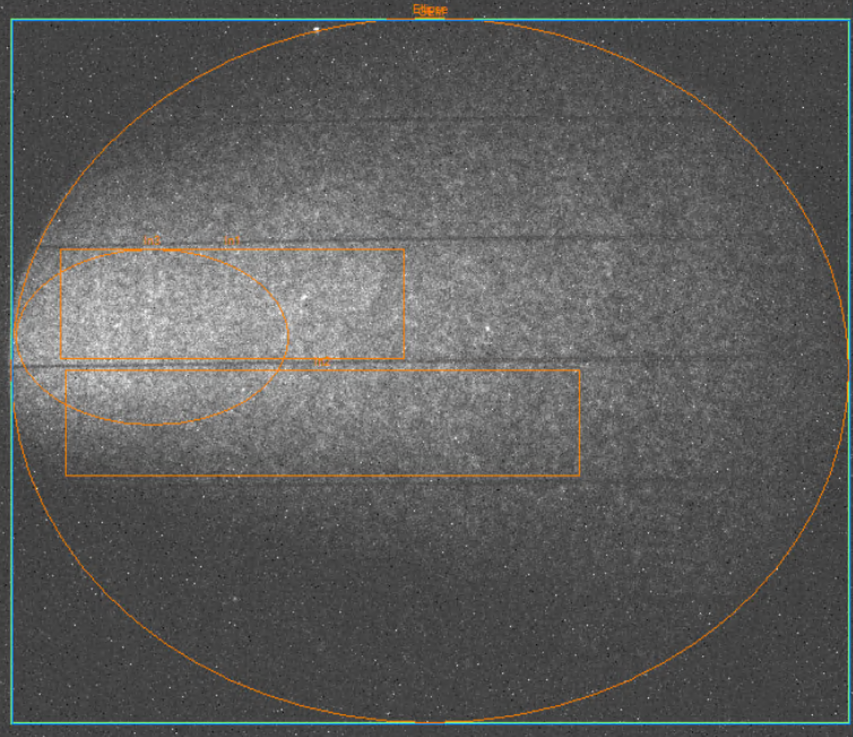}
	\caption{Example of 1 s exposure picture taken with the sCMOS camera with superimposed four regions the total light was evaluated from.}
	\label{fig:longexplemon}
\end{figure}
\begin{figure}[!t] 
	\centering
	\includegraphics[width=0.9\linewidth]{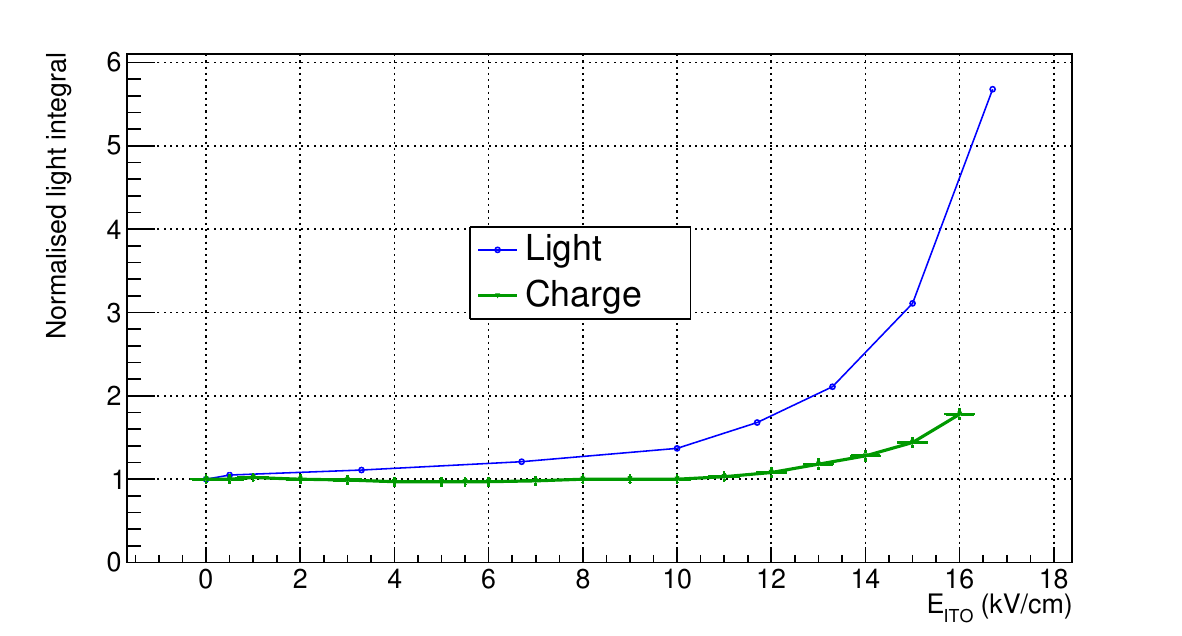}
	\caption{Comparison of the light output and of the charge measured on the ITO glass in LEMOn as a function of the induction field E$_{ITO}$ in He:CF$_4$ 60/40 at 1000 mbar.}
	\label{fig:light_charge_lemon}
\end{figure}

An example of a 1 s exposure sCMOS image acquired by LEMOn exposed to the $^{55}$Fe source is shown in Figure \ref{fig:longexplemon}, with four different regions (two elliptical and two rectangular) highlighted in orange. The light yield in the LEMOn data is evaluated by calculating the number of counts seen by all the sCMOS pixels in each region, after having subtracted the noise pixel by pixel exploiting images acquired in absence of source and with the GEM turned off (i.e. \emph{pedestal} runs). This number is normalised to the measurement with null induction field applied and the relative increase is averaged among the four regions.
Figure \ref{fig:light_charge_lemon} shows the comparison of the relative increase of the light output and of the extra charge produced in the process (see Section \ref{subsec:lemon_charge} for details on how the charge is evaluated) as a function of the induction field in He:CF$_4$ 60/40 at 1000 mbar, explicitly demonstrating the different rate of increase of the two quantities. The light enhancement measured with LEMOn and shown in Figure \ref{fig:light_charge_lemon} is consistent with the results reported in \cite{bib:EL_cygno} when one considers the errata corrige to the induction gap dimension claimed in the \cite{bib:EL_cygno} paper (actual: 2.5 mm, instead of the claimed 3 mm).
\subsection{GEMs electrodes current analysis}
\label{subsec:lemon_charge}
During the amplification processes a large amount of electrons and ions are generated close to the bottom of the holes of each GEM and drifted away in opposite directions by the electric fields applied. Considering the Shockley-Ramo theorem \cite{bib:Ramo,bib:Shockley}, the instantaneous current induced on the GEM electrodes depends on the amount of charge in motion (both ions and electrons), on their velocity and on a complicated function of the electric field along the charge path from its generation to the point of collection. The significant difference in ion and electron drift velocity leads to an average charge collection time of the order of $\mu$s for the firsts and few ns for the seconds.
The signal is induced as soon as a charge gets in motion and lasts until it is collected by an electrode. It is important to notice how a current signal is induced also on electrodes not collecting any charge due to the electron and ion motion. In this case, the signal is bipolar and its overall integral sums to zero. As a consequence, an infinite integration of the current signal allows to correctly measure a signal dependent solely on the actual charge collected by an electrode.

In absence of induction field E$_{ITO}$, all CYGNO prototypes are operated with the GEM3 bottom electrode at ground, since no charge needs to be collected with the optical readout approach, a configuration that causes the field lines to disorderly close on this. Immediately after the multiplication inside the holes of the last GEM amplification, an electric signal is induced on its electrodes. The upper electrode is responsible for the collection of the majority of the ions coming from the last step of multiplication, while the bottom one of the electrons. When the induction field E$_{ITO}$ is turned on, the field lines gets straightened and begin to close on the ITO glass rather than on the bottom GEM3 electrode. In this configuration, the ions are expected to be mostly collected on the top of GEM3, while the electrons will be shared between the bottom of the GEM3 and the ITO glass. When the electric field inside the gap is large enough to generate charge amplification, these additional electrons are collected on the ITO, while the newly generated ions are shared between the other electrodes, with the large majority of them being collected by the top and bottom ones of the GEM3.
\begin{figure}[!t] 
	\centering
	\includegraphics[width=0.9\linewidth]{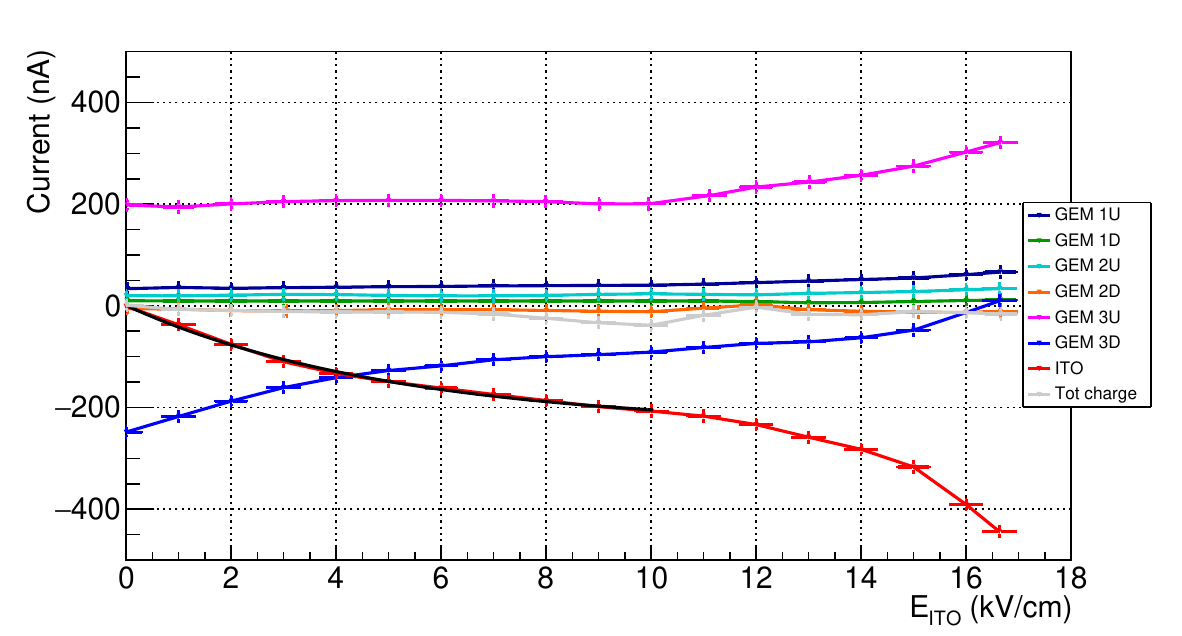}
	\caption{Currents measured in LEMOn as a function of the induction field E$_{ITO}$ for all the six electrodes of the amplification stage plus the ITO glass, where \emph{U} and \emph{D} represent respectively the upper and the bottom electrode of each GEM, and the total charge (in gray) is the sum of all the components. The black line represents the exponential fit described in Equation \ref{eq:fit}.}
	\label{fig:Lemoncharge}
\end{figure}
Given the LEMOn setup, the currents measured in this context can be considered as equivalent to an infinite integration of the signals induced on the GEM electrodes, causing only the electrons and ions actually collected to be relevant for the signal. The large $^{55}$Fe source activity employed guarantees that enough charge is collected by the electrodes, resulting in a current signal well above the sensitivity of the instrument. Figure \ref{fig:Lemoncharge} shows the continuous current measured in LEMOn as a function of the induction field for all the six electrodes of the amplification stage plus the ITO glass, and the total charge (in gray) is the sum of all the components.

The sum of the charge of all electrodes is always consistent with zero and mostly flat, as expected from general arguments by a well grounded electrical circuit. Similarly, the ITO glass and the GEM3D split between them (with proportions depending on the induction field) the current induced by the charges generated in the last amplification stage and moving in the induction gap. When no field is applied, the ITO sees a null current and all the 250 nA are collected by the GEM3D. The current measured on the upper electrode of the last GEM amplification, GEM3U, displays a constant behaviour up to about 10 kV/cm, where an increase starts. This \emph{breaking} point behaviour at 10 kV/cm is shared with the ITO and GEM3D measurements, which show an inflection point at the same value.
In order to evaluate the actual relative increment of measured charge with respect to null induction field, the charge sharing between GEM3D and the ITO glass need to be properly taken into account. To do this, the ITO current measured between 0 and 10 kV/cm is fitted with the function (black line in Figure \ref{fig:Lemoncharge}):
\begin{equation}
\label{eq:fit}
I_{ITO} = a+e^{b+cE_{ITO}}
\end{equation}
and the following parameters are obtained $a= (-240 \pm 20)$ nA, $b= (5.47 \pm 0.08)$ , $c= (-0.20 \pm 0.05)$ cm/kV. The fitted $a$, which represents the asymptote of the exponential function, is nicely consistent with the -250 nA value measured on the GEM3D due to the collection of all the electrons generated in the last GEM when no induction field is applied. In order to properly evaluate the actual charge generated in the gap, the ITO data are then normalised to the value at 10 kV/cm after having subtracted the fitted  $I_{ITO}$ function. These data are shown as a function of the induction field E$_{ITO}$ in Figure \ref{fig:light_charge_lemon} in comparison to the light output relative enhancement (illustrated in Section \ref{subsec:lemon_light}), explicitly displaying the different derivative in increase of the two quantities. This increment above 10 kV/cm in the measured charge is attributed to the generation of a small additional amount of charges right below GEM3 holes, which, nonetheless, can not account for the entire increase of the light output. 


\section{MANGO Experimental setup} \label{sec:MANGO}	
\label{sec:mango_det}
In order to expand and extend the results presented in \cite{bib:EL_cygno} and further validated in Section \ref{sec:lemon}, the MANGO prototype, illustrated in details in Section \ref{sec:mango}, is employed. This small prototype is used to be able to modify the GEM thicknesses and stacking option, in addition to the gas mixture (since no 20 $\times$ 24 cm$^2$ GEMs are available with thickness different from the standard 50 $\mu$m). In addition to the described setup, at a distance $\Delta$z$=3$ mm a metallic mesh from an ATLAS MicroMegas \cite{bib:Micromegas} (30 $\mu$m diameter metallic wires at 50 $\mu$m pitch, resulting in a transparency of $\sim 0.55 $) is placed in order to induce an electric fields below the electrode of the last GEM amplification, as shown in Figure \ref{fig:tpcinternsketch}. Recently, the metallic mesh was replaced with an ITO glass similar to the one employed in LEMOn with a larger transparency (0.9), showing that the results do not depend on the structure employed to produce the additional electric field after the last GEM amplification. As in LEMOn, the region between the last GEM amplification plane and the mesh is defined as the \emph{induction} region, and the electric field applied inside it the \emph{induction} field (E$_{Mesh}$). The drift gap measures 0.8 cm and the detector is operated with 1 kV/cm drift field, a configuration that guarantees a uniform electric field in the drift region without the need of a field cage.
\subsection{Datasets}
\label{subsubsec:mango_daq}	
A $\sim$ 480 kBq $^{55}$Fe X-rays source is employed to generate a 5.9 keV electrons in the MANGO active gas volume. The relative low source activity allows the reconstruction of each single $^{55}$Fe track in the sCMOS images (acquired with 0.5 s exposure), as discussed in Section \ref{sec:mango_light} and shown in Figure \ref{fig:waveformGEM}.

A systematic study of the performances of different He:CF$_4$ ratios in the gas mixture (60/40, 70/30) and different GEM thicknesses and stacking option is performed, since a 10 $\times$ 10 cm$^2$ GEM can be rather easily manufactured and purchased and are available with thicknesses different from the standard 50 $\mu$m. Two types of GEMs are employed: a thin 50 $\mu$m GEM with 70 $\mu$m radius holes and 140$\mu$m pitch (henceforth called "t") and a thicker 125 $\mu$m GEM with 175 $\mu$m radius holes and 350 $\mu$m pitch (henceforth called "T"). All the measurements are performed at the atmospheric pressure at Laboratori Nazionali del Gran Sasso (LNGS), which corresponds to (900 $\pm$ 7) mbar, being located at roughly 1000 m a.s.l.. Table \ref{tab:datataking} shows a summary of the different gas mixtures and GEM configurations explored with MANGO, and Table \ref{tab:app} the voltages applied to the various combinations of GEMs structures. 
\begin{table}[!t]
	\centering
	\begin{adjustbox}{max width=1.01\textwidth}
		\begin{tabular}{|c|c|c|c|c|c|c|c|}
			\hline
			
			\multirow{2}{*}{\large{Ampl stage}} & \multirow{2}{*}{\large{Tag}} & \multicolumn{2}{c|}{\large{Gas mixture (He:CF$_4$)}}  &\multicolumn{4}{c|}{\large{Study}}    \\ \cline{3-8} 
			& & 60/40 & 70/30 &  Gain & Energy Res &  Diffusion & E$_{Mesh}$ \\ \hline \hline
			Triple thin GEM & ttt & $\checkmark$  &  & $\checkmark$ & $\checkmark$ & $\checkmark$ & $\checkmark$ \\ \hline
			Double thick GEM & TT & $\checkmark$   & $\checkmark$ & $\checkmark$ & $\checkmark$ & $\checkmark$ & $\checkmark$ \\ \hline
			1 thick and 1 thin GEM & Tt   & $\checkmark$ & $\checkmark$ & $\checkmark$ & $\checkmark$ & $\checkmark$ & $\checkmark$ \\ \hline
		\end{tabular}
	\end{adjustbox}
	\caption{Table summarising the gas mixtures and GEMs configurations tested with the MANGO prototype.}
	\label{tab:datataking}
\end{table}\begin{table}[!t]
\centering
\begin{adjustbox}{max width=1.01\textwidth}
	\begin{tabular}{|c|c|c|c|c|}
		\hline
		\large{Config} 	& $V_{GEM1}$ (V) & $V_{GEM2}$ (V) & $V_{GEM3}$ (V)  & $V_{GEM}$ for E$_{Mesh}$ studies (V)\\ \hline \hline
		ttt 60/40 & 400-435  & 400-435  & 400-435 & 400+400+400=1200\\ 
		TT 60/40 & 770-780  & 470-520 & n.a & 775+490=1265\\ 
		Tt 60/40 & 740-780 & 400-435 & n.a & 770+400=1170\\ \hline
		TT 70/30 & 700-715 & 500 & n.a & 700+490=1190\\ 
		Tt 70/30 & 660-720 & 350-395 & n.a & 700+385=1085\\ \hline
	\end{tabular}
\end{adjustbox}
\caption{Table summarising the voltages applied to the various combinations of GEMs structure tested with the MANGO prototype. Each column shows the the range of voltages employed for each GEM. The definition of the configurations follows the ones in Table \ref{tab:datataking}}
\label{tab:app}
\end{table}	
\section{MANGO sCMOS images analysis}
\label{sec:mango_light}
\begin{figure}[t] 
	\centering
	\includegraphics[width=0.41\linewidth]{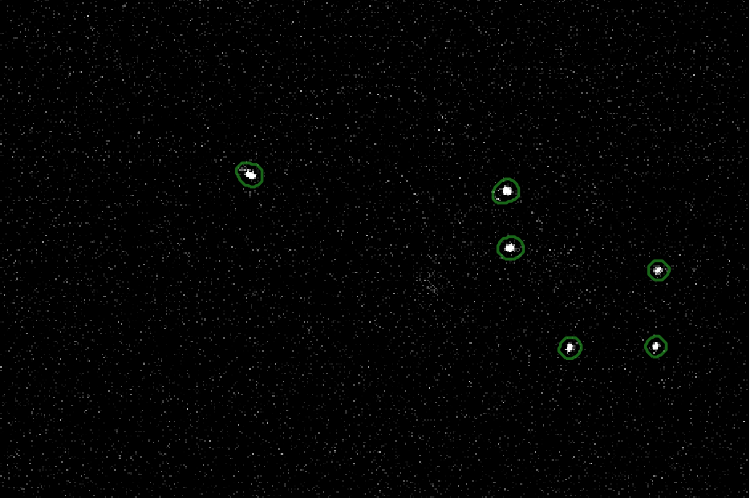} 
	\includegraphics[width=0.56\linewidth]{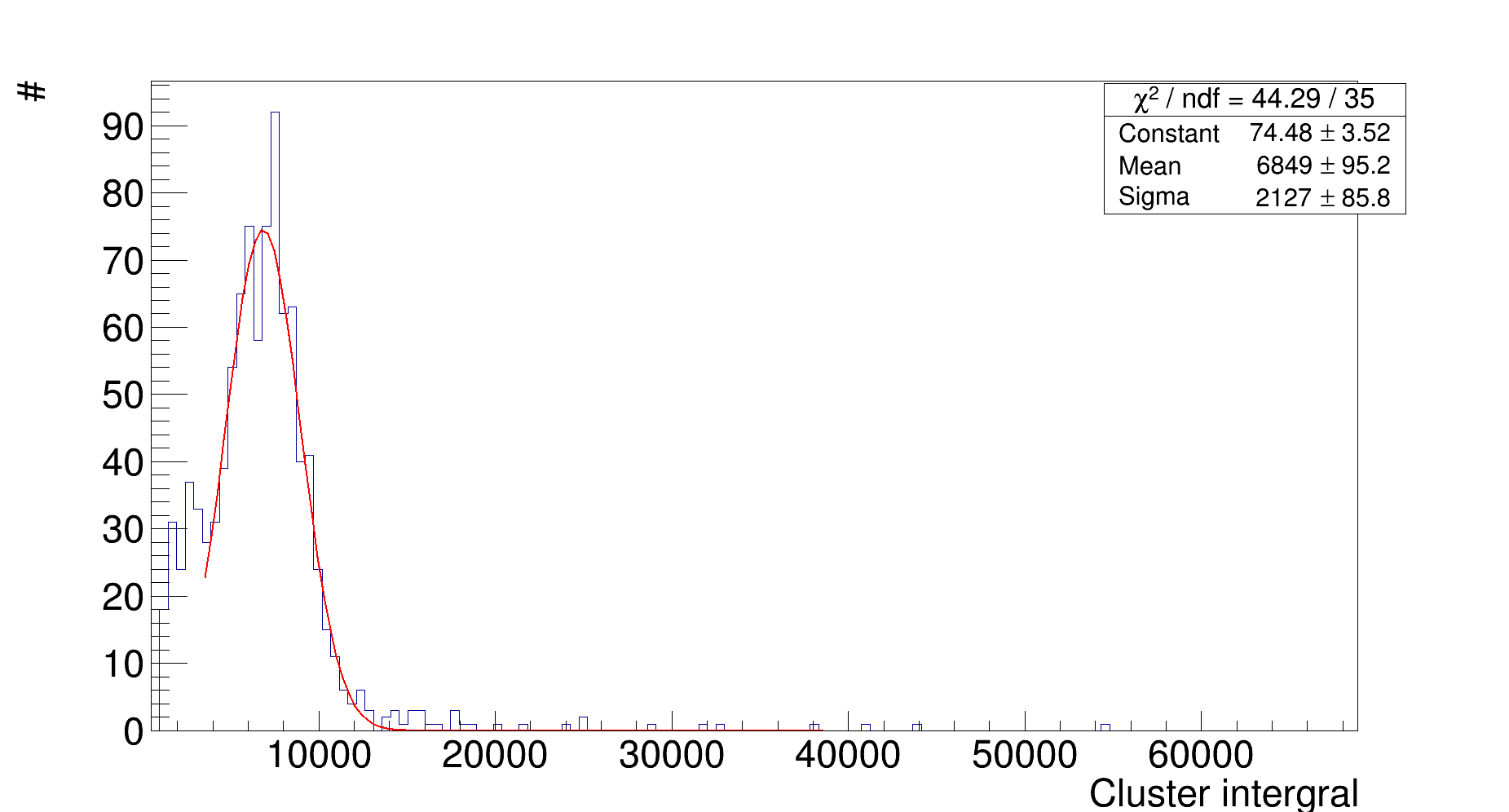}
	\caption{Example of $^{55}$Fe signals: on the left, an image acquired by the sCMOS camera in MANGO with a \emph{Tt} GEM configuration, He:CF$_4$ 60/40 gas mixture and 6 kV/cm induction field, where the $^{55}$Fe clusters are individually identified by the CYGNO reconstruction algorithm \cite{Baracchini:2020iwg,bib:cygnoIDBscan}; on the right, example of $^{55}$Fe light spectrum with superimposed the Gaussian fit from the same configuration.}
	\label{fig:waveformGEM}
\end{figure}

Figure \ref{fig:waveformGEM} shows, on the left panel, an example of 5.9 keV signals generated by the $^{55}$Fe X-ray  source  in MANGO as seen by the sCMOS camera. The images acquired are analysed with reconstruction code described and optimised in Chapter \ref{chap4}, employing the Chan-Vese algorithm, and the \fe spot dimensions are determined with the algorithm illustrated in Section \ref{sec:difffe}.

The distance travelled in the gas by the primary electrons created by the interaction with the $^{55}$Fe X-rays is small enough that the signal imaged by the sCMOS camera appears round, (see Section \ref{sec:difffe}). For this reason, a pure sample of \fe events is selected by requiring the ratio of the minor over the major axis of an ellipse containing the cluster (defined as \emph{slimness}) to be larger than 0.7, to reject long straight tracks from cosmic rays or short curly tracks from natural radioactivity. In addition, a geometrical cut is applied in order to analyse only tracks from the innermost region of the detector within a box of 900 $\times$ 900 pixels around the centre, to avoid effects due to field distortions.
For each of the found clusters satisfying the selection requirement described above, the energy deposited is calculated from the sum of the content of all the pixels belonging to the track (\textit{Integral}), after having subtracted the camera noise pixel by pixel. In addition, the dimension of the $^{55}$Fe round spot encodes the effect of the electron diffusion during the drift from its production to the detection point. Due to the very small drift gap of 0.8 cm in MANGO and the value of transverse diffusion of about 100 $\frac{\mu m}{\sqrt{cm}}$ at the drift field of operation \cite{Amaro_2022}, the spot dimension is dominated by the contribution of the amplification stage rather than diffusion during drift. Therefore, the analysis of the $^{55}$Fe spot dimension provides important information on the intrinsic diffusion due to the GEMs employed and the choice of stacking. The intrinsic diffusion of the amplification stage is estimated utilising the algorithm discussed in Section \ref{sec:difffe}.

The light output, the energy resolution and the diffusion of $^{55}$Fe induced events are studied for each of the configurations illustrated in Table \ref{tab:datataking} both as a function of the voltage applied across the GEMs, V$_{GEM}$, (i.e. the charge gain) and the intensity of the induction field after the last GEM amplification plane E$_{Mesh}$, and the results are discussed in the following Sections.
\subsection{Light yield as a function of the charge gain}
\label{subsec:gain}
\begin{figure}[!t] 
	\centering
	\includegraphics[width=1.05\linewidth]{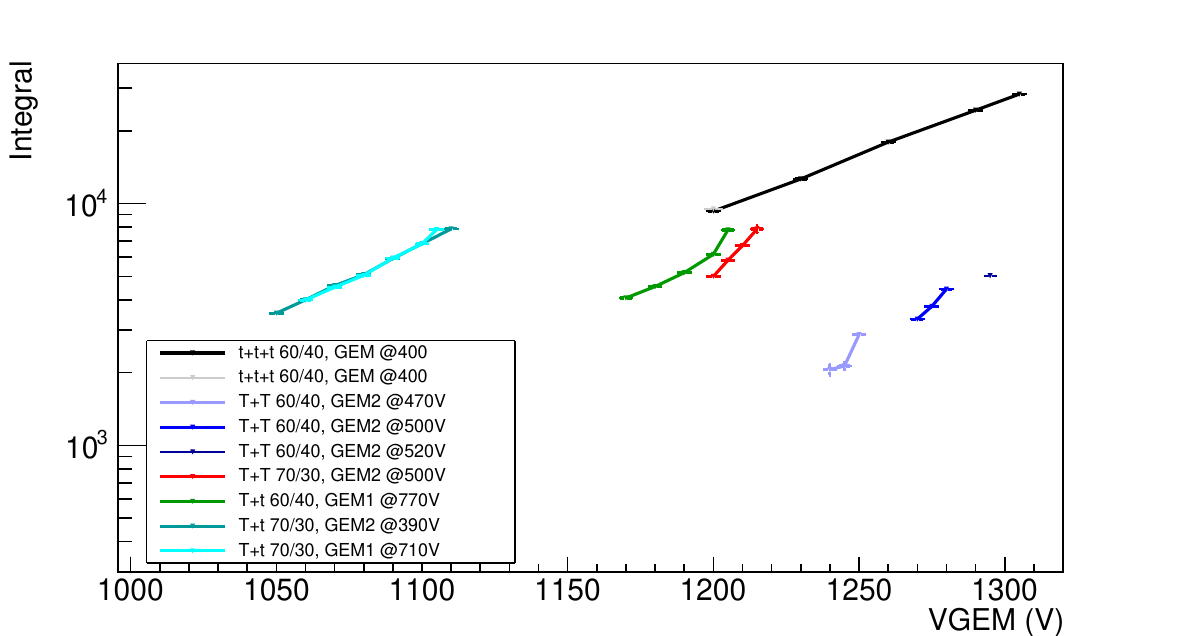}
	\caption{Gain scan summarising plot. The light integrals obtained by the $^{55}$Fe analysis are shown as a function of the total sum of the voltage applied across the GEMs. Different colours represent the various amplification and gas mixture combinations.}
	\label{fig:gain}
\end{figure}
The study of the dependence of the light yield on the voltage applied across the GEM electrodes V$_{GEM}$ (which effectively defines the charge gain of the detector) without adding any field to the induction region is presented in this Section. The values chosen for V$_{GEM}$ (and shown in Table \ref{tab:app}) depend, on the lower end, on the minimum voltage that allows the signal to be visible in the sCMOS images, and, on the larger end, on the voltages that are stable enough to keep the rate of sparks lower than 0.2 Hz. The light spectrum of the selected events is modelled with a Gaussian function and the fitted mean is taken as the light integral. An example of a fitted $^{55}$Fe light spectrum is shown in Figure \ref{fig:waveformGEM} right panel. The light integrals obtained by the $^{55}$Fe analysis for all the configuration of Table \ref{tab:datataking} are shown as a function of the total sum of the voltage applied across the GEMs in Figure \ref{fig:gain}.
\begin{table}[!t]
	\centering
	\begin{adjustbox}{max width=1.01\textwidth}
		\begin{tabular}{|c|c|c|c|c|c|c|}
			\hline
			
			\large{Ampl stage} & \large{Color on plot} & A'  &$\sigma_{A'}$ & B' [1/V] & $\sigma_{B'}$ [1/V] & Avg B' [1/V]    \\ \hline \hline
			ttt 60/40 & Black & -3.7 & 0.3 & 0.0107 & 0.0003 & 0.0107 $\pm$ 0.0003 \\ \hline
			Tt 60/40 & Green & -8 & 1 & 0.0140 & 0.0009 &  \\
			Tt 70/30 & Cyan & -7.3 & 0.7 & 0.0147 & 0.0006 & 0.0139 $\pm$  0.0004 \\ 
			Tt 70/30 & Dark Green & -5.8 & 0.6 & 0.0133 & 0.0005 &  \\  \hline
			TT 60/40  & Blue & -28& 10 &  0.029 & 0.008 & \multirow{2}{*}{0.030 $\pm$ 0.004} \\ 
			TT 70/30 & Red & -27 & 7 & 0.030 & 0.006 &  \\ \hline
		\end{tabular}
	\end{adjustbox}
	\caption{Table summarising the results of the fit with Equation \ref{eq:gainsimpl} to the data sets in Fig.\ref{fig:gain}. All data relative to the $TT$ configuration at 60/40 (i.e. all the data with blue shade colours) were fitted together.}		
	\label{tab:gain}
\end{table}

Since in this MANGO configuration the light is produced only in the electron avalanche amplification process happening within the GEMs, it is possible to interpret the results in Figure  \ref{fig:gain} in terms of detector charge gain. From the general description of the electron avalanche processes \cite{bib:tom,bib:Aoyama,bib:Williams,bib:Diethorn}, the reduced gain $\Gamma$ can be expressed as:
\begin{equation}
\label{eq:gain}
\Gamma = \frac{ln(G)}{n_g pt} = A \left(\frac{V_{GEM}}{n_g pt}\right)^m exp\left(-B\left(\frac{n_g pt}{V_{GEM}}\right)^{1-m}\right)
\end{equation}
with $G$ the gain, $n_g$ the number of GEMs used in the amplification stage, $p$ the gas pressure, $t$ thickness of the GEM, $V_{GEM}$ total voltage applied to the GEMs and $m$, $A$, $B$ free parameters. In particular, $m$ is constrained between 0 and 1 and depends on the gas \cite{bib:tom,bib:Aoyama}. For gain scans that do not span over a large range of voltages, $m$ can be approximated to 1, resulting in the more widely used expression of the gain in a gas detector:
\begin{equation}
ln(G) = A'+B'V_{GEM}
\label{eq:gainsimpl}
\end{equation}  
Equation \ref{eq:gainsimpl} can be used to fit all the data sets in Figure \ref{fig:gain}, and the fit results are listed in Table \ref{tab:gain}. 
The fit results show how each group of GEM stacking configuration (i.e. \emph{ttt}, \emph{Tt} and \emph{TT}) display the same gain slope (B' parameter) independently of the gas mixture used. For this reasons the average B' is also shown in last column of Table \ref{tab:gain}. Conversely, the He to CF$_4$ ratio influences the voltage on the GEMs needed to reach the same gain, with larger helium content requiring lower voltages. The total light output achievable is of the same order of magnitude once the same amplification structure is considered, with the only exception being the \emph{TT} at 60/40. While the increase of helium results beneficial in terms of a lower amplification voltage, it significantly increases the frequency of sparks and cascade instabilities.\\
The larger gain and light output is achieved with the three thin GEM configurations \emph{ttt}, with integral values on average $\sim$ 3 times larger than \emph{Tt} and up to $\sim$ 5 than \emph{TT}. As expected, having more planes of GEMs grants higher amplification.
It is also interesting to notice that for the \emph{Tt} set at 70/30, the scans are taken varying the voltage of the thick or the thin GEM alternatively. The light results of these sets is perfectly consistent as the points overlap each other nicely. This is consistent with the expectation of the gain dependence only on the total voltage applied across the GEMs, other than the stability of the detector during the data taking.\\
\begin{table}[!t]
	\centering
	\begin{tabular}{|c|c|c|c|c|c|}
		\hline
		Config & Colour & [0] $\frac{1}{torr\cdot cm}$ &$\sigma_{[0]}$ $\frac{1}{torr\cdot cm}$ & [1] $\frac{1}{torr\cdot cm\cdot V}$ &  $\sigma_{[1]}$ $\frac{1}{torr\cdot cm\cdot V}$     \\ \hline \hline
		ttt 60/40 & Black & -0.36 & 0.14 & 0.00106 & 0.00011 \\ \hline
		Tt 60/40 & Green & -0.7 & 0.2 & 0.0012 & 0.0004 \\
		Tt 70/30 & Cyan & -0.6 & 0.2 & 0.0012 &  0.0003 \\ 
		Tt 70/30 & Dark Green & -0.49 & 0.19 & 0.0011 & 0.0002  \\ \hline
		TT 60/40  & Blue & -1.6 & 0.9 &  0.0017 & 0.0007  \\ 
		TT 70/30 & Red & -1.6 & 1.0 & 0.0018 & 0.0006 \\ \hline
	\end{tabular}
	\caption{Table summarising the results of the linear fit of the reduced light gain of all the configurations as a function of the V$_{GEM}$, following Equation \ref{eq:gamm_vsvoltsigma}.}
	\label{tab:fit_gamma_vgem}
\end{table}

An interesting way to compare the different GEM amplification stacking options is by analysing their reduced gain $\Gamma$ as a function of reduced field $\Sigma$ generated inside the GEMs holes, which is the parameter effectively characterising the development of the electron avalanche. $\Sigma$ can be defined as:
\begin{equation}
\label{eq:sigma}
\Sigma = \frac{V_{GEM}}{n_g p t}
\end{equation}
Since, as discussed above, in parameter space of gains tested $m$ can be approximated to 1, the reduced gain can be written as:
\begin{equation}
\label{eq:gamm_vsvoltsigma}
\Gamma = A_0+ B_0\Sigma= A_0+ \frac{B_0}{pn_gt}V_{GEM}.
\end{equation}
This is a simple mathematical recombination of the terms in play in order to highlight the dependence on the number of GEMs and their applied voltages. Indeed,  when comparing Equation \ref{eq:gainsimpl} with Equation \ref{eq:gamm_vsvoltsigma}, it is valid that $A'n_gpt=A_0$ and $B_0=B'$. The results of the fits of the data presented in Figure \ref{fig:gain} with a function $$\Gamma = [0]+ [1]V_{GEM},$$ are summarised in Table \ref{tab:fit_gamma_vgem}.
Once the terms proportional to $V_{GEM}$ are expressed in terms of the number and thicknesses of the GEMs, their fitted values result highly consistent among all the configurations and gas mixture employed.\\

A more global analysis of the 60/40 mixture data aimed at characterising Equation \ref{eq:gain} for a wide spread of reduced fields, $\Sigma$, is described in Appendix \ref{appD}.
\subsection{Enhanced light yield with accelerated electrons}
\label{subsec:el_light}
Given the importance of enhancing the light yield for optically readout TPCs as discussed in Section \ref{subsec:dirdet}, the effect of the introduction of a strong electric field in the induction region is studied in details this Section, expanding on the results presented in \cite{bib:EL_cygno} and in Section \ref{sec:lemon}. As argued in Section \ref{subsec:cyg_gas}, it is theoretically possible to induce the emission of visible photon by means of CF$_4$ fragmentation intro neutral CF$_3^*$ without the generation of additional electrons. In order to investigate this possibility, an  induction field E$_{Mesh}$ is applied in the induction region below the last GEM electrode and this effect is studied for all the GEM stacking configurations and gas mixtures illustrated in Table \ref{tab:datataking}.
\begin{figure}[t] 
	\centering
	\includegraphics[width=1\linewidth]{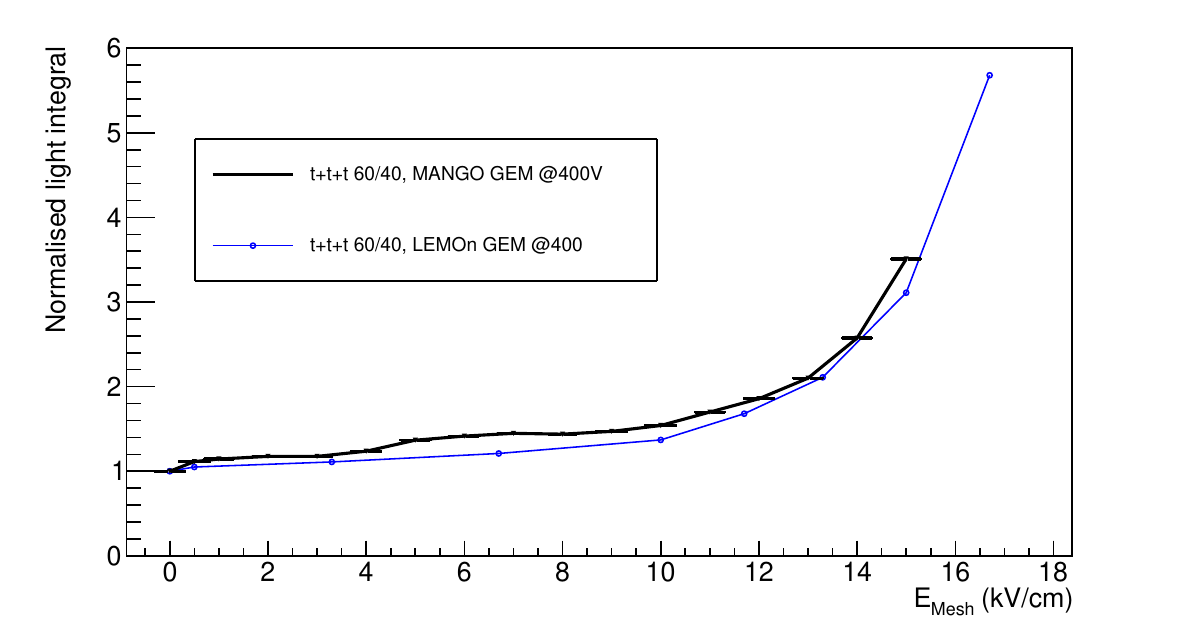}
	\caption{Relative increase of light integral for the \emph{ttt} configuration in MANGO and LEMOn.The two data sets are manifestly highly consistent with each other and with the measurements presented in \cite{bib:EL_cygno}, robustly confirming the results presented in Section \ref{subsec:lemon_light}. }
	\label{fig:ellightttt}
\end{figure}
In Figure \ref{fig:ellightttt} the relative increase in light yield with respect to no induction field applied is shown in black for a \emph{ttt} configuration with He:CF$_4$ 60/40 and 400 V applied on each GEM as a function of the induction field E$_{Mesh}$. The relative light increase measured with the LEMOn detector and discussed in Section \ref{sec:lemon} is superimposed in blue. The two trends are strongly consistent with each other and show the same features, clearly demonstrating that the light yield enhancement does not dependent on a single detector characteristics, but it is actually related to a physical phenomenon happening within the setup structure. It is important to stress that the two detectors employ identical voltages applied to the GEMs, but with different absolute gain, since LEMOn is located at LNF at about 150 m a.s.l., while MANGO at the LNGS at 1000 m a.s.l., highlighting the independence of the relative light output growth from the absolute gain of the detector. The two detectors moreover employ different structures to generate the induction field E$_{Mesh}$ (a metallic mesh in MANGO and a ITO glass in LEMOn) demonstrating that, once the transparency of these structures is properly taken into account, the light yield amplification results independent from this feature. 

The dependence of the light gain on the induction field is common to all the stacking configuration and gas mixtures studied, and can be split into three regions. Initially, as soon as the field is turned on, there is a boost in the light output of about 10\%. Afterwards, from 0.5 kV/cm  up to breaking point $E_b$ between 7 kV/cm and 10 kV/cm, the light grows linearly with E$_{Mesh}$, and beyond it, the light yield increase becomes exponential. This breaking point, where the increase changes from linear to exponential, is observed to depend on the gas mixture, being about  $E_b$ = 10 kV/cm for 60/40, and $E_b$ = 8 kV/cm for 70/30. Despite the difference in the type of analysis of the two data sets (see Section \ref{sec:lemon}), this behaviour is consistently shared also between the MANGO and the LEMOn data. The features of each of these regions is discussed in details in the following.\\

\paragraph{Region between 0 and 0.5 kV/cm} The enhancement generated by a field between 0 kV/cm and 0.5 kV/cm can be explained by the fact that typical MANGO (and LEMOn and all CYGNO prototypes) operation foresees the bottom of the last GEM electrode to be put to ground, to minimise the overall HV needed to be applied since no charge needs to be collected with an optical readout, as discussed in Section \ref{sec:lemon}. This implies that the field lines, typically closing disorderly on the lower electrode of the GEM, get straightened and become more stable within this additional small electric field, resulting in a slight light yield increase. This hypothesis is confirmed by simulation discussed in Section \ref{sec:maxwell}.

\paragraph{Linear region between 0.5 kV/cm and E$_b$} In the region between 0.5 kV/cm and a breaking point $E_b$, the light yield increase appears linearly proportional to the raise in induction field E$_{Mesh}$. Starting from the arguments presented in Section \ref{subsec:gain}, it is speculated that the induction field effectively linearly increases the reduced field $\Sigma$ inside the GEMs holes. To include a contribution from the $E_{Mesh}$, an additional term to Equation \ref{eq:gamm_vsvoltsigma} can be added as:
\begin{equation}
\label{eq:sigma_ext}
\Sigma = \frac{1}{p}\left(\frac{V_{g}}{n_gt}+\alpha E_{Mesh}\right) ,
\end{equation}
\begin{figure}[t] 
	\centering
	\includegraphics[width=0.49\linewidth]{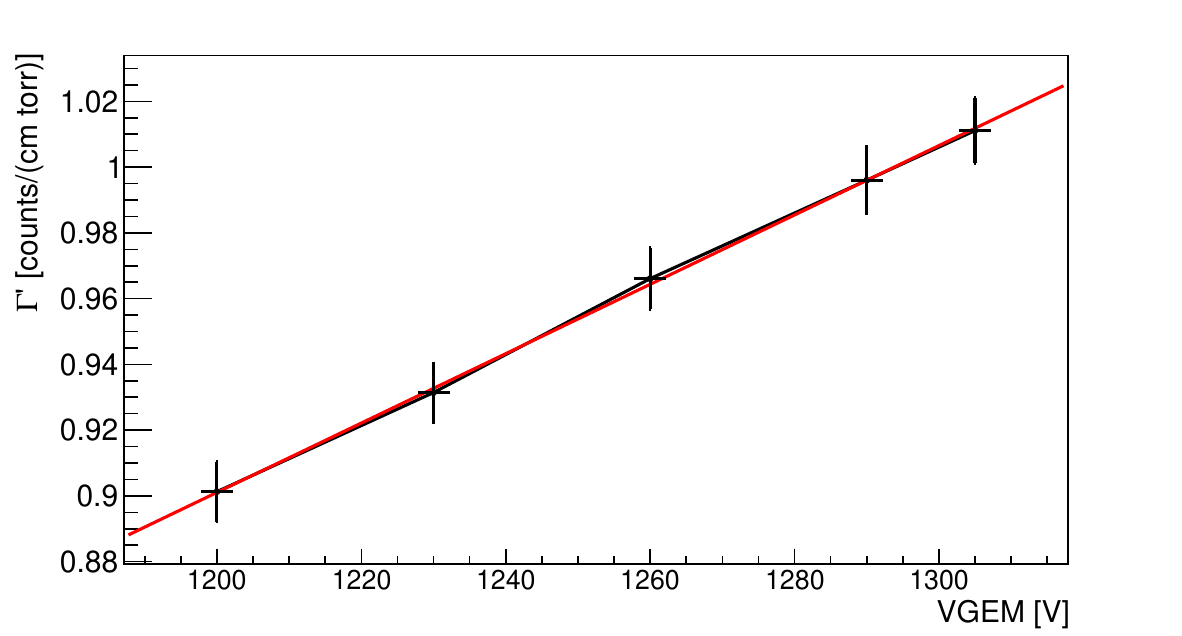}
	\includegraphics[width=0.49\linewidth]{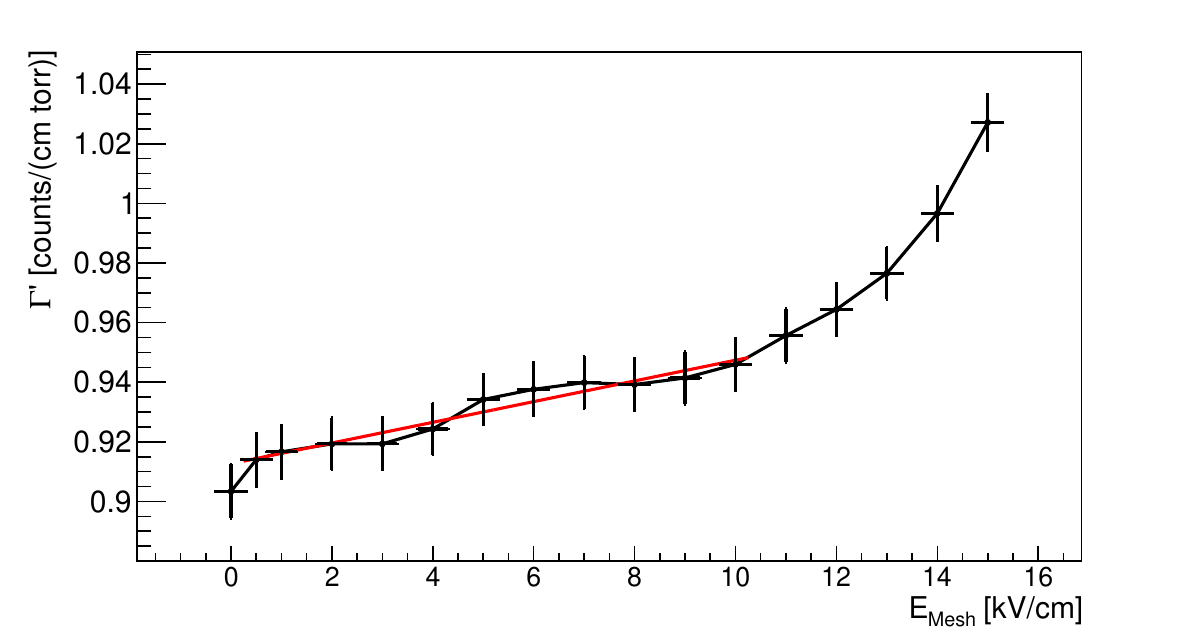}
	\caption{On the left, the reduced light gain as a function of V$_{GEM}$ with a linear fit superimposed. On the right, the reduced light gain is expressed as a function of $E_{Mesh}$ with a linear fit superimposed in the region below 10 kV/cm.}
	\label{fig:fitEh_gain}
\end{figure}
\begin{table}[!t]
	\centering
	\begin{adjustbox}{max width=1.01\textwidth}
		\begin{tabular}{|c|c|c|c|c|c|c|c|}
			\hline
			
			Conf  & [0] $\frac{1}{torr\cdot cm}$ &$\sigma_{[0]}$ $\frac{1}{torr\cdot cm}$ & [1] $\frac{1}{torr\cdot kV}$ &  $\sigma_{[1]}$ $\frac{1}{torr\cdot kV}$ & Cond. C & $\alpha$  & $\sigma_{\alpha}$  \\ \hline \hline
			ttt 60/40  & 0.912 & 0.005 & 0.0036 & 0.0010 & $\checkmark$ & 0.23 & 0.06\\ \hline
			Tt 60/40  & 0.747 & 0.009 & 0.0010 & 0.0009 & $\checkmark$ & 0.05 & 0.04 \\
			Tt 70/30  & 0.704 & 0.007 & 0.0029 & 0.0015  & $\checkmark$ & 0.14 & 0.07 \\ \hline
			TT 60/40   & 0.48 & 0.01 &  0.0025 & 0.0013 & $\checkmark$ & 0.06 & 0.03\\ 
			TT 70/30  & 0.505 & 0.004 & 0.0016 & 0.0010 & $\checkmark$ & 0.04 & 0.03\\ \hline
		\end{tabular}
	\end{adjustbox}
	\caption{Table summarising the results of the linear fit of the reduced light gain as a function of the $E_{Mesh}$, when V$_{GEM}$ is fixed. Cond. C is $\checkmark$ if that condition is fulfilled.}
	\label{tab:fit_gamma_mesh}
\end{table}
with $\alpha$ the coefficient of proportionality of $E_{Mesh}$. Therefore, the dependence of the reduced gain $\Gamma$ on the V$_{GEM}$ and $E_{Mesh}$ when m=1 can be expressed as:
\begin{equation}
\label{eq:gamm_vsmesh}
\Gamma = A_0+ \frac{B_0}{pn_gt}V_{GEM}+\frac{B_0\alpha }{p}E_{Mesh}
\end{equation}
When $E_{Mesh}$ is zero, Equation \ref{eq:gamm_vsvoltsigma} is recovered. On the contrary, if V$_{GEM}$ is fixed and the $E_{Mesh}$ is increased, a linear increase of $\Gamma$ is expected. Figure \ref{fig:fitEh_gain} shows the reduced gain $\Gamma$ for \emph{ttt} configuration with He:CF$_4$ 60/40 on the left as a function of the $V_{GEM}$, with $E_{Mesh}=0$, and on the right as a function of $E_{Mesh}$ with $V_{GEM} = 400$, displaying the linear dependence of the light yield on both quantities. Hence, a linear fit of the reduced gain $\Gamma$ as a function of $E_{Mesh}$ can be performed between 0.5 kV/cm and $E_b$ for each configuration under study with the function
\begin{equation}\label{eq:gamma_mesh}
\Gamma = [0] + [1] E_{Mesh},
\end{equation}
to measure the dependence on $E_{Mesh}$ of the light yield enhancement.
If the assumption that $E_{Mesh}$ contributes linearly to increase the effective field inside the GEMs holes is correct, the fitted $[0]$ term has to be equal to $A+\frac{B}{pn_gt}V_{GEM,0}$, where V$_{GEM,0}$ is the sum of the voltages applied to the GEMs. This is defined as \emph{Condition C}. The first order term $[1]$ of the linear fit allows to estimate the parameter $\alpha$, which defines the proportionality of the light increase to $E_{Mesh}$. Table \ref{tab:fit_gamma_mesh} shows the result of the fit with Equation \ref{eq:gamma_mesh} to all the configuration considered in this study, together with \emph{Condition C} and the $\alpha$ proportionality parameter. \emph{Condition C} is always verified, demonstrating a clear comprehension of the contribution of $E_{Mesh}$ to the linear part of light yield increase.\\

\begin{figure}[!t] 
	\centering
	\includegraphics[width=1\linewidth]{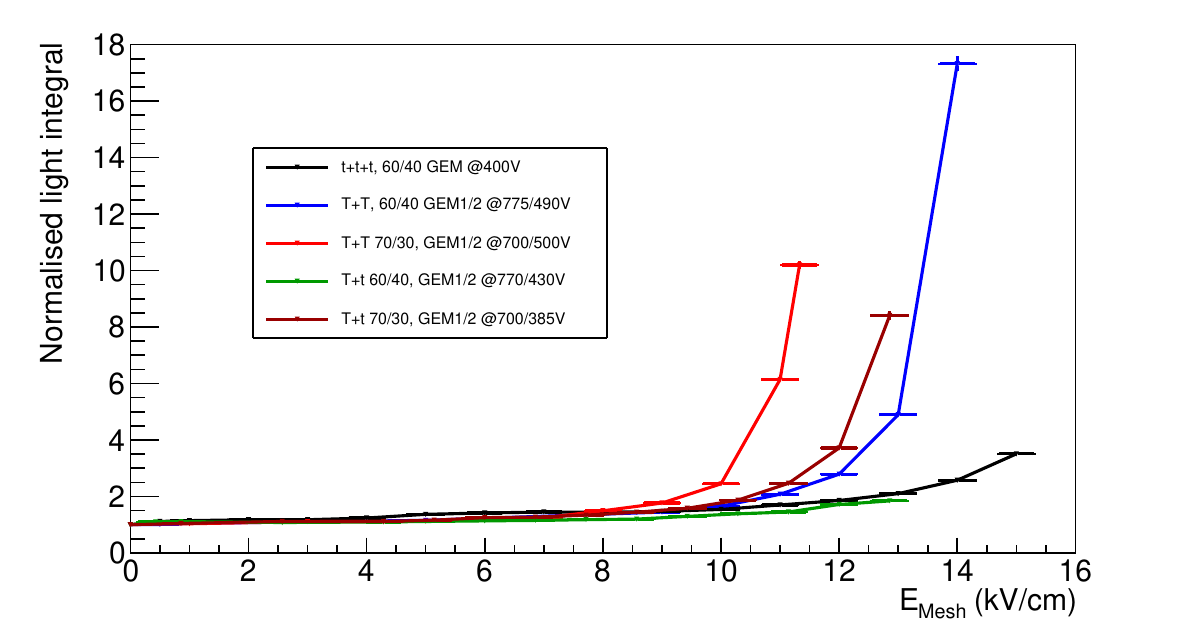}
	\caption{Relative increase of light output as a function of the induction field for all the GEMs stacking configurations studied with MANGO.}
	\label{fig:elall_light}
\end{figure}

\paragraph{Exponential region above E$_b$} The light yield enhancement beyond $E_b$ clearly stops being consistent with a linear increase, allowing to conclude that above this value a new phenomenon comes into play. In order to properly study this feature, the fitted function from Equation \ref{eq:gamma_mesh} is subtracted from the data in the entire range for each configuration for the following discussion. The relative light yield increase resulting after this subtraction is shown in Figure \ref{fig:elall_light}.
All the curves display very similar behaviours, including the three different light enhancement regions discussed above, where the differences between gas mixtures and amplification structures is highlighted by the different breaking points $E_b$ and different exponential rises.
The subtraction of the linear $E_{Mesh}$ enhancement allows to directly compare the different configurations in the entire range of the data by employing a modified expression to describe the increase of the light yield as:
\begin{equation}
\label{eq:fitel_light}
a+b \cdot e^{cE_{Mesh}-d}
\end{equation}
where $a$ represents the normalisation with respect to operation with null induction field E$_{Mesh}$, $b$ is the intensity of the exponential component, $c$ is the slope of the exponential, and $d$ is proportional to a shift in E$_{Mesh}$ field. The ratio between $d$ and $c$ returns the field value $E_{b}$ where the exponential growth starts, highly consistent with assumption used to evaluate the linear growth in the above discussion. 
Table \ref{tab:elbreak} summarises the parameters obtained by fitting the different data sets with Equation \ref{eq:fitel_light}. 
\begin{table}[!t]
	\centering
	\begin{adjustbox}{max width=1.01\textwidth}
		\begin{tabular}{|c|c|c|c|c|c|c|c|c|c|c|}
			\hline
			
			\large{Config} & $a$ & $\sigma_a$ & $b$ & $\sigma_b$ & $c$ [cm/kV] & $\sigma_c$ [cm/kV] & $d$ & $\sigma_d$ & \small{E$_{b}$ [kV/cm]} & \small{$\sigma_{E_b}$ [kV/cm]} \\ \hline  \hline
			ttt 60/40 & 0.99 & 0.02 & 0.04 & 0.02 & 0.8 & 0.1 & 8.2 & 0.6 & $9.8$  &  $1.5$ \\ 
			TT 60/40 & 0.99 & 0.03 & 0.09 & 0.03 & 1.1 & 0.1 & 10 & 1 &$9.5$  &  $1.2$ \\ 
			Tt 60/40 & 1.00 & 0.02 & 0.08 & 0.03 & 0.8 & 0.2 & 8 & 1 & $10$  &  $2$ \\ 
			Avg. 60/40 & & & & & & & & & $9.7$ & $0.8$ \\ \hline 
			TT 70/30 & 1.00 & 0.02 & 0.3 & 0.2 & 1.2 & 0.1 & 10 & 2 & $8.5$  &  $1.3$ \\ 
			Tt 70/30 & 0.99 & 0.02 & 0.17 & 0.08 & 0.86 & 0.03 & 7.6 & 0.9 & $8.8$  &  $0.9$ \\ 
			Avg. 70/30 & & & & & & & & & $8.7$ & $0.7$ \\ \hline  
		\end{tabular}
	\end{adjustbox}
	\caption{Table summarising the result of the fit with Equation \ref{eq:fitel_light} to the data of Figure \ref{fig:elall_light}, where the field $E_{b}$ represents the value at which the exponential light increase growth starts.}
	\label{tab:elbreak}
\end{table}
The results obtained show how the larger the helium concentration in the mixture, the smaller the E$_{Mesh}$ value needed to start the process responsible of the exponential growth of the light increase. This is consistent with what observed in Section \ref{subsec:gain} in terms of lower voltage across the GEM needed to start the electron avalanche responsible for the charge gain for higher helium fractions. For what concerns the exponential slope $c$, the \emph{TT} configurations appear to display a stronger boost of the light yield with respect to the other configurations, once the gas mixture is chosen. This feature will be discussed in more details in Section \ref{sec:disc} also in view of the simulations that will be presented in Section \ref{sec:maxwell}.
\subsection{Energy resolution}
\label{subsec:res}
\begin{figure}[!t] 
	\centering
	\includegraphics[width=1\linewidth]{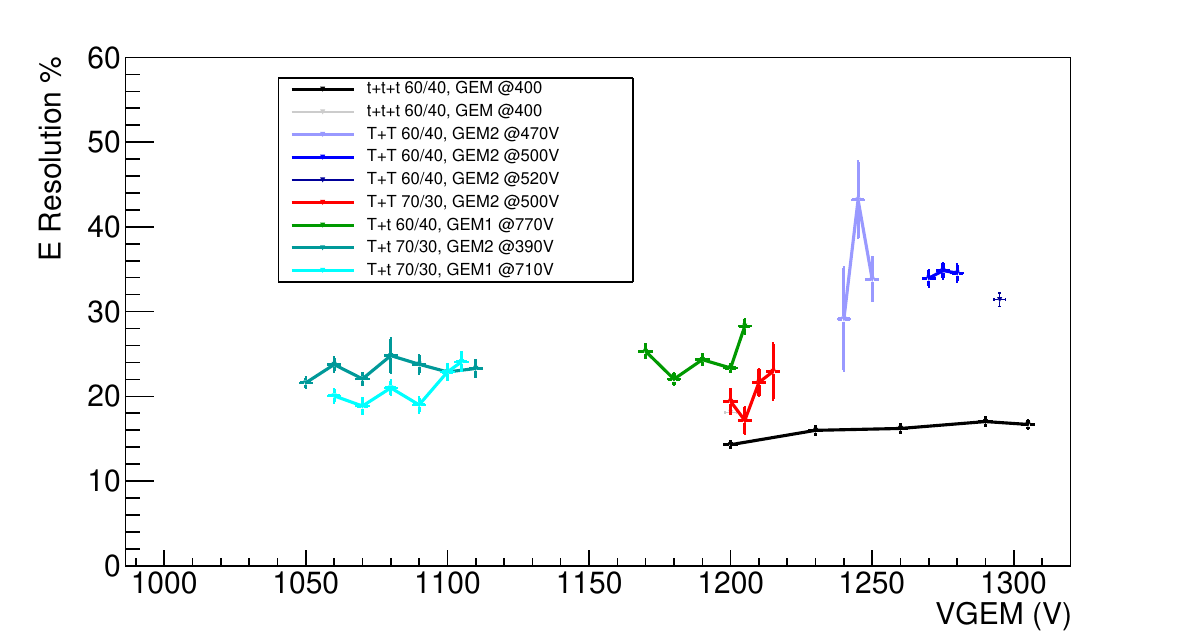}
	\caption{Energy resolution for the different amplification stages as a function of the sum of the  voltages applied to the GEMs with a null induction field. Different colours represent the various amplification and gas mixture combinations.}
	\label{fig:gaineres}
\end{figure}
The energy resolution at 5.9 keV is evaluated as the ratio of the sigma and the mean value of the Gaussian fit of the $^{55}$Fe spectrum. The results as a function of the different GEM configurations and voltages with a null induction field applied are shown in Figure \ref{fig:gaineres}. The energy resolution appears to strongly depend on the GEM configuration used, spanning from 15\% up to 35\%, but seems unaffected by the gas mixture used.\\

In general, for gains larger than 10$^3$ and if the avalanche in each GEM hole can be considered independent from one another \cite{bib:tom,bib:Alkhazov,Knoll}, the energy resolution $\sigma_E$ depends on the energy released in the event and on the gain according to:
\begin{equation}
\label{eq:resdet}
\sigma^2_E= \left(\frac{\sigma_{n_0}}{n_0}\right)^2 +\frac{1}{n_0} \left(\frac{\sigma_A}{\bar{A}}\right)^2 = \left(\frac{F}{n_0}\right) +\frac{1}{n_0} \left(\frac{\sigma_A}{\bar{A}}\right)^2
\end{equation}
with $n_0$ the number of primary electrons freed in the ionisation process, $\bar{A}$ the average avalanche gain, $\sigma_{n_0/A}$ the relative fluctuations, and $F$ the Fano factor \cite{Knoll}. Given the typical operating gains of the CYGNO prototypes \cite{bib:roby} of the order 10$^4$-10$^5$, the assumption can be considered valid. The first term of the right-hand side of Equation \ref{eq:resdet} is proportional to the Fano term which depends on the Fano factor of the gas (typically smaller than 1) and on the number of primary electrons generated. Since the same \fe source is employed for all the data taking, the Fano factor is foreseen to be constant. The different helium content tested modifies the number of primary electrons, but a variation below 3\% is expected between 60/40 and 70/30 mixture, making the effect on the Fano term negligible with respect to the uncertainties of the measurement.\\
Conversely, the second term of Equation \ref{eq:resdet} contains the fluctuation of the avalanche gain. While $\sigma_A$ is hard to parametrise, it is possible to show that this is inversely proportional to $\chi$, defined as \cite{bib:tom,bib:Alkhazov,bib:Schlumbohm}:
\begin{equation}
\label{eq:chilength}
\chi \propto \alpha \left(\frac{n_g p t}{V_{GEM}}\right) = A \Sigma^{m-1} e^{-B\Sigma^{m-1}}
\end{equation}
\begin{figure}[!t] 
	\centering
	\includegraphics[width=0.6\linewidth]{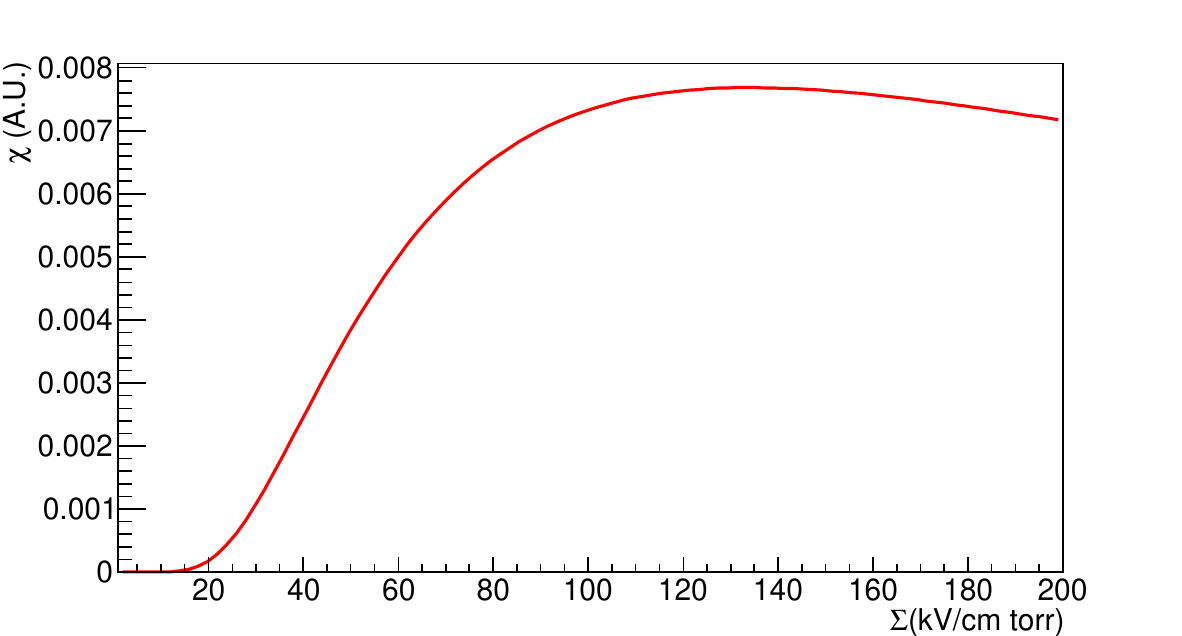}
	\caption{$\chi$ function using the parameters found in the fit of the gain from Table \ref{tab:tomplot}.}
	\label{fig:chi}
\end{figure}
being $\alpha$ the Townsend coefficient, thus related to the gain, and $\Sigma$, $m$, $A$, $B$, $p$, $n_g$, and $t$ the same as for Equation \ref{eq:gain}. This $\chi$ is proportional to the minimum distance an electron would need to travel to initiate an avalanche, and a higher value is related to a lower fluctuation in the gain \cite{bib:Schlumbohm}. If $A$, $B$ and $m$ values are taken from the global gain analysis fit discussed in Appendix \ref{appD}  and displayed in the first row of Table \ref{tab:tomplot}, namely 2.8, 134, 0 respectively, the $\chi$ results in the function plotted in Figure \ref{fig:chi}.
As a consequence, a larger reduced field and gain can induce smaller fluctuations in the avalanches. This is consistent with the results of Figure \ref{fig:gaineres}, where the configuration with larger reduced field (\textit{ttt}) has the better energy resolution. 
As for what concerns the resolution as a function of the increasing voltage across the GEMs, both the gain and the reduced fields are increasing. However, with the very high gains employed, the effect on the avalanches due to the reduction of $\bar{A}$ are expected to be negligible. Akin, as discussed in Section \ref{subsec:gain}, the range of reduced field scanned while varying the voltage across the GEMs is quite small, therefore having a negligible effect on the energy resolution.\\
It should be noted that the possible uncertainties and fluctuations given by the reconstruction code are also not included, but under study by the collaboration. Nevertheless, they are common to all the data taking and are not expected to modify the conclusion of this analysis.\\
\begin{figure}[!t] 
	\centering
	\includegraphics[width=1\linewidth]{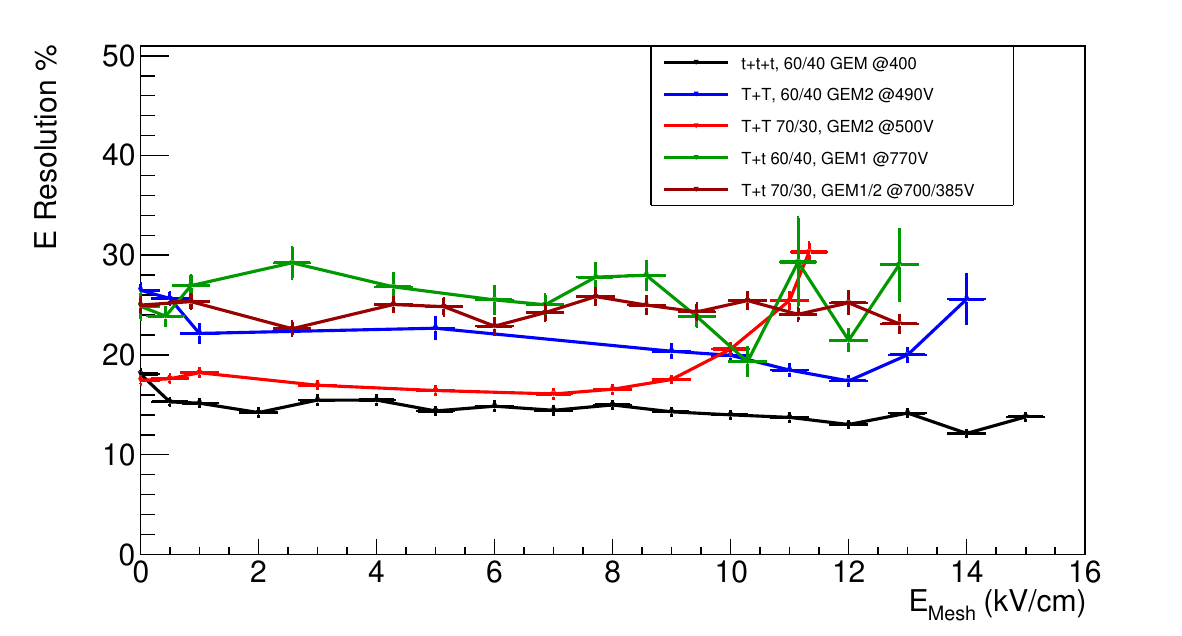}
	\caption{Energy resolution for the data sets with applied induction fields E$_{Mesh}$ as a function of E$_{Mesh}$.}
	\label{fig:el_res}
\end{figure}

The energy resolution as a function of the induction fields E$_{Mesh}$ is shown in the Figure \ref{fig:el_res}.
In this case, the energy resolution appears to remain constant independently from the light yield increase induced by the E$_{Mesh}$ field. This is in line with the hypothesis underlying this entire study, that is possible to amplify the light output of a gas detector without relevant additional charge gain. In fact, if the light yield enhancement were due to a pure additional electron avalanches generated in the induction gap within reduced electric field much smaller than those present in the GEM holes, the gain fluctuations would increase resulting in a worsening of the energy resolution. The exceptions to what just described are the \emph{TT} configurations at high fields. In this case the energy resolution is noticeably worsening with strong induction field, above the $E_b$, following again an exponential growth.
\subsection{Diffusion within the amplification stage}
\label{subsec:spot}
\begin{figure}[!t] 
	\centering
	\includegraphics[width=1\linewidth]{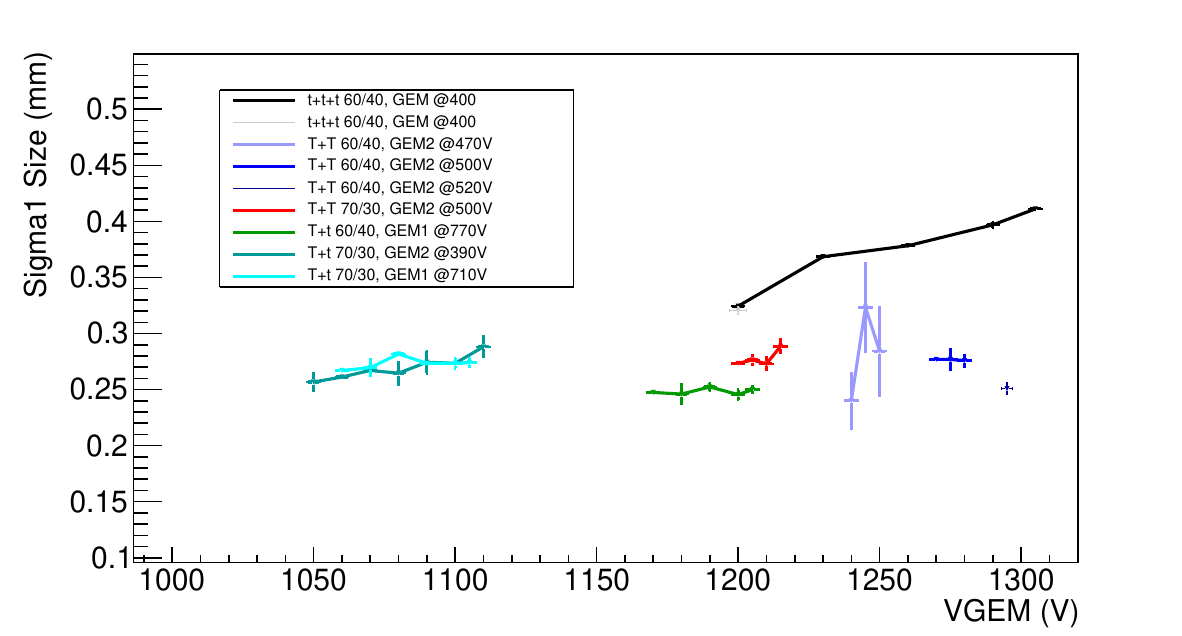}
	\caption{Primary sigma (averaged from the x and y projections) which represents the diffusion of the amplification structure as a function of the sum of the voltages across the GEMs. Different colours represent the various amplification and gas mixture combinations.}
	\label{fig:diffsig1}
\end{figure}
As illustrated in Section \ref{sec:difffe} and Section \ref{sec:mango_light}, a proper analysis of the $^{55}$Fe spot size allows to measure the contribution of the amplification stage to the total track diffusion. Figure \ref{fig:diffsig1} shows the primary sigma (averaged from the x and y projections) as a function of the voltage applied to the GEMs for the different setups of amplification. Most of the two GEMs structures are performing better than the three one, supporting the intuitive assumption that each stage of amplification contributes with an independent term to the overall diffusion. Indeed, the double GEM configurations diffusion is roughly 2/3 of the triple one, further backing up this argument.  
The \emph{Tt} stacking configurations perform better than the \emph{TT} with the same gas mixture, in line with the expectation that the granularity of the GEM closer to the sCMOS sensor sets the maximum achievable space resolution if larger than the camera pixels. Since a \emph{t} GEM has a pitch of 140 $\mu$m with respect to the 350 $\mu$m of the \emph{T}, the effective pixel size of the MANGO setup of 49 $\times$ 49 $\mu$m$^2$ results in fact sensitive to this feature. \\
Another interesting detail is that, while the triple GEM configuration diffusion linearly depends on the voltage applied to the GEM, the double ones display a much less significant (in some cases non-existent) increase. More recent data acquired with MANGO (but not discussed in this work) seems to suggest that the very large charge gain achieved with \emph{ttt}, coupled with the very small holes dimensions of these GEMs, are generating space charge effects that could be further worsening the diffusion in the amplification stage. This hypothesis is under study by the CYGNO collaboration.
\begin{figure}[!t] 
	\centering
	\includegraphics[width=1\linewidth]{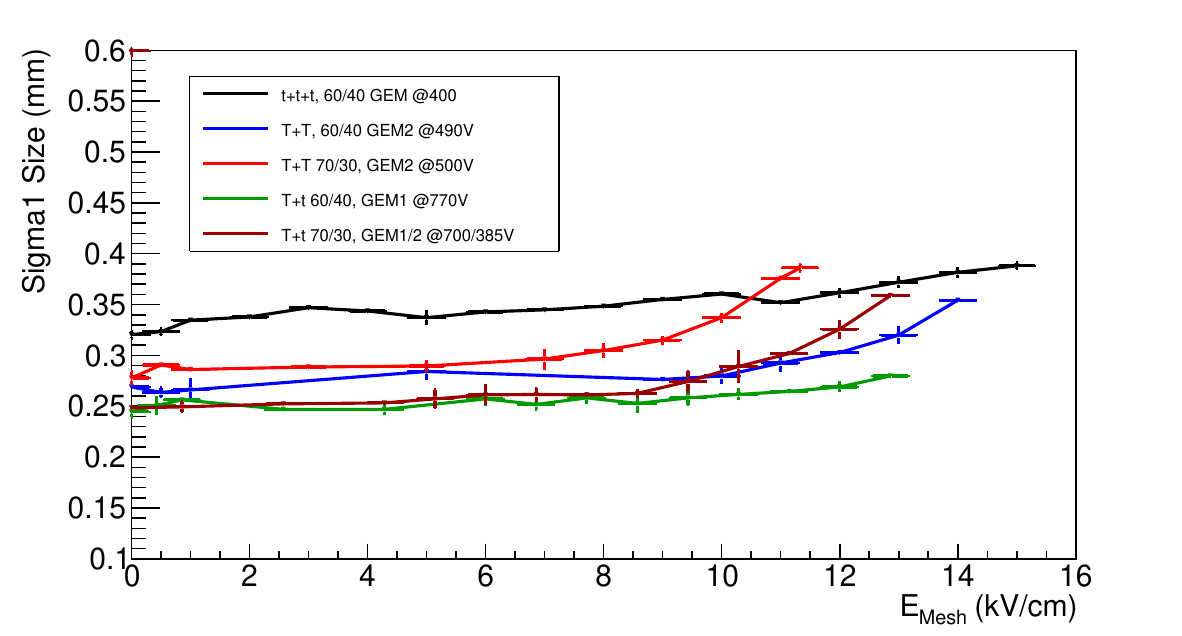}
	\caption{Amplification stage diffusion as a function of the $E_{Mesh}$  induction field.}
	\label{fig:el_size}
\end{figure}
The smaller diffusion measured with the \emph{Tt} configuration with respect to the \emph{TT} further demonstrates that the method developed to evaluate the diffusion within the GEMs and illustrated in Section \ref{sec:difffe} is independent from the light yield, having the first configuration a larger light output than the second.\\
\begin{figure}[!t] 
	\centering
	\includegraphics[height=4.1 cm]{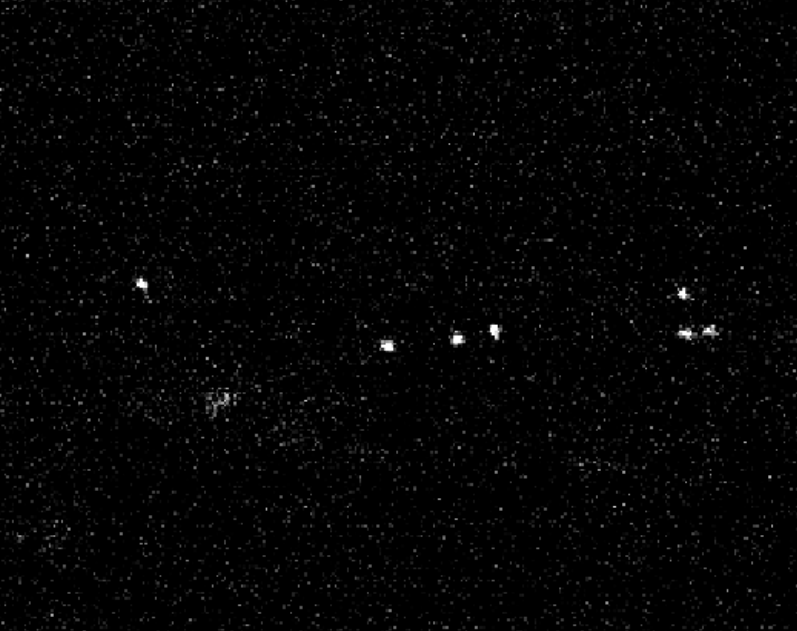}
	\includegraphics[height=4.1 cm]{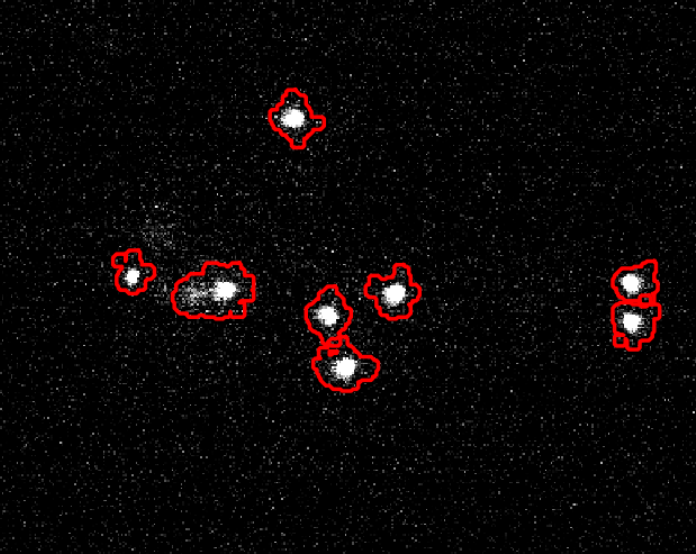}
	\caption{Raw images of the $^{55}$Fe data taking with the \emph{TT} amplification structure and 60/40 of He:CF$_4$ gas mixture. On the left a picture with the E$_{Mesh}= 0$ kV/cm, while on the right the field is 11 kV/cm.}
	\label{fig:TTlargespots}
\end{figure}

The diffusion at the amplification stage (which now includes the induction gap) is further investigated as a function of the $E_{Mesh}$ field and shown in Figure \ref{fig:el_size}. These results further confirm the assumption that no large amount of charge is generated by the electrons travelling in the induction gap, which would otherwise result also in a larger spread of the additional light generated. Since the sCMOS camera is focused on the last GEM electrode (and could not be focused on a volume, but only on a plane), the overall final effect results in a modest blur only slightly affecting the $^{55}$Fe spot size. Indeed, the \emph{ttt} diffusion at the maximum applied voltage on the GEM (light increase of a factor 3.0 with respect to 1200 V) is about 6\% larger than the diffusion at the maximum induction field E$_{Mesh}$ (light increase of a factor 3.5 with respect to null induction field at 1200 V).
The data show the same type of trend as the energy resolution and light yield, with a general change in behaviour after the $E_b$ fields in Table \ref{tab:elbreak} are applied. It can be noted that the \emph{TT} configurations are the ones experiencing again the worse increase in the diffusion among the tested amplification structures.  The increase in dimension is visible directly on the raw images as the example of the \emph{TT} 60/40 in Figure \ref{fig:TTlargespots}.
The \emph{ttt} and \emph{Tt} spot size dimensions are generally less affected by the increment in induction field, with a growth of less than 20\% at the largest E$_{Mesh}$ values tested, suggesting that this phenomenon is affecting the two kinds of GEM differently.
\section{Maxwell simulation}
\label{sec:maxwell}
\begin{figure}[!t] 
	\centering
	\includegraphics[width=0.47\linewidth]{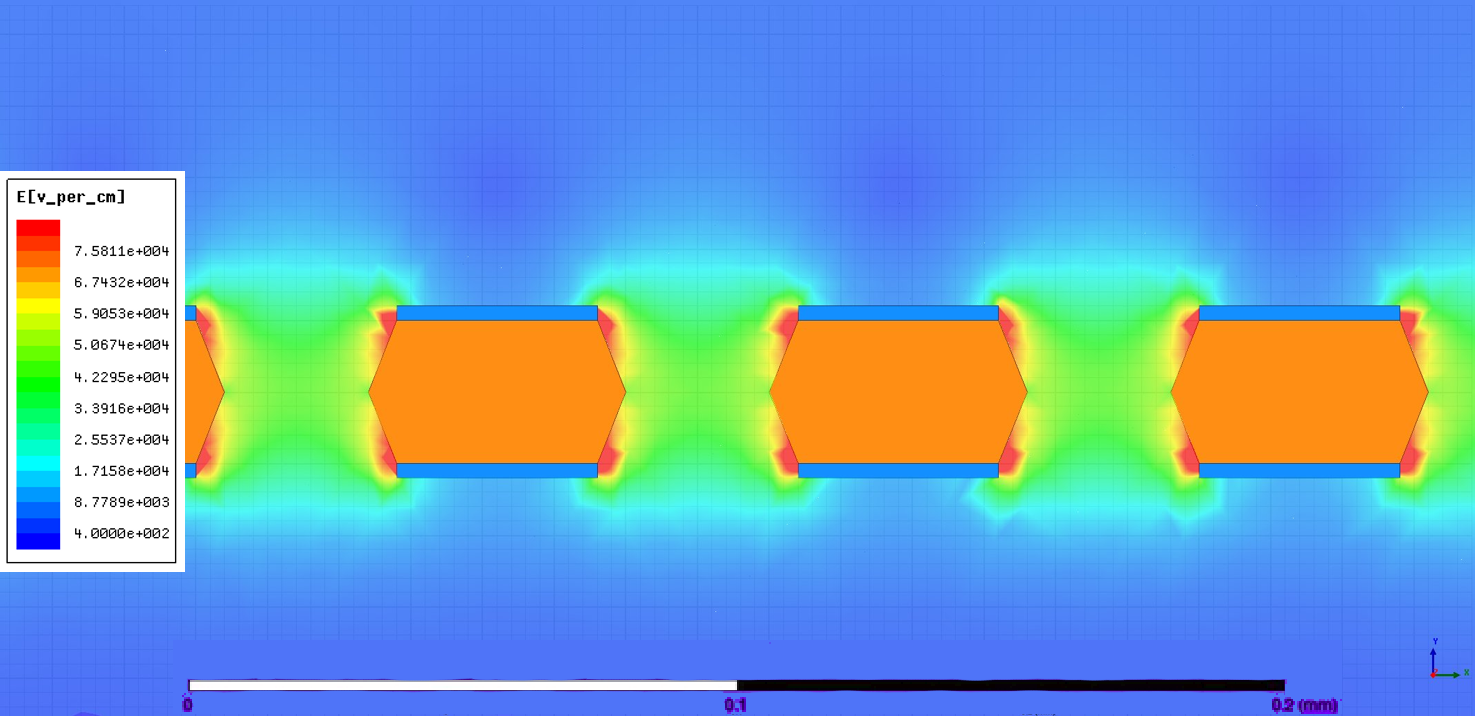}
	\includegraphics[width=0.43\linewidth]{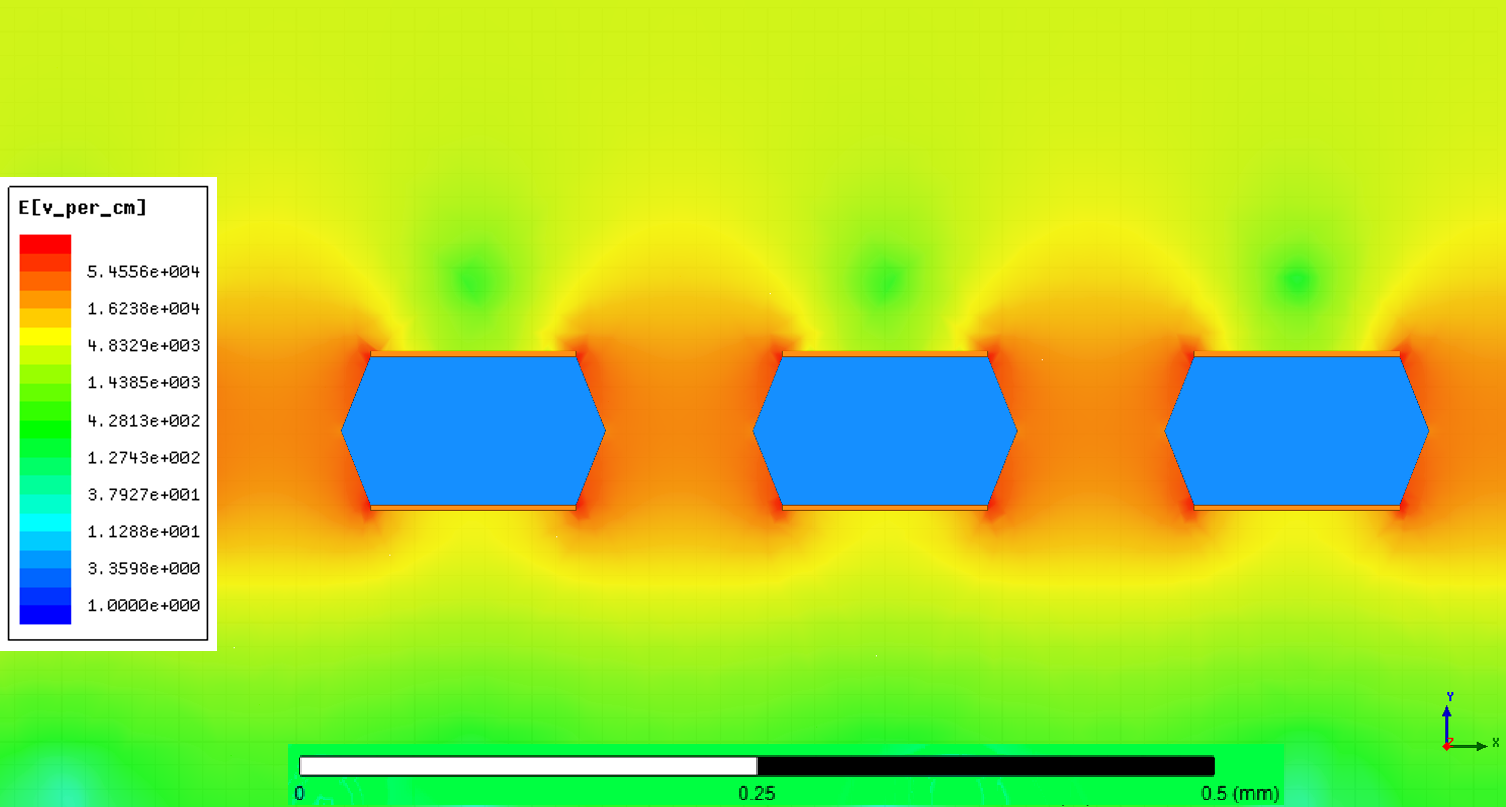}
	\caption{Examples of the 2D electric field maps generated by the Ansys Maxwell program. The vertical axis in the figure corresponds to the drift direction. The colour scale represents the intensity of the field, with red being the highest one. On the left, the detailed structure of the GEM holes for one thin GEM with 400 V applied across the GEM, 1000 V applied to the mesh and a transfer field of 0 kV/cm above the GEM. On the right, the same for a thick one with 490 V applied across the GEM, 1000 V applied to the mesh and a transfer field of 0 kV/cm above the GEM.}
	\label{fig:scheme}
\end{figure}
\begin{figure}[!t] 
	\centering
	\includegraphics[width=0.47\linewidth]{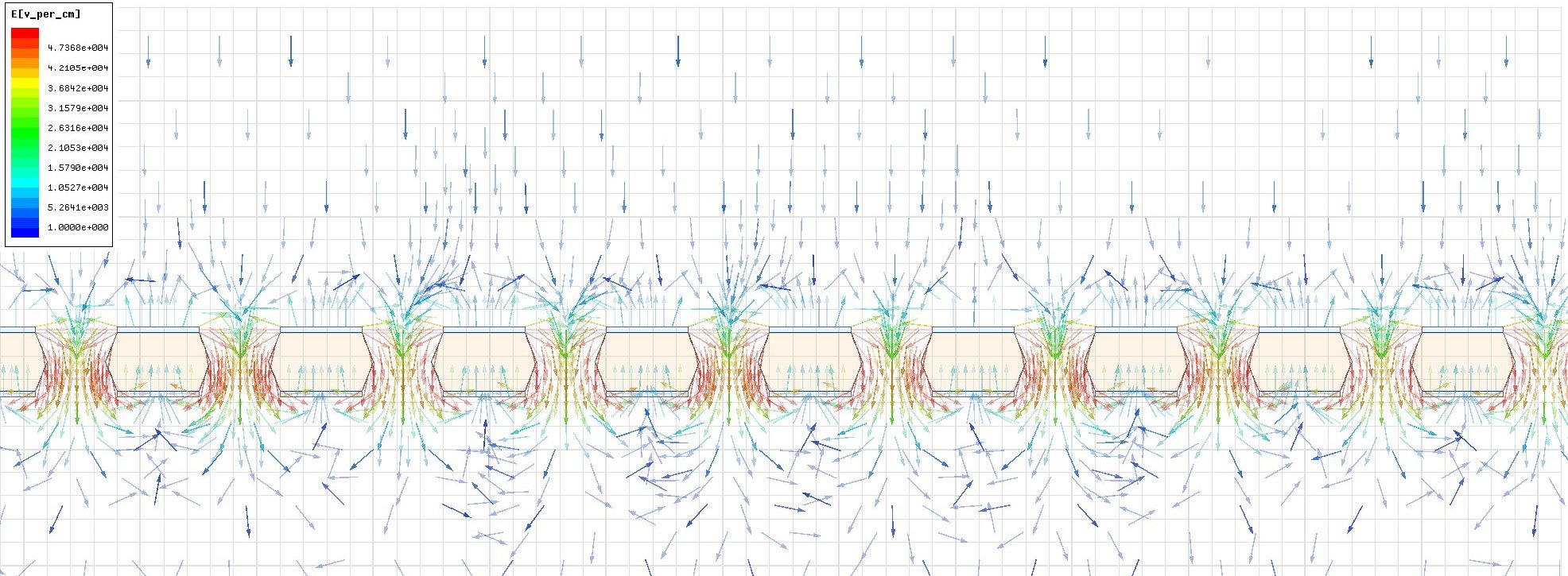}
	\includegraphics[width=0.43\linewidth]{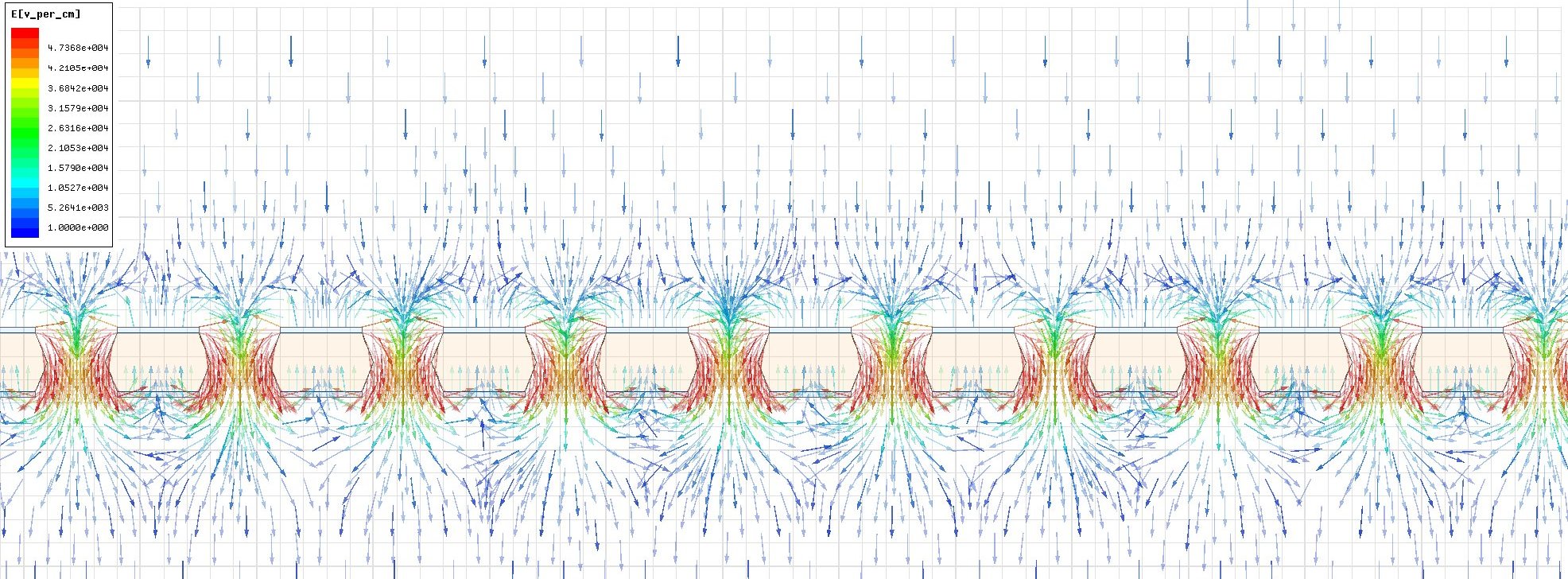}
	\caption{Examples of the 2D electric field line maps generated by the Ansys Maxwell program. The vertical axis in the figure corresponds to the drift direction. The line colour scale represents the intensity of the field, with red being the highest one. On the left, the detailed structure of the GEM holes when no induction field is applied, whilst on the right the same for 1 kV/cm of induction field. It is clearly visible how the field lines are much more ordered and straight towards the induction gap (bottom of the plot) in the right example than in the left one, as a result of the induction field addition.}
	\label{fig:flines}
\end{figure}
In order to deepen the understanding and comprehension of the light yield enhancement due to the additional induction field $E_{Mesh}$, the electric fields characteristics inside, above and below the GEMs holes are investigated through a simulation. The study is performed with Ansys Maxwell 15\footnote{\url{https://www.ansys.com/products/electronics/ansys-maxwell}}, a commercial software that allows to solve the electromagnetic equations and to obtain the electric field configuration of a specific geometry, among other features. 
A generalised simulation of the whole MANGO detector was initially performed with a coarse granularity in order to verify that the drift field and the induction field were constant and uniform. This confirmed the existence of a region few centimetres from the border where the above mentioned fields are uniform, and that the electric fields close to the holes of GEM3 were independent from the voltage configuration of GEM1 and GEM2. Therefore, to study in detail the influence of the induction field in the nearby of the GEM3 holes, only a single $t$ or $T$ 10 $\times$ 10 cm$^2$ GEM foil (representing GEM3), including all its holes with proper dimensions and conic shape, together with the metallic mesh is simulated in 2D (to minimise CPU time and since the geometry can be assumed to possess a cylindrical symmetry) with the actual applied voltages used in the data (see Table \ref{tab:datataking}) and discussed in the following. The granularity and accuracy of the simulation were increased up to a point where the electric field values reached a constant asymptote, not to have the result influenced by numerical errors.

Close to the GEM hole structure, the field results disuniform both above and below the GEM holes. The spatial scale of these irregularities covers a region of roughly 40 $\mu$m (100 $\mu$m) above and below a thin (thick) GEM holes, as shown in Figure \ref{fig:scheme} on the left (right). From the field lines evaluation, in the example of the $t$ GEM, displayed in Figure \ref{fig:flines}, it is possible to verify the assumption discussed in Section \ref{subsec:el_light}, for which the presence of the induction field would straighten the field lines below the GEM. Figure \ref{fig:flines} shows on the left the field lines in case no induction field is applied, whilst on the right a small 1 kV/cm induction field is present. The straightening of the field lines is clearly visible confirming the previous hypothesis.\\

\begin{figure}[!t] 
	\centering
	\includegraphics[width=0.9\linewidth]{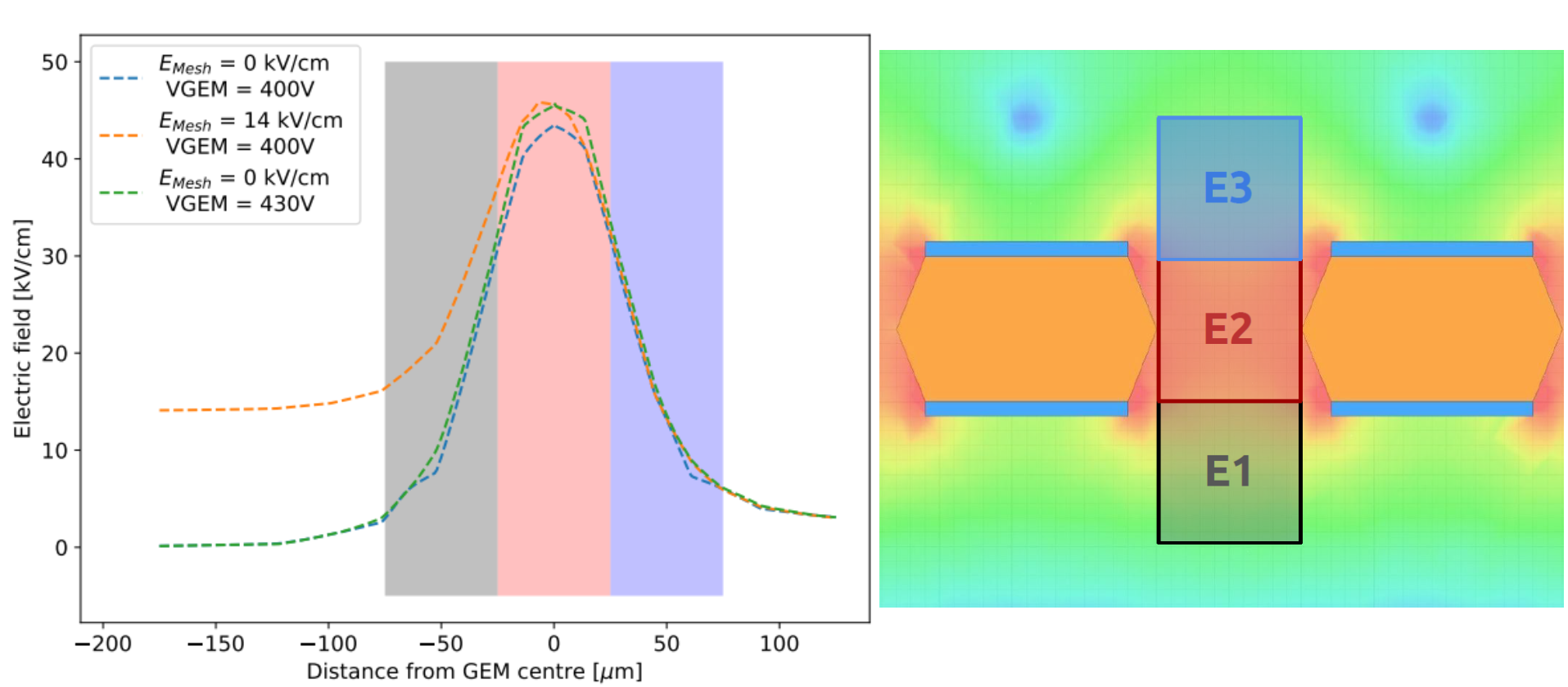}
	\caption{On the left, the profile of the electric field along the direction orthogonal to the GEM plane which passes through a $t$ GEM hole. The x-axis coordinate refers to the distance from the centre of the GEM hole, positive for above the GEM hole, negative for below, i.e. towards the induction gap. Three voltage configurations are shown described by the legend. Three regions are  highlighted in grey (E1), red (E2) and blue (E3) which are described in the text. On the right, a detail of the schematics of the thin GEM simulation with superimposed the same three regions E1, E2 and E3 described in the text.}
	\label{fig:profthin}
\end{figure}

\begin{figure}[!t] 
	\centering
	\includegraphics[width=0.49\linewidth]{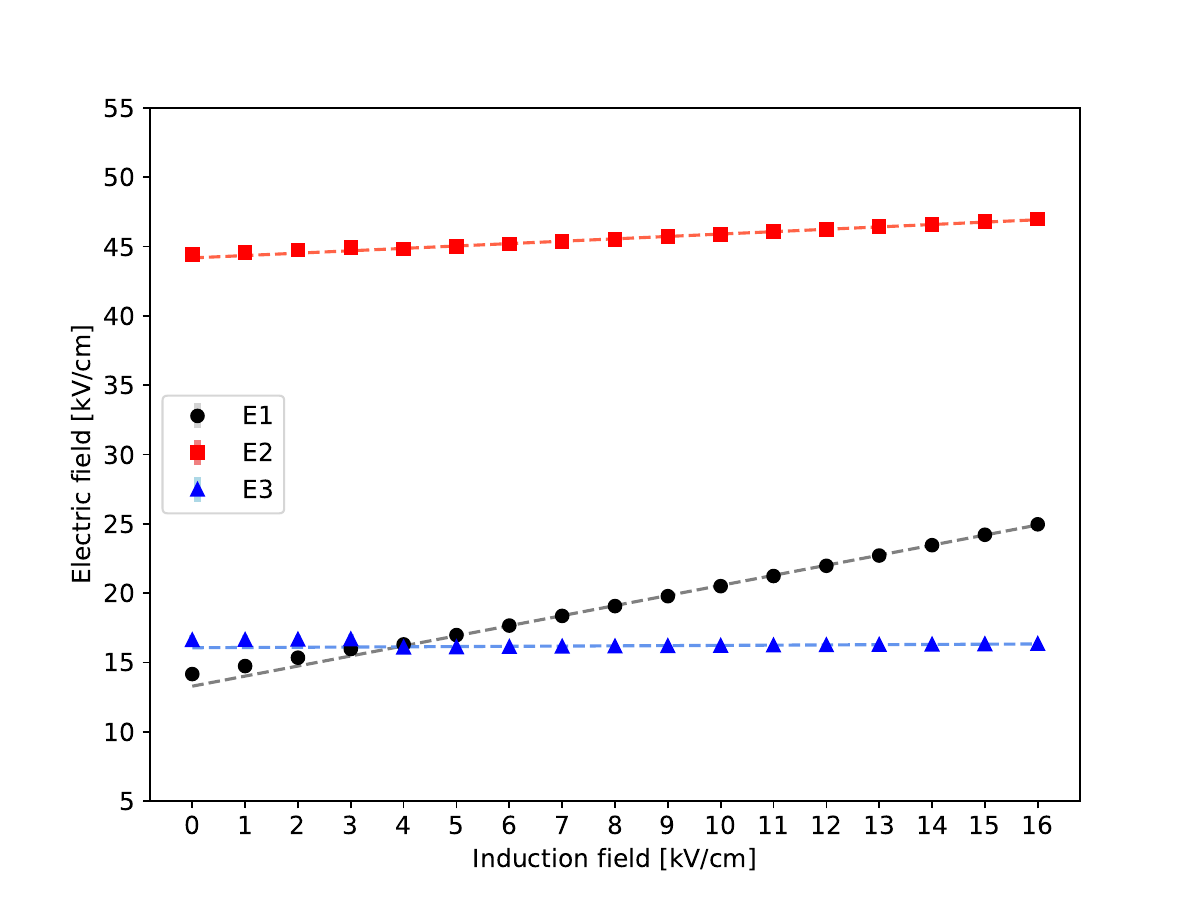}
	\includegraphics[width=0.49\linewidth]{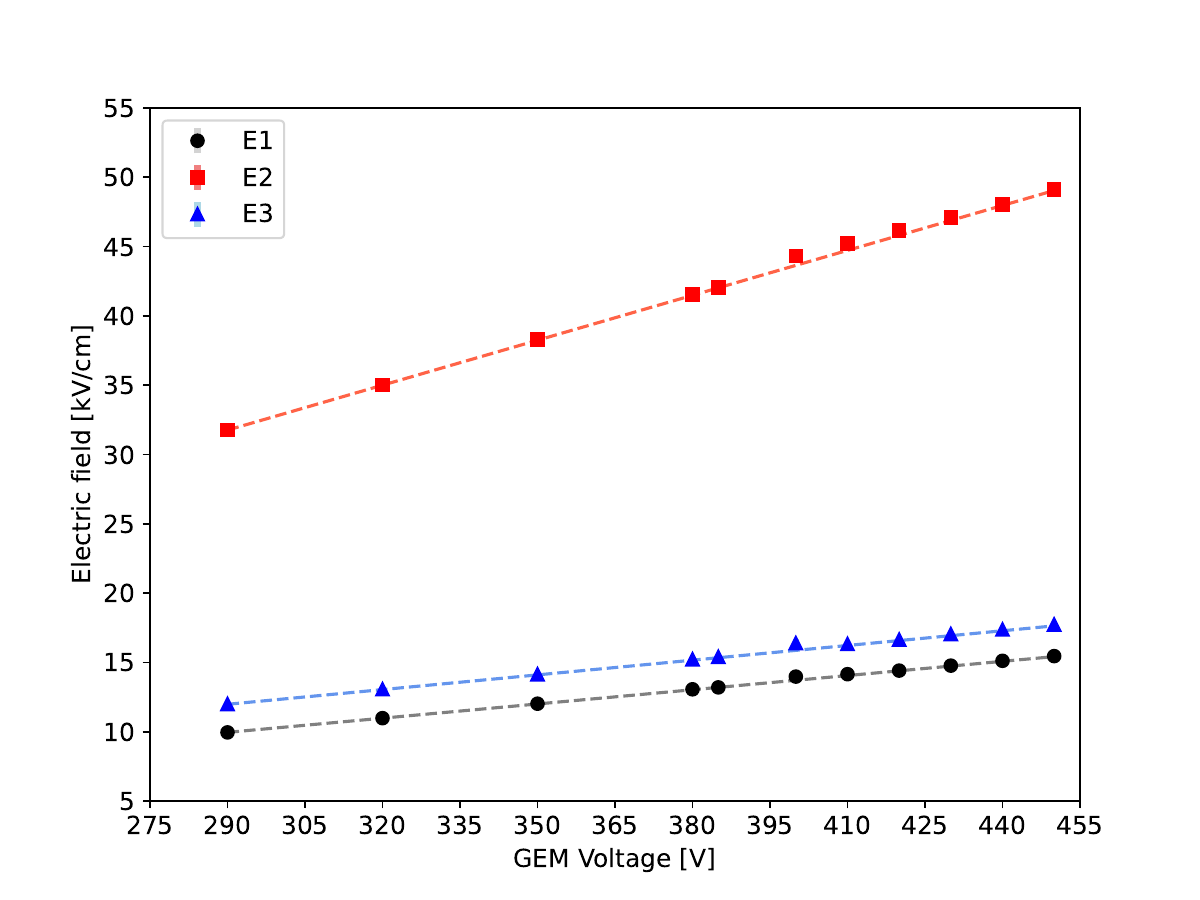}
	\caption{The simulated electric field in the three regions next to the GEM hole are displayed as a function of the induction field $E_{Mesh}$ on the left and as a function of V$_{GEM}$ on the right for a thin GEM geometry.}
	\label{fig:tttsim}
\end{figure}

The profile of the electric field in the direction orthogonal to the GEM plane which passes through a $t$ GEM hole is shown in Figure \ref{fig:profthin} on the left panel. The x-axis coordinate refers to the distance from the centre of the GEM hole, positive for above the GEM hole, negative for below, i.e. towards the induction gap. Three voltage configurations are shown: in blue 400 V across the GEM and no induction field is present, in green the voltage across the GEM is increased by 30 V, and finally in orange the voltage across the GEM is 400 but 14 kV/cm of induction field are present.
When no induction field is applied, the peak of the field is found at the zero coordinate, exactly at the centre of the GEM. The field symmetrically drops as the distance increases. Enlarging the voltage across the GEM affects the maximum field reached inside the GEM, but leaves the shape of the field profile untouched. Instead, when a strong induction field is added, not only the peak of the field inside the GEM hole increases, but the structure of the profile towards the induction gap changes. In particular, the slope of the decrease of the field is milder resulting in a stronger field close to the centre and edge of the GEM hole. In order to quantify this effect as a function of the voltage across the GEM and the induction field, the average value of the electric field is calculated in three different regions, highlighted in gray, red and blue in Figure \ref{fig:profthin} left. The region E2 is a square of 50 $\times$ 50 $\mu$m$^2$ centred in the centre of the GEM foil, to characterise the field inside the GEM hole. The regions E3 and E1 are taken adjacent to E2, with the same dimensions, respectively above and below E2. Figure \ref{fig:profthin} shows on the right panel the three regions in the Maxwell schematics.
The electric field in the three regions is simulated as a function of the induction field with a fixed 400 V across the GEM and as a function of V$_{GEM}$ with no induction field. The results are displayed in Figure \ref{fig:tttsim} respectively on the left and right panel. For each value of $E_{Mesh}$ or V$_{GEM}$, the electric field simulated by Maxwell is averaged inside each box and the obtained mean is further averaged over 20 adjacent holes. Linear fits are performed on the sets of data and are summarised in Table \ref{tab:tttsim} for the induction field dependence, as $A_{t,E}+ B_{t,E}E_{Mesh}$, and for the V$_{GEM}$ one, as $A_{t,V}+B_{t,V}V_{GEM}$.
The results remark that increasing the voltage across the GEM modifies the field in all the three regions, symmetrically in E1 and E3, and with larger intensity in E2. Instead, the addition of the induction field augments the field in E1 more strongly than in E2, while E3 is untouched. This allows to conclude that the increase in light output can not be explained by a modification of the GEM transparency due to the addition of the induction field below it.
The electric field inside the GEM hole (E2 region) increases linearly with both the induction field and $V_{GEM}$, consistently with the experimental results presented in Section \ref{subsec:el_light}. This simulation further confirms the assumptions discussed in the same Section that the induction field contributes linearly to the reduced field present inside the GEM holes (see Equation \ref{eq:gamm_vsmesh}) and confirm the hypotheses on the light production mechanism in this region. However, the change from linear to exponential light increment is obtained in the E1 region where the values reach fields above 20 kV/cm, considered high enough to start new amplification. If the configuration where $V_{GEM}=$ 400 V and $E_{Mesh}=$14.0 kV/cm is compared with the one in which $V_{GEM}=$ 430 V and $E_{Mesh}=$0 kV/cm (two operating conditions which produce extremely similar light yield), it can be observed that the increase in the E1 field with respect to nominal operating conditions ($V_{GEM}=$ 400 V and $E_{Mesh}=$0 kV/cm) is a factor ten larger in the first case compared with the second. Conversely, the increase in E2 is only a factor 2 larger when $V_{GEM}$ is increased, with respect to a raise in $E_{Mesh}$. It also has to be noted that while this is an average value of the field, Figure \ref{fig:profthin} shows that the closer one gets to the GEM, the larger the field. Both Figure \ref{fig:profthin} and Figure \ref{fig:tttsim} also demonstrate that the high intensity of the electric field in the region E1 is peculiar to the introduction of the induction field, thus it is not present in the regular operation of a GEM.\\
\begin{table}[!t]
	\centering
	\begin{adjustbox}{max width=1.01\textwidth}
		\begin{tabular}{|c|c|c|c|c|}
			\hline
			Fit parameter & $A_{t,E}$ (kV/cm) & $B_{t,E}$ & $A_{t,V}$ (kV/cm) & $B_{t,V}$ (kV/V cm) \\
			\hline \hline
			E1 region     & 13.28 $\pm$ 0.03 & 0.727 $\pm$  0.003 & 0.043 $\pm$ 0.002 & 0.034 $\pm$ 0.001\\
			E2 region     & 44.18 $\pm$ 0.03 & 0.172 $\pm$  0.003 & 0.042 $\pm$ 0.006 & 0.108 $\pm$ 0.001\\
			E3 region     & 16.25 $\pm$ 0.06 & 0.004 $\pm$  0.005 & 1.72 $\pm$ 0.01 & 0.035 $\pm$ 0.001\\
			\hline
		\end{tabular}
	\end{adjustbox}
	\caption{Result of a linear fits with the function $A_{t,E}+ B_{t,E}E_{Mesh}$ and $A_{t,V}+B_{t,V}V_{GEM}$ to the electric fields in E1, E2 and E3 regions as simulated with Maxwell for a $t$ GEM. }
	\label{tab:tttsim}
\end{table}
\begin{figure}[!t] 
	\centering
	\includegraphics[width=0.65\linewidth]{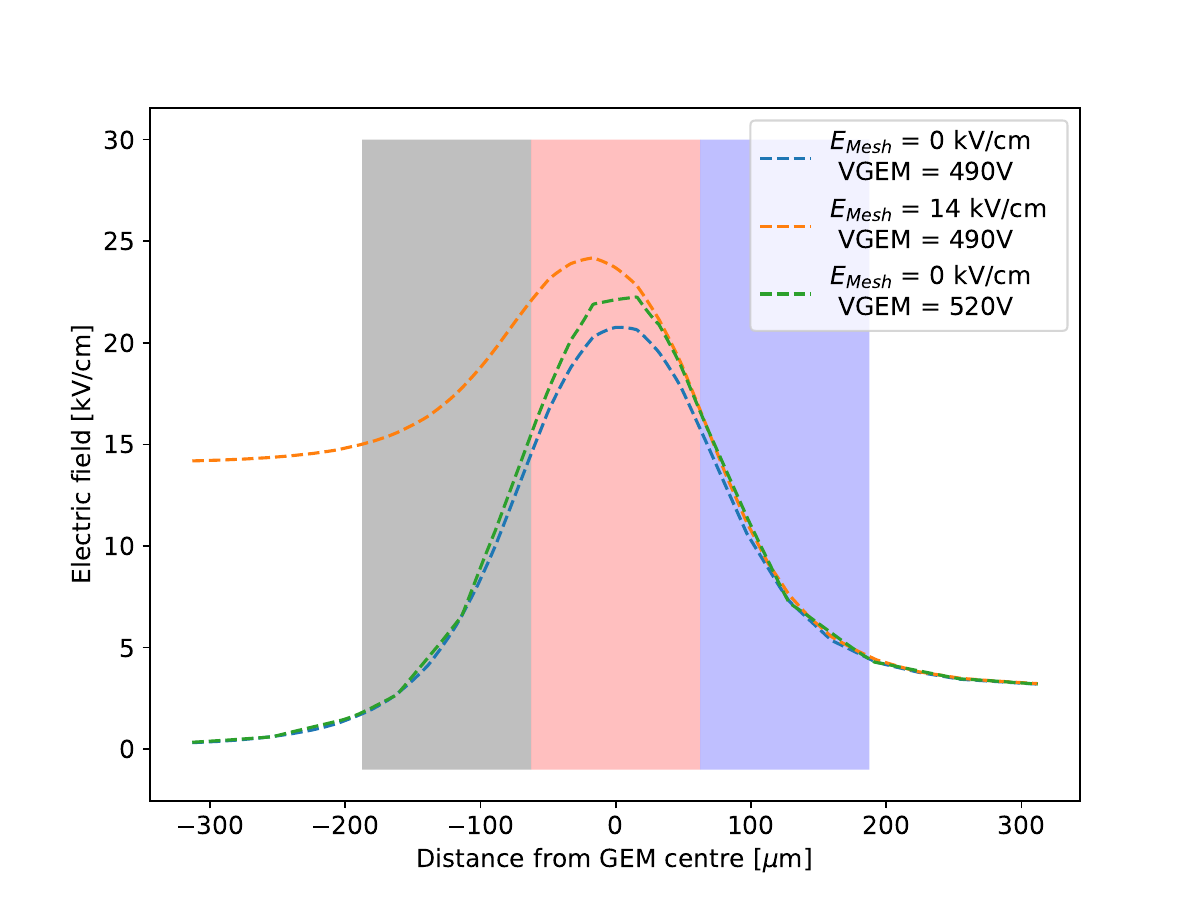}
	\caption{Profile of the electric field along the direction orthogonal to the GEM plane which passes through a $T$ GEM hole. The x-axis coordinate refers to the distance from the centre of the GEM hole, positive for above the GEM hole, negative for below, i.e. towards the induction gap. Three voltage configurations are shown described by the legend. Three regions are  highlighted in grey (E1), red (E2) and blue (E3) which are described in the text.}
	\label{fig:profthick}
\end{figure}

The same type of simulation and analysis is performed on a $T$ GEM. The reference voltage across the $T$ GEM is taken as 490 V in order to be consistent with the $E_{Mesh}$ studies described in Table \ref{tab:app}. Figure \ref{fig:profthick} shows the profile of the electric field along the direction orthogonal to the GEM plane which passes through a $T$ GEM hole. The modification of the electric field structure, with respect to the standard 490 V applied across the GEM, are obtained by raising the voltage of 30 V or by introducing 14 kV/cm in the induction gap. Akin to the $t$ GEM, not only the maximum field reached inside the GEM increases, but the slope of the field towards the induction gap decreases.  Due to the geometry, both the internal part of the GEM hole and the area below it are wider and the distortion seems to have a larger relative impact than in the $t$ case. To quantify the influence of the variation of the electric fields as a function of the voltage across the GEM and the induction field, three regions are defined in the 2D space of the simulation the average value of the electric field is calculated from, exactly as for the $t$. In order to adapt to a larger $T$ GEM the regions are selected with an area of 125 $\times$ 125 $\mu$m$^2$.

\begin{figure}[!t] 
	\centering
	\includegraphics[width=0.49\linewidth]{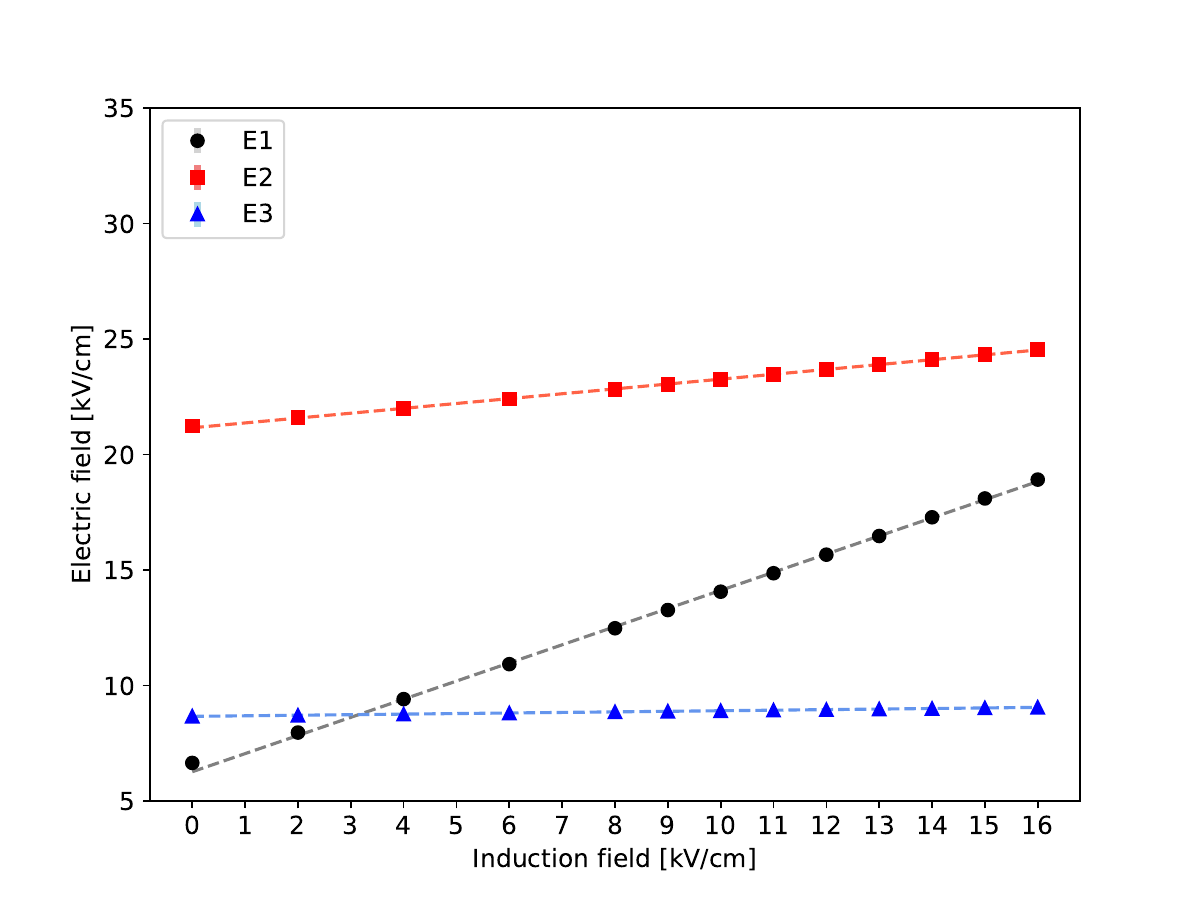}
	\includegraphics[width=0.49\linewidth]{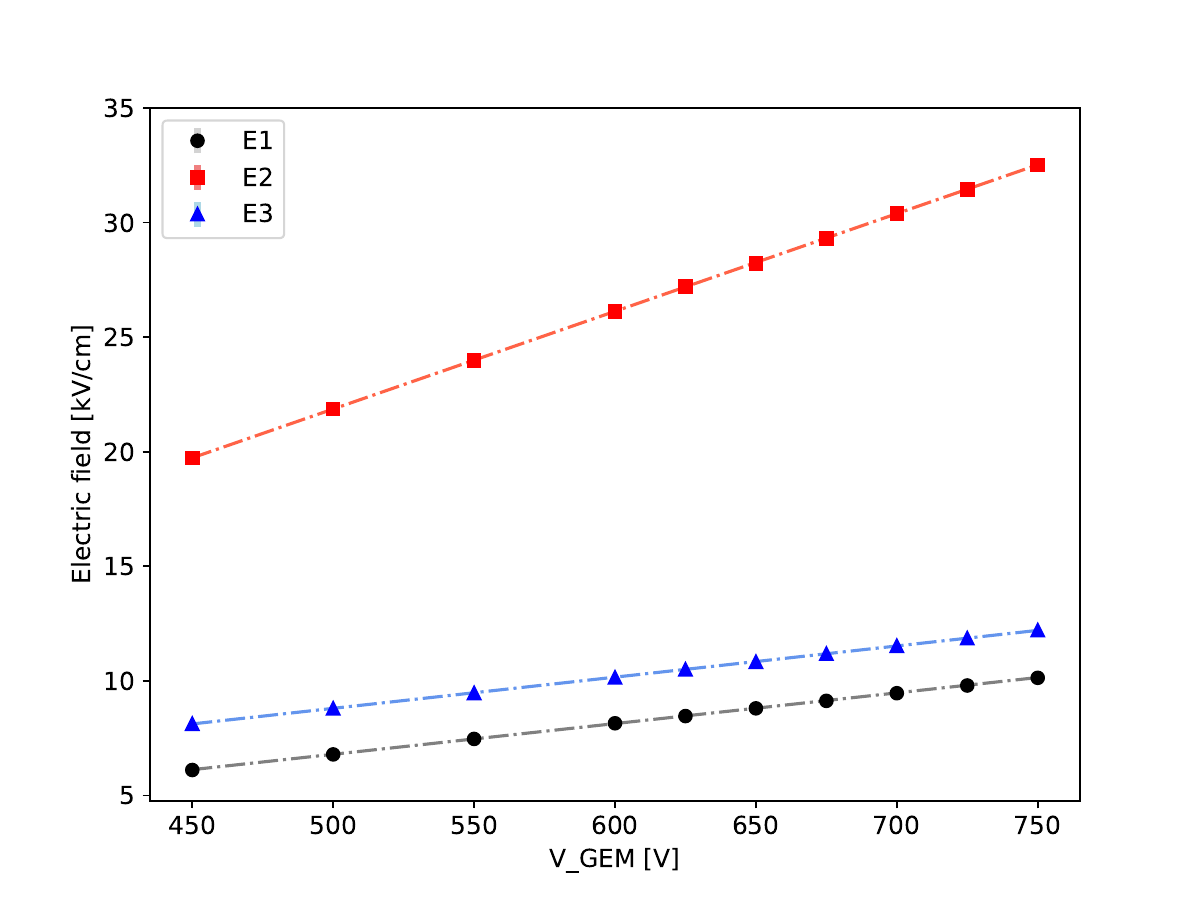}
	\caption{The simulated electric field in the three regions next to the GEM hole are displayed as a function of the induction field $E_{Mesh}$ on the left and as a function of V$_{GEM}$ on the right for a thin GEM geometry.}
	\label{fig:TTsim}
\end{figure}

The results of the average field in the E1, E2, and E3 regions as a function of the induction field and V$_{GEM}$ are displayed in Figure \ref{fig:TTsim} respectively on the left and right panel. Linear fits are performed on the sets of data and are summarised in Table \ref{tab:TTsim} for the induction field dependence, as $A_{T,E}+ B_{T,E}E_{Mesh}$, and for the V$_{GEM}$ one, as $A_{T,V}+B_{T,V}V_{GEM}$.
The results for the $T$ are coherent to the one attained for the $t$. The electric fields are generally lower in the case of the $T$ GEM, as expected from the geometry of the problem, and the values obtained in the region E2 of $\sim$ 20 kV/cm confirm that these fields allow amplification processes, as can be seen from the Figure \ref{fig:gain} and  Table \ref{tab:app}. Nonetheless, it can be noted that the slope of the increase of the field is larger for the $T$ than the $t$ one.\\

The results of these simulations show that, for all the configurations considered, the induction field influences linearly the electric field inside the GEM holes which, in turn, is expected to linearly affect the reduced gain of the amplification structure. This argument matches perfectly with the assumption of Equation \ref{eq:sigma_ext} and with the observation of exemplified in the right panel of Figure \ref{fig:fitEh_gain}.
\begin{table}[!t]
	\centering
	\begin{adjustbox}{max width=1.01\textwidth}
		\begin{tabular}{|c|c|c|c|c|}
			\hline
			Fit parameter & $A_{T,E}$ (kV/cm) & $B_{T,E}$ & $A_{T,V}$ (kV/cm) & $B_{T,V}$ (kV/V cm) \\
			\hline \hline
			E1 region     & 6.27 $\pm$ 0.02 & 0.785 $\pm$  0.002 & 0.09 $\pm$ 0.06 & 0.0134 $\pm$ 0.0005\\
			E2 region     & 21.16 $\pm$ 0.03 & 0.210 $\pm$  0.001 & 0.54 $\pm$ 0.06 & 0.042 $\pm$ 0.001\\
			E3 region     & 8.66 $\pm$ 0.02 & 0.002 $\pm$  0.006 & 2.00 $\pm$ 0.03 & 0.0136 $\pm$ 0.0005\\
			\hline
		\end{tabular}
	\end{adjustbox}
	\caption{Result of a linear fits with the function $A_{T,E}+ B_{T,E}E_{Mesh}$ and $A_{T,V}+B_{T,V}V_{GEM}$ to the electric fields in E1, E2 and E3 regions as simulated with Maxwell for a $T$ GEM. }
	\label{tab:TTsim}
\end{table}

The combination of these simulations with the results obtained in Section \ref{sec:lemon} and in Section \ref{sec:mango_light} suggests that when the induction field is raised beyond E$_b$, the reduced field in the bottom region of the GEM hole and few micrometers below it is high enough to produce a simultaneous (but not directly proportional) amplification of light and charge, effectively enlarging and shifting the actual amplification region of this configuration. In the $T$ GEM configuration, $E_{Mesh}$ appears to generate a larger region of additional amplification, with electric fields closer to the ones in the centre of the GEM hole. Once a single gas mixture is considered, this is consistent with the larger light output of the \textit{TT} configuration visible in Figure \ref{fig:elall_light} and characterised by a larger $c$ parameter in Table \ref{tab:elbreak}, which represented the slope of the exponential in the relative light increase as a function of $E_{Mesh}$. Moreover, the larger size is also compatible with the faster degradation in diffusion for the \textit{TT} configurations observed in Figure \ref{fig:el_size} with respect to the ones with a $t$ on the bottom.\\
\begin{figure}[!t] 
	\centering
	\includegraphics[width=0.5\linewidth]{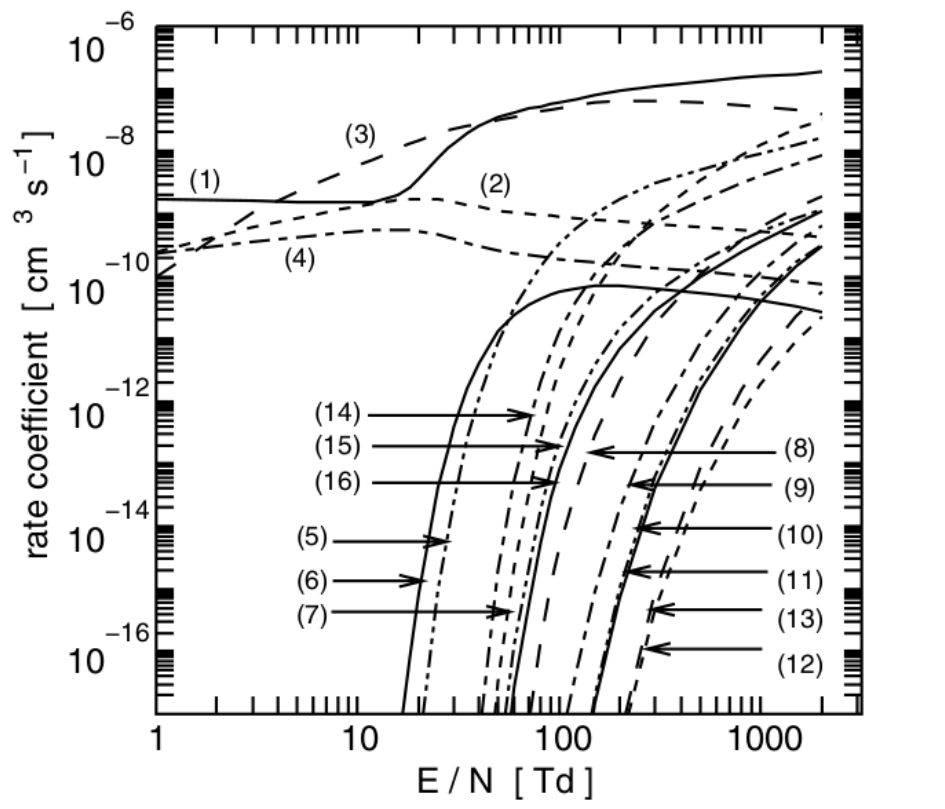}
	\caption{Rate coefficient of the different excitations (2-5), attachment (6), ionisation (7-13), dissociation (14-16) for CF$_4$ gas interaction with electrons as a function of the reduced electric field taken from \cite{bib:Kurihara}.}
	\label{fig:el_sezurto}
\end{figure}
The larger production of light with respect to charge can be explained by the fact that the energy threshold of the production of light through the fragmentation of the CF$^*_3$ has a lower threshold than the ionisation one. Figure \ref{fig:el_sezurto} shows the rate coefficient of the different excitations (2-5), attachment (6), ionisation (7-13), dissociation (14-16) for CF$_4$ gas interaction with electrons as a function of the reduced electric field taken from \cite{bib:Kurihara}. In particular, the process 14 refers to the neutral fragmentation responsible for the production of visible light and possesses a smaller cross section threshold than the ionisation ones. As the induction field introduces a region where amplification is possible, but with lower field strength than inside a GEM hole,  the light-producing process results favoured in terms of cross section with respect to charge production. The cross section presented in Figure \ref{fig:el_sezurto} are of pure CF$_4$, while the gas mixture studied contains large amounts of helium. The cross section for these mixtures are not found in literature, thus no conclusive assessment on the nature of the light to ratio production can be done. Yet, since the primary ionisation energy of He is nearly double the one of CF$_4$, it is fair to assume that the threshold of ionisation and fragmentation of CF$_4$ do not change significantly from this picture for the gas mixtures under study.

This result is suggesting that the innovative amplification strategy illustrated in this study effectively enhances the light amplification potentialities of the region below the GEM holes, while minimising the additional simultaneous charge production.

\section{Discussion}
\label{sec:disc}
In this study different combinations of GEMs and a varying amount of helium in the gas mixture were analysed with respect to their light yield properties, energy resolution and diffusion. The 70/30 mixture results in higher probability of discharges which makes the detector unstable and prone to failure while not providing any particular improvement with respect to the standard 60/40. When only the voltage across the GEM is increased, the \emph{ttt} configuration returns the better energy resolution and light yield as expected, but at the cost of a worse diffusion, which also depended on the voltage applied. The \emph{Tt} configuration has a lower light yield, around 3 times less, and guarantees a lower diffusion, with a reduction of 30\%. The \emph{TT} behaves in the middle for what concerns the diffusion, but possesses a light yield about 5 times lower than the \textit{ttt}.\\
\begin{table}[!t]
	\centering
	\begin{adjustbox}{max width=1.01\textwidth}
		\begin{tabular}{|c|c|c|c|c|}
			\hline
			\multicolumn{2}{|c|}{} & Integral & E res (\%) & Diff [$\mu$m] \\
			\hline \hline
			\multirow{3}{*}{\textit{ttt}} & min &  9510 $\pm$ 40 & 16.0 $\pm$ 0.3 & 320 $\pm$ 4\\
			& max V$_{GEM}$ & 28400 $\pm$ 110 & 16.6 $\pm$ 0.3 & 412 $\pm$ 5\\
			& max $E_{Mesh}$ & 33500 $\pm$ 140 & 13.8 $\pm$ 0.3 & 388 $\pm$ 5\\ \hline
			\multirow{3}{*}{\textit{TT}} & min &  3410 $\pm$ 20 & 28.0 $\pm$ 1.5 & 260 $\pm$ 3\\
			& max V$_{GEM}$ &  5090 $\pm$ 30 & 31.0 $\pm$ 0.6 & 255 $\pm$ 3\\
			&  max $E_{Mesh}$ &  58800 $\pm$ 300 & 25.7 $\pm$ 0.5 & 356 $\pm$ 5\\ \hline
			\multirow{3}{*}{\textit{Tt}} & min & 4600 $\pm$ 30 & 25.2 $\pm$ 0.5 & 245 $\pm$ 3\\
			& max V$_{GEM}$ &  7700 $\pm$ 40 & 27.8 $\pm$ 0.5 & 245 $\pm$ 3\\
			& max $E_{Mesh}$ &  11800 $\pm$ 50 & 26.8 $\pm$ 0.5 & 280 $\pm$ 4\\ \hline
		\end{tabular}
	\end{adjustbox}
	\caption{Summary for the three configurations at 60/40 (\textit{ttt}, \textit{Tt} and \textit{TT}) of the \textit{integral}, the energy resolution and the intrinsic diffusion in three scenarios: the minimum voltage applied to the GEMs with no induction field ("min"), the maximum voltages applied to the GEMs with no induction field ("max V$_{GEM}$"), and finally the minimum voltage applied to the GEMs with the maximum induction field ("max $E_{Mesh}$")}
	\label{tab:check2}
\end{table}

The introduction of a strong electric field  below the last GEM amplification plane enhances the light gain without affecting heavily the diffusion and the energy resolution. This innovative way of utilising the GEM with this specific gas mixture allows to improve the performances of the diverse GEM stacks employed. Table \ref{tab:check2} summarises for the three configurations at 60/40 (\textit{ttt}, \textit{Tt} and \textit{TT}) the \textit{integral} (proportional to the light yield), the energy resolution and the intrinsic diffusion in three scenarios: the minimum voltage applied to the GEMs with no induction field ("min"), the maximum voltages applied to the GEMs with no induction field ("max V$_{GEM}$"), and finally the minimum voltage applied to the GEMs with the maximum induction field ("max $E_{Mesh}$"). It can be noted that the addition of the induction field always allows to reach a light output larger than what is possible employing the GEMs in the standard way. Moreover, the larger light yield is accompanied by a similar, if not better energy resolution. For the \textit{ttt} configuration the intrinsic diffusion is also improved when the induction field is employed in place of the increase of the voltage applied to the GEMs. Conversely, for the \textit{TT} and \textit{Tt} ones, the diffusion worsens. This can be explained by the fact that the intrinsic diffusion of the \textit{ttt} was shown to be dependent on the applied voltage on the GEMs, differently from the \textit{TT} and \textit{Tt} cases (see Figure \ref{fig:el_res}).\\
\begin{table}[!t]
	\centering
	\begin{adjustbox}{max width=1.01\textwidth}
		\begin{tabular}{|c|c|c|c|c|c|}
			\hline
			\multicolumn{2}{|c|}{} & $E_{Mesh}$ [kV/cm] & Integral & E res (\%) & Diff [$\mu$m] \\
			\hline \hline
			\multirow{3}{*}{$0$} & \textit{ttt} & 0 $\pm$ 0 & 9510 $\pm$ 40 & 16.0 $\pm$ 0.3 & 320 $\pm$ 4\\
			& \textit{TT} & 12 $\pm$ 0.3 & 9420 $\pm$ 40 & 17.4 $\pm$ 0.4 & 302 $\pm$ 4\\
			& \textit{Tt} & 11.1 $\pm$ 0.3 & 9360 $\pm$ 40 & 27 $\pm$ 0.5 & 264 $\pm$ 3\\ \hline
			\multirow{3}{*}{$1$} & \textit{ttt} & 3 $\pm$ 0.3 & 11300 $\pm$ 50 & 15.5 $\pm$ 0.3 & 347 $\pm$ 5\\
			& \textit{TT} & 12.3 $\pm$ 0.4 & 11300 $\pm$ 50 & 17.9 $\pm$ 0.4 & 307 $\pm$ 4\\
			& \textit{Tt} & 12.3 $\pm$ 0.4 & 11300 $\pm$ 50 & 25.0 $\pm$ 0.5 & 273 $\pm$ 4\\ \hline
			\multirow{3}{*}{$2$} & \textit{ttt} & 15 $\pm$ 0.3 & 33500 $\pm$ 140 & 13.8 $\pm$ 0.3 & 388 $\pm$ 5\\
			& \textit{TT} & 14 $\pm$ 0.3 & 58800 $\pm$ 300 & 25.7 $\pm$ 0.5 & 356 $\pm$ 5\\
			& \textit{Tt} & 12.8 $\pm$ 0.2 & 11830 $\pm$ 50 & 26.8 $\pm$ 0.5 & 280 $\pm$ 4\\ \hline
		\end{tabular}
	\end{adjustbox}
	\caption{Summary, for each configuration, of the induction field, the integral, the energy resolution and intrinsic diffusion in three different scenarios. Scenario "0" refers to when the three configurations have the same light output and the \textit{ttt} has $E_{Mesh}=0$. Scenario "1" refers to when the three configurations have the highest light output equal to each other. Finally, scenario "2" refers to when the maximum induction field is applied to each configuration.}
	\label{tab:check1}
\end{table}

The performances of the different stacking options in the 60/40 gas mixture can be compared among each other to find the best solution. Table \ref{tab:check1} summarises, for each configuration, the induction field, the integral, the energy resolution and intrinsic diffusion in three different scenarios. Scenario $0$ refers to when the three configurations have the same light output and the \textit{ttt} has $E_{Mesh}=0$. Scenario $1$ refers to when the three configurations have the highest light output equal to each other. Finally, scenario $2$ refers to when the maximum induction field is applied to each configuration.\\
Scenarios $0$ and $1$ represent similar setups for which it is possible to demonstrate that the introduction of the induction field allows the 2 GEM stacks to attain a light output identical to the 3 GEM stack with similar energy resolution (only for the \textit{TT}) and reduced intrinsic diffusion.\\
Scenario $2$ directly compares the maximum light output achievable by the different configurations. The largest area generated under the $T$ GEMs thanks to the strong induction field permits to reach light gain dramatically high, maintaining a smaller intrinsic diffusion than the standard \textit{ttt}. The light output of \textit{TT} exceeds by a factor 2 the maximum achievable with standard operation of the  \textit{ttt} GEM stack, keeping the energy resolution at 5.9 keV below 30\% and intrinsic diffusion around 350 $\mu$m. \\
The analyses performed stress that depending on the experimental need each configuration excels in one of the variable studied. The \textit{ttt} has an excellent energy resolution thanks to the high gain and reduced field intensity, the \textit{Tt} always has the smallest intrinsic diffusion well below 300 $\mu$m, and finally the \textit{TT} allows to reach the largest light yields. In the context of a directional DM experiment, as discussed in Section \ref{subsubsec:dirposdisc}, the largest impact on the sensitivity comes from the energy threshold. CYGNO's expected energy threshold of 0.5 keV (see Section \ref{subsec:cyg_lim}) could be further lowered by the increase in light yield down to almost 0.25 keV, opening the possibility to search for \W DM around 0.5 \Gevc with the He target. At the same time, if more importance is given to the directional capabilities, such as HT recognition and angular resolution, the reduced diffusion of the \textit{Tt} can help improving the topology of the data for the imaging of the recoil tracks. For the above mentioned reasons, this study is deemed extremely important for the development of future recoiling imaging experiments.

\chapter{Negative Ions Drift operation with optical readout}
\label{chap6}
One of the main drawback of large gaseous TPCs is that the drift of the electronic cloud to the amplification plane typically deteriorates the intrinsic properties of the original ionisation trail generated by the track interacting with the gas due to diffusion. In particular, in the directional DM searches, the loss of information on the original topology spoils the evaluation of the direction and the HT recognition of the recoil track, fundamental variables in the directional quest for DM (see Section \ref{sec:directional}). The contribution to diffusion are mainly two: one happening during the drift in the sensitive volume of the TPC, whilst the other during the amplification processes. The former is directly dependent on the properties of the gas mixture chosen. A well known method to reduce the transverse diffusion (the one in the orthogonal direction to the drift field) is to apply magnetic fields \cite{Blum_rolandi}. Yet, in the context of DM search, the size of the experiments would require magnet with too big dimensions, along with the fact that the longitudinal diffusion would be worsened. Instead, the possibility of reducing the diffusion with the addition of a small percentage of SF$_6$ to exploit the advantages of a negative ion drift (NID) operation is analysed in this Chapter. 
This allows to construct much more compact detectors with longer drift distances,  with improved tracking performances with respect to electron drift (ED) operation,  better volume scaling and reduced background contamination. Typically, NID is operated at very low pressures below 100 Torr, with few examples of close to or exactly atmospheric pressure \cite{bib:nitec,LIGTENBERG2021165706}. In this thesis, for the fist time ever a NID operation at atmospheric pressure with a optical readout is presented. This is achieved by adding a small percentage of \SF 1.6\% to the standard CYGNO 60/40 He:CF$_4$ gas mixture. The data were acquired with the MANGO prototype illustrated in Section \ref{sec:mango}, and a great reduction of the transversal diffusion is demonstrated both along the drift and at the amplification stage.\\
Section \ref{sec:sf6} is dedicated to the description of the NID operation principles and of its current knowledge and state of the art. Section \ref{sec:sf6chem} describes the properties of the \SF gas in the context of its employment as a NID gas. Section \ref{sec:nidatm} is dedicated to the experimental setup and data taking of NID at atmospheric pressure, while Section \ref{sec:nidkeg} focuses on the diffusion studies of the NID at slightly below atmospheric pressure. Finally, Section \ref{sec:discunid} presents the discussion in light of the measurements shown in this Chapter.
\section{Negative ion drift operation}
\label{sec:sf6}
The introduction of an electronegative species in the gas mixture can change the properties of the gas, effectively modifying the conventional operation of a TPC \cite{NID_2_Martoff2000355,Ohnuki:2000ex}. In particular, free electrons can be very easily captured by these electronegative dopants at very short distances, about 10-100 $\mu$m, generating negative ions.  Due to the applied electric field, the anions drift to the anode and act as the effective track image carriers instead of electrons. Upon reaching the anode, the intense electric fields allow their additional electron to be released so that they can initiate a standard electron avalanche. This modified operation of the TPC brings two main advantages: a reduction in the diffusion coefficient and the possibility to fiducialise the detector.
\paragraph{Reduced diffusion}
One of the major advantages of the NID operation is the reduced diffusion during drift. In fact, the diffusion coefficient is proportional to the square root of the average energy of the drifting particle \cite{Blum_rolandi,Iontransport1}. Electrons are much lighter than the neutral component of the gas by orders of magnitude, which causes a non efficient momentum transfer with them. As a result, when electrons collide with ions, they have their direction randomised, but keep most of the energy acquired from the electric field in the time between two subsequent collisions. This  random part of the field energy is the dominant component in their total energy budget,  much larger than the thermal one, unless extremely low fields are applied. On the contrary, the anion masses are comparable to the neutral component of the gas which allows them to thermalise, efficiently dispersing the energy acquired from the drift field. This way, the diffusion can be reduced to the thermal limit \cite{Blum_rolandi} achieving dispersions of orders of magnitude less \cite{DRIFT:2014bny}. The approximate formula describing the diffusion width $\sigma$ of an electronic cloud which travelled a distance $L$ under the effect of an electric field $E$ is \cite{Blum_rolandi}\\
\begin{equation}
\label{eq:limterm}
\sigma^2 = \frac{2k_BTL}{eE},
\end{equation}
with $k_B$ the Boltzmann constant, $T$ the temperature in kelvin, $e$ the fundamental electric charge.
The reduction in diffusion allows to strongly improve the directional capabilities and the background rejection as the topology of the tracks results less spoiled by diffusion. The maximum  drift length, which can be employed in a TPC when tracking of low energy particles is required, is clearly directly correlated to the diffusion properties. Indeed, at large distances, the smearing effect induced becomes so strong that all types of tracks will result in blurred large spots. NID operation can, therefore, allow the use of much longer drift distances with respect to ED without significant degradation of tracking properties. Diffusions of the order of about 70-80 $\mu$m/$\sqrt{\rm{cm}}$ can be achieved with NID \cite{Ohnuki:2000ex,NID_2_Martoff2000355} compared with $\sim$ 600 $\mu$m/$\sqrt{\rm{cm}}$ measured for conventional ED like Ar:CH$_4$ (90/10) \cite{PDG}.
\paragraph{Fiducialisation}
\begin{figure*}[t]
	\centering
	\includegraphics[width=0.8\linewidth]{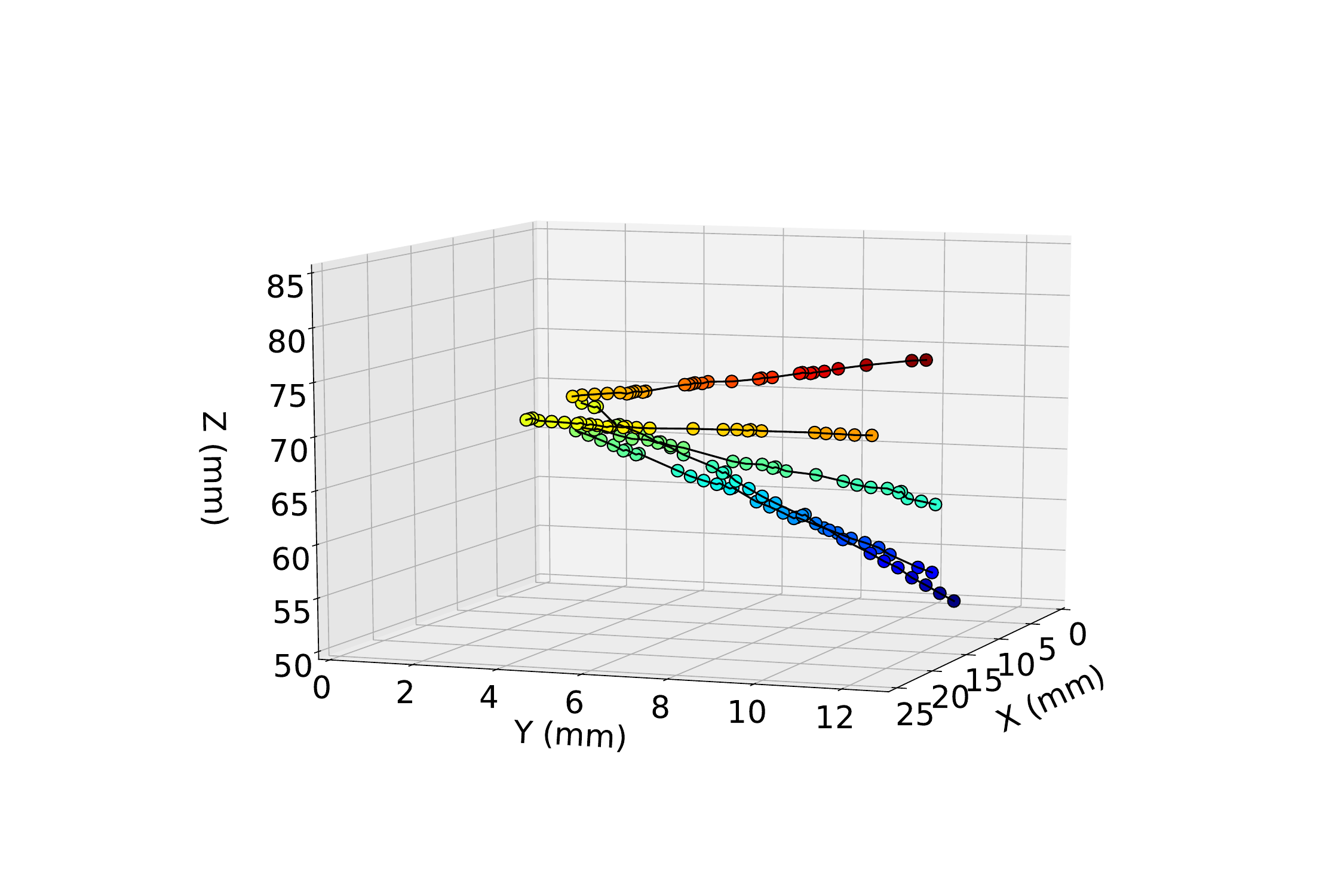}
	\caption{Reconstructed 3D alpha particles tracks. The z coordinate was obtained by the measurements of the time delay between minority carriers and SF$_6^-$ species. Figure taken from \cite{Ikeda:2020pex}}
	\label{fig:fiduc}
\end{figure*}
The fiducialisation is the ability localise the three-dimensional position of a recoil in a TPC, as discussed in Section \ref{subsec:background}. In TPC with segmented readout, the most difficult position to be measured is the one along the drift direction (z). With a NID operation, it was demonstrated that more than one species of negative ion carrier, with different masses, can be produced during the attachment of the primary electrons \cite{mincarry}. Among these different species one is responsible for the majority of the charge transport, and therefore the others are called \textit{minority carriers}. Since anion mobility depends on the mass and all the different species are generated in proximity of the original position of the primary charges, the difference in time of arrival of different anions effectively provides a measurement of the position of the event along z. The z coordinate can be estimated as \cite{Phan2016TheNP,Ikeda:2020pex}:
\begin{equation}
\label{eq:fidu}
z = \frac{v_mv_M}{v_m+v_M}\Delta T
\end{equation}
with $v_m$ the velocity of the minority carrier, $v_M$ the one for the major negative ion species, and $\Delta T$ the difference of time of arrival of the two.
\subsection{State of the art}
\label{subsec:stateart}
The studies on the possibility of employing NID operation in TPCs to reduce the diffusion were pioneered by Martoff, Snowden, Ohnuki and Spooner in 2000, when CS$_2$ electronegative gas was added to xenon or argon in a MultiWire Proportional Chamber (MWPC) \cite{NID_2_Martoff2000355}. These gas mixtures were operated at low pressure, 40 Torr, and drift velocities of the order of cm/ms were found, as expected for an ionic speed. The diffusion coefficient reported in this paper were extremely low, below 100 $\mu$m/$\sqrt{\text{cm}}$ for drift fields above 235 V/cm. This demonstrated a reduced diffusion with respect to other electron drift gas mixtures. Indeed, a He:CF$_4$ 60/40 cold gas mixture employed by CYGNO at a quite high 1 kV/cm, for example, has diffusion coefficient of $\sim$ 100 $\mu$m/$\sqrt{\text{cm}}$ (see Chapter \ref{chap3}), and Ar:CH$_4$ mixture reaches 600 $\mu$m/$\sqrt{\text{cm}}$ at atmospheric pressure and similar drift fields\cite{BIAGI1989716,PDG}. Subsequent studies on the negative ion properties of CS$_2$ in combination with various gases, such as He, CF$_4$ and Ar at pressures of 40 Torr with MWPC, confirmed the ionic nature of the charge carriers with reduced diffusion \cite{Ohnuki:2000ex,ALNER2004644,Martoff_2005,NID_3_Snowden}. The NID operation with CS$_2$ was also employed in the DRIFT experiment permitting to achieve zero background operation thanks to the fiducialisation, as discussed in Section \ref{subsec:dirdet}. The NID operation was also demonstrated at close to atmospheric pressure in mixture with helium in MWPCs \cite{Martoff_2005,Dion}, and very recently at 1030 mbar with a gas mixture of Ar:iC$_4$H$_{10}$:CS$_2$ with a GridPix charge readout \cite{LIGTENBERG2021165706}. Electron attachment to CS$_2$ was also proved to produce minority carriers, which helped DRIFT experiment achieving a full 3D fiducialisation \cite{mincarry}. Some studies were performed utilising nitromethane (CH$_3$NO$_2$) in combination with CO$_2$ as a NID carrier at low pressure ($\sim$ 70 Torr) \cite{Dion}.\\

The \SF gas is a well known electronegative gas that is mostly used in gaseous detector for its electronegativity to capture electrons in order to contain large discharges or streamer avalanches, like in Resistive Plate Chambers (RPCs) \cite{Knoll}. Thanks to its electronegativity and large fluorine content, it is an interesting candidate for the NID operation in the direct DM search context. NID operation with \SF was only recently demonstrated in 2016 \cite{Phan2016TheNP}. Pure \SF at pressures between 20 and 40 Torr was used in a TPC with 400 $\mu$m thick GEM with charge readout and negative ion operation was successfully obtained, also demonstrating the feasibility of fiducialisation with minority carrier transport. This achievement is extremely important as \SF is inert and non-toxic, while the CS$_2$ is highly toxic. Studies of the pure \SF at 20 Torr were performed by the NEWAGE collaboration with GEMs and charge readout demonstrated fiducialisation with minority carriers with a  130 $\mu$m resolution \cite{Ikeda:2020pex} for alpha particles, as shown in Figure \ref{fig:fiduc}. The NID operation with \SF was also proved with triple thin GEMs and TimePix2 charge readout \cite{bib:nitec}. In these studies, when used pure, it was operated at 75, 100 and  150 Torr and the reduced mobilities were measured independent of the pressure. \SF was also utilised in a mixture of Ar:CO$_2$:\SF (192:85:93) Torr and He:CF$_4$:\SF (360/240/10) Torr. Despite the small amount of dopant,  NID operation was demonstrated and the mobility of the negative ion carriers was observed larger than in the case of pure \SF. 
\subsubsection{Application to CYGNO}
In the context of the CYGNO experiment, the reduction of the diffusion is aimed at maximising the performances in the determination of the directional parameters. The NID is envisioned as a possible way to achieve it (see Section \ref{sec:future:CYGNO}). As the charge gain and optical properties of He:CF$_4$ are fundamental for the correct operation of the detector, only a very small percentage of electronegative gas is added, with the goal of introducing the NID operation without modifying the other gas characteristics. Therefore, a concentration of 1.6\% of \SF gas is introduced, conform to the measurements performed in \cite{bib:nitec} which demonstrated a NID operation with He:CF$_4$. In the following Section, the principal characteristics of the \SF gas and its effect on the diffusion are described.
\section{SF$_6$ properties}
\label{sec:sf6chem}
When an ionising particle crosses the sensitive volume of a TPC, electron and ions pairs are generated due to the energy deposited. In presence of even small quantity of SF$_6$ gas, these electrons get captured very rapidly, within a micron from the generation position at low pressure (below 100 Torr) and even travelling less distance at higher pressures \cite{SF6xsec,SF6cose1,SF6cose2,SF6cose3,SF6cose4,SF6cose5,SF6cose6,SF6cose7,SF6cose8,SF6cose9}. The capture process generates an excited state of the SF$_6$ negative ion (SF$_6^{-*}$) which stabilises by radiation or collision. The principal phenomena occurring to a SF$_6^{-*}$ are listed below:
\begin{equation}
\label{eq:autodetachment}
\text{SF}_6^{-*} \rightarrow \text{SF}_6 + e^-  \hspace{1.5 cm} \text{(auto-detachment)}
\end{equation}
\begin{equation}
\label{eq:stabil_coll}
\text{SF}_6^{-*} + \text{SF}_6 \rightarrow \text{SF}_6^- + \text{SF}_6  \hspace{1.5 cm} \text{(collisional stabilisation)}
\end{equation}
\begin{equation}
\label{eq:autodisso}
\text{SF}_6^{-*} \rightarrow \text{SF}_5^- + F  \hspace{1.5 cm} \text{(auto-dissociation)}
\end{equation}
The auto-detachment, process \ref{eq:autodetachment}, results in the release of the electron and is counterproductive with respect to the desire of drifting negative ion in the TPC. The mean time of auto-detachment was measured in different experimental environments which returned controversial results with the mean time ranging from tens of $\mu$s up to ms \cite{SF6autodet,SF6autodet2,HARLAND197129,Christophorou1971AtomicAM,CHRISTOPHOROU197855,CHRISTOPHOROU1984477,KLOTS197661,R_W_Odom_1975,FOSTER1975479,SF6xsec}. On the other hand, the average time between collisions of molecules in a gas during drift is in the scale of ns. Thus, the stabilisation by collision, process \ref{eq:stabil_coll}, is much more likely to happen, independently from which of the measured time of auto-detachment is considered.
\begin{figure*}[t]
	\centering
	\includegraphics[width=0.4\linewidth]{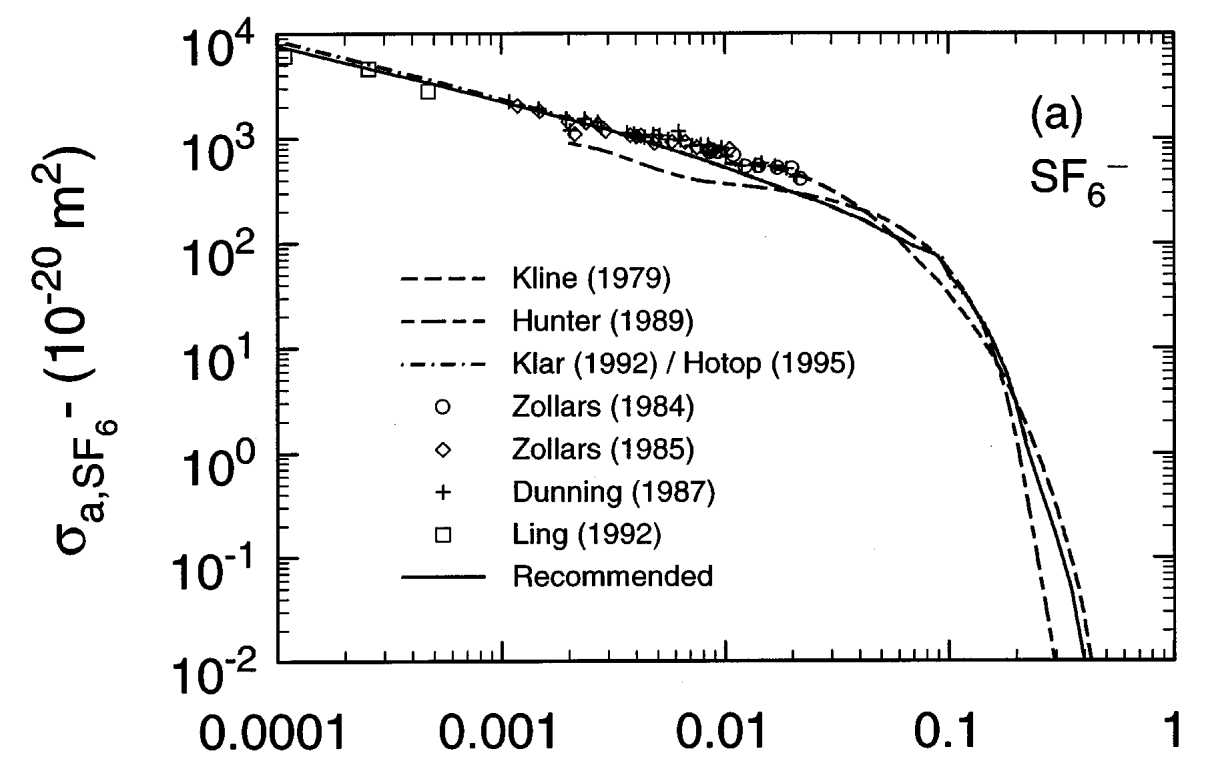}
	\includegraphics[width=0.4\linewidth]{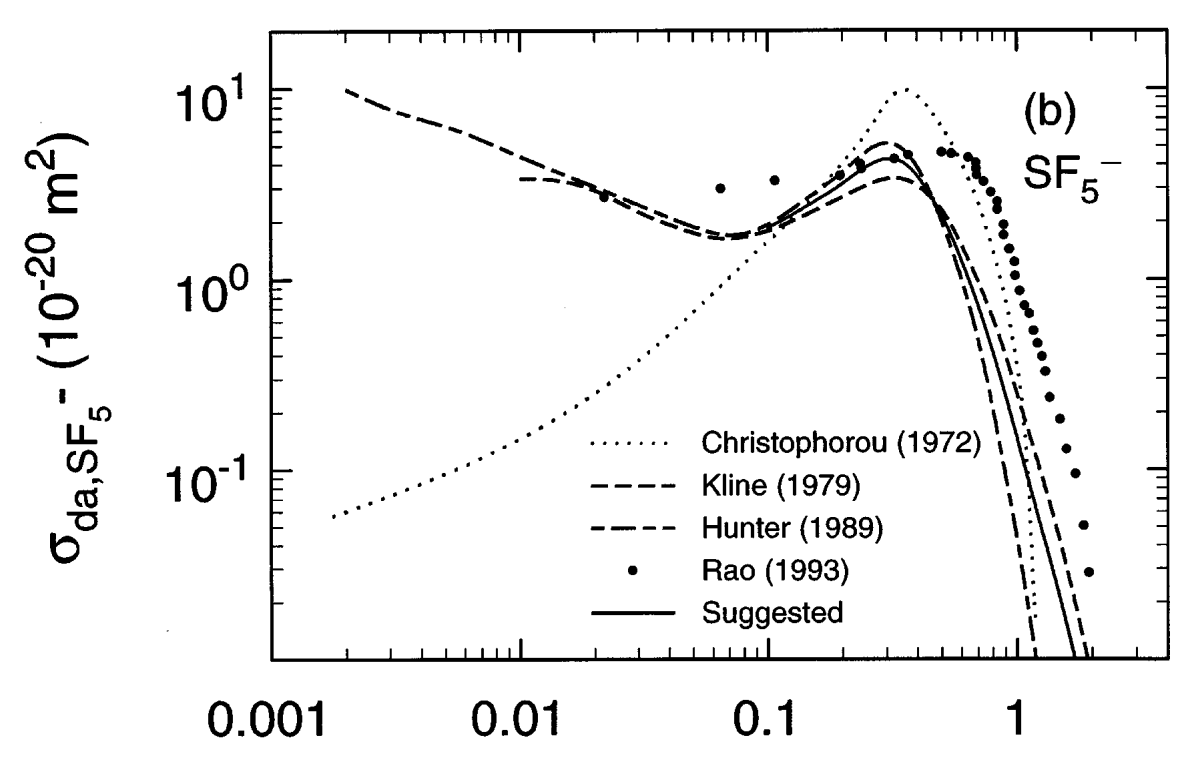}
	\caption{Cross section of the production of SF$_6^-$ (left) and SF$_5^-$ (right) after attachment as a function of the electron energy in eV. Figures taken from \cite{SF6xsec}}
	\label{fig:crosssecsf6}
\end{figure*}
The process \ref{eq:autodisso} also generates a negative ion species, but with a different mass. This is very useful as the SF$_5^-$ can also be utilised as a negative ion carrier. Figure \ref{fig:crosssecsf6} shows the cross section of the collisional stabilisation (on the left) and of the auto-dissociation (on the right) as a function of the electron energy in eV before the attachment \cite{SF6xsec}. It can be deduced that, especially with low energy collisions, the SF$_6^-$ species is the most abundant. With electric drift fields of the order of hundreds of V/cm, the energy acquired by the electrons before being absorbed is expected to be below tenths of eV. Nevertheless, a non negligible amount of SF$_5^-$ can be produced when the energy of the electron rises significantly, as in the case of higher electric drift fields. As the production of SF$_5^-$ is suppressed with respect to the SF$_6^-$ at the typical drift fields employed in TPC detectors, in the context of NID operation, they are known as minority carriers. During the attachment process and the motion of these ions, there are other phenomena which can complicate the transport of the charge, such as the double auto-detachment with the production of SF$_4^-$ or formation of clusters of SF$_6$ molecules that surround the negative species \cite{Patterson1}. Nevertheless, these have a much lower occurrence than the above mentioned and represent second order effects on the signal formation.\\

After the creation of negative ions, these objects thousand times heavier than electrons drift towards the amplification stage under the pull of the electric field of the TPC. When the anion species reaches the amplification stage, it is necessary to extract the electrons from the anion in order to start the amplification processes, as the large mass of the ion prevents the electric field to accelerate it enough to generate further ionisation. The detachment processes were studied in \cite{sf6detach,SF6xsec} and it was found that the one with the largest cross section to remove the extra electron from a stabilised SF$_6^-$ is by collision with other species. The collisional detachment of Equation \ref{eq:coldetsf6} is measured to have a very high energy threshold of 90 eV in the centre of mass of the collision, requiring very strong collisions. 
\begin{equation}
\label{eq:coldetsf6}
\text{SF}_6^{-} + \text{SF}_6 \rightarrow \text{SF}_6 + \text{SF}_6 + e 
\end{equation}
However, the collision with SF$_6$ can generate different negative ion species whose interaction can also produce electron detachment. Some of these processes are listed below:
\begin{equation}
\label{eq:creasf5}
\text{SF}_6^{-} + \text{SF}_6 \rightarrow \text{SF}_6 + \text{SF}_5^{-} +  \text{F}
\end{equation}
\begin{equation}
\label{eq:coldetsf5}
\text{SF}_5^{-} + \text{SF}_6 \rightarrow \text{SF}_5 + \text{SF}_6 + e 
\end{equation}
\begin{equation}
\label{eq:creaf}
\text{SF}_6^{-} + \text{SF}_6 \rightarrow \text{F}^- + \text{SF}_5 + \text{SF}_6 
\end{equation}
\begin{equation}
\label{eq:coldetf}
\text{F}^{-} + \text{SF}_6 \rightarrow \text{F} + \text{SF}_6 +  e 
\end{equation}
The cross section for the collisional production of these other negative ion species was found be much below ($\mathcal{O}$(10) eV in the centre of mass of the collision). Particularly interesting is the process of Equation \ref{eq:coldetf} for which the detachment of the electron from fluorine negative ion has a quite small threshold of about 8 eV in the centre of mass. Hence, the simplest mechanism to free the electrons from the SF$_6^-$ is considered via collisional detachment via the production of fluorine negative ion. The detachment process with the smallest threshold require extremely intense electric fields, above 375 Td \cite{SF6xsec}. With very large voltage potential across the GEMs, the intense electric field generated in the holes is strong enough to induce the detachment processes. This way, gains of the order of 10$^{3-4}$ can be achieved also with NID operation \cite{NEWAGE1,Phan2016TheNP}. The efficiency of the extraction of the electrons in the presence of strong electric fields is under study in the community \cite{NEWAGE1}.
\section{NID at atmospheric pressure}
\label{sec:nidatm}
The detector employed for the NID operation studies in a CYGNO-like detector is the MANGO prototype, described in Section \ref{sec:mango}. The drift gap was set to 5 cm, the maximum which can fit the gas-tight acrylic internal vessel, by employing a 5 cm field cage composed of 1 mm silver wires encased in 0.5 cm thick polycarbonate field rings with 7.4 cm internal diameter at 1 cm pitch, enclosed on one side by a cathode made by a thin copper layer deposited on a PCB and on the other by the GEMs amplification stage. The wires are connected by 1 G$\Omega$ resistors to partition the HV provided by an external CAEN N1570\footnote{\url{https://www.caen.it/products/n1570/}} power supply in order to maintain a uniform drift field. The detector is operated at the atmospheric pressure of LNGS (900 $\pm$ 7) mbar ((684 $\pm$ 5) Torr) with the He:CF$_4$ 60/40 gas mixture (ED) and the He:CF$_4$:SF$_6$ 59/39.4/1.6 NID mixture, following the results of \cite{bib:nitec}, in continuous flux mode. An  $^{241}$Am source producing 5.485 MeV alpha particles is placed in between the field cage rings to generate tracks inside the detector active volume. The PMT and the output of the charge sensitive preamplifier (CAEN A422A\footnote{\url{https://www.caen.it/products/a422a/}} with 300 $\mu$s decay time) connected to the bottom electrode of GEM3 are acquired through a highly performing Teledyne LeCroy oscilloscope\footnote{\url{https://cdn.teledynelecroy.com/files/pdf/hdo6000b-datasheet.pdf}}. The Orca Fusion images are collected by the custom Hamamatsu software Hokawo 3.0 in free running mode and with 0.5 s exposure. The two sets of data are analysed independently, the
first to provide an estimate of the negative ion gas mixture drift velocity and mobility, and the second to compare the light yields between ED and NID. The results are reported respectively in
Section \ref{subsec:driftmobatm} and Section \ref{subsec:gainatm}.
\subsection{Drift velocity and mobility measurements}
\label{subsec:driftmobatm}
\begin{figure}[!t] 
	\centering
	\includegraphics[width=0.95\linewidth]{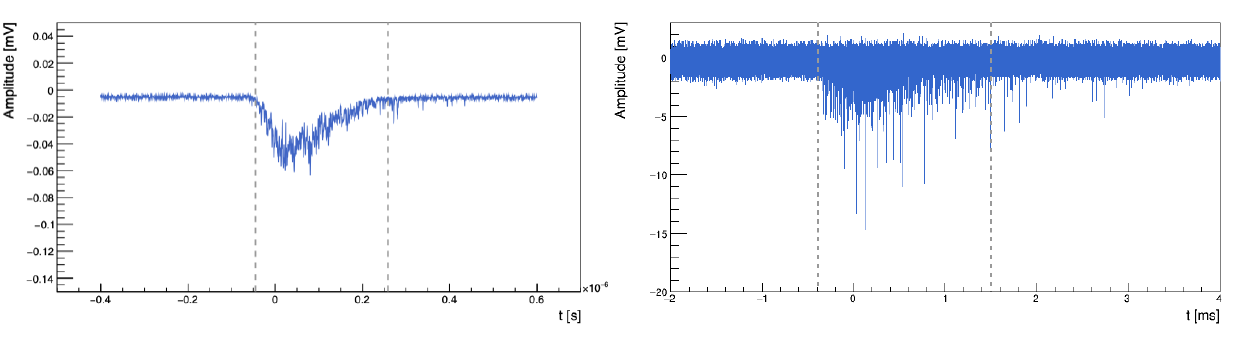}
	\caption{Example of PMT waveforms representing tilted alpha tracks obtain with the ED (left) and NID (right) mixtures. The vertical dashed lines on the left panel represent beginning and end of the PMT signal, as estimanted with the algorithm described in the text. The difference in time scale between the two waveforms remarks the different regime of drift velocity of the two gas mixtures.}
	\label{fig:pmt_wf}
\end{figure}
To measure drift velocity and mobility of the gas mixtures of interest, the PMT signal induced by alpha particles was studied placing the $^{241}$Am source at 2.5 cm distance from the GEMs. The drift fields tested are 300 V/cm, 400 V/cm, 550 V/cm and 700 V/cm.  The GEMs are operated at 310 V each in the ED case and at 550/545/540 V for GEM1/GEM2/GEM3 for the NID case.\\
In order to measure the drift velocity, since in the MANGO setup it not possible to obtain any information on the instant in which an alpha particle is ionising the gas, the time extension $\Delta T$ of the PMT waveforms is exploited. Indeed, $\Delta T$ is the product of the drift velocity of the charges with the extent of the alpha track in the direction of the drift field, $\Delta z$. By measuring $\Delta T$ with ED gas mixture, it is possible to calculate the $\Delta z$ travelled because the drift velocity is known (see Section \ref{sec:conc_goalCYGNO}). Without moving the source and evaluating the $\Delta T$  He:CF$_4$:SF$_6$ 59/39.4/1.6 mixture, the  NID drift velocity can be evaluated.\\

Figure \ref{fig:pmt_wf} shows on the left an example of a tilted alpha tracks PMT waveform for ED, where the dashed lines represent the start and end of the signal, as defined in the following discussion. The ED signals are acquired with a sampling of 400 ps/pt with a negative edge threshold of -40 mV on the PMT waveform amplitude. The signal is a continuous electric waveform with a typical time extent of hundreds of ns. When the track is not parallel to the GEM plane a side of the waveform is more intense, displaying the Bragg peak of the alpha particle.\\
Figure \ref{fig:pmt_wf} shows on the right an example of a tilted alpha tracks waveforms for NID. To the writer's knowledge, this is the first time that a NID signal is observed by means of a PMT.  Differently from the ED one, the fast response of the PMT combines with the slow drift velocity of NID anions to produce a much sparse signal. These NID signals consist in a series of thousands of small peaks, with amplitudes between 5 and 30 mV and time lengths of a handful of nanoseconds, extending over several ms (depending on the drift field applied) spaced each other hundreds of ns or even $\mu$s. Each of these small peaks is believed to represent the arrival of one or few primary ionisation clusters at the amplification plane, whose time distance due to the slow anion drift velocity allows to identify individually. While the study of this topic goes beyond the scope of this analysis, it is important to notice how this feature could enable the cluster counting, an analysis technique which can offer superior energy resolution and particle identification performances\cite{FISCHLE1991202}, especially in He-based detectors \cite{CATALDI1997458,Chiarello_2017}. Since each of the primary cluster signals closely resemble the one caused by single photoelectron background, a simple threshold-based trigger applied on the PMT would result ineffective to acquire NID waveforms. Thus, the trigger conditions chosen and the analysis procedures vary significantly between the ED and NID dataset. For this reason, NID events are recorded by employing a 250 mV positive edge threshold to the GEM preamplifier signal output which, in case of an alpha track, is generated long and continuous thanks to the long decay time of the preamplifier, permitting to simply trigger on it.
\subsubsection{PMT waveform analysis}
\label{subsubsec:omtanal}
The time duration of the PMT ED waveform is estimated by defining the start (the end) of the signal where the absolute value of the measured amplitude goes above (below) 3 times the waveform noise RMS, where the latter is evaluated from the first 100 bins of the waveform which do not contain any signal. From the average time extension of the ED PMT waveform $\Delta T$ and the simulated drift velocity $v_{drift}$ for He:CF$_4$ 60/40 at 900 mbar (see Section \ref{sec:conc_goalCYGNO}), the $\Delta Z$ extent covered by the alpha tracks in the drift direction in this setup is extracted from the formula:
\begin{equation}
\Delta Z = v_{drift} \times \Delta T
\end{equation}
which is found to be (0.7 $\pm$ 0.2) cm.\\

The analysis of the NID waveforms results in a significantly more complicated task due to the  difficulty in defining the beginning and the end of the signal, since each peak resembles noise and their spacing is of the order of $\mu$s. For this reason, a preliminary original algorithm was developed to analyse PMT waveform signals, which is planned to be improved for future and further analyses. The time duration of the single peaks composing the NID signals requires a very high sampling rate of GS/s from the oscilloscope in order to correctly detect and sample them. However, with this sampling rate, and the irregular and discrete nature of the NID signals, the majority of the samples saved by the oscilloscope contains electronic noise. The amplitude RMS of the electronic noise is evaluated on the first 500 bins of the waveform, which are expected to not contain any signal. From the original waveforms, only peaks above 6 times the RMS, equivalent to $\sim$ 2.5 mV, are retained for the following analyses. The peaks resulting from this selection are used to build a rebinned waveform, with a total of 150 bins over a 10 ms time extension. On this rebinned histogram, the beginning (end) of the signal are defined when two consecutive bins are above (below) 10 mV. The systematic uncertainties associated to this analysis strategy are evaluated by varying the number of bins of the rebinned histogram from 150 to 250 and the threshold on the two consecutive bins from 5 mV to 15 mV. 
\subsubsection{Results}
\label{subsubsec:result}
\begin{figure}[!t] 
	\centering
	\includegraphics[width=0.49\linewidth]{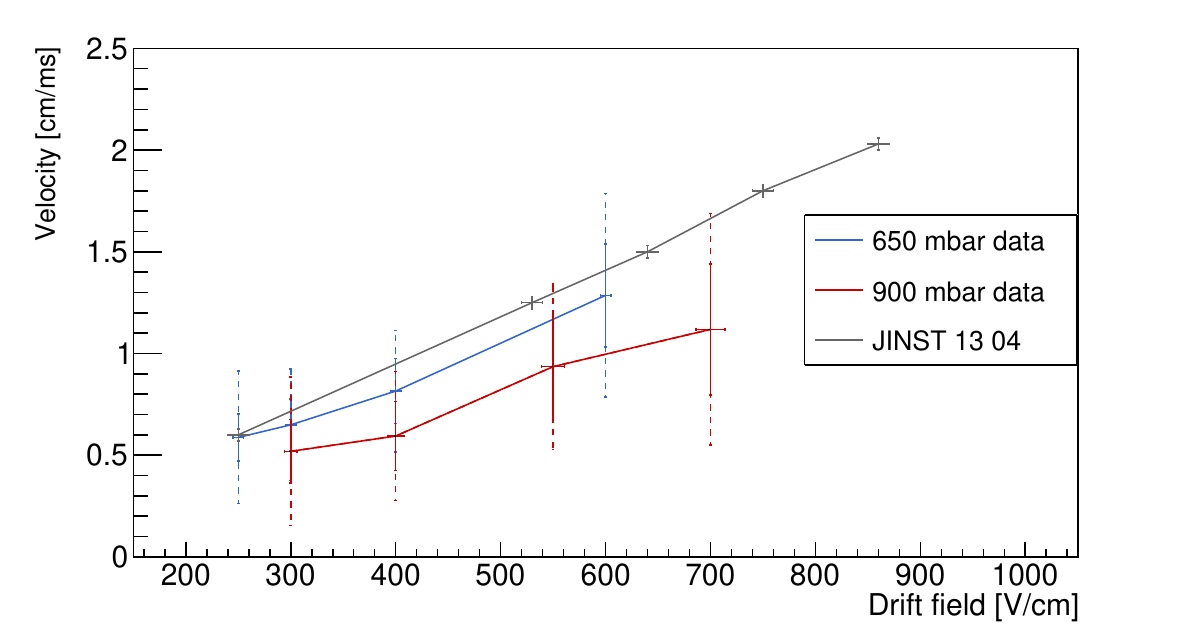}
	\includegraphics[width=0.49\linewidth]{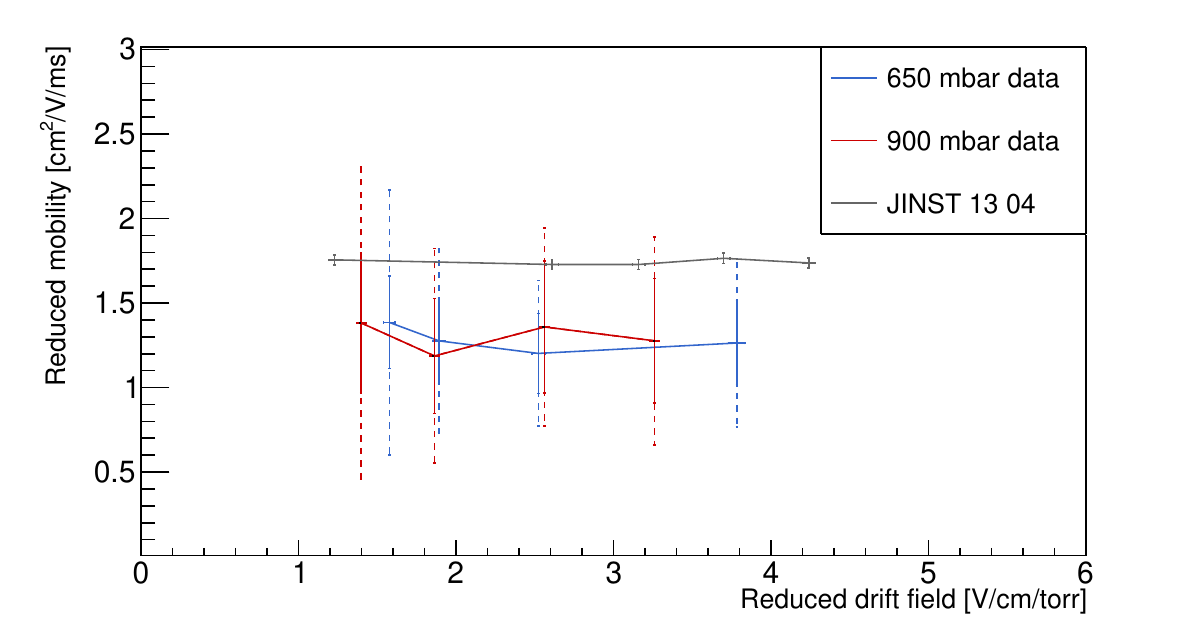}
	\caption{On the left, NID drift velocity as a function of the drift field. Both data sets at atmospheric pressure and 650 mbar (see Section \ref{subsec:driftmobkeg}) and the results extracted from \cite{bib:nitec} (JINST 13 04 in the legend) are shown in red, blue and grey respectively. On the right, NID reduced mobility as a function of the reduced drift field.Both data sets at atmospheric pressure and 650 mbar (see Section \ref{subsec:driftmobkeg}) and the results extracted from \cite{bib:nitec} (JINST 13 04 in the legend) are shown in red, blue and grey respectively. In both panels, the solid line error bars represent the statistical errors, while the dashed the systematics evaluated as illustrated in the text.}
	\label{fig:mob_meas}
\end{figure}
From the average measured time extension of NID signal, given the known $\Delta Z$ extent of the alpha setup evaluated from the ED data, the NID drift velocity as a function of the applied drift field can be extracted. These are shown for the 4 applied drift fields in left panel of Figure \ref{fig:mob_meas}. As the systematic uncertainties, evaluated as discussed above, may not be related to the standard error theory, the full analysis to obtain the drift velocity is repeated in the different conditions of binning and threshold described to estimate the systematic uncertainties. From the distribution of the drift velocities obtained, a confidence interval of 68\% is used to estimate the systematics. At 250 V/cm, when the tracks are more spread in time, they reach 60\%, while drops around 40\% at higher fields. These large values are somewhat expected as the analysis procedure is preliminary. Nevertheless, the $\mathcal{O}$(1) cm/ms measured drift velocities demonstrate unequivocally that the He:CF$_4$:SF$_6$ 59/39.4/1.6  gas mixture at 900 mbar induces NID operations. The results from an identical analysis performed with data taken at 650 mbar are also shown in Figure \ref{fig:mob_meas} and point at the same conclusions. These data will be described in Section \ref{subsec:driftmobkeg}.\\
The reduced mobility evaluated from these velocities is compared with the results published in \cite{bib:nitec} on a He:CF$_4$:SF$_6$ 360:240:10 Torr gas mixture measured with pixel charge readout in right panel of Figure \ref{fig:mob_meas}, showing the consistency of the three measurements, also considering the large systematic uncertainties. The larger uncertainties of the measurement presented in this thesis with respect to the results of \cite{bib:nitec} can be explained with the methodology utilised. In fact, the latter exploited a beam of particles whose operation provided the absolute time in which the particles ionise the gas, and the velocity was estimated from the time delay between that instant and the moment the signal was detected at the amplification stage. This has a much better precision than relying on the extent covered by the alpha tracks in the drift direction, but it was not possible to use this technique with MANGO.

\begin{figure}[!t] 
	\centering
	\includegraphics[width=0.8\linewidth]{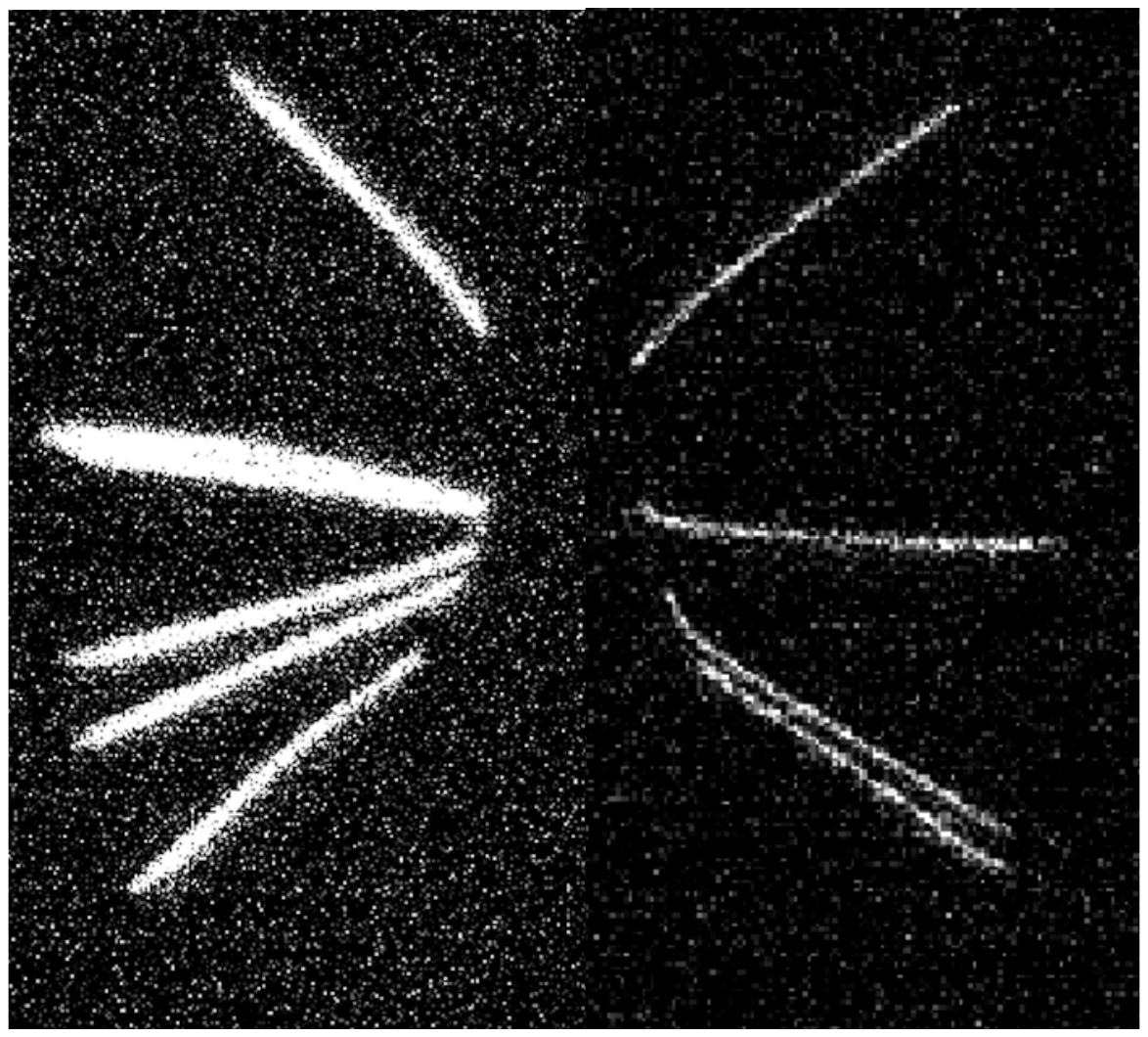}
	\caption{Example of a 0.5 s sCMOS image acquired with MANGO exposed to the $^{241}$Am source placed at 2.5 cm from the GEMs plane and operated with He:CF$_4$ 60/40 (left) and He:CF$_4$:SF$_6$ 59/39.4/1.6 (right) at atmospheric LNGS pressure (900 mbar).}
	\label{fig:alphas}
\end{figure}
\subsection{sCMOS images analysis}
\label{subsec:gainatm}
An example of a 0.5 s sCMOS image acquired with MANGO exposed to the $^{241}$Am source placed at 2.5 cm from the GEM plane and operated with He:CF$_4$ 60/40 (left) and He:CF$_4$:SF$_6$ 59/39.4/1.6 (right) at atmospheric LNGS pressure (900 mbar) is shown in Figure \ref{fig:alphas}, where the difference in diffusion suffered by the alpha tracks while drifting in the two gas mixtures appears evident despite the similar light yield conditions (as will be detailed in the following).

\begin{figure}[!t] 
	\centering
	\includegraphics[width=0.47\linewidth]{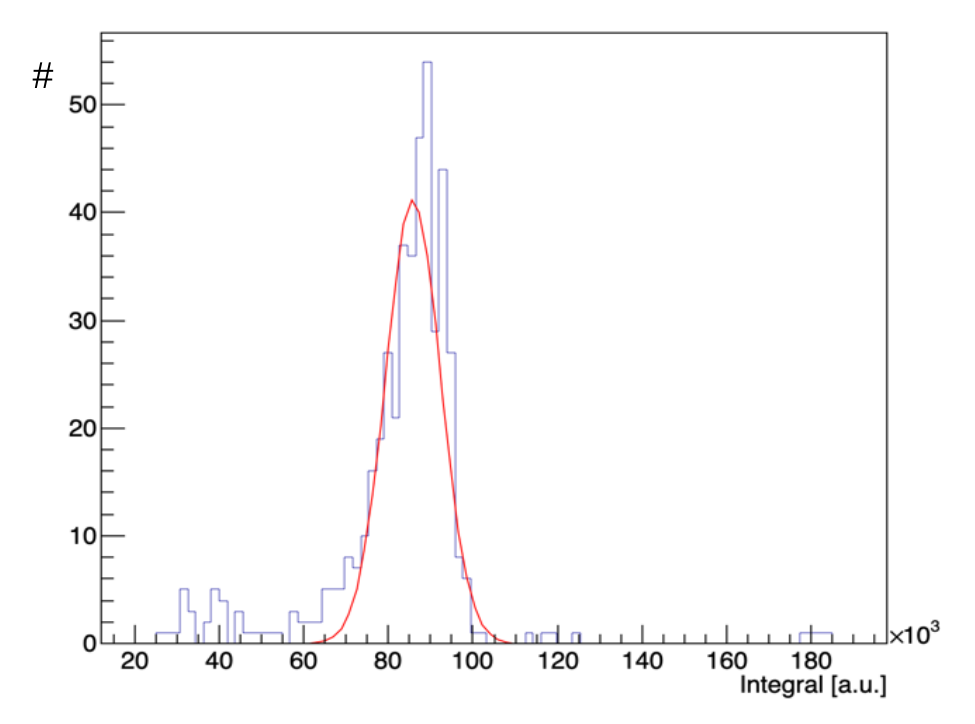}
	\includegraphics[width=0.45\linewidth]{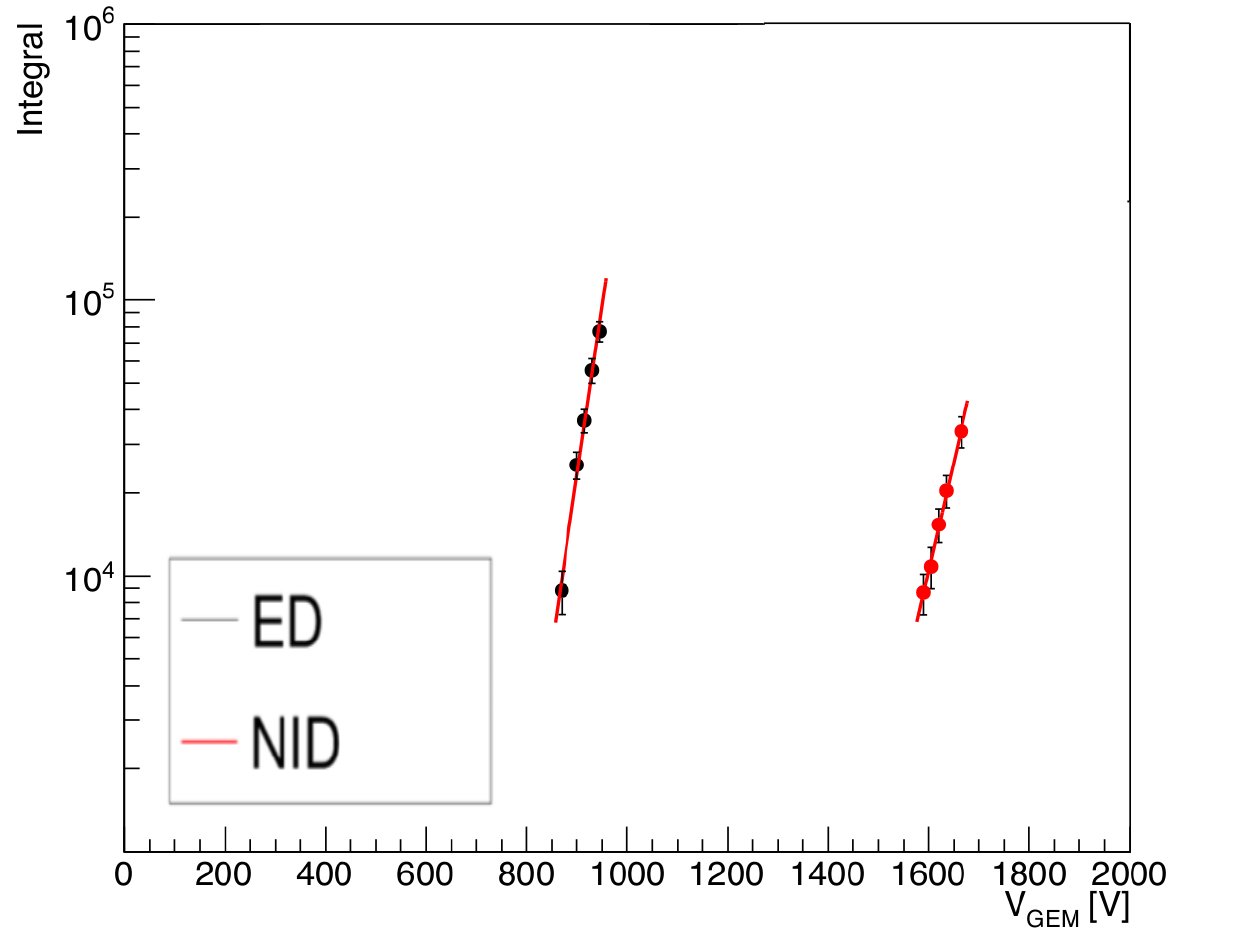}
	\caption{On the left, the distribution the sum of the content of the pixels belonging to each cluster (\textit{integral}) representing an alpha track which survived the selection cuts described in the text. A Gaussian fit is superimposed. On the right, the average light integral as a function of the total voltage applied on the GEMs V$_{GEM}$ for ED in red and NID in black at 2.5 cm drift distance, with 1000 V/cm and 400 V/cm drift field applied respectively.}
	\label{fig:gain_scan}
\end{figure}
To evaluate the gain achievable with the NID gas mixture a scan in voltages was performed  at the shortest drift distance allowed by the setup, i.e. 2.5 cm, in order to minimise the contribution of attachment or any other effect resulting in charge loss during drift. The sCMOS images are analysed with the reconstruction code described in Chapter \ref{chap4}. Alpha tracks are then selected by requiring the track length to be larger than 1.2 cm and the ratio of the track width over length to be less than 0.3, to reject noise and mis-reconstructed events. For each of the found clusters satisfying the selection requirements described above, the light yield, proportional to the energy deposited, is evaluated by fitting with a Gaussian function the distribution the sum of the content of the pixels belonging to each cluster (\textit{integral}). An example of the just mentioned distribution for the ED gas mixture with the voltage across the GEMs of 315 V is displayed in Figure \ref{fig:gain_scan} on the left panel.
\begin{table}[!t]
	\centering
	\vspace{0.3 cm}
	\begin{tabular}{|c|c|c|}
		\hline
		Gas Mixture	& a & b [1/V]\\ 
		\hline \hline
		ED & -4 $\pm$ 1 & 0.020 $\pm$ 0.001\\
		NID & -20 $\pm$ 2 & 0.018 $\pm$ 0.002\\
		\hline
	\end{tabular}
	\vspace{0.3 cm}
	\caption{Results of the fit to the data in Figure \ref{fig:gain_scan} with the function $ e^{a+bx}$.}
	\label{tab:expo}
\end{table}
Figure \ref{fig:gain_scan} shows, on the right panel, the measured averaged integral as a function of the sum of the applied voltages on the three GEMs, V$_{GEM}$, for ED in black and NID in red, with respectively 1000 V/cm and 400 V/cm drift field. The voltages necessary for the NID are larger than the one for the ED as expected from the reasoning of Section \ref{sec:sf6chem}. Due to this larger voltage requirement, the power supply maximum voltage applied to the cathode does not allow to surpass 700 V/cm in the drift region. As no light output difference is found between 400 and 700 V/cm, the former is chosen not to stress the power supply. The two sets of data are fitted with an exponential function $e^{a+bV_{GEM}}$ and the fit results are displayed in Table \ref{tab:expo}. The fitted exponential slope result consistent between the ED and the NID gas mixtures. The measurement also demonstrates that it is possible to attain amplification with the NID gas mixture similar to what can be achieved with ED, even though at a much larger voltage on the GEM.\\
A rough estimation of the absolute gain attained in this configuration can be performed exploiting the analysis of Chapter \ref{chap5} in Section \ref{subsec:gain}. The light gain of the \fe as a function of the sum of the voltages across the GEMs of the He:CF$_4$ mixture in the triple GEM configuration was measured and was fitted with $\ln(G)=A'+B'V_{GEM}$ whose results can be found in Table \ref{tab:gain}. The fit on the data extended from 1200 V up to 1305 V. In the assumption that the same function is correctly representing the gain with the same parameters at voltages below 1200 V, the light yield of an \fe signal at 900 V can be estimated as $\sim$ 4 10$^2$, 25 times less than the one at 1200 V. The absolute gain at atmospheric pressure for the same gas mixture was measured 3.3 10$^5$ at 1200 V in \cite{bib:roby}. Therefore, the at 900 V it can be estimated to be $\sim$ 10$^4$. In the further assumption that the NID mixture produces similar amount of photons per secondary electron as the ED mixture, it can be concluded that a gain of roughly 10$^4$ was reached during the measurements. While it is possible that the amplification drops faster for lower voltages than what described by the fitting curve of the data presented in this work, it is also known to be higher at lower pressure \cite{Blum_rolandi}. Therefore, the gain achieved in this NID mixture can be approximately believed to be around 1 10$^4$. Gas gains in \SF NID gas mixture of $\sim$ 2-3 10$^{3}$ were also measured in other works \cite{Phan2016TheNP,Ikeda:2020pex}, though the operative pressures were more than 10 times lower. If further measurements confirm the present findings, this could be one of the largest gains ever achieved at high pressure for a NID \SF-based gas mixture.
\section{Modified MANGO setup for diffusion studies}
\label{sec:nidkeg}
\begin{figure}[!t]
	\centering
	\includegraphics[height=6.2 cm]{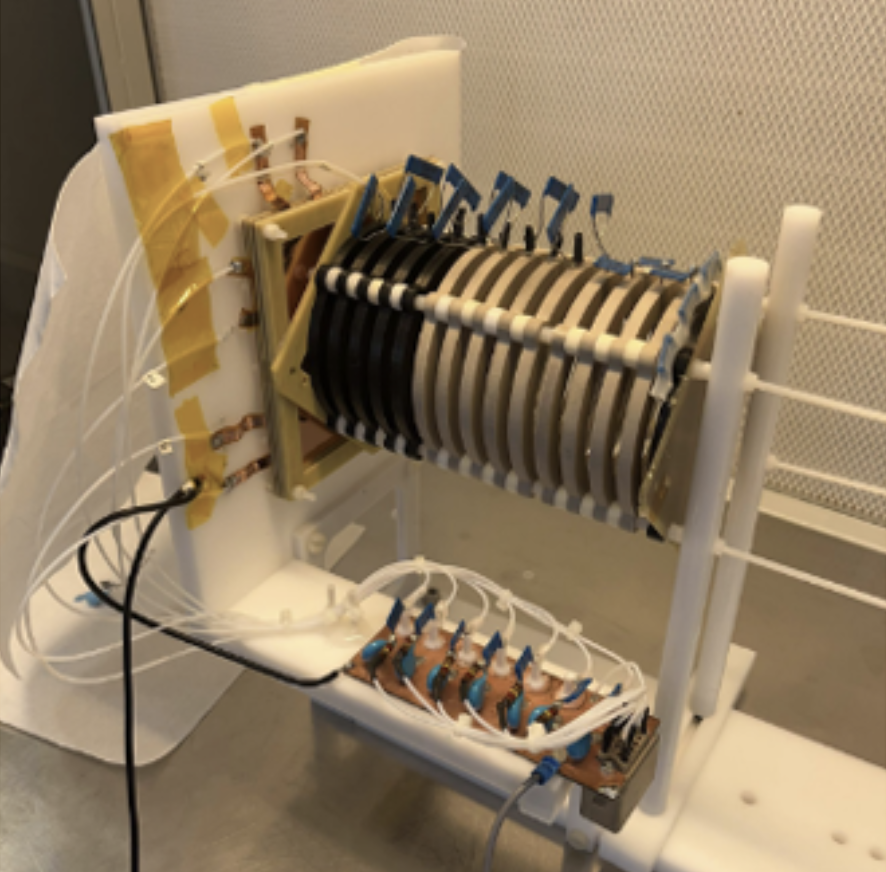}
	\includegraphics[height=6.2 cm]{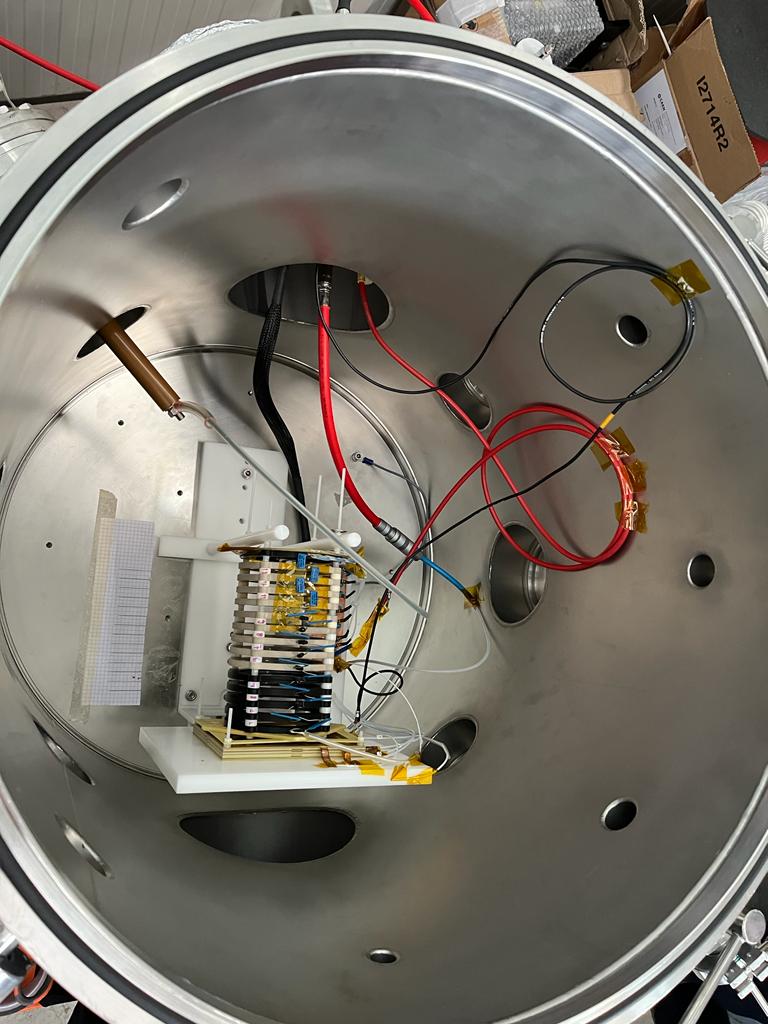}
	\caption{Pictures of the updated MANGO setup. On the left, a detail of the GEM stack equipped with a 15 cm long field cage and the structural support, while on the right, the same structure inserted in the 150 l stainless steel vacuum vessel.}
	\label{fig:mangosketch2}
\end{figure}

In order to carry out measurements of the diffusion during drift, the maximum 5 cm drift length available in the standard MANGO configuration results too small. Hence, the same amplification stage was adapted to another detector structure, whose picture is shown in Figure \ref{fig:mangosketch2} . In this configuration MANGO is renamed MANGOk (MANGO in a keg).
The amplification is always provided by a stack of triple thin 50 $\mu$m GEMs with 10 $\times$ 10 cm$^2$ active area, 2 mm spacing  and a transfer field of 2.5 kV/cm in between and powered  exactly as before. The TPC is equipped with a 15 cm long field cage, built exactly as the 5 cm long one employed for the measurements in the previous Section. The TPC structure is installed in a 150 l stainless steel vacuum vessel.

MANGOk is equipped with the same optical readout comprised of the identical photomultiplier tube and sCMOS camera. These photon detectors are installed outside the vacuum vessel and coupled to the MANGOk amplification plane through a quartz window with a 90\% transparency. The ORCA-Fusion is positioned at $(26.6 \pm 0.4)$ cm from the last GEM and focused on it. Within this scheme, the camera images an area of 14.1 $\times$ 14.1 cm$^2$, resulting in an effective pixel size of 61 $\times$ 61  $\mu$m$^2$. 

As the photo-sensors need to be positioned at a larger distance from the last GEM amplification plane, the solid angle coverage drops of a factor 3. In order to obtain a light yield at the sCMOS sensor high enough to be able to measure diffusion at large distances also for weak electric drift fields, the operative pressure is reduced to (650 $\pm$ 1) mbar ((494.0 $\pm$ 0.7 ) Torr) to increase the gain. The same ED and NID gas mixtures are employed for the measurement. After sealing, for each measurement the vessel is pumped down to less than 0.1 mbar with a dry scroll vacuum pump and then filled with the chosen gas mixture. The GEMs are operated at 310 V each in the ED case and at 535/530/525 V for GEM1/GEM2/GEM3 for the NID case. The operating GEM voltages are chosen in order to obtain a comparable light yield for the two mixtures also within this setup, as it will be shown in Section \ref{subsec:edcrosskeg}. 

The $^{241}$Am source was used akin to the measurements described in Section \ref{sec:nidatm}, but this time the source was further collimated to reduce the spread of outgoing direction of the alpha particles. The acquisition and trigger strategy is identical to the one illustrated in Section \ref{sec:nidatm}.
\subsection{Drift velocity and mobility measurements}
\label{subsec:driftmobkeg}
For the measurement of the drift velocity and reduced mobility, the source was placed at 3.5 cm distance from the GEMs and collimated it in order to produce in the gas volume tilted tracks, as parallel as possible to the drift direction, in order to maximise the extension of the tracks along this coordinate.  The drift fields applied are 250 V/cm, 300 V/cm, 400 V/cm and 600 V/cm. The analysis of the PMT waveforms is performed following the same procedure as in Section \ref{subsec:driftmobatm}. The average $\Delta Z$ found in this configuration is (1.5 $\pm$ 0.3) cm and the results of the drift velocity and reduced mobility can be found in Figure \ref{fig:mob_meas}, on the left and right respectively. The statistical uncertainties are slightly smaller with respect to the previous measurements thanks to the more consistent track length because of the collimation of the source. The results confirm all the conclusions derived in Section \ref{subsec:driftmobatm}.
\subsection{sCMOS images analysis for diffusion parameters estimation}
\label{subsec:diffkeg}
In order to evaluate the diffusion parameters of the gas mixtures employed, data are acquired at six different drift distances (2.5, 3.5, 4.5, 6.5, 9.5 and 12.5 cm from the GEMs) each with varying drift field (150 V/cm, 200 V/cm, 250 V/cm, 300 V/cm, 350 V/cm, 400 V/cm and 600 V/cm) for both the ED and NID gas mixtures at 650 mbar. The source is positioned in order to produce alpha tracks populating the plane as perpendicular as possible to the drift direction. 

\begin{figure}[!t] 
	\centering
	\includegraphics[width=0.9\linewidth]{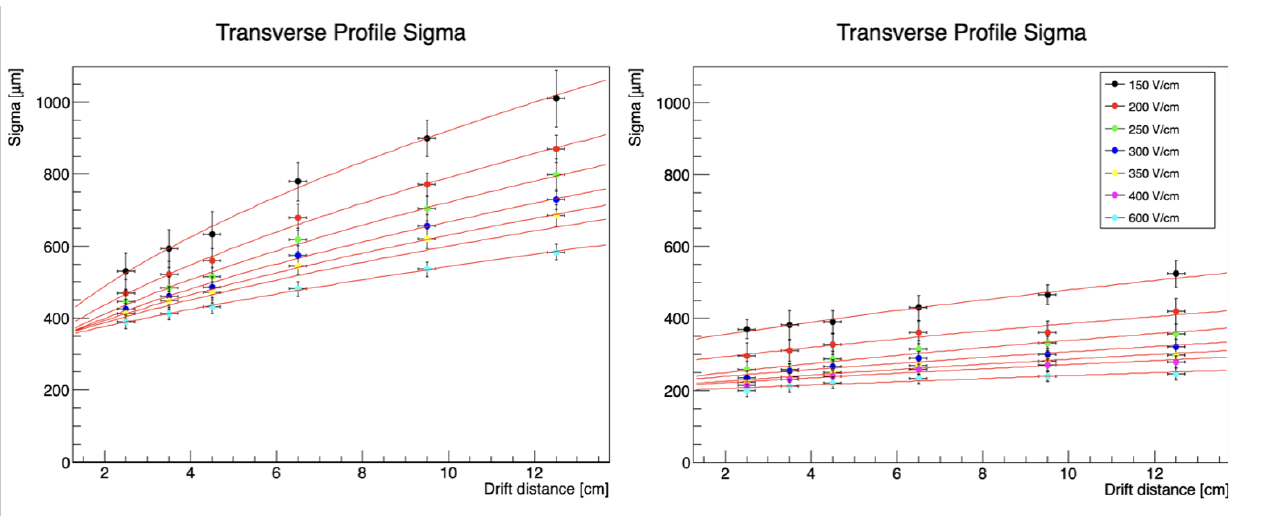}
	\caption{Estimation of the diffusion as a function of the drift distance for different applied drift fields. ED is shown on the left and NID on the right with superimposed the fit performed with Equation \ref{eq:diff}. The legend shows in colours the different drift fields applied for both plots.}
	\label{fig:diff}
\end{figure}

\begin{figure}[t] 
	\centering
	\includegraphics[width=0.9\linewidth]{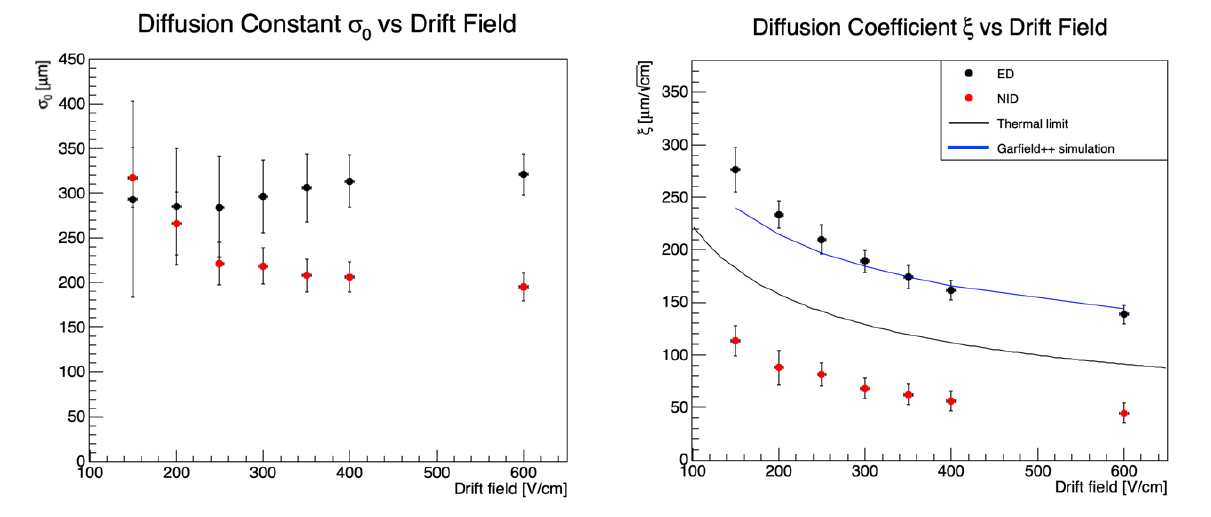}
	\caption{Fitted $\sigma_0$ (left)  and  $\xi$ (right) from Equation \ref{eq:diff} as a function of the applied drift field for ED (black) and NID (red). In right panel the expected thermal behaviour in black and the Garfield++ simulation of the ED gas mixture in blue are also shown.}
	\label{fig:sigma_vs_drift}
\end{figure}

The evaluation of the diffusion from the alpha particle tracks is computed following the analysis algorithm described in Section \ref{sec:diffalpha}. Figure \ref{fig:diff} displays the measured diffusion as a function of the drift distance and the applied drift field for ED on the left and NID on the right. The measured $\sigma_{meas}$ are fitted with
\begin{equation}
\label{eq:diff}
\sigma_{meas} = \sqrt{\sigma_0^2 + \xi^2 L}
\end{equation}
where $\sigma_0$ represents the constant diffusion suffered in the GEM amplification plane, the original spread of the alpha track and everything which does not depend on the drift distance, while $\xi$ the diffusion coefficient, and $L$ the drift distance from the GEM \cite{Blum_rolandi}. The results of the fit with Equation \ref{eq:diff} are reported in Table \ref{tab:fit}. The dependence of fitted values of $\sigma_0$ and $\xi$ on the applied drift field is shown respectively in left and right panel of Figure \ref{fig:sigma_vs_drift} for ED in black and NID in red. In addition, in the right panel of Figure  \ref{fig:sigma_vs_drift} the thermal expected behaviour is displayed in black and the diffusion for the ED mixture studied as simulated by the Garfield++\footnote{\url{https://gitlab.cern.ch/garfield/garfieldpp}} software in blue.
The latter nicely matches the measured diffusion of the ED mixture.
\begin{table}[!t]
	\centering
	\vspace{0.3 cm}
	\begin{tabular}{|c|c|c|c|c|}
		\hline
		Drift field [V/cm] 	& $\sigma_0^{ED}$ [$\mu$m] & $\xi^{ED}$ [$\mu$m/$\sqrt{\text{cm}}$] & $\sigma_0^{NID}$ [$\mu$m] & $\xi^{NID}$ [$\mu$m/$\sqrt{\text{cm}}$] \\  
		\hline \hline
		150 & 300 $\pm$ 100 & 280 $\pm$ 20 & 320 $\pm$ 30 & 110 $\pm$ 10 \\
		200 & 290 $\pm$ 60 &  230 $\pm$ 10 & 260 $\pm$ 30 & 90 $\pm$ 20 \\
		250 & 290 $\pm$ 60 & 210 $\pm$ 10 & 220 $\pm$ 20 & 81 $\pm$ 10 \\
		300 & 300 $\pm$ 40 & 190 $\pm$ 10 & 220 $\pm$ 20 & 68 $\pm$ 10 \\
		350 & 300 $\pm$ 40 & 170 $\pm$ 10 & 210 $\pm$ 20 & 60 $\pm$ 10 \\
		400 & 310 $\pm$ 30 & 160 $\pm$ 10 & 210 $\pm$ 20 & 56 $\pm$ 9 \\
		600 & 320 $\pm$ 20 & 140 $\pm$ 10 & 200 $\pm$ 20 & 45 $\pm$ 10 \\
		\hline
	\end{tabular}
	\vspace{0.3 cm}
	\caption{Table summarising the fit with Equation \ref{eq:diff} to the data shown in Figure \ref{fig:diff} as a function of the applied drift fields.}
	\label{tab:fit}
\end{table}
Both ED and NID diffusion coefficients $\xi$ decrease with the increase of the drift field as expected following the well known 1/$\sqrt{E}$ dependence \cite{Knoll,Blum_rolandi}. While the measured ED diffusion is, as expected from general arguments and from simulation, far from the thermal limit, the NID mixture is characterised by a very small diffusion coefficient $\xi$, $\le$ 50 $\mu$m/$\sqrt{cm}$ at 600 V/cm drift field. Another interesting feature is displayed in left panel of Figure  \ref{fig:sigma_vs_drift}, which shows that the constant diffusion term $\sigma_0$ due to the GEMs amplification plane is significantly reduced with NID operation compared to ED as well, of about a factor 1/3 from and average of 300 $\mu$m to 200 $\mu$m, and result independent (as expected) on the applied drift field. Overall, the NID operations reported in this paper demonstrate the possibility to reduce the diffusion during drift to 45 $\mu$m/$\sqrt{cm}$. This finding is compatible with the smallest diffusion coefficient ever measured in a gas detector, to the writer's knowledge, obtained in \cite{NID_2_Martoff2000355} with a 40 Torr CS$_2$-based gas mixture.
\subsection{Setup crosscheck}
\label{subsec:edcrosskeg}
\begin{figure}[!t] 
	\centering
	\includegraphics[width=0.95\linewidth]{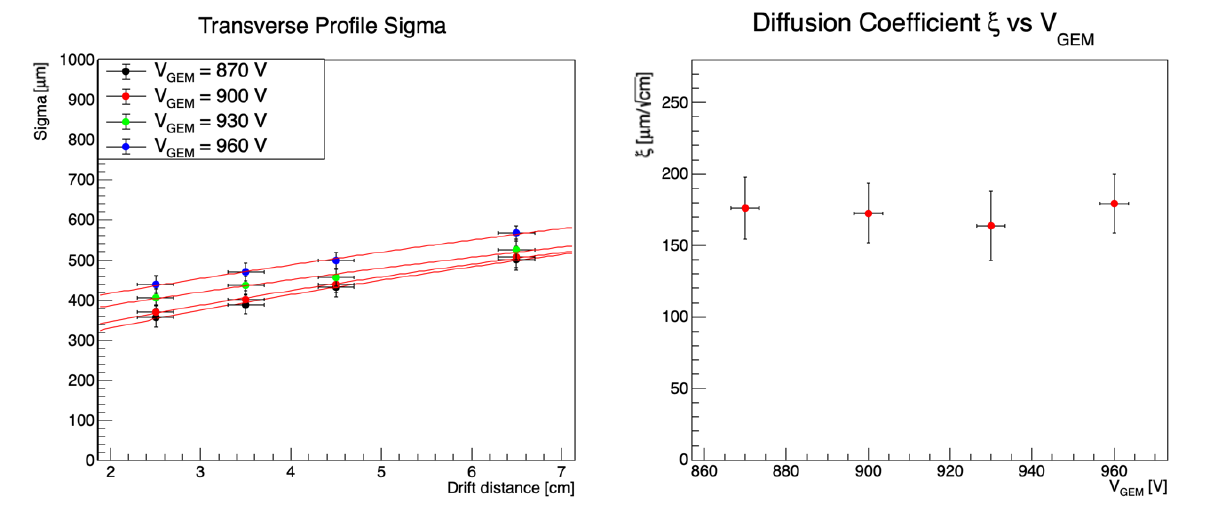}
	\caption{On the left, measured diffusion in ED as a function of the drift distance with 400 V/cm drift field and different voltages $V_{GEM}$ applied on the GEMs (see legend). On the right, fitted $\xi$ from Equation \ref{eq:diff} as a function of $V_{GEM}$ for the data in the left panel.}
	\label{fig:sigma_vs_vgem}
\end{figure}

In order to further validate the analysis methodology, the experimental apparatus, and the results presented, the diffusion for the ED mixture is studied as a function of the applied voltage on the GEMs. The goal of this cross-check is to demonstrate that the analysis approach is not affected by the total light (and charge) produced in the GEM amplification plane, even at higher gains than the chosen operating voltage for the measurements discussed so far. In order to show this, the drift field was fixed to 400 V/cm and the voltage applied on the GEM varied from 290 V to 320 V on each GEM (for an overall voltage spanning from 870 V to 960 V). Figure \ref{fig:sigma_vs_vgem} shows on the left the measured diffusion for the ED mixture at 400 V/cm as a function of the drift distance with 4 different voltages applied on the GEMs $V_{GEM}$, with superimposed a fit with Equation \ref{eq:diff}. Right panel of Figure \ref{fig:sigma_vs_vgem} displays the fitted diffusion coefficient $\xi$ as a function of V$_{GEM}$, clearly showing no dependence at all on the gas gain (hence light produced), further demonstrating the robustness of the previous measurements.\\

\begin{figure}[!t] 
	\centering
	\includegraphics[width=0.9\linewidth]{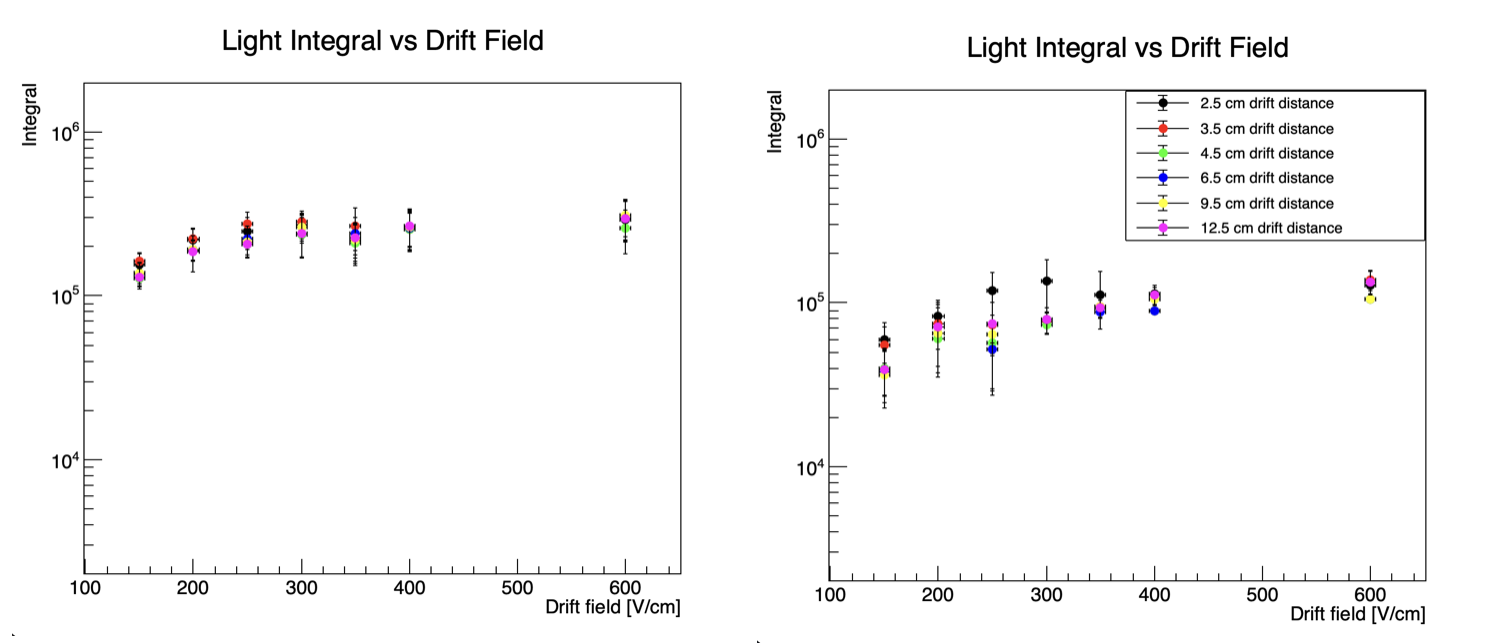}
	\caption{Light integral as a function of the applied drift, on the left for ED and on the right for NID gas mixture. The legend displays the drift distance of each measurement for both plots.}
	\label{fig:int_vs_field}
\end{figure}

Furthermore, the light yield of ED and NID was studied at the nominal V$_{GEM}$ operating voltages employed in the MANGOk setup and data taking (see beginning of Section \ref{subsec:diffkeg}) as a function of the applied drift field at various distances, from 2.5 cm up to 12.5 cm, to confirm that the diffusion measurements do not suffer from systematic associated to drift lengths and drift fields. In Figure \ref{fig:int_vs_field} this is shown on the left for ED and on the right for NID gas mixture. These measurements demonstrate that the ED and NID operating conditions results in comparable light yield and that it is overall independent of other parameters varied.
\section{Discussion}
\label{sec:discunid}
The results illustrated in this Chapter are the first demonstration of the feasibility of NID operation with optical readout at LNGS atmospheric pressure (900 mbar) and the first measurement of the transversal diffusion of NID for a He:CF$_4$:SF$_6$ 59/39.4/1.6 mixture at 650 mbar, representing a decisive breakthrough for directional DM searches and low diffusion TPC applications, not only with optical readout.
\\
While the diffusion parameters found for the ED gas mixture are perfectly consistent with expectation and gas simulations, the NID measurements fall well below the expected thermal limit. Although puzzling, it is not the first time this feature has been observed. In the Ar:CS$_2$:CH$_{4}$ and Xe:CS$_2$ mixtures studied in \cite{NID_2_Martoff2000355}, a transverse diffusion below the thermal limit is actually measured, contrary to what reported in the text by the authors. By using the values reported in Table 1 of \cite{NID_2_Martoff2000355} and the known drift distance of the measurement, effective temperatures between 90 K and 130 K can be in fact evaluated from the data acquired at 23.5 kV/m (with even lower temperature at higher drift fields). The authors of \cite{NID_2_Martoff2000355} recognise in fact that the trend of the diffusion with increasing drift fields appears to fall more rapidly than the $1/\sqrt{E}$ dependence expected from transport properties in gas beyond 23.5 kV/m. It must also be noted that the measurements reported in \cite{NID_2_Martoff2000355} suffer from large systematics due to the experimental technique employed, that results in uncertainties up to 80\%.\\
In \cite{Ohnuki:2000ex}, NID transverse diffusion in pure CS$_2$ at 40 Torr is compared to Ar:CS$_2$ mixtures with varying Ar ratios at the same pressure. The relative effective temperatures evaluated from data result mostly consistent with thermal behaviour, although hinting that the diffusion is decreasing with increasing Ar concentration, with a fitted T$_{eff}$ = (240 $\pm$ 50) K for Ar:CS$_2$ 75\%:25\%. The authors of \cite{Ohnuki:2000ex} argue that, having the CS$_2$ anion a larger mass than Ar, it might tend to preserve its direction of motion after a collision with the neutral gas, possibly violating one of the assumptions underlying Equation \ref{eq:limterm}.\\
The results presented in \cite{Dion} show indication of a reduction of the longitudinal diffusion of CS$_2$ negative ion species when employed together with large amount of helium as buffer gas, as if the He overall effect is to \emph{cool down} the CS$_2$ anions motion. These measurements can not be directly numerically compared to the ones presented in this work, being them related to the longitudinal diffusion, which is known from experiments \cite{LIGTENBERG2021165706,NID_3_Snowden} and calculation \cite{Iontransport1,Iontransport2,FERRARI19969,Dion} to be always larger than the transversal one due to the electric field anisotropy \cite{Blum_rolandi}. Nonetheless, the observed trend results similar to the transverse diffusion behaviour as a function of increasing Ar concentration of \cite{Ohnuki:2000ex}.
It is important to notice how only longitudinal diffusion measurements of SF$_6$ NID at 40 Torr exists \cite{Phan2016TheNP}, that result nicely consistent with thermal behaviour (even though the constant contribution to diffusion due to the amplification stage is not taken into account in this estimation). Nonetheless, for the reasons discussed above, this can not be directly compared to the work presented here, that employs SF$_6$ in a very small concentration, together with He as buffer gas in large ratio at (nearly) atmospheric pressure.

The theory of ions transport in gas \cite{Iontransport1,Iontransport2,FERRARI19969} results quite ineffective in predicting the expected transport properties of the gas mixture under investigation in this work. Indeed, when an ion swarm moves in a gas subject to an electric field, its transport properties can in principle be determined from scratch from the stationary and homogeneous version of the appropriate Boltzmann (or Fokker-Planck) equation. The problem with this approach is that the Boltzmann collision integral, except for very special and simple cases, cannot be exactly evaluated, and approximation and limits needs to be taken for each peculiar case. While several approximated theories exist, all rely on specific assumptions or limits which can not be considered directly applicable to the case presented.
In particular, the molecular nature of the SF$_6$ and CF$_4$ gases (that implies many processes in which energy can be efficiently dissipated), combined with extremely large difference in mass between the SF$_6$ negative ion and the helium (of the order of 36) clearly complicates the description of the motion of the negative ion in the collisions during the drift. The collaboration is working with the goal of finding a complete theoretical explanation of the experimental findings, but the possibility that a large disparity in mass can lead to diffusion even below the thermal limit is appealing to say the least. 
In conclusion, the measurement is considered solid and robust and, while it is not possible to directly cross-check the values with existing measurements, the trend of a below thermal diffusion with NID species heavier than the neutral gas they drift in is found in other studies.\\

A systematic study of the gain and directional performances while varying the ratio of the gas constituents and with different amplification structure is planned for the future, in order to increment the light gain and further investigate the outstanding diffusion properties of He:CF$_4$:SF$_6$ mixtures. Possible amplification structures to be tested are thicker GEMs, such as combination of structures presented in Chapter \ref{chap5}, COBRA structures, a modification of regular GEMs \cite{Amaro_2010kh}, and ThickGEM-multiwire hybrid \cite{EZERIBE2021164847}. A test with charge readout through Timepix3 \cite{LIGTENBERG2021165706} and possibly He:SF$_6$ gas mixtures with varying SF$_6$ concentration is also envisioned in the near future, to further explore the possibility that the measured diffusion below the thermal limit depends on the ratio of the drifting ions with respect to the neutral gas molecules.\\

It is important to stress how the avalanche gain achieved with triple thin GEMs and the NID He:CF$_4$:SF$_6$ mixture at 900 mbar, estimated in Section \ref{subsec:gainatm} of $\mathcal{O}$(10$^4$), is already large enough to grant $\mathcal{O}$(1) keV energy threshold for charge readouts as illustrated in \cite{bib:cygnus}, and to allow single ion full detection efficiency, as discussed in \cite{LIGTENBERG2021165706}.
While the solid angle reduction for optical readout still limits the energy threshold obtained in this first tested configuration, the work reported in this Chapter represents only a first step in the investigation of the potentialities of NID mixtures for optical readout. In addition, the fast development of sCMOS cameras sensors (see Section \ref{subsubsec:CMOS}) allows to envisage the possibility to improve of about a factor ten the performances reported in this Chapter with the use of the latest Orca Quest Hamamatsu model, and possibly lower the threshold even further with updated and more advanced versions. 

A reduction of a factor 3 in the drift diffusion (from 130 of ED down to 45 $\mu$m/$\sqrt{\text{cm}}$ of NID) together with 100 $\mu$m less diffusion in the amplification stage, in the context of CYGNO would permit to increment the drift length of 3 times with improved tracking capabilities, directionality and HT recognition, compactness and scalability.

The groundbreaking results presented in this Chapter, combined with the recent advances in the directional TPC field illustrated in Section \ref{sec:directional}, unconditionally demonstrate that a large scale directional experiment with competitive sensitivity for both Spin Dependent and Spin Independent couplings appears today not only technologically feasible, but also the only viable option to venture into the Neutrino Fog and to eventually establish beyond any doubt the nature of DM.

\chapter{Directionality relevance in the dark matter searches}
\label{chap7}
The motion of the Sun around the centre of the Galaxy generates for an observer on Earth an apparent wind of DM particles coming from the Cygnus constellation, as discussed in Chapter \ref{chap2}. This induces a clear directional dependence of the nuclear recoils caused by the elastic scattering of \W,  which no form of background can mimic (see Section \ref{sec:directional}). Directional detectors employ state of the art technologies to follow an innovative path for the direct search of DM that can exploit this peculiar signature. In Chapter \ref{chap2} the advantages of directional searches with respect to traditional experiments were illustrated in terms of discrimination of backgrounds (see Section \ref{subsubsec:dirlimits} and Section  \ref{subsubsec:dirnufog}), capability of positive claim of discovery (Section \ref{subsubsec:dirposdisc}) and potential to provide constraint on DM properties and perform DM astronomy (Section \ref{subsubsec:dirastronomydm}).
In this Chapter, these potentialities are further elaborated and detailed within the expected performances of an experiment based on the CYGNO approach in terms of DM exclusion limits and identification of DM properties. In particular, in Section \ref{sec:limitsCygno} a 30 m$^3$ CYGNO experiment expected sensitivity to SI and SD coupling, obtained by exploiting the directional distribution of the observed events, is discussed. Section \ref{sec:discr_2models} describes the capability of the directional search to provide discrimination between two DM models that displays very similar nuclear recoil energy spectra, with significant differences in the angular distribution.
\section{WIMP dark matter limits with CYGNO detector}
\label{sec:limitsCygno}
In the quest for the direct detection of DM, when no significance for \W induced recoils over the expected background is found, it is only possible to set limits in the cross section versus \W mass parameter space. Due to the nature of the DM halo and \W  velocity distribution, the angular distribution of NRs results much more efficient than the energy spectrum in discriminating signal from background (expected to be isotropic, see Section \ref{sec:directional}). The CYGNO project timeline (see Section \ref{sec:timeline}) foresees the development of a $\mathcal{O}$(30) m$^3$ experiment for directional DM searches and solar neutrino measurement (see PHASE\_2 in Section \ref{sec:future:CYGNO}). In order to evaluate the potentialities of CYGNO-30 to set exclusion limits in the SI and SD parameter space, a statistical analysis based on the Bayesian approach was developed. The next Sections are dedicated to briefly describe the basics of Bayesian inference (Section \ref{subsec:bayesian}), how this can be applied to simulated angular distributions of measured background events in CYGNO experiment to define a 90\% C.I. on the observed number of signal events (Section \ref{subsec:limit_signmod}-\ref{subsec:limit_likelihood}) and how from this a sensitivity limit on the DM masses versus SI and SD coupling can be extracted (Section \ref{subsec:limits}).
\subsection{Bayesian statistical approach}
\label{subsec:bayesian}
For a rigorous statistical analysis, the two main approaches applied in the diverse fields of physics are the frequentist and the Bayesian \cite{bib:Baxter_2021,bib:ROSZKOWSKI200910,bib:arina2014bayesian}.
The principal difference between the two resides in the meaning of the obtained results, where the Bayesian probability distributions can be referred to the true value of the variable of interest, for example the actual SI cross section of \W with a nucleon. This is particularly relevant in cases the expected value of interest is very close to the boundary of an unphysical region, as for example in DM searches when the expected number of events induced by \Ws is close to zero. In the Bayesian approach, the boundary information can be naturally taken into account in the calculation by inserting them in the prior probabilities, which represent the knowledge prior to the execution of the experiment (see below in the text). Instead, the evaluation of the frequentist limits strongly depends on the data acquired and, despite the prescription provided by the \textit{unified} interval methodology \cite{PDG}, the limit can sometimes be in the unphysical region. While this is not technically wrong, it requires careful communication of the results \cite{PDG}. Moreover, with Bayesian statistics, it is possible to obtain probability distributions on the real value of the variables under study, directly expressing a degree of belief. In this sense, probability actually expresses the chance of a parameter of being in a certain range. Hence, for the computation of the CYGNO experiment limits, the Bayesian approach is preferred due the simpler and more robustness of the limit evaluation close to the physical boundary of the variables.\\

A detailed review of the Bayesian statistical approach can be found in \cite{bib:Gelman,bib:DAgostini}. Its foundation lies in the application of the Bayes' theorem:
\begin{equation}
\label{eq:Bayes0}
p(A|B)=\frac{p(B|A)p(A)}{p(B)}
\end{equation}
meaning that the probability of an event A, given the occurrence of B is equal to the probability of B given A times the probability of A normalised to the total probability of B.

Let assume that a generic set of n independent quantities $\vec{x}$ are used to describe the state of a system or the outcome of a measurement. $p(\vec{x})$ is defined as the probability density function of the $\vec{x}$ variables for which the following characteristics are valid:
\begin{itemize}
	\item $p(\vec{x})d\vec{x}$ is the normalised probability of the system to find the vector ${x_1,x_2,..,x_n}$ in a multidimensional infinitesimal volume $\{[x_1,x_1+dx_1],[x_2,x_2+dx_2],...,[x_n,x_n+dx_n]\}$.
	\item $\int_{D}^{}p(\vec{x})d\vec{x}=1$, where $D$ is the domain of $p$.
\end{itemize}
$p(\vec{x})$ is interpreted as the proper infinitesimal probability of the variable $\vec{x}$. For example, when considering a mono-dimensional problem, if $\int_{0}^{10}f(x)dx=0.4$ it means that the probability of finding  $x$ between 0 and 10 is 40\%. In fact, a probability distribution function $p(\vec{x})$ can be assigned to a set of measurements $\vec{x}$, or to any model and parameter which is of interest to the specific analysis. Thus, since all of them can be considered connected to a probability distribution, they follow the rules of probability.\\
For each model that is expected to represent the data, a vector $\vec{a}$ of all the parameters defining such model can be defined. These can be divided in parameters fundamental for the description of the underlying physics, $\vec{\mu}$, and in nuisance parameters  $\vec{\theta}$ which are useful to describe the system and the data, but are not of interest for for the output of the statistical inference. For example, in the direct DM search context, the $\vec{\mu}$ can represent the expected number of  signal and background events, while the $\vec{\theta}$ can include detector effects and resolution of measurable quantities. The Bayesian approach allows to compute the probability of any model given a certain amount of information related to it, exploiting the following revisited form of the Bayes' theorem:
\begin{equation}
\label{eq:Bayes}
p(\vec{\mu},\vec{\theta}\vert \vec{x},H) = \frac{p(\vec{x}\vert\vec{\mu},\vec{\theta},H)\pi(\vec{\mu},\vec{\theta}\vert H) }{ \int_{\Omega}\int_{D}p(\vec{x}\vert \vec{\mu},\vec{\theta},H)\pi(\vec{\mu},\vec{\theta}\vert H)d\vec{\mu} d\vec{\theta} } 
\end{equation}

In Equation (\ref{eq:Bayes}), the following notation is used:
\begin{itemize}
	\item $\vec{\mu}$ is the vector of the free parameters of interest, representing the fundamental elements of the model under test;
	\item $\vec{\theta}$ is the vector of nuisance parameters, necessary to describe theoretical assumptions and experimental conditions that can affect the results. They can be not completely known and may depend on prior probability distributions;
	\item $\vec{x}$ is the vector representing the data set. It can be made of actual experimental data or simulated data;
	\item $H$ is the hypothesis under test;
	\item $\Omega$ is the nuisance parameters space.
	\item $D$ is the parameters space of $\vec{\mu}$.
	\item $p(\vec{\mu}\vert\vec{x})$ is the posterior probability function for the parameters $\vec{\mu}$, given $\vec{x}$, meaning the probability distribution on the real values of $\vec{\mu}$ in the light of the observations performed;
	\item $p(\vec{x}\vert\vec{\mu},\vec{\theta},H)$ is the likelihood function $\Likeli(\vec{\mu},\vec{\theta},H)$, describing how the observed data are likely to be produced by a particular set of parameters ($\vec{\mu}$, $\vec{\theta}$ and the hypothesis under test $H$);
	\item $\pi(\vec{\mu})$ is the prior probability of a parameter. This includes the expectations of the parameters as well as constraints and knowledge previously obtained from other experiments;
	
\end{itemize}
It is relevant to remark the importance of the prior probabilities $\pi(y)$. They represent the probability distribution of $y$ that express one's beliefs prior to the execution of the experiment. It contains all the previous knowledge available on the variables. This term, which never appears in the frequentist approach, permits the posterior probabilities on the real value of the parameter of interest to be attained. Indeed, it is the one that, by means of the Bayes' theorem, allows to obtain a posterior probability function, a probability distribution, on the model itself or on the true value of a quantity of interest. Instead, in a frequentist framework, the lack of the prior probability only permits to gain information on of the likelihood of the outcome of an experiment, without any direct connection to the physical quantities underneath.\\
The posterior distribution dependence on the nuisance parameters can be eventually removed by marginalising on $\vec{\mu}$ distributions as:
\begin{equation}
\label{eq:Bayes2}
p(\vec{\mu},\vert \vec{x},H) =  \int_{\Omega}p(\vec{\mu},\vec{\theta}\vert \vec{x},H) d\vec{\theta} 
\end{equation}
The experimental upper bound of a set of parameters of interest $\vec{\mu}$ whose true values are estimated  to be consistent with to zero, can be evaluated as the 90\% credible interval (C.I.) of the posterior. In the Bayesian approach, the upper bound is a statement on the true value of the parameter of interest. The quantity of 90\% C.I. has to be interpreted as the value below which the true values of $\vec{\mu}$ are believed to lie at 90\% probability level, given the present experimental information. Considering a mono-dimensional example where the parameter of interest is $\mu_1$, the standard and robust procedure to estimate the C.I. is defined as follows \cite{PDG,bib:DAgostini}:
\begin{equation}
\label{eq:CI}
\mu_1(90\% CI): \int_{0}^{\mu_1(90\%CI)} p(\mu_1\vert\vec{x},H)d\mu_1=0.9
\end{equation}
where $p(\mu_1\vert\vec{x},H)$ is the posterior probability marginalised over the nuisance parameters.\\

\subsubsection{Application to the CYGNO-30 exclusion limits}
\label{subsubsec:exclproc}
In the context of the evaluation of the expected exclusion limit for a 30 m$^3$ CYGNO-like detector, the Bayesian inference technique can be exploited to attain the upper bound limit on the number of DM events $\mu_s$ for a given WIMP mass, when $N_{evt}$ events are observed, which are compatible with the $\mu_b$ expected background events. The \W mass values chosen for the analysis spanned from the minimum detectable, which is a function of the energy threshold (see  Section \ref{subsubsec:limit_gasmix}), up to 100 GeV/c$^2$ for the SI and 500 GeV/c$^2$ for the SD, as at these large masses it is expected a loss of efficiency in exposure due to the light target employed, characteristic of the CYGNO gas mixture. The maximum mass studied for the SD coupling is larger than the one for SI. In the SD \W to proton limits the leading experiment is PICO \cite{bib:Amole_2019} which employs fluorine target and extends their limits up to 500 \Gevc. Conversely, in the SI ones, at masses above 100 \Gevc, xenon-based experiments dominate with orders of magnitude lower cross section bounds than the fluorine can permit due to the $A^2$ term. Thus, the masses studied are limited to 100 \Gevc.\\
Monte Carlo (MC) techniques are employed to simulate fake experiments the posterior probability of $\mu_s$ is evaluated from. The limit evaluation discussed in this thesis is based on different background scenarios (see Section \ref{subsec:limit_backmod} for details). For each possible value of these, the actual number of events in the pseudo-data spectrum is randomly extracted from a Poisson distribution of average $\mu_b$. For each of these events, a direction is also randomly sampled from the background angular distribution, and a resolution Gaussian smearing is applied. In the context of the application to CYGNO, these events are used to fill a histogram which represents the outcome of the simulated measurement. The likelihood function of the detected events is fitted on the data and the posterior probability is evaluated from this with the use of JAGS package (Just Another Gibbs Sampler)\cite{bib:Plummer}, after having marginalised on the nuisance parameters. The JAGS software allows to straightforwardly implement the Bayesian network constructed and calculates the integral through the numerical method based on the Markov Monte Carlo Chain via an algorithm called Gibbs Sampling. From the posterior probability the 90\% C.I. on $\mu_s$ is computed utilising Equation \ref{eq:CI}. In order to avoid suffering from any underfluctuation of the background (as undersampling), 500 data samples are simulated for each scenario, and the average result is taken as the final value.

A CYGNO-like detector is sensitive to the energy of recoils, along with the direction, which  provides an additional, very powerful handle, to discriminate signal from background events and to characterise DM. Indeed, energy and angle of recoil are bound by the kinematic of the elastic scattering and provide orthogonal information on the event nature. Measuring both enhances the power of the analysis. Nevertheless, as from the discussion of Section \ref{sec:directional}, the directional information is anticipated to grant better discrimination capability. Therefore, in the current study, for the sake of simplicity, only the angular information of the recoils is considered.\\
In the following Sections, the ingredients necessary to perform the estimation of the upper bound limit are described, from the detector response to the signal and background modelling and the likelihood function.
\subsection{Detector response}
\label{subsec:detecresp}
\subsubsection{Angular resolution}
Since CYGNO’s approach directional capabilities are still under evaluation, for this study, an angular resolution of 30$^{\circ}$ in the whole detectable energy range is assumed. This choice is backed up by experimental measurement in the 50–400 keV$_{\rm{nr}}$ range by the NEWAGE experiment \cite{Nakamura:2012zza} and by recent simulation within the CYGNUS context \cite{bib:cygnus} (details on the difference between keV$_{\rm{nr}}$ and keV$_{\rm{ee}}$ are presented in Section \ref{subsubsec:limit_gasmix}). Full head tail (HT) recognition down to the energy threshold is furthermore assumed, for the sake of simplicity. While it is known that 100\% HT efficiency can represent an optimistic scenario for energies below 10 keV$_{\rm{nr}}$, 30$^{\circ}\times$30$^{\circ}$ are deemed to be a pessimistic expectation for higher energy NRs  in the CYGNO context. In the fake experiments simulation, events are extracted and analysed with an angular distribution in 2D Galactic coordinates, to maximise the sensitivity to the NR anisotropy.
\subsubsection{Energy region of interest}
\label{subsubsec:ROI}
The energy region of interest (ROI) of the CYGNO experiment, defined as $\left[E_{thr},E_{max}\right]$, is an important parameter as it not only determines the range where DM is searched, but also affects the angular distribution of the recoils (see next Section). \\
The determination of the upper energy limit follows the kinematics of the WIMP scatter for every element of the gas mixture. As a matter of fact, the escape velocity of the Galaxy, $v_{esc}$, limits the maximum energy of recoil induced to a specific target element as discussed in Section \ref{sec:WIMPparadigm}. The maximum energy was described in Equation \ref{eq:maxenergyrec} here reported:
\begin{equation}
\label{eq:maxErecoil}
E_{max}= \frac{1}{2}m_{\chi} r (v_{lab}\cos\gamma +v_{esc})^2 
\end{equation}
where $r$ is the efficiency in the recoil momentum transfer which is maximised when the DM mass $m_{\chi}$ and the mass of the nucleus $m_A$ are equal, $v_{lab}$ is the velocity of the laboratory with respect to the Galactic Rest Frame, and $\gamma$ is the angle between the direction of the recoil and the opposite one to the motion of the laboratory.
For each direction, target and DM mass, the upper energy threshold can be adjusted so that all the recoils detectable are included. In a range of \W masses from 100 \Mevc up to 500 \Gevc, and considering the elements of H, He, C, and F present in the CYGNO gas mixture (see Chapter \ref{chap3}), this upper energy limit never exceeds 180 keV$_{\rm{nr}}$. Since NRs at these energy are not expected to exceed a $\mathcal{O}$(10) mm length, they can be considered to be always contained inside the CYGNO PHASE 2 active gas volume (see Figure \ref{fig:sim}). Moreover, the background rejection capabilities improve by orders of magnitude with energies above 20 keV$_{\rm{ee}}$, as seen in Section \ref{sec:directional}, so no drawback is foreseen when the energy range is extended at those values.\\
The choice of the low energy threshold for the simulated events has a very important role in the determination of the shape of the angular distribution. Section \ref{sec:LIME} and Section \ref{sec:future:CYGNO} illustrate how an energy threshold of 1 keV$_{\rm{ee}}$ was evaluated on CYGNO prototypes by means of a \fe source, which can be expected to improve to 0.5 keV$_{\rm{ee}}$ thanks to the improved sensitivity of the ORCA Quest camera (and future developments in sCMOS sensors). Thus, these two scenarios are considered in the following. Thus, for the study both the energy thresholds are considered as a conservative and realistic scenarios.
\subsection{Signal model}
\label{subsec:limit_signmod}
The double differential rate of the \W induced recoils as a function of the recoil energy and angle was calculated in Section \ref{subsec:ratecalculation}. Thus, the angular distribution of the recoils, named $\frac{dD}{d\cos\gamma}$,  for a single target $A$ and normalised to 1 is:
\begin{eqnarray}
\label{eq:rateWIMP2}
\frac{dD}{d\cos\gamma}=\alpha_0\int_{E_{thr}}^{E_{max}}S(E)\left(e^{-\frac{\left(\frac{\sqrt{2m_AE}}{2\mu_A}-v_{lab}\cos\gamma\right)^2}{v_p^2}} - e^{-\frac{v_{esc}^2}{v_p^2}}\right)dE,
\end{eqnarray}
with $\alpha_0$ the normalisation factor $v_p$ the velocity dispersion in the SHM, $m_A$ the mass of the target, and $\mu_A$ its reduced mass with the \W mass. The distribution clearly depends on the \W mass which is unknown. Since it is of interest to calculate a upper limit for each \W and not for a average one, the analysis and the calculation of the angular distribution is repeated for each of the \W masses considered.\\
The energy ROI directly affects the range of integration the distribution is obtained from, and therefore its shape, and its choice was discussed in the previous Section. 
The $E_{max}$ value depends on the target mass, the WIMP mass, the escape and lab velocities, as well as the angle $\gamma$ (see Equation \ref{eq:maxErecoil}). In turn, the actual E$_{thr}$ (and therefore the ROI) results different for each target, since it is affected by the quenching of the visible ionisation energy (see Section \ref{subsubsec:limit_gasmix}). As a result, the angular distribution changes for each target element considered and needs to be computed separately.
\begin{figure}[t] 
	\centering
	\includegraphics[width=0.9\linewidth]{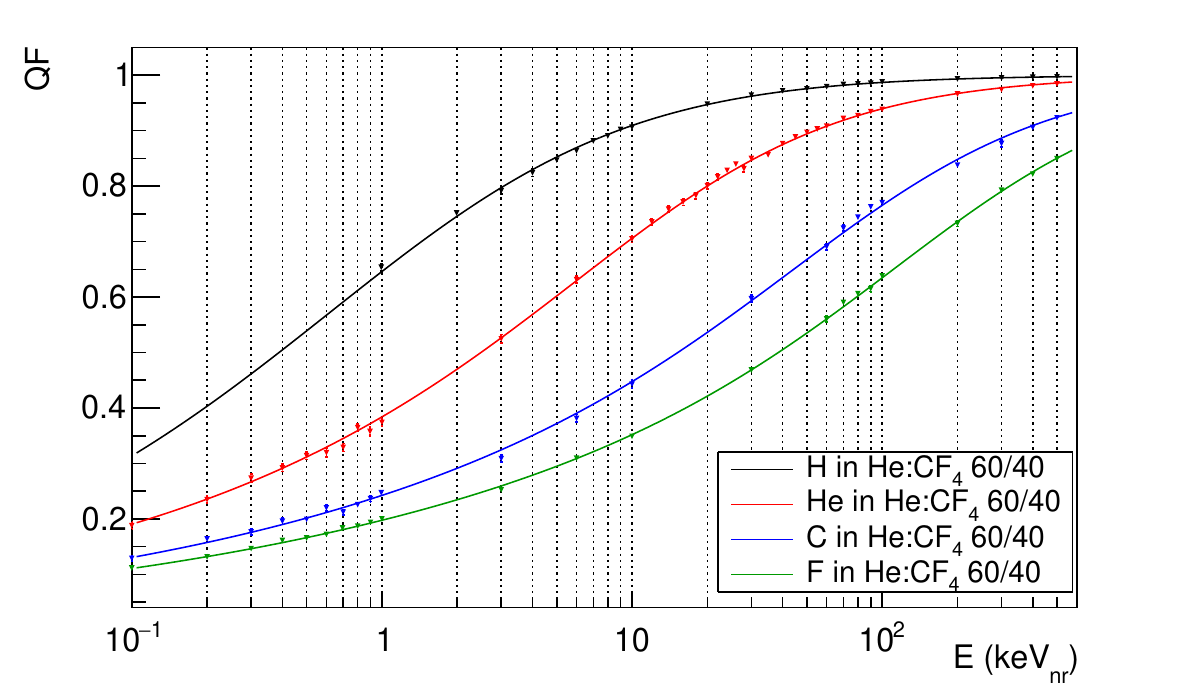}
	\caption{Quenching factor as a function of the nuclear recoil energy for the various target elements of the gas mixture simulated with the SRIM software. The fitting functions are the representation of the quenching factor behaviour from the Equation \ref{eq:QF}.}
	\label{fig:QF}
\end{figure}

\subsubsection{Composition of the gas mixture}
\label{subsubsec:limit_gasmix}
The gas mixtures considered for the future performances of the CYGNO detector are the standard He:CF$_4$ 60/40 and the possible He:CF$_4$:iC$_{4}$H$_{10}$ 58/40/2 both at 1 atmosphere, following the arguments of Chapter \ref{chap3}.  These gas mixtures contain different elements for each of which kinematics and momentum transfer characteristics change. Thus, contribution of every element in the gas mixture needs to be correctly taken into account when evaluating the signal model.
\paragraph{Quenching factor}
\begin{table}[t]
	\begin{center}
		\begin{adjustbox}{max width=1.01\textwidth}
			\begin{tabular}{|c|c|c|c|c|}
				\hline
				& \multicolumn{2}{c|}{1 keV$_{\text{ee}}$} & \multicolumn{2}{c|}{0.5 keV$_{\text{ee}}$} \\ \cline{2-5}
				& $E_{thr,\text{nr}}$ (keV$_\text{nr}$) & Min DM mass (GeV/c$^2$) &  $E_{thr,\text{nr}}$ (keV$_\text{nr}$) & Min DM mass (GeV/c$^2$)\\ \hline \hline
				H & 1.4 & 0.5 & 0.8 & 0.3\\ \hline
				He & 2.1 & 1.0 & 1.2 & 0.7\\ \hline
				C & 3.1 & 1.9 & 1.8 & 1.4\\ \hline
				F & 3.8 & 2.5 & 2.2 & 1.9\\ \hline
			\end{tabular}
		\end{adjustbox}
		\caption{Summary for all the elements and the energy thresholds, of the effective threshold in nuclear recoil energy and the corresponding minimum DM mass detectable.}
		\label{tab:ethres}
	\end{center}
\end{table}
It is relevant to underline that NRs deposit energy in a different way than ERs, especially at the low $\mathcal{O}$(1) keV energies of interest for direct DM searches. Indeed, electron kinetic energy is efficiently transferred to the gas ionisation, while low energy NRs dissipate energy in a medium through inelastic Coulomb interactions with atomic electrons, referred to as electronic energy losses, and elastic scattering in the screened electric field of the nuclei, referred to as nuclear energy losses. The ionisation quenching factor (QF) is defined as the fraction of the kinetic energy released through ionization by a recoil in a medium. In this context, it is possible to introduce the observed nuclear recoil energy, E$_{ee}$ in electronvolts electron equivalent (eV$_{ee}$) as the nuclear recoil energy that is measured by ionisation, that is related to the total kinetic nuclear recoil energy E$_{nr}$ expressed in eV$_{nr}$ as E$_{ee}$ = QF E$_{nr}$.
The estimation of the QF is performed by the collaboration by means of a SRIM simulation for each element of the gas mixtures. Figure \ref{fig:QF} shows the simulated QF as a function of the NR energy for the four elements under investigation. The coloured fits follow the Equation \ref{eq:QF} which describes its expected behaviour \cite{bib:QFpaper}.
\begin{equation}
\label{eq:QF}
QF = \frac{a\left(E_{\text{nr}}+bE_{\text{nr}}^c\right)}{1+a\left(E_{\text{nr}}+bE_{\text{nr}}^c\right)}
\end{equation}
Since the QF differs for each nucleus, the effective threshold on the NR energy, $E_{thr,\text{nr}}$, and the angular distribution will be affected differently for each target element. The effective energy threshold for each element defines the minimum WIMP mass each of them can be sensitive to. Indeed, if in Equation \ref{eq:maxErecoil} $E_{max}$ is replaced with $E_{thr,\text{nr}}$ and $\cos\gamma=1$, it is possible to calculate the minimum DM mass $m_{\chi}$ for which a certain element is able to recoil with an energy the detector is sensitive to. Table \ref{tab:ethres} summarises for each element and  detectable energy threshold, the effective threshold in nuclear recoil energy and the corresponding minimum detectable DM mass.
\begin{figure}[t] 
	\centering
	\includegraphics[width=0.48\textwidth]{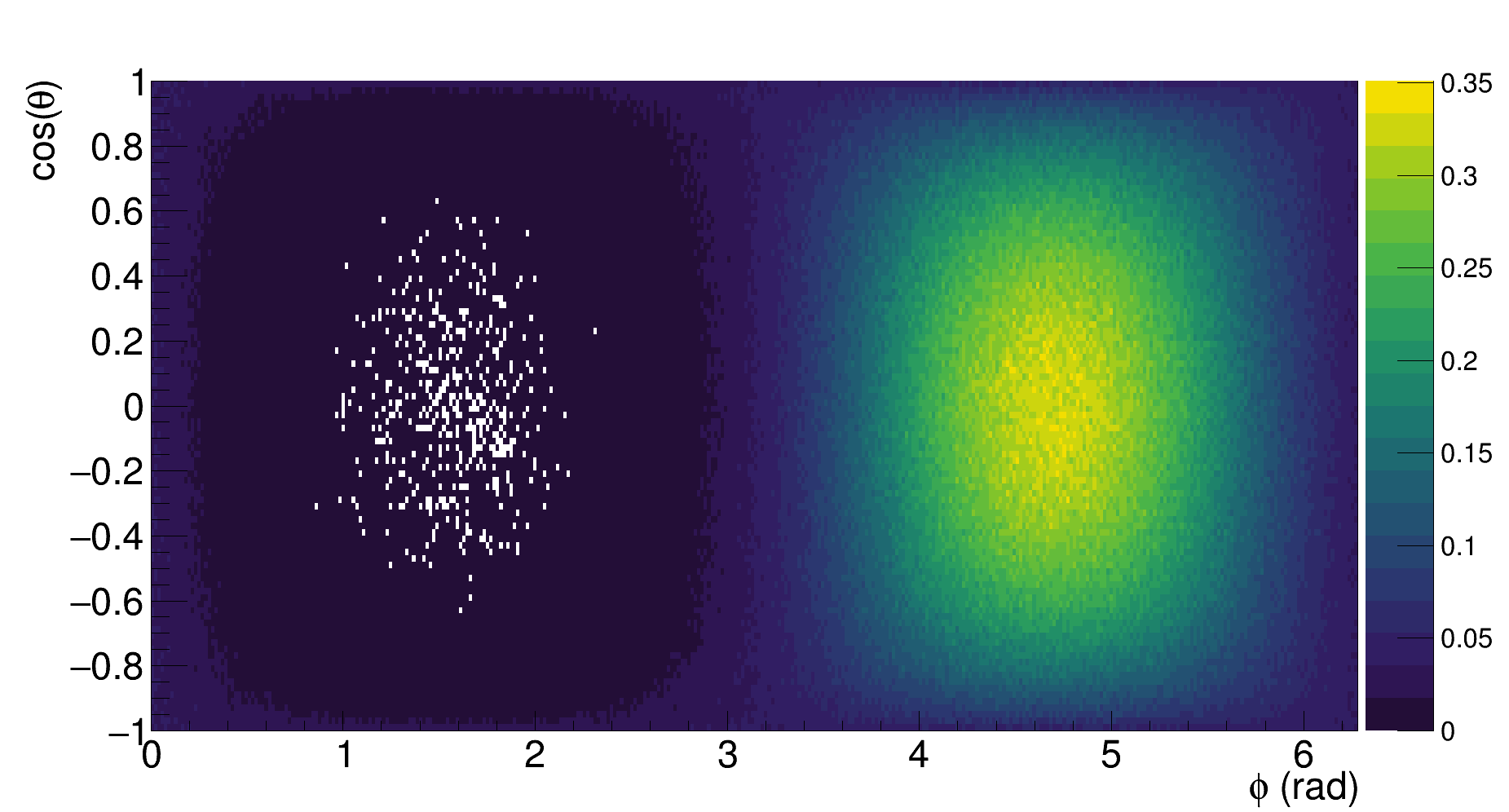} 
	\includegraphics[width=0.48\textwidth]{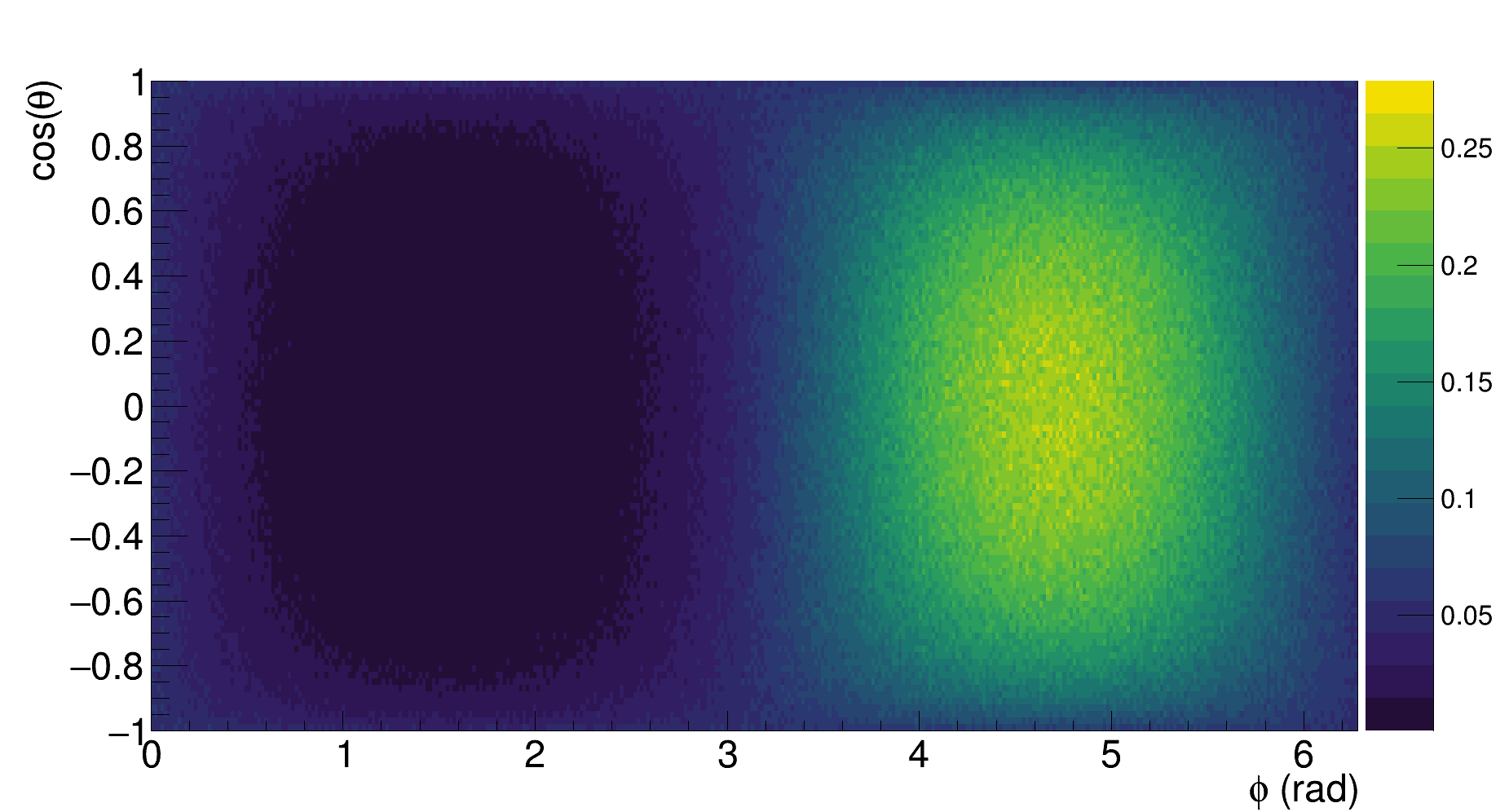}
	\caption{Two examples of the angular distribution of recoils due to DM in Galactic coordinates, obtained by Monte Carlo simulations. Left: helium recoils induced by 10 GeV/c$^2$ DM. Right: fluorine recoils induced by 100 GeV/c$^2$ DM.}
	\label{fig:exemp_WIMP}
\end{figure}
It is clear how the reduction in energy threshold can push the experiment  sensitivity towards lower DM masses. Similarly, lighter targets have a better momentum transfer with low mass \Ws and higher QF, which increase the recoil energy and consequently allows to probe lower DM masses. It is important to notice that the standard CYGNO gas mixture He:CF$_4$ 60/40 can be sensitive to \W masses below the GeV/c$^2$, even without the addition of the hydrocarbon component, but with a sufficiently low energy threshold.\\
With the $E_{thr,\text{nr}}$ determined, the angular distributions for each element can be computed and Figure \ref{fig:exemp_WIMP}  shows an example of angular NR distribution in Galactic coordinate for He on the left and F on the right, with 1 keV$_{\rm{ee}}$ detectable energy threshold induced by 10 GeV/c$^2$ WIMP (on the left) and 100 GeV/c$^2$ (on the right).
\paragraph{Element interaction probability}
In order to correctly take into account the response of a multi-target detector to the \W interaction and sum the contribution of the different elements consistently,  it is important to consider the different probability of WIMP-nucleon interaction and NR detection for each element. The ratio of the number of expected recoils for one element over the total number of recoils for all elements is deemed to be a good estimation of the above mentioned probabilities.\\
Equation \ref{eq:rateWIMP} expresses the double differential rate of nuclear recoils in angle and energy per unit of mass with a single recoiling species in the SI assumption. From this, the total number of events for a single monoatomic molecule $i$ can be obtained integrating in the energy and angle and by multiplying by the exposure of the detector:
\begin{equation}
\label{eq:totnumSIonetarget}
N_{DMevt,i} = tV\frac{P}{P_{atm}}\frac{T_0}{T}\rho_i\frac{ N_0}{A_{mol,i}}\frac{2\rho_0\sigma_{n,SI}}{m_{\chi}^2r_i}\frac{\mu_{A,i}^2}{\mu_n^2}A_i^2 I_{i}^{E\gamma}(m_{\chi},E_{thr,i}),
\end{equation}
with:
\begin{itemize}
	\item $t$ the time of exposure;
	\item $V$ the volume of the detector;
	\item $P$ the working pressure of the gas;
	\item $P_{atm}$ the atmospheric pressure;
	\item $T$ the working temperature expressed in Kelvin;
	\item $T_0$ the temperature of 0 degrees Celsius expressed in Kelvin;
	\item  $\rho_i$ the gas density at atmospheric pressure and 0 degrees Celsius;
	\item $N_0$ the Avogadro number;
	\item $A_{mol,i}$ the molar mass of the gas;
	\item $\rho_0$ the local DM density (Section \ref{sec:WIMPparadigm});
	\item $\mu_{A,i}$ the reduced mass between the \W mass and the mass of the nucleus $A$ defined as in Equation \ref{eq:red_mass};
	\item $\mu_{n}$ the reduced mass between the \W mass and the mass of the nucleon;
	\item $A_i$ the atomic mass of the nucleus;
	\item $I_{i}^{E\gamma}(m_{\chi},E_{thr,i})$ the velocity distribution integrated in the velocity, energy and angle after the Radon transformation (see Section \ref{subsec:ratecalculation}).
\end{itemize}
The last term of Equation \ref{eq:totnumSIonetarget} is defined as:
\begin{eqnarray}
\label{eq:integralvel}
I^{E\gamma}(m_{\chi},E_{thr}) = &\bigint_{E_{thr}}^{E_{max}}dE\bigint_{-1}^1d\cos\gamma S(E)\pi \frac{v_p^3}{v_{lab}}\alpha'\times& \\ &\left(e^{-\frac{\left(\frac{\sqrt{2m_AE}}{2\mu_A}-v_{lab}\cos\gamma\right)^2}{v_p^2}} - e^{-\frac{v_{esc}^2}{v_p^2}}\right)&, \nonumber
\end{eqnarray}
The total number of events for a gas mixture can hence be written summing Equation \ref{eq:totnumSIonetarget} over the different elements as:
\begin{equation}
\label{eq:totnumSImoretargets}
N_{DMevt} = tV\frac{P}{P_{atm}}\frac{T_0}{T}\sum_{i}^{n_{mol}}\sum_{j}^{n_{el,i}}\rho_ik_i\frac{ N_0}{A_{mol,i}}N_{at,i,j}\frac{2\rho_0\sigma_{n,SI}}{m_{\chi}^2r_j}\frac{\mu_{A,j}^2}{\mu_n^2}A_j^2 I_{j}^{E\gamma}(m_{\chi},E_{thr,j}),
\end{equation}
where $k_i$ is the percentage of content of the molecule $i$ in the gas, $N_{at,i,j}$ the number of atoms $j$ in the molecule $i$, $n_{el,i}$ the total number of elements in the gas $i$, and $n_{mol}$ the number of molecules composing the gas mixture. 
All the terms depending on the gas can be rearranged in a single quantity $F_{i,j}$ as:
\begin{equation}
\label{eq:Fij}
F_{i,j} = \rho_ik_i\frac{N_{at,i,j}}{A_{mol,i}}\frac{2}{r_j}\mu_{A,j}^2A_j^2 I_{j}^{E\gamma}(m_{\chi},E_{thr,j}).
\end{equation}
\begin{figure}[t]
	\centering
	\includegraphics[width=0.7\textwidth]{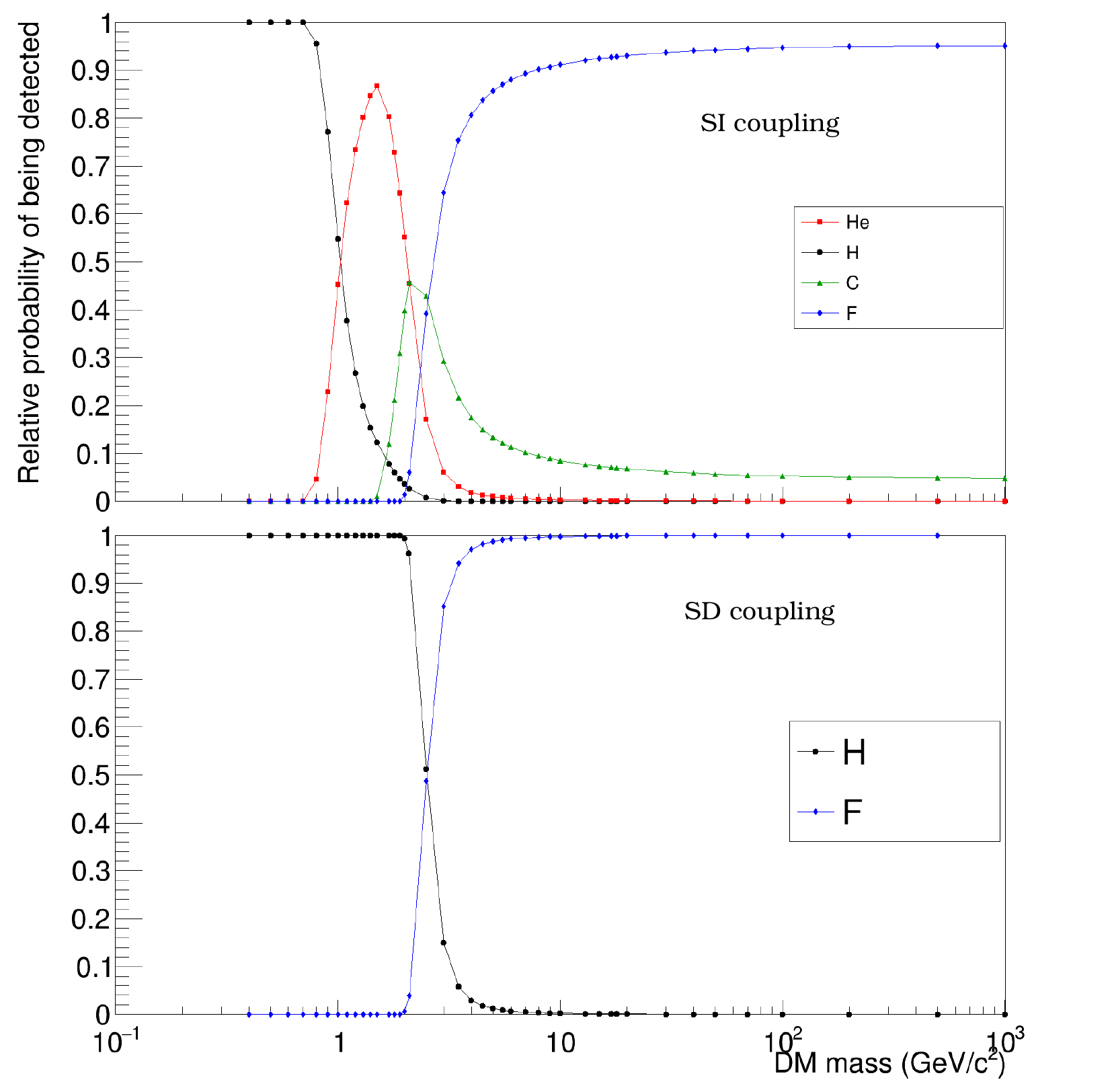} 
	\caption{The probability of each element of being hit and detected as a function of the DM mass, if the energy threshold for the detection is 0.5 keV$_{\text{ee}}$. The lower masses are dominated by the lighter element, while above 6 GeV/c$^2$ fluorine is the most probable.}
	\label{fig:limit_prob}
\end{figure}
The probability of each element $X$ to be hit by a \W and detected by the experimental setup can hence be defined as:
\begin{equation}
\label{eq:limit_proba_elem}
P_{X}=\frac{N_{DMevt,X}}{N_{DMevt}}= \frac{\sum_{i}^{n_{mo}}F_{i,X}}{\sum_{i}^{n_{mo}}\sum_{j}^{n_{el,i}}F_{i,j}}.
\end{equation}
This estimation of the probability includes all the dependencies on the DM halo model, the nuclear cross section, the quenching factors and other gas properties, while, at the same time, factorises the unknown SI nucleon-WIMP cross section $\sigma_{n,SI}$.  
Similar calculation can be performed for the SD coupling, for which the relative function $F_{i,j,SD}$ can be written as:
\begin{equation}
\label{eq:FSD}
F_{i,j,SD} = \rho_ik_i\frac{ N_{at,i,j}}{A_{mol,i}}\frac{2}{r_j}\mu_A^2\frac{4\left\langle S_p\right\rangle(J_j+1)}{3J_j} I_{j}^{E\gamma}(m_{\chi},E_{thr,j}).
\end{equation}
Figure \ref{fig:limit_prob} shows the probability of each element of being hit and detected as a function of the DM mass both for SI (on top) and SD (on bottom), with an energy threshold for the detection of 0.5 keV$_{\text{ee}}$.

The region of the DM velocity distribution (and hence DM masses) accessible to detection is limited at lower values by the energy threshold and at higher values by $v_{esc}$. 
Due of their light masses, hydrogen and helium have the largest probability to be detected for WIMP masses of a couple of GeV/c$^2$. At higher DM masses, when the window of integration on the velocity distribution is quite large, the A$^2$ cross section enhancement for the SI coupling and the large abundance for the SD dominate, making fluorine the most probable detectable element.
\subsubsection{Signal event prior probability}
\label{subsubsec:prior}
A uniform distribution 0 to 1000 is chosen as the prior on the expected number of signal event, given that articulated signal prior probabilities cannot be assumed without risking biases, because the actual cross section of DM with protons is unknown. Indeed, events per year is a non-negative defined variable, and due to current limits in the DM community, it is hardly believable that more than 1000 events per year would be produced an O(30) m$^3$ CYGNO detector.
\subsection{Background model}
\label{subsec:limit_backmod}
The background event angular distribution can be reasonably assumed flat in Galactic coordinate. Indeed, any local source of background which is not already isotropic with respect to the detector, once transformed into Galactic coordinates, will be diluted and smeared thanks to the motion of the Earth around its own axis, and will result in a mostly flat distribution (see Section \ref{sec:directional}). 
Since the phenomena inducing NRs and indistinguishable ER backgrounds in the detector differently from a WIMP interaction do not affect their flat angular distribution, the probability of each element to recoil due to backgrounds is assumed identical.
\subsubsection{Background event prior probability}
\label{subsubsec:priorback}
The number of expected background events for CYGNO Phase\_2 cannot easily be predicted at this stage of the project, since it will depend on the outcome of LIME underground installation, CYGNO-04 performances, and the possible improvements in terms of material radioactivity minimisation and background rejection discussed in Sections \ref{sec:LIME} and \ref{sec:future:CYGNO}. 

However, considering LIME background simulations (Section \ref{sec:LIME}), the preliminary results on Machine Learning techniques for particle identification (which already demonstrated a nearly order of magnitude improvement over the performances illustrated in Section \ref{subsec:cyg_backrej}) and the reduced contamination of the materials that will be employed for the detector construction, a tentative estimation returned about 10$^{1}$-10$^{2}$ events per year in the range 1-20 keV$_{\text{ee}}$ for CYGNO-04 (see Section \ref{subsec:cygno04}). By simply scaling for the volume, a factor 100 increase can be expected for CYGNO-30 with respect to CYGNO-04. At the same time, given the fast development of sCMOS sensors (see Table \ref{tab:hama}) and the lesson learned with CYGNO-04, it is reasonable to expect to further improve the experimental technique and reduce the backgrounds for a 30 m$^3$ experiment. For these reasons, three different scenarios are considered in this study, with  $\mu_b$ equal to 100, 1000 and 10000 events per year. A Possonian distribution of the number of events for each scenario is employed as prior. 

\subsection{Likelihood}
\label{subsec:limit_likelihood}
When no events can be recognised as WIMP induced recoils with the desired significance, the upper limit on the $\mu_s$ number of events expected from signal can be evaluated as the 90\% C.I. of the posterior. In the determination of the credible interval to evaluate the limit sensitivity for a CYGNO-30 experiment, the posterior probability on the $\mu_s$ signal events is estimated in the hypothesis, $H_1$, that both background and signal are present in the data sample. 
As discussed in Section \ref{subsec:detecresp}, the analysis is performed in Galactic coordinates in order to maximise the anisotropy of the WIMP-induced NRs with respect to backgrounds. For each background scenario, the actual number of events is randomly extracted from a Poissonian distribution and a direction is assigned to each, randomly sampling the background angular distribution. After applying a Gaussian smearing to account for the resolution, a histogram representing the measured event direction in Galactic coordinates is filled, with its binning reflecting the angular resolution (bin size double the resolution value). 
For the C.I. evaluation, the likelihood function in Equation \ref{eq:Bayes} is profiled along the angular distribution as:
\begin{equation}
\label{eq:likelihood_cygno}
\Likeli(\vec{x}|\mu_s,\mu_b,H_1)=(\mu_b+\mu_s)^{N_{evt}}e^{-(\mu_b+\mu_s)  }\prod_{i=1}^{N_{\text{bins}}} \left[ \left( \frac{\mu_b}{\mu_b+\mu_s}P_{i,b}+ \frac{\mu_s}{\mu_b+\mu_s}P_{i,s}\right)^{n_i}\frac{1}{n_i!}\right]
\end{equation}
with:
\begin{itemize}
	\item $N_{evt}$ is the total number of events of the data sample;
	\item $i$ is the index representing the bin of the histogram in the 2D angular Galactic coordinates;
	\item $n_i$ is the number of events occurring in the $i$th bin;
	\item $\mu$ is the number of expected events due to WIMP-induced recoil ($\mu_s$) or background ($\mu_b$), given a certain WIMP mass;
	\item $P_{i,x}$ is the probability of a single event to end up in the $i$th bin, according to x hypothesis (background $b$ or signal $s$).
	
\end{itemize}
The likelihood function is obtained as the product of a term which represents the Poissonian distribution of having measured $N_{evt}$ when $\mu_b+\mu_s$ are expected, and the profiled term on the angular distribution of the recoils. 

The profile term is written in a event-binned form, which means that the function is calculated as the independent probability of having a certain amount of events happening in each bin. The marginalised probability $P_{i,x}$ in each bin contains the information coming from the original theoretical angular distribution, the probability of each element to recoil and the probability of an event to migrate from one bin to another due to resolution effects, and is expressed as:
\begin{equation}
\label{eq:pmargin}
P_{i,x}=\sum_{j=i}^{N_{adjacent}}\left[P^{\text{migrate}}_{j\rightarrow i}\sum_{k}^{n_{el}}P_{k,j,x}^{theo}P_{k,x}^{el}\right]
\end{equation}
with:
\begin{itemize}
	\item $P_{k,j,x}^{theo}$ the theoretical probability of an event of the element $k$ to end up in the $j$th bin following the hypothesis $x$. This is the one directly calculated from the spectra obtained in Section \ref{subsec:limit_signmod}.
	\item $P_{k,x}^{el}$ is the probability of the element $k$ of recoiling as evaluated in Section \ref{subsubsec:limit_gasmix}.
	\item $P^{\text{migrate}}_{j\rightarrow i}$ is probability of an event to migrate from the $j$th bin to the $i$th due to resolution effect.
	\item the sum runs over all the bins adjacent to the $i$th one ($i$th included) in the histogram. 
\end{itemize}
The probability of migration is calculated in the approximation of a Gaussian smearing effect of the resolution. The evaluation of $P^{\text{migrate}}_{x,j\rightarrow i}$ in the more general 2D case, assuming that the resolution in two independent variables are themselves independent, can be estimated from:
\begin{equation}
\label{eq:migration1}
P^{\text{migrate}}_{x,j\rightarrow i}(\alpha,\eta)=\int_{i_{a\alpha}}^{i_{b\alpha}}d\alpha\int_{i_{a\eta}}^{i_{b\eta}}d\eta\int_{j_{a\bar{\alpha}}}^{j_{b\bar{\alpha}}}d\bar{\alpha}\int_{j_{a\bar{\eta}}}^{j_{b\bar{\eta}}}d\bar{\eta}\frac{1}{\sqrt{2\pi}\sigma_\eta}e^{-\frac{(\eta-\bar{\eta})^2}{2\sigma_\eta^2}} \frac{1}{\sqrt{2\pi}\sigma_{\alpha}}e^{-\frac{(\alpha-\bar{\alpha})^2}{2\sigma_{\alpha}^2}}f_x(\bar{\alpha},\bar{\eta})
\end{equation}
with $\alpha$ and $\eta$ the two independent variables, $\bar{\alpha}$ and $\bar{\eta}$ the original value of the two independent variables before any smearing that ended up in the $j$th bin, $a$ and $b$ the extremes of the $i$th or $j$th bin, $f_x(\bar{\alpha},\bar{\eta})$ the probability spectrum from model x, and $ \sigma_\eta $ and $ \sigma_{\alpha} $ the experimental resolutions of the variables. Since for the Gaussian distributions is valid that $\sigma_{\alpha} \sim (i_{b\alpha}-i_{a\alpha})$ and the dimension of the bins are built as described before, the calculation can be reduced by fixing the mean value of the Gaussian smearing functions to the central value of the bin\footnote{Assumption verified with numerical integration} as:

$$
P^{\text{migrate}}_{x,j\rightarrow i}(\alpha,\eta)=\int_{\alpha_{ci}-\sigma_{ci}}^{\alpha_{ci}+\sigma_{ci}}d\alpha\int_{\eta_{ci}-\sigma_{ci}}^{\eta_{ci}+\sigma_{ci}}d\eta\frac{1}{\sqrt{2\pi}\sigma_{\eta_{cj}}}e^{-\frac{(\eta-\eta_{cj})^2}{2\sigma_{\eta_{cj}}^2}} \frac{1}{\sqrt{2\pi}\sigma_{\alpha_{cj}}}e^{-\frac{(\alpha-\alpha_{cj})^2}{2\sigma_{\alpha_{cj}}^2}}\cdot$$
$$\cdot\int_{j_{a\bar{\alpha}}}^{j_{b\bar{\alpha}}}d\bar{\alpha}\int_{j_{a\bar{\eta}}}^{j_{b\bar{\eta}}}d\bar{\eta}f_x(\bar{\alpha},\bar{\eta})=$$

$$= \int_{\alpha_{ci}-\sigma_{ci}}^{\alpha_{ci}+\sigma_{ci}}d\alpha\int_{\eta_{ci}-\sigma_{ci}}^{\eta_{ci}+\sigma_{ci}}d\eta\frac{1}{\sqrt{2\pi}\sigma_{\eta_{cj}}}e^{-\frac{(\eta-\eta_{cj})^2}{2\sigma_{\eta_{cj}}^2}} \frac{1}{\sqrt{2\pi}\sigma_{\alpha_{cj}}}e^{-\frac{(\alpha-\alpha_{cj})^2}{2\sigma_{\alpha_{cj}}^2}}\cdot P_{j,x}^{theo}$$

where the subscript $cj$ ($ci$) means evaluated in the centre of the $j$th ($i$th) bin, and $P_{j,x}^{theo}$ is the result of the sum on the elements $k$ of Equation \ref{eq:pmargin}. Thus, in this approximation, which is perfectly valid in the current analysis, the probability of migration is model independent and can be expressed by:
\begin{equation}
\label{eq:migration2.1}
P^{\text{migrate}}_{j\rightarrow i}(\alpha,\eta)=\int_{\alpha_{ci}-\sigma_{ci}}^{\alpha_{ci}+\sigma_{ci}}d\alpha\int_{\eta_{ci}-\sigma_{ci}}^{\eta_{ci}+\sigma_{ci}}d\eta\frac{1}{\sqrt{2\pi}\sigma_{\eta_{cj}}}e^{-\frac{(\eta-\eta_{cj})^2}{2\sigma_{\eta_{cj}}^2}} \frac{1}{\sqrt{2\pi}\sigma_{\alpha_{cj}}}e^{-\frac{(\alpha-\alpha_{cj})^2}{2\sigma_{\alpha_{cj}}^2}}
\end{equation}
For a more simple example of a 1D  angular case, $ P^{\text{migrate}}_{i\rightarrow i}=0.683 $ and $ P^{\text{migrate}}_{i+1\rightarrow i}=0.159 $. A numerical integration was performed for a couple of \W masses and target of the full Equation \ref{eq:migration1} and of the approximated version and resulted consistent within a factor 10$^{-4}$.\\

As both background and signal can be present at once, the probabilities from the two spectra are weighted on the expected number of signal and background events. 
The parameters of interest of the analysis are $\mu_b$ and $\mu_s$, which will be fitted on the data set to find their best approximation with respect to the data sample and on which the posterior probability will be computed. In the simulation of the expected performances, $\mu_b$ can be used as a control variable that has to be consistent with the known simulated background intensity, whilst in a real analysis it can assess the intensity of the background. Instead, $\mu_s$ is directly related to the WIMP-nucleon elastic cross section, useful to evaluate the limits.
\subsection{Effect of directionality on the limits}
\label{subsec:effect_dir_limit}
Following the procedure described in Section \ref{subsubsec:exclproc}, for each \W mass,  the posterior probability at 90\% C.I. of the number of WIMP-induced recoil $\mu_s$ is computed and averaged to obtain the final result, following Equation \ref{eq:CI}.
\begin{figure}[t] 
	\centering
	\includegraphics[width=0.9\textwidth]{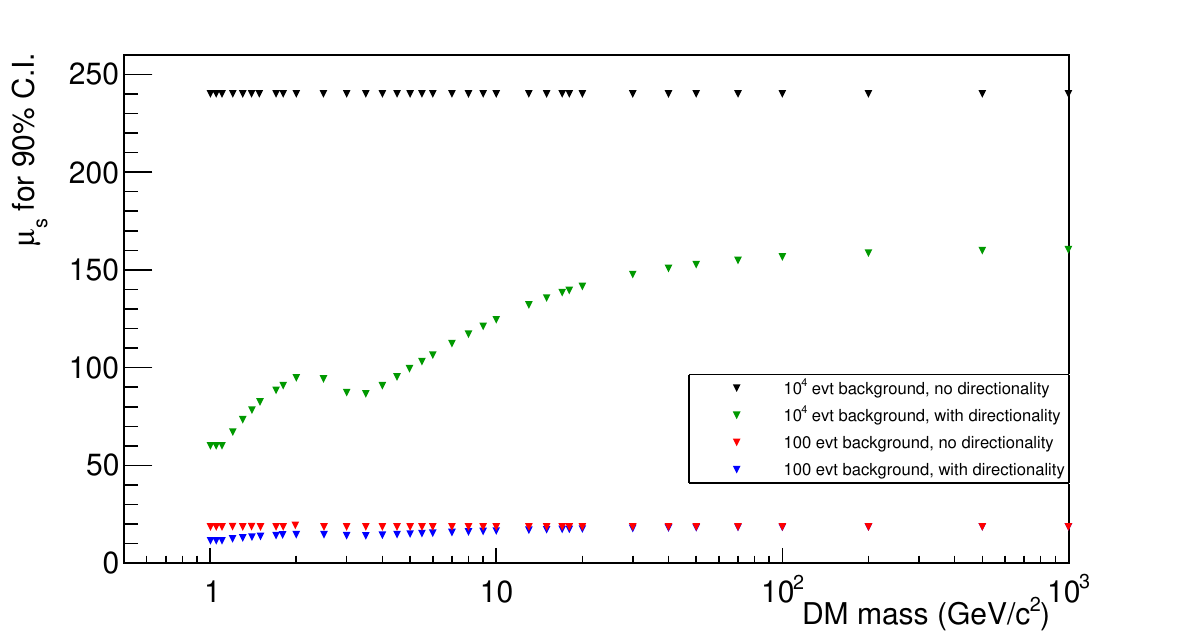} 
	\caption{The $\mu_{s,90\%}$ corresponding to the 90\% percentile of the posterior probability of $\mu_s$ for two background configurations for the SI coupling and 1 keV$_{\text{ee}}$ energy threshold. The green and blue points are obtained employing the likelihood of Equation \ref{eq:likelihood_cygno} which includes the profiling on the angular information. The black and red points are attained using the likelihood of Equation \ref{eq:likelihood_cygno_nodir} which corresponds to a simple counting experiment with no energy nor angular information.}
	\label{fig:CI_nodir}
\end{figure}
In order to explicitly show the power of directionality in setting limits on WIMP-nucleon scattering, the same analysis is repeated in the assumption that the experiment under evaluation can only measure the number of detected events, with no energy or additional information. This can be achieved by replacing the likelihood described in Section \ref{subsec:limit_likelihood} with a function which only takes into account the Poissonian statistics of counting, written as:
\begin{equation}
\label{eq:likelihood_cygno_nodir}
\Likeli(\vec{x}|\mu_s,\mu_b,H_1)=\frac{(\mu_b+\mu_s)^{N_{evt}}}{N_{evt}!}e^{-(\mu_b+\mu_s)}
\end{equation}
While this is a simplified assumption, given that, typically, the energy distribution of the detected events is available, the background spectrum highly depends on the exact materials and shielding employed in the experiment and results therefore difficult to predict with precision at this stage of the project development. Thus, the comparison of the directional performances is carried out with respect to a simple counting experiment.\\

The pure counting statistical analysis is performed for a 1 keV$_{\rm{ee}}$ threshold and with the most optimistic and pessimistic background scenarios (namely, 100 and 10000 events) to highlight the differences with respect to directional detection.
Figure \ref{fig:CI_nodir} shows the 90\% C.I. on the number of signal events $\mu_s$ as a function of the DM mass when the direction information is taken into account and when it is not.  The upper limit on $\mu_s$ achievable by directional searches is always lower or equal than the corresponding value for pure counting experiment. Figure \ref{fig:CI_nodir} highlights another important feature of directional searches: the largest  the number of background events detected, the stronger the improvement  due to the use of the angular distribution in setting upper bounds on the signal. This results evident when considering that a smaller background statistical sample can more easily resemble a signal dipole structure due to fluctuation only, compared to a large number of events that will distribute more evenly in Galactic coordinates. This is an important result since even in the presence of large, irreducible backgrounds, a directional detector will be able to set improved exclusion limits with increasing exposure. The presence of minima at 1 and 2.5 GeV/c$^2$ can be explained by the fact that there the helium and fluorine have their threshold of DM mass sensitivity. Indeed, close to their sensitivity threshold, only a very small fraction of the fastest WIMPs can induce a detectable recoil, driving the angular distribution to become very peaked and thus easier to discriminate. Due to the energy threshold chosen and the kinematic between heavy \Ws (above $\sim$ 50 \Gevc) and fluorine nuclei, the angular distributions of the recoils have their dipole structure more spread, reducing slightly the discrimination power with respect to low \W masses. Thus, while the range of the improvement due to the angular distribution information varies, the enhancement of the discrimination is dramatic for large backgrounds. 
\subsection{CYGNO-30 C.I. 90\%}
\label{subsec:limits}
From the estimations of the upper bound on the \W induced events $\mu_s$ from the 90\% C.I., it is possible to evaluate the upper bound on the WIMP-nucleon elastic cross section employing Equation \ref{eq:totnumSImoretargets} and adopting $N_{DMevt}=\mu_{s,90\%}$.
\subsubsection{CYGNO-30 Spin Independent limit}
\label{subsubsec:limit_SI}
\begin{figure}[!t] 
	\centering
	\includegraphics[width=0.8\textwidth]{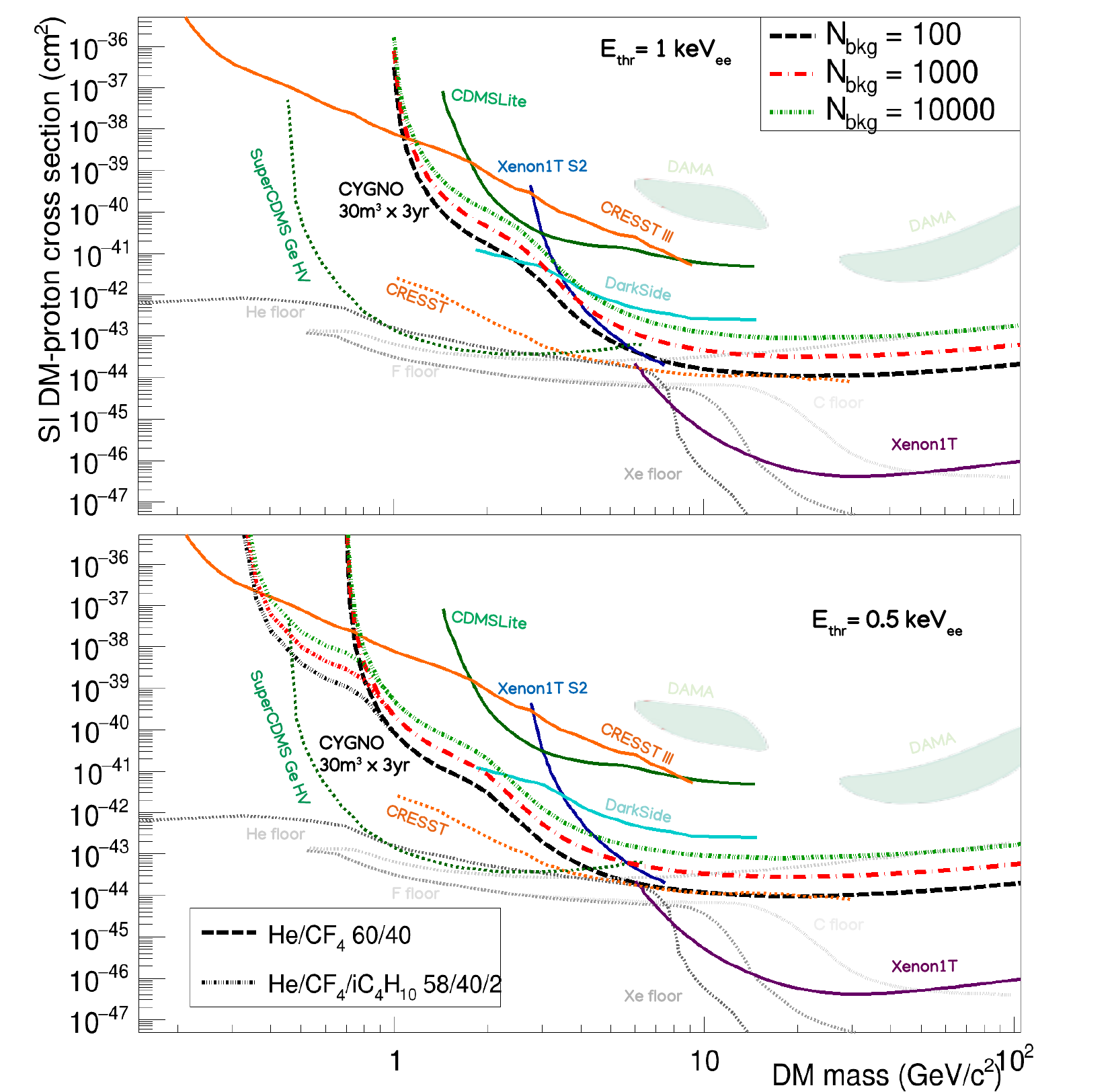} 
	\caption{Spin-independent 90\% C.I. for WIMP-nucleon cross section for 30 m$^3$ CYGNO detector for 3 years of exposure with different background level assumptions and an operative threshold of 1 keV$_{\text{ee}}$ (\textbf{top} plot) and 0.5 keV$_{\text{ee}}$ (\textbf{bottom} plot). The dashed curves correspond to a He:CF$_4$ 60/40 gas mixture with N$_{bkg}$ = 100 (black), 1000 (red) and 10000 (dark green). The dotted curves show the sensitivity for a He:CF$_4$:isobutane 58/40/2 mixture. Current bounds from Xenon1T (violet) \cite{Aprile:2018dbl}, Xenon1T S2 analysis (blue) \cite{bib:Aprile_2019}, DarkSide (cyan) \cite{bib:Agnes_2018}, CRESST III (orange) \cite{bib:Mancuso:2020gnm} and CDMSLite (green) \cite{bib:Agnese_2018} are also shown. The densely dotted curves show the future expected limits of SuperCDMS Ge (green) \cite{SuperCDMS:2016wui} and CRESST (orange) \cite{Willers:2017vae}. The light gray regions denote DM constraints as evaluated from the NR annual modulation reported by the analysis of the DAMA data \cite{bib:Savage_2009}, while the different gray curves show the neutrino background levels for various targets \cite{Boehm:2018sux}.}
	\label{fig:SI}
\end{figure}
Figure \ref{fig:SI} shows in the top part the expected SI limits for a 30 m$^3$ CYGNO experiment for a 3-year exposure with 1 keV$_{\text{ee}}$ energy threshold. The improvements achievable by lowering the threshold to 0.5 keV$_{\text{ee}}$ are shown in the bottom of Figure \ref{fig:SI}, together with the results achievable with the addition of 2\% of isobutane, as discussed in Section \ref{sec:future:CYGNO}.

The shape of the limit mirrors the the probability of detection of each element shown in Figure \ref{fig:limit_prob}. There is a kink on the curve at around 0.7 GeV/c$^2$ corresponding to the transition from hydrogen-dominated to helium-dominated recoils, and at 3 GeV/c$^2$, from helium to fluorine-dominated recoils through carbon. The carbon percentage on the total gas mixture (8\%) is too low to produce a visible effect on the curve.\\
Figure \ref{fig:SI} shows how all the scenarios considered in this sensitivity evaluation will be able to probe regions in WIMP masses versus cross section planes not yet explored, therefore significantly contributing to future DM searches for low WIMP masses. While it is true that the expected reach of future SuperCDMS \cite{SuperCDMS:2016wui}, CRESST \cite{Willers:2017vae}, Darkside low-mass \cite{Darksidelowmass} and NEWS-G experiments may be able to cover these regions, all of these will be realised through modes of operation that strongly reduce (if not even completely give up) tools for background discrimination. Strict requirements on the radio purity of materials and the estimation of the expected backgrounds are necessary to accomplish this goal. As a matter of fact, any observed signal in this region by these experiments will be difficult to interpret unambiguously as a DM signal, whilst CYGNO's exploitation of the directional properties of the recoils constitutes a crucial and decisive test for a positive identification of DM.

Considering the discussion on the expected backgrounds for CYGNO-04 in Section \ref{subsubsec:priorback} and the foreseen one year of livetime, the sensitivity of PHASE\_1 can be approximately inferred by scaling of a factor 225 (75 for the active volume ratio times 3 for the number of years) the CYGNO-30 limits of Figure \ref{fig:limit_prob}. In this simplified scenario, CYGNO-04 can result competitive with the current most stringent limit at low 1-5 GeV/c$^2$ WIMP masses in the best background scenario. In addition, it will be able to completely explore with robust directionality capabilities the regions corresponding to the DAMA/LIBRA modulation observation, even in the most pessimistic background scenario.
\subsubsection{CYGNO-30 Spin Dependent limit}
\label{subsubsec:limit_SD}
\begin{figure}[!t]
	\centering
	\includegraphics[width=0.8\textwidth]{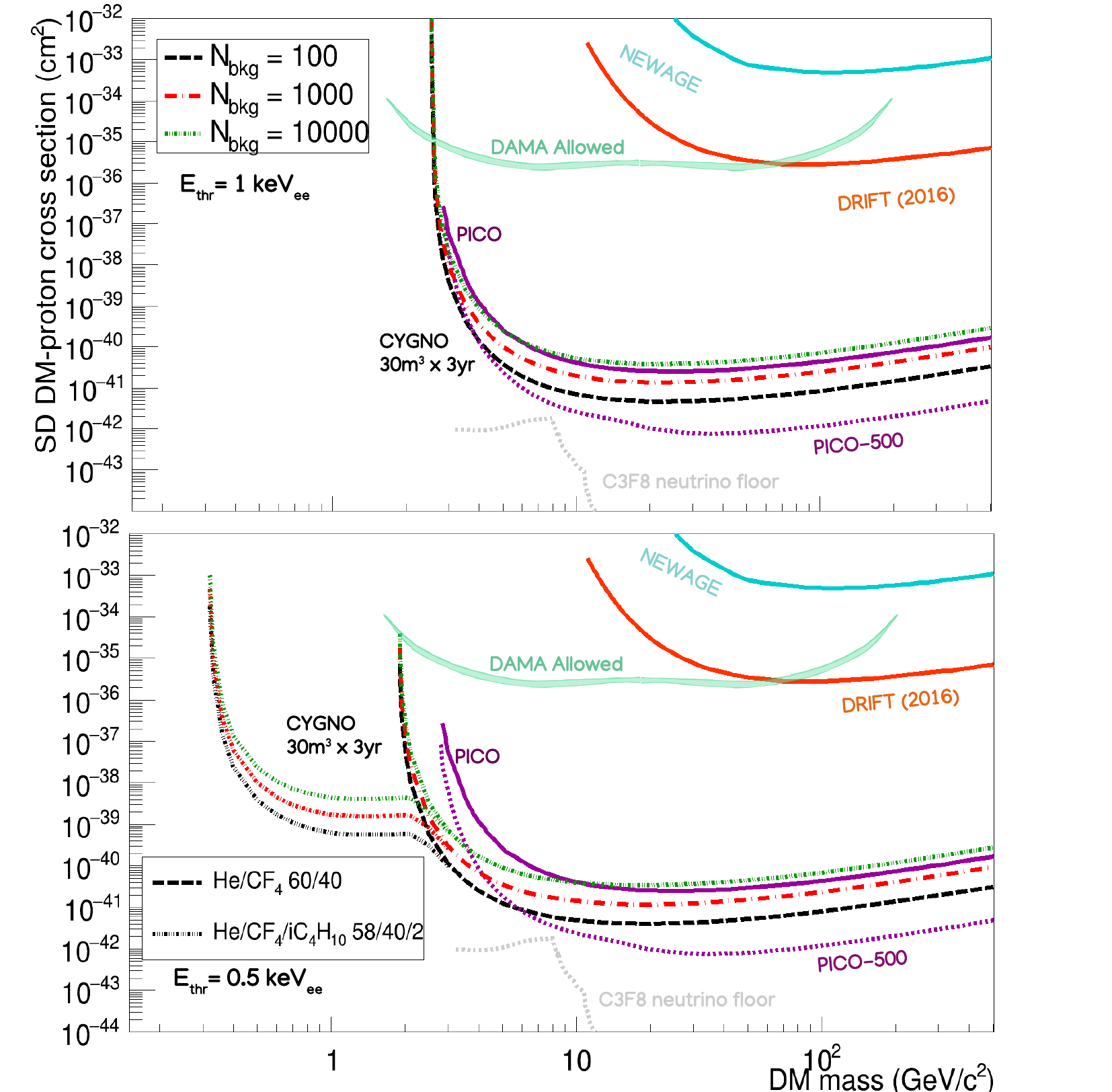} 
	\caption{Spin-dependent 90\% C.I. for WIMP--proton cross sections for 30 m$^3$ CYGNO detector for 3 years of exposure with different background level assumptions and an operative threshold of 1 keV$_{\text{ee}}$ (\textbf{top} plot) and 0.5 keV$_{\text{ee}}$ (\textbf{bottom} plot). The dashed curves correspond to a He:CF$_4$ 60/40 gas mixture with N$_{bkg}$ = 100 (black), 1000 (red) and 10000 (dark green).  The dotted curves show the sensitivity for a He:CF$_4$:isobutane 58/40/2 mixture. Current bounds from PICO (purple) \cite{bib:Amole_2019}, DRIFT (orange) \cite{bib:drift2}, and NEWAGE (cyan) \cite{bib:yakabe2020limits} are also shown. The allowed region estimated from the NR annual modulation reported by analysis on  DAMA data is denoted by the light green band \cite{bib:Savage_2004}. The~light gray dotted line representing the neutrino floor for C$_3$F$_8$ is also taken from PICO \cite{bib:Amole_2019}.} 
	\label{fig:SD}
\end{figure}
As stated in Section \ref{sec:WIMPparadigm}, fluorine and hydrogen permit CYGNO to be efficiently sensitive to SD coupling. Figure \ref{fig:SD} shows in the top part the expected SD limits for a 30 m$^3$ CYGNO experiment for a 3-year exposure with 1 keV$_{\text{ee}}$ energy threshold. The improvements achievable with an operating threshold of 0.5 keV$_{\text{ee}}$ are shown in the bottom of Figure \ref{fig:SD}, together with the results obtained adding 2\% of isobutane, as discussed in Section \ref{sec:future:CYGNO}.
The strong coupling, the large amount of fluorine and the light nature of hydrogen (in the assumption of the addition of isobutane to the gas mixture) allow CYGNO-30 to be able to explore wide regions not yet excluded by the PICO experiment in the low background scenario.
The PICO experiment, which possesses the strongest sensitivity among all the existing and planned experiments exploring SD coupling, is based on an energy threshold approach. Hence, also in this context, a confirmation of the Galactic origin of the detected signal is lacking for PICO while is available for CYGNO. 

In the context of directional detectors, it is relevant to note how the CYGNO project is expected to strongly outperform all other directional competitors discussed in Section \ref{subsubsec:dirgas}. By the scaling argument illustrated at the end of Section \ref{subsubsec:limit_SI}, it can be easily inferred from Figure \ref{fig:SD} how already CYGNO-04 will be able to improve of nearly two orders of magnitude current directional limit on SD coupling, fully covering the contour region identified by the NR annual modulation measured by DAMA/LIBRA.\newpage
\section{Discrimination between dark matter models}
\label{sec:discr_2models}
As illustrated in Section \ref{subsubsec:dirastronomydm}, directional DM detectors possess not only the ability to make a positive claim of detection, but are also able to identify the nature of DM  with far greater efficiency than their non-directional counterparts.  To explicitly quantify this feature,  directional and non-directional searches are compared in the capability to  discriminate between the classic WIMP scenario and a second considerably different DM model able to produce very similar detector response in terms of NR energy spectrum, but with significant differences in the angular distribution. The alternative paradigm considered here is a specific realisation of a dark sector that couples to the SM through a dark photon, developed by the authors of \cite{DeRocco_2019,DeRocco_2019_2}. In this model, light DM candidates of MeV/c$^2$ masses can be produced by Supernovae and emerge from them with semirelativistic momenta, large enough to be detected in existing and proposed WIMP detectors. For these reasons this model  will be henceforth referred to as SuperNova dark matter (\SN). The larger velocity and smaller mass allows \SN particles to perfectly fit in the purpose of this analysis (see more details in Section \ref{subsec:SNDM}). The minimum number of detected signal events needed to discriminate between WIMPs and SNDM for two toy experimental setup (a directional and a non-directional one) is evaluated in the following in order to explicitly demonstrate the power of the angular spectrum over the energy distribution to probe the DM nature.
\subsection{Supernova dark matter model}
\label{subsec:SNDM}
\subsubsection{Flux of \SN}
\label{subsec:fluxSNDM}
In this Section, it is shown that the production by supernovae of an approximately constant-in-time but highly anisotropic flux of semirelativistic DM is a generic feature of DM models on the \Mevc scale with suitably strong coupling to the Standard Model (SM).

As was first pointed out in \cite{DeRocco_2019_2}, Galactic supernovae can produce a flux of hot \Mevc-scale DM at Earth that is roughly constant in time. Note that while \cite{DeRocco_2019_2} treats a specific example model, this is in fact a general feature of models with new degrees of freedom on the \Mevc scale. Since the temperature of a core-collapse supernova (SN) can reach upwards of 30 MeV, supernovae will produce a large thermal flux of particles with masses up to hundreds of MeV (at which point the flux becomes heavily Boltzmann-suppressed). If the new degree of freedom is coupled sufficiently strongly to the SM, then it becomes diffusively trapped within the protoneutron star (PNS), remaining in thermal contact with the SM bath out to some radius (the "energy sphere" $r_E$). Akin to the SN neutrinos, the DM flux will then approximate blackbody emission from this sphere with a temperature set by the local temperature at the energy sphere. (See \cite{DeRocco_2019_2} for further details.)

For sufficiently massive particles ($m_X > T(r_E)$, with $m_X$ the mass of the new particle), the escaping flux will be semirelativistic with a velocity distribution approximately Maxwell-Boltzmann at the temperature of the energy sphere. The Maxwell-Boltzmann distribution exhibits a roughly order-one spread in velocities (i.e. $\sim75\%$ of the distribution lies between $v = 0.5\bar{v}$ and $v = 1.5\bar{v}$ where $\bar{v}\approx \sqrt{T/m_X}$ is the average speed). This spread in velocities leads to a spread in arrival times of these particles at Earth. For a supernova a distance $d$ away, the spread in arrival time is $\Delta t \approx (d/\bar{v})\delta v$ with $\delta v \equiv \frac{\Delta v}{v} \approx 1$.

Consider a supernova that occurs in the centre of the Milky Way ($\sim 3000$ light years away). The DM flux will be produced over $\sim 10$ seconds (the cooling timescale of the PNS), but the arrival on Earth of the bulk of the flux will take place over $\sim 3000(c/\bar{v})$ years. Since Type II supernovae are predicted to occur in the Galaxy at a rate of roughly 2 per century \cite{Beacom_SN}, the emissions of $\gtrsim 100$ SN will all be overlapping at Earth at any  given time. These overlapping emissions produce what can be called the "diffuse galactic flux" of hot dark matter produced in supernovae (SNDM).

While the term ``diffuse'' is used to indicate that this flux is approximately constant in time, it is \textit{not} isotropic. It is instead very strongly peaked towards the centre of the Galaxy, which is where the majority of supernovae take place. To quantify this, the double-exponential profile of \cite{Adams_2013} is utilised for the core-collapse SN density rate in our galaxy:
\begin{equation}
\frac{dn_{SN}}{dt} = A e^{-r/R_d} e^{-|z|/H}
\end{equation}
with $R$ the galactocentric radius and $z$ the height above the Galactic mid-plane. Earth sits at $R_E = 8.7$ kpc and $z_E = 24$ pc. The parameter values are provided by \cite{Adams_2013} for Type II SN: $R_d = 2.9$ kpc, $H =$ 95 pc. The SN rate is normalised to two SN per century.

\begin{figure*}[!t]
	\centering
	\includegraphics[width=0.9\textwidth]{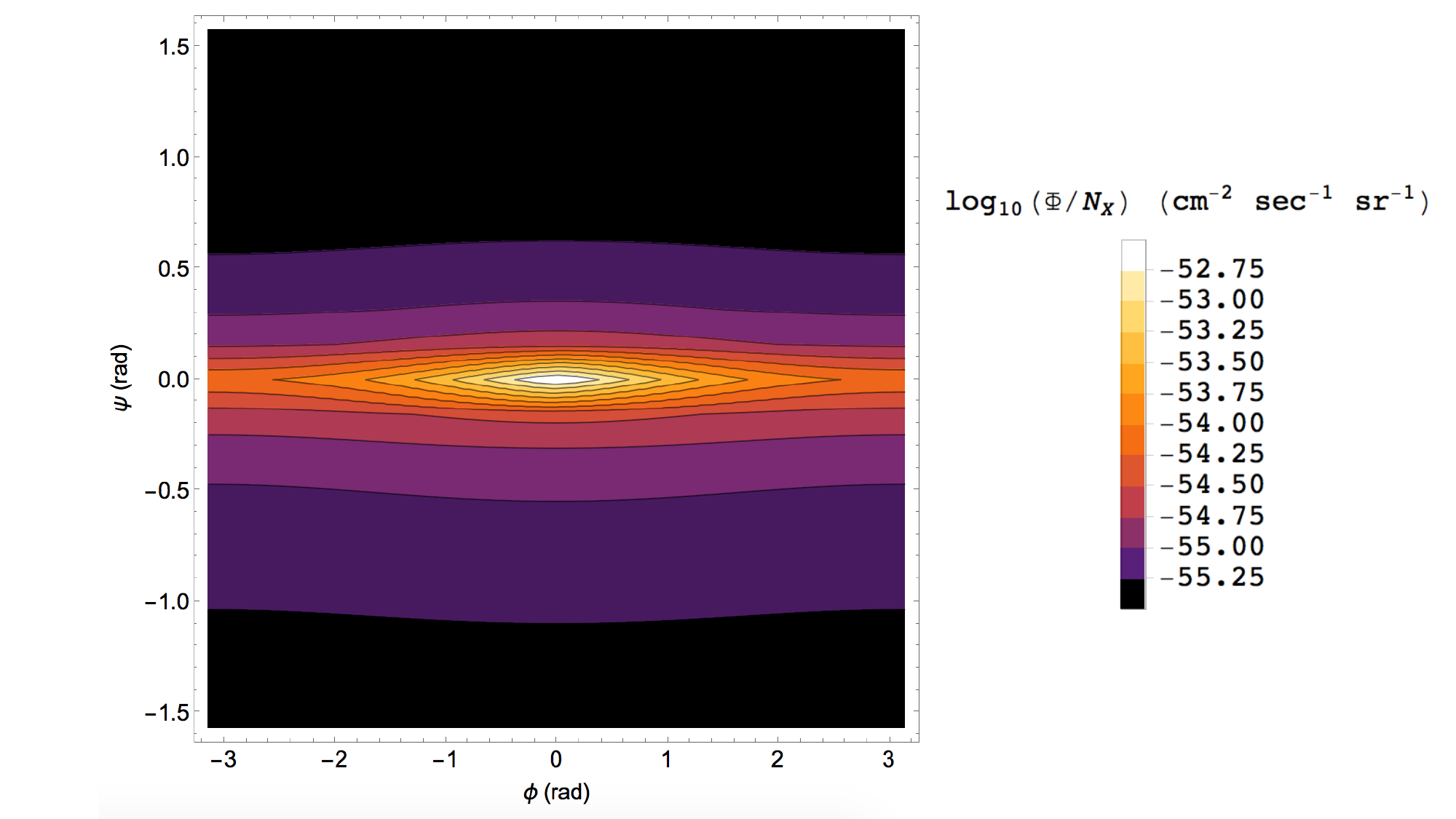}
	\caption{Sky map for the flux of light DM produced by Galactic SNe. The scale has been normalized by $N_X$, the total number of DM particles produced in a single supernova. It is evident that the increased rate of SNe in the Galactic centre results in a large flux from that region. Note that the expected flux looking directly out of the plane of the Galaxy is four orders of magnitude smaller than the flux coming from the Galactic centre.}
	\label{fig:skymap}
\end{figure*}

The flux along a given line of sight is computed by performing the following integration:
\begin{multline}
\Phi(\psi, \phi) = \\
N_{X}\int_0^{\infty} \frac{dn_{SN}}{dt}\biggm\lvert_{\left\{\begin{subarray}\\r = \sqrt{r_E^2+(l\cos\psi)^2-2r_E(l\cos\psi\cos\phi)},\\
	z = z_E + l\sin\psi\end{subarray}\right\}}~dl
\end{multline}
with $\psi$ and $\phi$ the polar and azimuthal angles respectively and $N_X$ the total number of SNDM particles emitted by a single SN. In Figure \ref{fig:skymap}, the result of this computation is displayed with $N_X$ divided out, as it simply adjusts the normalisation.
It can be seen that the flux of DM due to supernovae is strongly peaked towards the Galactic centre. The flux from this region exceeds that of the flux coming from directly out of the plane by four orders of magnitude.

Note that for this entire discussion, the focus was on supernovae occurring within the Galaxy. It is natural to ask about whether there is an isotropic contribution to this flux due to extragalactic supernovae, which is, for example, how the diffuse supernova neutrino background is formed \cite{Beacom_SN}. However, the neutrinos are all travelling at $c$, which means that they experience no time-spreading effect. For massive particles, the time-spreading effect becomes so large on extragalactic scales that the corresponding flux, even integrated out to high redshift, is subdominant to the Galactic flux. Another line of sight integral can be performed as specified by \cite{Beacom_SN}:
\begin{equation}
\Phi = N_X\int_0^{\infty} R_{SN}(z)\left|\frac{dt}{dz}\right|~dz
\end{equation}
where $R_{SN}(z)$ is the redshift-dependent Type II supernova rate (taken from \cite{Beacom_SN}) and $(\frac{dt}{dz})^{-1} = H_0 (1+z) \sqrt{ \Omega_{\Lambda} + \Omega_{m} (1+z)^3}$. Performing this integral and dividing by $N_X$ as above, it is valid that $\log_{10}(\Phi/N_X) \approx -56.3$. Note that this is an order of magnitude below even the weakest line of sight for the Galactic contribution (directly out of the plane). As a result, the isotropic extragalactic contribution is ignored for the remainder of the study.

To summarise the preceding discussion, the production of an anisotropic constant flux of high-momentum particles by Galactic SNe is a generic feature of any DM model with mass $\mathcal{O}(10)-\mathcal{O}(100)$ \Mevc and coupling to the SM sufficient to diffusively trap the particle within the PNS out to $\sim 10-20$ km.
\subsubsection{Scattering kinematics and \W comparison}
\label{subsec:SNDMkin}
In the standard WIMP picture, WIMPs have masses of $\mathcal{O}(10) - \mathcal{O}(100)$ \Gevc \cite{Schumann_2019}, three orders of magnitude greater than the SNDM discussed in the previous Section. However, direct detection experiments built to target WIMPs are actually sensitive to SNDM as well. This is due to the simple fact that the cosmological abundance of WIMPs travel at the Galactic virial velocity ($v\approx 10^{-3} c$) while the SNDM travels at some order-one fraction of $c$. The semi-relativistic velocity of the SNDM compensates for its lighter mass and results in a similar momentum to a cold \Gevc-scale WIMP.

Equation \ref{eq:relativ_kinematics} displays the momentum of a recoiling target nucleus in a DM detector with no assumptions that the momentum transfer is non-relativistic. In the case of \Mevc-scale DM travelling semirelativistically, it is valid that $m_A \gg m_{\chi}, p_0$. Hence:
\begin{equation}
\label{eq:q_SNDM}
q \approx 2 p_0 \cos\theta_r.
\end{equation}
It can immediately be noted that combining Equation \ref{eq:q_SNDM} with the equivalent for \Ws (Equation \ref{eq:nonrelativ_kinematics}), it can be deduced that a nucleus will have the same recoil momentum if hit by a \W or a \SN particle if 
\begin{equation}
\label{eq:SNDM_WIMP_equiv}
p_0=v\mu_A
\end{equation}
with $v$ being the velocity of the incoming \W.

This poses a significant challenge. One would wish to be able to discriminate between a cold \Gevc-scale WIMP and hot SNDM if some signal were to be detected. Unfortunately, since $\mu_A v$ in the WIMP scenario and $p_0$ in the SNDM scenario can be of comparable order, even if one were able to generate large statistics on the recoil energy distribution in a detector, it would be very difficult to discriminate between the two models. Differences will show up in the recoil spectrum due to differences in the shape of the incoming momentum spectrum, but both are approximately Boltzmann and there is considerable uncertainty on the WIMP velocity distribution \cite{RevModPhys.85.1561} and momentum spectrum of the SNDM that could limit discrimination. Just how difficult it is to discriminate these models with energetic information alone will be quantified in Section \ref{sec:results}.

The best discriminator is the fact that, as will be explicitly shown in the following Sections, the SNDM is highly anisotropic, with a steep peak towards the Galactic centre. WIMPs, in contrast, appear to originate from the Cygnus constellation (see Section \ref{sec:directional}). This means that the signals from SNDM and from a cosmological abundance of WIMPs would be \textit{perpendicular}. In this way, directional detection can allow to discriminate these two populations with a very small number of events.
\subsubsection{\SN model}
\label{subsec:SNDMmod}
In order to evaluate the expected signal from the SNDM in a fiducial experiment (introduced in Section \ref{subsec:SNDMexp_par}), a specific example model of SNDM that produces the features discussed in Section \ref{subsec:SNDMkin} will be described. (This is the same model as used in \cite{DeRocco_2019_2}.) 

Namely, it is a dark sector with a Dirac fermion coupled to the SM via the four-fermion operator
\begin{equation}
\label{eq:interaction}
\mathcal{L} \supset \frac{e \epsilon g_d}{\Lambda^2} \bar \chi \gamma_\mu \chi J^\mu_\text{em}
\end{equation}
with $\chi$ the dark matter and $J^\mu_\text{em}$ the electromagnetic current of the SM. This can be viewed as the case wherein the dark sector is coupled to the SM by a dark photon with mass $\Lambda$ and kinetic mixing $\epsilon$ and a dark charge $g_d$. $\Lambda$ is taken as $\gg \mathcal{O}(10)$ MeV large such that it is not produced on-shell in the SN. The coupling is further parametrised to the SM by the variable $y\equiv \frac{\epsilon^2 g_d^2}{4\pi} \left(\frac{m_{\chi}}{\Lambda}\right)^4$ \cite{Izaguirre:2015yja}.

As a result of this choice of coupling term, the predominant interaction that keeps the dark sector thermally coupled to the SM is scattering with electrons. It is the radius at which this interaction ceases to become efficient that the temperature of the escaping DM is set (the "energy sphere" described in Section \ref{subsec:fluxSNDM}). This is defined formally by finding the radius $r_E$ at which the optical depth for this interaction ($\tau_E$) is 2/3:
\begin{equation}
\label{eq:opticaldepth}
\tau_E|_{r=r_E} \equiv \int_{r_E}^{\infty} \sqrt{\lambda_{\chi e}^{-1}(r)[\lambda_{\chi p}^{-1}(r) + \lambda_{\chi e}^{-1}(r)]}~dr = \frac{2}{3}
\end{equation}
where
$\lambda_{\chi p}$
and 
$\lambda_{\chi e}$
are the mean free path for DM scattering with protons and electrons respectively.

This condition simply provides a mathematical prescription to determine at what radius the SNDM decouples from the SM thermal bath. At radii $r < r_E$, the SNDM is undergoing rapid scatters off of electrons, which allows the SNDM population to stay in thermal equilibrium with the SM. However, at radii $r > r_E$, the electron density has dropped to the point at which the average SNDM particle will escape to infinity without undergoing any more electron scatters. The optical depth computed in Equation \ref{eq:opticaldepth} is roughly the number of scatters expected for a SNDM particle to undergo as it escapes. For this reason, when it drops below unity (or, more formally, 2/3 \cite{Raffelt:2001kv}), this simply means that SNDM particles emitted at that radius will not exchange energy with the SM bath and will not be thermally coupled to the SM. Instead, their temperature will be set at the last radius at which they \text{were} thermally-coupled to the SM, which is just the definition of the "energy sphere". So, as stated previously, it is the local temperature of the SM bath at this energy sphere that sets the temperature of the escaping SNDM flux, which in turn sets the momentum spectrum of incident SNDM at an Earth-based detector.\footnote{Note that the proton scattering appears in the formula as the SNDM is still undergoing diffusive scatters off of the protons that do \textit{not} exchange energy even once the electron scatters have become inefficient.} For further details, see \cite{DeRocco_2019_2}.
\subsubsection{\SN recoil spectrum }
\label{subsec:SNDMrate}
The spectra of the recoils induced by \SN particles can be computed starting from the same conditions in Equation \ref{eq:rate2}:
\begin{equation}
\label{eq:rateSN1}
\frac{dR}{dq^2d\Omega} = \frac{N_0}{A}\frac{d\sigma}{dq^2d\Omega} v \frac{\rho_0}{m_{\chi}}f(\vec{v})d^3v = \frac{N_0}{A}\frac{\rho_0}{m_{\chi}} \frac{d\sigma}{dq^2}\frac{1}{2\pi}\int \delta\left(\cos\theta-\frac{q}{2p}\right) v f(\vec{v})d^3v,
\end{equation}
where the kinematic condition has been modified accordingly.\\
As the \SN particles are semi-relativistic, the most general representation of their momentum distribution in natural units is the Fermi-Dirac's:
\begin{equation}
\label{eq:SNDM_distr}
f(\vec{p})d^3p = f(p)d^3p = \frac{\beta}{e^{\frac{\sqrt{p^2+m^2}}{T}}+1}
\end{equation}
with $T$ the temperature of the thermal equilibrium reached inside the region in the SN collapse at the moment of decoupling, and $\beta$ a normalisation factor. As the interest is to express the momentum distribution at the laboratory, the gravitational effects these particles are subject to during the escape from the SN must not be neglected. As described in \cite{Deroccowork}, the gravitational redshift changes the particles momentum as follows:
\begin{equation}
\label{eq:SNDM_pstar}
p_*=\sqrt{\frac{p^2+2\Phi m^2}{1-2\Phi}}
\end{equation}
with $p_*$ the momentum on the surface of the star. $\Phi$ is the redshift parameter which depends on the star mass, being $\Phi \equiv \int_{r_E}^{\infty}\frac{m_{enc}(r)}{r}dr$, where $r_E$ is the energy sphere and $m_{enc}$ is the mass enclosed in a radius $r$. As a consequence not all particles can escape the SN core, but a minimum momentum is required. As a consequence, the modified distribution function of the \SN particles momentum which arrive on Earth is
\begin{equation}
\label{eq:SNDM_pstardistr}
f_*(p)=
\begin{cases}
p<0 & 0\\
p>0  &\frac{1}{e^{\frac{\sqrt{\frac{p^2+m^2}{1-2\Phi}}}{T}}+1}
\end{cases}
\end{equation}

Due to the high degree of anisotropy in the angular distribution discussed above, the flux of particles is approximated as being produced from a point source at the Galactic centre. In addition, since the semi-relativistic speed of the \SN particles is much larger than the velocity of the Sun, no transformation of coordinates is required as the laboratory is still with respect to the \SN particles. Applying all these considerations into Equation \ref{eq:rateSN1}, the spectra of \SN induced recoils is:
\begin{equation}
\label{eq:rateSNDM}
\frac{dR}{dEdcos\theta}\propto \frac{1}{(1-2\Phi)^{3/2}}\int S(E)\delta(cos\theta-\frac{\sqrt{2m_AE}}{2p})\sqrt{\frac{p^2+2\Phi m_{\chi}^2}{p^2+m_{\chi}^2}}f_*(p)dp
\end{equation}
For a more detailed calculation the reader is sent to Appendix \ref{app:calcrateSNDM}.
\subsection{Frequentist statistical approach}
\label{subsec:frequentist}
For the discrimination of two models of DM, the standard frequentist approach with the likelihood ratio test is opted for because very robust and reliable.\\
The frequentist inference statistics is well described in text books as \cite{bib:young_smith_2005,bib:Taylor}.\\
Let assume a generic set of n independent quantities $\vec{x}$ are used to describe the state of a system or the outcome of a measurement. $f(\vec{x})$ is defined as the probability density function for which the following characteristics are valid:
\begin{itemize}
	\item $f(\vec{x})d\vec{x}$ is the normalised occurrence of the state to have the vector ${x_1,x_2,..,x_n}$ in a multidimensional infinitesimal volume $\{[x_1,x_1+dx_1],[x_2,x_2+dx_2],...,[x_n,x_n+dx_n]\}$.
	\item $\int_{D}^{}f(\vec{x})d\vec{x}=1$, where $D$ is the domain of the $f$.
\end{itemize}
The normalised occurrence of an event in the first item can be interpreted as the probability of the state to be found in that particular region of the parameter space, but this approach is not interested into the probability of an event, but on its possible repetition. For example, when considering a mono-dimensional problem, if $\int_{0}^{10}f(x)dx=0.4$ it means that if the same experiment were repeated an infinite amount of times, the outcome of the measurement would be 40\% of the times in the range between 0 and 10, but states nothing about the probability of the underlying phenomenon.\\
Considering experimental measurements it is very common to define the likelihood function $\Likeli=\Likeli(\vec{x},\vec{a},H)$, which is a function that represents the likelihood of the data set with a theoretical model that is expected to describe system considered. $\vec{x}$ is the vector representing the data, $\vec{a}$ the one representing parameters of the model and $H$ is the hypothesis under which $\vec{a}$ depicts the model. In general, by definition, the higher the likelihood the more the data agree with the model.\\
\begin{figure}[t] 
	\centering
	\includegraphics[width=0.8\linewidth]{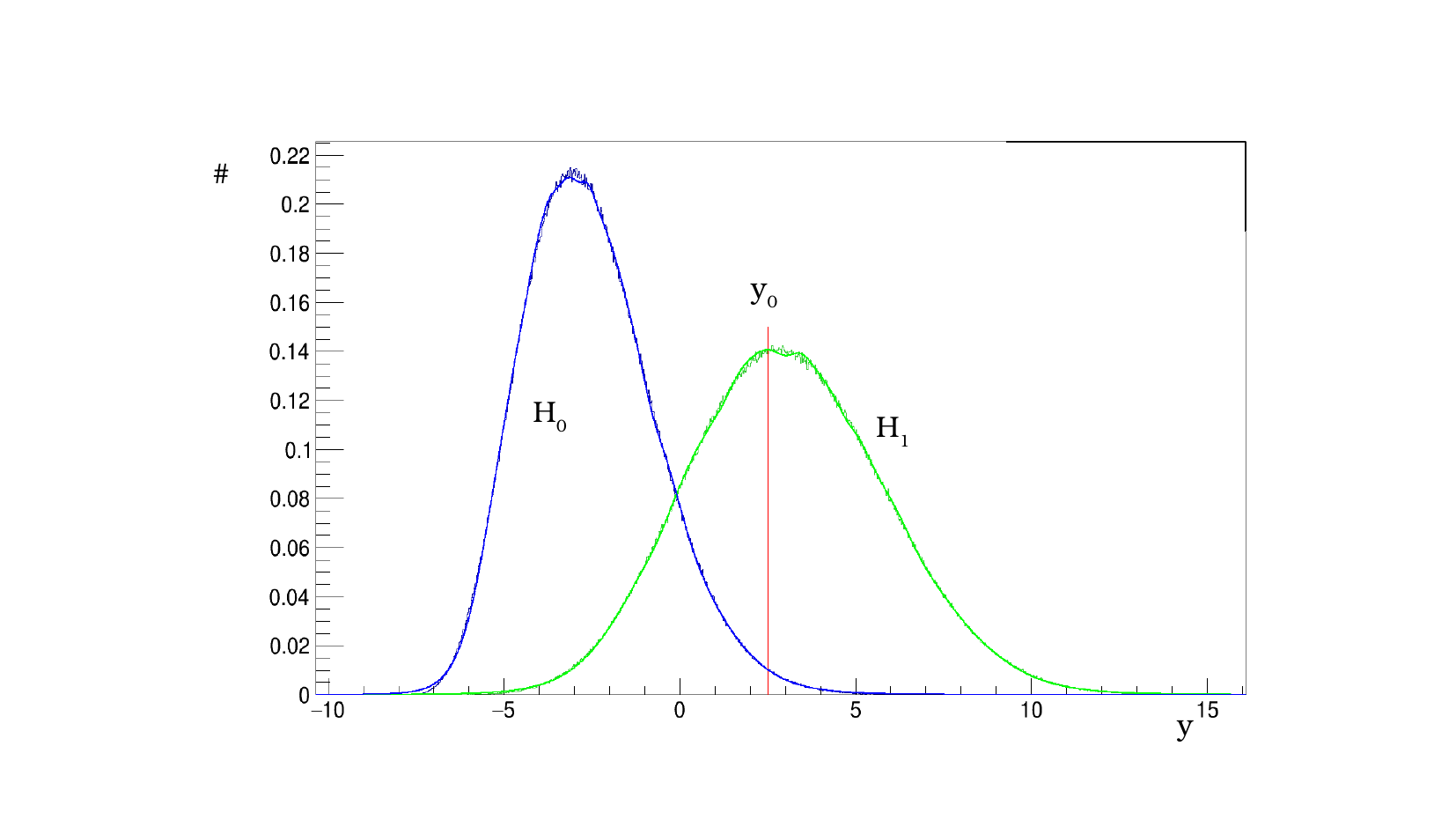}
	\caption{Mock example of the possible density distributions of a discriminating variable $y$ in case of two hypotheses $H_0$ and $H_1$. The distribution in blue is the one that follows the hypothesis $H_0$, while the green one follows $H_1$. The red line represents the $y_0$, value chosen as the discrimination.}
	\label{fig:exfreq}
\end{figure}

When trying to discriminate between two different models, once a set of events is measured, the two hypotheses are named $H_0$ and $H_1$. The analysis consists in finding a variable (or a set of variables) $y=y(\vec{x})$ so that its density distributions for the two models, $g_{H_0}(y)$ and $g_{H_1}(y)$, are as separated as possible, i.e. with the least part the domain of y for which both $g_{H_0}$ and $g_{H_1}$ are different from zero. Once the distributions are set, a value of $y=y_0$ determines the separation between the acceptance of hypothesis 0 and 1. Figure \ref{fig:exfreq} represents a mock example of what described up to now. The blue distribution is the $g_{H_0}$, the green one is $g_{H_1}$, the red vertical line is $y_0$. If the measurement $\vec{x}$ will result in $y<y_0$ then the preferred model given the data will be $H_0$. 

Considering the example in Figure \ref{fig:exfreq}, the following definitions can be described:
\begin{itemize}
	\item Error Type I: $a_I=\int_{y_0}^{+\infty}g_{H_0}(y)dy$, the probability of choosing $H_1$, when $H_0$ is the truth. This determines the significance of a test.
	\item   Error Type II: $a_{II}=\int_{-\infty}^{y_0}g_{H_1}(y)dy$, the probability of choosing $H_0$, when $H_1$ is the truth. 1-$a_{II}$ defines the power of a test.
\end{itemize}
In literature, one of the most robust and easiest to evaluate variable is the likelihood ratio test defined as:
\begin{equation}
\label{eq:likelihodd_ratio}
\lambda=\frac{\Likeli_{H_1}}{\Likeli_{H_0}}
\end{equation}
with $\Likeli_{H_0}$ the likelihood function evaluated on the data that follows the assumptions of the hypothesis $H_0$, and $\Likeli_{H_1}$ the likelihood function evaluated on the data that follows the assumptions of the hypothesis $H_1$. Since the logarithmic operation does not add maxima or minima to a function, very often the loglikelihood ratio test is utilised. It is important to remember that $\Likeli_{H_0}$ and $\Likeli_{H_1}$ both depend on the parameters of the model chosen.
\subsection{Experimental parameters}
\label{subsec:SNDMexp_par}
To perform the analysis, it is necessary to introduce some experimental parameters typical of the DM direct detectors such as energy resolution, energy threshold, and target. To keep the argument of this comparison general enough, the experiments taken into consideration need to have an energy measurement capability and a DM mass sensitivity in a similar range. 
\paragraph{Target material} Given what has been discussed in the previous Sections, a liquid-Xe dual phase TPC and a gaseous TPC with He:CF$_4$ at 1 atm will be considered as benchmark experimental models in this study. The choice of the former is dictated by the observation that this is the leading approach in WIMP searches above 10 \Gevc masses and represent the largest existing realisations of a DM detector. It should be noted that similar experiments employing liquid argon, while still currently limited to 50 kg mass \cite{Agnes:2018fwg}, have demonstrated improved capabilities for ER/NR discrimination with respect to Xe-based detectors and are working on the realisation of a 50-tonne detector with timelines comparable to the xenon approach \cite{Aalseth:2017fik}. Nonetheless, the simplified approach employed in this analysis to the problem assumes zero ER background and, being based on the experimentally measured energy profile of the events, can be easily scaled between Xe and Ar by taking into account the differences in the momentum transferred to the nuclei due to the different masses.

Although inherently challenging, gaseous TPCs potentially provide the best architecture and the best observables for directional dark matter detection. Gaseous TPCs can detect the full 3D electron cloud created by an ionisation event in the active gas and were thoroughly described in Section \ref{subsubsec:dirgas}.

He:CF$_4$ is chosen as TPC gas mixture since at the moment it is the only mixture with simultaneous sensitivity to both spin-independent and spin-dependent couplings that has been demonstrated to have good tracking capability even at 1 atm \cite{bib:cygnoIDBscan}. Moreover, it is interesting to show how very light targets like He are particularly useful not only to explore low WIMP masses, but also in the discrimination between WIMPs and other models like the one discussed in this study, due to helium atoms' high sensitivity to the transferred momentum. For the benchmark TPC full 3D tracking capability is assumed, including track sense determination (head-tail). 
\begin{figure}[!t]
	\centering
	\includegraphics[width=0.65\textwidth]{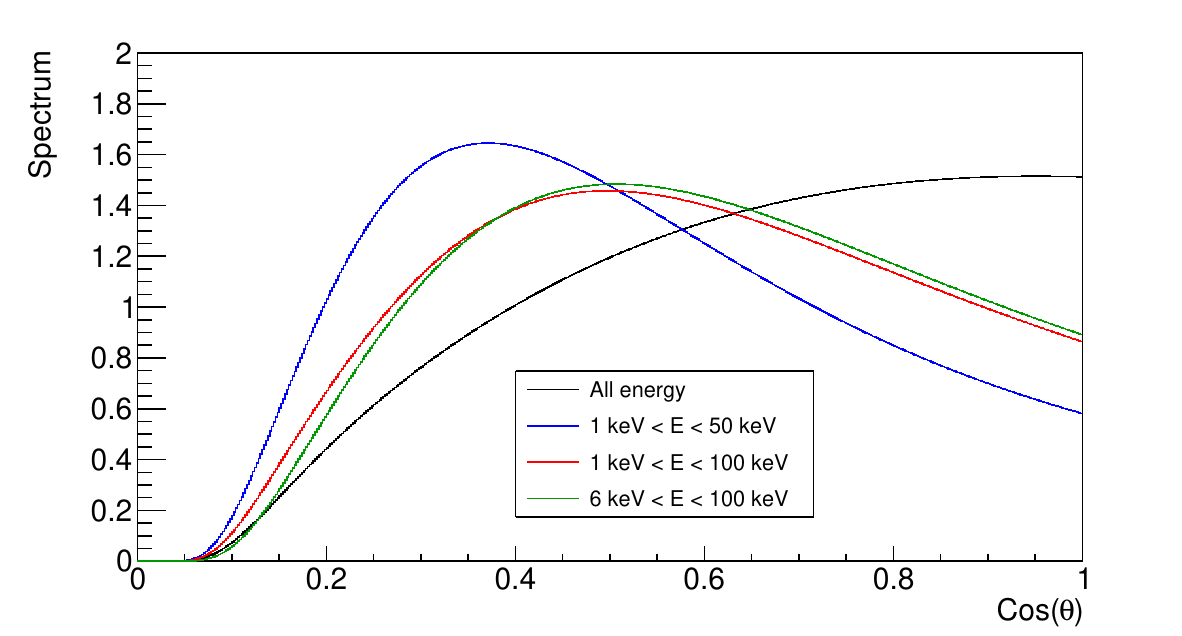}
	\caption{Angular distributions of the fluorine recoils for an example of SNDM model (more details on the calculation of the spectra in Section \ref{subsec:sn_wimp_spectra}).  The different colour represents diverse energy ranges the distribution is evaluated from. The effect of the energy range on the final output is clearly visible as it completely changes the shape.}
	\label{fig:sn_wimpkinemeff}
\end{figure}
\paragraph{Energy range} It is interesting to note that typical WIMP DM searches have not only a lower energy threshold, but also an upper energy bound on the Region Of Interest (ROI). The reason for this can be discerned from Figure 1 of \cite{Aprile:2018dbl}, where the expected Xe nuclear recoil energy spectra for different WIMP masses are shown together with the experimental detection efficiency and energy ROI selection. As can be seen, the upper ROI limit is chosen to nearly match the maximum possible nuclear recoil energy for a 200 \Gevc WIMP mass. However, it is important to  note that the selection on the energy region can strongly affect the shape of the angular recoil distributions as shown in Figure \ref{fig:sn_wimpkinemeff}. The latter displays the one dimensional angular distributions of the fluorine recoils for an example of SNDM model. The different colours represent diverse energy ranges the distribution is evaluated from. The effect of the energy range on the final output is clearly visible.

Given that the goal of this analysis is to evaluate the capability of DM detectors tailored for WIMP searches to discriminate between WIMPs and models such as the one discussed in the previous Sections, a ROI for the Xe-based detector of [4.9, 40,9] keV$_{\text{nr}}$ is employed\cite{bib:Aprile:2019bbb}. For the gas TPC, since no real underground detectors have been operated yet with such configurations, the lower energy threshold is extrapolated to be 5.9 keV$_{\text{nr}}$ for both He and F recoils from the measurements reported in \cite{bib:cygnoIDBscan}, and the results of simulation discussed in \cite{Vahsen:2020pzb}. For the upper energy thresholds, the same assumptions of \cite{Aprile:2018dbl} are utilised and it is extrapolated to be 100 keV$_{\text{nr}}$ for both He and F recoils.

While the chosen lower energy thresholds do not necessarily represent the lowest thresholds achieved by these experimental techniques, values at which electron recoil discrimination is still significantly effective are adopted, since it is decided to work under the assumption of zero background.

\paragraph{Energy resolution} Similarly, the experimental energy resolution are extracted from measurements on actual data. In particular, the Xe-based detector energy resolution dependence is assumed to follow the relation $\sigma_{E}(E) = a/\sqrt{E} + b$ with $a$ and $b$ taken from Table III of \cite{Aprile:2017xxh}. For the gas TPC, the function shown in \cite{Vahsen:2014fba} is adopted describing the relative gain (and therefore energy) resolution as $\sigma_{G}(G) = \sqrt{d^2 + c^2/E}$ with d = (1.94$\pm$0.07), and c = (22.3$\pm$1.5) $\sqrt{\rm{keV}}$, with the constraint to reproduce the 18$\%$ energy resolution at 5.9 keV$_{\text{ee}}$ reported by the He:CF$_4$ detector in \cite{bib:fe55} and 2$\%$ above 50 keV$_{\text{ee}}$ as in typical gas detectors \cite{Vahsen:2014fba}. 

\paragraph{Angular resolution} Gaseous TPC angular resolution is constrained at very low energies (below about 50 keV$_{\text{nr}}$) mainly by multiple scattering, straggling, and diffusion during drift. Given that no measurements exist of angular resolutions from 3D DM TPCs in a realistic regime (in terms of underground operation of detectors of $\mathcal{O}$(1) m$^3$ dimensions), and to be as general as possible, the analysis is performed with a wide range of possible resolutions, spanning from 2$^{\circ}$ to 45$^{\circ}$. The former is an almost perfect angular resolution which represents the ideal case when all angular information is available to the experiment, neglecting the aforementioned diffusion, straggling, and multiple scattering. The latter reflects a scenario of low resolution where a hemisphere in Galactic coordinates is split into a handful of distinct bins. This last assumption is backed up by measurements in the 50-400 keV$_{\text{nr}}$ range by the NEWAGE experiment \cite{Nakamura:2012zza} and is consistent with the simulation studies presented in \cite{Vahsen:2020pzb}.

\paragraph{Background} The fiducial experimental scenarios are assumed to have perfect background rejection. The interest is \textit{not} in the detection of a signal, but in the subsequent discrimination between two models after a discovery. In order not to obfuscate this point, henceforth, the capability of the fiducial experimental setups to fully reject any other background sources through typical analysis techniques is considered perfect. As a corollary of this, it will be far from the interests of this analysis to compare the exposures of the experiments and the specific cross sections of the different models, and instead cross sections and exposures that make discrimination maximally difficult will be selected.

Nonetheless, it is important to stress that, while the details of the background are different for every experimental setup, the energy spectrum of backgrounds in direct detection experiments often highly resembles the spectrum expected from the signal, while the angular distribution does not due to the general isotropy of the background sources or clear directional point source (see Section \ref{sec:directional}).
\subsection{Construction of the energy and angular distributions}
\label{subsec:sn_wimp_spectra}
\begin{figure*}[t]
	\centering
	\includegraphics[width=0.45\textwidth]{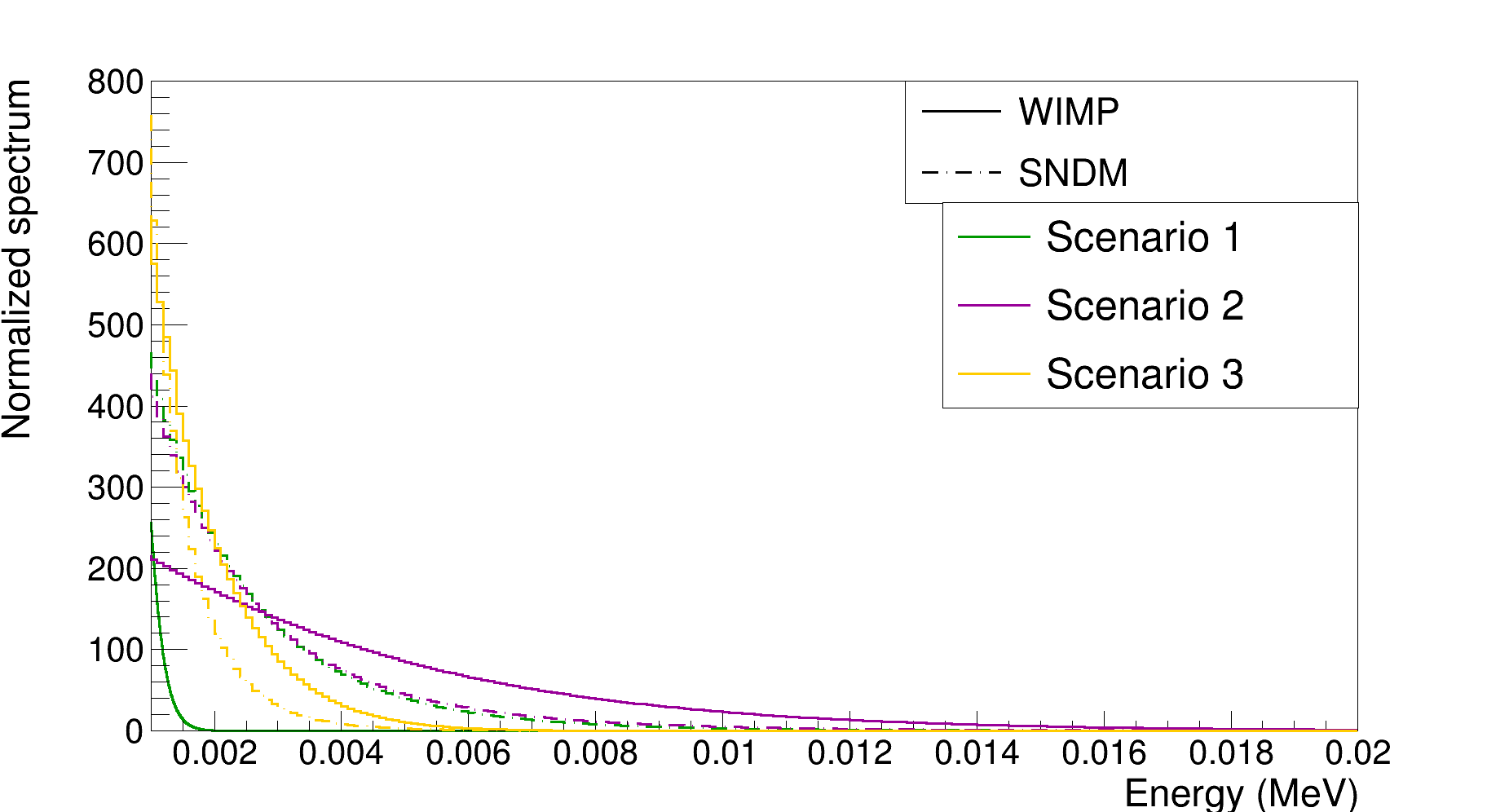}
	\includegraphics[width=0.45\textwidth]{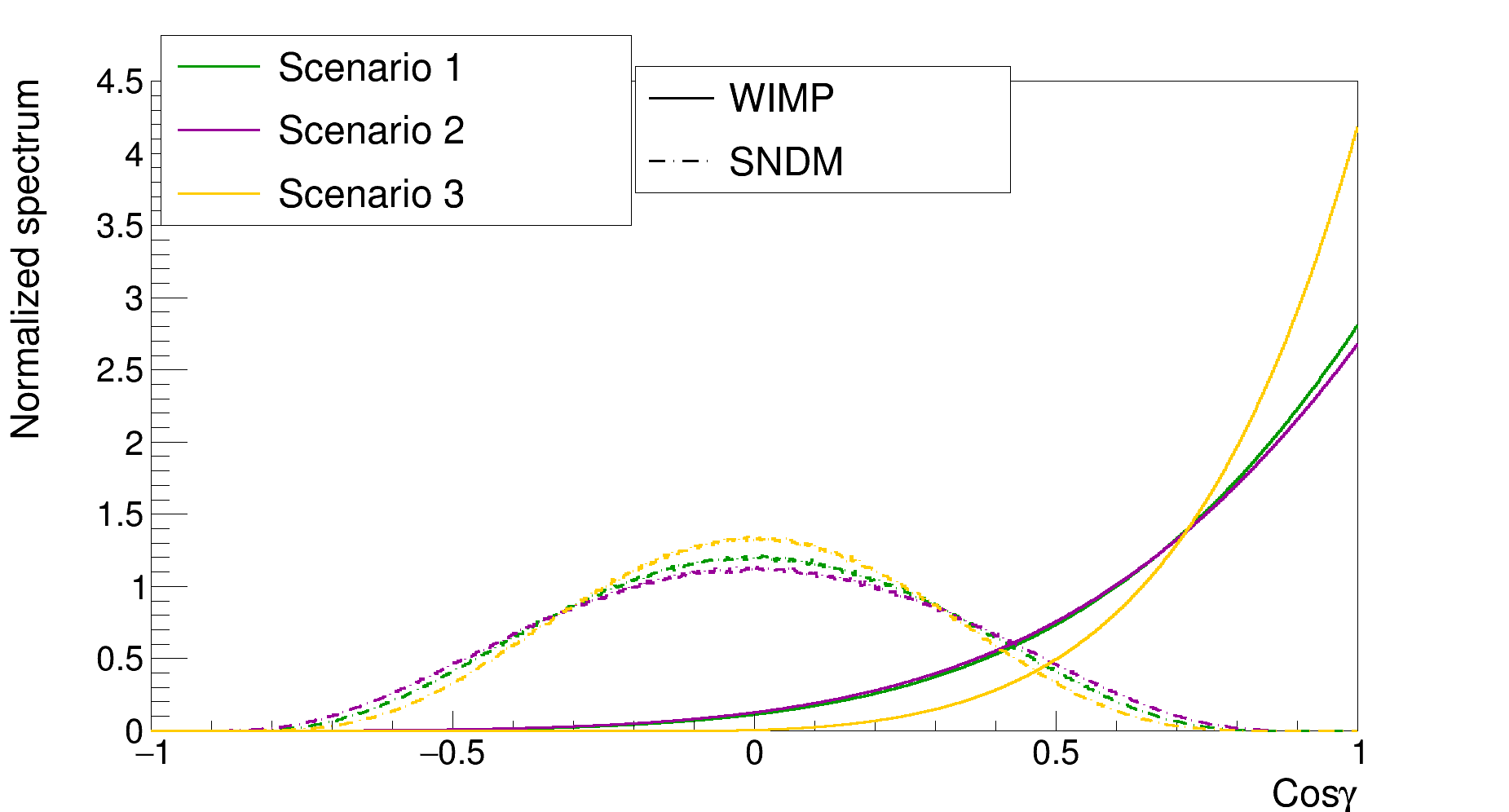}
	\includegraphics[width=0.45\textwidth]{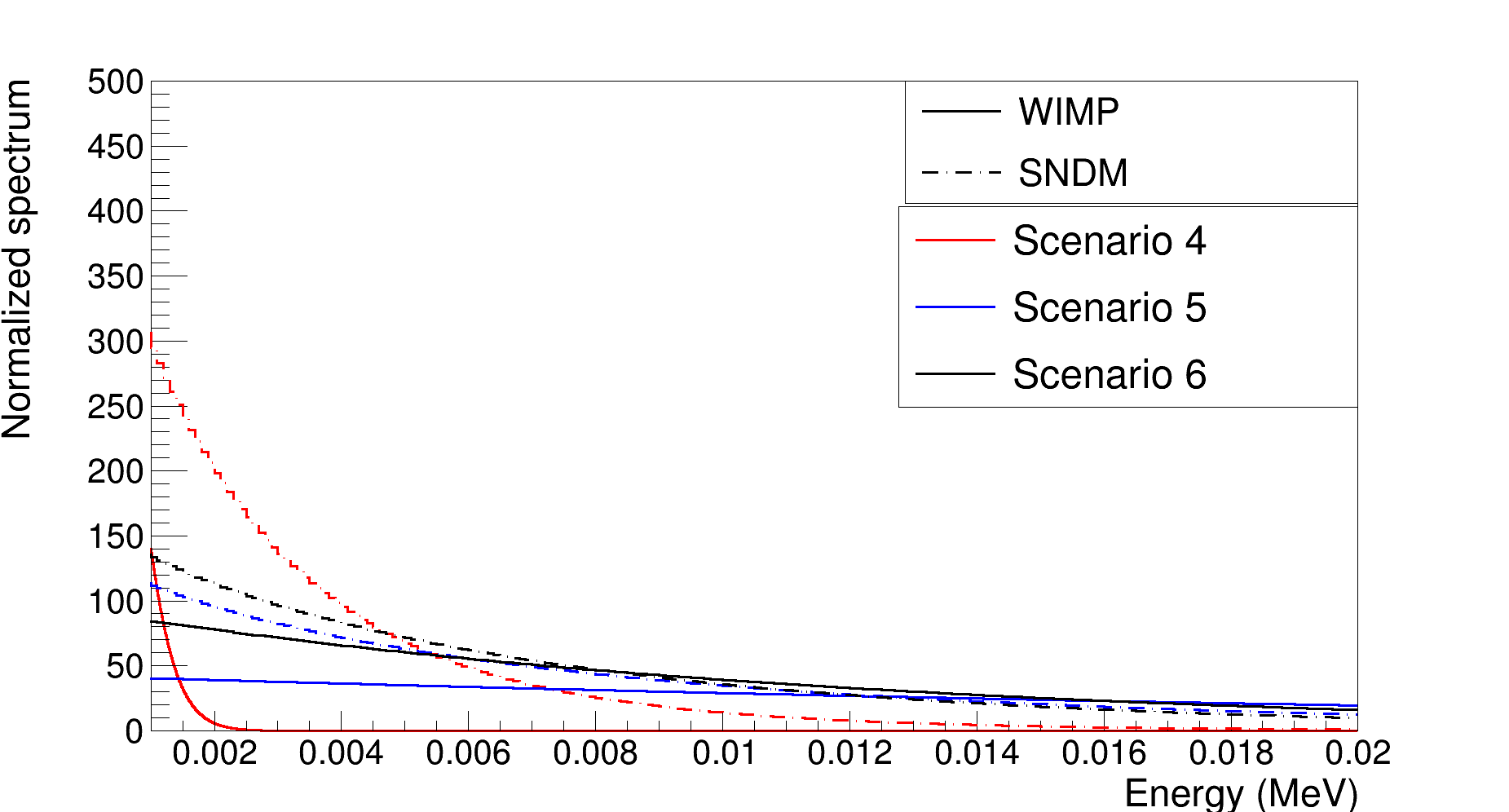}
	\includegraphics[width=0.45\textwidth]{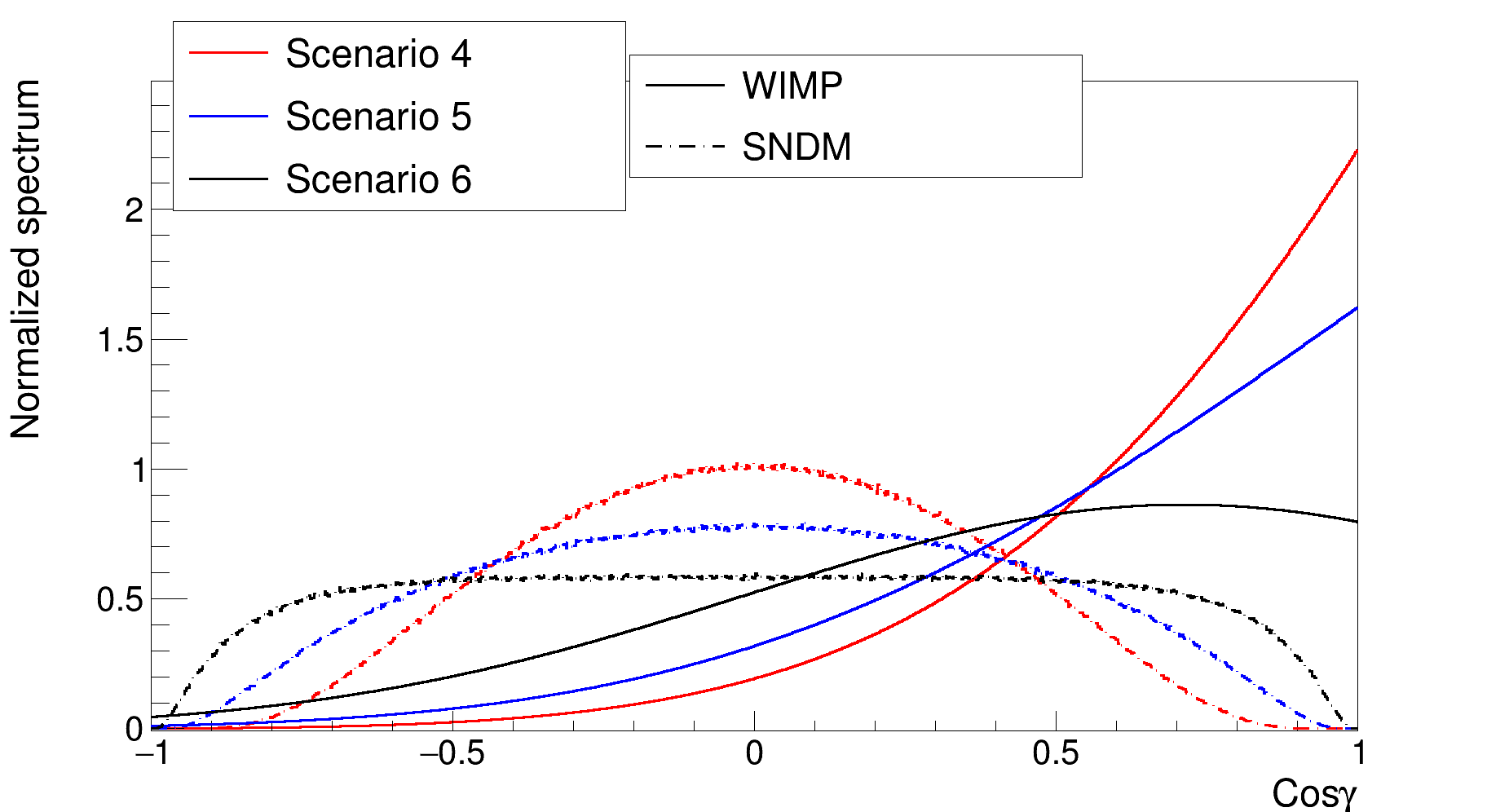}
	\caption{Energy (left) and 1D angular recoil spectra (right) for 10 \Gevc (top) and 100 \Gevc (bottom) WIMP masses for the six scenarios considered. (See Table \ref{tab:sn_wimp_cases}.) The angular spectra are displayed as a function of the angle $\gamma$ between the recoil direction and the one opposite to the motion of the Sun, as in Section \ref{sec:WIMPparadigm}. Solid lines denote the WIMP spectrum while dashed lines denote the SNDM spectrum; colour denotes the scenario in question. Note the similarity in energetic spectral shape for WIMPs and SNDM of the same scenario (owing to the choice of WIMP and SNDM masses to deposit similar energy in nuclear recoils) and the net difference in angular spectrum by virtue of the roughly perpendicular arrival directions.}
	\label{fig:sn_wimp_cases}
\end{figure*}
In this Section, the various scenarios considered in the study are presented as test cases for discrimination between a WIMP and SNDM signal in fiducial experimental setups.

Since the argument of this analysis is that a WIMP signal and SNDM signal are very difficult to distinguish through solely the nuclear recoil energy spectra, pairs of WIMP masses and SNDM scenarios (mass, SNDM-SM coupling $y$, corresponding temperature at escape $T$, and associated redshift factor $\Phi$) are considered so that they will produce a similar energy deposition in a DM detector. 
As shown in Section \ref{subsec:SNDMkin}, in order to obtain this, the WIMP $\mu_A v$ needs to match the SNDM $p_0$ such that the WIMP and SNDM have comparable momentum transfer to the target nucleus. An example of lighter (10 \Gevc) and heavier (100 \Gevc) WIMP mass are chosen to compare to various SNDM scenarios. Since the energy and angular recoil distribution are proportional to the WIMP-nucleus cross section, this is allowed to scale freely to best match the energy spectrum produced by the SNDM. As per the fiducial experimental choices discussed in Section \ref{subsec:SNDMexp_par}, $^4$He, $^{19}$F, and $^{131}$Xe are elected as target elements and six different scenarios are considered for comparison, where a "scenario" is a choice of WIMP mass, SNDM mass and coupling, and target nucleus. These scenarios are listed in Table \ref{tab:sn_wimp_cases}. It is believed that these choices are a good representation of the various cases that may be encountered in terms of experimental approaches, target materials, and DM parameter space. 

\begin{figure*}[!t]
	\centering
	\includegraphics[width=0.35\textwidth]{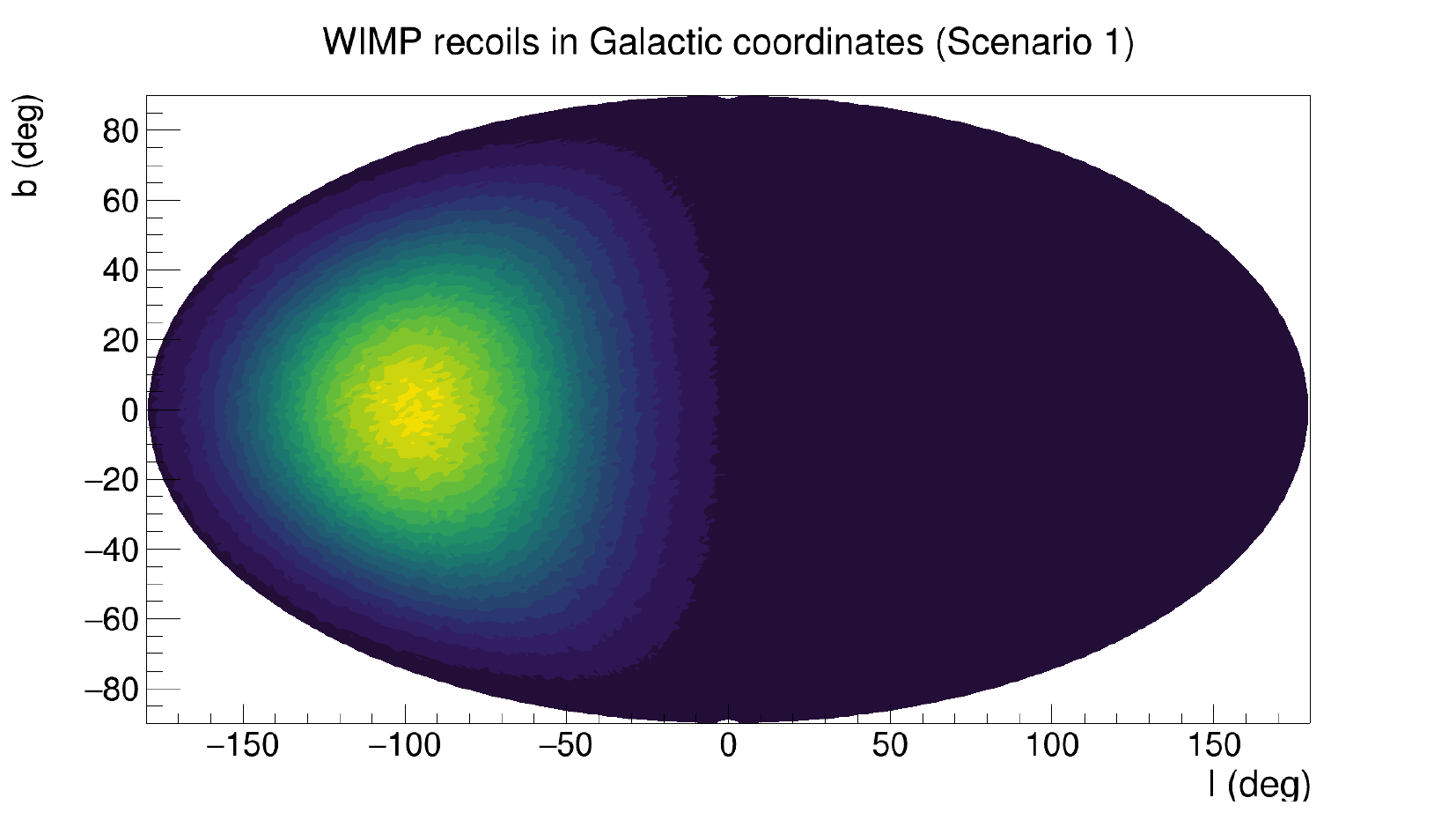}
	\includegraphics[width=0.35\textwidth]{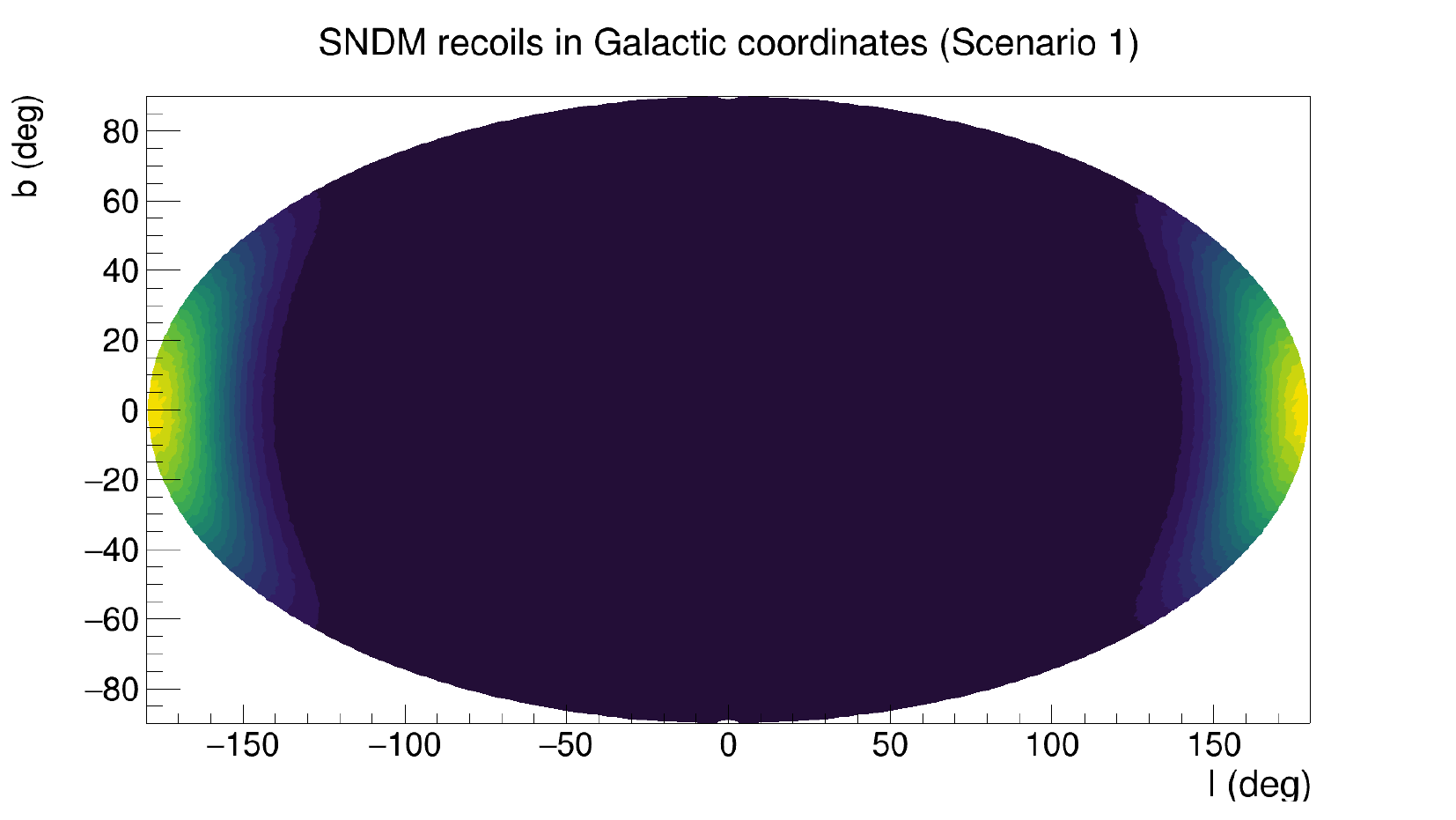}
	\includegraphics[width=0.35\textwidth]{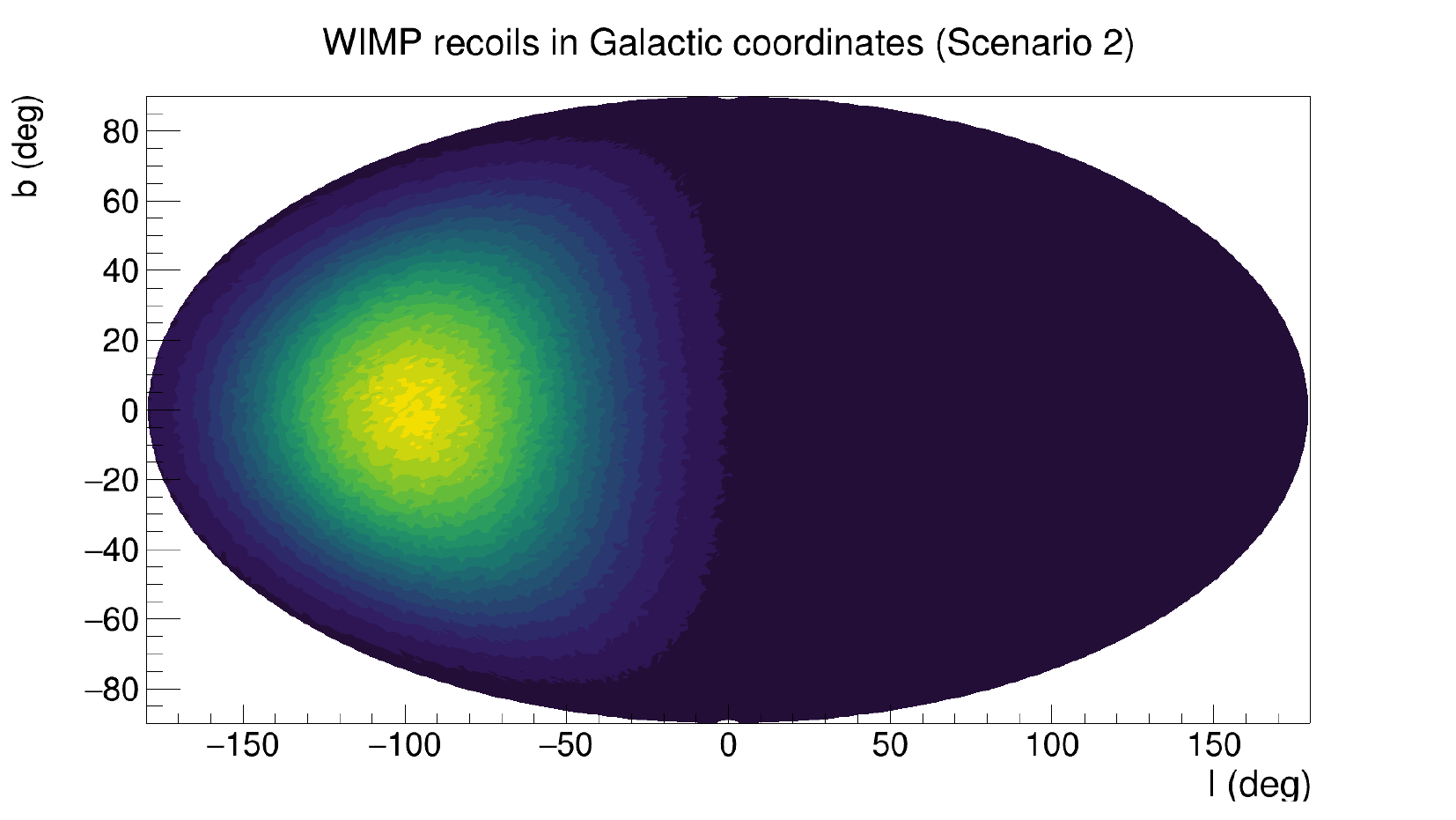}
	\includegraphics[width=0.35\textwidth]{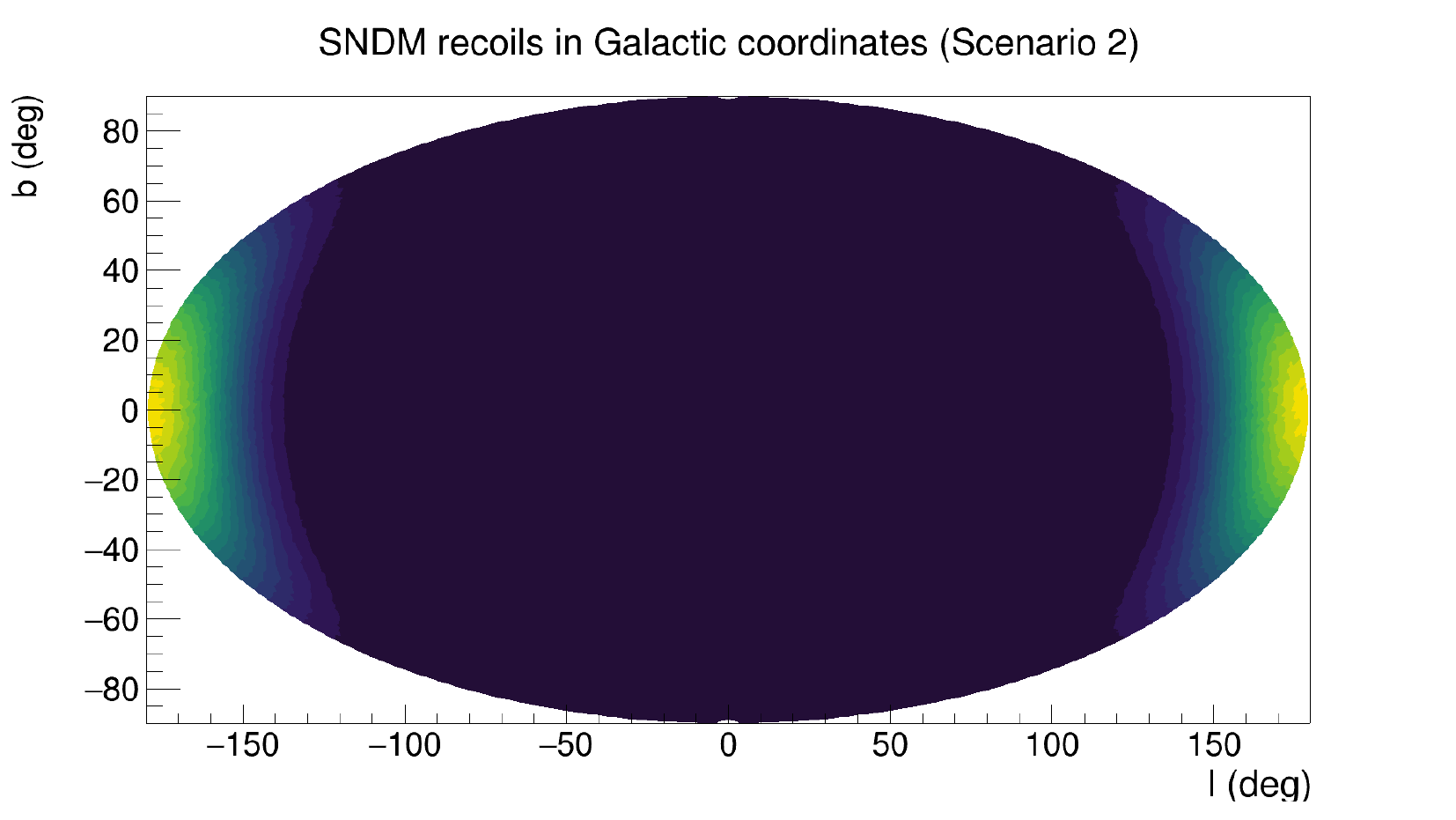}
	\includegraphics[width=0.35\textwidth]{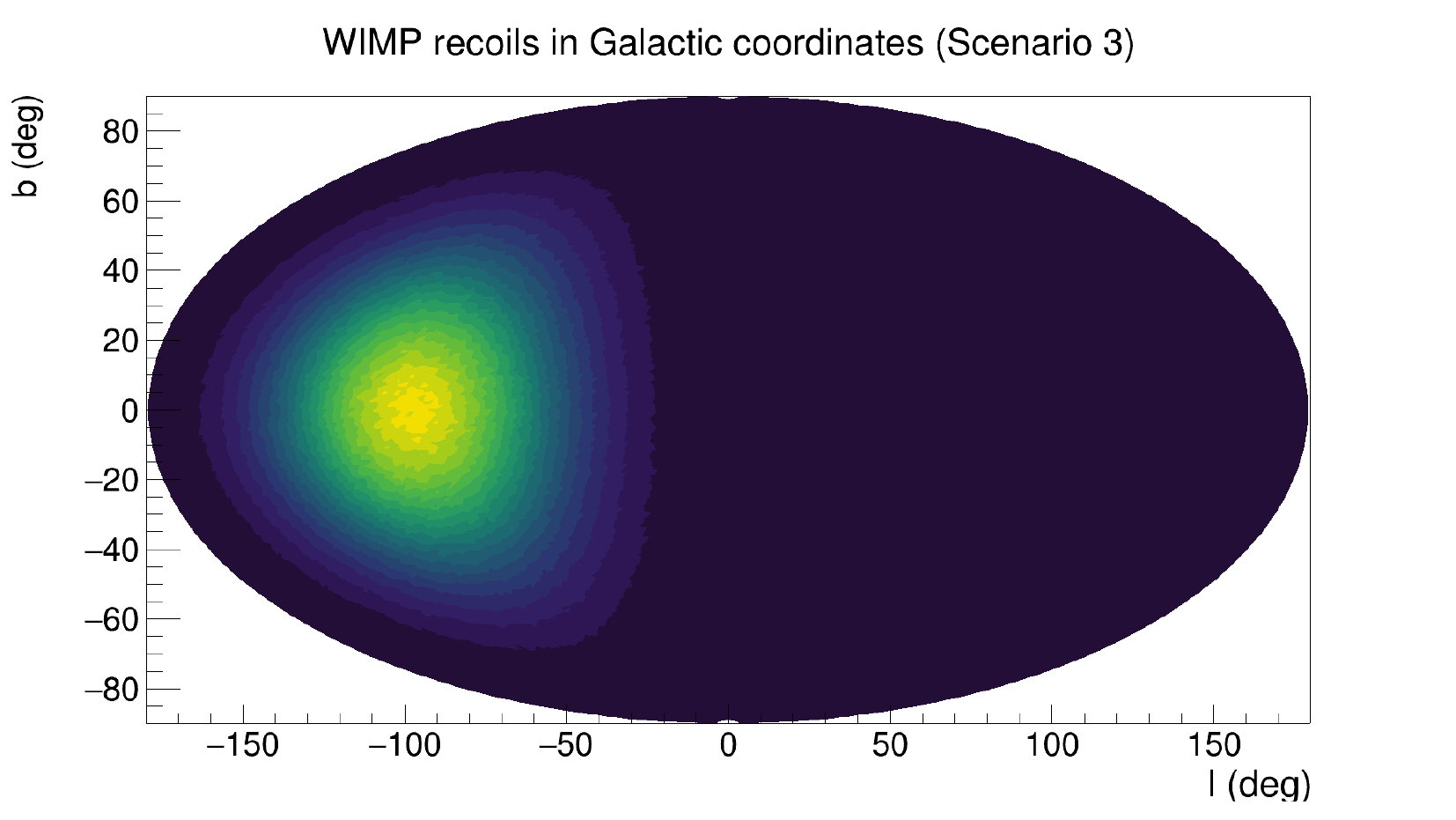}
	\includegraphics[width=0.35\textwidth]{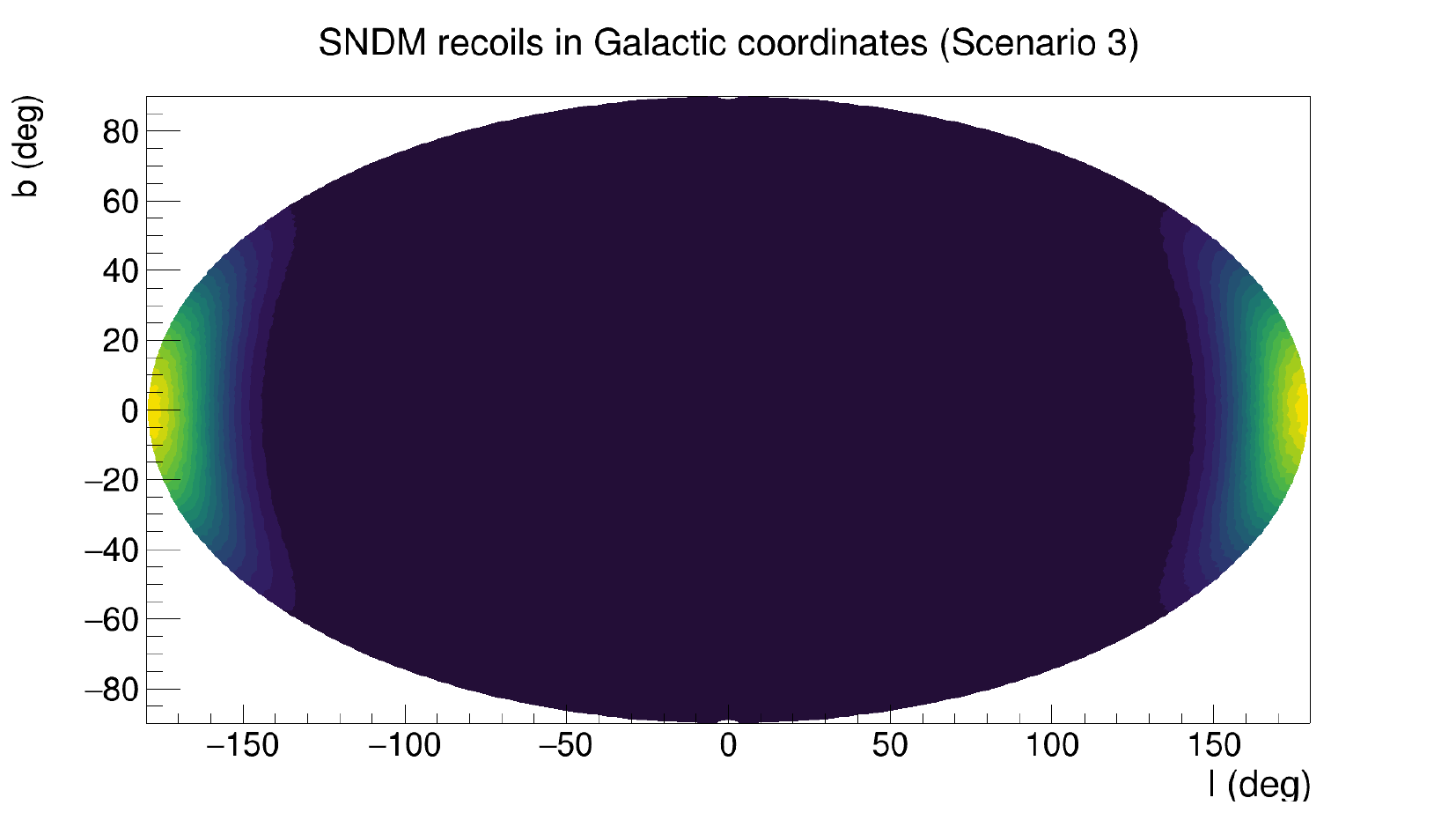}
	\includegraphics[width=0.35\textwidth]{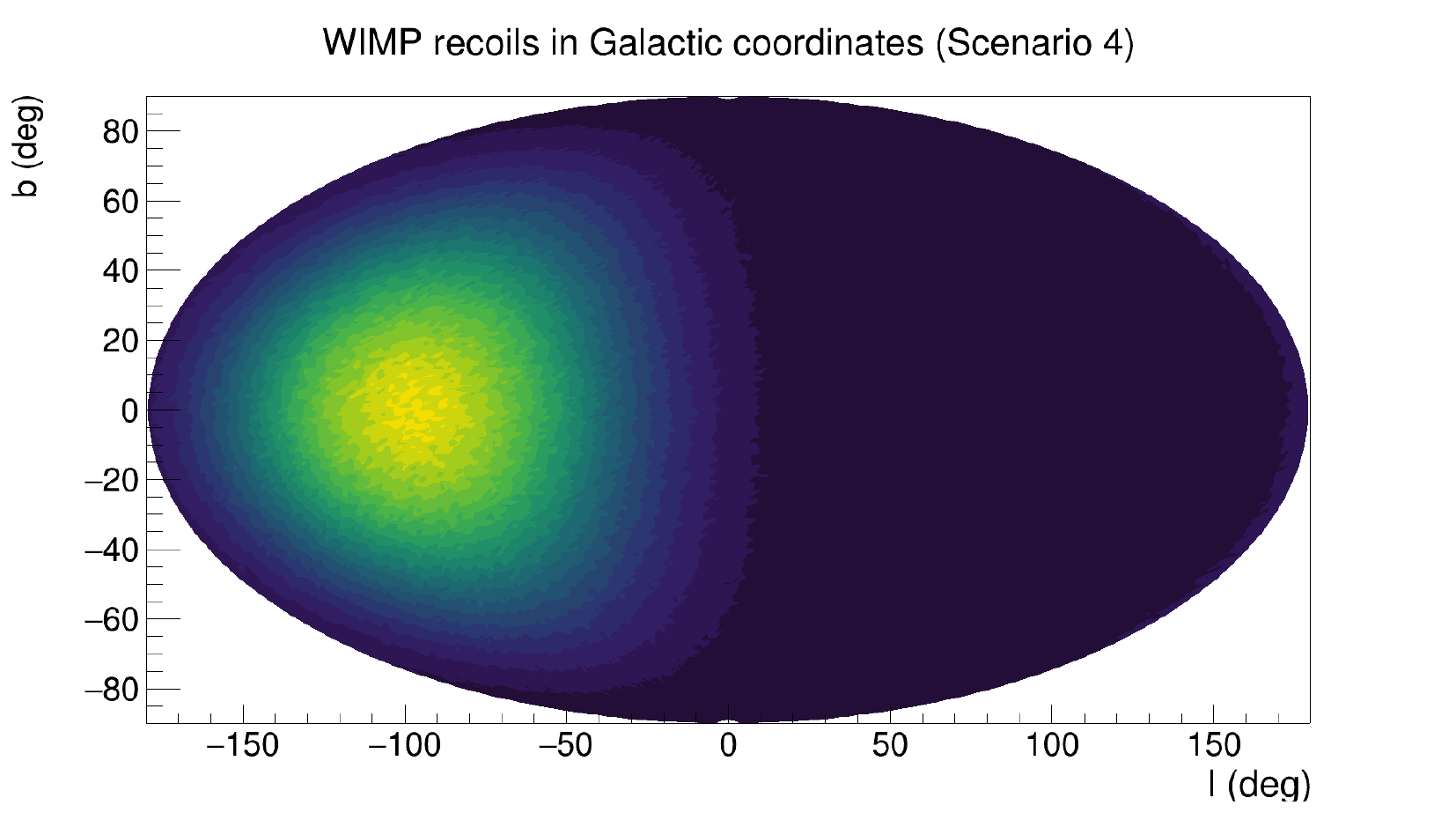}
	\includegraphics[width=0.35\textwidth]{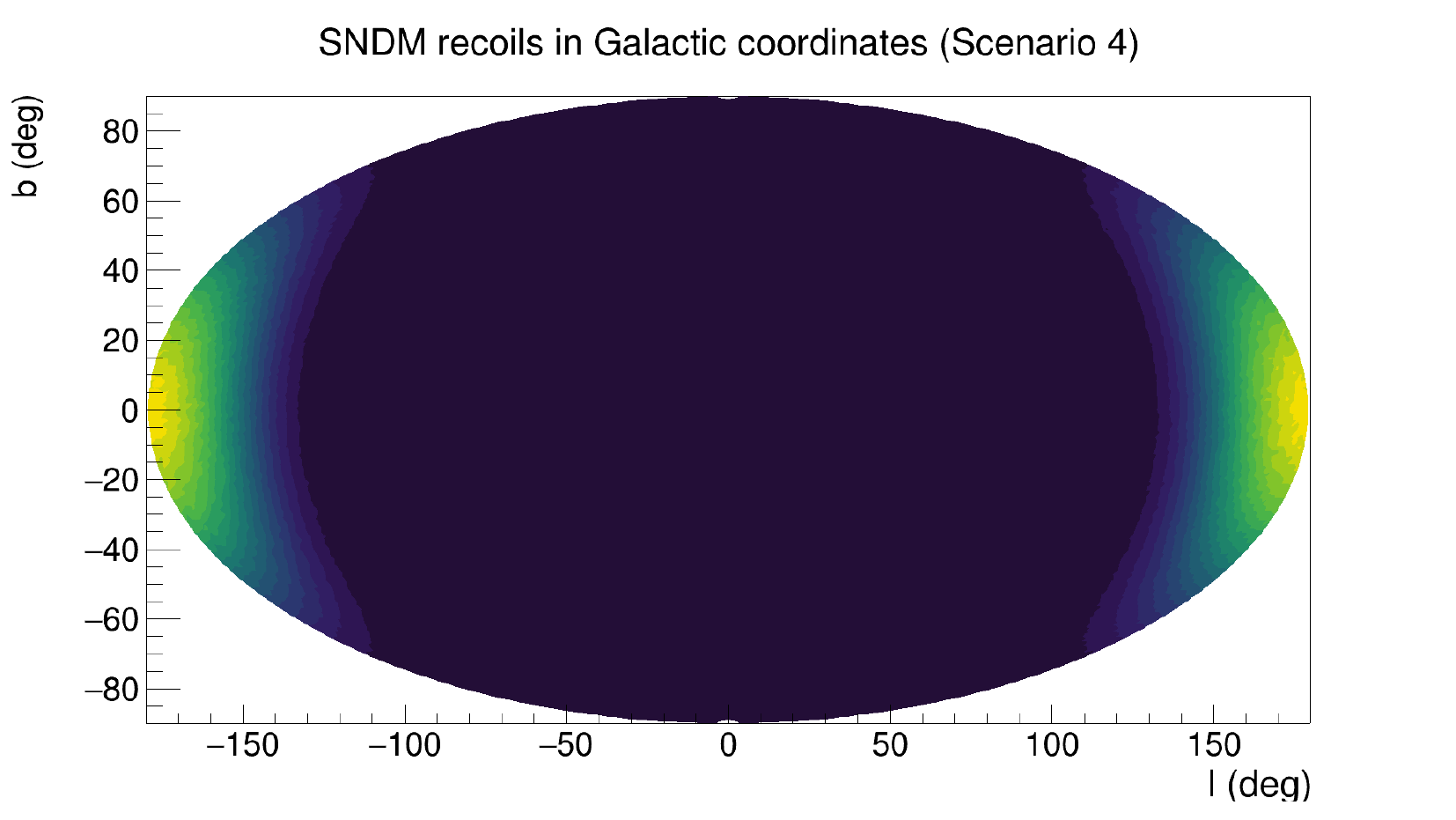}
	\includegraphics[width=0.35\textwidth]{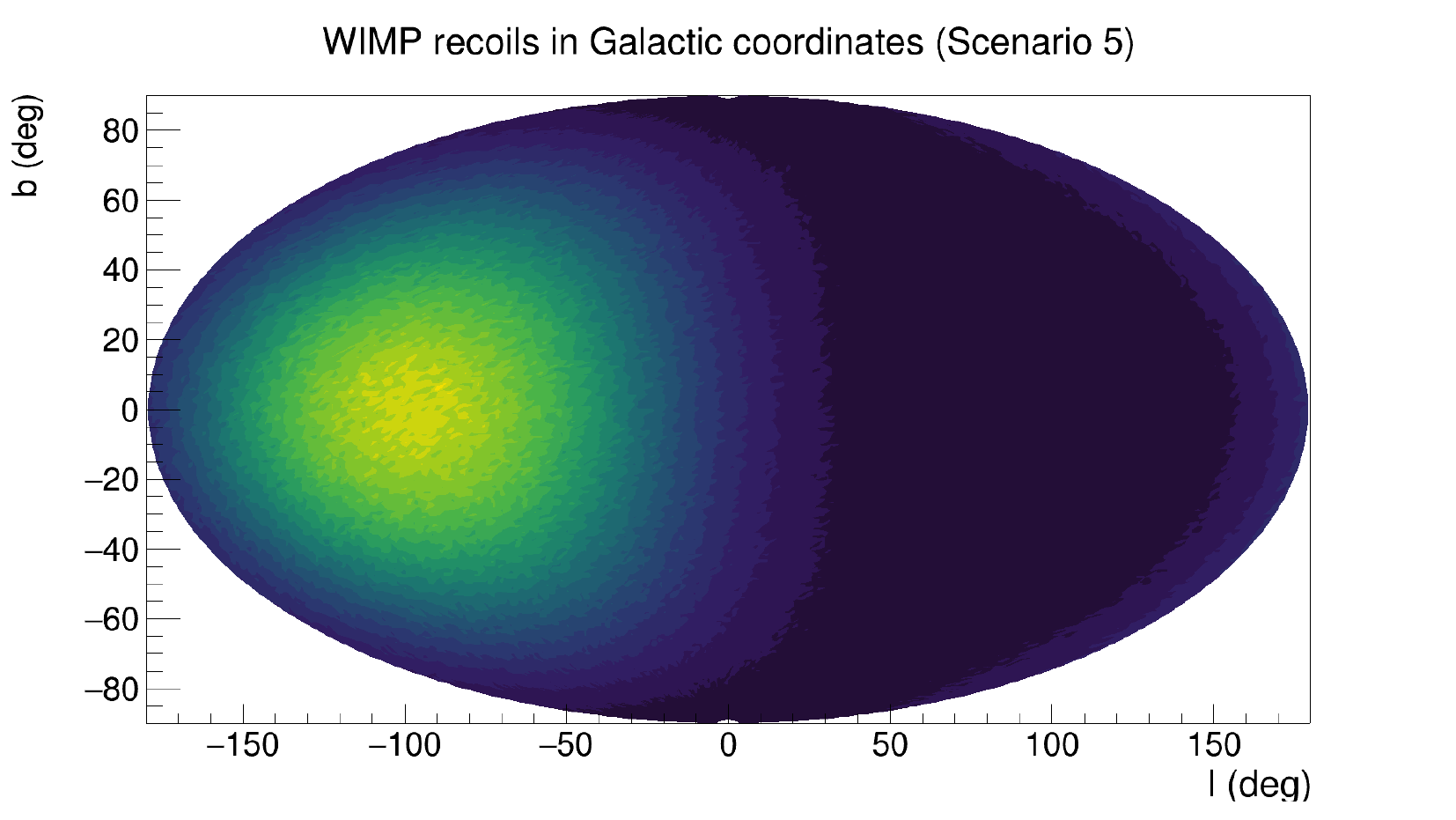}
	\includegraphics[width=0.35\textwidth]{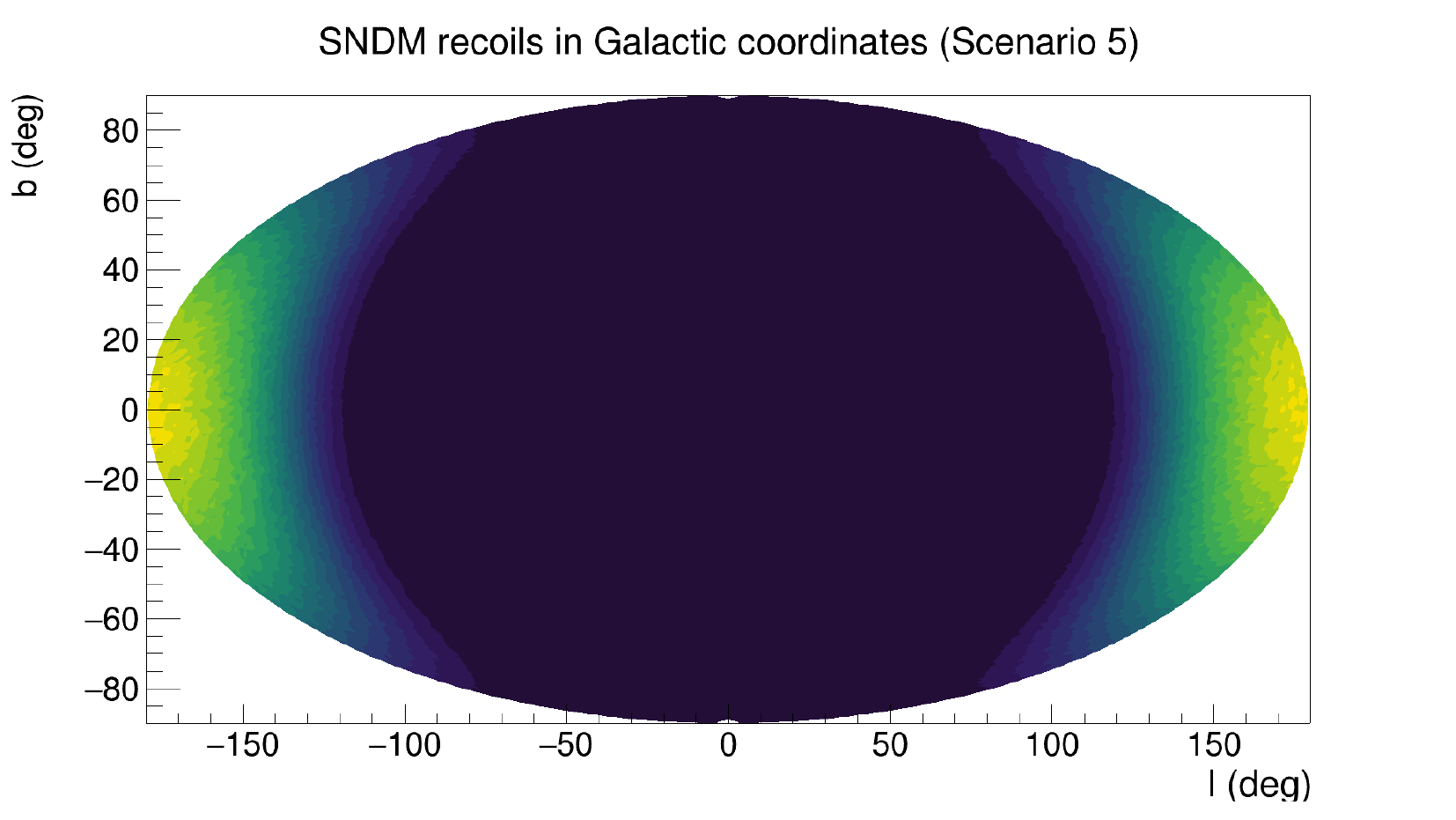}
	\includegraphics[width=0.35\textwidth]{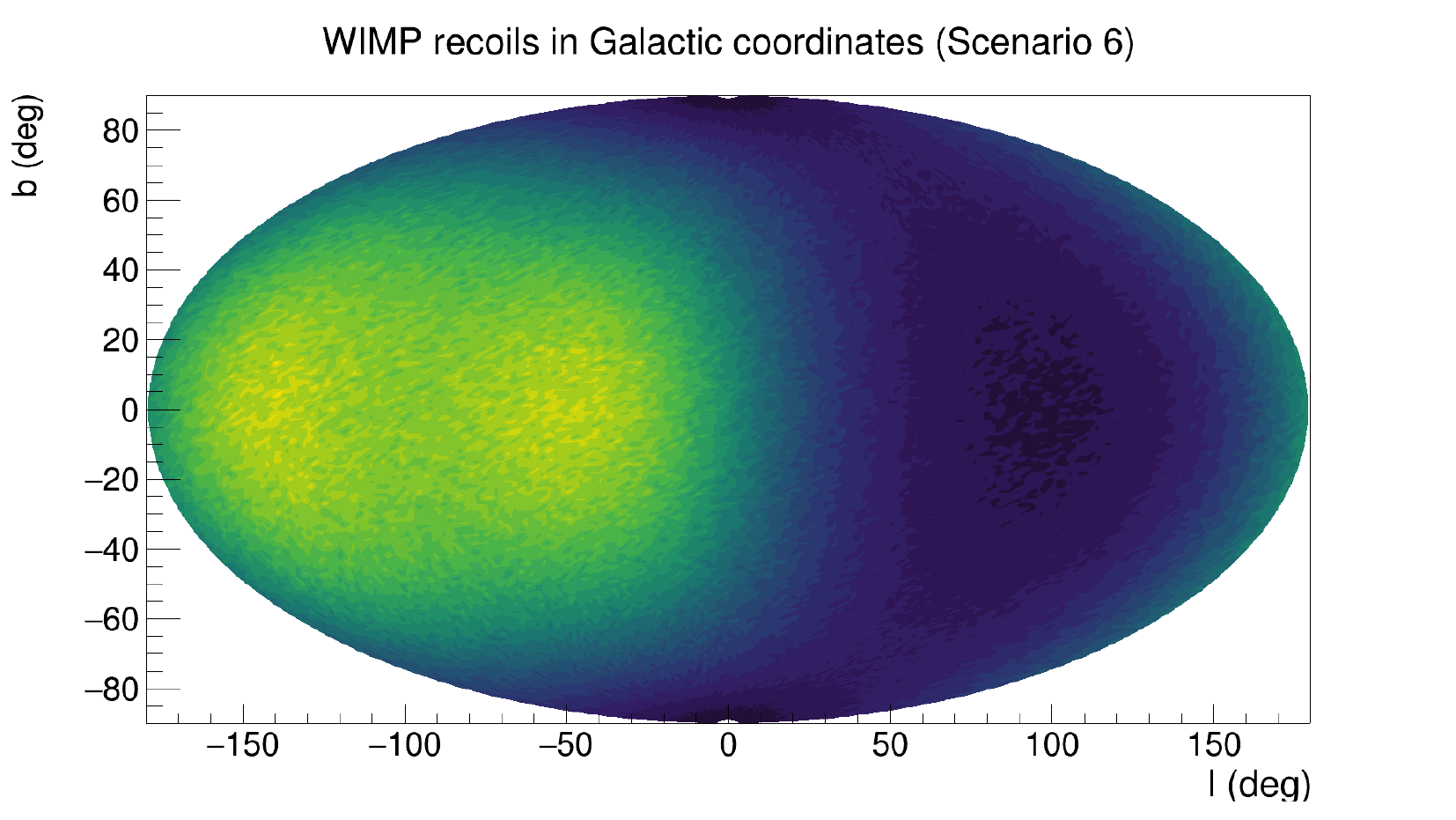}
	\includegraphics[width=0.35\textwidth]{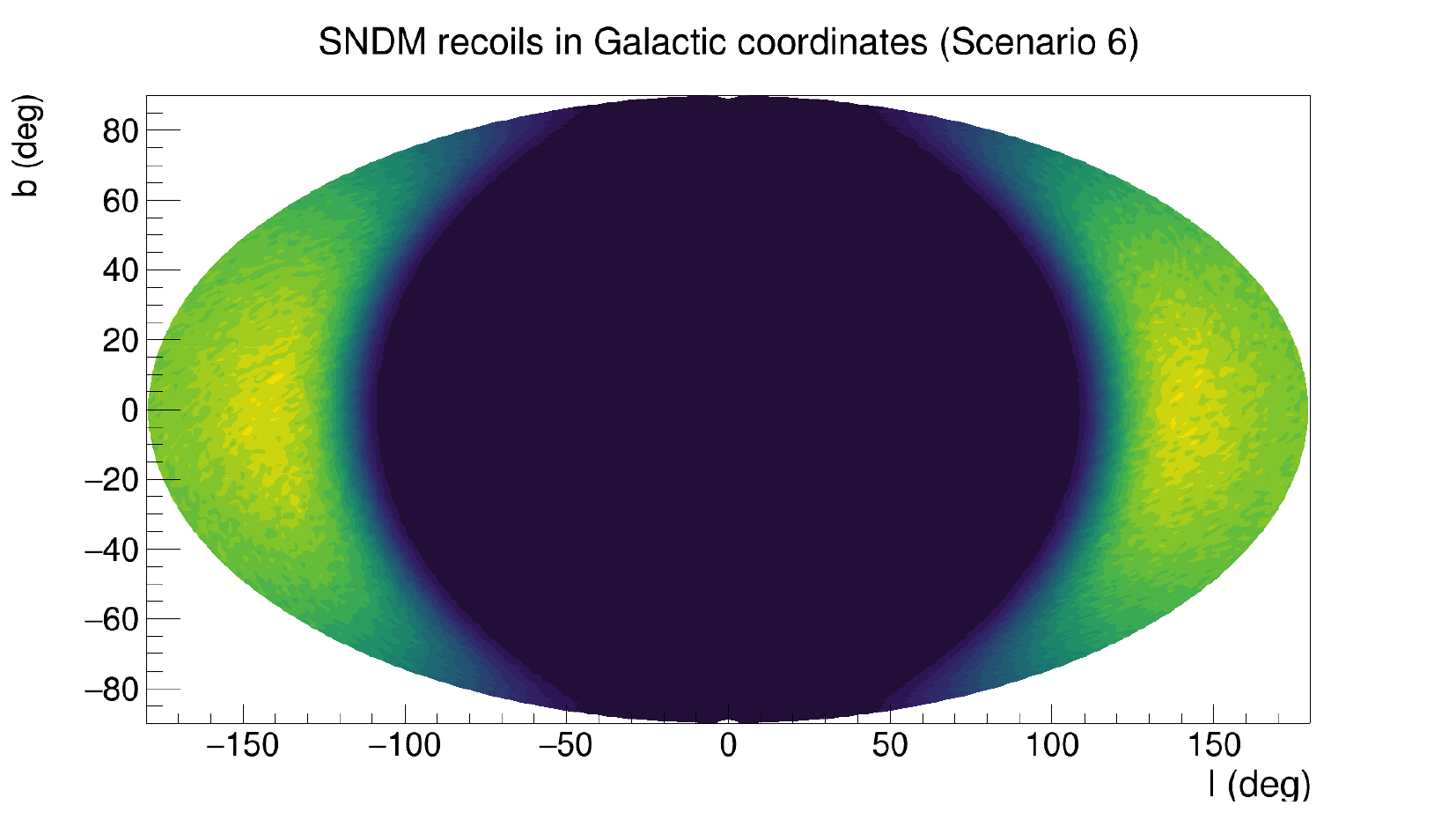}
	\caption{Comparison of the angular distribution of nuclear recoils in Galactic coordinates from WIMP on the left and SN DM interactions on the right for the six scenarios considered (1 to 6 from top to bottom), where the colour scale indicates the recoils density.}
	\label{fig:sn_wimp_2d}
\end{figure*}

Figure \ref{fig:sn_wimp_cases} shows the energy (left) and 1D angular (right) distributions of nuclear recoils for the six scenarios considered, divided into 10 \Gevc WIMP scenarios (top) and 100 \Gevc scenarios (bottom). These spectra are shown after having applied the cut on the energy ROI, as discussed in Section \ref{subsec:SNDMexp_par}. Here $\gamma$ is the angle between the recoiling nucleus and the one opposite to the laboratory velocity in the lab frame, as from Equation \ref{eq:rateWIMP2}. The full 2D angular spectra are shown in Figure \ref{fig:sn_wimp_2d}.

As is evident in the plots, the energy spectra possess very similar shapes for a given scenario, as expected by the choice of WIMP mass and SNDM parameters. The angular distributions, on the contrary,  demonstrate a dramatic difference in shape due to the approximately perpendicular arrival directions of the WIMP and SNDM. It is this difference that allows the angular spectra to discriminate between the two models with very few events.

\begin{table*}
	\begin{center}
		\begin{adjustbox}{max width=1.01\textwidth}
			\begin{tabular}{|c|c|c|c|c|c|c|}
				\hline
				Scenario & Target & WIMP Mass [\Gevc]  & SNDM Mass [\Mevc]  & $T$ [MeV] & $\log_{10} y$  & $\Phi$   \\ 
				\hline
				\hline
				1 & $^4$He & 10  & 5& 0.31 & -13.3 & 0.006 \\ 
				2 &$^{19}$F & 10 & 7 & 1.0 & -14.3 & 0.02 \\ 
				3 &$^{131}$Xe & 10 & 9 & 1.6 & -14.6 & 0.03 \\ 
				4 &$^4$He & 100 & 5 & 0.52 & -14.0 & 0.01 \\ 
				5 &$^{19}$F & 100 & 14 & 3.0  & -15.0 & 0.07 \\ 
				6 &$^{131}$Xe & 100 & 38 & 13.4 & -16.0 & 0.1\\
				\hline
			\end{tabular}
		\end{adjustbox}
		\caption{Various signal scenarios for comparison. The target nucleus, WIMP mass, SNDM mass, SNDM-SM coupling $y$, temperature of SNDM particles at the energy sphere set by this $y$, and the redshift factor $\Phi$ from this energy sphere are presented. The WIMP masses and SNDM parameters are chosen in order to have comparable momentum transfer to the nuclei such that energetic discrimination is difficult.}
		\label{tab:sn_wimp_cases}
	\end{center}
\end{table*}

\newpage
\subsection{Likelihood}
\label{subsec:frequentistlikeli}
The methodology employs a profiled likelihood ratio test, a modification of the likelihood ratio test. Indeed, in regular likelihood ratio tests, the likelihood function depends only on the expected number of events of a determined hypothesis. In the studied case, it will be absolutely identical for the two models as the number of DM detected events $\mu_n$ is chosen a priori. Instead, the profiled version modifies the function in order to take into account the dependence of the $\mu_n$ on a measurable quantity which derives from the model under test. In the analysis, the number of signal events needed in order to distinguish the SNDM signal from the WIMP signal is estimated assuming only one measurable quantity is detectable and therefore included in the profiling of the likelihood.\\

In order to do this, $\mu_n$ events are randomly extracted according to the energy and angular spectra defined in Section \ref{subsec:SNDMrate} and \ref{subsec:limit_signmod} and shown in Figure \ref{fig:sn_wimp_cases}. With these, a histogram is filled to represent the outcome of the measurement performed by an experiment, after a proper Gaussian smearing of the extracted quantity following the expected experimental resolutions illustrated in Section \ref{subsec:SNDMexp_par}. The range of the histogram allows to take into account the ROI discussed in the Section \ref{subsec:SNDMexp_par} and the bin sizes are chosen to be twice the $\sigma$ resolution evaluated in the centre of the bin itself (i.e. if $x_0$ is the centre of the bin, the bin range will go from $x_0-\sigma_{res}(x_0)$ to $x_0+\sigma_{res}(x_0)$). 
Given this set of simulated data, the likelihood function under two hypotheses is calculated: WIMP signal (hypothesis $H_0$) and SNDM signal (hypothesis $H_1$). The likelihood is evaluated as a simple multinomial PDF as follows:
\begin{equation}
\label{eq:likelihood}
\Likeli_{y|x}=\mu_{n}!\prod_{i=1}^{N_{\text{bins}}} \left[ \left( \sum_{j=i}^{N_{\text{adjacent}}} P^{\text{migrate}}_{j\rightarrow i}P_{j|x} \right)^{n_i}\frac{1}{n_i!}\right]
\end{equation}

where
\begin{itemize}
	\item $ x$ denotes the assumed hypothesis: H$_0$ (WIMP) or H$_1$ (SNDM)
	\item $ y$ denotes the Monte Carlo true analytic spectral the data are generated from: $W$ (WIMP) or $S$ (SNDM)
	\item $ n_i $ is the number of signal events in the $i$th bin
	\item the product runs over all the bins of the histogram of the experiment (the term becomes irrelevant if $ n_i=0 $)
	\item $ P_{j|x} $ is the probability of an event in the $j$th bin under hypothesis $x$
	\item $P^{\text{migrate}}_{j\rightarrow i} $ is the probability of an event that occurred in the $j$th bin migrates to the $i$th bin due to resolution effects, which captures the effect of spectral smearing due to imperfect resolution
	\item the sum runs over all the bins adjacent to the $i$th one ($i$th included) in the histogram. (Note: for the energy histograms both adjacent and next-to-adjacent bins are considered, corresponding to a total of 5 bins in the sum; for the angular histograms only truly adjacent bins are considered, corresponding to 3 bins in the 1D case and 9 bins in the 2D case).
\end{itemize}

The likelihood is written in a event-binned form, which means that the function is calculated as the independent probability of having a certain amount of events happening in each bin. The probability in each bin is not estimated as a Poissonian distribution, but directly from the probability distribution given by the energy and angular spectra.\\
The evaluation of $ P^{\text{migrate}}_{j\rightarrow i}$ is calculated as in Section \ref{subsec:limit_likelihood}.

Once the likelihood is evaluated, the likelihood ratio variable is calculated as
\begin{equation}
\label{eq:likeliratio}
\lambda_{y}=\frac{\Likeli
	_{y|H_1}}{\Likeli_{y|H_0}}
\end{equation}
with $y=W,S$. The operation is repeated 5$\cdot$10$^5$ times to obtain the probability distributions of the likelihood ratios $f_W(\lambda_W)$ and $f_S(\lambda_S)$. 
The $\mu_n$ events that are considered enough to distinguish between the two models is found when both the probabilities of committing either a type I or type II error are 5$\%$ or less to ensure symmetry between the two hypotheses.
It is important to note that $\mu_n$ is never chosen smaller than 3, as this is the minimum number of events needed to distinguish a Poisson fluctuation from zero background.
\subsection{Results}
\label{sec:results}
\begin{figure*}[t]
	\centering
	\includegraphics[width=0.49\textwidth]{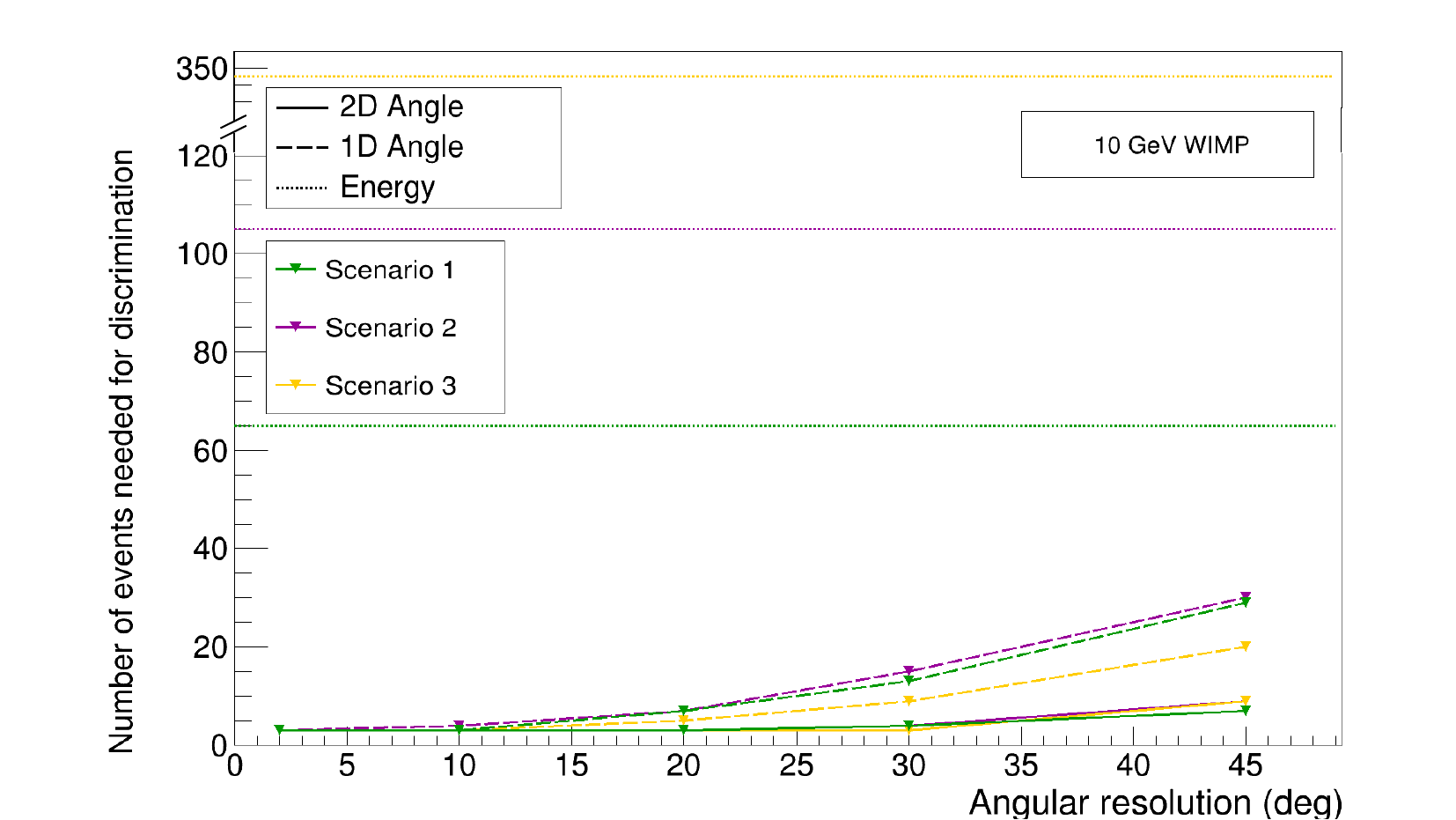}
	\includegraphics[width=0.49\textwidth]{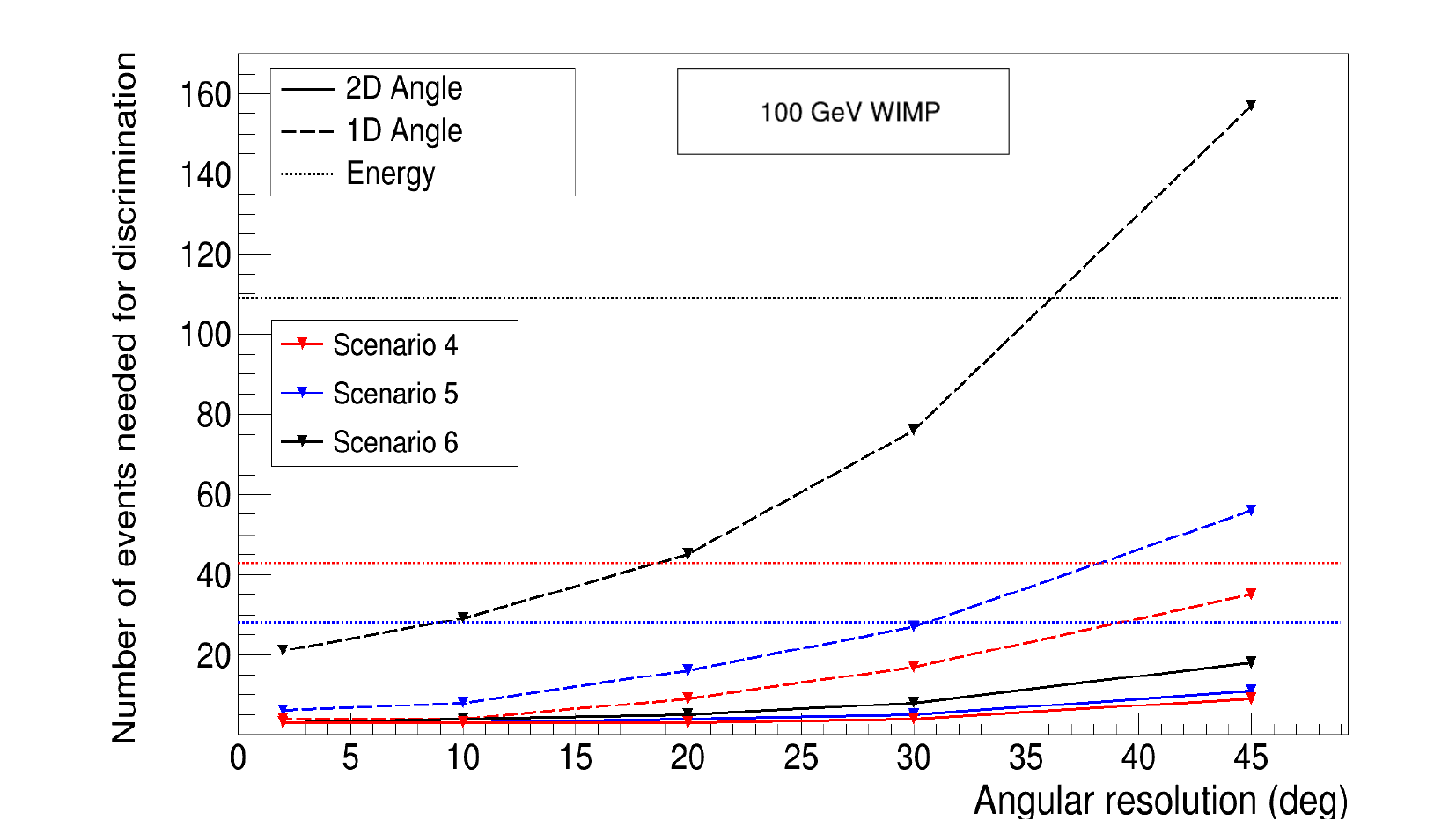}
	\caption{The average number of events necessary for discriminating between a WIMP and SNDM signal in the fiducial experimental setups for the various scenarios in Table \ref{tab:sn_wimp_cases}, plotted as a function of angular resolution. (n.b. For the 2D spectrum, the horizontal axis has units of deg$~\times~$deg for each point, e.g. ``30'' denotes $30^{\circ} \times 30^{\circ}$.) Results for energy spectra are horizontal as they possess no dependence on angular resolution. It is clear that angular information allows model discrimination with far fewer average signal events than when using solely the energy spectra. While the specific numeric values are dependent on the choice of fiducial experimental setup, the point remains that directional detectors can provide an order of magnitude improvement in discrimination capability for a large selection of realistic angular resolutions.}
	\label{fig:relall}
\end{figure*}

Following the procedure described in the previous Section, the number of detected events inside the ROI needed to distinguish between the WIMP and SNDM models are shown in Figure \ref{fig:relall} for the six scenarios considered as a function of angular resolution. The average number of events to distinguish the models using purely the energy spectra is displayed as a dotted line (horizontal, as there is no angular resolution dependence), the number of events for using only the 1D angular spectrum as a dashed line, and the number of events using only 2D angular spectrum as a solid line. It is germane to stress the choice of the symmetry of the two types of error during the analysis. This way, no hypothesis is preferred in advance just because it is named $H_0$ instead of $H_1$.\\
The order of magnitude of improvement the angular information provides it clearly visible. Even in the worst case scenario with 45$^{\circ}$ angular resolution, for which only a handful of bins are available for the determination of the angular distribution of the detected events, the angular information is still far more powerful at discriminating the two models than the energy information. As expected, a full 2D angular detection performs much better than its projection on a single axis, as it provides more information on the original shape of the recoil distribution. This allows model discrimination with more than one order of magnitude fewer events compared to using the energy spectra, especially when a good resolution is available or if the kinematics are particularly favourable, as for the scenario 3, where the SNDM and WIMP distributions are strongly peaked in different directions.

As expected, the angular resolution significantly affects the discriminating power. Indeed, a worse angular resolution allows the events to migrate away from their expected position, effectively smearing the original shape of the recoil distributions. Moreover, a worse resolution leads to a decrease in the number of bin available which furthermore dilutes the angular distribution of the data. For the 1D angular spectra, the number of events needed to discriminate seems to double roughly every 10$^{\circ}$ lost in resolution, resulting in a strong degradation of the information from a resolution of 30$^{\circ}$ or worse. For the 2D angle, worsening the resolution has a lesser effect, with a clear degradation of the discrimination power noticeable only at resolutions worse than $30^{\circ}\times30^{\circ}$. Moreover, the relative increase in number of events is less intense than the 1D case.

Another interesting feature appearing in the right-hand plot of Figure \ref{fig:relall} is that energy distributions tuned to be similar in terms of total momentum transfer can still exhibit differences in shape that allow better discrimination than with 1D angular information alone, as for the case of $^{19}$F target with a WIMP mass of 100 GeV/c$^2$ (scenario 5). A similar effect is happening also for the $^{131}$Xe target with 100 GeV/c$^2$ WIMP mass (scenario 6), but in this case this is due to the well-known ringing effect \cite{Bozorgnia_2012} that results in more similar angular distributions. Nonetheless, it is important to notice that this happens only in the case of very poor angular resolution, and is not present when this is improved beyond 30$^{\circ}$. This underlines the fact that such a feature is only an artifact of this simplistic analysis wherein the angular and energy distributions are not used together, with the proper correlations induced by the kinematics, as a real experiment would do. Clearly, a combination of the energetic and angular information furnishes an even better discrimination power and is bound to perform equally or better than when a single information is available by construction.\\

Overall, the study shows that directionality and the angular distribution can be very powerful tools for the discrimination of different model, such as the example of WIMP and SNDM. Even when few events are available, a directional detector would retain a strong advantage with respect to experiments only able to access the energetic information. In addition to this, in agreement with the discussion of Section \ref{sec:directional}, while directionality can improve the limits setting, its true power resides in the DM astronomy and positive identification of DM. Indeed, the studies present in this Chapter demonstrated how the upper bound limits improved by a factor less than 10, while the discrimination of DM models reached improvements with respect to the energy measurements up to two orders of magnitude.	
\chapter*{Conclusions}
\addcontentsline{toc}{chapter}{Conclusion}
\label{concl}
In the last decades, the discoveries of the Higgs boson and gravitational waves demonstrated an excellent agreement between experimental results and the most important leading theoretical frameworks, the Standard Model (SM) of particle physics and General Relativity. However, simultaneously, several cosmological and astrophysical observations indicate that the current knowledge of the fundamental particles and forces of the Universe is still incomplete. The leading cosmological model, the $\Lambda$CDM, supports the existence of a new constituent of the Universe, usually referred to as dark matter (DM) since all the observation of its presence are related to gravitational probes. Some of the most established hypotheses postulate that DM is comprised of non relativistic Weakly Interactive Massive Particles (\Ws) with masses from $\mathcal{O}$(1) \Gevc to $\mathcal{O}$(1)TeV/c$^2$, which weakly couple to the SM particles. Indeed, a stable, weakly-interacting particle in thermal equilibrium with the early Universe would be able to reproduce the observed relic DM density.\\
\Ws are expected to populate the Milky Way, enveloping it in a DM halo. In the assumption that DM can interact other than gravitationally with the SM, the collision between \Ws and regular matter on Earth would produce detectable recoils with energies up to 100 keV. The direct detection experiments aim at revealing DM induced recoils of nuclei with extremely low rates, below 1 event per kilogram year, by exposing large amount of instrumented mass. Due to the low expected rate, the direct detection of \Ws is considered a search for very rare events and therefore challenging requirements on the experimental background are unavoidable. Classical background minimisation techniques are deep underground operation (to suppress cosmic rays), use of radio-pure components (to avoid natural radioactivity) and active or passive shielding of the detector.
The motion of the Sun and Earth around the centre of the Galaxy induces an apparent wind of \Ws roughly coming from the Cygnus constellation. This produces a strong directional dependence in the distribution of the recoils, which cannot be shared by any type of background, also thanks to the rotation of the Earth around its own axis. The access to the angular distribution of the recoils is an extremely powerful mean to set improved exclusion limits even in presence of large background, but especially to be able to positively claim DM discovery and study its characteristics.\\
The CYGNO project plans to build a large, $\mathcal{O}$(30) m$^3$, directional detector for rare event searches, such as DM. The detector is based on a gaseous time projection chamber filled with a He:CF$_4$ gas mixture, in order to be sensitive to both spin-depedent and spin-independent \W interactions. A gaseous TPC detector is intrinsically sensitive to the topology of the recoil tracks, which allows to measure their directional characteristics. CYGNO will be equipped with an amplification stage composed of three 50 $\mu$m gas electron multipliers (GEMs) and will be optically readout with photomultiplier tubes and scientific CMOS cameras. The latter allows to image large readout areas with high sensitivity while maintaining extremely good granularity which, coupled to the high segmentation of the GEMs, grants the capability to retrieve the topological information of the recoils down to $\mathcal{O}$(1) keV energies.\\

The work presented in this thesis was carried out in the context of the CYGNO experiment with the intention to maximise the light yield (to lower the energy threshold), minimise the diffusion during drift (to improve tracking capabilities) and explicitly show the potential of the directionality in different case studies.\\

By the arguments described in Chapter \ref{chap2}, the energy threshold is one of the most important parameters for a direct detection experiment as it directly determines the lower limit of \W masses the experiment can probe. While the CYGNO optical approach permits to image large areas reducing the number of readout detectors necessary, at the same time significantly diminishes the solid angle coverage, lowering the total number of photons collected. This, in turn, affects the minimum energy detectable. In Chapter \ref{chap5}, the possibility of adding a strong electric field below the last GEM (\textit{induction field}) to enhance the production of photons was thoroughly studied. Previous investigations, published in \cite{bib:EL_cygno}, suggested that the light production could be increased with little to no extra charge generation. Employing CYGNO prototypes, the phenomenon was thoroughly analysed increasing the electric field intensity with respect to the publication, and combining it with variation of the standard triple 50 $\mu$m thick GEM ($t$) amplification structure. Thicker 125 $\mu$m GEMs ($T$) were utilised in pairs or in combination with another $t$ one and the performances were meticulously compared by measuring the light output, the energy resolution and the intrinsic diffusion of the amplification structures. Supported by Ansys Maxwell simulations of the electric field, it was found that the induction field acts upon the structure of the electric field of the closest GEM, generating a region below the GEM holes where the intensity of the electric field can reach a 10 times increment with respect to when no field is added. This feature allows to improve the light yield. A production of photons larger than charge was observed thanks to rigorous charge measurements and the reason is attributed to the peculiar fragmentation thresholds, characteristic of the CF$_4$, which require less energy to emit visible photons than to ionise the molecule. The addition of this induction field proved to outperform the same amplification structure when pushed to the breakdown limit both in light yield and intrinsic diffusion, while keeping similar energy resolutions. Each type of structure combined with the induction field was found to excel in the measurement of a specific observable: a triple $t$ stack had the best energy resolution, a double $T$ one had the best light yield, and a $Tt$ one had the lowest intrinsic diffusion.\\
The possibility to enhance the light yield by almost a factor 2 with respect to the maximum achievable by the standard CYGNO amplification stage opens the possibility to strongly reduce the energy threshold while preserving all the other characteristics. Combined with the advancement of the optical sensors, exemplified in Table \ref{tab:hama}, this could lead to a recoil imaging directional detector with a threshold of a couple of hundreds of eV capable of probing DM masses down to 100 \Mevc. Such an improvement of the performances of this kind of amplification stage and its versatility can be exploited in other fields. For example, in X-ray polarimetry, the energies under study are of $\sim$ 2$-$8 keV, but the determination of the direction and the topology of the track are extremely important \cite{xray}. In this sense, the reduced intrinsic diffusion of the $Tt$ configuration of about 100 $\mu$m, with respect to the standard triple $t$ while losing only a factor 2 of light yield, poses itself as excellent possibility. This innovative way of employing the GEM amplification structure in a He:CF$_4$ mixture results therefore extremely relevant for several optical TPC applications, even beyond DM searches.\\

In a TPC, the primary electron diffusion degrades the tracking capabilities of the detector. In the directional DM field, this aspect becomes even more relevant, since the goal is to characterise the topology of low energy short recoil tracks in order to reconstruct the original direction and its vectorial or axial determination, i.e. \textit{head-tail} recognition (see Chapter \ref{chap2}). The addition of electronegative molecules in the gas mixture can permit the primary electrons to be captured by them within few hundreds of $\mu$m, letting these anions drift and transport the charge. This Negative Ion Drift (NID) operation was pioneered with studies with CS$_2$ and \SF and demonstrated the possibility to reduce the diffusion to the thermal limit. In Chapter \ref{chap6}, 1.6\% of \SF was introduced in the CYGNO gas mixture and for first time ever to the writer's knowledge NID operation was successfully obtained at pressures close to 1 bar with an optical readout. Measurements with alpha particles of the mobility of the charge carriers in the gas demonstrated to be consistent with previous charge-based analyses of the same gas mixture. Gas gains of roughly 10$^4$ were achieved, which are in line with the values found in literature. Systematic measurements of the transverse diffusion showed that extremely low diffusion coefficients can be attained, down to 45 $\mu$m/$\sqrt{\text{cm}}$, one of the lowest ever reached. In the context of the CYGNO experiment, this achievement is dramatically relevant, because, as soon as larger gains or sCMOS sensors with improved sensitivity are attainable, it will be possible to widen the drift region with similar or better tracking performances. The impact of such modification is a game changer in this field as larger drift region will make a CYGNO module more compact while keeping the same volume, improving the scalability for larger detectors and reducing radioactive contamination from the surfaces. In addition, the low diffusion parameters found and the relative high gain of the mixture for a charge readout-based  detector, demonstrate the potential feasibility of the CYGNUS project, the most ambitious in the directional DM field.\\

The relevance of the directionality in the DM search was described extensively in Chapter \ref{chap2}. The power of the positive discovery granted by the directional detectors makes their construction fundamental even in case a non-directional experiment claimed a discovery. In this work, the power of directionality was analysed in two exemplifying cases. A generic Bayesian-based statistical analysis tool was developed with the goal of evaluating the 90\% C.I. upper limit of a directional-based detector. The case studied focused on the expected performances of a CYGNO-like detector of $\mathcal{O}$(30) m$^3$. In case no statistical significance is reached for DM signal, the upper limit on the SI cross section can be as low as 10$^{-43}$ cm$^2$ at 5 \Gevc, in the assumption of 100 events per year of background and 0.5 keV$_{\text{ee}}$ energy threshold, while in the SD to proton coupling the expected limits are competitive with the projection of the PICO-500 experiment, the most sensitive in this field. Particular attention was given to the advantages of the inclusion of directionality in the limit estimation which turned out to improve the evaluation, with respect to pure counting experiments, by a factor between 1.5 and 5  depending on WIMP masses and background level. In addition, the power of directionality in the discrimination between DM models was examined in a toy analysis where the number of events needed to discriminate between the \W model and a dark fermion-based one was computed for different measurable quantities. The study showed that, if the recoil energy induced by the DM candidate of the two different models is similar, measuring the direction of the recoils can lower by up to 2 orders of magnitude the number of DM events which are needed to discriminate the two scenarios. As a consequence, if a xenon-based detector found events inconsistent with the background expectations, extremely large exposure would be necessary to determine the \W nature of the particle that induced them. This stresses once more the fundamental role of directional detectors in the DM field.\\

In conclusion, the importance of the directional detectors for DM searches was exemplified by employing rigorous statistical tools. In the context of the CYGNO experiment, the expected performances for a $\mathcal{O}$(30) m$^3$ detector were evaluated highlighting the role of directionality, and R\&D studies on the amplification structure and gas diffusion were carried out. The results of the two latter studies open the possibility for very large low-diffusion low-energy threshold gaseous detector experiments with optical readout. The impact of the research, though, is not limited to DM matter searches, but can positively impact gaseous imaging detectors. 

\appendixpage
\appendix
\chapter{Velocity integral calculation}
\label{app:calcrateWIMP}
In this appendix the calculation of the velocity integral is performed with more details than in the text of Chapter \ref{chap2}.\\
Starting from Equation \ref{eq:rate_I}, the velocity integral can be written as:
\begin{equation}
\label{eq:rate_I2}
I= \frac{1}{2\pi}\int \delta\left(v\cos\theta-\frac{q}{2\mu}\right)f(\vec{v})d^3v .
\end{equation}
The term $v\cos\theta$ can be written as the scalar product of the vector $\vec{v}$ of the \W velocity with a unitary vector $\hat{w}$ which has the direction of the recoiling nucleus.
\begin{equation}
\label{eq:rate_I3}
I= \frac{\alpha}{2\pi}\int \delta\left(\vec{v}\cdot \hat{w}-\frac{q}{2\mu}\right)v^2e^{-\frac{v^2}{v_p^2}}dvd\cos\epsilon d\lambda .
\end{equation}
where the $f(\vec{v})$ has been replaced by its representation in Equation \ref{eq:fv} and $d^3v$ is expanded in spherical coordinates. Following \cite{Gondolo_2002}, the Radon transform $\hat{f}(w,\hat{w})$ can be defined as:
\begin{equation}
\label{eq:radontrans}
\hat{f}(w,\hat{w}) \equiv \int \delta\left(\vec{v}\cdot \hat{w} - w\right)f(\vec{v})d^3v
\end{equation} 
The integral is written in a frame of reference comoving with Galaxy, thus to correctly take calculate the integral in the laboratory frame a change of coordinates is needed. A pure translation of the velocity $\vec{v}$ in $\vec{v}+\vec{v_{lab}}$ is performed. Thanks to the properties of the Radon transform \cite{Gondolo_2002}, the result of the change in coordinates can be written as:
\begin{equation}
\label{eq:rate_I4}
I= \frac{\alpha}{2\pi}\int \delta\left(\vec{v}\cdot \hat{w}+\vec{v}_{lab}\cdot \hat{w} - \frac{q}{2\mu}\right)v^2e^{-\frac{v^2}{v_p^2}}dvd\cos\epsilon d\lambda,
\end{equation}
as $\vec{v}_{lab}$ is taken as $-v_{lab}\hat{u_z}$ where $\hat{u_z}$ is tangent to the circular orbit of the Sun but in the opposite sense than the motion of the Sun. The scalar products can be made explicit:
\begin{equation}
\label{eq:rate_I5}
I= \frac{\alpha}{2\pi}\int \delta\left(v\cos\epsilon+v_{lab}\cos\gamma - \frac{q}{2\mu_A}\right)v^2e^{-\frac{v^2}{v_p^2}}dvd\cos\epsilon d\lambda,
\end{equation}
Here $\gamma$ is the angle between the recoil and the direction of $\vec{v}_{lab}$, while $\epsilon$ is the angle between the recoil and the \W impinging direction. It is important to stress that $\epsilon$ and $\theta$ are indeed representing the same angle, but the physical interpretation changes. $\theta$ is the angle between the two aforementioned directions when the DM particle is fixed and the recoil direction is left free to vary. Instead, in this integral the recoil direction is fixed, but the \W is now changing the arrival direction while integrating on the 3D velocity space. In conclusion, the angles are the same, but are renamed due to physical meaning.\\
Now, integrating the angular part and considering Dirac's delta as the derivative of the Heaviside's Theta:
\begin{equation}
\label{eq:rate_I6}
I= \alpha\int \Theta\left(v+v_{lab}\cos\gamma - \frac{q}{2\mu_A}\right)ve^{-\frac{v^2}{v_p^2}}dv =\alpha \int_{\frac{q}{2\mu_A}>v_{lab}\cos\gamma}^{\infty}ve^{-\frac{v^2}{v_p^2}}dv
\end{equation}
and finally,
\begin{equation}
\label{eq:rate_I7}
I= \frac{\alpha v_p^2}{2} e^{-\frac{\left(\frac{\sqrt{2m_AE}}{2\mu_A}-v_{lab}\cos\gamma\right)^2}{v_p^2}},
\end{equation}
In this calculation the escape velocity was not included. Following \cite{Gondolo_2002}, and using the complete version of $f(\vec{v})$:
\begin{equation}
\label{eq:fvcomplete}
f(\vec{v}) = \alpha' e^{-\frac{v^2}{v_p^2}}\Theta(v_{esc}-\vert\vec{v}+\vec{v_{lab}}\vert),
\end{equation}
the integral turns out to be:
\begin{equation}
\label{eq:rate_Ifinal}
I= \frac{\alpha' v_p^3}{2v_{lab}}\left( e^{-\frac{\left(\frac{\sqrt{2m_AE}}{2\mu_A}-v_{lab}\cos\gamma\right)^2}{v_p^2}}- e^{-\frac{v_{esc}^2}{v_p^2}}\right)\Theta\left(\cos\gamma -\frac{\frac{\sqrt{2m_AE}}{2\mu_A}-v_{esc}}{v_{lab}}\right),
\end{equation}
\chapter{Global gain analysis}
\label{appD}
\begin{figure}[!t] 
	\centering
	\includegraphics[width=1\linewidth]{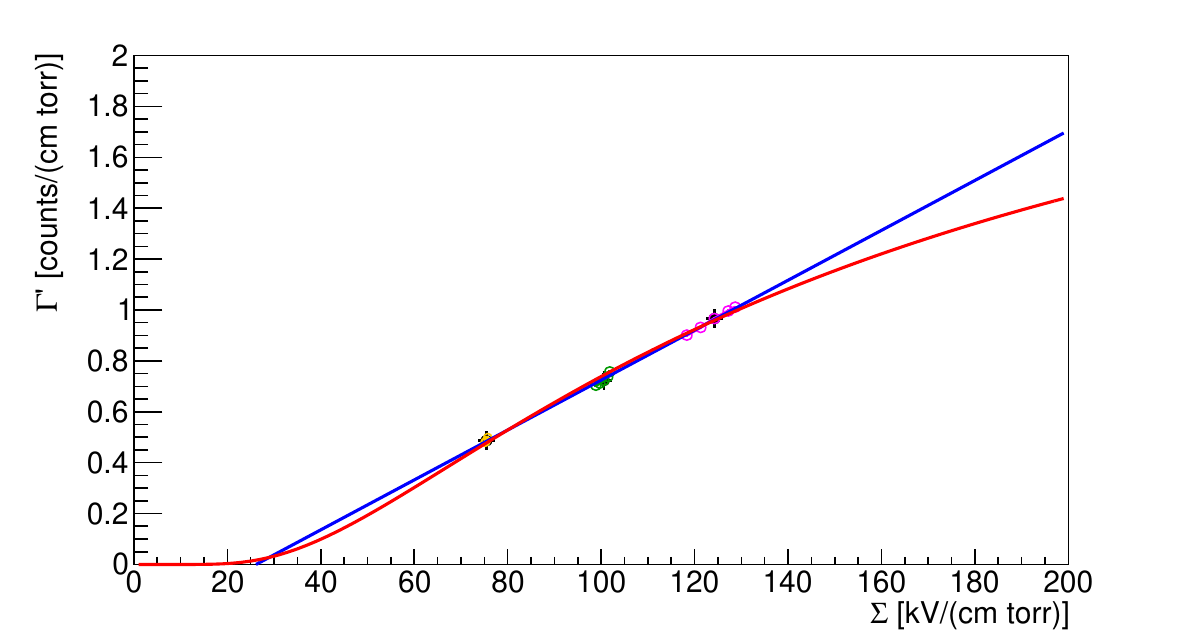}
	\caption{The reduced light gain as a function of the reduced field. The data points are obtained with the average value of gain and field of the three different configurations of amplification at 60/40 of He:CF$_4$. The two fitting curves represent the extreme function obtained from Equation \ref{eq:gainbis}.}
	\label{fig:tomplot}
\end{figure}

In Section \ref{subsec:gain}, Equation \ref{eq:gain}, which expresses the reduced gain $\Gamma$ as a function of the reduced field $\Sigma$, was obtained from \cite{bib:tom,bib:Aoyama} and is here reported for simplicity:
\begin{equation}
\label{eq:gainbis}
\Gamma = \frac{ln(G)}{n_g pt} = A \left(\frac{V_{GEM}}{n_g pt}\right)^m exp\left(-B\left(\frac{n_g pt}{V_{GEM}}\right)^{1-m}\right),
\end{equation}
with $n_g$ the number of GEMs used in the amplification stage, $p$ the gas pressure, $t$ thickness of the GEM, $V_{GEM}$ total voltage applied to the GEMs and $m$, $A$, $B$ free parameters. In addition, the reduced field $\Sigma$ is defined as:
\begin{equation}
\label{eq:redf}
\Sigma = \frac{V_{GEM}}{n_g pt}
\end{equation}
In Section \ref{subsec:gain}, the parameter $m$ is always approximated with the value of 1 as the gain scans do not span over a large range of voltages, hence of reduced fields. However, to have a wider view on the correctness of the amplification modelling, it is interesting to compare the data taken with the same gas mixture of He:CF$_4$ at 60/40 and different stacking option (see Table \ref{tab:app}). This way, due to the change in thickness and number of GEMs, the reduced field varies much more than when only the voltage across the GEMs is modified. The average reduced gain and field obtained averaging the results of the three different amplification configurations are displayed in Figure \ref{fig:tomplot}, with errors calculated from the propagation of the uncertainties on the $^{55}$Fe peaks and on the working pressure.  In this context, the reduced gain $\Gamma$ can be written as a function of the reduced field $\Sigma$ as (for $m=1$, akin to Equation \ref{eq:gamm_vsvoltsigma}):

\begin{equation}
\Gamma = A_0 +B_0\Sigma = \frac{(\Sigma-\Sigma_0)}{D}
\label{eq:gamma_m1}
\end{equation}	
and (for $m=0$):
\begin{equation}
\Gamma = A_2 e^{-\frac{B}{\Sigma}} 
\label{eq:gamma_m0}
\end{equation}

where $\Sigma_0$, $D$, $A_2$ and $B$ are the free parameters. The fits with the right-hand side of Equation \ref{eq:gamma_m1} (in blue) and Equation \ref{eq:gamma_m0} (in red) to the average reduced gain of the three different GEM configurations are shown in Figure \ref{fig:tomplot} and the fitted parameters are reported in Table \ref{tab:tomplot}.
Both curves well represent the data points, showing that the reduced fields explored in this study do not allow to determine $m$ for this gas mixtures and GEM configurations. This result is consistent with the expected operation of the GEM amplification structures, as discussed in the work presented in \cite{bib:tom}.
\begin{table}[!t]
	\centering
	\begin{tabular}{|c|c|c|c|}
		\hline
		
		\large{$m$} & \large{Par0} & \large{Par1}  & \large{$\chi^2/NdF$}    \\ \hline
		0 & $2.8 \pm 0.3$  &  $134 \pm 15$ &  0.5/1   \\ \hline
		1 & $26 \pm 7$ & $102 \pm 10$ &  0.07/1   \\ \hline  
	\end{tabular}
	\caption{Table summarising the results of the fit with Equation \ref{eq:gamma_m0} and Equation \ref{eq:gamma_m1} to the data in Fig.\ref{fig:tomplot}. For $m=0$ \{Par0,Par1\} = \{$A_2$ [1/(Torr cm)],$B$ [V/(Torr cm)]\}, while for $m=1$  \{Par0,Par1\} = \{$\Sigma_0$ [V/(Torr cm)],$D$ [V] \}, with the corresponding unit of measurements.}
	\label{tab:tomplot}
\end{table}	

\chapter{SNDM differential spectrum calculation}
\label{app:calcrateSNDM}
In this appendix the derivation of the double differential spectrum of the \SN induced nuclear recoils is calculated. Equation \ref{eq:rateSN1} displays the double differential rate in momentum transfer and solid angle. Utilising the general Fermi's Golden rule to make explicit the differential cross section it is obtained:
\begin{equation}
\label{eq:rateSN2}
\frac{dR}{dq^2d\Omega} = \frac{N_0}{A}n_{SNDM} \frac{1}{2\pi}\int \frac{\sigma_{SNDM,A}}{4p^2} \delta\left(\cos\theta-\frac{q}{2p}\right) v f(\vec{v})d^3v,
\end{equation}
with $n_{SNDM}$ the number density of \SN particles close to the laboratory location and $\sigma_{SNDM,A}$ is the cross section between the dark fermion and the nucleaus of mass number $A$. The value of the cross section is highly model dependent. According to the discussion in \cite{DeRocco_2019_2}, the temperature $T$ of thermal equilibrium and the coupling factor $y$ between the dark fermion and the SM particles are related each other again depending on the specific model. As the purpose of this calculation in the context of the thesis is to compute the shape of the energetic and angular spectra for a \SN particle (see Section \ref{sec:discr_2models}) the overall multiplicative constants are irrelevant for the evaluation. Since the velocity $v$ is linked to the momentum $p$ relativistically, the velocity distribution can be rewritten in terms of momenta. Moreover, the integral of the velocity $v$ has to be performed on the modified momentum distribution due to the gravitational redshift as in Equation \ref{eq:SNDM_pstardistr} and \ref{eq:SNDM_pstar}. As a result the following relation is obtained:
\begin{equation}
\label{eq:rateSN3}
\frac{dR}{dq^2d\cos\theta} 
\propto \int S(q)\frac{1}{p^2} \delta\left(\cos\theta-\frac{q}{2p}\right) \frac{p/m_{\chi}}{\sqrt{1+(p/m_{\chi})^2}} p_*^2f_*(p_*)dp_*d\Omega_p,
\end{equation}
Now, it is possible to integrate on the angular part in the phase space and on the azimuthal angle of the recoil. In addition to this, writing the $p_*$ terms as a function of the $p$ at the laboratory:
\begin{equation}
\label{eq:rateSN4}
\frac{dR}{dEd\cos\theta} 
\propto \frac{1}{(1-2\Phi)^{3/2}}\int S(E) \delta\left(\cos\theta-\frac{\sqrt{2m_AE}}{2p}\right) \sqrt{\frac{p^2+2\Phi m_{\chi}^2}{p^2+m_{\chi}^2}}f_*(p)dp,
\end{equation}
which is used in Equation \ref{eq:rateSNDM}. If one wishes to continue the calculation to derive the single differential rates, $p$ can be made the principle variable of the Dirac's delta, $$\delta\left(\cos\theta-\frac{\sqrt{2m_AE}}{2p}\right)= \delta\left(p-\frac{\sqrt{2m_AE}}{2\cos\theta}\right)\frac{p}{\cos\theta}$$
and the integral on $p$ solved exploiting the properties of the delta function.
\chapter{Schneider Xenon Lens technical sheet}\label{appC}
\includepdf[pages={1-2}]{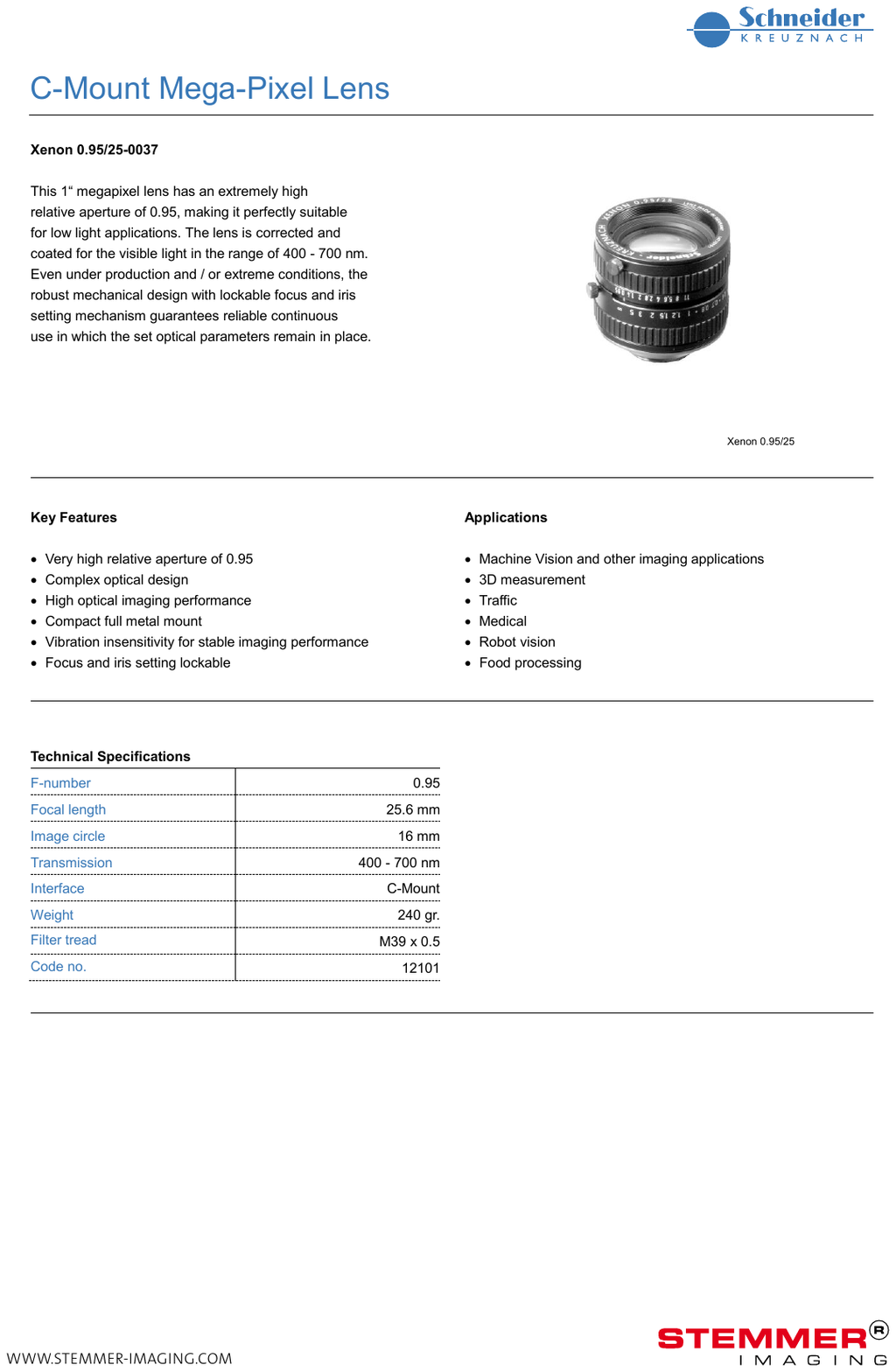}

\backmatter
\bibliographystyle{customthesis}
\bibliography{Reference.bib}

\begin{thebibliography}{100}
\expandafter\ifx\csname urlstyle\endcsname\relax
  \providecommand{\doi}[1]{doi:\discretionary{}{}{}#1}\else
  \providecommand{\doi}{doi:\discretionary{}{}{}\begingroup
  \urlstyle{rm}\Url}\fi

\bibitem{bib:Plank_2020}
PLANCK.
\newblock \emph{Planck 2018 results}.
\newblock \emph{Astronomy: Astrophysics}, \textbf{641} (2020):A6.
\newblock \doi{10.1051/0004-6361/201833910}.

\bibitem{Frenk_2012}
C.~Frenk, S.~White.
\newblock \emph{Dark matter and cosmic structure}.
\newblock \emph{Annalen der Physik}, \textbf{524(9-10)} (2012):507--534.
\newblock \doi{10.1002/andp.201200212}.

\bibitem{bib:Bertone:2004pz}
G.~Bertone, et~al.
\newblock \emph{Particle dark matter: evidence, candidates and constraints}.
\newblock \emph{Physics Reports}, \textbf{405(5)} (2005):279--390.
\newblock \doi{https://doi.org/10.1016/j.physrep.2004.08.031}.

\bibitem{Strigari_2013}
L.~E. Strigari.
\newblock \emph{Galactic searches for dark matter}.
\newblock \emph{Physics Reports}, \textbf{531(1)} (2013):1--88.
\newblock \doi{10.1016/j.physrep.2013.05.004}.

\bibitem{Begeman_1991}
K.~G. Begeman, A.~H. Broeils, R.~H. Sanders.
\newblock \emph{{Extended rotation curves of spiral galaxies: dark haloes and
  modified dynamics}}.
\newblock \emph{Monthly Notices of the Royal Astronomical Society},
  \textbf{249(3)} (1991):523--537.
\newblock \doi{10.1093/mnras/249.3.523}.

\bibitem{sanders_2010}
R.~H. Sanders.
\newblock \emph{The Dark Matter Problem: A Historical Perspective}.
\newblock Cambridge University Press (2010).
\newblock \doi{10.1017/CBO9781139192309}.

\bibitem{Zwicky_1937}
F.~{Zwicky}.
\newblock \emph{{On the Masses of Nebulae and of Clusters of Nebulae}}.
\newblock \emph{apj}, \textbf{86} (1937):217.
\newblock \doi{10.1086/143864}.

\bibitem{Smith_1936}
S.~{Smith}.
\newblock \emph{{The Mass of the Virgo Cluster}}.
\newblock \emph{apj}, \textbf{83} (1936):23.
\newblock \doi{10.1086/143697}.

\bibitem{Bertone_2018}
G.~Bertone, D.~Hooper.
\newblock \emph{History of dark matter}.
\newblock \emph{Reviews of Modern Physics}, \textbf{90(4)} (2018).
\newblock \doi{10.1103/revmodphys.90.045002}.

\bibitem{Bandiera:2003qgt}
R.~Bandiera, R.~Maiolino, eds.
\newblock \emph{{Proceedings, 21st Texas Symposium on Relativistic Astrophysics
  (Texas in Tuscany)}: {Florence, Italy, December 9-13, 2002}}. World
  Scientific, River Edge, N.J. (2003).
\newblock \doi{10.1142/5373}.

\bibitem{GRWald}
R.~M. Wald.
\newblock \emph{{General Relativity}}.
\newblock Chicago Univ. Pr., Chicago, USA (1984).
\newblock \doi{10.7208/chicago/9780226870373.001.0001}.

\bibitem{Clowe_2006}
D.~Clowe, M.~Brada{\v{c}}, A.~H. Gonzalez, M.~Markevitch, S.~W. Randall,
  C.~Jones, D.~Zaritsky.
\newblock \emph{A Direct Empirical Proof of the Existence of Dark Matter}.
\newblock \emph{The Astrophysical Journal}, \textbf{648(2)} (2006):L109--L113.
\newblock \doi{10.1086/508162}.

\bibitem{Huterer_2010}
D.~Huterer.
\newblock \emph{Weak lensing, dark matter and dark energy}.
\newblock \emph{General Relativity and Gravitation}, \textbf{42(9)}
  (2010):2177--2195.
\newblock \doi{10.1007/s10714-010-1051-z}.

\bibitem{Wegg_2016}
C.~Wegg, O.~Gerhard, M.~Portail.
\newblock \emph{{MOA}-{II} Galactic microlensing constraints: the inner Milky
  Way has a low dark matter fraction and a near maximal disc}.
\newblock \emph{Monthly Notices of the Royal Astronomical Society},
  \textbf{463(1)} (2016):557--570.
\newblock \doi{10.1093/mnras/stw1954}.

\bibitem{Kolb:1990vq}
E.~W. Kolb, M.~S. Turner.
\newblock \emph{{The Early Universe}}, vol.~69 (1990).
\newblock \doi{10.1201/9780429492860}.

\bibitem{OZER1987776}
M.~Özer, M.~Taha.
\newblock \emph{A model of the universe free of cosmological problems}.
\newblock \emph{Nuclear Physics B}, \textbf{287} (1987):776--796.
\newblock \doi{https://doi.org/10.1016/0550-3213(87)90128-3}.

\bibitem{Arbey_2021}
A.~Arbey, F.~Mahmoudi.
\newblock \emph{Dark matter and the early Universe: A review}.
\newblock \emph{Progress in Particle and Nuclear Physics} (2021):103865.
\newblock \doi{10.1016/j.ppnp.2021.103865}.

\bibitem{ELWright}
E.~L. Wright.
\newblock \emph{Theoretical Overview of Cosmic Microwave Background Anisotropy}
  (2003).
\newblock \doi{10.48550/ARXIV.ASTRO-PH/0305591}.

\bibitem{Sparre_2015}
M.~Sparre, C.~C. Hayward, V.~Springel, M.~Vogelsberger, S.~Genel, P.~Torrey,
  D.~Nelson, D.~Sijacki, L.~Hernquist.
\newblock \emph{The star formation main sequence and stellar mass assembly of
  galaxies in the Illustris simulation}.
\newblock \emph{Monthly Notices of the Royal Astronomical Society},
  \textbf{447(4)} (2015):3548--3563.
\newblock \doi{10.1093/mnras/stu2713}.

\bibitem{Springel_2005}
V.~Springel, S.~D.~M. White, A.~Jenkins, C.~S. Frenk, N.~Yoshida, L.~Gao,
  J.~Navarro, R.~Thacker, D.~Croton, J.~Helly, J.~A. Peacock, S.~Cole,
  P.~Thomas, H.~Couchman, A.~Evrard, J.~Colberg, F.~Pearce.
\newblock \emph{Simulations of the formation, evolution and clustering of
  galaxies and quasars}.
\newblock \emph{Nature}, \textbf{435(7042)} (2005):629--636.
\newblock \doi{10.1038/nature03597}.

\bibitem{nu2GC}
T.~{Ishiyama}, M.~{Enoki}, M.~A.~R. {Kobayashi}, R.~{Makiya}, M.~{Nagashima},
  T.~{Oogi}.
\newblock \emph{{The $\nu^{2}$GC simulations: Quantifying the dark side of the
  universe in the Planck cosmology}}.
\newblock \emph{pasj}, \textbf{67(4)} (2015):61.
\newblock \doi{10.1093/pasj/psv021}.

\bibitem{White}
S.~D.~M. {White}, C.~S. {Frenk}, M.~{Davis}.
\newblock \emph{{Clustering in a neutrino-dominated universe}}.
\newblock \emph{apjl}, \textbf{274} (1983):L1--L5.
\newblock \doi{10.1086/184139}.

\bibitem{Milgrom_2015}
M.~Milgrom.
\newblock \emph{{MOND} theory}.
\newblock \emph{Canadian Journal of Physics}, \textbf{93(2)} (2015):107--118.
\newblock \doi{10.1139/cjp-2014-0211}.

\bibitem{Famaey_2012}
B.~Famaey, S.~S. McGaugh.
\newblock \emph{Modified Newtonian Dynamics ({MOND}): Observational
  Phenomenology and Relativistic Extensions}.
\newblock \emph{Living Reviews in Relativity}, \textbf{15(1)} (2012).
\newblock \doi{10.12942/lrr-2012-10}.

\bibitem{Mcgaugh_1}
S.~McGaugh.
\newblock \emph{Testing Galaxy Formation and Dark Matter with Low Surface
  Brightness Galaxies} (2021).
\newblock \doi{10.48550/ARXIV.2103.05003}.

\bibitem{Milgrom_2020}
M.~Milgrom.
\newblock \emph{{MOND} vs. dark matter in light of historical parallels}.
\newblock \emph{Studies in History and Philosophy of Science Part B: Studies in
  History and Philosophy of Modern Physics}, \textbf{71} (2020):170--195.
\newblock \doi{10.1016/j.shpsb.2020.02.004}.

\bibitem{Skordis_2021}
C.~Skordis, T.~Z{\l}o{\'{s} }nik.
\newblock \emph{New Relativistic Theory for Modified Newtonian Dynamics}.
\newblock \emph{Physical Review Letters}, \textbf{127(16)} (2021).
\newblock \doi{10.1103/physrevlett.127.161302}.

\bibitem{Angus_1}
G.~W. Angus.
\newblock \emph{{Is an 11 eV sterile neutrino consistent with clusters, the
  cosmic microwave background and modified Newtonian dynamics?}}
\newblock \emph{Monthly Notices of the Royal Astronomical Society},
  \textbf{394(1)} (2009):527--532.
\newblock \doi{10.1111/j.1365-2966.2008.14341.x}.

\bibitem{McGaugh_1999}
S.~S. McGaugh.
\newblock \emph{Distinguishing between Cold Dark Matter and Modified Newtonian
  Dynamics: Predictions for the Microwave Background}.
\newblock \emph{The Astrophysical Journal}, \textbf{523(2)} (1999):L99--L102.
\newblock \doi{10.1086/312274}.

\bibitem{Zeldovich}
Y.~B. {Zel'dovich}, I.~D. {Novikov}.
\newblock \emph{{The Hypothesis of Cores Retarded during Expansion and the Hot
  Cosmological Model}}.
\newblock \emph{sovast}, \textbf{10} (1967):602.

\bibitem{Hawking}
S.~Hawking.
\newblock \emph{{Gravitationally Collapsed Objects of Very Low Mass}}.
\newblock \emph{Monthly Notices of the Royal Astronomical Society},
  \textbf{152(1)} (1971):75--78.
\newblock \doi{10.1093/mnras/152.1.75}.

\bibitem{Coc_2017}
A.~Coc, E.~Vangioni.
\newblock \emph{Primordial nucleosynthesis}.
\newblock \emph{International Journal of Modern Physics E}, \textbf{26(08)}
  (2017):1741002.
\newblock \doi{10.1142/s0218301317410026}.

\bibitem{Zyla:2020zbs}
P.~Zyla, et~al.
\newblock \emph{{Review of Particle Physics}}.
\newblock \emph{PTEP}, \textbf{2020(8)} (2020):083C01.
\newblock \doi{10.1093/ptep/ptaa104}.

\bibitem{Carr_2020}
B.~Carr, F.~Kühnel.
\newblock \emph{Primordial Black Holes as Dark Matter: Recent Developments}.
\newblock \emph{Annual Review of Nuclear and Particle Science}, \textbf{70(1)}
  (2020):355--394.
\newblock \doi{10.1146/annurev-nucl-050520-125911}.

\bibitem{Chapline}
G.~F. Chapline.
\newblock \emph{Cosmological effects of primordial black holes}.
\newblock \emph{Nature}, \textbf{253} (1975).
\newblock \doi{https://doi.org/10.1038/253251a0}.

\bibitem{Billard_2022}
J.~Billard, M.~Boulay, S.~Cebri{\'{a} }n, L.~Covi, G.~Fiorillo, A.~Green,
  J.~Kopp, B.~Majorovits, K.~Palladino, F.~Petricca, L.~R. (chair),
  M.~Schumann.
\newblock \emph{Direct detection of dark matter APPEC committee report}.
\newblock \emph{Reports on Progress in Physics}, \textbf{85(5)} (2022):056201.
\newblock \doi{10.1088/1361-6633/ac5754}.

\bibitem{Green_2021}
A.~M. Green, B.~J. Kavanagh.
\newblock \emph{Primordial black holes as a dark matter candidate}.
\newblock \emph{Journal of Physics G: Nuclear and Particle Physics},
  \textbf{48(4)} (2021):043001.
\newblock \doi{10.1088/1361-6471/abc534}.

\bibitem{PrimordialBH}
P.~Villanueva-Domingo, O.~Mena, S.~Palomares-Ruiz.
\newblock \emph{A Brief Review on Primordial Black Holes as Dark Matter}.
\newblock \emph{Frontiers in Astronomy and Space Sciences}, \textbf{8} (2021).
\newblock \doi{10.3389/fspas.2021.681084}.

\bibitem{limitsSM}
J.~Ellis.
\newblock \emph{Limits of the Standard Model} (2002).
\newblock \doi{10.48550/ARXIV.HEP-PH/0211168}.

\bibitem{Lee_2021}
H.~M. Lee.
\newblock \emph{Lectures on physics beyond the Standard Model}.
\newblock \emph{Journal of the Korean Physical Society}, \textbf{78(11)}
  (2021):985--1017.
\newblock \doi{10.1007/s40042-021-00188-x}.

\bibitem{Fan_2013}
J.~Fan, A.~Katz, L.~Randall, M.~Reece.
\newblock \emph{Double-Disk Dark Matter}.
\newblock \emph{Physics of the Dark Universe}, \textbf{2(3)} (2013):139--156.
\newblock \doi{10.1016/j.dark.2013.07.001}.

\bibitem{Lovell_2012}
M.~R. Lovell, V.~Eke, C.~S. Frenk, L.~Gao, A.~Jenkins, T.~Theuns, J.~Wang,
  S.~D.~M. White, A.~Boyarsky, O.~Ruchayskiy.
\newblock \emph{The haloes of bright satellite galaxies in a warm dark matter
  universe}.
\newblock \emph{Monthly Notices of the Royal Astronomical Society},
  \textbf{420(3)} (2012):2318--2324.
\newblock \doi{10.1111/j.1365-2966.2011.20200.x}.

\bibitem{1992ApJ}
O.~E. {Gerhard}, D.~N. {Spergel}.
\newblock \emph{{Dwarf Spheroidal Galaxies and the Mass of the Neutrino}}.
\newblock \emph{APJ}, \textbf{389} (1992):L9.
\newblock \doi{10.1086/186336}.

\bibitem{PhysRevD.33.1585}
R.~J. Scherrer, M.~S. Turner.
\newblock \emph{On the relic, cosmic abundance of stable, weakly interacting
  massive particles}.
\newblock \emph{Phys. Rev. D}, \textbf{33} (1986):1585--1589.
\newblock \doi{10.1103/PhysRevD.33.1585}.

\bibitem{Bernal_2017}
N.~Bernal, M.~Heikinheimo, T.~Tenkanen, K.~Tuominen, V.~Vaskonen.
\newblock \emph{The dawn of {FIMP} Dark Matter: A review of models and
  constraints}.
\newblock \emph{International Journal of Modern Physics A}, \textbf{32(27)}
  (2017):1730023.
\newblock \doi{10.1142/s0217751x1730023x}.

\bibitem{Feng_2010}
J.~L. Feng.
\newblock \emph{Dark Matter Candidates from Particle Physics and Methods of
  Detection}.
\newblock \emph{Annual Review of Astronomy and Astrophysics}, \textbf{48(1)}
  (2010):495--545.
\newblock \doi{10.1146/annurev-astro-082708-101659}.

\bibitem{Fox:20196D}
P.~J. Fox.
\newblock \emph{{TASI Lectures on WIMPs and Supersymmetry}}.
\newblock \emph{PoS}, \textbf{TASI2018} (2019):005.
\newblock \doi{10.22323/1.333.0005}.

\bibitem{Roszkowski_2018}
L.~Roszkowski, E.~M. Sessolo, S.~Trojanowski.
\newblock \emph{{WIMP} dark matter candidates and searches{\textemdash}current
  status and future prospects}.
\newblock \emph{Reports on Progress in Physics}, \textbf{81(6)} (2018):066201.
\newblock \doi{10.1088/1361-6633/aab913}.

\bibitem{Andriot_2017}
D.~Andriot, G.~L. G{\'{o} }mez.
\newblock \emph{Signatures of extra dimensions in gravitational waves}.
\newblock \emph{Journal of Cosmology and Astroparticle Physics},
  \textbf{2017(06)} (2017):048--048.
\newblock \doi{10.1088/1475-7516/2017/06/048}.

\bibitem{PDG}
P.~D. Group.
\newblock \emph{{Review of Particle Physics}}.
\newblock \emph{Progress of Theoretical and Experimental Physics},
  \textbf{2020(8)} (2020).
\newblock \doi{10.1093/ptep/ptaa104}.
\newblock 083C01.

\bibitem{Appelquist_2001}
T.~Appelquist, H.-C. Cheng, B.~A. Dobrescu.
\newblock \emph{Bounds on universal extra dimensions}.
\newblock \emph{Physical Review D}, \textbf{64(3)} (2001).
\newblock \doi{10.1103/physrevd.64.035002}.

\bibitem{Servant_2003}
G.~Servant, T.~M. Tait.
\newblock \emph{Is the lightest Kaluza{\textendash}Klein particle a viable dark
  matter candidate?}
\newblock \emph{Nuclear Physics B}, \textbf{650(1-2)} (2003):391--419.
\newblock \doi{10.1016/s0550-3213(02)01012-x}.

\bibitem{Peccei:1977hh}
R.~D. Peccei, H.~R. Quinn.
\newblock \emph{{CP Conservation in the Presence of Instantons}}.
\newblock \emph{Phys. Rev. Lett.}, \textbf{38} (1977):1440--1443.
\newblock \doi{10.1103/PhysRevLett.38.1440}.

\bibitem{Abel:2020pzs}
C.~Abel, et~al.
\newblock \emph{{Measurement of the Permanent Electric Dipole Moment of the
  Neutron}}.
\newblock \emph{Phys. Rev. Lett.}, \textbf{124(8)} (2020):081803.
\newblock \doi{10.1103/PhysRevLett.124.081803}.

\bibitem{Pendlebury:2015lrz}
J.~M. Pendlebury, et~al.
\newblock \emph{{Revised experimental upper limit on the electric dipole moment
  of the neutron}}.
\newblock \emph{Phys. Rev. D}, \textbf{92(9)} (2015):092003.
\newblock \doi{10.1103/PhysRevD.92.092003}.

\bibitem{PhysRevLett.40.279}
F.~Wilczek.
\newblock \emph{Problem of Strong $P$ and $T$ Invariance in the Presence of
  Instantons}.
\newblock \emph{Phys. Rev. Lett.}, \textbf{40} (1978):279--282.
\newblock \doi{10.1103/PhysRevLett.40.279}.

\bibitem{di_Cortona_2016}
G.~G. di~Cortona, E.~Hardy, J.~P. Vega, G.~Villadoro.
\newblock \emph{The {QCD} axion, precisely}.
\newblock \emph{Journal of High Energy Physics}, \textbf{2016(1)} (2016).
\newblock \doi{10.1007/jhep01(2016)034}.

\bibitem{axions_astro}
G.~Raffelt.
\newblock \emph{AXIONS IN ASTROPHYSICS AND COSMOLOGY} (1995).
\newblock \doi{10.48550/ARXIV.HEP-PH/9502358}.

\bibitem{Sikivie_2008}
P.~Sikivie.
\newblock \emph{Axion Cosmology}.
\newblock In \emph{Lecture Notes in Physics}. Springer Berlin Heidelberg
  (2008), pp. 19--50.
\newblock \doi{10.1007/978-3-540-73518-2_2}.

\bibitem{Mass__2002}
E.~Mass{\'{o} }, F.~Rota, G.~Zsembinszki.
\newblock \emph{Axion thermalization in the early universe}.
\newblock \emph{Physical Review D}, \textbf{66(2)} (2002).
\newblock \doi{10.1103/physrevd.66.023004}.

\bibitem{Garcia_Irastorza_2022}
I.~G. Irastorza.
\newblock \emph{An introduction to axions and their detection}.
\newblock \emph{{SciPost} Physics Lecture Notes} (2022).
\newblock \doi{10.21468/scipostphyslectnotes.45}.

\bibitem{Raffelt_2007}
G.~G. Raffelt.
\newblock \emph{Axions—motivation, limits and searches}.
\newblock \emph{Journal of Physics A: Mathematical and Theoretical},
  \textbf{40(25)} (2007):6607.
\newblock \doi{10.1088/1751-8113/40/25/S05}.

\bibitem{nuoscillation}
M.~H. e.~a. Ahn.
\newblock \emph{Measurement of neutrino oscillation by the K2K experiment}.
\newblock \emph{Phys. Rev. D}, \textbf{74} (2006):072003.
\newblock \doi{10.1103/PhysRevD.74.072003}.

\bibitem{Dodelson_1994}
S.~Dodelson, L.~M. Widrow.
\newblock \emph{Sterile neutrinos as dark matter}.
\newblock \emph{Physical Review Letters}, \textbf{72(1)} (1994):17--20.
\newblock \doi{10.1103/physrevlett.72.17}.

\bibitem{Hall_2010}
L.~J. Hall, K.~Jedamzik, J.~March-Russell, S.~M. West.
\newblock \emph{Freeze-in production of {FIMP} dark matter}.
\newblock \emph{Journal of High Energy Physics}, \textbf{2010(3)} (2010).
\newblock \doi{10.1007/jhep03(2010)080}.

\bibitem{SHROCK1982359}
R.~E. Shrock.
\newblock \emph{Electromagnetic properties and decays of Dirac and Majorana
  neutrinos in a general class of gauge theories}.
\newblock \emph{Nuclear Physics B}, \textbf{206(3)} (1982):359--379.
\newblock \doi{https://doi.org/10.1016/0550-3213(82)90273-5}.

\bibitem{darksector}
A.~Jim, et~al.
\newblock \emph{Dark Sectors 2016 Workshop: Community Report} (2016).
\newblock \doi{10.48550/ARXIV.1608.08632}.

\bibitem{Fabbrichesi_2021}
M.~Fabbrichesi, E.~Gabrielli, G.~Lanfranchi.
\newblock \emph{The Physics of the Dark Photon}.
\newblock Springer International Publishing (2021).
\newblock \doi{10.1007/978-3-030-62519-1}.

\bibitem{PhysRevD.70.083501}
K.~Sigurdson, M.~Doran, A.~Kurylov, R.~R. Caldwell, M.~Kamionkowski.
\newblock \emph{Dark-matter electric and magnetic dipole moments}.
\newblock \emph{Phys. Rev. D}, \textbf{70} (2004):083501.
\newblock \doi{10.1103/PhysRevD.70.083501}.

\bibitem{Boehm_2004}
C.~Boehm, T.~A. En{\ss}lin, J.~Silk.
\newblock \emph{Can annihilating dark matter be lighter than a few {GeVs}?}
\newblock \emph{Journal of Physics G: Nuclear and Particle Physics},
  \textbf{30(3)} (2004):279--285.
\newblock \doi{10.1088/0954-3899/30/3/004}.

\bibitem{DeRocco_2019}
W.~DeRocco, P.~W. Graham, D.~Kasen, G.~Marques-Tavares, S.~Rajendran.
\newblock \emph{Observable signatures of dark photons from supernovae}.
\newblock \emph{Journal of High Energy Physics}, \textbf{2019(2)} (2019).
\newblock \doi{10.1007/jhep02(2019)171}.

\bibitem{Conrad2014}
J.~Conrad.
\newblock \emph{Indirect Detection of WIMP Dark Matter: a compact review}
  (2014).
\newblock \doi{10.48550/ARXIV.1411.1925}.

\bibitem{Buchmueller_2017}
O.~Buchmueller, C.~Doglioni, L.-T. Wang.
\newblock \emph{Search for dark matter at colliders}.
\newblock \emph{Nature Physics}, \textbf{13(3)} (2017):217--223.
\newblock \doi{10.1038/nphys4054}.

\bibitem{Battaglieri_2022}
M.~Battaglieri, M.~Bond{\'{\i} }, A.~Celentano, P.~Cole, M.~D. Napoli, R.~D.
  Vita, L.~Marsicano, N.~Randazzo, E.~Smith, M.~Spreafico, M.~Wood.
\newblock \emph{Dark matter search with the {BDX}-{MINI} experiment}.
\newblock \emph{Physical Review D}, \textbf{106(7)} (2022).
\newblock \doi{10.1103/physrevd.106.072011}.

\bibitem{snowall}
J.~Cooley, et~al.
\newblock \emph{Report of the Topical Group on Particle Dark Matter for
  Snowmass 2021} (2022).
\newblock \doi{10.48550/ARXIV.2209.07426}.

\bibitem{Gondolo_2002}
P.~Gondolo.
\newblock \emph{Recoil momentum spectrum in directional dark matter detectors}.
\newblock \emph{Physical Review D}, \textbf{66(10)} (2002).
\newblock \doi{10.1103/physrevd.66.103513}.

\bibitem{LEWIN199687}
J.~Lewin, P.~Smith.
\newblock \emph{Review of mathematics, numerical factors, and corrections for
  dark matter experiments based on elastic nuclear recoil}.
\newblock \emph{Astroparticle Physics}, \textbf{6(1)} (1996):87--112.
\newblock \doi{https://doi.org/10.1016/S0927-6505(96)00047-3}.

\bibitem{SCHNEE_2011}
R.~W. Schnee.
\newblock \emph{{Introduction} {to} {dark} {matter} {experiments}}.
\newblock In \emph{Physics of the Large and the Small}. {WORLD} {SCIENTIFIC}
  (2011).
\newblock \doi{10.1142/9789814327183_0014}.

\bibitem{Standard_halo}
J.~{Binney}, S.~{Tremaine}.
\newblock \emph{{Galactic Dynamics: Second Edition}} (2008).

\bibitem{bib:Baxter_2021}
D.~Baxter, et~al.
\newblock \emph{Recommended conventions for reporting results from direct dark
  matter searches}.
\newblock \emph{The European Physical Journal C}, \textbf{81(10)} (2021).
\newblock \doi{10.1140/epjc/s10052-021-09655-y}.

\bibitem{Reid_2019}
M.~J. Reid, K.~M. Menten, A.~Brunthaler, X.~W. Zheng, T.~M. Dame, Y.~Xu, J.~Li,
  N.~Sakai, Y.~Wu, K.~Immer, B.~Zhang, A.~Sanna, L.~Moscadelli, K.~L.~J. Rygl,
  A.~Bartkiewicz, B.~Hu, L.~H. Quiroga-Nu{\~{n} }ez, H.~J. van Langevelde.
\newblock \emph{Trigonometric Parallaxes of High-mass Star-forming Regions: Our
  View of the Milky Way}.
\newblock \emph{The Astrophysical Journal}, \textbf{885(2)} (2019):131.
\newblock \doi{10.3847/1538-4357/ab4a11}.

\bibitem{Eilers_2019}
A.-C. Eilers, D.~W. Hogg, H.-W. Rix, M.~K. Ness.
\newblock \emph{The Circular Velocity Curve of the Milky Way from 5 to 25 kpc}.
\newblock \emph{The Astrophysical Journal}, \textbf{871(1)} (2019):120.
\newblock \doi{10.3847/1538-4357/aaf648}.

\bibitem{bib:Mayet_2016zxu}
F.~Mayet, et~al.
\newblock \emph{A review of the discovery reach of directional Dark Matter
  detection}.
\newblock \emph{Physics Reports}, \textbf{627} (2016):1--49.
\newblock \doi{10.1016/j.physrep.2016.02.007}.

\bibitem{Piffl_2014}
T.~Piffl, C.~Scannapieco, J.~Binney, M.~Steinmetz, R.-D. Scholz, M.~E.~K.
  Williams, R.~S. de~Jong, G.~Kordopatis, G.~Matijevi{\v{c} },
  O.~Bienaym{\'{e}}, J.~Bland-Hawthorn, C.~Boeche, K.~Freeman, B.~Gibson,
  G.~Gilmore, E.~K. Grebel, A.~Helmi, U.~Munari, J.~F. Navarro, Q.~Parker,
  W.~A. Reid, G.~Seabroke, F.~Watson, R.~F.~G. Wyse, T.~Zwitter.
\newblock \emph{The {RAVE} survey: the Galactic escape speed and the mass of
  the Milky Way}.
\newblock \emph{Astronomy \& Astrophysics}, \textbf{562} (2014):A91.
\newblock \doi{10.1051/0004-6361/201322531}.

\bibitem{Gaia_escape}
G.~{Monari}, B.~{Famaey}, I.~{Carrillo}, T.~{Piffl}, M.~{Steinmetz}, R.~F.~G.
  {Wyse}, F.~{Anders}, C.~{Chiappini}, K.~{Jan{\ss}en}.
\newblock \emph{{The escape speed curve of the Galaxy obtained from Gaia DR2
  implies a heavy Milky Way}}.
\newblock \emph{aap}, \textbf{616} (2018):L9.
\newblock \doi{10.1051/0004-6361/201833748}.

\bibitem{Navarro_1996}
J.~F. Navarro, C.~S. Frenk, S.~D.~M. White.
\newblock \emph{The Structure of Cold Dark Matter Halos}.
\newblock \emph{The Astrophysical Journal}, \textbf{462} (1996):563.
\newblock \doi{10.1086/177173}.

\bibitem{refId0}
{Polukhina, Natalia}, {Starkov, Nikolai}.
\newblock \emph{New experiment for WIMP direct search (NEWSdm)}.
\newblock \emph{EPJ Web Conf.}, \textbf{191} (2018):02023.
\newblock \doi{10.1051/epjconf/201819102023}.

\bibitem{PhysRevD.37.2703}
K.~Griest.
\newblock \emph{Effect of the Sun's gravity on the distribution and detection
  of dark matter near the Earth}.
\newblock \emph{Phys. Rev. D}, \textbf{37} (1988):2703--2713.
\newblock \doi{10.1103/PhysRevD.37.2703}.

\bibitem{Froborg_2020}
F.~Froborg, A.~R. Duffy.
\newblock \emph{Annual modulation in direct dark matter searches}.
\newblock \emph{Journal of Physics G: Nuclear and Particle Physics},
  \textbf{47(9)} (2020):094002.
\newblock \doi{10.1088/1361-6471/ab8e93}.

\bibitem{triaxialhalo}
M.~Kuhlen, M.~Vogelsberger, R.~Angulo.
\newblock \emph{Numerical Simulations of the Dark Universe: State of the Art
  and the Next Decade} (2012).
\newblock \doi{10.48550/ARXIV.1209.5745}.

\bibitem{Gaia}
M.~Perryman, K.~Zioutas.
\newblock \emph{Gaia, Fundamental Physics, and Dark Matter} (2021).
\newblock \doi{10.48550/ARXIV.2106.15408}.

\bibitem{halosim1}
M.~Vogelsberger, A.~Helmi, V.~Springel, S.~D.~M. White, J.~Wang, C.~S. Frenk,
  A.~Jenkins, A.~Ludlow, J.~F. Navarro.
\newblock \emph{{Phase-space structure in the local dark matter distribution
  and its signature in direct detection experiments}}.
\newblock \emph{Monthly Notices of the Royal Astronomical Society},
  \textbf{395(2)} (2009):797--811.
\newblock \doi{10.1111/j.1365-2966.2009.14630.x}.

\bibitem{Pillepich_2014}
A.~Pillepich, M.~Kuhlen, J.~Guedes, P.~Madau.
\newblock \emph{THE DISTRIBUTION OF DARK MATTER IN THE MILKY WAY’S DISK}.
\newblock \emph{The Astrophysical Journal}, \textbf{784(2)} (2014):161.
\newblock \doi{10.1088/0004-637X/784/2/161}.

\bibitem{Purcell_2009}
C.~W. Purcell, J.~S. Bullock, M.~Kaplinghat.
\newblock \emph{THE DARK DISK OF THE MILKY WAY}.
\newblock \emph{The Astrophysical Journal}, \textbf{703(2)} (2009):2275.
\newblock \doi{10.1088/0004-637X/703/2/2275}.

\bibitem{Kurylov_2004}
A.~Kurylov, M.~Kamionkowski.
\newblock \emph{Generalized analysis of the direct weakly interacting massive
  particle searches}.
\newblock \emph{Physical Review D}, \textbf{69(6)} (2004).
\newblock \doi{10.1103/physrevd.69.063503}.

\bibitem{Tovey_2000}
D.~Tovey, R.~Gaitskell, P.~Gondolo, Y.~Ramachers, L.~Roszkowski.
\newblock \emph{A new model-independent method for extracting spin-dependent
  cross section limits from dark matter searches}.
\newblock \emph{Physics Letters B}, \textbf{488(1)} (2000):17--26.
\newblock \doi{10.1016/s0370-2693(00)00846-7}.

\bibitem{Radon_dean}
S.~Deans.
\newblock \emph{{The Radon Transform and Some of Its Applications}}.
\newblock John Wiley \& Sons Ltd. (1983).

\bibitem{Radon_radon}
J.~Radon.
\newblock \emph{$\ddot{U}$ber die Bestimmung von Funktionen durch ihre
  Integralwerte l$\ddot{a}$ngs gewisser Mannigfaltigkeiten}.
\newblock \emph{Akad. Wiss.}, \textbf{262(69)} (1917).

\bibitem{RevModPhys.85.1561}
K.~Freese, M.~Lisanti, C.~Savage.
\newblock \emph{Colloquium: Annual modulation of dark matter}.
\newblock \emph{Rev. Mod. Phys.}, \textbf{85} (2013):1561--1581.
\newblock \doi{10.1103/RevModPhys.85.1561}.

\bibitem{Monroe_2007}
J.~Monroe, P.~Fisher.
\newblock \emph{Neutrino backgrounds to dark matter searches}.
\newblock \emph{Physical Review D}, \textbf{76(3)} (2007).
\newblock \doi{10.1103/physrevd.76.033007}.

\bibitem{DEBICKI2009429}
Z.~Dębicki, K.~Jędrzejczak, J.~Karczmarczyk, M.~Kasztelan, R.~Lewandowski,
  J.~Orzechowski, J.~Szabelski, M.~Szeptycka, P.~Tokarski.
\newblock \emph{Thermal neutrons at Gran Sasso}.
\newblock \emph{Nuclear Physics B - Proceedings Supplements}, \textbf{196}
  (2009):429--432.
\newblock \doi{https://doi.org/10.1016/j.nuclphysbps.2009.09.084}.
\newblock Proceedings of the XV International Symposium on Very High Energy
  Cosmic Ray Interactions (ISVHECRI 2008).

\bibitem{osti_15215}
D.~G. Madland, E.~D. Arthur, G.~P. Estes, J.~E. Stewart, M.~Bozoian, R.~T.
  Perry, T.~A. Parish, T.~H. Brown, T.~R. England, W.~B. Wilson, W.~S.
  Charlton.
\newblock \emph{SOURCES 4A: A Code for Calculating (alpha,n), Spontaneous
  Fission, and Delayed Neutron Sources and Spectra}.
\newblock \doi{10.2172/15215}.

\bibitem{Kudryavtsev2008}
V.~A. Kudryavtsev, L.~Pandola, V.~Tomasello.
\newblock \emph{Neutron- and muon-induced background in underground physics
  experiments}.
\newblock \emph{The European Physical Journal A}, \textbf{36(2)}
  (2008):171--180.
\newblock \doi{10.1140/epja/i2007-10539-6}.

\bibitem{Westerdale_2016}
S.~Westerdale, E.~Shields, F.~Calaprice.
\newblock \emph{A prototype neutron veto for dark matter detectors}.
\newblock \emph{Astroparticle Physics}, \textbf{79} (2016):10--22.
\newblock \doi{10.1016/j.astropartphys.2016.01.005}.

\bibitem{Schumann_2019}
M.~Schumann.
\newblock \emph{Direct detection of {WIMP} dark matter: concepts and status}.
\newblock \emph{Journal of Physics G: Nuclear and Particle Physics},
  \textbf{46(10)} (2019):103003.
\newblock \doi{10.1088/1361-6471/ab2ea5}.

\bibitem{osti_1056763}
E.~Aguayo~Navarrete, R.~T. Kouzes, J.~L. Orrell, T.~J. Berguson, A.~T. Greene.
\newblock \emph{Estimation of Cosmic Induced Contamination in Ultra-low
  Background Detector Materials}.
\newblock \doi{10.2172/1056763}.

\bibitem{Billard_2014}
J.~Billard, E.~Figueroa-Feliciano, L.~Strigari.
\newblock \emph{Implication of neutrino backgrounds on the reach of next
  generation dark matter direct detection experiments}.
\newblock \emph{Physical Review D}, \textbf{89(2)} (2014).
\newblock \doi{10.1103/physrevd.89.023524}.

\bibitem{Xenon_future}
J.~e.~a. Aalbers.
\newblock \emph{A Next-Generation Liquid Xenon Observatory for Dark Matter and
  Neutrino Physics} (2022).
\newblock \doi{10.48550/ARXIV.2203.02309}.

\bibitem{Vahsen:2020pzb}
S.~Vahsen, et~al.
\newblock \emph{{CYGNUS: Feasibility of a nuclear recoil observatory with
  directional sensitivity to dark matter and neutrinos}} (2020).

\bibitem{O_Hare_2021}
C.~A.~J. O'Hare.
\newblock \emph{New Definition of the Neutrino Floor for Direct Dark Matter
  Searches}.
\newblock \emph{Physical Review Letters}, \textbf{127(25)} (2021).
\newblock \doi{10.1103/physrevlett.127.251802}.

\bibitem{bib:Snowmass2022}
G.~Krnjaic, N.~Toro, A.~Berlin, B.~Batell, N.~Blinov, L.~Darme,
  P.~DeNiverville, P.~Harris, C.~Hearty, M.~Hostert, K.~J. Kelly, D.~McKeen,
  S.~Trojanowski, Y.~D. Tsai.
\newblock \emph{A Snowmass Whitepaper: Dark Matter Production at
  Intensity-Frontier Experiments} (2022).
\newblock \doi{10.48550/ARXIV.2207.00597}.

\bibitem{Ohare2}
C.~A.~J. O'Hare, et~al.
\newblock \emph{Recoil imaging for directional detection of dark matter,
  neutrinos, and physics beyond the Standard Model} (2022).
\newblock \doi{10.48550/ARXIV.2203.05914}.

\bibitem{O_Hare_2015}
C.~A. O'Hare, A.~M. Green, J.~Billard, E.~Figueroa-Feliciano, L.~E. Strigari.
\newblock \emph{Readout strategies for directional dark matter detection beyond
  the neutrino background}.
\newblock \emph{Physical Review D}, \textbf{92(6)} (2015).
\newblock \doi{10.1103/physrevd.92.063518}.

\bibitem{Coarasa_2019}
I.~Coarasa, J.~Amar{\'{e}}, S.~Cebri{\'{a}}n, C.~Cuesta, E.~Garc{\'{\i}}a,
  M.~Mart{\'{\i}}nez, M.~A. Oliv{\'{a}}n, Y.~Ortigoza, A.~O.
  de~Sol{\'{o}}rzano, J.~Puimed{\'{o}}n, A.~Salinas, M.~L. Sarsa, P.~Villar,
  J.~A. Villar.
\newblock \emph{{ANAIS}-112 sensitivity in the search for dark matter annual
  modulation}.
\newblock \emph{The European Physical Journal C}, \textbf{79(3)} (2019).
\newblock \doi{10.1140/epjc/s10052-019-6733-4}.

\bibitem{Bernabei:2022xgg}
R.~Bernabei, et~al.
\newblock \emph{{Recent Results from DAMA/LIBRA and Comparisons}}.
\newblock \emph{Moscow Univ. Phys. Bull.}, \textbf{77(2)} (2022):291--300.
\newblock \doi{10.3103/S0027134922020138}.

\bibitem{G_Gerbier_2008}
G.~Gerbier, (forthe EDELWEISS~Collaboration).
\newblock \emph{Status of the EDELWEISS-II experiment}.
\newblock \emph{Journal of Physics: Conference Series}, \textbf{120(4)}
  (2008):042017.
\newblock \doi{10.1088/1742-6596/120/4/042017}.

\bibitem{Willers:2017vae}
M.~Willers, et~al.
\newblock \emph{{Direct dark matter search with the CRESST-III experiment -
  status and perspectives}}.
\newblock \emph{J. Phys. Conf. Ser.}, \textbf{888(1)} (2017):012209.
\newblock \doi{10.1088/1742-6596/888/1/012209}.

\bibitem{Angloher_2016}
G.~Angloher, P.~Carniti, L.~Cassina, L.~Gironi, C.~Gotti, A.~Gütlein,
  D.~Hauff, M.~Maino, S.~S. Nagorny, L.~Pagnanini, G.~Pessina, F.~Petricca,
  S.~Pirro, F.~Pröbst, F.~Reindl, K.~Schäffner, J.~Schieck, W.~Seidel.
\newblock \emph{The {COSINUS} project: perspectives of a {NaI} scintillating
  calorimeter for dark matter search}.
\newblock \emph{The European Physical Journal C}, \textbf{76(8)} (2016).
\newblock \doi{10.1140/epjc/s10052-016-4278-3}.

\bibitem{SuperCDMS:2016wui}
R.~Agnese, et~al.
\newblock \emph{{Projected Sensitivity of the SuperCDMS SNOLAB experiment}}.
\newblock \emph{Phys. Rev. D}, \textbf{95(8)} (2017):082002.
\newblock \doi{10.1103/PhysRevD.95.082002}.

\bibitem{KUZNIAK2016340}
M.~Kuźniak, et~al.
\newblock \emph{DEAP-3600 Dark Matter Search}.
\newblock \emph{Nuclear and Particle Physics Proceedings}, \textbf{273-275}
  (2016):340--346.
\newblock \doi{https://doi.org/10.1016/j.nuclphysbps.2015.09.048}.
\newblock 37th International Conference on High Energy Physics (ICHEP).

\bibitem{ABE201378}
K.~Abe, et~al.
\newblock \emph{Light WIMP search in XMASS}.
\newblock \emph{Physics Letters B}, \textbf{719(1)} (2013):78--82.
\newblock \doi{https://doi.org/10.1016/j.physletb.2013.01.001}.

\bibitem{Aprile_2022}
E.~Aprile, et~al.
\newblock \emph{Search for New Physics in Electronic Recoil Data from
  {XENONnT}}.
\newblock \emph{Physical Review Letters}, \textbf{129(16)} (2022).
\newblock \doi{10.1103/physrevlett.129.161805}.

\bibitem{Aalseth:2017fik}
C.~E. Aalseth, et~al.
\newblock \emph{{DarkSide-20k: A 20 tonne two-phase LAr TPC for direct dark
  matter detection at LNGS}}.
\newblock \emph{Eur. Phys. J. Plus}, \textbf{133} (2018):131.
\newblock \doi{10.1140/epjp/i2018-11973-4}.

\bibitem{bib:Amole_2019}
C.~Amole, M.~Ardid, I.~Arnquist, D.~Asner, D.~Baxter, E.~Behnke, M.~Bressler,
  B.~Broerman, G.~Cao, C.~Chen, et~al.
\newblock \emph{Dark matter search results from the complete exposure of the
  PICO-60 C3F8 bubble chamber}.
\newblock \emph{Physical Review D}, \textbf{100(2)} (2019).
\newblock \doi{10.1103/physrevd.100.022001}.

\bibitem{Buttazzo_2020}
D.~Buttazzo, P.~Panci, N.~Rossi, A.~Strumia.
\newblock \emph{Annual modulations from secular variations: relaxing {DAMA}?}
\newblock \emph{Journal of High Energy Physics}, \textbf{2020(4)} (2020).
\newblock \doi{10.1007/jhep04(2020)137}.

\bibitem{Davis_2014}
J.~H. Davis, C.~McCabe, C.~B{\oe}hm.
\newblock \emph{Quantifying the evidence for dark matter in {CoGeNT} data}.
\newblock \emph{Journal of Cosmology and Astroparticle Physics},
  \textbf{2014(08)} (2014):014--014.
\newblock \doi{10.1088/1475-7516/2014/08/014}.

\bibitem{Messina_2020}
A.~Messina, M.~Nardecchia, S.~Piacentini.
\newblock \emph{Annual modulations from secular variations: not relaxing
  {DAMA}?}
\newblock \emph{Journal of Cosmology and Astroparticle Physics},
  \textbf{2020(04)} (2020):037--037.
\newblock \doi{10.1088/1475-7516/2020/04/037}.

\bibitem{Damaanalysis}
G.~Adhikari, et~al.
\newblock \emph{An induced annual modulation signature in COSINE-100 data by
  DAMA/LIBRA's analysis method} (2022).
\newblock \doi{10.48550/ARXIV.2208.05158}.

\bibitem{Adhikari_2021}
G.~Adhikari, et~al.
\newblock \emph{Strong constraints from {COSINE}-100 on the {DAMA} dark matter
  results using the same sodium iodide target}.
\newblock \emph{Science Advances}, \textbf{7(46)} (2021).
\newblock \doi{10.1126/sciadv.abk2699}.

\bibitem{PhysRevLett.118.251302}
D.~S. Akerib, et~al.
\newblock \emph{Limits on Spin-Dependent WIMP-Nucleon Cross Section Obtained
  from the Complete LUX Exposure}.
\newblock \emph{Phys. Rev. Lett.}, \textbf{118} (2017):251302.
\newblock \doi{10.1103/PhysRevLett.118.251302}.

\bibitem{LZ}
J.~Aalbers, et~al.
\newblock \emph{First Dark Matter Search Results from the LUX-ZEPLIN (LZ)
  Experiment} (2022).
\newblock \doi{10.48550/ARXIV.2207.03764}.

\bibitem{Takahashi_2020}
F.~Takahashi, M.~Yamada, W.~Yin.
\newblock \emph{{XENON}1T Excess from Anomaly-Free Axionlike Dark Matter and
  Its Implications for Stellar Cooling Anomaly}.
\newblock \emph{Physical Review Letters}, \textbf{125(16)} (2020).
\newblock \doi{10.1103/physrevlett.125.161801}.

\bibitem{Adari_2022}
P.~Adari, et~al.
\newblock \emph{{EXCESS} workshop: Descriptions of rising low-energy spectra}.
\newblock \emph{{SciPost} Physics Proceedings}, \textbf{(9)} (2022).
\newblock \doi{10.21468/scipostphysproc.9.001}.

\bibitem{Green_2007}
A.~M. Green, B.~Morgan.
\newblock \emph{Optimizing {WIMP} directional detectors}.
\newblock \emph{Astroparticle Physics}, \textbf{27(2-3)} (2007):142--149.
\newblock \doi{10.1016/j.astropartphys.2006.10.006}.

\bibitem{COPI199943}
C.~J. Copi, J.~Heo, L.~M. Krauss.
\newblock \emph{Directional sensitivity, WIMP detection, and the galactic
  halo}.
\newblock \emph{Physics Letters B}, \textbf{461(1)} (1999):43--48.
\newblock \doi{https://doi.org/10.1016/S0370-2693(99)00830-8}.

\bibitem{Billard_2012}
J.~Billard, F.~Mayet, D.~Santos.
\newblock \emph{Assessing the discovery potential of directional detection of
  dark matter}.
\newblock \emph{Physical Review D}, \textbf{85(3)} (2012).
\newblock \doi{10.1103/physrevd.85.035006}.

\bibitem{Vahsen_2021}
S.~E. Vahsen, C.~A. O'Hare, D.~Loomba.
\newblock \emph{Directional Recoil Detection}.
\newblock \emph{Annual Review of Nuclear and Particle Science}, \textbf{71(1)}
  (2021):189--224.
\newblock \doi{10.1146/annurev-nucl-020821-035016}.

\bibitem{Lee_2012}
S.~K. Lee, A.~H. Peter.
\newblock \emph{Probing the local velocity distribution of {WIMP} dark matter
  with directional detectors}.
\newblock \emph{Journal of Cosmology and Astroparticle Physics},
  \textbf{2012(04)} (2012):029--029.
\newblock \doi{10.1088/1475-7516/2012/04/029}.

\bibitem{Kavanagh_2016}
B.~J. Kavanagh, C.~A. O'Hare.
\newblock \emph{Reconstructing the three-dimensional local dark matter velocity
  distribution}.
\newblock \emph{Physical Review D}, \textbf{94(12)} (2016).
\newblock \doi{10.1103/physrevd.94.123009}.

\bibitem{O_Hare_2014}
C.~A. O'Hare, A.~M. Green.
\newblock \emph{Directional detection of dark matter streams}.
\newblock \emph{Physical Review D}, \textbf{90(12)} (2014).
\newblock \doi{10.1103/physrevd.90.123511}.

\bibitem{gaiasausage}
N.~W. Evans, C.~A.~J. O'Hare, C.~McCabe.
\newblock \emph{SHM$^{++}$: A Refinement of the Standard Halo Model for Dark
  Matter Searches in Light of the Gaia Sausage} (2018).
\newblock \doi{10.48550/ARXIV.1810.11468}.

\bibitem{Kruijssen_2018}
J.~M.~D. Kruijssen, J.~L. Pfeffer, M.~Reina-Campos, R.~A. Crain, N.~Bastian.
\newblock \emph{The formation and assembly history of the Milky Way revealed by
  its globular cluster population}.
\newblock \emph{Monthly Notices of the Royal Astronomical Society},
  \textbf{486(3)} (2018):3180--3202.
\newblock \doi{10.1093/mnras/sty1609}.

\bibitem{Naidu_2020}
R.~P. Naidu, C.~Conroy, A.~Bonaca, B.~D. Johnson, Y.-S. Ting, N.~Caldwell,
  D.~Zaritsky, P.~A. Cargile.
\newblock \emph{Evidence from the H3 Survey That the Stellar Halo Is Entirely
  Comprised of Substructure}.
\newblock \emph{The Astrophysical Journal}, \textbf{901(1)} (2020):48.
\newblock \doi{10.3847/1538-4357/abaef4}.

\bibitem{Myeong_2017}
G.~C. Myeong, N.~W. Evans, V.~Belokurov, N.~C. Amorisco, S.~E. Koposov.
\newblock \emph{Halo substructure in the {SDSS}--Gaia catalogue: streams and
  clumps}.
\newblock \emph{Monthly Notices of the Royal Astronomical Society},
  \textbf{475(2)} (2017):1537--1548.
\newblock \doi{10.1093/mnras/stx3262}.

\bibitem{Battat_2016}
J.~Battat, et~al.
\newblock \emph{Readout technologies for directional {WIMP} Dark Matter
  detection}.
\newblock \emph{Physics Reports}, \textbf{662} (2016):1--46.
\newblock \doi{10.1016/j.physrep.2016.10.001}.

\bibitem{SEKIYA_2005}
H.~Sekiya, M.~Minowa, Y.~Shimuzu, W.~Suganuma, Y.~Inoue.
\newblock \emph{{Dark} {matter} {search} {with} {direction} {sensitive}
  {scintillator}}.
\newblock In \emph{Neutrino Oscillations and Their Origin}. {WORLD}
  {SCIENTIFIC} (2005).
\newblock \doi{10.1142/9789812701824_0049}.

\bibitem{Belli_2020}
P.~Belli, R.~Bernabei, F.~Cappella, V.~Caracciolo, R.~Cerulli, N.~Cherubini,
  F.~A. Danevich, A.~Incicchitti, D.~V. Kasperovych, V.~Merlo, E.~Piccinelli,
  O.~G. Polischuk, V.~I. Tretyak.
\newblock \emph{Measurements of ZnWO$_4$ anisotropic response to nuclear
  recoils for the {ADAMO} project}.
\newblock \emph{The European Physical Journal A}, \textbf{56(3)} (2020).
\newblock \doi{10.1140/epja/s10050-020-00094-z}.

\bibitem{carbonnano}
L.~M. Capparelli, G.~Cavoto, D.~Mazzilli, A.~D. Polosa.
\newblock \emph{Directional Dark Matter Searches with Carbon Nanotubes} (2014).
\newblock \doi{10.48550/ARXIV.1412.8213}.

\bibitem{Nygren_2013}
D.~R. Nygren.
\newblock \emph{Columnar recombination: a tool for nuclear recoil directional
  sensitivity in a xenon-based direct detection WIMP search}.
\newblock \emph{Journal of Physics: Conference Series}, \textbf{460(1)}
  (2013):012006.
\newblock \doi{10.1088/1742-6596/460/1/012006}.

\bibitem{Nakamura_2018}
K.~Nakamura, S.~Ban, M.~Hirose, A.~Ichikawa, Y.~Ishiyama, A.~Minamino,
  K.~Miuchi, T.~Nakaya, H.~Sekiya, S.~Tanaka, K.~Ueshima.
\newblock \emph{Angular dependence of columnar recombination in high pressure
  xenon gas using time profiles of scintillation emission}.
\newblock \emph{Journal of Instrumentation}, \textbf{13(07)}
  (2018):P07015--P07015.
\newblock \doi{10.1088/1748-0221/13/07/p07015}.

\bibitem{NAKA2013519}
T.~Naka, T.~Asada, T.~Katsuragawa, K.~Hakamata, M.~Yoshimoto, K.~Kuwabara,
  M.~Nakamura, O.~Sato, T.~Nakano, Y.~Tawara, G.~{De Lellis}, C.~Sirignano,
  N.~D'Ambrossio.
\newblock \emph{Fine grained nuclear emulsion for higher resolution tracking
  detector}.
\newblock \emph{Nuclear Instruments and Methods in Physics Research Section A:
  Accelerators, Spectrometers, Detectors and Associated Equipment},
  \textbf{718} (2013):519--521.
\newblock \doi{https://doi.org/10.1016/j.nima.2012.11.106}.
\newblock Proceedings of the 12th Pisa Meeting on Advanced Detectors.

\bibitem{Shiraishi_2023}
T.~Shiraishi, S.~Akamatsu, T.~Naka, T.~Asada, G.~D. Lellis, V.~Tioukov,
  G.~Rosa, R.~Kobayashi, N.~D'Ambrosio, A.~Alexandrov, O.~Sato.
\newblock \emph{Environmental sub-MeV neutron measurement at the Gran Sasso
  surface laboratory with a super-fine-grained nuclear emulsion detector}.
\newblock \emph{Physical Review C}, \textbf{107(1)} (2023).
\newblock \doi{10.1103/physrevc.107.014608}.

\bibitem{Ariga2020}
A.~Ariga, T.~Ariga, G.~De~Lellis, A.~Ereditato, K.~Niwa.
\newblock \emph{Nuclear Emulsions}.
\newblock Springer International Publishing, Cham (2020), pp. 383--438.
\newblock \doi{10.1007/978-3-030-35318-6_9}.

\bibitem{DeLellis:20227s}
G.~De~Lellis.
\newblock \emph{{Nuclear Emulsions for WIMP Search with a directional
  measurement}}.
\newblock \emph{PoS}, \textbf{EPS-HEP2021} (2022):157.
\newblock \doi{10.22323/1.398.0157}.

\bibitem{Newdm1}
S.~A. {Gorbunov}, N.~S. {Konovalova}.
\newblock \emph{{New Experiment NEWSdm for Direct Searches for Heavy Dark
  Matter Particles}}.
\newblock \emph{Physics of Atomic Nuclei}, \textbf{83(1)} (2020):83--91.
\newblock \doi{10.1134/S1063778820010056}.

\bibitem{Alexandrov2019}
A.~Alexandrov, G.~De~Lellis, V.~Tioukov.
\newblock \emph{A Novel Optical Scanning Technique with an Inclined Focusing
  Plane}.
\newblock \emph{Scientific Reports}, \textbf{9(1)} (2019):2870.
\newblock \doi{10.1038/s41598-019-39415-8}.

\bibitem{Betzig2006-rv}
E.~Betzig, G.~H. Patterson, R.~Sougrat, O.~W. Lindwasser, S.~Olenych, J.~S.
  Bonifacino, M.~W. Davidson, J.~Lippincott-Schwartz, H.~F. Hess.
\newblock \emph{Imaging intracellular fluorescent proteins at nanometer
  resolution}.
\newblock \emph{Science}, \textbf{313(5793)} (2006):1642--1645.

\bibitem{plasmonreso}
J.~J. Mock, M.~Barbic, D.~R. Smith, D.~A. Schultz, S.~Schultz.
\newblock \emph{Shape effects in plasmon resonance of individual colloidal
  silver nanoparticles}.
\newblock \emph{The Journal of Chemical Physics}, \textbf{116(15)}
  (2002):6755--6759.
\newblock \doi{10.1063/1.1462610}.

\bibitem{Alexandrov2020}
A.~Alexandrov, T.~Asada, G.~De~Lellis, A.~Di~Crescenzo, V.~Gentile, T.~Naka,
  V.~Tioukov, A.~Umemoto.
\newblock \emph{Super-resolution high-speed optical microscopy for fully
  automated readout of metallic nanoparticles and nanostructures}.
\newblock \emph{Scientific Reports}, \textbf{10(1)} (2020):18773.
\newblock \doi{10.1038/s41598-020-75883-z}.

\bibitem{Di_Marco_2016}
N.~{Di Marco}, on~behalf of~the NEWS Collaboration.
\newblock \emph{NEWS: Nuclear Emulsions for WIMP Search}.
\newblock \emph{Journal of Physics: Conference Series}, \textbf{718(4)}
  (2016):042018.
\newblock \doi{10.1088/1742-6596/718/4/042018}.

\bibitem{Golovatiuk_2021}
A.~Golovatiuk.
\newblock \emph{Directional Dark Matter Search with the NEWSdm experiment}.
\newblock \emph{Journal of Physics: Conference Series}, \textbf{2156(1)}
  (2021):012044.
\newblock \doi{10.1088/1742-6596/2156/1/012044}.

\bibitem{bib:tpc1}
J.~N. Marx, D.~R. Nygren.
\newblock \emph{{The Time Projection Chamber}}.
\newblock \emph{Phys. Today}, \textbf{31N10} (1978):46--53.
\newblock \doi{10.1063/1.2994775}.

\bibitem{bib:tpc2}
D.~R. Nygren.
\newblock \emph{{The Time Projection Chamber: A New 4 pi Detector for Charged
  Particles}}.
\newblock \emph{eConf}, \textbf{C740805} (1974):58.

\bibitem{bib:tpc3}
W.~B. Atwood, et~al.
\newblock \emph{{Performance of the ALEPH time projection chamber}}.
\newblock \emph{Nucl. Instrum. Meth. A}, \textbf{306} (1991):446--458.
\newblock \doi{10.1016/0168-9002(91)90038-R}.

\bibitem{Knoll}
G.~Knoll.
\newblock \emph{{Radiation Detection and Measurement (4th ed.)}}.
\newblock John Wiley, Hoboken, NJ (2010).

\bibitem{SAULI20162}
F.~Sauli.
\newblock \emph{The gas electron multiplier (GEM): Operating principles and
  applications}.
\newblock \emph{Nuclear Instruments and Methods in Physics Research Section A:
  Accelerators, Spectrometers, Detectors and Associated Equipment},
  \textbf{805} (2016):2--24.
\newblock \doi{https://doi.org/10.1016/j.nima.2015.07.060}.
\newblock Special Issue in memory of Glenn F. Knoll.

\bibitem{DUNE}
R.~Acciarri, et~al.
\newblock \emph{Long-Baseline Neutrino Facility (LBNF) and Deep Underground
  Neutrino Experiment (DUNE) Conceptual Design Report, Volume 4 The DUNE
  Detectors at LBNF} (2016).
\newblock \doi{10.48550/ARXIV.1601.02984}.

\bibitem{Blum_rolandi}
W.~Blum, L.~Rolandi, W.~Riegler.
\newblock \emph{{Particle detection with drift chambers}}.
\newblock Particle Acceleration and Detection, ISBN = 9783540766834 (2008).
\newblock \doi{10.1007/978-3-540-76684-1}.

\bibitem{Martoff_2005}
C.~Martoff, R.~Ayad, M.~Katz-Hyman, G.~Bonvicini, A.~Schreiner.
\newblock \emph{Negative ion drift and diffusion in a {TPC} near 1 bar}.
\newblock \emph{Nuclear Instruments and Methods in Physics Research Section A:
  Accelerators, Spectrometers, Detectors and Associated Equipment},
  \textbf{555(1-2)} (2005):55--58.
\newblock \doi{10.1016/j.nima.2005.08.103}.

\bibitem{LEWIS201581}
P.~Lewis, S.~Vahsen, I.~Seong, M.~Hedges, I.~Jaegle, T.~Thorpe.
\newblock \emph{Absolute position measurement in a gas time projection chamber
  via transverse diffusion of drift charge}.
\newblock \emph{Nuclear Instruments and Methods in Physics Research Section A:
  Accelerators, Spectrometers, Detectors and Associated Equipment},
  \textbf{789} (2015):81--85.
\newblock \doi{https://doi.org/10.1016/j.nima.2015.03.024}.

\bibitem{bib:Antochi_2021}
V.~Antochi, et~al.
\newblock \emph{Performance of an optically read out time projection chamber
  with ultra-relativistic electrons}.
\newblock \emph{Nuclear Instruments and Methods in Physics Research Section A:
  Accelerators, Spectrometers, Detectors and Associated Equipment},
  \textbf{999} (2021):165209.
\newblock \doi{10.1016/j.nima.2021.165209}.

\bibitem{mincarry}
D.~P. Snowden-Ifft.
\newblock \emph{Discovery of multiple, ionization-created CS2 anions and a new
  mode of operation for drift chambers}.
\newblock \emph{Review of Scientific Instruments}, \textbf{85(1)}
  (2014):013303.
\newblock \doi{10.1063/1.4861908}.

\bibitem{Battat_2021}
T.~D. collaboration.
\newblock \emph{Improved sensitivity of the DRIFT-IId directional dark matter
  experiment using machine learning}.
\newblock \emph{Journal of Cosmology and Astroparticle Physics},
  \textbf{2021(07)} (2021):014.
\newblock \doi{10.1088/1475-7516/2021/07/014}.

\bibitem{DRIFT:2014bny}
J.~B.~R. Battat, et~al.
\newblock \emph{{First background-free limit from a directional dark matter
  experiment: results from a fully fiducialised DRIFT detector}}.
\newblock \emph{Phys. Dark Univ.}, \textbf{9-10} (2015):1--7.
\newblock \doi{10.1016/j.dark.2015.06.001}.

\bibitem{Battat_2016_HT}
J.~B.~R. Battat, E.~Daw, A.~Ezeribe, J.-L. Gauvreau, J.~L. Harton, R.~Lafler,
  E.~R. Lee, D.~Loomba, A.~Lumnah, E.~Miller, F.~Mouton, A.~Murphy, S.~Paling,
  N.~Phan, M.~Robinson, S.~Sadler, A.~Scarff, F.~S. II, D.~Snowden-Ifft,
  N.~Spooner.
\newblock \emph{First measurement of nuclear recoil head-tail sense in a
  fiducialised {WIMP} dark matter detector}.
\newblock \emph{Journal of Instrumentation}, \textbf{11(10)}
  (2016):P10019--P10019.
\newblock \doi{10.1088/1748-0221/11/10/p10019}.

\bibitem{NEWAGE1}
T.~Ikeda, K.~Nakamura, T.~Shimada, R.~Yakabe, T.~Hashimoto, H.~Ishiura,
  T.~Nakamura, H.~Ito, K.~Ichimura, K.~Abe, K.~Kobayashi, T.~Tanimori, H.~Kubo,
  A.~Takada, H.~Sekiya, A.~Takeda, K.~Miuchi.
\newblock \emph{{Direction-sensitive dark matter search with the low-background
  gaseous detector NEWAGE-0.3b”}}.
\newblock \emph{Progress of Theoretical and Experimental Physics},
  \textbf{2021(6)} (2021).
\newblock \doi{10.1093/ptep/ptab053}.
\newblock 063F01.

\bibitem{Hashimoto_2018}
T.~Hashimoto, K.~Miuchi, K.~Nakamura, R.~Yakabe, T.~Ikeda, R.~Taishaku,
  M.~Nakazawa, H.~Ishiura, A.~Ochi, Y.~Takeuchi.
\newblock \emph{Development of a low-alpha-emitting $\mathrm{\mu}$-{PIC} for
  {NEWAGE} direction-sensitive dark-matter search}.
\newblock In \emph{{AIP} Conference Proceedings}. Author(s) (2018).
\newblock \doi{10.1063/1.5019004}.

\bibitem{bib:yakabe2020limits}
R.~Yakabe, K.~Nakamura, T.~Ikeda, H.~Ito, Y.~Yamaguchi, R.~Taishaku,
  M.~Nakazawa, H.~Ishiura, T.~Nakamura, T.~Shimada, T.~Tanimori, H.~Kubo,
  A.~Takada, H.~Sekiya, A.~Takeda, K.~Miuchi.
\newblock \emph{First limits from a 3d-vector directional dark matter search
  with the NEWAGE-0.3b' detector} (2020).

\bibitem{Ikeda:2020pex}
T.~Ikeda, T.~Shimada, H.~Ishiura, K.~D. Nakamura, T.~Nakamura, K.~Miuchi.
\newblock \emph{{Development of a negative ion micro TPC detector with SF$_6$
  gas for the directional dark matter search}}.
\newblock \emph{JINST}, \textbf{15(07)} (2020):P07015.
\newblock \doi{10.1088/1748-0221/15/07/P07015}.

\bibitem{VAHSEN201595}
S.~Vahsen, M.~Hedges, I.~Jaegle, S.~Ross, I.~Seong, T.~Thorpe, J.~Yamaoka,
  J.~Kadyk, M.~Garcia-Sciveres.
\newblock \emph{3-D tracking in a miniature time projection chamber}.
\newblock \emph{Nuclear Instruments and Methods in Physics Research Section A:
  Accelerators, Spectrometers, Detectors and Associated Equipment},
  \textbf{788} (2015):95--105.
\newblock \doi{https://doi.org/10.1016/j.nima.2015.03.009}.

\bibitem{Amaro_2022}
F.~D. Amaro, E.~Baracchini, L.~Benussi, S.~Bianco, C.~Capoccia, M.~Caponero,
  D.~S. Cardoso, G.~Cavoto, A.~Cortez, I.~A. Costa, R.~J. da~Cruz~Roque,
  E.~Dan{\'{e}}, G.~Dho, F.~D. Giambattista, E.~{Di Marco}, G.~G. di~Cortona,
  G.~D'Imperio, F.~Iacoangeli, H.~P.~L. J{\'{u}}nior, G.~S.~P. Lopes,
  A.~da~Silva Lopes~J{\'{u}}nior, G.~Maccarrone, R.~D.~P. Mano, M.~Marafini,
  R.~R.~M. Gregorio, D.~J.~G. Marques, G.~Mazzitelli, A.~G. McLean, A.~Messina,
  C.~M.~B. Monteiro, R.~A. Nobrega, I.~F. Pains, E.~Paoletti, L.~Passamonti,
  S.~Pelosi, F.~Petrucci, S.~Piacentini, D.~Piccolo, D.~Pierluigi, D.~Pinci,
  A.~Prajapati, F.~Renga, F.~Rosatelli, A.~Russo, J.~M.~F. dos Santos,
  G.~Saviano, N.~J.~C. Spooner, R.~Tesauro, S.~Tomassini, S.~Torelli.
\newblock \emph{The {CYGNO} Experiment}.
\newblock \emph{Instruments}, \textbf{6(1)} (2022):6.
\newblock \doi{10.3390/instruments6010006}.

\bibitem{Riffard_2016}
Q.~Riffard, D.~Santos, O.~Guillaudin, G.~Bosson, O.~Bourrion, J.~Bouvier,
  T.~Descombes, J.-F. Muraz, L.~Lebreton, D.~Maire, P.~Colas, I.~Giomataris,
  J.~Busto, D.~Fouchez, J.~Brunner, C.~Tao.
\newblock \emph{{MIMAC} low energy electron-recoil discrimination measured with
  fast neutrons}.
\newblock \emph{Journal of Instrumentation}, \textbf{11(08)}
  (2016):P08011--P08011.
\newblock \doi{10.1088/1748-0221/11/08/p08011}.

\bibitem{Tao:2020muh}
Y.~Tao, et~al.
\newblock \emph{{Dark Matter Directionality Detection performance of the
  Micromegas-based $\mu$TPC-MIMAC detector}}.
\newblock \emph{Nucl. Instrum. Meth. A}, \textbf{1021} (2022):165412.
\newblock \doi{10.1016/j.nima.2021.165412}.

\bibitem{BATTAT20146}
J.~B. Battat, C.~Deaconu, G.~Druitt, R.~Eggleston, P.~Fisher, P.~Giampa,
  V.~Gregoric, S.~Henderson, I.~Jaegle, J.~Lawhorn, J.~P. Lopez, J.~Monroe,
  K.~A. Recine, A.~Strandberg, H.~Tomita, S.~Vahsen, H.~Wellenstein.
\newblock \emph{The Dark Matter Time Projection Chamber 4Shooter directional
  dark matter detector: Calibration in a surface laboratory}.
\newblock \emph{Nuclear Instruments and Methods in Physics Research Section A:
  Accelerators, Spectrometers, Detectors and Associated Equipment},
  \textbf{755} (2014):6--19.
\newblock \doi{https://doi.org/10.1016/j.nima.2014.04.010}.

\bibitem{Ahlen_2011}
S.~Ahlen, J.~Battat, T.~Caldwell, C.~Deaconu, D.~Dujmic, W.~Fedus, P.~Fisher,
  F.~Golub, S.~Henderson, A.~Inglis, A.~Kaboth, G.~Kohse, R.~Lanza, A.~Lee,
  J.~Lopez, J.~Monroe, T.~Sahin, G.~Sciolla, N.~Skvorodnev, H.~Tomita,
  H.~Wellenstein, I.~Wolfe, R.~Yamamoto, H.~Yegoryan.
\newblock \emph{First dark matter search results from a surface run of the 10-L
  {DMTPC} directional dark matter detector}.
\newblock \emph{Physics Letters B}, \textbf{695(1-4)} (2011):124--129.
\newblock \doi{10.1016/j.physletb.2010.11.041}.

\bibitem{bib:cygnus}
S.~E. Vahsen, et~al.
\newblock \emph{{CYGNUS: Feasibility of a nuclear recoil observatory with
  directional sensitivity to dark matter and neutrinos}} (2020).

\bibitem{bib:garfield1}
R.~Veenhof.
\newblock \emph{{Garfield, a drift chamber simulation program}}.
\newblock \emph{Conf.\ Proc.\ C}, \textbf{9306149} (1993):66--71.

\bibitem{bib:garfield2}
R.~Veenhof.
\newblock \emph{{GARFIELD, recent developments}}.
\newblock \emph{Nucl.\ Instrum.\ Meth.\ A}, \textbf{419} (1998):726--730.
\newblock \doi{10.1016/S0168-9002(98)00851-1}.

\bibitem{bib:Margato_2013}
L.~M.~S. Margato, A.~Morozov, M.~M. F.~R. Fraga, L.~Pereira, F.~A.~F. Fraga.
\newblock \emph{Effective decay time of {CF}$_4$ secondary scintillation}.
\newblock \emph{Journal of Instrumentation}, \textbf{8(07)}
  (2013):P07008--P07008.
\newblock \doi{10.1088/1748-0221/8/07/p07008}.

\bibitem{bib:Fraga:2003uu}
M.~Fraga, F.~Fraga, S.~Fetal, L.~Margato, R.~Marques, A.~Policarpo.
\newblock \emph{The GEM scintillation in He–CF4, Ar–CF4, Ar–TEA and
  Xe–TEA mixtures}.
\newblock \emph{Nuclear Instruments and Methods in Physics Research Section A:
  Accelerators, Spectrometers, Detectors and Associated Equipment},
  \textbf{504(1)} (2003):88--92.
\newblock \doi{https://doi.org/10.1016/S0168-9002(03)00758-7}.
\newblock Proceedings of the 3rd International Conference on New Developments
  in Photodetection.

\bibitem{bib:Morozov_2012}
A.~Morozov, L.~M.~S. Margato, M.~M. F.~R. Fraga, L.~Pereira, F.~A.~F. Fraga.
\newblock \emph{Secondary scintillation in CF$_4$: emission spectra and photon
  yields for {MSGC} and {GEM}}.
\newblock \emph{Journal of Instrumentation}, \textbf{7(02)}
  (2012):P02008--P02008.
\newblock \doi{10.1088/1748-0221/7/02/p02008}.

\bibitem{bib:Kurihara}
M.~Kurihara, Z.~L. Petrovic, T.~Makabe.
\newblock \emph{Transport coefficients and scattering cross-sections for plasma
  modelling in {CF}$_4$-Ar mixtures: a swarm analysis}.
\newblock \emph{Journal of Physics D: Applied Physics}, \textbf{33(17)}
  (2000):2146--2153.
\newblock \doi{10.1088/0022-3727/33/17/309}.

\bibitem{bib:StabilityCygno}
E.~Baracchini, et~al.
\newblock \emph{Stability and detection performance of a {GEM}-based Optical
  Readout {TPC} with He/CF$_4$ gas mixtures}.
\newblock \emph{Journal of Instrumentation}, \textbf{15(10)}
  (2020):P10001--P10001.
\newblock \doi{10.1088/1748-0221/15/10/p10001}.

\bibitem{bib:roby}
R.~Campagnola.
\newblock \emph{{Study and optimization of the light-yield of a triple-GEM
  detector }}.
\newblock Ph.D. thesis, Sapienza University of Rome, CERN-THESIS-2018-027
  (2018. http://cds.cern.ch/record/2313231).

\bibitem{bib:geant}
S.~Agostinelli, et~al.
\newblock \emph{{GEANT4: A Simulation toolkit}}.
\newblock \emph{Nucl. Instrum. Meth. A}, \textbf{506} (2003):250--303.
\newblock \doi{10.1016/S0168-9002(03)01368-8}.

\bibitem{Tripathy:2021puz}
A.~Tripathy, P.~K. Sahu, S.~Swain, S.~Sahu.
\newblock \emph{{Measurement of ion backflow fraction in GEM detectors}}.
\newblock \emph{Nucl. Instrum. Meth. A}, \textbf{1013} (2021):165596.
\newblock \doi{10.1016/j.nima.2021.165596}.

\bibitem{bib:gem}
F.~Sauli.
\newblock \emph{{GEM: A new concept for electron amplification in gas
  detectors}}.
\newblock \emph{Nucl. Instrum. Meth. A}, \textbf{386} (1997):531--534.
\newblock \doi{10.1016/S0168-9002(96)01172-2}.

\bibitem{bib:lemon_btf}
V.~C. Antochi, G.~Cavoto, I.~A. Costa, E.~D. Marco, G.~D'Imperio,
  F.~Iacoangeli, M.~Marafini, A.~Messina, D.~Pinci, F.~Renga, C.~Voena,
  E.~Baracchini, A.~Cortez, G.~Dho, L.~Benussi, S.~Bianco, C.~Capoccia,
  M.~Caponero, G.~Maccarrone, G.~Mazzitelli, A.~Orlandi, E.~Paoletti,
  L.~Passamonti, D.~Piccolo, D.~Pierluigi, F.~Rosatelli, A.~Russo, G.~Saviano,
  S.~Tomassini, R.~A. Nobrega, F.~Petrucci.
\newblock \emph{A GEM-based Optically Readout Time Projection Chamber for
  charged particle tracking} (2020).

\bibitem{MONTEIRO201218}
C.~Monteiro, L.~Fernandes, J.~Veloso, C.~Oliveira, J.~{dos Santos}.
\newblock \emph{Secondary scintillation yield from GEM and THGEM gaseous
  electron multipliers for direct dark matter search}.
\newblock \emph{Physics Letters B}, \textbf{714(1)} (2012):18--23.
\newblock \doi{https://doi.org/10.1016/j.physletb.2012.06.066}.

\bibitem{opticalcharpak}
G.~Charpak, W.~Dominik, N.~Zaganidis.
\newblock \emph{Optical imaging of the spatial distribution of beta-particles
  emerging from surfaces.}
\newblock \emph{Proceedings of the National Academy of Sciences},
  \textbf{86(6)} (1989):1741--1745.
\newblock \doi{10.1073/pnas.86.6.1741}.

\bibitem{PMT}
H.~Photonics.
\newblock \emph{Photomultiplier Tubes}.
\newblock Hamamatsu (2000).

\bibitem{CMOStech}
B.~Razavi.
\newblock \emph{Design of Analog CMOS Integrated Circuits}.
\newblock McGraw-Hill Education (2000).

\bibitem{PPD}
E.~R. Fossum, D.~B. Hondongwa.
\newblock \emph{A Review of the Pinned Photodiode for CCD and CMOS Image
  Sensors}.
\newblock \emph{IEEE Journal of the Electron Devices Society}, \textbf{2(3)}
  (2014):33--43.
\newblock \doi{10.1109/JEDS.2014.2306412}.

\bibitem{CCDopt}
D.~Vorobiev, A.~Irwin, Z.~Ninkov, K.~Donlon, D.~Caldwell, S.~Mochnacki.
\newblock \emph{Direct measurement of the Kepler Space Telescope CCD's
  intra-pixel response function} (2019).
\newblock \doi{10.48550/ARXIV.1909.12248}.

\bibitem{Zhan_2017}
H.~Zhan, X.~Zhang, L.~Cao.
\newblock \emph{Intrapixel effects of {CCD} and {CMOS} detectors}.
\newblock \emph{Journal of Instrumentation}, \textbf{12(04)}
  (2017):C04010--C04010.
\newblock \doi{10.1088/1748-0221/12/04/c04010}.

\bibitem{CCDnoise}
A.~K. Boyat, B.~K. Joshi.
\newblock \emph{A Review Paper: Noise Models in Digital Image Processing}
  (2015).
\newblock \doi{10.48550/ARXIV.1505.03489}.

\bibitem{CCDcmosnoise}
M.~Konnik, J.~Welsh.
\newblock \emph{High-level numerical simulations of noise in CCD and CMOS
  photosensors: review and tutorial} (2014).
\newblock \doi{10.48550/ARXIV.1412.4031}.

\bibitem{CCDhubble}
L.~Pritchard, et~al.
\newblock \emph{{STIS Instrument Handbook}} (2022).

\bibitem{LSST}
N.~Petitdidier.
\newblock \emph{LSST: Characterization of the CCD sensors}.
\newblock Master's thesis, KTH Engineering Sciences, Stockholm, SW (2015).

\bibitem{vignetting}
A.~Rowlands.
\newblock \emph{Physics of Digital Photography}.
\newblock 2053-2563. IOP Publishing (2017).
\newblock \doi{10.1088/978-0-7503-1242-4}.

\bibitem{bib:EL_cygno}
E.~Baracchini, et~al.
\newblock \emph{First evidence of luminescence in a He/CF$_4$ gas mixture
  induced by non-ionizing electrons}.
\newblock \emph{Journal of Instrumentation}, \textbf{15(08)}
  (2020):P08018--P08018.
\newblock \doi{10.1088/1748-0221/15/08/p08018}.

\bibitem{CORRADI200796}
G.~Corradi, F.~Murtas, D.~Tagnani.
\newblock \emph{A novel High-Voltage System for a triple GEM detector}.
\newblock \emph{Nuclear Instruments and Methods in Physics Research Section A:
  Accelerators, Spectrometers, Detectors and Associated Equipment},
  \textbf{572(1)} (2007):96--97.
\newblock \doi{https://doi.org/10.1016/j.nima.2006.10.166}.
\newblock Frontier Detectors for Frontier Physics.

\bibitem{bib:Micromegas}
F.~Kuger.
\newblock \emph{Micromesh-selection for the ATLAS New Small Wheel Micromegas
  detectors}.
\newblock \emph{Journal of Instrumentation}, \textbf{11(11)} (2016):C11043.
\newblock \doi{10.1088/1748-0221/11/11/C11043}.

\bibitem{bib:jinst_orange1}
M.~Marafini, V.~Patera, D.~Pinci, A.~Sarti, A.~Sciubba, E.~Spiriti.
\newblock \emph{{High granularity tracker based on a Triple-GEM optically read
  by a CMOS-based camera}}.
\newblock \emph{JINST}, \textbf{10(12)} (2015):P12010.
\newblock \doi{10.1088/1748-0221/10/12/P12010}.

\bibitem{bib:fe55}
I.~A. Costa, et~al.
\newblock \emph{Performance of optically readout {GEM}-based {TPC} with a
  $^{55}$Fe source}.
\newblock \emph{Journal of Instrumentation}, \textbf{14(07)}
  (2019):P07011--P07011.
\newblock \doi{10.1088/1748-0221/14/07/p07011}.

\bibitem{bib:GEMoptical}
Marafini, et~al.
\newblock \emph{Study of the Performance of an Optically Readout Triple-GEM}.
\newblock \emph{IEEE Transactions on Nuclear Science}, \textbf{65(1)}
  (2018):604--608.
\newblock \doi{10.1109/TNS.2017.2778503}.

\bibitem{dbscan1996}
M.~Ester, H.-P. Kriegel, J.~Sander, X.~Xu.
\newblock \emph{A Density-based Algorithm for Discovering Clusters a
  Density-based Algorithm for Discovering Clusters in Large Spatial Databases
  with Noise}.
\newblock In \emph{Proceedings of the Second International Conference on
  Knowledge Discovery and Data Mining}, KDD'96. AAAI Press (1996), pp.
  226--231.

\bibitem{gac}
V.~Caselles, R.~Kimmel, G.~Sapiro.
\newblock \emph{{Geodesic Active Contours}}.
\newblock \emph{International Journal of Computer Vision}, \textbf{22}
  (1997):61--79.

\bibitem{mgac}
{P. Marquez-Neila and L. Baumela\ and L. Alvarez}.
\newblock \emph{A Morphological Approach to Curvature-Based Evolution of Curves
  and Surfaces}.
\newblock \emph{IEEE Transactions on Pattern Analysis and Machine
  Intelligence}, \textbf{36(1)} (2014):2--17.

\bibitem{Randomforest}
G.~Louppe.
\newblock \emph{Understanding Random Forests: From Theory to Practice} (2014).
\newblock \doi{10.48550/ARXIV.1407.7502}.

\bibitem{Belli1989DeepUN}
P.~Belli, R.~Bernabei, S.~D'angelo, M.~P. de~Pascale, L.~Paoluzi, R.~Santonico,
  N.~Taborgna, N.~Iucci, G.~Villoresi.
\newblock \emph{Deep underground neutron flux measurement with large BF3
  counters}.
\newblock \emph{Il Nuovo Cimento A (1965-1970)}, \textbf{101} (1989):959--966.

\bibitem{Haffke2011BackgroundMI}
M.~Haffke, L.~Baudis, T.~Bruch, A.~Ferella, T.~M. Undagoitia, M.~Schumann,
  Y.-F. Te, A.~van~der Schaaf.
\newblock \emph{Background measurements in the Gran Sasso Underground
  Laboratory}.
\newblock \emph{Nuclear Instruments and Methods in Physics Research Section A:
  Accelerators, Spectrometers, Detectors and Associated Equipment},
  \textbf{643(1)} (2011):36--41.
\newblock \doi{10.1016/j.nima.2011.04.027}.

\bibitem{Bruno2019FluxMO}
G.~Bruno, W.~Fulgione.
\newblock \emph{Flux measurement of fast neutrons in the Gran Sasso underground
  laboratory}.
\newblock \emph{The European Physical Journal C} (2019).

\bibitem{bib:drift1}
E.~Daw, et~al.
\newblock \emph{{The DRIFT Directional Dark Matter Experiments}}.
\newblock \emph{EAS Publ. Ser.}, \textbf{53} (2012):11--18.
\newblock \doi{10.1051/eas/1253002}.

\bibitem{Battat:2015rna}
J.~B.~R. Battat, et~al.
\newblock \emph{{Reducing DRIFT Backgrounds with a Submicron Aluminized-Mylar
  Cathode}}.
\newblock \emph{Nucl. Instrum. Meth. A}, \textbf{794} (2015):33--46.
\newblock \doi{10.1016/j.nima.2015.04.070}.

\bibitem{Castel_2019}
J.~Castel, S.~Cebri{\'{a}}n, I.~Coarasa, T.~Dafni, J.~Gal{\'{a}}n, F.~J. Iguaz,
  I.~G. Irastorza, G.~Luz{\'{o}}n, H.~Mirallas, A.~O. de~Sol{\'{o}}rzano,
  E.~Ruiz-Ch{\'{o}}liz.
\newblock \emph{Background assessment for the {TREX} dark matter experiment}.
\newblock \emph{The European Physical Journal C}, \textbf{79(9)} (2019).
\newblock \doi{10.1140/epjc/s10052-019-7282-6}.

\bibitem{Antochi_2018}
V.~Antochi, E.~Baracchini, G.~Cavoto, E.~D. Marco, M.~Marafini, G.~Mazzitelli,
  D.~Pinci, F.~Renga, S.~Tomassini, C.~Voena.
\newblock \emph{Combined readout of a triple-GEM detector}.
\newblock \emph{Journal of Instrumentation}, \textbf{13(05)} (2018):P05001.
\newblock \doi{10.1088/1748-0221/13/05/P05001}.

\bibitem{BOREXINO:2018ohr}
M.~Agostini, et~al.
\newblock \emph{{Comprehensive measurement of $pp$-chain solar neutrinos}}.
\newblock \emph{Nature}, \textbf{562(7728)} (2018):505--510.
\newblock \doi{10.1038/s41586-018-0624-y}.

\bibitem{Boehm:2018sux}
C.~B\oe{}hm, D.~G. Cerde\~no, P.~A.~N. Machado, A.~Olivares-Del~Campo,
  E.~Perdomo, E.~Reid.
\newblock \emph{{How high is the neutrino floor?}}
\newblock \emph{JCAP}, \textbf{01} (2019):043.
\newblock \doi{10.1088/1475-7516/2019/01/043}.

\bibitem{bib:bertuzzo}
E.~Bertuzzo, F.~F. Deppisch, S.~Kulkarni, Y.~F. Perez~Gonzalez,
  R.~Zukanovich~Funchal.
\newblock \emph{{Dark Matter and Exotic Neutrino Interactions in Direct
  Detection Searches}}.
\newblock \emph{JHEP}, \textbf{04} (2017):073.
\newblock \doi{10.1007/JHEP04(2017)073}.

\bibitem{Seguinot:1992zu}
J.~Seguinot, T.~Ypsilantis, A.~Zichichi.
\newblock \emph{{A High rate solar neutrino detector with energy
  determination}}.
\newblock \emph{Conf. Proc. C}, \textbf{920310} (1992):289--313.

\bibitem{Arpesella:1996uc}
C.~Arpesella, C.~Broggini, C.~Cattadori.
\newblock \emph{{A possible gas for solar neutrino spectroscopy}}.
\newblock \emph{Astropart. Phys.}, \textbf{4} (1996):333--341.
\newblock \doi{10.1016/0927-6505(95)00051-8}.

\bibitem{Soffitta:2012hx}
P.~Soffitta, F.~Muleri, S.~Fabiani, E.~Costa, R.~Bellazzini, A.~Brez,
  M.~Minuti, M.~Pinchera, G.~Spandre.
\newblock \emph{{Measurement of the position resolution of the Gas Pixel
  Detector}}.
\newblock \emph{Nucl. Instrum. Meth. A}, \textbf{700} (2013):99--105.
\newblock \doi{10.1016/j.nima.2012.09.055}.

\bibitem{medianfilter}
Y.~Dong, S.~Xu.
\newblock \emph{A New Directional Weighted Median Filter for Removal of
  Random-Valued Impulse Noise}.
\newblock \emph{IEEE Signal Processing Letters}, \textbf{14(3)}
  (2007):193--196.
\newblock \doi{10.1109/LSP.2006.884014}.

\bibitem{Baracchini:2020iwg}
E.~Baracchini, et~al.
\newblock \emph{{A density-based clustering algorithm for the CYGNO data
  analysis}}.
\newblock \emph{JINST}, \textbf{15(12)} (2020):T12003.
\newblock \doi{10.1088/1748-0221/15/12/T12003}.

\bibitem{chanvese}
P.~Getreuer.
\newblock \emph{{Chan-Vese Segmentation}}.
\newblock \emph{{Image Processing On Line}}, \textbf{2} (2012):214--224.
\newblock \url{https://doi.org/10.5201/ipol.2012.g-cv}.

\bibitem{Mumfordshah}
D.~Mumford, J.~Shah.
\newblock \emph{Optimal approximations by piecewise smooth functions and
  associated variational problems}.
\newblock \emph{Communications on Pure and Applied Mathematics}, \textbf{42(5)}
  (1989):577--685.
\newblock \doi{https://doi.org/10.1002/cpa.3160420503}.

\bibitem{root}
R.~Brun, F.~Rademakers.
\newblock \emph{ROOT — An object oriented data analysis framework}.
\newblock \emph{Nuclear Instruments and Methods in Physics Research Section A:
  Accelerators, Spectrometers, Detectors and Associated Equipment},
  \textbf{389(1)} (1997):81--86.
\newblock \doi{https://doi.org/10.1016/S0168-9002(97)00048-X}.
\newblock New Computing Techniques in Physics Research V.

\bibitem{bib:Ramo}
S.~Ramo.
\newblock \emph{Currents Induced by Electron Motion}.
\newblock \emph{Proceedings of the IRE}, \textbf{27(9)} (1939):584--585.
\newblock \doi{10.1109/JRPROC.1939.228757}.

\bibitem{bib:Shockley}
W.~Shockley.
\newblock \emph{{Currents to conductors induced by a moving point charge}}.
\newblock \emph{J. Appl. Phys.}, \textbf{9(10)} (1938):635--636.
\newblock \doi{10.1063/1.1710367}.

\bibitem{bib:cygnoIDBscan}
E.~Baracchini, et~al.
\newblock \emph{Identification of low energy nuclear recoils in a gas time
  projection chamber with optical readout}.
\newblock \emph{Measurement Science and Technology}, \textbf{32(2)}
  (2020):025902.
\newblock \doi{10.1088/1361-6501/abbd12}.

\bibitem{bib:tom}
T.~Thorpe, S.~Vahsen.
\newblock \emph{Avalanche gain and its effect on energy resolution in GEM-based
  detectors}.
\newblock \emph{Nuclear Instruments and Methods in Physics Research Section A:
  Accelerators, Spectrometers, Detectors and Associated Equipment}
  (2022):167438.
\newblock \doi{https://doi.org/10.1016/j.nima.2022.167438}.

\bibitem{bib:Aoyama}
T.~Aoyama.
\newblock \emph{Generalized gas gain formula for proportional counters}.
\newblock \emph{Nuclear Instruments and Methods in Physics Research Section A:
  Accelerators, Spectrometers, Detectors and Associated Equipment},
  \textbf{234(1)} (1985):125--131.
\newblock \doi{https://doi.org/10.1016/0168-9002(85)90817-4}.

\bibitem{bib:Williams}
A.~Williams, R.~Sara.
\newblock \emph{Parameters affecting the resolution of a proportional counter}.
\newblock \emph{The International Journal of Applied Radiation and Isotopes},
  \textbf{13(5)} (1962):229--238.
\newblock \doi{https://doi.org/10.1016/0020-708X(62)90115-1}.

\bibitem{bib:Diethorn}
W.~Diethorn.
\newblock \emph{A methane proportional counter system for natural radiocarbon
  measurements.}
\newblock Master's thesis, Carnegie Inst. of Tech., Pittsburgh, PA (1956).

\bibitem{bib:Alkhazov}
G.~Alkhazov.
\newblock \emph{Statistics of electron avalanches and ultimate resolution of
  proportional counters}.
\newblock \emph{Nuclear Instruments and Methods}, \textbf{89} (1970):155--165.
\newblock \doi{https://doi.org/10.1016/0029-554X(70)90818-9}.

\bibitem{bib:Schlumbohm}
H.~Schlumbohm.
\newblock \emph{Zur Statistik der Elektronenlawinen im ebenen Feld. III}.
\newblock \emph{Zeitschrift für Physik}, \textbf{151(5)} (1958):563--576.
\newblock \doi{10.1007/BF01338427}.

\bibitem{bib:nitec}
E.~Baracchini, G.~Cavoto, G.~Mazzitelli, F.~Murtas, F.~Renga, S.~Tomassini.
\newblock \emph{{Negative Ion Time Projection Chamber operation with SF$_6$ at
  nearly atmospheric pressure}}.
\newblock \emph{JINST}, \textbf{13(04)} (2018):P04022.
\newblock \doi{10.1088/1748-0221/13/04/P04022}.

\bibitem{LIGTENBERG2021165706}
C.~Ligtenberg, M.~{van Beuzekom}, Y.~Bilevych, K.~Desch, H.~{van der Graaf},
  F.~Hartjes, K.~Heijhoff, J.~Kaminski, P.~Kluit, N.~{van der Kolk}, G.~Raven,
  J.~Timmermans.
\newblock \emph{On the properties of a negative-ion TPC prototype with GridPix
  readout}.
\newblock \emph{Nuclear Instruments and Methods in Physics Research Section A:
  Accelerators, Spectrometers, Detectors and Associated Equipment},
  \textbf{1014} (2021):165706.
\newblock \doi{https://doi.org/10.1016/j.nima.2021.165706}.

\bibitem{NID_2_Martoff2000355}
C.~Martoff, D.~Snowden-Ifft, T.~Ohnuki, N.~Spooner, M.~Lehner.
\newblock \emph{Suppressing drift chamber diffusion without magnetic field}.
\newblock \emph{Nuclear Instruments and Methods in Physics Research Section A:
  Accelerators, Spectrometers, Detectors and Associated Equipment},
  \textbf{440(2)} (2000):355--359.
\newblock \doi{https://doi.org/10.1016/S0168-9002(99)00955-9}.

\bibitem{Ohnuki:2000ex}
T.~Ohnuki, D.~P. Snowden-Ifft, C.~J. Martoff.
\newblock \emph{{Measurement of carbon disulfide anion diffusion in a TPC}}.
\newblock \emph{Nucl. Instrum. Meth. A}, \textbf{463} (2001):142--148.
\newblock \doi{10.1016/S0168-9002(01)00222-4}.

\bibitem{Iontransport1}
\emph{Kinetic Theory of Mobility and Diffusion: Sections 5.1 – 5.2}, chap.~5.
\newblock John Wiley \& Sons, Ltd (1988), pp. 137--193.
\newblock \doi{https://doi.org/10.1002/3527602852.ch5a}.

\bibitem{Phan2016TheNP}
N.~S. Phan, R.~Lafler, R.~J. Lauer, E.~R. Lee, D.~Loomba, J.~A. Matthews, E.~H.
  Miller.
\newblock \emph{The novel properties of SF6 for directional dark matter
  experiments}.
\newblock \emph{Journal of Instrumentation}, \textbf{12} (2016):P02012 --
  P02012.

\bibitem{BIAGI1989716}
S.~Biagi.
\newblock \emph{A multiterm Boltzmann analysis of drift velocity, diffusion,
  gain and magnetic-field effects in argon-methane-water-vapour mixtures}.
\newblock \emph{Nuclear Instruments and Methods in Physics Research Section A:
  Accelerators, Spectrometers, Detectors and Associated Equipment},
  \textbf{283(3)} (1989):716--722.
\newblock \doi{https://doi.org/10.1016/0168-9002(89)91446-0}.

\bibitem{ALNER2004644}
G.~Alner, H.~Araujo, R.~Ayad, A.~Bewick, J.~Burwell, M.~Carson, C.~Chen,
  J.~Davies, J.~Dawson, T.~Gamble, M.~Garcia, J.~Heidecker, A.~Howard,
  M.~Katz-Hyman, W.~Jones, J.~Kirkpatrick, V.~Kudryavtsev, T.~Lawson,
  V.~Lebedenko, M.~Lehner, J.~Lewin, P.~Lightfoot, I.~Liubarsky, R.~Lüscher,
  J.~McMillan, C.~Martoff, B.~Morgan, J.~Mulholland, G.~Nicklin, T.~Ohnuki,
  S.~Paling, A.~Petkov, R.~Preece, J.~Quenby, J.~Roberts, M.~Robinson,
  E.~Rykoff, A.~Sanders, D.~Snowden-Ifft, N.~Smith, P.~Smith, N.~Spooner,
  T.~Sumner, D.~Tovey, N.~Villaume, R.~Walker.
\newblock \emph{The DRIFT-I dark matter detector at Boulby: design,
  installation and operation}.
\newblock \emph{Nuclear Instruments and Methods in Physics Research Section A:
  Accelerators, Spectrometers, Detectors and Associated Equipment},
  \textbf{535(3)} (2004):644--655.
\newblock \doi{https://doi.org/10.1016/j.nima.2004.06.172}.

\bibitem{NID_3_Snowden}
D.~P. Snowden-Ifft, J.~L. Gauvreau.
\newblock \emph{{High Precision Measurements of Carbon Disulfide Negative Ion
  Mobility and Diffusion}}.
\newblock \emph{Rev. Sci. Instrum.}, \textbf{84} (2013):053304.
\newblock \doi{10.1063/1.4803004}.

\bibitem{Dion}
M.~Dion.
\newblock \emph{{The study of electronegative gases for Time Projection
  Chambers}}.
\newblock Ph.D. thesis, Temple University (2009).

\bibitem{SF6xsec}
L.~G. Christophorou, J.~K. Olthoff.
\newblock \emph{Electron Interactions With SF6}.
\newblock \emph{Journal of Physical and Chemical Reference Data},
  \textbf{29(3)} (2000):267--330.
\newblock \doi{10.1063/1.1288407}.

\bibitem{SF6cose1}
H.~G. Crompton~R.W.
\newblock \emph{Thermal Electron Attachment to SF$_6$ and CFCl$_3$}.
\newblock \emph{Australian Journal of Physics}, \textbf{36} (1983):15--30.
\newblock \doi{10.1071/PH830015}.

\bibitem{SF6cose2}
Z.~L. Petrovic, R.~W. Crompton.
\newblock \emph{Thermal-electron attachment to SF6 at room temperature and
  500K}.
\newblock \emph{Journal of Physics B: Atomic and Molecular Physics},
  \textbf{18(13)} (1985):2777.
\newblock \doi{10.1088/0022-3700/18/13/024}.

\bibitem{SF6cose3}
T.~M. Miller, A.~E.~S. Miller, J.~F. Paulson, X.~Liu.
\newblock \emph{Thermal electron attachment to SF4 and SF6}.
\newblock \emph{The Journal of Chemical Physics}, \textbf{100(12)}
  (1994):8841--8848.
\newblock \doi{10.1063/1.466738}.

\bibitem{SF6cose4}
P.~G. Datskos, L.~G. Christophorou, J.~G. Carter.
\newblock \emph{Erratum: Effect of temperature on the attachment of slow (less
  than 1 eV) electrons to CH3Br [J. Chem. Phys. 97, 9031 (1992)]}.
\newblock \emph{The Journal of Chemical Physics}, \textbf{99(5)}
  (1993):4238--4238.
\newblock \doi{10.1063/1.466078}.

\bibitem{SF6cose5}
L.~Christophorou, J.~Olthoff, D.~Green.
\newblock \emph{Gases for Electrical Insulation and Arc Interruption: Possible
  Present and Future Alternatives to Pure SF<sub>6</sub>} (1997).
\newblock \doi{https://doi.org/10.6028/NIST.tn.1425}.

\bibitem{SF6cose6}
P.~Spanel, S.~Matejcik, D.~Smith.
\newblock \emph{The varying influences of gas and electron temperatures on the
  rates of electron attachment to some selected molecules}.
\newblock \emph{Journal of Physics B: Atomic, Molecular and Optical Physics},
  \textbf{28(14)} (1995):2941.
\newblock \doi{10.1088/0953-4075/28/14/015}.

\bibitem{SF6cose7}
M.~Braun, S.~Barsotti, S.~Marienfeld, E.~Leber, J.~M. Weber, M.~W. Ruf,
  H.~Hotop.
\newblock \emph{High resolution study of anion formation in low-energy electron
  attachment to SF6 molecules in a seeded supersonic beam}.
\newblock \emph{The European Physical Journal D - Atomic, Molecular, Optical
  and Plasma Physics}, \textbf{35(2)} (2005):177--191.
\newblock \doi{10.1140/epjd/e2005-00230-6}.

\bibitem{SF6cose8}
A.~A. Viggiano, T.~M. Miller, J.~F. Friedman, J.~Troe.
\newblock \emph{Low-energy electron attachment to SF6. III. From thermal
  detachment to the electron affinity of SF6}.
\newblock \emph{The Journal of Chemical Physics}, \textbf{127(24)}
  (2007):244305.
\newblock \doi{10.1063/1.2804764}.

\bibitem{SF6cose9}
R.~L. Merlino, S.-H. Kim.
\newblock \emph{Measurement of the electron attachment rates for SF6 and C7F14
  at Te=0.2 eV in a magnetized Q machine plasma}.
\newblock \emph{The Journal of Chemical Physics}, \textbf{129(22)}
  (2008):224310.
\newblock \doi{10.1063/1.3039078}.

\bibitem{SF6autodet}
D.~Edelson, J.~E. Griffiths, K.~B. McAfee.
\newblock \emph{Autodetachment of Electrons in Sulfur Hexafluoride}.
\newblock \emph{The Journal of Chemical Physics}, \textbf{37(4)}
  (1962):917--918.
\newblock \doi{10.1063/1.1733191}.

\bibitem{SF6autodet2}
R.~Compton, L.~Christophorou, R.~Huebner.
\newblock \emph{Temporary negative ion resonances in benzene and benzene
  derivatives}.
\newblock \emph{Physics Letters}, \textbf{23(11)} (1966):656--658.
\newblock \doi{https://doi.org/10.1016/0031-9163(66)90211-3}.

\bibitem{HARLAND197129}
P.~Harland, J.~Thynne.
\newblock \emph{Decomposition of the metastable SF4$^-$ ion}.
\newblock \emph{Inorganic and Nuclear Chemistry Letters}, \textbf{7(1)}
  (1971):29--31.
\newblock \doi{https://doi.org/10.1016/0020-1650(71)80116-2}.

\bibitem{Christophorou1971AtomicAM}
L.~G. Christophorou.
\newblock \emph{Atomic and Molecular Radiation Physics} (1971).

\bibitem{CHRISTOPHOROU197855}
L.~Christophorou.
\newblock \emph{The Lifetimes of Metastable Negative Ions}.
\newblock vol.~46 of \emph{Advances in Electronics and Electron Physics}.
  Academic Press (1978), pp. 55--129.
\newblock \doi{https://doi.org/10.1016/S0065-2539(08)60411-4}.

\bibitem{CHRISTOPHOROU1984477}
L.~Christophorou, D.~McCorkle, A.~Christodoulides.
\newblock \emph{6 - Electron Attachment Processes}.
\newblock In L.~Christophorou, ed., \emph{Electron–Molecule Interactions and
  their Applications}. Academic Press (1984), pp. 477--617.
\newblock \doi{https://doi.org/10.1016/B978-0-12-174401-4.50011-0}.

\bibitem{KLOTS197661}
C.~E. Klots.
\newblock \emph{Rate constants for unimolecular decomposition at threshold}.
\newblock \emph{Chemical Physics Letters}, \textbf{38(1)} (1976):61--64.
\newblock \doi{https://doi.org/10.1016/0009-2614(76)80255-2}.

\bibitem{R_W_Odom_1975}
R.~W. Odom, D.~L. Smith, J.~H. Futrell.
\newblock \emph{A study of electron attachment to SF6 and auto-detachment and
  stabilization of SF6-}.
\newblock \emph{Journal of Physics B: Atomic and Molecular Physics},
  \textbf{8(8)} (1975):1349.
\newblock \doi{10.1088/0022-3700/8/8/026}.

\bibitem{FOSTER1975479}
M.~S. Foster, J.~Beauchamp.
\newblock \emph{Chemical consequences of electron scavenging by SF6 in
  radiolysis experiments}.
\newblock \emph{Chemical Physics Letters}, \textbf{31(3)} (1975):479--481.
\newblock \doi{https://doi.org/10.1016/0009-2614(75)85067-6}.

\bibitem{Patterson1}
P.~L. Patterson.
\newblock \emph{Mobilities of Negative Ions in SF6}.
\newblock \emph{The Journal of Chemical Physics}, \textbf{53(2)}
  (1970):696--704.
\newblock \doi{10.1063/1.1674047}.

\bibitem{sf6detach}
Y.~Wang, R.~L. Champion, L.~D. Doverspike, J.~K. Olthoff, R.~J. Van~Brunt.
\newblock \emph{Collisional electron detachment and decomposition cross
  sections for SF$_6$, SF$_5$, and F$^-$ on SF6 and rare gas targets}.
\newblock \emph{The Journal of Chemical Physics}, \textbf{91(4)}
  (1989):2254--2260.
\newblock \doi{10.1063/1.457033}.

\bibitem{FISCHLE1991202}
H.~Fischle, J.~Heintze, B.~Schmidt.
\newblock \emph{Experimental determination of ionization cluster size
  distributions in counting gases}.
\newblock \emph{Nuclear Instruments and Methods in Physics Research Section A:
  Accelerators, Spectrometers, Detectors and Associated Equipment},
  \textbf{301(2)} (1991):202--214.
\newblock \doi{https://doi.org/10.1016/0168-9002(91)90460-8}.

\bibitem{CATALDI1997458}
G.~Cataldi, F.~Grancagnolo, S.~Spagnolo.
\newblock \emph{Cluster counting in helium based gas mixtures}.
\newblock \emph{Nuclear Instruments and Methods in Physics Research Section A:
  Accelerators, Spectrometers, Detectors and Associated Equipment},
  \textbf{386(2)} (1997):458--469.
\newblock \doi{https://doi.org/10.1016/S0168-9002(96)01164-3}.

\bibitem{Chiarello_2017}
G.~Chiarello, C.~Chiri, G.~Cocciolo, A.~Corvaglia, F.~Grancagnolo, M.~Panareo,
  A.~Pepino, F.~Renga, G.~Tassielli, C.~Voena.
\newblock \emph{Application of the Cluster Counting/Timing techniques to
  improve the performances of high transparency Drift Chamber for modern HEP
  experiments}.
\newblock \emph{Journal of Instrumentation}, \textbf{12(07)} (2017):C07021.
\newblock \doi{10.1088/1748-0221/12/07/C07021}.

\bibitem{Iontransport2}
\emph{Kinetic Theory of Mobility and Diffusion: Section 5.3}, chap.~5.
\newblock John Wiley \& Sons, Ltd (1988), pp. 193--224.
\newblock \doi{https://doi.org/10.1002/3527602852.ch5b}.

\bibitem{FERRARI19969}
L.~Ferrari.
\newblock \emph{Diffusion coefficients of ions in lighter gases in an electric
  field}.
\newblock \emph{Chemical Physics}, \textbf{206(1)} (1996):9--34.
\newblock \doi{https://doi.org/10.1016/0301-0104(96)00013-4}.

\bibitem{Amaro_2010kh}
F.~D. Amaro, C.~Santos, J.~F. C.~A. Veloso, A.~Breskin, R.~Chechik, J.~M.~F.
  dos Santos.
\newblock \emph{{The Thick-COBRA: A New Gaseous Electron Multiplier for
  Radiation Detectors}}.
\newblock \emph{JINST}, \textbf{5} (2010):P10002.
\newblock \doi{10.1088/1748-0221/5/10/P10002}.

\bibitem{EZERIBE2021164847}
A.~Ezeribe, C.~Eldridge, W.~Lynch, R.~{Marcelo Gregorio}, A.~Scarff,
  N.~Spooner.
\newblock \emph{Demonstration of ThGEM-multiwire hybrid charge readout for
  directional dark matter searches}.
\newblock \emph{Nuclear Instruments and Methods in Physics Research Section A:
  Accelerators, Spectrometers, Detectors and Associated Equipment},
  \textbf{987} (2021):164847.
\newblock \doi{https://doi.org/10.1016/j.nima.2020.164847}.

\bibitem{bib:ROSZKOWSKI200910}
L.~Roszkowski, R.~{Ruiz de Austri}, J.~Silk, R.~Trotta.
\newblock \emph{On prospects for dark matter indirect detection in the
  Constrained MSSM}.
\newblock \emph{Physics Letters B}, \textbf{671(1)} (2009):10--14.
\newblock \doi{https://doi.org/10.1016/j.physletb.2008.11.061}.

\bibitem{bib:arina2014bayesian}
C.~Arina.
\newblock \emph{Bayesian analysis of multiple direct detection experiments}
  (2014).

\bibitem{bib:Gelman}
A.~G. e~al.
\newblock \emph{{Bayesian Data Analysis}} (2013).
\newblock \doi{https://doi.org/10.1201/b16018}.

\bibitem{bib:DAgostini}
G.~D'Agostini.
\newblock \emph{{Bayesian reasoning in data analysis: A critical introduction}}
  (2003).

\bibitem{bib:Plummer}
M.~Plummer.
\newblock \emph{JAGS: A Program for Analysis of Bayesian Graphical Models Using
  Gibbs Sampling}.
\newblock \emph{Proceedings of the 3rd International Workshop on Distributed
  Statistical Computing (DSC 2003)} (2003):1--10.

\bibitem{Nakamura:2012zza}
K.~Nakamura, et~al.
\newblock \emph{{Low pressure gas study for a direction-sensitive dark matter
  search experiment with MPGD}}.
\newblock \emph{JINST}, \textbf{7} (2012):C02023.
\newblock \doi{10.1088/1748-0221/7/02/C02023}.

\bibitem{bib:QFpaper}
J.~Lindhard, V.~Nielsen, M.~Scharff, P.~V. Thomsen.
\newblock \emph{INTEGRAL EQUATIONS GOVERNING RADIATION EFFECTS. (NOTES ON
  ATOMIC COLLISIONS, III)}.
\newblock \emph{Kgl. Danske Videnskab., Selskab. Mat. Fys. Medd.}

\bibitem{Aprile:2018dbl}
E.~Aprile, et~al.
\newblock \emph{{Dark Matter Search Results from a One Ton-Year Exposure of
  XENON1T}}.
\newblock \emph{Phys. Rev. Lett.}, \textbf{121(11)} (2018):111302.
\newblock \doi{10.1103/PhysRevLett.121.111302}.

\bibitem{bib:Aprile_2019}
E.~Aprile, J.~Aalbers, F.~Agostini, M.~Alfonsi, L.~Althueser, F.~Amaro,
  V.~Antochi, E.~Angelino, F.~Arneodo, D.~Barge, et~al.
\newblock \emph{Light Dark Matter Search with Ionization Signals in XENON1T}.
\newblock \emph{Physical Review Letters}, \textbf{123(25)} (2019).
\newblock \doi{10.1103/physrevlett.123.251801}.

\bibitem{bib:Agnes_2018}
P.~Agnes, I.~Albuquerque, T.~Alexander, A.~Alton, G.~Araujo, D.~Asner, M.~Ave,
  H.~Back, B.~Baldin, G.~Batignani, et~al.
\newblock \emph{Low-Mass Dark Matter Search with the DarkSide-50 Experiment}.
\newblock \emph{Physical Review Letters}, \textbf{121(8)} (2018).
\newblock \doi{10.1103/physrevlett.121.081307}.

\bibitem{bib:Mancuso:2020gnm}
M.~Mancuso, et~al.
\newblock \emph{{Searches for Light Dark Matter with the CRESST-III
  Experiment}}.
\newblock \emph{J. Low Temp. Phys.}, \textbf{199(1-2)} (2020):547--555.
\newblock \doi{10.1007/s10909-020-02343-3}.

\bibitem{bib:Agnese_2018}
R.~Agnese, A.~Anderson, T.~Aralis, T.~Aramaki, I.~Arnquist, W.~Baker,
  D.~Balakishiyeva, D.~Barker, R.~Basu~Thakur, D.~Bauer, et~al.
\newblock \emph{Low-mass dark matter search with CDMSlite}.
\newblock \emph{Physical Review D}, \textbf{97(2)} (2018).
\newblock \doi{10.1103/physrevd.97.022002}.

\bibitem{bib:Savage_2009}
C.~Savage, G.~Gelmini, P.~Gondolo, K.~Freese.
\newblock \emph{Compatibility of DAMA/LIBRA dark matter detection with other
  searches}.
\newblock \emph{Journal of Cosmology and Astroparticle Physics},
  \textbf{2009(04)} (2009):010–010.
\newblock \doi{10.1088/1475-7516/2009/04/010}.

\bibitem{Darksidelowmass}
C.~E. Aalseth, et~al.
\newblock \emph{{Future Dark Matter Searches with Low-Radioactivity Argon, by
  The Global Argon Dark Matter Collaboration}}.
\newblock \emph{Input to the European Particle Physics Strategy Update}.

\bibitem{bib:drift2}
J.~B.~R. Battat, et~al.
\newblock \emph{{Low Threshold Results and Limits from the DRIFT Directional
  Dark Matter Detector}}.
\newblock \emph{Astropart. Phys.}, \textbf{91} (2017):65--74.
\newblock \doi{10.1016/j.astropartphys.2017.03.007}.

\bibitem{bib:Savage_2004}
C.~Savage, P.~Gondolo, K.~Freese.
\newblock \emph{Can WIMP spin dependent couplings explain DAMA data, in light
  of null results from other experiments?}
\newblock \emph{Physical Review D}, \textbf{70(12)} (2004).
\newblock \doi{10.1103/physrevd.70.123513}.

\bibitem{DeRocco_2019_2}
W.~DeRocco, P.~W. Graham, D.~Kasen, G.~Marques-Tavares, S.~Rajendran.
\newblock \emph{Supernova signals of light dark matter}.
\newblock \emph{Physical Review D}, \textbf{100(7)} (2019).
\newblock \doi{10.1103/physrevd.100.075018}.

\bibitem{Beacom_SN}
J.~F. Beacom.
\newblock \emph{The Diffuse Supernova Neutrino Background}.
\newblock \emph{Annual Review of Nuclear and Particle Science}, \textbf{60(1)}
  (2010):439--462.
\newblock \doi{10.1146/annurev.nucl.010909.083331}.

\bibitem{Adams_2013}
S.~M. Adams, C.~S. Kochanek, J.~F. Beacom, M.~R. Vagins, K.~Z. Stanek.
\newblock \emph{OBSERVING THE NEXT GALACTIC SUPERNOVA}.
\newblock \emph{The Astrophysical Journal}, \textbf{778(2)} (2013):164.
\newblock \doi{10.1088/0004-637X/778/2/164}.

\bibitem{Izaguirre:2015yja}
E.~Izaguirre, G.~Krnjaic, P.~Schuster, N.~Toro.
\newblock \emph{{Analyzing the Discovery Potential for Light Dark Matter}}.
\newblock \emph{Phys. Rev. Lett.}, \textbf{115(25)} (2015):251301.
\newblock \doi{10.1103/PhysRevLett.115.251301}.

\bibitem{Raffelt:2001kv}
G.~G. Raffelt.
\newblock \emph{{Muon-neutrino and tau-neutrino spectra formation in
  supernovae}}.
\newblock \emph{Astrophys. J.}, \textbf{561} (2001):890--914.
\newblock \doi{10.1086/323379}.

\bibitem{Deroccowork}
E.~Baracchini, W.~DeRocco, G.~Dho.
\newblock \emph{Discovering supernova-produced dark matter with directional
  detectors}.
\newblock \emph{Physical Review D}, \textbf{102(7)} (2020).
\newblock \doi{10.1103/physrevd.102.075036}.

\bibitem{bib:young_smith_2005}
G.~A. Young, R.~L. Smith.
\newblock \emph{Bibliography}.
\newblock Cambridge Series in Statistical and Probabilistic Mathematics.
  Cambridge University Press (2005), p. 218–222.
\newblock \doi{10.1017/CBO9780511755392.013}.

\bibitem{bib:Taylor}
J.~R. Taylor.
\newblock \emph{An Introduction to Error Analysis: The Study of Uncertainties
  in Physical Measurements}.
\newblock University Science Books, 2 sub ed. (1996).

\bibitem{Agnes:2018fwg}
P.~Agnes, et~al.
\newblock \emph{{DarkSide-50 532-day Dark Matter Search with Low-Radioactivity
  Argon}}.
\newblock \emph{Phys. Rev.}, \textbf{D98(10)} (2018):102006.
\newblock \doi{10.1103/PhysRevD.98.102006}.

\bibitem{bib:Aprile:2019bbb}
E.~Aprile, et~al.
\newblock \emph{{XENON1T Dark Matter Data Analysis: Signal Reconstruction,
  Calibration and Event Selection}}.
\newblock \emph{Phys. Rev.}, \textbf{D100(5)} (2019):052014.
\newblock \doi{10.1103/PhysRevD.100.052014}.

\bibitem{Aprile:2017xxh}
E.~Aprile, et~al.
\newblock \emph{{Signal Yields of keV Electronic Recoils and Their
  Discrimination from Nuclear Recoils in Liquid Xenon}}.
\newblock \emph{Phys. Rev.}, \textbf{D97(9)} (2018):092007.
\newblock \doi{10.1103/PhysRevD.97.092007}.

\bibitem{Vahsen:2014fba}
S.~Vahsen, M.~Hedges, I.~Jaegle, S.~Ross, I.~Seong, T.~Thorpe, J.~Yamaoka,
  J.~Kadyk, M.~Garcia-Sciveres.
\newblock \emph{{3-D tracking in a miniature time projection chamber}}.
\newblock \emph{Nucl. Instrum. Meth. A}, \textbf{788} (2015):95--105.
\newblock \doi{10.1016/j.nima.2015.03.009}.

\bibitem{Bozorgnia_2012}
N.~Bozorgnia, G.~B. Gelmini, P.~Gondolo.
\newblock \emph{Ring-like features in directional dark matter detection}.
\newblock \emph{Journal of Cosmology and Astroparticle Physics},
  \textbf{2012(06)} (2012):037–037.
\newblock \doi{10.1088/1475-7516/2012/06/037}.

\bibitem{xray}
E.~Costa.
\newblock \emph{General History of X-Ray Polarimetry in Astrophysics} (2022).
\newblock \doi{10.48550/ARXIV.2209.08181}.

\end{thebibliography}

\end{document}